\documentclass[a4paper, 11pt, oneside]{Thesis}  
\graphicspath{{./Figures/}}  

\usepackage[square, numbers, comma, sort&compress]{natbib}  
\usepackage{hypernat}
\usepackage[T1]{fontenc}
\usepackage{lmodern} 
\usepackage{graphicx}
\usepackage{bm}
\usepackage{bbm}
\usepackage{color}
\usepackage{xcolor}
\usepackage{footnote}
\usepackage{epstopdf}
\usepackage{widetext}
\usepackage[Lenny]{fncychap}

\usepackage[toc,page]{appendix}
\usepackage{hyperref}
\usepackage{here} 
\usepackage[applemac]{inputenc}
\usepackage{verbatim}  
\usepackage[square]{natbib}  
\usepackage{vector}  
\newcommand{\myparallel}{{\mkern3mu\vphantom{\perp}\vrule depth 0pt\mkern2mu\vrule depth 0pt\mkern3mu}}
\usepackage{dsfont}
\hypersetup{urlcolor=RawSienna,colorlinks=true}  

\begin{document}
\frontmatter	  

\title{Hybrid superconductor-semiconductor nanowire junctions as useful platforms to study Majorana bound states}
   \authors{
            Jorge Cayao
            }
\addresses  {\groupname\\\deptname\\\univname}  
\date {Madrid, May 2016}
\subject    {}

\maketitle

\newpage{\
\thispagestyle{empty}}
\newpage{\
\thispagestyle{empty}}

\setstretch{1}  

\fancyhead{}  
\rhead{\thepage}  
\lhead{}  

\pagestyle{fancy}  

\Committee{

\centerline{\Large
Prof. Rosa L\'{o}pez Gonzalo} 
\vskip 12mm
\centerline{\Large Prof. Alfredo Levy Yeyati} 
\vskip 12mm
\centerline{\Large Prof. Karsten Flensberg} 
\vskip 12mm
\centerline{\Large Prof. Pascal Simon} 
\vskip 12mm
\centerline{\Large Dr. Sebastian Bergeret} 
}
\begin{center}
\vskip -12mm
{\Large
{\bf Thesis Advisor:} Dr. Ram\'on Aguado Sola}
\end{center}

\clearpage  

\pagestyle{empty}  

\null\vfill

\vfill\vfill\vfill\vfill\vfill\vfill\null
\clearpage  


\acknowledgements{

First and foremost I would like to express my profound gratitude to my advisor, Ram\'{o}n Aguado, for his enormous support, guidance and patience during these years. 
As well, my eternal thanks to Pablo San-Jos\'{e} and Elsa Prada for helping me  throughout all these years and guiding me to finish my projects.
Without the continual assistance of all of them, this work would not get much farther than to this page. 
Additionally, I would like to thank Peter Sta\v{n}o for his friendly support, hospitality and supervision during my research visits to Bratislava and RIKEN.

I acknowledge the funding of the Spanish Research Council (CSIC) through the JAE-Predoc Program during these four years and also of the Spanish Ministry of Economy and Innovation through the grant FIS2012-33521.
}
\clearpage  

\dedicatory{
In memory of my beloved brother Luciano
}
\clearpage
\newpage{\
\thispagestyle{empty}}


\abstract{
\subsection*{Abstract}

One of the most promising platforms for one-dimensional topological superconductivity is based on semiconducting nanowires with strong spin-orbit coupling (SOC), 
where $s$-wave superconductivity is induced by proximity effect and an external Zeeman field drives the system into the topological superconducting 
phase with Majorana bound states (MBSs) at the end of the wire. During last years this idea has led to a great number of important experiments in 
hybrid superconductor-semiconductor systems, where the main signature is an emergent zero-bias peak (ZBP) in the differential conductance as the Zeeman field is increased.

This thesis focuses on the study of hybrid superconductor-semiconductor junctions made of semiconducting nanowires with Rashba SOC. In the first part, 
we introduce the emergence of one-dimensional topological superconductivity, and then we show details on modeling one-dimensional hybrid junctions made of semiconducting nanowires with SOC.
Afterwards, we fully analyse the Andreev spectrum and study phase-biased transport, which exhibit non-trivial signatures in the topological phase with MBSs.
In the second part, we focus on transport in a voltage-biased short superconductor-normal-superconductor (SNS) junction made of semiconducting nanowires with SOC as the applied Zeeman field drives the system into the topological superconducting phase. 
We show that the dissipative multiple Andreev reflection (MAR) current at different junction transparencies exhibits unique features related to topology such as gap inversion, 
the formation of MBSs and fermion-parity conservation. 
In the third part, we carried out a detailed study of helicity and confinement in long SNS junctions based on semiconducting nanowires. The main conclusion in this part is that 
a long junction with a helical normal section but still in the topologically trivial regime, supports a low-energy sub-gap spectrum consisting of multiple zero-energy crossings 
that smoothly evolve towards MBSs as the junction becomes topological.  
 In the fourth part, we investigate a novel approach to engineer MBSs in non-topological superconducting wires, where we propose to create a sufficiently transparent normal-superconductor (NS) junction on a Rashba wire with a topologically trivial superconducting side and a helical normal side.
 We finish with an analysis of screening properties of proximitized nanowires with SOC and Zeeman fields within linear response theory, relevant for experiments trying to measure Majorana bound states and their non-trivial overlap.
Along this thesis, we emphasise the importance of employing hybrids superconductor-semiconductor nanowire junctions towards the unambiguously detection
of MBSs beyond zero-bias anomalies. 
\vfil
{\bf Keywords: } Topological superconductivity, Majorana bound states, Andreev bound states, hybrid NS and SNS junctions, spin-orbit coupling, Zeeman interaction, $s$-wave superconductivity
}

\newpage{\
\thispagestyle{empty}}


\resumen{
\subsection*{Resumen}

Una de las plataformas m\'{a}s prometedoras para superconductividad topol\'{o}gica en una dimensi\'{o}n est\'{a} basada
en nanohilos semiconductores con fuerte acoplamiento de esp\'{i}n-\'{o}rbita (AEO), donde se induce superconductividad de onda $s$ v\'{i}a efecto de proximidad y un campo Zeeman 
externo conduce al sistema hacia la fase topol\'{o}gica con estados ligados de Majorana (ELM) en los bordes del nanohilo. Durante los \'{u}ltimos a\~{n}os, esta idea ha dado lugar a un gran 
n\'{u}mero de experimentos en sistemas h\'{i}bridos superconductor-semiconductor, donde la principal caracter\'{i}stica is la emergencia de un pico a voltaje cero en la conductancia 
diferencial a medida que el campo magn\'{e}tico es aumentado.

Esta tesis se centra en el estudio de uniones h\'{i}bridas superconductor-semiconductor hechas de nanohilos semiconductores con AEO de tipo Rashba.
En la primera parte, introducimos la emergencia de superconductividad topol\'{o}gica en una dimensi\'{o}n, y luego mostramos como modelar uniones h\'{i}bridas
en una dimensi\'{o}n hechas de nanohilos semiconductores con AEO. 
Despu\'{e}s, realizamos un an\'{a}lisis detallado del espectro de Andreev y del transporte dependiente de la fase, el cual exhibe caracter\'{i}sticas no triviales in la fase 
topol\'{o}gica con ELM.
En la segunda parte, nos centramos en transporte dependiente del voltaje en una union corta superconductor-normal-superconductor (SNS) hecha de nanohilos semiconductores con 
AEO, a medida que el campo Zeeman aplicado conduce al sistema hacia la fase topol\'{o}gica superconductora.
Mostramos que la corriente disipativa de las m\'{u}ltiples reflexiones de Andreev (MAR) a diferentes transparencias de la uni\'{o}n exhibe propiedades relacionadas a la topolog\'{i}a, 
tales como la inversi\'{o}n del gap, la formaci\'{o}n de ELM y la conservaci\'{o}n de la paridad fermi\'{o}nica.
En la tercera parte, hacemos un estudio detallado de helicidad y confinamiento en uniones largas SNS basadas en nanohilos semiconductores. 
La conclusi\'{o}n m\'{a}s importante de esta parte es que una uni\'{o}n larga con una regi\'{o}n helical, pero todav\'{i}a en el r\'{e}gimen topologicamente trivial, 
contiene un espectro de baja energ\'{i}a dentro del gap que consiste en cruces m\'{u}ltiples en energ\'{i}a cero que evolucionan suavemente hacia ELM a  medida que la uni\'{o}n se hace topol\'{o}gica.
En la cuarta parte, investigamos un nuevo enfoque para dise\~{n}ar ELM en nanohilos superconductores no topol\'{o}gicos, donde proponemos crear una 
uni\'{o}n normal-superconductor (NS) suficientemente transparente en un nanohilo de tipo Rashba, con una parte en la fase topol\'{o}gicamente trivial y la otra en la fase helical normal.
Terminamos con un an\'{a}lisis de propiedades de apantallamiento en nanohilos superconductores con AEO y campo Zeeman dentro de 
la teor\'{i}a de respuesta lineal, relevante para experimentos que intentan medir ELM y su solapamiento no trivial.
A lo largo de esta tesis, hacemos hincapi\'{e} en la importancia del emplear uniones h\'{i}bridas superconductor-semiconductor de nanohilos para la detecci\'{o}n sin 
ambig\"{u}edades de ELM mas all\'{a} de la anomal\'{i}a a voltaje cero.

\vfil
{\bf Palabras clave: } Superconductividad topol\'{o}gica, estados ligados de Majorana, estados ligados de Andreev, uniones h\'{i}bridas NS y SNS, acoplamiento de sp\'{i}n-\'{o}bita, interacci\'{o}n de Zeeman, superconductividad de onda $s$
}

\newpage{\
\thispagestyle{empty}}






\clearpage  












\setstretch{1}  

\pagestyle{fancy}  

    \renewcommand*\contentsname{{\bf Contents}}
\lhead{\emph{ Contents}}  
\tableofcontents  

\newpage{\
\thispagestyle{empty}}

\clearpage

        \renewcommand*\listfigurename{\bf List of Figures}
\lhead{\emph{List of Figures}}  
\listoffigures  


\setstretch{1.5}  
\clearpage  
\lhead{\emph{Abbreviations}}  
\listofsymbols{ll}  
{
\textbf{MBS} & \textbf{M}ajorana \textbf {B}ound \textbf {S}tate \\
\textbf{SOC} & \textbf{S}pin-\textbf {O}rbit \textbf {C}oupling \\
\textbf{ABS} & \textbf{A}ndreev \textbf {B}ound \textbf{S}tate \\
\textbf{NS} & \textbf{N}ormal metal-\textbf {S}uperconductor \\
\textbf{SNS} & \textbf{S}uperconductor-\textbf {N}ormal metal-\textbf{S}uperconductor \\
\textbf{MLL} & \textbf{M}ajorana \textbf {L}ocalization \textbf {L}ength \\
\textbf{NW} & \textbf{N}ano-\textbf {W}ire \\
\textbf{TS} & \textbf {T}opological \textbf{S}uperconductor \\
\textbf{EP} & \textbf{E}xceptional \textbf {P}oint \\
\textbf{ZBA} & \textbf{Z}ero \textbf {B}ias \textbf{A}nomaly \\
\textbf{ZBP} & \textbf{Z}ero \textbf {B}ias \textbf {P}eak \\
\textbf{MAR} & \textbf{M}ultiple \textbf{A}ndreev \textbf {R}eflection \\
\textbf{RPA} & \textbf{R}andom \textbf{P}hase \textbf {A}pproximation \\
\textbf{S-matrix}& \textbf{S}cattering \textbf{M}atrix \\
\textbf{GF}& \textbf{G}reen's \textbf{F}unction \\
\textbf{RGF}& \textbf{R}ecursive  \textbf{G}reen's \textbf{F}unction \\
\textbf{Qubit} &\textbf{Q}uantum \textbf{B}it\\
\textbf{2DEG} &\textbf{T}wo-\textbf{D}imensional \textbf{E}lectron \textbf{G}as\\
\textbf{1D} &\textbf{O}ne-\textbf{D}imensional\\
\textbf{QHS} &\textbf{Q}uantum \textbf{H}all \textbf{S}tate\\
\textbf{IQHS} &\textbf{I}nteger \textbf{Q}uantum \textbf{H}all \textbf{S}tate\\
\textbf{FQHS} &\textbf{F}ractional \textbf{Q}uantum \textbf{H}all \textbf{S}tate\\
\textbf{QSHS} &\textbf{Q}uantum \textbf{S}pin  \textbf{H}all \textbf{S}tate\\
}

\setstretch{1}  

\pagestyle{empty}  

\addtocontents{toc}{\vspace{1em}}  

\newpage{\
\thispagestyle{empty}}

\mainmatter	  
\pagestyle{fancy}  


\chapter{{\bf Introduction}} 
\label{Chapter01}
\lhead{Chapter \ref{Chapter01}. \emph{Introduction: topological superconductivity and Majorana bound states}} 



 


\section{Overview of the research in the field}
The discovery of new materials based on condensed matter systems led to a large number of technological applications mainly because it was possible to identify, 
characterise and classify matter in different states. In all these states, matter is formed by a substantial number of constituents ordered in different phases that 
correspond to different internal structures or orders, which are associated with symmetries. 
 
The common framework to describe and classify such emergent phases at the quantum level is based on Landau's Fermi liquid theory \cite{landau} and the Ginzburg-Landau theory 
of symmetry breaking \cite{landau2}. In Landau's  Fermi liquid  theory, the low-energy excitations  (near the  Fermi surface) in an interacting Fermi liquid can be considered as 
non-interacting quasiparticles with renormalised properties such as mass, velocity, etc. However, such picture is not longer valid at low temperatures because some systems 
are unstable towards phases that are characterized by a local order parameter. 
At high temperature entropy dominates and leads to a disordered state, while at low temperature energy dominates and leads to an ordered state. Therefore, an ordered phase 
appears at low temperature when the system spontaneously loses one of the symmetries present at high temperatures: namely, phase transitions, described by Ginzburg-Landau 
theory  \cite{landau}, occur where some of the symmetries presented in the system are broken.
Hence, based on a classification made according to the concept of symmetry breaking, there exist many possible arrangements that originate states such crystalline solids, which break translation symmetry; liquid crystals, which break rotational but not translational symmetry; magnets, which break time-reversal symmetry and rotational symmetry of the spin space; superconductors, which break the more subtle gauge symmetry leading to novel phenomena such as flux quantisation and Josephson effects, among others.

During the last decades there has been a bunch of theoretical and experimental discoveries, which do not fit into the above picture.
These new phases of matter do not break any symmetry but instead exhibit fundamental properties robust to smooth changes in materials parameters that 
do not change unless the system experiments a quantum phase transition. Such phases, referred to as \emph{topological phases}, do not have a local order parameter, but they rather posses a 
so-called topological order parameter  \cite{tkkn,xiao}. Consequently, a description based on the Ginzburg-Landau theory of symmetry-breaking fails. 

In mathematics, \emph{topology} studies whether objects can be deformed smoothly into each other, without creating a hole in the deformation process \cite{nakahara}. 
The concept of topological invariance, and therefore topological invariants, was introduced to classify different geometrical objects into broad classes. For instance, 
2D surfaces are classified by their number of holes. 
In condensed matter physics, on the other hand, one considers Hamiltonians with an energy gap separating the ground state  from the excited states, where a smooth deformation is 
defined as a change in the Hamiltonian that does not close the energy gap. If two gapped Hamiltonians can be continuously transformed into each other without closing the energy gap, 
then it is said that these two systems are topologically equivalent \cite{Hasan:RMP10}. These classes are distinguished by a topological invariant called the Chern 
number \cite{nakahara}. 
This Chern number is a well defined integer in the absence of band crossings (gapped subbands). 
It can therefore only change value when subbands cross, or reconnect, as an external parameter is varied.
The process where two gapped Hamiltonians cannot be continuously transformed  into each other without closing the energy gap,  implies a change in the topological invariant, 
the Chern number, and such process is called \emph{topological phase transition}. 
A fundamental consequence of the topological classification of gapped band structures is the existence of gapless conducting states at interfaces where the 
topological invariant changes \cite{Hasan:RMP10,Qi:RMP11}.

In the late 80's the discoveries of the Integer Quantum Hall effect (IQHE) \cite{Pepper} and later on its fractional (FQHE) counterpart \cite{Tsui} 
were well described by the previous ideas. 
These experiments investigated the motion of electrons confined to two dimensions and exposed to a strong perpendicular magnetic field at low temperatures.
The experiments reported that 
this state is characterised by an energy gap between the ground state and the excited states,
 a quantised Hall conductance, $\nu e^{2}/h$ and a vanishing longitudinal resistance, where integer values of  $\nu$ stands for the IQHE and fractional values of $\nu$ 
 for the FQHE. In the quantum Hall state the bulk of the two-dimensional sample is insulating and the electric current is carried along the edge of the 
 sample, only. The flow of this unidirectional current does not exhibit dissipation and leads to the quantisation of the Hall conductance. 
 Astonishingly, the value of the Hall conductance is determined by the ratio of two fundamental constants, and therefore is independent of any properties of the material 
 being measured, 
 disorder or other macroscopic details. 
 Afterwards, it was demonstrated that the quantisation of the Hall conductance, is related to the Chern number \cite{tkkn}, and therefore the Hall state is a result of a 
 topological phase. 
As discussed above, a consequence of the topological classification of gapped band structures is the existence of gapless conducting states at the interfaces where the 
topological invariant changes \cite{Hasan:RMP10,Qi:RMP11}. The electronic states between the integer Hall effect and vacuum are \emph{chiral}  in the sense that they propagate in one direction only along the edge.
Physically, the Chern number determines the number of these chiral states, which propagate along the edge of the sample. Due to this chirality, these chiral states are robust against 
disorder, due to the absence of counter-propagating modes in which to backscatter, and carry electric current without dissipation \cite{Hasan:RMP10,Qi:RMP11}.

Later, the two-dimensional quantum spin Hall effect (QSHE) in HgTe/CdTe quantum wells \cite{bergv,Zhang2}
and its three-dimensional counterpart in bismuth chalcogenides \cite{haij,has33} were predicted and discovered. The QSH state is the first example of the so-called two-dimensional topological insulator.  
Unlike the Hall effect described before, which breaks time-reversal 
symmetry, these new quantum states (QSHE) belong to a new class of materials called topological insulators, which are invariant under time-reversal and in which spin-orbit coupling 
plays a key role  \cite{Hasan:RMP10,Qi:RMP11}. 
 The QSH state is invariant under time-reversal, has a charge excitation gap in the 2D bulk, but has topologically protected 1D gapless edge states that lie inside the bulk 
 insulating gap \cite{tkkn, xiao, PhysRevLett.95.226801}. 
 Unlike the QHS, the edge states in the QSHS are distinct: two states in with opposite spin polarization counter propagate at a given edge \cite{PhysRevLett.95.226801,PhysRevLett.95.146802,PhysRevLett.96.106401,PhysRevB.73.045322}. For this reason, 
 these states are also known as \emph{helical}, where the spin is correlated with the direction of motion \cite{PhysRevLett.96.106401}. In this case, time-reversal symmetry prevents the helical edge states from 
 backscattering \cite{Hasan:RMP10,Qi:RMP11}.
 The spectrum of a QSHS cannot be smoothly deformed into another of a topologically 
 trivial insulator without helical states, thus representing a new topologically distinct state of matter.
These topological insulators can be characterized by a $Z_{2}$ topological invariant, which is determined from the band 
structure \cite{PhysRevLett.95.146802,PhysRevB.73.045322}. The helical states in the QSHS can be viewed as two copies of chiral edge states of the QHS related by time-reversal symmetry.

The search and study of topological phases in superconductors started even before topological insulators \cite{volovikbook}, 
but it was until the classification of topological insulators that similar ideas were used to topologically classify superconductors 
\cite{roya, PhysRevB.78.195125, kitaev09, PhysRevLett.102.187001}.
Indeed, there is a direct analogy between insulators and superconductors because the equations that describe quasiparticles in a superconductor or superfluids like He$^{3}$, 
the so-called Bogoliubov-de Gennes (BdG) equations, have a similar mathematical formulation as the Dirac equation for topological insulators, with the superconducting gap 
corresponding to the band gap of the insulator. 
In superconductors, additionally to the time-reversal symmetry in topological insulators, one finds charge-conjugation or electron-hole symmetry. These two symmetries and the 
product of the two of them (chiral or sublattice symmetry) led to the classification 
of topological superconductors \cite{PhysRevB.78.195125}. As in topological insulators, here we also find time-reversal invariant and breaking topological superconductors, 
where the former is classified \cite{roya, PhysRevB.78.195125, kitaev09, PhysRevLett.102.187001}
by a $Z_{2}$ invariant in 1D and 2D and a $Z$ invariant in 3D, while the latter are classified by an integer \cite{PhysRevB.61.10267} in a similar fashion as the Quantum Hall 
insulators \cite{tkkn}. It is important to mention here that when time-reversal is present the systems belong to the BDI class, while for time-reversal breaking  
to the D class \cite{roya,kitaev09}.
The ones that have attracted much attention during last years are the time-reversal breaking topological superconductors, mainly because they are related with non-Abelian statistics 
with potential application to topological quantum computation \cite{RevModPhys.80.1083}.

In 2D, the integer classification of topological superconductors is very similar to that of topological insulators. Indeed, a QHS with Chern number $N$ has $N$ chiral edge states, 
while a chiral superconductor with topological number $N$ has $N$ chiral Majorana edge states, which resemble the ones in the QHS but with electron-hole redundancy 
\cite{Hasan:RMP10,Qi:RMP11}. The simplest platform in 2D for a topological superconductor consists of a spinless superconductor with $p_{x}\pm ip_{x}$ symmetry 
\cite{Read:PRB00,Volovik:JL99}, while a spinless superconductor with $p$ symmetry is the simplest one in 1D \cite{kitaev}. 
These superconductors support topological phases that host chiral Majorana edge states propagating at the boundary of domain-walls (regions that break time-reversal invariance), 
while a vortex binds a stable Majorana zero mode \cite{volovik,PhysRevB.61.10267,Beenakker:11}. 
Similarly, in 1D  the system hosts Majorana zero modes bound at its boundaries with trivial (non-topological) regions (one Majorana bound state at each end of the  1D system). 
These modes were named after Ettore Majorana, who introduced a similar concept in the context of high energy physics, where a Majorana fermion is a fermionic particle that is 
identical to its own anti-particle \cite{majorana,wilcek}.
In condensed matter, superconductors offer a natural platform for studying these exotic modes, since non-degenerate quasiparticle excitations in superconductors at zero energy 
indeed exhibit such Majorana character \cite{PhysRevB.81.224515,Beenakker:11}, where a particle is identical to its anti-particle. 
These phenomena have attracted massive theoretical interest during last years mainly because these zero modes represent the simplest case of non-Abelian anyons with profound 
implications in topological quantum computation \cite{kitaev,RevModPhys.80.1083,stern}.
Although the pairing symmetry, $p$ in 1D and  $p_{x}\pm ip_{x}$ in 2D, of the required superconductors can rarely emerge intrinsically, there are some important cases where 
nature does it for us.
The first proposal was the fractional quantum Hall effect state at filling fraction $5/2$ \cite{PhysRevB.61.10267} and later experiment on Sr$_{2}$RuO$_{4}$ 
compound provided evidence as the best experimental candidate for topological superconductivity with $p_{x}\pm ip_{y}$ \cite{RevModPhys.75.657}, similar to the A-phase in a 
superfluid liquid He$^{3}$ \cite{sigrist,ishida}.
Despite the efforts, conclusive signatures of such exotic $p$-wave superconductivity are still missing.

Even though $p$-wave and $p_{x}\pm p_{y}$ pairings are not robust against disorder and thus scarce in nature, 
a number of platforms were proposed in order to engineer such non-trivial superconductivity. These ideas have attracted enormous attention because they are based on combining traditional well-known effects in condensed matter physics, where all existing proposals are based on $s$-superconductors, representing an advantage over intrinsic non-trivial superconductivity.
Indeed, Fu and Kane \cite{Fu:PRL08} proposed to proximitize the surface of a 3D topological insulator with an $s$-wave superconductor, where below the superconductor critical temperature, the high transparency of the contacts gives rise to proximity-induced superconductivity, thus generating a 2D topological superconductor. The surface of a topological insulator hosts a single Dirac cone. Then, for any chemical potential residing within the bulk gap there is only one single Fermi surface, since the Dirac cone is non-degenerate, and therefore the spinless regime can be achieved. Electrons along this Fermi surface are not spin-polarised, so $p_{x}\pm ip_{y}$ pairing can be effectively induced on the surface of the 3D topological insulator by a conventional $s$-wave superconductor via proximity effect. The key ingredient for inducing such non-trivial pairing is based on the strong spin-orbit interaction, an intrinsic property of topological insulators, which gives rise to spin-momentum locking.
Time-reversal breaking of any form will generate chiral Majorana edge states at the boundary between topologically superconducting and magnetically gapped regions in the surface of a 3D topological insulator. Moreover, a magnetic field creates a vortex on the surface, which can trap midgap states, and therefore bind a stable Majorana zero mode. As it was pointed out previously, these zero modes possess special properties that can be used in topological quantum computation \cite{kitaev,RevModPhys.80.1083,stern}.
In fact, a well separated pair of Majorana bound states forms a fermionic two-level system, a qubit, which can be either occupied or empty defining a non-local qubit. Remarkably, the quantum information in this qubit is stored non-locally leading to long coherence times, which is a necessary requirement for robust quantum computing \cite{kitaev,RevModPhys.80.1083,stern}. The topological protection in these systems relies on the presence of a non-zero gap for quasiparticle excitations \cite{RevModPhys.80.1083}. Following Fu and Kane's groundbreaking proposal described above, many authors pursued alternative 
approaches towards engineering two one dimensional, $p_{x}\pm ip_{y}$ and $p$, superconductors, where semiconductor-based proposals are actively investigated \cite{Sau:PRL10,Alicea:PRB10}.

Along these lines, nanowire research has acquired another flavor due to the possibility of creating one-dimensional counterparts of topological surface states. 
The simplest model for engineering topological superconductivity in one-dimension is based on spinless fermions, an idea developed by Kitaev \cite{kitaev}. 
It consists of a chain of spinless fermions that supports a topological phase with Majorana bound states at each end. 
Interestingly, it was theoretically shown that it is possible to engineer the Kitaev's model only by considering traditional ingredients such as s-wave superconductivity and a spin texture that can be provided by strong Rashba spin-orbit and Zeeman interactions \cite{PhysRevLett.105.077001, PhysRevLett.105.177002}, rotating Zeeman fields \cite{PhysRevB.82.045127}, or RKKY interaction \cite{PhysRevLett.111.186805}.
These one-dimensional structures are thought to possess several distinct advantages when it comes to fabrication and subsequent detection of the Majorana zero modes.
Here, the system becomes spinless by applying a Zeeman field that opens a gap, while strong spin-orbit interaction guarantees that there is a finite antiparallel spin component between opposite momenta within each electronic band and therefore the induced superconducting pairing opens a quasiparticle excitation gap.
In quantum wires, Majoranas occur either at the wire ends or at a domain wall between topological and non-topological regions of the wire.
Unlike their two-dimensional counterparts, Majorana bound states in nanowires do not require the presence of a vortex in the system, eliminating decoherence which arises from low lying vortex-core quasiparticle states. 
Most importantly, the topological superconducting phase with Majoranas can be reached by varying the chemical potential, which can be tuned using gates, or 
by increasing the external magnetic field. This tunability makes nanowires the most promising scheme for the detection of Majorana bound states in condensed matter systems. 
Although a number of experiments have been reported \cite{frolov,xu,Das:NP12,Finck:PRL13,Churchill:PRB13,Lee:13}, still new geometries and studies are needed.

In this thesis we  consider the platform based on nanowires with strong Rashba spin-orbit coupling and Zeeman interaction \cite{PhysRevLett.105.077001, PhysRevLett.105.177002}, and then
investigate hybrid superconductor-semiconductor junctions made up of these nanowires.

\newpage
\section{Fermionic and Majorana operators}
\label{sec11}
Systems of many-body particles are appropriately described within the second quantisation formalism, where states are denoted by $\ket{N_{1},N_{2},\cdots}$ and identified by the number of particles $N_{i}$ in each single-particle state $i$. For fermions, $N_{i}=1,0$ denotes an occupied or empty state. Now, we define the fermionic creation operator $c^{\dagger}_{i}$, for the single particle state, as an operator which increases $N_{i}$ by one if the state $i$ is empty, $c^{\dagger}_{i}\ket{0_{i}}=\ket{1_{i}}$, and give zero otherwise, $c^{\dagger}_{i}\ket{1_{i}}=0$. Likewise, we define the fermionic annihilation operator $c_{i}$, which decreases $N_{i}$ by one if the state $i$ is occupied, $c_{i}\ket{1_{i}}=\ket{0_{i}}$, and give zero otherwise, $c_{i}\ket{0_{i}}=0$. Any state can be constructed by successive applications of creation operators on the states with no particles, the vacuum states.
The occupation number operator is defined as $\hat{N}_{i}=c^{\dagger}_{i}c_{i}$, which measures the number of particles in state $i$.
Due to the antisymmetry of the fermionic state $\ket{N_{1},N_{2},\cdots}$, creation and annihilation operators fulfil the following anti-commutation relations \cite{mahan},
\begin{equation}
\begin{split}
\{c^{\dagger}_{i},c_{j}\}&=c^{\dagger}_{i}c_{j}+c_{j}c^{\dagger}_{i}=\delta_{ij}\,,\\
\{c^{\dagger}_{i},c_{j}^{\dagger}\}&=c^{\dagger}_{i}c_{j}^{\dagger}+c_{j}^{\dagger}c^{\dagger}_{i}=0\,,\\
\{c_{i},c_{j}\}&=c_{i}c_{j}+c_{j}c_{i}=0\,,
\end{split}
\end{equation}
where $\delta_{ij}$ is the Kronecker delta.
Any physical operator can be written in terms of the creation and annihilation operators we have defined above \cite{mahan}. 
Now, we introduce a new kind of operators based on a decomposition of a Dirac (complex) fermion in terms of two real operators, known as Majorana operators \cite{kitaev},
\begin{equation}
\label{fermionMajoTrans}
c_{j}=\frac{1}{2}\Big(\gamma_{j}^{A}+i\gamma_{j}^{B}\Big)\,,\quad c_{j}^{\dagger}=\frac{1}{2}\Big(\gamma_{j}^{A}-i\gamma_{j}^{B}\Big)\,,
\end{equation}
where a Majorana operator is usually understood as half of a normal fermionic operator. 
In fact, any fermion operator can be defined in terms of Majorana operators, and a description in terms of these new operators is usually helpful in systems where the number of particles is only conserved modulo $2$ \cite{kitaev}, as in superconducting systems. 

The inverse transformation of Eqs.\,(\ref{fermionMajoTrans}) gives us the Majorana operators,
\begin{equation}
\label{MajTrans}
\gamma_{j}^{A}=c_{j}+c_{j}^{\dagger}\,,\quad \gamma_{j}^{B}=i(c^{\dagger}-c_{j})\,.
\end{equation}
These new operators $\gamma$ satisfy the following algebra
\begin{equation}
\label{MajAlgebra}
\{\gamma_{i}^{A},\gamma_{j}^{B}\}=2\delta_{ij}\delta_{AB}\,,\quad \gamma_{j}=\gamma^{\dagger}_{j}\,,\quad \gamma_{j}^{2}=\gamma^{\dagger2}_{j}=1\,.
\end{equation}
Any fermionic operator that satisfies previous conditions is a Majorana fermion operator, where
the second expression indeed expresses the essence of a Majorana fermion: a particle created by the operator $\gamma^{\dagger}$ is identical to its antiparticle created by $\gamma$. The term \emph{Majorana} refers to the \emph{real} nature of such operators, as in the Majorana's representation of the Dirac's equation in particle physics \cite{majorana}. 
The condensed matter counterparts, considered here, however, are not connected with the Majorana's original idea and its application to neutrinos \cite{majorana}. Despite the little connection between these two views, the terminology is extensively used by the condensed matter community. If additionally, the Majorana operator $\gamma$ from Eq.\,(\ref{MajTrans}) commutes with the systems's Hamiltonian,
\begin{equation}
\label{Majzeromode}
[H,\gamma_{j}]=0\,,
\end{equation}
then it represents a Majorana zero mode (MZM) and because these zero modes emerge bound to defects it is also common to call them Majorana bound states (MBS) \cite{Alicea:RPP12,Beenakker:11,StanescuModel13,RevModPhys.87.137,RevModPhys.87.1037}. This phenomenon has no analogy in particle physics and it is in condensed matter physics that such Majorana zero modes have acquired an enormous interest due to their potential application in topological quantum computation \cite{kitaev}. Later, in Sec.\,\ref{KitaevModel}, we will describe in more detail these zero modes in one-dimensional systems.
For now, the main condition for the emergence of a Majorana fermion in condensed matter systems is: \emph{a particle being its own anti-particle}. 
Notice that Eqs.\,(\ref{MajAlgebra}) constitute a linear combination of creation and annihilation operators, thus a natural platform for investigating the emergence of these exotic physics includes superconducting systems as we will see later.
Previous discussion is rather ideal and in realistic physical systems, however, the condition given by Eq.\,(\ref{Majzeromode}) is hardly fulfilled, and in general one has,
\begin{equation}
\label{Majzeromode2}
[H,\gamma_{j}]\approx {\rm e}^{-\frac{x}{\xi}}\,,
\end{equation}
where $x$ represents the separation between two MBSs, and $\xi$ the correlation length associated with the Hamiltonian $H$. Notice that for long enough $x$, Eq.\,(\ref{Majzeromode2}) indeed implies Eq.\,(\ref{Majzeromode}), and the condition for a Majorana zero mode is fulfilled.

The fermion (occupation) number operator, in terms of the Majorana operators, read
\begin{equation}
\hat{N}_{i}=c^{\dagger}_{i}c_{i}=\frac{1}{2}\big(1+i\gamma_{j}^{A}\gamma_{j}^{B}\big)\,,
\end{equation}
then the expectation values of the number operator follow
\begin{equation}
\label{epepe}
\bra{j}c^{\dagger}_{j}c_{j}\ket{j}=
\begin{cases}
  1,   & \ket{j}\, \text{occupied }, \\
    0, & \ket{j}\,\text{empty}.
\end{cases}\,\rightarrow\quad
\bra{j}2c^{\dagger}_{j}c_{j}-1\ket{j}=
\bra{j}i\gamma_{j}^{A}\gamma_{j}^{B}\ket{j}=
\begin{cases}
  1,    & \ket{j}\, \text{occupied }, \\
    -1, & \ket{j}\,\text{empty}.
\end{cases}
\end{equation}
For superconducting systems, the particle number operator $\hat{N}_{j}$ does not commute with the Hamiltonian. Instead, we define the number parity operator as 
\begin{equation}
\label{hdhdh}
P=(-1)^{\hat{N}}=-i\gamma_{j}^{A}\gamma_{j}^{B}\,,
\end{equation} 
which anti-commutes with the Majorana operators $\{\gamma_{i},P\}=0$, and commutes with a Hamiltonian being quadratic in the fermionic creation and annihilation operators $[H,P]=0$. 
Thus, taking into account Eq.\,(\ref{epepe}),
\begin{equation}
\bra{j}-i\gamma_{j}^{A}\gamma_{j}^{B}\ket{j}=
\begin{cases}
  -1,    & \ket{j}\, \text{occupied }\rightarrow \text{odd parity} \\
    1, & \ket{j}\,\text{empty}\rightarrow \text{even parity}\,.
\end{cases}
\end{equation} 
which implies that the eigenstates of $H$ can be divided in states with even and odd parity.

\newpage
\section{Bogoliubov-de Gennes formalism for superconductivity}
\label{bdgformalism}
As we have pointed out previously introduction, superconductivity provides a natural platform for investigating Majorana physics in condensed matter systems. Thus, 
in this part we describe the Bogoliubov-de Gennes (BdG) formalism of the BCS\footnote{Named after John Bardeen, Leon Cooper, and John Robert Schrieffer (BCS) \cite{PhysRev.108.1175}} theory of superconductivity  \cite{PhysRev.108.1175}, which describes quasiparticle excitations in superconductors.
The BCS Hamiltonian emerges from Eq.\,(\ref{eq1}) by making the assumption that, at low temperatures even a weak attractive interaction\footnote{In conventional (singlet) superconductors, as the one discussed next, the attraction is due to an exchange of phonons. However, magnetic interactions, for instance, can also induce attraction between electrons favouring triplet pairing with a non-zero spin of the pair.} between electrons near the Fermi surface enables the formation of bound pairs of time-reversed states ($k,\uparrow$) and ($-k,\downarrow$), called Cooper pairs \cite{PhysRev.104.1189}. 
Then, considering only the terms decisive for superconductivity, we write down the so-called reduced Hamiltonian \cite{tinkham}
\begin{equation} 
\label{reducedH}
H=\sum_{k,\sigma}\xi_{k}\,c^{\dagger}_{k\sigma}c_{k\sigma}-\sum_{k,k'}V_{k,k'}\,c^{\dagger}_{k,\uparrow}c^{\dagger}_{-k,\downarrow}
c_{-k',\downarrow}c_{k',\uparrow}\,,
\end{equation}
where the first term is the free electron term, $\xi_{k}=\hbar^{2}k^{2}/2m-\mu$ with $m$ the effective electron's mass and $\hbar$ the Planck's constant, second term  describes the scattering of Cooper pairs with momenta $(k',-k')$ into another pair with momenta $(k,-k)$ with amplitude $V_{k,k'}$. 
In the previous Hamiltonian, we have introduced the second quantization representation, where
$c^{\dagger}_{k\sigma}$($c_{k\sigma}$) is the creation (annihilation) operator which creates (destroys) an electron with momentum $k$ and spin $\sigma=\uparrow,\downarrow$.
Exact analytical treatment of Eq.\,(\ref{reducedH}) is complicated and therefore it is usual to follow the mean-field approach. 
This is justified by taking into account the fact that the ground state of the BCS Hamiltonian is a coherent many-body state, and therefore the pair of operators such as $c_{-k',\downarrow}c_{k',\uparrow}$ can have non-zero expectation values in such ground state, unlike averaging to zero in a normal metal. Moreover, since now we deal with a huge number of particles, fluctuations about these expectation values are expected to be rather small, and therefore some really small terms are neglected. For details see Appendix\,\ref{BCS}. 
Then, after some algebra, we arrive at the Hamiltonian in the mean field approximation,
\begin{equation}
\label{mfaH0}
H_{MFA}=\sum_{k,\sigma}\xi_{k}\,c^{\dagger}_{k\sigma}c_{k\sigma}
+\sum_{k}\Big[\Delta_{k}^{*}\,c_{-k,\downarrow}c_{k,\uparrow}+\Delta_{k}\,c_{k,\uparrow}^{\dagger}c_{-k,\downarrow}^{\dagger}\Big]-\sum_{k}\Delta_{k}\average{c_{-k,\downarrow}c_{k,\uparrow}}\,,
\end{equation}
where we have defined $\Delta_{k}=-\sum_{k'}V_{k,k'}\average{c_{-k,\downarrow}c_{k,\uparrow}}$, which is in general complex and it represents the pairing potential or superconducting order parameter, and $\average{}$ is taken over the BCS ground state. The symmetry of $\Delta_{k}$ is intrinsic from the superconducting material, and it can give rise to superconductors with different properties.
See Appendix \ref{pairings} for a brief discussion on singlet and triplet superconducting pairings.
Here, we consider $s$-wave pairing, where electrons of different spin and momenta are paired together as given by Eq.\,(\ref{mfaH0}). From here on, we consider $\Delta_{k}=\Delta$ as for $s$-wave symmetry the pairing potential is independent of $k$.
Now, when studying superconducting systems, it is standard to  treat  electrons  and  holes  at  the  same footing. Therefore, it is appropriate to introduce a new set of spinors $\Psi_{k}$, where previous Hamiltonian can be written as, see Appendix for more details,
\begin{equation}
\label{mfaH}
H_{MFA}=\sum_{k}\Psi_{k}^{\dagger}\,H_{BdG}\,\Psi_{k}\,+\,\sum_{k}\Big[\xi_{k}-\Delta\average{c_{-k,\downarrow}c_{k,\uparrow}}\Big]\,,\quad 
\end{equation}
where  $\Psi_{k}=
\begin{pmatrix}
c_{k,\uparrow},
c_{k,\downarrow},
c_{-k,\uparrow}^{\dagger},
c_{-k,\downarrow}^{\dagger}
 \end{pmatrix}^{\dagger}$ is known as the Nambu spinor \cite{zagoskin}. 
Then, previous, and 
\begin{equation}
\label{hbdg}
H_{BdG}=\xi_{k}\tau_{z}\otimes\sigma_{0}-{\rm Re}(\Delta)\tau_{y}\otimes\sigma_{y}-{\rm Im}(\Delta)\tau_{x}\otimes\sigma_{y}
\,=\,
\begin{pmatrix}
\xi_{k}&0&0&\Delta\\
0&\xi_{k}&-\Delta&0\\
0&-\Delta^{*}&-\xi_{-k}&0\\
\Delta^{*}&0&0&-\xi_{-k}\\
\end{pmatrix}
\end{equation}
is the so-called Bogoliubov-de Gennes Hamiltonian \cite{DeGennes}, where $\tau_{i}$ and $\sigma_{i}$, $i=0,x,y,z$, are Pauli matrices acting on the electron-hole and spin subspaces\footnote{The matrices $\sigma_{i}$ and $\tau_{i}$ act on the spin and particle-hole degree of freedom, respectively:
\begin{displaymath}
\sigma_{0}=\begin{pmatrix}
1&0\\
0&1
\end{pmatrix}\,,\quad
\sigma_{x}=\begin{pmatrix}
0&1\\
1&0
\end{pmatrix}
\,,\quad
\sigma_{y}=\begin{pmatrix}
0&-i\\
i&0
\end{pmatrix}
\,,\quad
\sigma_{z}=\begin{pmatrix}
1&0\\
0&-1
\end{pmatrix}\,.
\end{displaymath}}. 
 Notice that we have considered the superconducting pairing as  $\Delta=|\Delta|\,{\rm e}^{i\varphi}={\rm Re}(\Delta)+i{\rm Im}(\Delta)$, where $\varphi$ is the superconducting phase.  
Moreover, due to  inversion asymmetry, $\xi_{k}=\xi_{-k}$, the spectrum is invariant under $k\rightarrow -k$, and the BdG Hamiltonian can be written as
\begin{equation} 
H_{BdG}=
\begin{pmatrix}
H_{0}& i\sigma_{y}\Delta\\
-i\sigma_{y}\Delta^{*}&-H_{0}^{*}
\end{pmatrix}\,,\quad
H_{0}=
\begin{pmatrix}
\xi_{k}&0\\
0&-\xi_{k}
\end{pmatrix}
\end{equation}
where $^{*}$ denotes the complex conjugation operation and in the following represented by the operator $K$, and $\xi_{k}=\hbar^{2}k^{2}/2m-\mu$.
Now, we aim at diagonalizing $H_{MFA}$, which is done by finding the eigenvalues of $H_{BdG}$, since
 the second term in Eq.\,(\ref{mfaH}) is constant and therefore can be neglected.
The eigenvalue problem, 
\begin{equation}
\label{BdGEqs}
H_{BdG}\Psi=E\Psi\,,
\end{equation} 
gives rise to the so-called Bogoliubov-de Gennes equations and $\Psi$ is a four component vector (two for electron and two for hole) \cite{zagoskin}. 
The Bogoliubov-de Gennes Hamiltonian, $H_{BdG}$, is Hermitian and acts on spinors $\Psi$, whose first half is formed out of annihilation operators of electrons and the second half out of creation operators of the same electrons. The second half with creation operators of the same electrons can be viewed as annihilation operators of an extra set of holes, so that we introduce a redundant description, where the amount of degrees of freedom are doubled in the system. 
The off-diagonal blocks of $H_{BdG}$ couple electrons and holes in opposite spins bands switched by the spin Pauli matrix $\sigma_{y}$ via the pairing potential $\Delta$ \cite{RevModPhys.87.1037}.

This relation between electrons and holes automatically imposes a symmetry on $H_{BdG}$: the so-called electron-hole or charge-conjugation symmetry. 
In the spinor basis given by $\Psi_{k}=
\begin{pmatrix}
c_{k,\uparrow},
c_{k,\downarrow},
c_{-k,\uparrow}^{\dagger},
c_{-k,\downarrow}^{\dagger}
 \end{pmatrix}^{\dagger}$, each eigenfunction $\Psi$ of $H_{BdG}$ at energy $E>0$ has a copy $\tau_{x}\Psi$ at $-E$, where the Pauli matrix $\tau_{x}$ switches electrons and holes \cite{RevModPhys.87.1037}.
The action of such symmetry converts an electron into a hole and vice versa and it is represented by an anti-unitary operator $P=\tau_{x}\otimes\sigma_{0}K$, where the Pauli matrix $\tau_{x}$ acts on the electron-hole subspaces, $\sigma_{0}$ on the spin sector and $K$ is the complex conjugation operator. Then, one can show that $H_{BdG}=-PH_{BdG}P^{-1}=-\tau_{x}H_{BdG}^{*}\tau_{x}$. Thus, due to the minus sign, the spectrum of $H_{BdG}$ must be symmetric around zero energy: for every vector $\Psi$ of $H_{BdG}$ with energy $E$, there is an electron-hole symmetric eigenvector $P\Psi$ with energy $-E$.
Notice that the Hamiltonian given by Eq.\,(\ref{hbdg}) can also be written in the alternative basis given by $\tilde{\Psi}_{k}=
\begin{pmatrix}
c_{k,\uparrow},
c_{k,\downarrow},
-c_{-k,\downarrow}^{\dagger},
c_{-k,\uparrow}^{\dagger}
 \end{pmatrix}^{\dagger}$. In this basis, however, the particle-hole symmetry equals $\tilde{P}=(\sigma_{y}\otimes\tau_{y})K$. Important, although the explicit form of the particle-hole operator depends on the spinor basis, both $P$ and $\tilde{P}$ are anti-unitary and square to $+1$ and the physics discussed here does not depend on it.

The eigenvalues of $H_{BdG}$, found after solving Eq.\,(\ref{BdGEqs}), represent the energy spectrum of quasiparticle excitations in a superconductor, known in the literature as Bogoliubov 
quasiparticles \cite{DeGennes,tinkham,zagoskin},
\begin{equation}
\label{BdGSC}
E_{k,\pm}=\pm\sqrt{\xi_{k}^{2}+|\Delta|^{2}}\,,
\end{equation}
where $\xi_{k}$ represent the dispersion of free electrons in a normal metal, while $\Delta$ represents the $s$-wave superconducting pairing potential, which couples electrons and holes.
After finding the eigenvalues of $H_{MFA}$, given by previous equations, one is now able to think
 that there is a new basis in which the mean-field Hamiltonian is diagonal, up to a constant term that we have dropped off, see second term in Eq.\,(\ref{mfaH}), and therefore we can write
\begin{equation} 
H_{MFA}\cong\sum_{k}
\begin{pmatrix}
\alpha_{k,\uparrow},\alpha_{k,\downarrow},\alpha_{-k,\uparrow}^{\dagger},\alpha_{-k,\downarrow}^{\dagger}
\end{pmatrix}^{\dagger}
\begin{pmatrix}
E_{1}&0&0&0\\
0&E_{2}&0&0\\
0&0&E_{3}&0\\
0&0&0&E_{4}\\
\end{pmatrix}
\begin{pmatrix}
\alpha_{k,\uparrow}\\
\alpha_{k,\downarrow}\\
\alpha_{-k,\uparrow}^{\dagger}\\
\alpha_{-k,\downarrow}^{\dagger}
\end{pmatrix}
\end{equation}
where $E_{1,2}=E_{k,+}$, $E_{3,4}=E_{k,-}$. The operators $\alpha^{\dagger}_{k,\uparrow}, \alpha^{\dagger}_{k,\downarrow}, \alpha^{\dagger}_{-k,\uparrow},\alpha^{\dagger}_{-k,\downarrow}$ are fermionic and create quasiparticles with energies $E_{k,+},E_{k,+},E_{k,-},E_{k,-}$, respectively.
The relation between operators $\alpha$ and $c$ is given by an unitary matrix formed by the eigenvectors of $H_{BdG}$, which gives rise to the Bogoliubov transformation\footnote{Bogoliubov-Valatin transformation in the Russian literature, although along this thesis we will refer to as Bogoliubov transformation, only.}, see Appendix \ref{bogotrans} for further details on the derivation,
\begin{equation}
\label{bogomaintex}
\begin{split}
\alpha_{k,\uparrow}&=u_{k}c_{k,\uparrow}+v_{k}c_{-k,\downarrow}^{\dagger} \,,\\
\alpha_{k,\downarrow}&= u_{k}c_{k,\downarrow}+v_{k}c_{-k,\uparrow}^{\dagger} \,,\\
\alpha_{-k,\uparrow}^{\dagger}&=-v_{k}c_{k,\downarrow}+u_{k}c_{-k,\uparrow}^{\dagger} \,,\\
\alpha_{-k,\downarrow}^{\dagger}&= -v_{k}c_{k,\uparrow}+u_{k}c_{-k,\downarrow}^{\dagger}\,,
\end{split}
\end{equation}
where $u_{k}$ and $v_{k}$ are the so-called coherence factors and given by,
\begin{equation}
u_{k}=\frac{1}{\sqrt{2}}\sqrt{1+\frac{\xi_{k}}{E_{k}} }\,, \quad v_{k}=\frac{1}{\sqrt{2}}\sqrt{1-\frac{\xi_{k}}{E_{k}}}\,,
\end{equation}
and satisfy $|u_{k}|^{2}+|v_{k}|^{2}=1$. The quasiparticle operators given by Eqs.\,(\ref{bogomaintex}) are not independent and indeed one notices that $\alpha_{k,\uparrow(\downarrow),E_{k,+}}$ is connected to $\alpha_{-k,\downarrow(\uparrow),E_{k,-}}^{\dagger}$, leading to only two independent quasiparticle operators, due to the redundant description we have introduced with the Bogoliubov-de Gennes description. 

\begin{figure} 
  \centering\includegraphics[width=.9 \columnwidth]{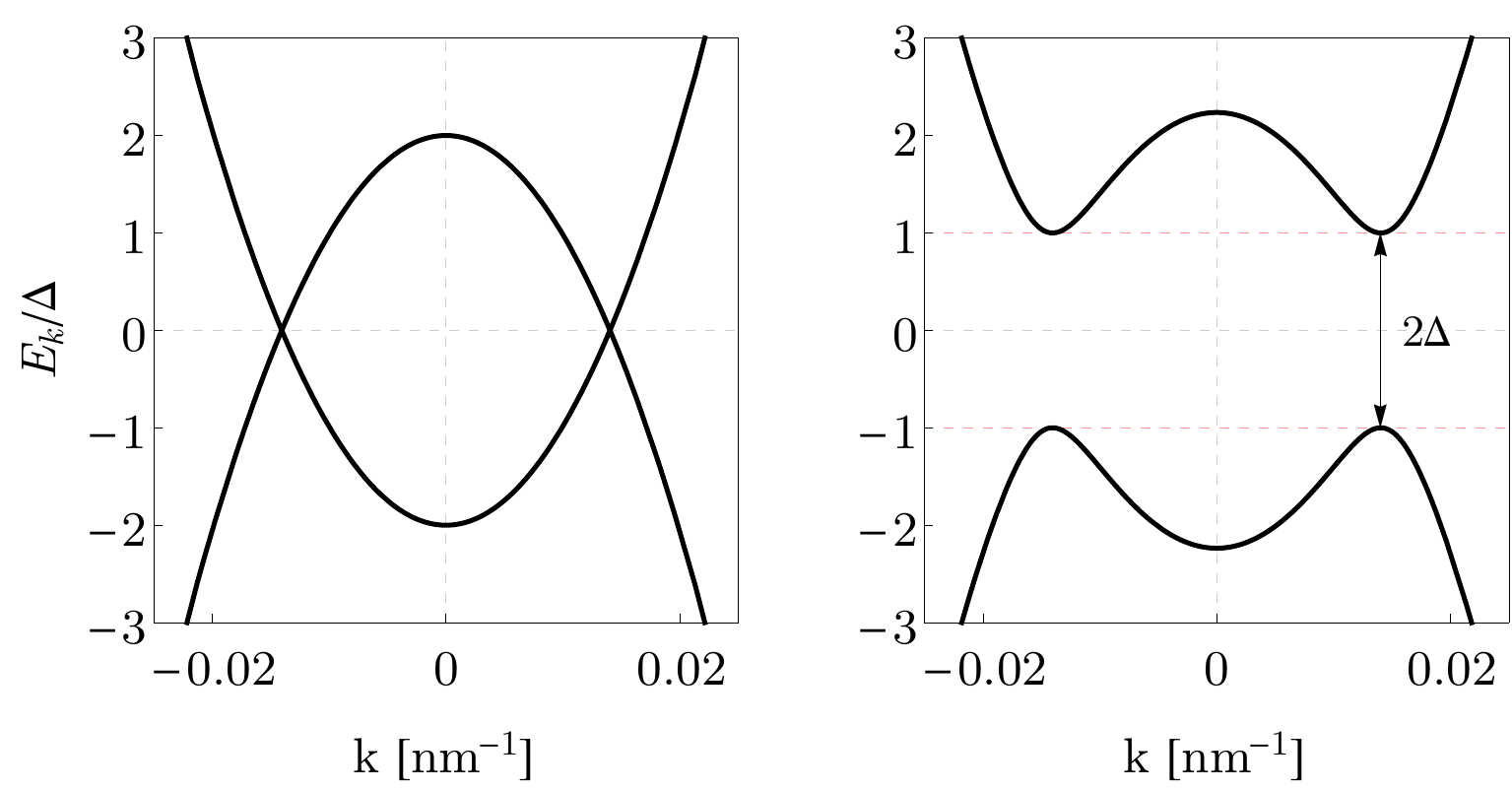} 
     \caption[]{ Energy spectrum in a superconductor from Eq.\,(\ref{BdGSC}). Left panel shows the bands for $\Delta=0$, while the right panel for $\Delta=0.25$\,meV. The chemical potential is set to $\mu=0.5$\,meV, and $m=0.015\,m_{e}$, where $m_{e}$ is the free electron mass.
}
   \label{fig1:BdG}
\end{figure}
The spectrum shown in Fig.\,\ref{fig1:BdG}, given by Eq.\,(\ref{BdGSC}), is two-fold degenerate due to spin, so that in principle there are four bands. This degeneracy can be lifted by applying a Zeeman magnetic field for instance.
At zero pairing potential, $\Delta=0$, the spectrum consists of two parabolas for electrons and two for holes. Superconductivity opens an energy gap of $2\Delta$ at the Fermi points in the spectrum,
giving rise to energy bands, which are a mixture of electron and hole.
This can be further seen from Eqs.\,(\ref{bogomaintex}), where quasiparticles created (destroyed) by operators $\alpha_{k,\sigma}^{\dagger}$ ($\alpha_{k,\sigma}$) are a linear combination of electron and hole $c$ operators with opposite spins and represent the elementary excitations in a superconductor.  Notice that the spectrum shown in Fig.\,\ref{fig1:BdG} is similar to the one of a band insulator with fine-tuned particle-hole symmetry, with the sole difference that quasiparticles in a superconductor are mixture of electrons and holes, as given by Eqs.\,(\ref{bogomaintex}).

The energy $\Delta$ is the lowest single-particle excitation energy in the superconducting state. Since particles are paired to form Cooper pairs in the ground state, it is not possible to excite individual quasiparticles with energy $E_{k}$, but rather one must break such pairs and excite them to the sea of Bogolibov quasiparticles represented by the band $E_{k}$. The minimum of energy needed for breaking a Cooper pair corresponds to $2\Delta$, and it determines the pairing energy. Cooper pairs maintain their correlation within a length called the coherence length, and following the BCS theory it is defined as \cite{DeGennes,tinkham,zagoskin}
\begin{equation}
\xi_{sc}\equiv \frac{\hbar v_{F}}{\pi \Delta(0)}\,,
\end{equation}
where $v_{F}$ is the Fermi velocity in the superconductor and $\Delta(0)$ is the zero temperature superconducting pairing\footnote{In general, we have defined: $\Delta_{k}=-\sum_{k'}V_{k,k'}\average{c_{-k,\downarrow}c_{k,\uparrow}}$. This equation can be written as $\Delta_{k}=-\sum_{k'}V_{k,k'}\frac{\Delta_{k'}}{2E_{k'}}{\rm tanh}\big(\frac{E_{k'}}{2\kappa T}\big)$, which constitutes a self-consistency equation for the pairing potential $\Delta_{k}$ and it is known as the BCS gap equation \cite{DeGennes,tinkham,zagoskin}. Within the BCS approach, $V_{k,k'}=V$, singlet pairing, which do no depend on momentum $\Delta_{k}=\Delta_{k'}=\Delta$. Then $\frac{1}{V}=\sum_{k}\frac{1}{2E_{k}}{\rm tanh}\big(\frac{E_{k}}{2\kappa T}\big)$, where $E_{k}$ given by Eq.\,(\ref{BdGSC}). From this equation one can calculate the dependence on the temperature of pairing potential $\Delta(T)$. Then, taking the zero temperature limit one can show that in the weak-coupling limit $\Delta(0)=2\hbar\omega_{D}{\rm exp}(-2/VN_{0})$, where $\hbar\omega_{D}$ is the Debye energy that characterises the cutoff of the phonon spectrum, and $N_{0}$ is the density of states at the Fermi level for electrons of one spin projection. Now, considering that at the critical temperature, $T_{c}$, the pairing potential vanishes $\Delta=0$, one can show that $\frac{2\Delta(0)}{\kappa T_{c}}=3.52$ holds for conventional low-$T_{c}$ superconductors.}. In previous definition, we have assumed zero temperature and a superconductor without impurities. In alloys with a mean-free path $l<\xi_{sc}$, the coherence length follows $\bar{\xi}_{sc}=\sqrt{\xi_{sc}l}$ \cite{tinkham}. 
\begin{figure} 
  \centering\includegraphics[width=.6\columnwidth]{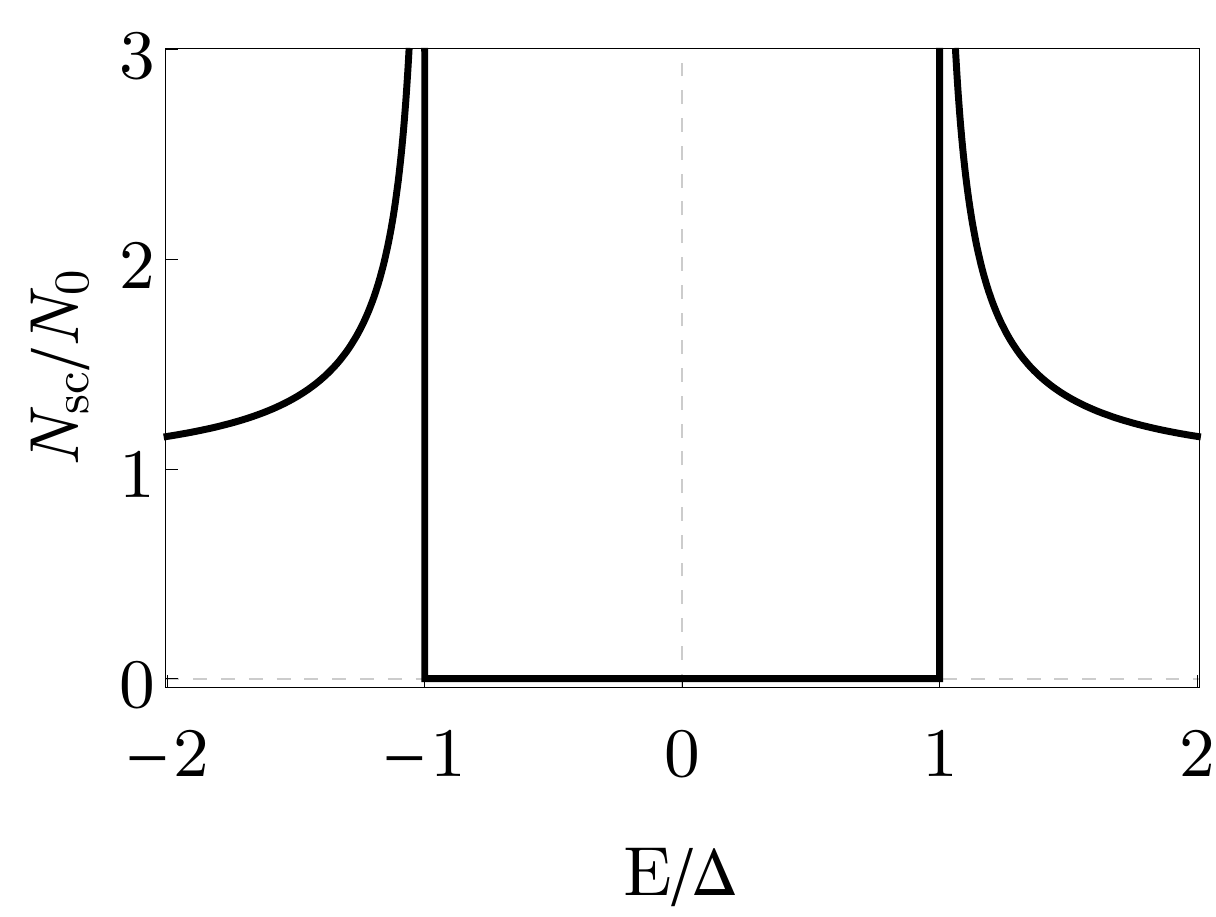} 
     \caption[Density of states for a $s$-wave superconductor]{ Density of states as a function of energy for a $s$-wave superconductor. 
     Superconductivity opens a gap of $2\Delta$, which is not allowed for energy states due to the $s$-wave nature of the pairing potential.}
   \label{fig2:DOSSC}
\end{figure}
Another special result of the BCS theory was on the density of states in the superconductor: $N_{sc}(E)=dN/dE=N_{0}d\xi/dE$, where $N_{0}$ is the density of states at the Fermi energy in the normal state assumed constant within $\pm\Delta$. Then, according to Eq.\,(\ref{BdGSC}), $N_{sc}(E)=|E|/\sqrt{E^{2}-|\Delta|^{2}}$ for $E>|\Delta|$ exhibits divergences at the gap edges $E=\pm \Delta$ giving rise to the van Hove singularities of the superconducting spectrum\footnote{These peaks are also known as coherence peaks.}, and zero within the gap of width $2\Delta$, as shown in Fig.\,\ref{fig1:BdG} \cite{DeGennes,tinkham,zagoskin}. The fact that there are no states within $2\Delta$ is based on the nature of our $s$-wave superconductor and supported by the Anderson's theorem\footnote{states that an $s$-wave superconductor does not host states within the gap at zero magnetic field \cite{Anderson-theorem}.}\cite{Anderson-theorem}

\begin{figure} 
  \centering\includegraphics[width=.99\columnwidth]{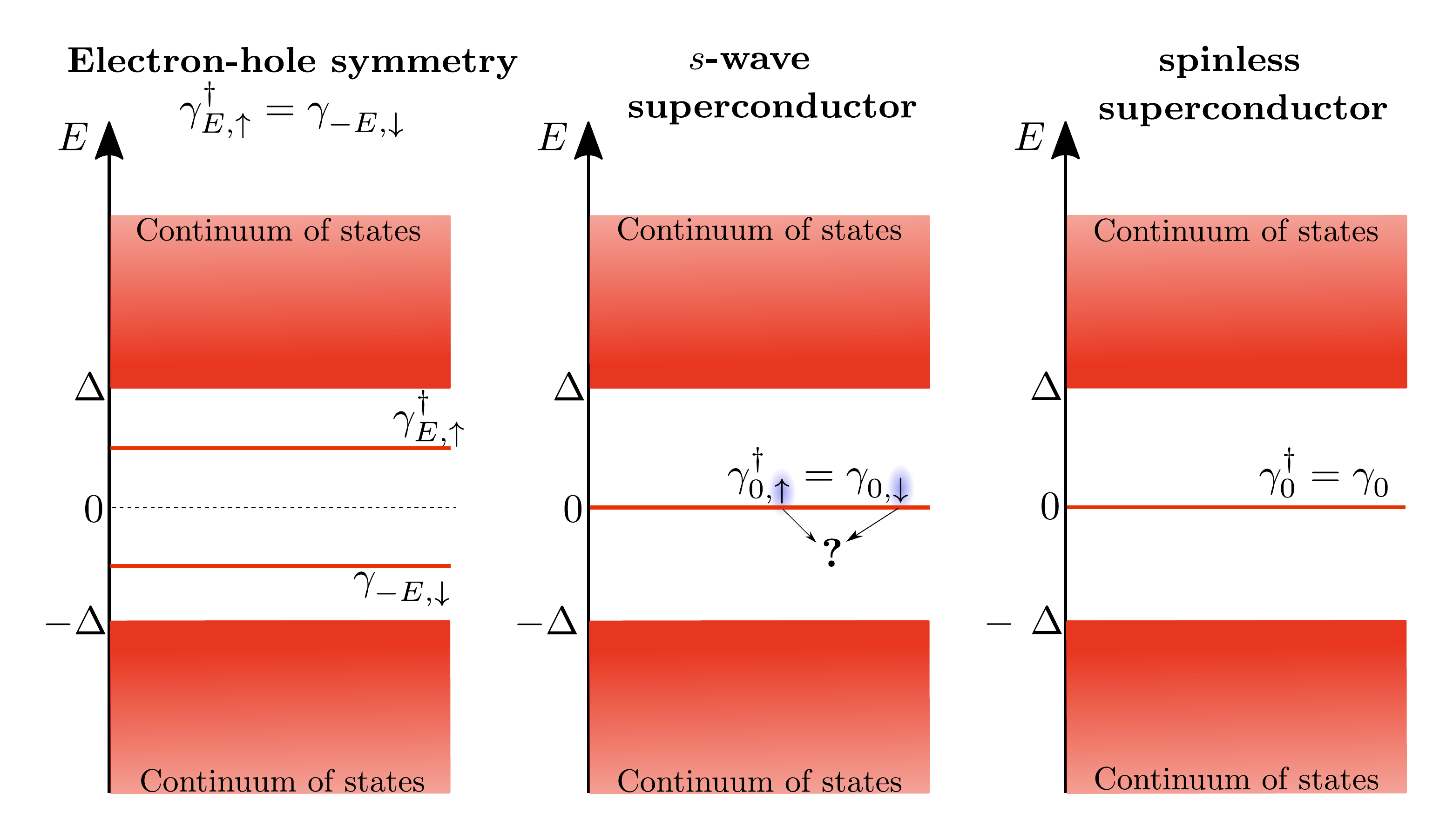} 
     \caption[Sketch of the need of spinless superconductors for MBSs]{(Color online) Sketch of the need of spinless superconductors for MBSs. (Left) Solutions of the Bogoliubov-de Gennes come in pairs, which are related by electron-hole symmetry. (Middle) In a $s$-wave superconductor, spin degeneracy does not allow to have robust zero modes. (Right) In 1D, the simplest systems which allow non-degenerate zero modes are spinless superconductors, $p$-wave superconductors. }
   \label{fig3:BdG}
\end{figure}

As we have already discussed, the eigenvalues of the Bogoliubov-de Gennes Hamiltonian come in pairs due to electron-hole symmetry, and it is schematically shown in Fig.\,\ref{fig3:BdG}. Thus, in a superconductor, the creation of a quasiparticle with energy $E$ is identical to the annihilation of a quasiparticle with energy $-E$. 
This idea can be seen in the quasiparticle operators given by Eqs.\,(\ref{bogomaintex}), where $\alpha_{k,\uparrow(\downarrow),E_{k,+}}^{\dagger}=\alpha_{-k,\downarrow(\uparrow),E_{k,-}}$.

We have pointed out in Sec.\,\ref{sec11} that superconducting systems are natural platforms for 
investigating the emergence of Majorana physics in condensed matter systems.
The main condition we described was based on a particle being its own anti-particle. This property is represented in terms of Majorana operators by the relation $\gamma=\gamma^{\dagger}$.
As already discussed, electron-hole symmetry in $s$-wave superconductors introduces somehow a similar condition, $\gamma_{E,\uparrow}^{\dagger}=\gamma_{-E,\downarrow}$. 
The condition for having a particle which is its own antiparticle must therefore occur at zero energy. However, the problem with $s$-wave superconductors is evidently the spin due to spin degeneracy, and therefore the zero energy state is not robust.
Thus, it is not always obvious how to impose this condition in trivial superconductors, such as $s$-wave, due to spin degeneracy. This is the reason why spinless or spin-polarised superconductors are required, and in the next section we discuss the simplest proposal in one dimension.

\clearpage
\newpage
\section{One-dimensional topological superconductivity}
\label{KitaevModel}
\begin{figure} 
   \includegraphics[width=1 \columnwidth]{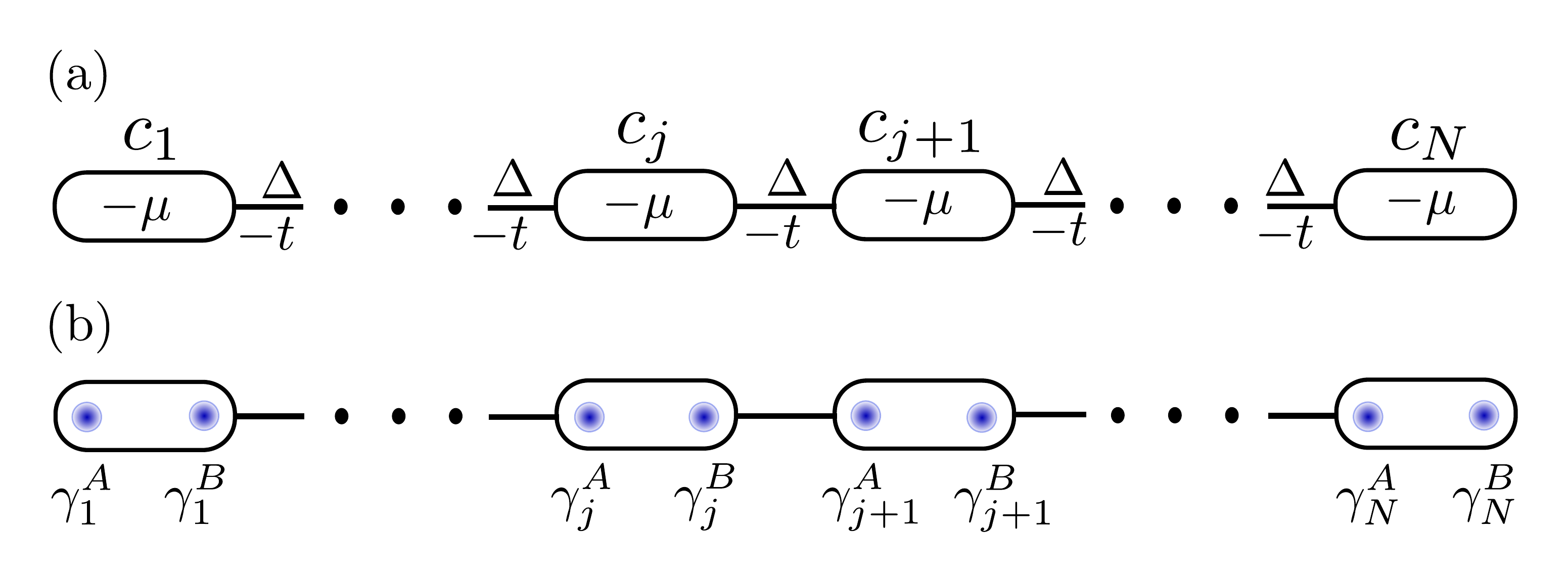} 
     \caption[Kitaev model]{(Color online) Kitaev model. (a) One-dimensional chain of spinless fermions $c_{j}$ with $N$ sites given by Eq.\,(\ref{kiatev0}), where $t$ and $\Delta$ represent the nearest-neighbor hopping and pairing amplitudes, respectively. (b) Each fermionic site can be decomposed into two Majorana fermions $\gamma_{j}^{A}$ and $\gamma_{j}^{A}$ following Eqs.\,(\ref{fermionMajoTrans}).
}
   \label{fig:kitaev}
\end{figure}

In this part we introduce the Kitaev's proposal \cite{kitaev}, a model to engineer Majorana bound states in one-dimensional quantum wires based on $p$-wave superconductivity. It describes a 1D system of fermions with the same spin that can be viewed as spin-polarized or spinless.
The model consists of a chain with N sites, where each site can be empty or occupied by a fermion
\begin{equation}
\label{kiatev0}
H=-\mu\sum_{j=1}^{N}\Big(c^{\dagger}_{j}c_{j}-\frac{1}{2}\Big)\,+\,\sum_{j=1}^{N-1}\Big[- t\,\big(c^{\dagger}_{j}c_{j+1}+c^{\dagger}_{j+1}c_{j}\big)\,+\,\Delta\,c_{j}c_{j+1}\,+\,\Delta^{*}\,c^{\dagger}_{j+1}c^{\dagger}_{j}\Big]\,,
\end{equation}
where $\mu$ represents the onsite energy, $t$ is the nearest-neighbor hopping amplitude and $\Delta$ is the superconducting pairing potential between nearest neighbors sites (assumed real for   now) and with lattice spacing $a$, see Fig.\,\ref{fig:kitaev}(a). 
The operator $c^{\dagger}_{j}(c_{j})$ creates (destroys) a spinless fermion at site $j$. Fermions obey the following anti-commutation relations
\begin{equation}
\{c_{i}^{\dagger},c_{j}^{\dagger}\}=\{c_{i},c_{j}\}\,=\,0\,,\quad \{c_{i}^{\dagger},c_{j}\}=c_{i}^{\dagger}c_{j}+c_{j}c_{i}^{\dagger}=\delta_{ij}\,.
\end{equation}
For a better understanding of how Majorana zero modes emerge, let us first consider the situation of a chain with open boundary conditions, and then write the Eq.\,(\ref{kiatev0}) in the Majorana basis. 

\begin{figure} 
   \includegraphics[width=1 \columnwidth]{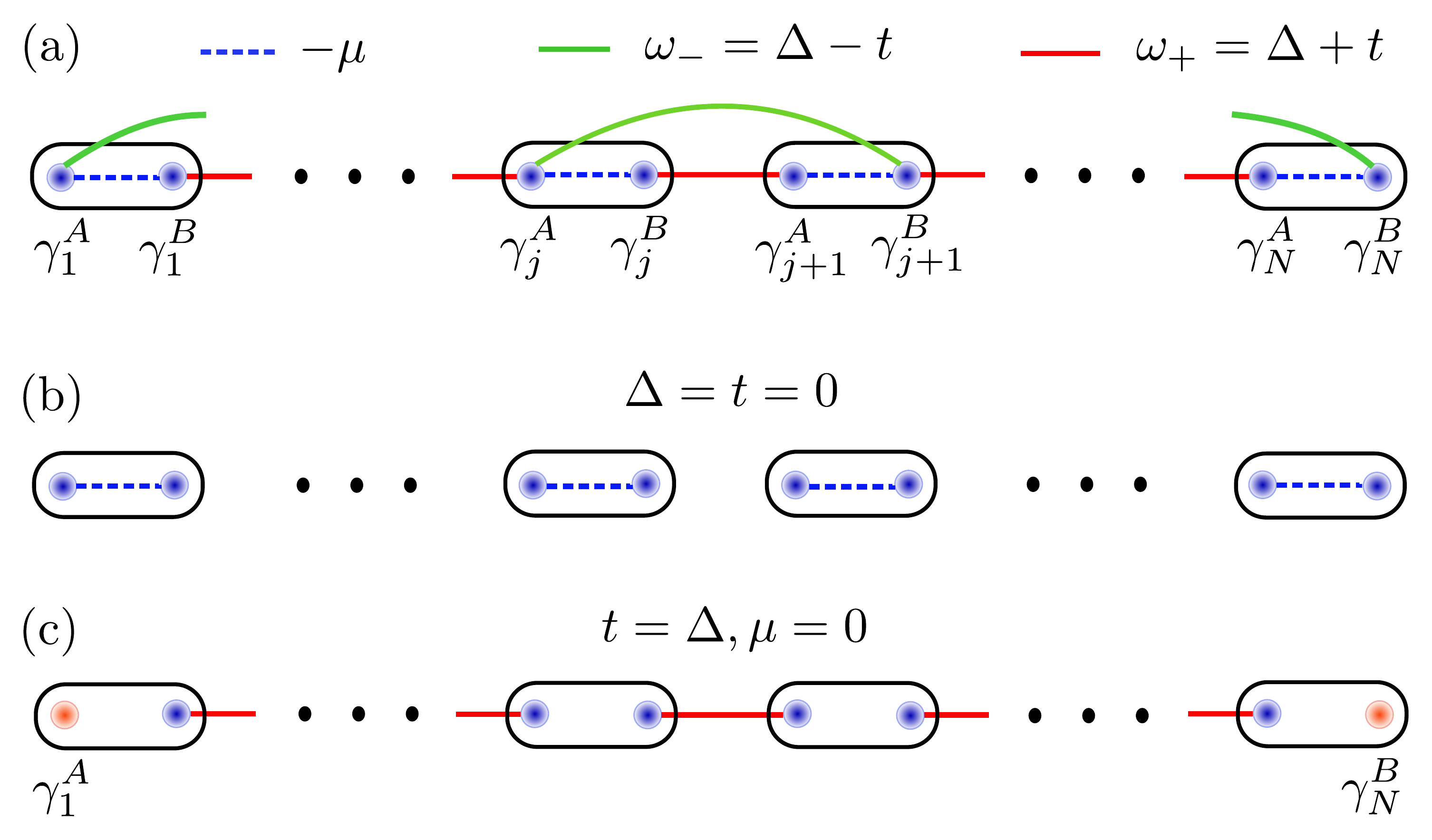} 
     \caption[Kitaev model in the Majorana basis and its trivial and topological phases]{(Color online) (a) Kitaev model in the Majorana basis. Representation of the N-site chain of spinless fermions in the Majorana basis, Eq.\,(\ref{kitaev0a}). Notice that $\mu$, $\omega_{-}$ and $\omega_{+}$ represent hopping amplitudes between Majorana fermions of the same fermionic site $j$, between the first Majorana of site j with the second Majorana of site $j+1$, and between the second Majorana of site $j$ with the first Majorana of site $j+1$, respectively. Kitaev model: two phases. (b) Trivial phase: Majorana operators of the same physical site are coupled to form a Dirac fermion, Eq.\,(\ref{kitaev0b}). (c) Topological phase: Majorana operators on adjacent lattice sites are coupled leaving two unpaired Majorana fermions at the end of the chain, Eq.\,(\ref{kitaev0c}). 
}
   \label{fig:kitaev2}
\end{figure}

As we have introduced in Sec.\,\ref{sec11}, any fermion operator can be defined in terms of two new operators, Eqs.\,(\ref{fermionMajoTrans}) and their inverse Eqs.\,(\ref{MajTrans}), known as Majorana operators, see Fig.\,\ref{fig:kitaev}(b). Although sometimes it does not lead to any novelty but rather it complicates the problem, we will show that in this situation such decomposition leads to interesting physics.
Therefore, the Hamiltonian given by Eq.\,(\ref{kiatev0}) in terms of these new operators reads (see Appendix \ref{AppA0} for more details on the derivation)
\begin{equation}
\label{kitaev0a}
H=-\frac{i\mu}{2}\sum_{j=1}^{N}\gamma^{A}_{j}\gamma^{B}_{j}+\frac{i}{2}\sum_{j=1}^{N-1}\Big[\omega_{+}\gamma^{B}_{j}\gamma^{A}_{j+1}+ \omega_{-}\gamma^{A}_{j}\gamma^{B}_{j+1}\Big]\,,
\end{equation}
where $\mu$, $\omega_{-}=\Delta-t$ and $\omega_{+}=\Delta+t$ represent hopping amplitudes between Majorana fermions of the same fermionic site $j$, between the first Majorana of site $j$ with the second Majorana of site $j+1$, and between the second Majorana of site $j$ with the first Majorana of site $j+1$, respectively. Fig.\,\ref{fig:kitaev2} shows the situation we have described here. 
The Hamiltonian given by Eq\,(\ref{kitaev0a}) exhibit different interesting properties depending on the values of the system parameters, $\mu$, $\Delta$ and $t$. Such properties belong to the emergence of different quantum phases.
Now, we illustrate the difference between the topological and trivial phases by looking at two special limits.

\textbf{The trivial phase:} For $t=\Delta=0$, the Hamiltonian given by Eq.\,(\ref{kitaev0a}) reads
\begin{equation}
\label{kitaev0b}
H=-\frac{i\mu}{2}\sum_{j=1}^{N}\gamma^{A}_{j}\gamma^{B}_{j}\,.
\end{equation}
Previous sum tells us that Majorana operators from the same physical site are paired together to form a fermion and the ground state is given by all fermion states empty ($\mu<0$) or occupied ($\mu>0$). See Fig.\,\ref{fig:kitaev2}(b). At this point nothing special happens and the system is topologically trivial.

\textbf{The topological phase:} On the other hand, an interesting situation happens when $\omega_{-}=0$ and $\mu=0$. Indeed, for the former, Eq.\,(\ref{kitaev0a}) acquires the following form,
\begin{equation}
\label{kitaev0c}
H=it\sum_{j=1}^{N-1}\gamma^{B}_{j}\gamma^{A}_{j+1}\,=\,i\Delta\Big[\gamma^{B}_{1}\gamma^{A}_{2}+\gamma^{B}_{2}\gamma^{A}_{3}+\cdots+\gamma^{B}_{N-1}\gamma^{A}_{N}\Big]\,.
\end{equation}
Remarkably, 
 $H$ does not contain operators $\gamma_{1}^{A}$ and $\gamma_{N}^{B}$ and only Majorana operators on adjacent lattice sites are coupled, as sketched in Fig.\,\ref{fig:kitaev2}(c). These operators that do not appear in the Hamiltonian represent zero-energy Majorana modes localized at the ends of the chain. Since such operators are not coupled to any Majorana operator, they commute with the Hamiltonian 
$[H,\gamma_{1}^{A}]=[H,\gamma_{N}^{B}]=0$.

Now, we can define a new set of fermionic operators that involve only Majorana operators included in the Hamiltonian given by 
Eq.\,(\ref{kitaev0a}),
\begin{equation}
\begin{split}
d_{j}&=\frac{1}{2}\Big(\gamma_{j}^{B}+i\gamma_{j+1}^{A} \Big)\,,\quad d_{j}^{\dagger}=\frac{1}{2}\Big(\gamma_{j}^{B}-i\gamma_{j+1}^{A} \Big)\,.
\end{split}
\end{equation}
Hence, in terms of these new operators, Eq.\,(\ref{kitaev0a}) reads
\begin{equation}
\label{fermioMan}
H=2t\sum_{j=1}^{N-1}\Big(d_{j}^{\dagger}d_{j}-\frac{1}{2} \Big)\,,
\end{equation}
which can be further written as
\begin{equation}
H=2t\sum_{j=1}^{N-1}d_{j}^{\dagger}d_{j}-t(N-1)\,,
\end{equation}
with ground state energy $E_{0}=-t(N-1)$ for $t>0$.

One realises that the two unpaired Majorana operators residing at the ends of the chain, $\gamma_{1}^{A}$ and $\gamma_{N}^{B}$, can be fused into an ordinary fermion operator, hence
\begin{equation}
\label{newopMaj}
f=\frac{1}{2}\Big(\gamma_{1}^{A}+i\gamma_{N}^{B}\Big)\,,\quad f^{\dagger}=\frac{1}{2}\Big(\gamma_{1}^{A}-i\gamma_{N}^{B}\Big)\,.
\end{equation} 
This fermion is delocalised with contributions from both ends of the chain, and since it is absent in Eq.\,(\ref{fermioMan}) to occupy its quasiparticle state requires zero energy. 
One can check that the space formed by all the ground states $\ket{\psi}$ of $H$ can be written using the operators $f$ and $f^{\dagger}$ defined above. The occupation number operator is  $n=f^{\dagger}f=(1+i\gamma_{1}^{A}\gamma_{N}^{B})/2$ and can be used to label the ground state, which exhibits a two-fold degeneracy arising from the two possible occupancies $n=0,1$ of the ordinary fermion state $f$. The two ground states indeed are $\ket{0}$, which satisfies $f\ket{0}=0$ and the other ground state can be defined as $\ket{1}=f^{\dagger}\ket{0}$. One notices that $\braket{1}{0}=0$, implying that $\ket{1}$ and $\ket{0}$ are two different ground states and that indeed the ground state degeneracy of the Kitaev's model in the topological regime is two-fold. 

The parity state defines a degenerate two-level system, a qubit, so that it can be used to encode quantum information \cite{kitaev}.
 Since the definition of such number operator state takes into account the zero energy fermion operator made of two Majorana operators at the end of the chain, the fermion state cannot be measured with any local measurement on one of the bound states at the end of the chain: such parity state can be accessed by a joint measurement of the two Majoranas. This is the reason why it is considered that the information in such a quit is stored non-locally \cite{kitaev}.

Up to this part, we have discussed the two distinct phases that exhibits the Kitaev's model: a trivial phase and the topological phase with unpaired Majorana zero modes located at the end of the chain. 
Now, in order to investigate the properties of the superconducting bulk without perturbations from the ends of the chain and to fully describe the two phases we have studied previously, we assume that the chain forms a closed loop with periodic boundary conditions. This requires to add an extra term in the Hamiltonian given by Eq.\,(\ref{kiatev0})
\begin{equation}
-t\Big[c_{N}^{\dagger}c_{N+1}+c_{N+1}^{\dagger}c_{N}\Big]+\Delta c_{N}c_{N+1}+\Delta^{*}c_{N+1}^{\dagger}c_{N}^{\dagger}\,.
\end{equation}
Hence, the Hamiltonian given by Eq.\,(\ref{kiatev0}), reads
\begin{equation}
\label{kiatev0x}
H=-\mu\sum_{j=1}^{N}\Big(c^{\dagger}_{j}c_{j}-\frac{1}{2}\Big)\,+\,\sum_{j=1}^{N}\Big[- t\,\big(c^{\dagger}_{j}c_{j+1}+c^{\dagger}_{j+1}c_{j}\big)\,+\,\Delta\,c_{j}c_{j+1}\,+\,\Delta^{*}\,c^{\dagger}_{j+1}c^{\dagger}_{j}\Big]\,.
\end{equation}
Notice that the addition of the extra term implies that the second sum runs till site $N$ and allows interactions between sites $N$ and $N+1$. The consideration of periodic boundary conditions is achieved by requiring that sites $1$ and $N+1$ correspond to the same site, so that  $c_{1}=c_{N+1}$ and $c_{1}^{\dagger}=c_{N+1}^{\dagger}$. Indeed, this is the closed loop assumption we have made. It is thus appropriate to consider the Hamiltonian given by Eq.\,(\ref{kiatev0x}) in momentum space $k$; and since electrons and holes are involved in the problem, it is useful to rewrite the momentum Hamiltonian in Nambu space, see App.\,\ref{AppA0} for more details,
\begin{equation}
H=\frac{1}{2}\sum_{k}\psi_{k}^{\dagger}H_{BdG}\psi_{k}+\frac{1}{2}\sum_{k}(-2t\cos ka)\,,\quad 
\psi_{k}\,=\,\begin{pmatrix}
c_{k}\\
c_{-k}^{\dagger}
 \end{pmatrix}\,,
\end{equation}
where $\psi_{k}$ defines a Nambu operator and $H_{BdG}$ is the so-called Bogoliubov-de Gennes Hamiltonian,
\begin{equation}
H_{BdG}=\xi_{k}\tau_{z}+\Delta_{k}\tau_{y}=\mathbf{h}\cdot\mathbf{\tau}\,,
\end{equation}
where $\mathbf{h}=(0,\Delta_{k},\xi_{k})$, $\Delta_{k}=-2\Delta\sin ka$, $\xi_{k}=-\mu-2t\cos ka$, and $\mathbf{\tau}=(\tau_{x},\tau_{y},\tau_{z})$.
The energy spectrum of $H_{BdG}$ is then given by
\begin{equation}
\label{quasipKiatev}
E_{k,\pm}=\pm\sqrt{(\mu+2t\cos ka)^{2}+4\Delta^{2}\sin^{2}ka}\,.
\end{equation}
For $\Delta\neq0$, the energy gap closes when both elements inside the square root  vanish simultaneously: $\sin ka=0$ and $\mu+2t\cos ka=0$. The former vanishes at isolated points $k=0$ and $k=\pi/a$. Hence, for $k=0$, $\mu=-2t$, while for $k=\pi/a$, $\mu=2t$. 
Previous can be understood by considering that the superconducting pairing $\Delta_{k}$ has $p$-wave nature, being an odd function in $k$, thereby Cooper pairs are prohibited to form at $k=0$ or $k=\pi/a$. This closing of the energy gap represent a special kind of phase transition and it is called topological phase transition for reasons that will be understood below.
The two lines $\mu=\pm 2t$ separate two gapped superconducting phases, $|\mu|<|2t|$ and $|\mu|>|2t|$, that are connected only by making zero the gap of the energy spectrum. 
\begin{figure} 
  \begin{center} 
  \includegraphics[width=.9\columnwidth]{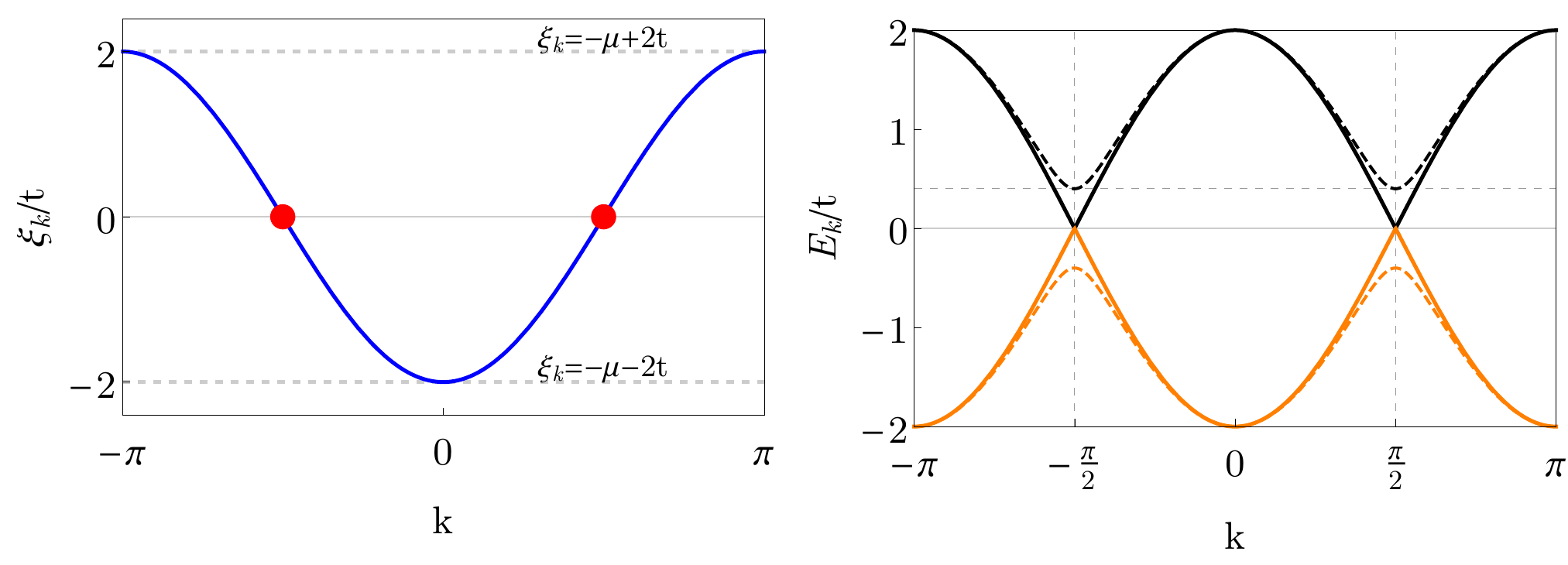} 
     \caption[ Energy dispersion of the normal system in the Kitaev's model]{(Color online) (left) 
   Energy dispersion of the normal system in the Kitaev's model $\xi_{k}=-\mu-2t\cos ka$ with $a=1$.
     The two horizontal dashed lines represent a closure of the energy gap when the chemical potential is tuned to $\mu=-2t$ for $k=0$ and $\mu=2t$ for $k=\pm \pi$. Within the region $-2t<\mu<2t$ one finds an odd number of pairs of Fermi points (red filled circles) for $-\pi<k<\pi$, where the topological invariant $\nu=-1$ leads to a clear signature of the topological superconducting phase with Majorana zero modes located at the ends of the chain. Outside of such region, the system is in the non-topological phase.
      (right) Quasiparticle excitation spectrum given by Eq.\,(\ref{quasipKiatev}) with $\mu=0$, where $\Delta=0$ (solid curves) and $\Delta=0.2 t$ (dashed curves). Notice that at $k=\pm\pi/2$ the superconducting pairing opens a gap $2\Delta$ (dashed horizontal line). 
}
   \label{fig:kitaev4}
\end{center}
\end{figure}

When $\Delta$ is much smaller than the relevant energy scales in the problem, these two phases can be distinguished by means of topological invariants as briefly discussed in the introduction. In the case of one-dimensional superconductors such topological invariant is the Majorana number \cite{kitaev} $M=(-1)^{\nu}$, where $\nu$ represents the number of pairs of Fermi points in the Brillouin zone of the normal system ($\Delta=0$). Remarkably, when restoring superconductivity ($\Delta\neq0$) the system becomes a topological superconductor with Majorana zero modes located at the ends of the system: odd number of pairs of Fermi points indicates the emergence of the topological phase, while even of the trivial one. 
Notice that this topological invariant cannot be changed in a continuous way without closing the gap of the spectrum as discussed above.
For $|\mu|<|2t|$, the number of pairs of Fermi points is odd and therefore the Majorana number is $M=-1$. See Fig.\,\ref{fig:kitaev4}. It represents the existence of Majorana zero modes and confirms the discussion we have made in the geometry with open boundary conditions. On the other hand $|\mu|>|2t|$ is a trivial phase. We conclude this part by pointing out that along this thesis we will refer to these Majorana zero modes bounded at the ends of the chain as to \emph{Majorana bound states} (MBSs).

In general, for a small but non-zero $\mu$, the Majorana bound states are not really localise at the ends of the wire, but their wave-functions exhibit an exponential decay into the bulk of the wire, ${\rm e}^{-x/\xi}$, see Eq.\,(\ref{Majzeromode2}). The non-zero spatial overlap of the two Majorana wave-functions results in a non-zero energy splitting between the two Majorana states. For long  enough wires the splitting is very small that the two Majorana states can be considered to be degenerate. 
Moreover, the Majoranas can also split when the higher-energy states in the bulk
come very close to zero energy, hence the Majorana modes are protected as long as the bulk energy gap is finite. 
This follows from the particle-hole symmetry involved in the problem, where the spectrum has to be symmetric around zero energy. Therefore, trying to move the Majorana zero modes 
from zero energy individually is impossible, as it would violate particle-hole symmetry.

Therefore, we conclude that Majorana bound states at the end of the Kitaev's chain are protected by electron-hole symmetry, and by the absence of zero-energy excitations in the bulk of the wire, and not by fine tuning of the model's parameters.

\newpage
\section{Physical realization based on nanowires with Rashba spin-orbit coupling}
\label{Rashbawire}
\begin{figure} 
 \begin{center}
 \includegraphics[width=.8 \columnwidth]{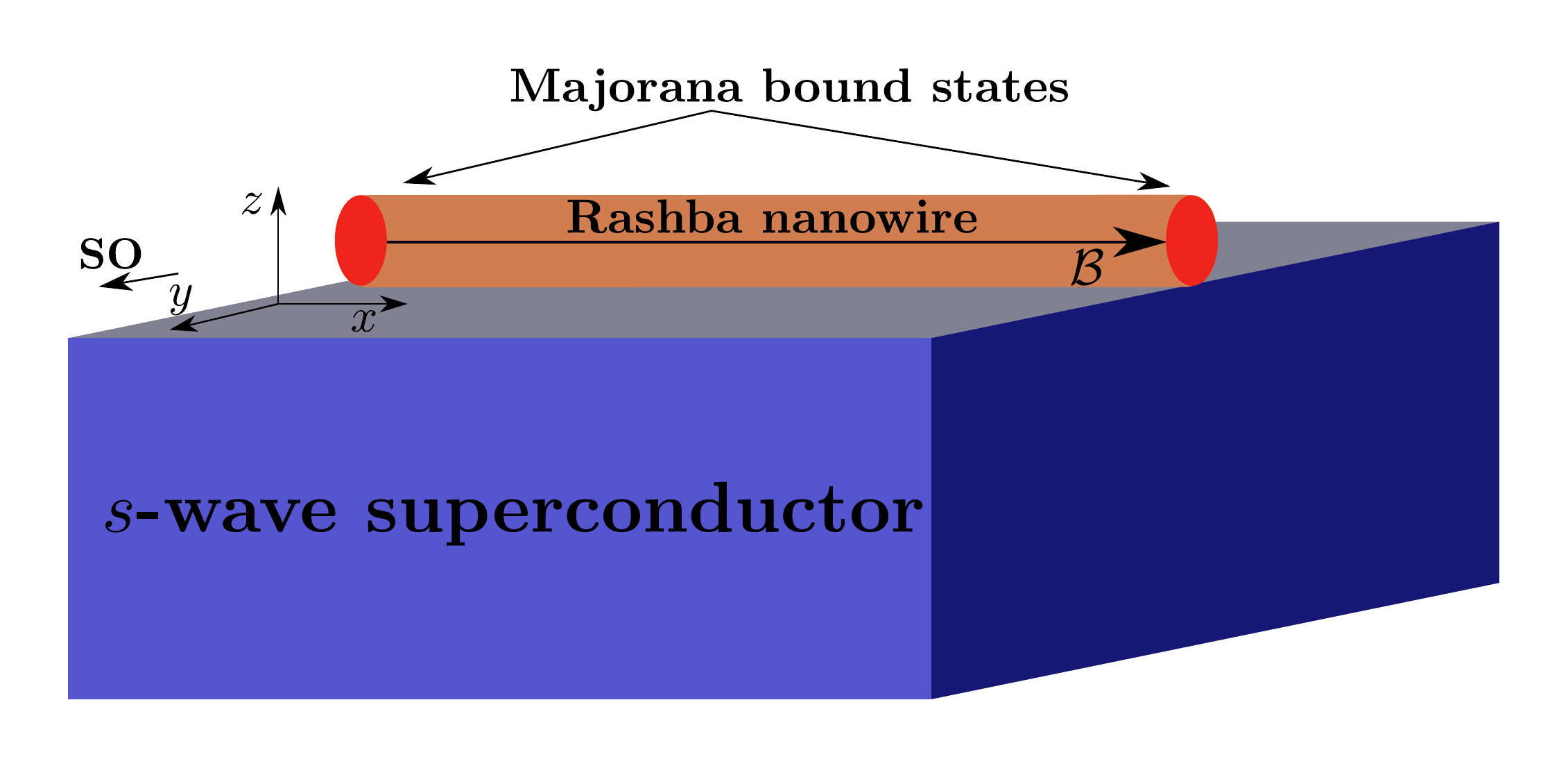} 
     \caption[Sketch of a 1D semiconducting nanowire placed on a s-wave superconductor]{(Color online) Sketch of a 1D semiconducting Rashba nanowire placed on a s-wave superconductor. A magnetic field $\mathcal{B}$ is applied along the wire $x$-axis, which is perpendicular to the spin-orbit field. These ingredients represent a solid platform for investigating the topological superconducting phase and therefore Majorana bound states, red filled circles, in 1D condensed matter systems.
}
   \label{fig:RashbaNWproposal}
   \end{center}
\end{figure}

A promising physical realisation for engineering topological superconductivity in one dimension based on the Kitaev's model is schematically shown in Fig.\, \ref{fig:RashbaNWproposal}. It involves one-dimensional semiconducting nanowires with strong spin-orbit coupling (SOC) (such as InSb, InAs), where conventional $s$-wave superconductivity is induced by proximity effect and an external Zeeman field, perpendicular to the spin-orbit axis, drives the system into the topological superconducting phase with Majorana bound states at the end of the wire \cite{PhysRevLett.105.177002,Lutchyn:PRL10}. Along this thesis, we will refer to a nanowire with Rashba SOC as to Rashba nanowire.
The idea is such that the normal system, with zero superconductivity, can support an odd number of pairs of Fermi points for a given chemical potential, similar to the normal system in the Kitaev's proposal, and therefore when switching on superconductivity a topological superconducting phase emerges hosting Majorana bound states at each end of the wire. 
To understand how the topological phase arises in this system, it is worth to firstly describe the normal state of the nanowire. 
Before going further we point out that for investigating topological superconductivity one really needs a spin texture an $s$-wave superconductivity. For the proposal we discuss along this thesis, the interplay of Rashba spin-orbit coupling and Zeeman interaction gives rise to the spin-texture, however, other systems as chains of magnetic atoms \cite{PhysRevB.84.195442,PhysRevLett.111.186805,PhysRevB.88.020407,PhysRevLett.111.147202,PhysRevB.88.155420} or topological insulators \cite{Fu:PRL08,Fu:PRB09,Hasan:RMP10,Qi:RMP11} can also lead to similar conclusions. The calculations we present along this chapter and in the rest of this thesis correspond to typical  parameters for InSb. For reviews see for instance \cite{Alicea:RPP12,Beenakker:11,StanescuModel13,RevModPhys.87.137,RevModPhys.87.1037}.
\subsection*{Nanowire with Rashba spin-orbit coupling and Zeeman interaction}
We consider a single channel nanowire in one-dimension\footnote{The radius of the wire
is small compared to the Fermi wavelength and there is a single 1D occupied mode.} with SOC and Zeeman interaction, whose Hamiltonian is given by 
\begin{equation}
\label{H0}
\mathcal{H}_{0}\,=\,\mathcal{H}_{kin}\,+\,\mathcal{H}_{SOC}\,+\,\mathcal{H}_{Z}\,,
\end{equation}
where the first, second and third terms are the kinetic, spin-orbit coupling and Zeeman Hamiltonians, respectively. We refer to $\mathcal{H}_{0}$ as to the for the normal system Hamiltonian to distinguish it from the superconducting one to be described next. The kinetic term reads
\begin{equation}
\label{kinH}
\begin{split}
\mathcal{H}_{kin}\,&=\,\sum_{\sigma}\int\,dx\,\psi_{\sigma}^{\dagger}(x)\left[ -\frac{\hbar^{2}\partial_{x}^{2}}{2m}-\mu\right]\psi_{\sigma}(x)\,,\\
\end{split}
\end{equation} 
where $m$ is the effective electron's mass in the nanowire and $\mu$ the chemical potential, which determines the filling of the nanowire. The spin-orbit coupling Hamiltonian is described by
\begin{equation}
\label{soH}
\begin{split}
\mathcal{H}_{SOC}&\,=\,\sum_{\sigma \sigma^{\prime}}\int dx\,\psi_{\sigma}^{\dagger}(x)\left[ \boldsymbol{\alpha}\cdot\boldsymbol{\sigma}\right]_{\sigma\sigma^{\prime}}\,\left[ -i\hbar\partial_{x}\right]\psi_{\sigma^{\prime}}(x)\,,\\
\end{split}
\end{equation}
where the spin direction is such that $\boldsymbol{\alpha}\cdot\boldsymbol{\sigma}=-\alpha_{R}\sigma_{y}/\hbar$, $\boldsymbol{\sigma}=(\sigma_{x},\sigma_{y},\sigma_{z})$ is the vector of Pauli matrices,
$\sigma=\uparrow,\downarrow$ denotes the spin direction along the $y$-axis, and $\alpha_{R}$ represents the strength of Rashba spin-orbit coupling.
The Zeeman Hamiltonian associated to the magnetic field $\mathcal{B}$ along the the nanowire $x$-axis, perpendicular to the spin-orbit axis,
\begin{equation}
\label{B1Hamil}
\begin{split}
\mathcal{H}_{Z}\,&=\,B\sum_{\sigma\sigma^{\prime}}\int dx\,\psi^{\dagger}_{\sigma}(x)\left[\sigma_{x}\right]_{\sigma\sigma^{\prime}}\psi_{\sigma^{\prime}}(x)\,,\\
\end{split}
\end{equation}
where $B=g\mu_{B}\mathcal{B}/2$ is the Zeeman energy, $\mathcal{B}$ is the applied magnetic field, $g$ is the wire's $g$-factor and $\mu_{B}$ the Bohr magneton. In the previous Hamiltonians, $\psi_{\sigma}$ represents the annihilation operator of an electron at position $x$ with spin $\sigma=\uparrow,\downarrow$. The fact that we require the Zeeman and the spin-orbit axes to be perpendicular will become clear later. For a discussion on the effects of a parallel field see Appendix \ref{appRashba}.
It is appropriate to introduce the Hamiltonian density $H_{0}$
\begin{equation}
\label{H0Rashba}
\mathcal{H}_{0}=\int dx\,\psi^{\dagger}(x)\,H_{0}\,\psi(x)\,,\quad 
\text{in the basis:}\,\quad \psi=\begin{pmatrix}
\psi_{\uparrow}\\
\psi_{\downarrow}
 \end{pmatrix}\,,
\end{equation}
where the Hamiltonian density reads
\begin{equation}
\label{H0Hamil}
H_{0}=\frac{p^{2}_{x}}{2m}-\mu-\frac{\alpha_{R}}{\hbar}\sigma_{y}p_{x}+B\sigma_{x}\,,
\end{equation}
and $p_{x}=-i\hbar\partial_{x}$ is the momentum operator. Typical values for InSb nanowires \cite{Mourik:S12} include the electron's effective mass $m=0.015m_{e}$ and the spin-orbit strength $\alpha_{R}=20$\,meVnm. Mostly along this thesis we will refer to Hamiltonian densities denoted by $H$ simply as to Hamiltonians without loss of generality.

The eigenvalues and eigenvectors of previous Hamiltonian, $H_{0}$, are found by solving the Schrodinger equation $H_{0}\Psi=E\Psi$. In terms of plane waves $\Psi_{k}(r)=\phi\,{\rm e}^{ikr}$, see Appendix \ref{appRashba} for details on the derivation, we get
\begin{equation}
\label{H0vEnVec}
\varepsilon_{k,\pm}=\xi_{k}\pm\sqrt{B^{2}+\alpha_{R}^{2}k^{2}}\,,\quad \Psi_{k,\pm}(r)=\phi_{\pm}(k)\,\frac{1}{\sqrt{L}}{\rm e}^{ikr}\,,\quad  \phi_{\pm}(k)=\frac{1}{\sqrt{2}}\begin{pmatrix}
\pm\gamma_{k}\\
1
 \end{pmatrix}\,,
\end{equation}
where $\xi_{k}=\frac{\hbar^{2}k^{2}}{2m}-\mu$, and $\gamma_{k}=\frac{(i\alpha_{R} k+B)}{\sqrt{B^{2}+\alpha_{R}^{2}k^{2}}}$.
\begin{figure} 
 \begin{center}
\includegraphics[width=.9 \columnwidth]{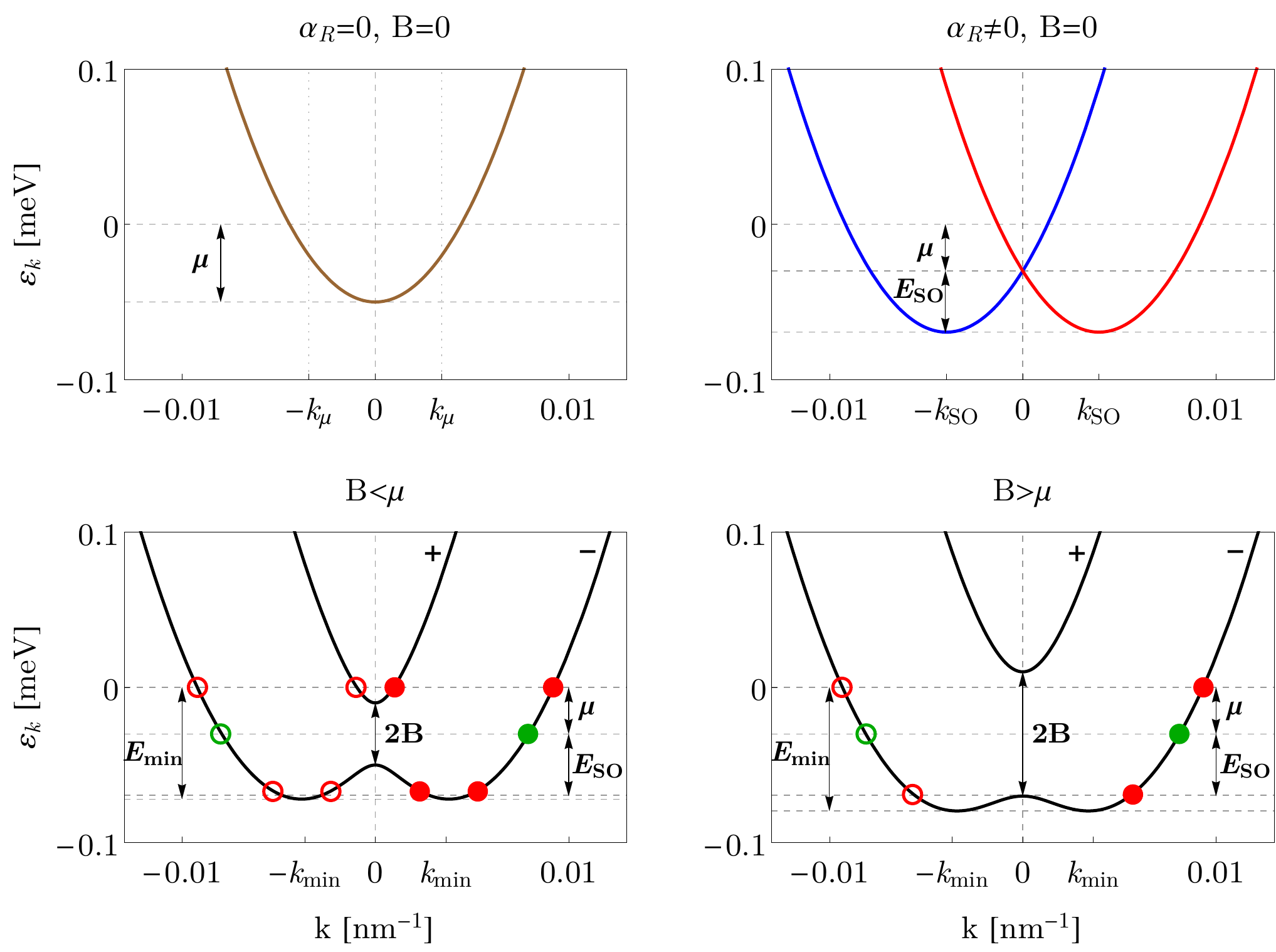} 
     \caption[Energy dispersion of a Rashba Nanowire]{(Color online) (left) 
  Energy dispersion for a Rashba Nanowire given by Eq.\,(\ref{H0vEnVec}). Top left panel at $\alpha_{R}=0$, $B=0$: the energy spectrum consists of two degenerate parabolas. Top right panel at $\alpha_{R}\neq0$, $B=0$: the parabolas are shifted by $k_{SO}$  and cross at $k=0$. 
  Bottom panels: any Zeeman field perpendicular to the SO axis, $B$, opens a gap of $2B$ at zero momentum $k=0$ lifting the spin degeneracy.
     When the chemical potential is tuned inside the Zeeman gap, the system supports an odd number of pairs of Fermi points (red filled circles). 
   Within the Zeeman gap, notice that for $B>\mu$ the system possesses two Fermi points per energy, which correspond to counter propagating states with different spins, and therefore this gap is also called helical gap.
      The spectrum in this case has local and global extremes at $k=0$ and $\pm k_{min}=\pm\sqrt{k_{SO}^{2}-k_{Z}^{4}/4k_{SO}^{2}}$. For the former, the corresponding energies are $-\mu\pm B$, while for the latter $E_{min}=-\mu-E_{SO}-B^{2}/4E_{SO}$. Parameters considered here are correspond to InSb nanowire: $\alpha_{R}=20$\,meVnm, $B=0.02$\,meV for bottom left and $B=0.04$\,meV for bottom right, $\mu=0.03$
}
   \label{fig:RashbaNW}
    \end{center}
\end{figure}

In Fig.\,\ref{fig:RashbaNW}, we present the energy dispersion for the nanowire with Rashba SOC and Zeeman interaction, which is 
given by Eq.\,(\ref{H0vEnVec}). The top left panel of Fig.\,\ref{fig:RashbaNW} shows the energy dispersion of the free electron Hamiltonian that consists of two superimposed parabolas, one for each spin. The value of the chemical potential $\mu$ is measured from the bottom of the band and determines the filling of the nanowire. The spin-orbit coupling shift the two parabolas, which cross at zero momentum $k=0$, by momenta $\pm k_{SO}=\pm m\alpha_{R}/\hbar^{2}$ and by energy $E_{SO}=m\alpha_{R}^{2}/2\hbar^{2}$, see top right panel in Fig.\,\ref{fig:RashbaNW}. The spin of the two electronic bands is aligned along $\pm y$. 
Up to this part, one can clearly notices that there is no possibility to mimic the topological phase when superconductivity is applied in a similar way as it was done in the Kitaev's approach, since for any $\mu$ there are two pairs of Fermi points which corresponds to an even number and thus to a trivial phase.
An external magnetic field, if perpendicular to the spin-orbit field, solves this issue (bottom panels).
Indeed, a Zeeman magnetic field $\mathcal{B}$ lifts the spin degeneracy at $k=0$ by removing the level crossing and opens a gap in the spectrum of $2B$ at zero momentum $k=0$. 
When the chemical potential $\mu$ is tuned to be inside the gap opened by the Zeeman field $|\mu|<B$, the system hosts an odd number of pairs of Fermi points (two Fermi points, red filled circles), thus only the lowest band is partially occupied and the nanowire behaves as spinless.
Therefore, the system can reach the topological superconducting phase by placing the nanowire on a $s$-wave superconductor. 
When the chemical potential lies within this anti-crossing gap, the system has two Fermi points, as opposed to four Fermi points for above 
or below this gap. This window is a \emph{helical gap}, since the two fermi points correspond to counter propagating states with different spins (the spin projection is locked
to momentum) and hence the name helical. Therefore, a nanowire with $B>\mu$ is helical and along this thesis we will refer to it as to \emph{helical nanowire}. 
As mentioned, when the chemical potential lies within this helical gap, the wire is spin polarised and it appears spinless. Turning on a $s$-wave pairing weakly compared to $B$, then effectively $p$-wave pairs states and therefore drives the system into the topological phase.
We will see that the helical regime plays an important role in normal transport in hybrid junctions (see Chap.\,\ref{Chap2}) and it is crucial towards the emergence of the topological superconducting phase as we will see later in this thesis.

Electrons in the semiconducting nanowire feel an effective superconducting pairing potential as a result of the so-called \emph{proximity effect} \cite{RevModPhys.36.225,Doh272}. 
To occur such effect, a good interface between the wire and superconductor should be made, so that electrons can tunnel between these two systems.
Within the BCS theory, $s$-wave superconducting pairing couples states with opposite 
momenta $k$ and spin and can be described by a phenomenological or reduced Hamiltonian
\begin{equation}
\label{hsc}
\mathcal{H}_{sc}=\int dx\Big[\mathbf{\Delta}\psi_{\uparrow}^{\dagger}(k)\psi_{\downarrow}^{\dagger}(-k)+\mathbf{\Delta}^{\dagger}\psi_{\downarrow}(-k)\psi_{\uparrow}(k)\Big]\,,
\end{equation}
where $\mathbf{\Delta}={\rm e}^{i\varphi}\Delta$ is the pairing potential, which is complex in general but for now we will consider it to be real, and $\varphi$ 
is the superconducting phase. 
The full system is now described by the sum of Eqs.\,(\ref{H0Rashba}) and \,(\ref{hsc}), 
\begin{equation}
\label{fullHamiltonian}
\mathcal{H}=\mathcal{H}_{0}+\mathcal{H}_{sc}=\mathcal{H}_{kin}+\mathcal{H}_{SOC}+\mathcal{H}_{Z}+\mathcal{H}_{sc}\,.
\end{equation}

Again, when dealing with superconducting systems, it is appropriate to make use of the Bogoliubov formalism described in Sec.\,\ref{bdgformalism}. Therefore,
the full Hamiltonian, $\mathcal{H}$, can be then written in Nambu space by defining new spinors
$\Psi_{k}$, so that 
\begin{equation}
\mathcal{H}\,=\,\frac{1}{2}\int dx\,\Psi^{\dagger}_{k}H_{BdG}\,\Psi_{k}\,,\quad\quad
 \Psi_{k}=
\begin{pmatrix}
\psi_{\uparrow}(k),
\psi_{\downarrow}(k),
\psi_{\uparrow}^{\dagger}(-k),
\psi_{\downarrow}^{\dagger}(-k))
\end{pmatrix}^{\dagger}
\end{equation}
where, 
\begin{equation}
\label{BdGqq}
\begin{split}
H_{BdG}&\,=\,\left(-\hbar^{2}\partial^{2}_{x}/2m-\mu\right)\tau_{z}\sigma_{0}\,-\,i\alpha_{R}\tau_{z}\sigma_{y}\,\partial_{x}+\,B\tau_{z}\sigma_{x}\,+\,\Delta\tau_{y}\sigma_{y}\,,
\end{split}
\end{equation}
is the Bogoliubov-de Gennes Hamiltonian. The Pauli matrices $\sigma$ and $\tau$  act in spin and electron-hole subspaces, respectively.  
The energy spectrum of Hamiltonian (\ref{BdGqq}) is then given by
\begin{equation}
\label{BandsNambu1}
E^{2}_{k,\pm}=\xi_{k}^{2}+(\alpha_{R} k)^{2}+B^{2}+\Delta^{2}\pm2\sqrt{B^{2}\Delta^{2}+\left[B^{2}+(\alpha_{R}k)^{2}\right]\xi_{k}^{2}}\,,
\end{equation}
where $\xi_{k}=\hbar^{2}k^{2}/2m-\mu$ is the free electron energy dispersion. 

\begin{figure} 
 \begin{center}
\includegraphics[width=.8 \columnwidth]{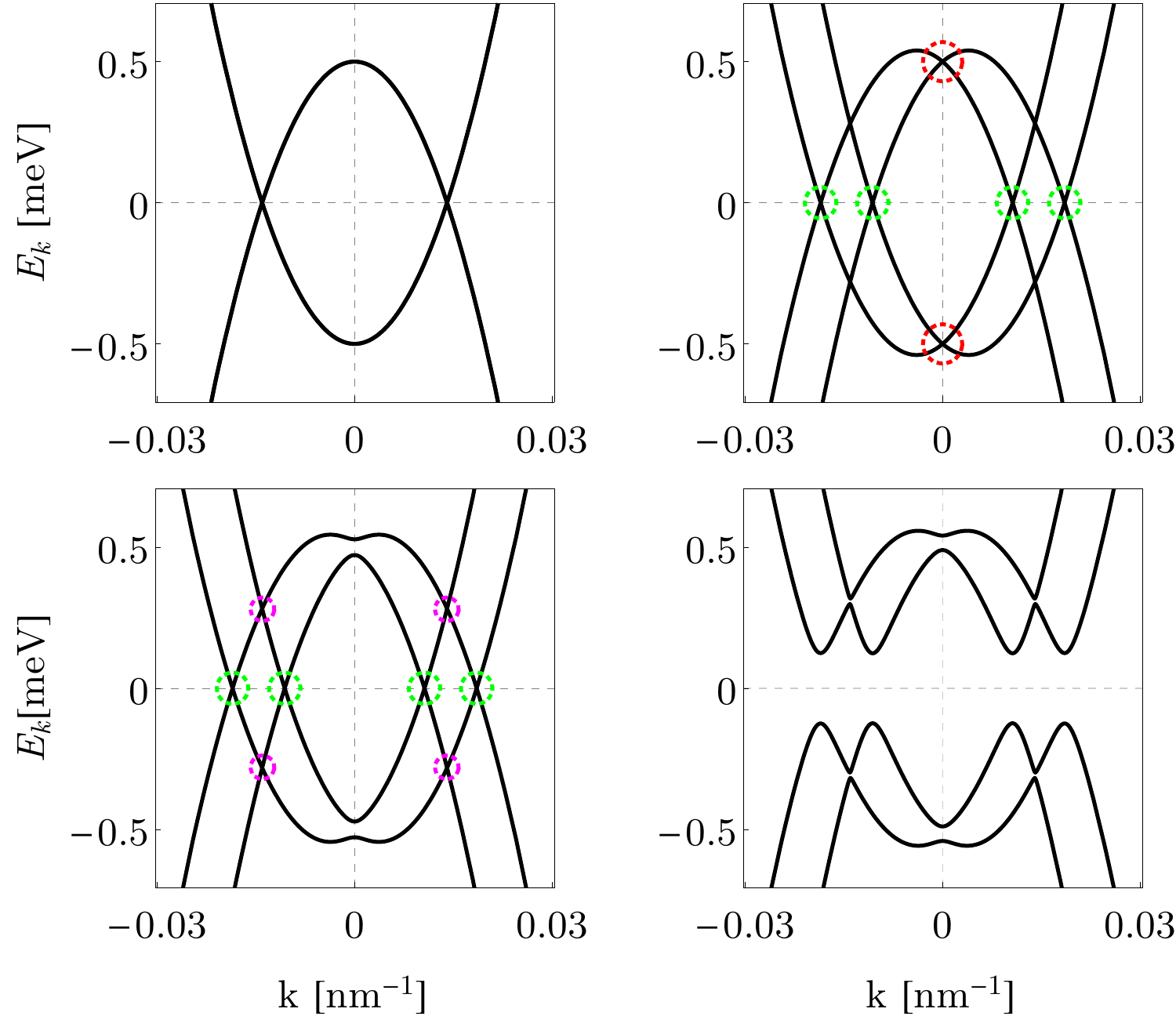} 
     \caption[Nambu energy dispersion of a Rashba Nanowire]{(Color online)
Nambu energy spectrum for a Rashba Nanowire given by Eq.\,(\ref{BandsNambu1}). Top row: (left ) Free electrons have a parabolic dispersion, while for holes it is an inverted parabola, (right) spin-orbit coupling shifts the bands, which cross at $k=0$. Bottom row: (left) the Zeeman field opens a gap, red circles,$\alpha_{R}\neq0$ and $B\neq0$. The green circles mark the four Fermi momenta points, given by Eqs.\,(\ref{fermpoints}). (right) The superconducting pairing opens gaps at the Fermi momenta (green dashed circles) and also at finite energy (magenta dashed circles). Parameters: $\alpha_{R}=20$\,meVnm, $B=0.03$\,meV, $\mu=0.5$\,meV.
}
   \label{fig:RashbaD0NW}
   \end{center}
\end{figure}

First, we analyse the role of superconductivity in the energy spectrum given by previous equation without analysing the emergence of the topological phase, see Fig.\,\ref{fig:RashbaD0NW}. The band spectrum, additionally to the normal regime, contains an inverted parabola (left-top panel) that corresponds to the kinetic Hamiltonian of a hole due to the Nambu description. Introducing spin-orbit, (right-top panel), such parabolas split and cross at $k=0$, while the Zeeman field opens a gap of $2B$ at $k=0$, as shown in Fig.\,\ref{fig:RashbaNW}.
(bottom panels) The superconducting pairing $\Delta$ modifies the energy spectrum and it opens a gap at each Fermi point encircled with dashed green circles (bottom row). Moreover, notice that the superconducting pairing opens gaps mixing different bands at finite energy (Magenta dashed circles). In the following we discuss in more detail the dependence of the gaps at $k_{F}$ on the Zeeman field.

Eq.\,(\ref{BandsNambu1}) allows us to investigate the evolution of the energy spectrum with the Zeeman field and the emergence of MBSs.
It is, however, instructive to write down the Hamiltonians given by Eqs.\,(\ref{H0Rashba}) and (\ref{hsc}) in the basis constructed from Eqs.\,(\ref{H0vEnVec}) as follows \cite{Alicea:PRB10,Alicea:RPP12},
 \begin{equation}
 \label{eq111}
 \psi(k)=\phi_{-}(k)\psi_{-}(k)+\phi_{+}(k)\psi_{+}(k)\,,
 \end{equation}
 where $\psi_{\pm}$ are operators that annihilates states in the upper/lower bands at momentum $k$ with energy $\varepsilon_{k,\pm}$ and $\phi_{\pm}$ 
 the respective normalized wave-functions calculated previously and given by Eq.\,(\ref{H0vEnVec}). 
 From here on we will refer to this basis as to \emph{helical}.
 We can decompose previous equations into the two spinor components, for details see Appendix \ref{appRashba}, thus
 \begin{equation}
 \label{basis1}
   \psi_{\uparrow}(k)=\frac{1}{\sqrt{2}}\left[-\gamma_{k}\psi_{-}(k)+\gamma_{k}\psi_{+}(k)\right]\,,\quad   \psi_{\downarrow}(k)=\frac{1}{\sqrt{2}}\left[\psi_{-}(k)+\psi_{+}(k)\right]\,.
 \end{equation}
In this basis, the Hamiltonian $\mathcal{H}_{0}$ is diagonal, see Appendix \ref{appRashba}.
By introducing Eqs.\,(\ref{basis1}) into Eqs.\,(\ref{H0Rashba}) and (\ref{hsc}), we get  
for the full Hamiltonian in this new basis, see Appendix \ref{appRashba},
\begin{equation}
\label{fullHhelical}
\begin{split}
\mathcal{H}&=
\int \frac{dk}{2\pi}
\big[
\varepsilon_{k,+}\psi_{+}^{\dagger}(k)\psi_{+}(k)
+
\varepsilon_{k,-}\psi_{-}^{\dagger}(k)\psi_{-}(k)
\big]\\
&+
\bigg[\frac{\Delta_{--}(k)}{2}\psi_{-}^{\dagger}(k)\psi_{-}^{\dagger}(-k)
+\frac{\Delta_{++}(k)}{2}\psi_{+}^{\dagger}(k)\psi_{+}^{\dagger}(-k)
+\Delta_{+-}(k)\psi_{+}^{\dagger}(k)\psi_{-}^{\dagger}(-k)+h.c
\bigg],
\end{split}
\end{equation}
where 
\begin{equation}
\label{pairings0}
\Delta_{--,++}(k)=\frac{\pm i\alpha_{R} k \Delta}{\sqrt{B^{2}+\alpha_{R}^{2}k^{2}}}\,,\quad \Delta_{+-}(k)=\frac{B \Delta}{\sqrt{B^{2}+\alpha_{R}^{2}k^{2}}}\,,
\end{equation}
represent the different pairing functions that arise in our nanowire due to the interplay of Rashba SOC and Zeeman interaction when placed on a $s$-wave superconductor. 
The first line of Eq.\,(\ref{fullHhelical}) is just the normal Rashba nanowire Hamiltonian, while in the second line, the first and second terms associated to $\Delta_{--,++}(k)$, describe pairing between states of the same $\mp$ band, while the third term associated to $\Delta_{+-}(k)$, represents pairing between states of different band. 
Moreover, it is important to notice that $\Delta_{--,++}(k)$ are odd functions of momentum $k$, while $\Delta_{+-}(k)$ is even. This implies that $\Delta_{+-}$ is an interband $s$-wave pairing, while $\Delta_{--,++}$ is an intraband $p$-wave pairing.
 Eq.\,(\ref{fullHhelical}) can be written in Nambu space, see Appendix \ref{fullhSC0},
 \begin{equation}
 \mathcal{H}=\frac{1}{2}\int \frac{dk}{2\pi}\Psi^{\dagger}(k)\,H_{BdG}\,\Psi(k)\,, \quad \Psi(k)=\begin{pmatrix}
\psi_{+}^{\dagger}(k),
\psi_{-}^{\dagger}(k),
\psi_{+}(-k),
\psi_{-}(-k)
\end{pmatrix}^{\dagger}
 \end{equation}
 where the Bogoliubov-de Gennes Hamiltonian reads
 \begin{equation}
 H_{BdG}=
 \begin{pmatrix}
 \varepsilon_{k,+}&0&\Delta_{++}(k)&\Delta_{+-}(k)\\
 0&\varepsilon_{k,-}&-\Delta_{+-}(k)&\Delta_{--}(k)\\
 \Delta_{++}^{\dagger}(k)&-\Delta_{+-}^{\dagger}(k)&-\varepsilon_{-k,+}&0\\
 \Delta_{+-}^{\dagger}(k)&\Delta_{--}^{\dagger}(k)&0&-\varepsilon_{-k,-}
 \end{pmatrix}\,.
 \end{equation}

 The eigenvalues of $H_{BdG}$ are given by
 \begin{equation}
 \label{SpectrumFull}
E_{\pm}^{2}(k)=|\Delta_{++}(k)|^{2}+\Delta_{+-}^{2}(k)+\frac{\varepsilon_{k,+}^{2}+\varepsilon_{k,-}^{2}}{2}
\pm|\varepsilon_{k,+}-\varepsilon_{k,-}|\sqrt{\Delta_{+-}^{2}(k)+\Big[\frac{\varepsilon_{k,+}+\varepsilon_{k,-}}{2}\Big]^{2}}\,,
\end{equation}
where $\varepsilon_{k,\pm}$ are given by Eq.\,(\ref{H0vEnVec}) and $\Delta_{++,--,+-}$ by Eqs.\,(\ref{pairings0}).

\begin{figure} 
 \begin{center}
 \includegraphics[width=.99 \columnwidth]{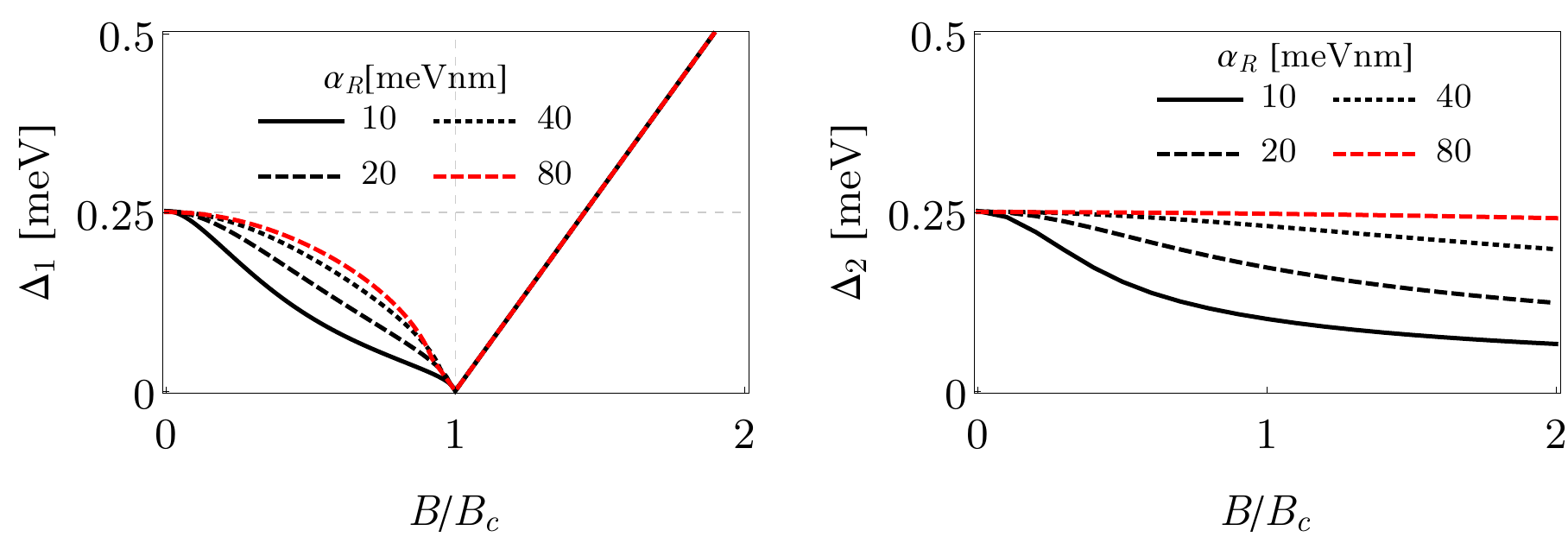} 
     \caption[Energy gaps of the Nambu Rashba nanowire]{(Color online) Energy gaps of the Nambu Rashba nanowire. Left and right panels: dependence of the gap at small, $k_{F,+}$, $\Delta_{1}$, and large momenta, $k_{F,-}$, $\Delta_{2}$, as function of the Zeeman field. Notice that the gap $\Delta_{1}$ closes at $B_{c}$, while $\Delta_{2}$ remains roughly constant only for strong spin-orbit coupling. Parameters: $\alpha_{R}=20$\,meVnm, $\Delta=0.25$\,meV and $\mu=0.5$\,meV.
}
   \label{gapsapp}
   \end{center}
\end{figure}

As we have already discussed, the $s$-wave pairing opens gaps in the energy spectrum given by Eq.\,(\ref{SpectrumFull}), as shown in Fig.\,(\ref{fig:RashbaD0NW}).
Now, we concentrate on the gaps opened at the Fermi momenta $k_{F,\pm}$. The energy band that experiments such gap opening is the lower band $E_{-}$. Therefore, we define the two gaps opened by $\Delta$ as
\begin{equation}
\Delta_{1}=E_{-}(k_{F,+})\,,\quad \Delta_{2}=E_{-}(k_{F,-})\,,
\end{equation}
where $E_{-}$ is the lower band given by Eq.\,(\ref{SpectrumFull}) and $k_{F,\pm}$ the Fermi momenta given by Eq.\,(\ref{fermpoints}). Notice that $\Delta_{1}$ corresponds to a low momentum gap $k_{F,+}$, while $\Delta_{2}$ to higher momentum $k_{F,-}$. Since the Fermi momenta depend on the spin-orbit, Zeeman interaction and chemical potential, the gaps will behave differently for increasing such parameters. Experimentally, however, it is more reliable to vary the Zeeman field  $B$ or chemical potential $\mu$ than the spin-orbit coupling strength $\alpha_{R}$. Thus, it is then natural to assume fixed SOC and vary $B$ or $\mu$. 
It is important to notice that the energy spectrum given by Eq.\,(\ref{SpectrumFull}) at $k=0$ reads
\begin{equation}
\label{topotrans}
E_{\pm}(k=0)=|B\pm\sqrt{\Delta^{2}+\mu^{2}}|\,.
\end{equation}
Remarkably, from Eq.\,(\ref{topotrans}), we observe that this lower band is zero, $E_{k,-}=0$, when Zeeman field reaches $\sqrt{\Delta^{2}+\mu^{2}}$, otherwise it is finite, then
\begin{equation}
\label{topotrans2}
E_{-}(k=0)=|B-B_{c}|\,,
\end{equation}
where 
\begin{equation}
\label{Bc}
B_{c}\equiv\sqrt{\Delta^{2}+\mu^{2}}
\end{equation}
is the critical field at which $E_{-}(k=0)=0$.
This point signals a phase transition between two gapped phases. 
It was shown that these two phases are topologically different and $E_{k,-}=0$ defines the topological phase transition into a topological superconducting phase with MBSs \cite{PhysRevLett.105.177002,Lutchyn:PRL10}. As in the Kitaev's model \cite{kitaev}, the topological invariant that distinguish these phases is the Majorana number $M=(-1)^{\nu}$, where $\nu$ is the number of pairs of Fermi points in the normal dispersion (see Fig.\,\ref{fig:RashbaNW}). The system host an odd number of pairs of Fermi points when $\mu$ is within the helical gap, $|\mu|<B$.
Fig.\,\ref{gapsapp} shows the behaviour of the two gaps as function of $B$ for different values of the spin-orbit coupling strength. Observe that the low momentum gap, $\Delta_{1}$, firstly decreases as the Zeeman field increases, reaching zero at $B=B_{c}$. By further increasing the Zeeman field, it increases linearly with $\Delta_{1}\approx|B-B_{c}|$. 
On the other hand, the higher momentum gap, $\Delta_{2}$, is strongly dependent on the spin-orbit coupling. Indeed, it decreases for a small values of $\alpha_{R}$ but it is finite for all relevant fields. However, notice that for strong spin-orbit coupling the gap $\Delta_{2}$ remains constant for all fields.
\begin{figure}[h] 
 \begin{center}
 \includegraphics[width=.9 \columnwidth]{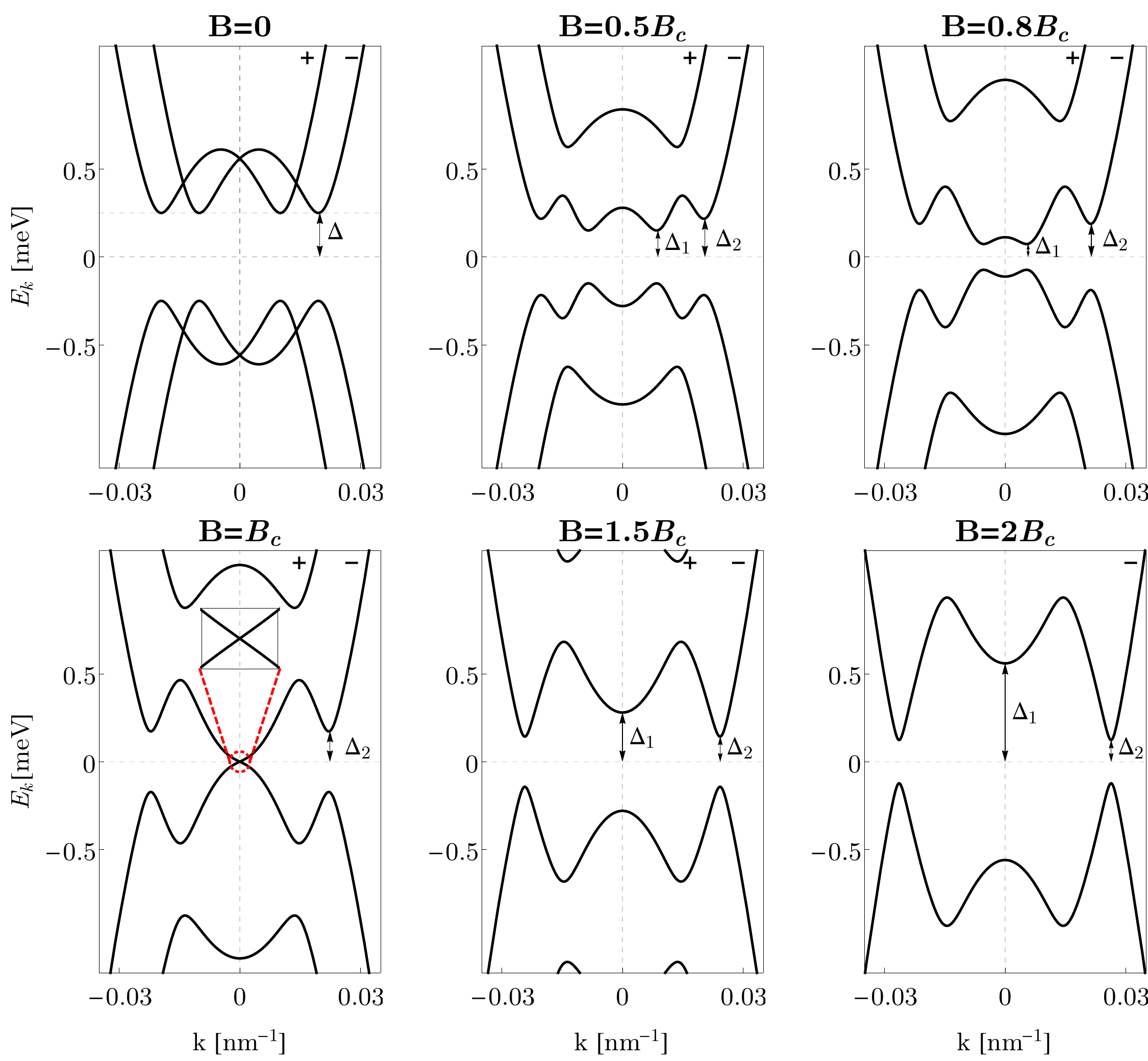} 
     \caption[Evolution of Nambu Rashba nanowire bands with the Zeeman field]{(Color online) Evolution of the Nambu Rashba nanowire bands with the Zeeman field. Parameters: $\alpha_{R}=20$\,meVnm, $\Delta=0.25$\,meV and $\mu=0.5$\,meV.
}
   \label{fig:fullRashbaNambu}
   \end{center}
\end{figure}

In Fig.\,\ref{fig:fullRashbaNambu}, we show the energy bands evolution given by Eq.\,(\ref{SpectrumFull}) as one increases the Zeeman field $B$.
At $B=0$, a non-zero superconducting pairing $\Delta$ opens a gap of $E_{-}(k_{F,\pm})$ at the Fermi points $\pm k_{F,\pm}$ given by Eqs.\,(\ref{fermpoints}) and marked with green circles in Fig.\,\ref{fig:RashbaD0NW}. Moreover, it modifies the gap at $k=0$ as seen in Eq.\,(\ref{topotrans}).
We have seen that the spin-orbit coupling splits NW states into two subbands of opposite helicity at $B=0$, see Figs.\,\ref{fig:RashbaNW} and \ref{fig:RashbaD0NW}. At finite $B$, these two subbands, which we label $+$ and $-$, have spins canted away from the SO axis. 
The s-wave pairing $\Delta$, expressed in the $\psi_{\pm}$ basis, takes the form of an intraband $p$-wave $\Delta_{++/--}(k)$, plus an interband $s$-wave pairing $\Delta_{+-}(k)$ \cite{Alicea:RPP12}, given by Eqs.\,(\ref{pairings0}). Without the latter, the problem decouples into two independent $p$-wave superconductors, while $\Delta_{+-}$ can be understood as a weak coupling between them.
Now, we identify two sectors, associated to the two subband denoted by $\pm$.
The energy gaps $\Delta_{1}$ and $\Delta_{2}$, defined by Eqs.\,(\ref{pairings0}), for each sector $\pm$ arise as soon as the Zeeman field is switched on and have a different dependence on the Zeeman field $B$: the former represents the energy gap at small momentum $k_{F,+}$, while the latter at high momentum $k_{F,-}$. As one increases the Zeeman field, 
$\Delta_{2}$ remains  roughly constant for strong spin-orbit coupling \cite{Prada:PRB12,Rainis:PRB13}, while $\Delta_{1}$ gets reduced as can be seen in Fig.\,\ref{fig:fullRashbaNambu} (top middle and top right panels).
Remarkably, as the Zeeman field approaches the critical field $B_{c}$, $\Delta_{1}$ vanishes (roughly) linearly, see Fig.\,\ref{gapsapp}. For $B=B_c$, $\Delta_{1}$ is zero and  becomes centered at $k=0$, Fig.\,\ref{fig:fullRashbaNambu} (bottom left panel).
By further increasing the Zeeman field $B$, $\Delta_{1}$ reopens and follows $\Delta_{1}= |B-B_c|$, the zero momentum energy of the lowest subband, see Fig.\,\ref{fig:fullRashbaNambu} (bottom middle and right panels). This closing and reopening of the energy gap $\Delta_{1}$, also known as gap inversion, signals a topological transition, induced by the effective removal of the $-$ sector away from the low-energy problem (see discussion in next paragraph).
Below $B_c$ the NW is composed of two spinless p-wave superconductors, and is therefore topologically trivial. 
Above $B_c$, $\Delta_-$ is no longer a p-wave gap, but rather a normal (Zeeman) spectral gap already present in the normal state, transforming the wire into a single-species $p$-wave superconductor with non-trivial topology. This phase contains MBSs, protected by the effective gap $\Delta_\mathrm{eff}=\mathrm{Min}(\Delta_1,\Delta_2)$, at the wire ends. Above a certain field $B_c^{(2)}$, the gap $\Delta_\mathrm{eff}$ saturates at $\Delta_2$ and the physics of superconducting helical edge states in spin-Hall insulators is recovered \cite{Fu:PRB09,Badiane:PRL11}, see Fig.\,\ref{gapsapp}. Indeed, in the following we show that for high Zeeman fields $B\gg B_{c}$, the Hamiltonian $\mathcal{H}$ given by Eq.\,(\ref{fullHhelical}) can be connected onto the Kitaev's model hosting Majorana bound states.

To have a further insight of the previous discussion, the system has to exhibit $p$-wave pairing symmetry according to the Kitaev's model. Thus,
it is convenient to project the system Hamiltonian onto the lower band $-$.
This is allowed because for reaching the topological phase one needs strong Zeeman field, then the upper band $+$, see Fig.\,\ref{fig:fullRashbaNambu}, can be removed from the low-energy problem. Therefore, we can write,
\begin{equation}
\label{fullHlowestband}
\begin{split}
\mathcal{H}&=
\int \frac{dk}{2\pi}\bigg[
\varepsilon_{k,-}\psi_{-}^{\dagger}(k)\psi_{-}(k)
+
\frac{\Delta_{--}(k)}{2}\psi_{-}^{\dagger}(k)\psi_{-}^{\dagger}(-k)
+\frac{\Delta_{--}^{\dagger}(k)}{2}\psi_{-}(-k)\psi_{-}(k)
\bigg],
\end{split}
\end{equation}
where the superconducting pairing potential, or order parameter, 
\begin{equation}
\Delta_{--}(k)=\frac{i\alpha_{R}k\Delta}{\sqrt{B^{2}+\alpha_{R}^{2}k^{2}}}
\end{equation}
 has $p$-wave symmetry.
 Now, we write previous Hamiltonian in the BdG form,
 \begin{equation}
 \mathcal{H}=\frac{1}{2}\int\frac{dk}{2\pi}\psi^{\dagger}(k) H_{BdG}\psi(k)\,,\quad 
 \psi(k)=\begin{pmatrix}
\psi_{-}(k)\\
\psi_{-}(-k)^{\dagger}
\end{pmatrix}\,,
 \end{equation}
 where
 \begin{equation}
 H_{BdG}=
 \begin{pmatrix}
\varepsilon_{k,-}&\Delta_{--}(k)\\
\Delta_{--}^{\dagger}(k)&-\varepsilon_{-k,-}\,
\end{pmatrix}\,,
 \end{equation}
 whose energy spectrum is given by
 \begin{equation}
 E_{k,-}=\pm\sqrt{\varepsilon_{k,-}^{2}+|\Delta_{--}(k)|^{2}}\,.
 \end{equation}
 Previous equation is in essence the energy spectrum of a $p$-wave superconductor, thus being in concordance with the Kitaev's model described in previous section. Therefore, it is not a coincidence that the model given by the Hamiltonian Eq.\,(\ref{fullHamiltonian}) indeed describes Majorana-like physics when $B>B_{c}$. 
 
 From the point of view of topology, the infinite, Zeeman polarised, single-channel semiconducting nanowire in proximity to a conventional $s$-wave superconductor, described in this section, belongs to the so called one-dimensional D class \cite{Altland:PRB97}, which has an invariant $\nu'$ that may be $\nu'=0$ (topologically trivial) or $\nu'=1$ (non-trivial).
This system undergoes a band topological transition from $\nu'=0$ to $\nu'=1$ when the Zeeman splitting $B$, perpendicular to the spin-orbit axis, exceeds a critical value $B_c=\sqrt{\mu_S^2+\Delta^2}$, where $\mu_S$ and $\Delta$ are the wire's Fermi energy and induced gap respectively. 

\begin{figure*}
\centering
\begin{minipage}[b]{.45\textwidth}
\centering
\includegraphics[width=.99\textwidth]{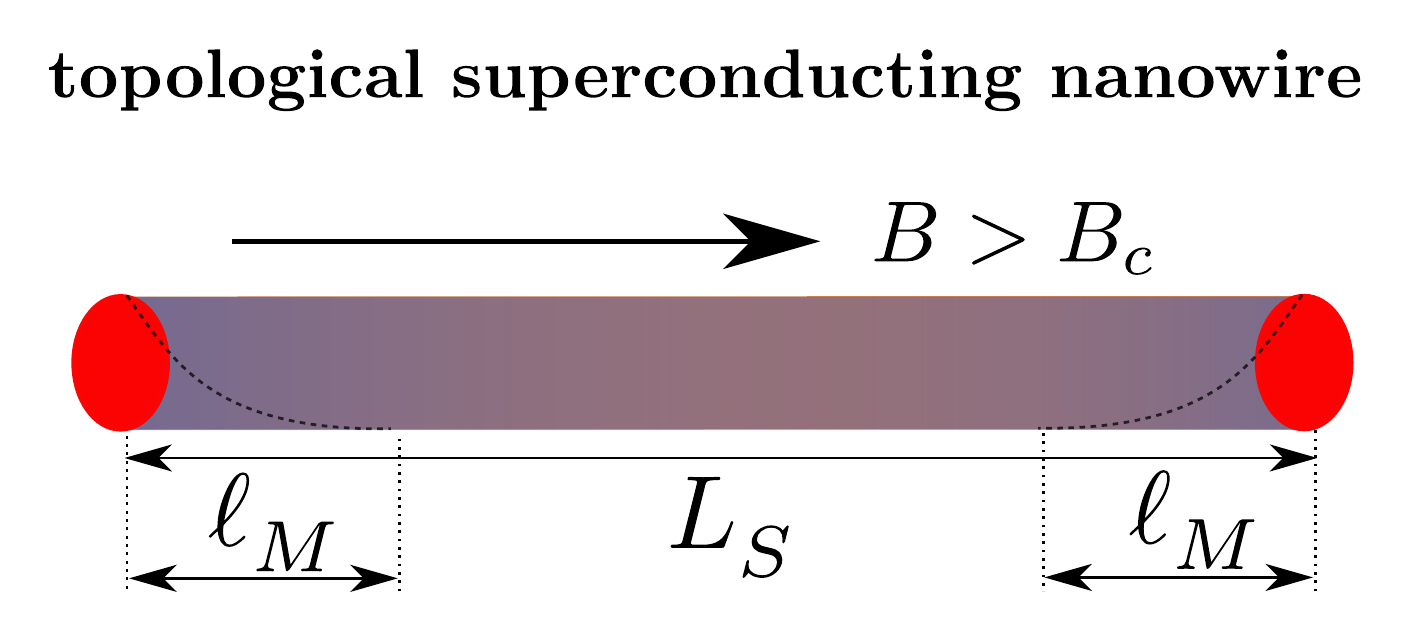} 
    \vspace{1.5cm}
\end{minipage}\quad
\begin{minipage}[b]{.45\textwidth}
\centering
\includegraphics[width=.99\textwidth]{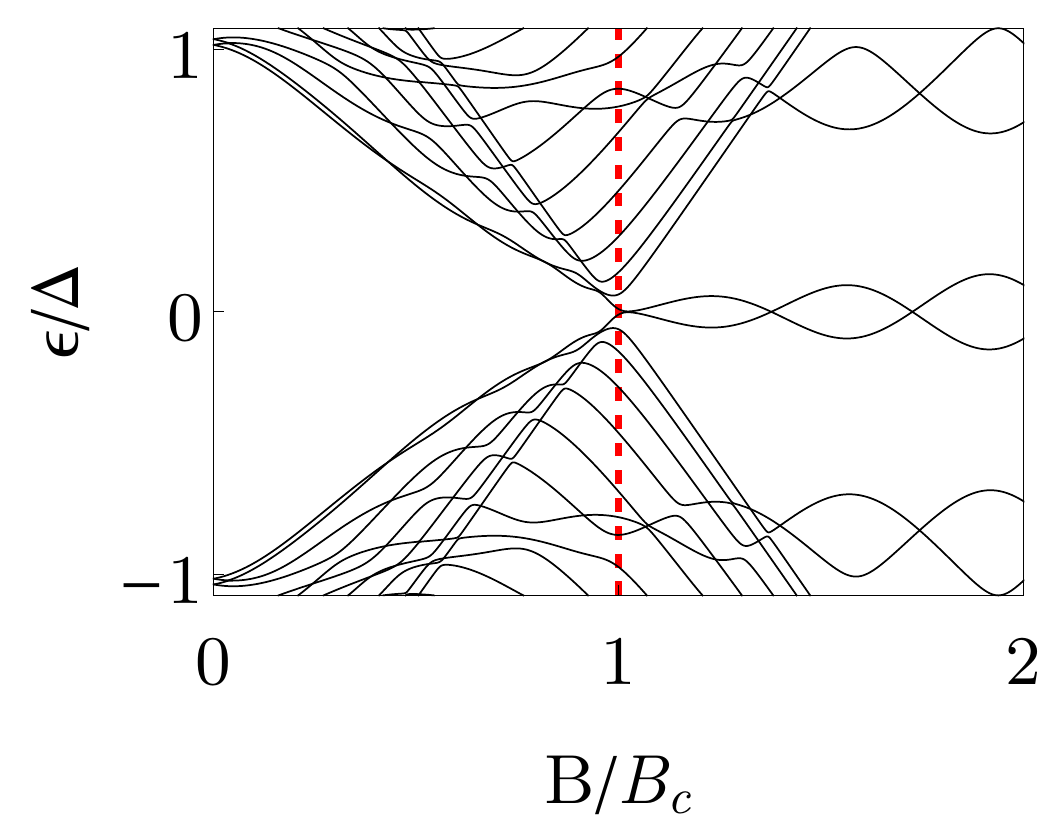} 
\end{minipage}
\caption[Topological superconducting wire and energy levels as function of B]{(Color online) Left: A finite-length semiconductor nanowire with Rashba SOC placed on a $s$-wave superconductor develops weakly overlapping Majorana bound states at its ends (red filled circles) under a Zeeman field exceeding the critical field $B_{c}$. The wave functions of the MBSs decay exponentially into the bulk of the superconductor and it is characterised by the Majorana localisation length $\ell_{M}$. Right: The spectrum of the wire in this regime has a pair of approximate zero modes, with a small splitting that oscillates with $B$ but decreases when increasing the length of the wire $L_{S}$ (see Sec.\,\ref{SNSRashbaNWs}). The vertical red dashed line marks the topological transition point.
Parameters: $\alpha_{R}=20$\,meVnm, $\mu=0.5$\,meV.}
\label{introSplitting}
\end{figure*}

 An interface between a $B>B_c$ semi-infinite wire (with $\nu'=1$) and vacuum ($\nu'=0$) binds a single subgap state \cite{PhysRevB.78.195125,PhysRevLett.109.150408}.
The state is pinned to zero energy by the particle-hole symmetry of the Bogoliubov-de Gennes description, locked strictly midway between electrons and holes; it satisfies the Majorana condition $\gamma=\gamma^\dagger$, and is hence known as a MBS. This robust pinning is broken when two MBSs are brought within a finite distance of each other, i.e. in a topological superconducting nanowire of finite length $L_{S}$, see left panel in Fig. \ref{introSplitting}. Both MBSs then overlap and hybridize into a single conventional fermion of finite energy, or Andreev bound state. Right panel in Fig. \ref{introSplitting} presents the evolution of the wire spectrum with $B$, showing the energy splitting of MBSs for $B>B_{c}$. The energy splitting can be small for long enough wires, though strictly speaking it is non-zero for any length \cite{Lim:PRB12,Prada:PRB12,Rainis:PRB13,DasSarma:PRB12}. The MBSs become merely quasi-stationary states, with their energy splitting representing a Rabi frequency at which one oscillates into the other. See Chap.\,\ref{SNSRashbaNWs} for details on the calculation of the spectrum presented in Fig.\,\ref{introSplitting}.
  
The wave-functions of the MBSs exponentially decay into the bulk of the superconducting wire  \cite{DasSarma:PRB12,PhysRevB.86.085408} and can be approximately write down for $x\gg \xi_{eff}$ as $\Psi(x)={\rm e}^{-x/\xi_{eff}}{\rm e}^{\pm ik_{F,eff}x}$, where $\xi_{eff}$ is the effective coherence length and $k_{F,eff}$ the effective Fermi wave vector associated with the zero-mode solution \cite{DasSarma:PRB12}. 
The relevant decay length characterising this overlap is the effective coherence length  $\xi_{eff}$ and we refer to it as to the Majorana localisation length denoted later by $\ell_{M}$, which tells us how well are MBSs localised at the ends of the wire.  For a finite spatial overlap between their wave-functions, the MBSs acquire a finite energy and they are not longer true zero modes, thus for $L_{S}\gg\xi_{eff}$ \cite{DasSarma:PRB12}
 \begin{equation}
 \label{esplitting}
 \Delta E\approx\hbar^{2}k_{F,eff}\frac{{\rm e}^{-2L_{S}/\xi_{eff}}}{m\,\xi_{eff}}\,{\rm cos}(k_{F,eff}\xi_{eff})\,,
 \end{equation}
 where $m$ is the effective electron's mass in the nanowire, and $L_{S}$ is the wire's length. 
 Notice that such energy splitting exhibits an oscillating behaviour in the system parameters $\mu$ and $B$ through $k_{F}$. We also notice that for sufficiently long wires, $L_{S}\gg2\ell_{M}$, the energy splitting can be very small so that it can be assumed to be zero $\Delta E\approx0$.

To conclude this section, we remark that we have described in detail the platform for engineering Majorana bound states in nanowires with strong Rashba spin-orbit coupling (SOC) in proximity to an $s$-wave superconductor, where an external Zeeman field, perpendicular to the spin-orbit axis, drives the system into the topological superconducting phase \cite{PhysRevLett.105.177002,Lutchyn:PRL10}. The topological phase transition is controlled by increasing the Zeeman field and the topological phase is determined by the presence of two Majorana bound states, one at each end of the wire, which can be splitted in energy or not depending on the relation between the length of the wire and the Majorana localisation length.

\section{Experimental signatures}
 Although recent experiments have focused on diverse theoretical proposals, in this section we aim at giving a brief description of the ones based on Rashba nanowires in proximity to $s$-wave superconductors  \cite{PhysRevLett.105.177002,Lutchyn:PRL10}. This platform has attracted serious attention mainly because all the ingredients are well-known phenomena in condensed matter. As we have explained in the two previous sections, according to theory, signatures Majorana physics in quantum wires implies to find Majorana bound states at each end of the wire. For the system to host such zero modes, there has to be a topological phase transition as the applied external Zeeman field increases, which is distinguishable by observing the closing and reopening of the gap. 
  
Following these ideas, a nanowire with Rashba SOC can be partially placed on a $s$-wave superconductor, where the portion in contact with the superconductor acquires superconducting correlations by the induced proximity effect and we refer to it as to the \emph{superconducting part} S, while the other portion of the wire does not contain superconductivity and therefore is refereed to as to the \emph{normal part} N.
When the applied Zeeman field exceeds the critical field, $B>B_{c}\equiv\sqrt{\mu^{2}+\Delta^{2}}$, the S part becomes topological superconducting and therefore host Majorana bound states emerge at its two ends, one at each end. 
This configuration gives rise to the \emph{hybrid} geometry known as NS junction, and offers a number of advantages over superconducting nanowires in that it allows for contacting the junction with gates and thus tune its chemical potential. Indeed, superconductor-semiconductor hybrid devices can be assembled from semiconductor nanowires individually contacted by superconductor electrodes. Below the superconductor critical temperature, the high transparency of the contacts gives rise to proximity-induced superconductivity. The nanowires form superconducting weak links operating as mesoscopic Josephson junctions with electrically tunable coupling \cite{Doh:S05}. This allows to study a wealth of fundamental physical phenomena in a tunable and well controlled manner. Now, the matter is how one should proceed in order to measure the presence of these MBSs or at least one of them in NS junctions.
 
 In a NS junction, the S part possesses a superconducting order parameter, which make the energy spectrum in S gapped, while N remains metallic. An electron traveling towards the NS interface is not transmitted since there are no states within the gap of S, but it is rather reflected as a hole. This process is known as Andreev reflection and will be discussed in more detail in Sec.\,\ref{introhybridNSSNS}.
The general relation between the conductance of a single mode NS junction and the Andreev reflection probability is given by $G=(2e^{2}/h)A$, where the factor of $2$ arises because the Andreev reflection of an electron into a hole doubles the current and $A$ is the Andreev reflection probability which depends on the nature of the interface \cite{PhysRevB.25.4515}. Since $B\neq0$, there is no time-reversal symmetry  and therefore Kramers degeneracy does not apply. The system still possesses particle-hole symmetry, which requires that any $A$ is two-fold degenerate (Beri degeneracy \cite{PhysRevB.79.245315}) with exceptions for $A=0$ and $A=1$, which may be non-degenerate. 
It was shown that a right-moving electron from N experiments an Andreev reflection probability from a MBS (a zero energy state within the gap) which is non-degenerate and pinned to unity, $A=1$, thus producing a quantised conductance of $(2e^{2}/h)$ at zero energy (the energy of the MBS), whereas without the MBS the conductance vanishes \cite{Law:PRL09,Flensberg:PRB10,Liu:PRL12,PhysRevB.63.144531,Beenakker:11}. On the other hand, all other Andreev reflected modes are two-fold degenerate and the conductance become $(4e^{2}/h)A$. Therefore, measuring the conductance at zero energy constitutes a powerful tool for detecting the presence or absence of MBSs in a NS hybrid junction when MBSs emerge at the ends of the S region. 

By applying electrostatic gates at the NS contact, one makes a tunnel junction between the N and S parts of the NS junction. Then, under an applied voltage bias $V$ across the junction, we can measure the tunneling current $I$ through this weak link. In the tunneling regime the differential conductance $dI/dV$ is proportional to the density of states at the end of the superconducting region, which is adjacent to the tunneling contact. The density of states, to a very good approximation, measures the conductance of a NS junction. Therefore, one expects that as the Zeeman field increases the differential conductance $dI/dV$ should trace the gap closing and reopening, which leads to the topological superconducting phase with a peak at $V=0$ of high $2e^{2}/h$. This peak is referred to as \emph{zero-bias peak} (ZBP) or \emph{zero-bias anomaly} (ZBA). 
During the last years, this approach was extensively pursued experimentally in hybrid NS junctions made of nanowires, looking for the zero-bias peak in the conductance across such junction when the S part becomes topological.

We point out that the observation of a zero-bias peak in the conductance is not the unique condition for testing MBSs in nanowires. Stronger evidence could be provided by the observation of non-Abelian interference (braiding) \cite{Nayak:RMP08}, or by transport in phase-sensitive superconductor-normal-superconductor (SNS) junctions. The latter approach, which typically involves the measurement of an anomalous ``fractional'' $4\pi$-periodic ac Josephson effect \cite{Kitaev:P01,Fu:PRB09,Kwon:EPJB03}, is much less demanding than performing braiding. Realistically, however, the fractional effect, detected through, e.g., the absence of odd steps in Shapiro experiments \cite{Kwon:EPJB03,Jiang:PRL11,Dominguez:PRB12,Rokhinson:NP12}, may be difficult to measure (dissipation is expected to destroy it in the steady state), or may even develop without relation to topology \cite{Sau:12b}. 
Although it has been shown that the $4\pi$ periodicity survives in the dynamics, such as noise and transients \cite{Badiane:PRL11, San-Jose:PRL12a,Pikulin:PRB12}, simpler experimental probes of MBSs are extremely desirable.
On the other hand, it has been proposed to employ T-junctions in order to test braiding statistics in one-dimension \cite{Alicea:NP11}, however, on the experimental side it still represents a major challenge than measuring the zero-bias peak in conductance. 

\begin{figure}[!ht]
\centering
\includegraphics[width=.5\textwidth]{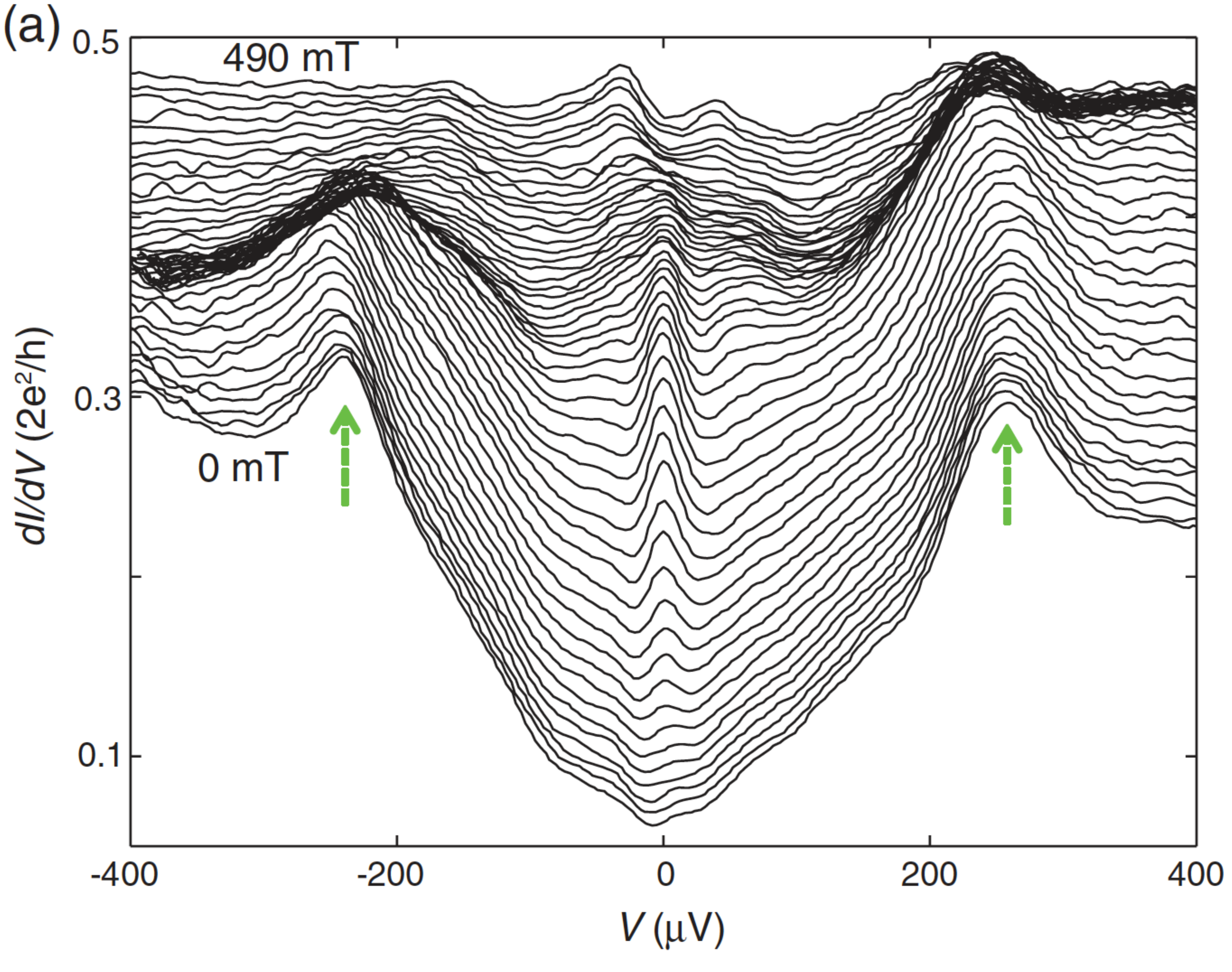} \\
\includegraphics[width=.5\textwidth]{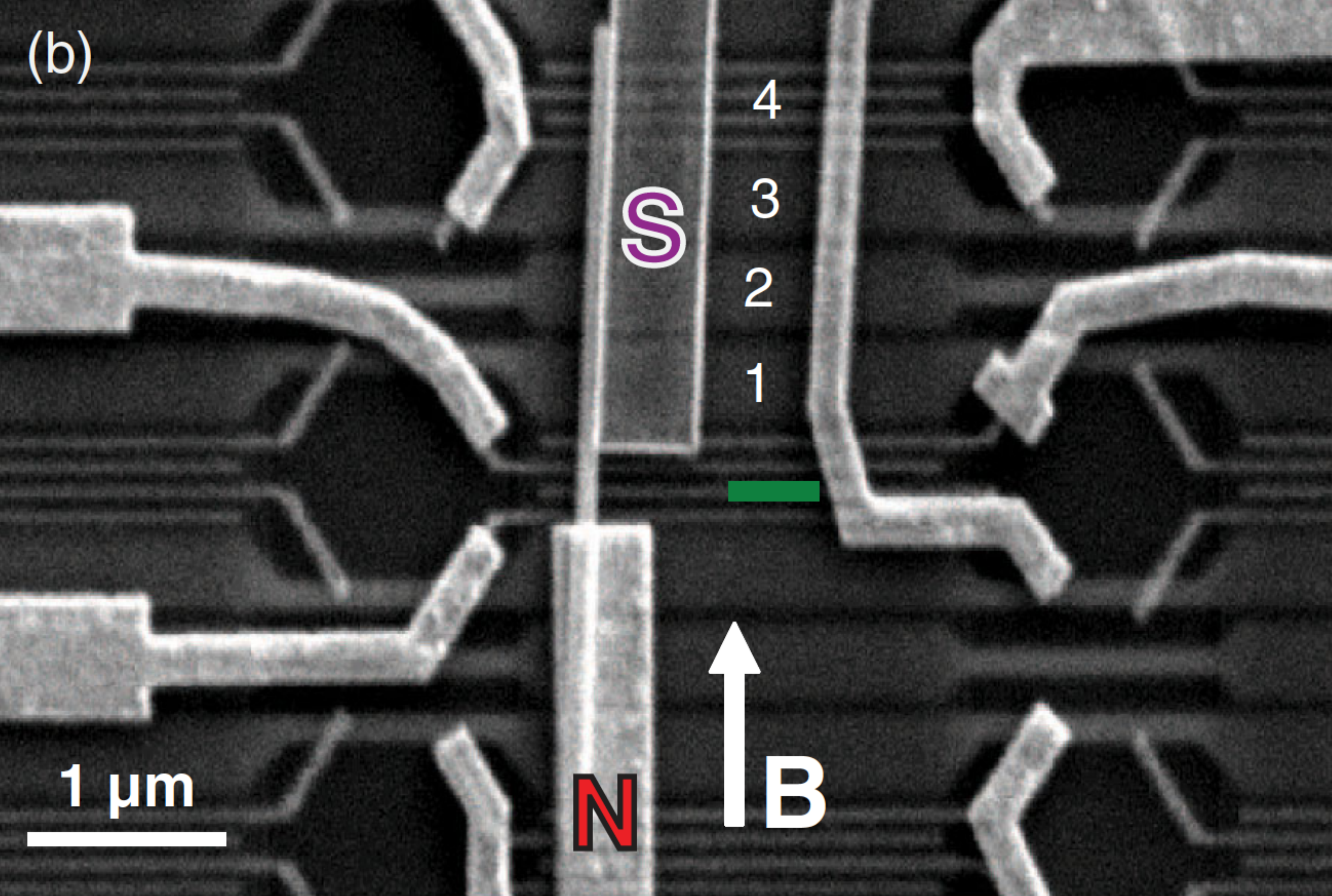} 
\caption[Experimental signatures of MBSs in nanowires]{(Color online) The Delft experiment. (a) Tunneling conductance $dI/dV$ versus $V$ for different values of the magnetic field, showing the induced superconducting gap and the emergence of the zero-bias peak of amplitude $0.05 (2e^{2}/h)$ when the magnetic field, perpendicular to the SO axis, is between 
$\sim100$ and $\sim400$\,mT. (b) Scanning electron microscope image of the device with normal (N) and superconducting (S) contacts attached to an InSb nanowire. Adapted from \cite{Mourik:S12}.}
\label{ZBPX.pdf}
\end{figure}

Regarding the materials, InAs \cite{Das:NP12} and InSb nanowires \cite{doi:10.1021/nl203846g} are known to have strong spin-orbit interaction, $\alpha_{R}=10$\,meVnm and $\alpha_{R}=20$\,meVnm, and large $g$-factor \cite{Das:NP12,doi:10.1021/nl901333a}. Indeed, the $g$-factor in bulk InAs and InSb is very large, $g\approx15$ and $g\approx50$, respectively, which makes such nanowires to accept high Zeeman fields as a result of the applied magnetic field.
Indeed, when a good proximity effect between the nanowire with an s-wave superconductor is generated, the large $g$-factors allow for exceptionally weak fields to drive the wire into a topological superconductor with MBSs. The strong spin-orbit coupling further allows this topological state to possess a relatively large gap that remains robust against disorder \cite{PhysRevB.83.184520}. A relatively good proximity effect was measured in both systems \cite{Doh272,Nishio:N11,Nilsson:NL12}, making them promising for the search of MBSs. 

Following these ideas, the first evidence was reported by the Delft group \cite{Mourik:S12}. 
In this experiment the geometry was similar to the NS junctions described above. 
They employed an InSb semiconductor quasi-one-dimensional wire with strong spin-orbit coupling partially deposited on a substrate equipped with gates and 
contacted with superconducting (niobium titanium nitride) and normal metal electrodes as shown in Fig.\,\ref{ZBPX.pdf}(b). 
The presence of the MBS at the end of the superconducting section adjacent to the NS interface of the wire was tested by measuring the tunnelling current $I$ through the weak link, created between the superconducting and normal electrodes, under an applied bias $V$.  In this setup, the differential conductance $dI/dV$ is proportional to the density of states in the superconducting end adjacent to the tunnelling contact. By increasing the magnetic field, a clear zero-bias peak was observed for a range of magnetic fields 
of $0.1$\,T$\leq B\leq400$\,mT, and disappearing for higher fields, see Fig.\,\ref{ZBPX.pdf}(a).
These results support the existence of MBSs in NS junctions, however, a number of features related to the emergence of MBSs are missing.
Indeed, the value of the zero-bias quantized conductance was smaller than the predicted, $0.05(2e^{2}/h$), and the experiment does not show the gap closing and 
reopening for entering into the topological phase, when such zero-bias peak emerge, as predicted by theory. Theoretical studies attribute this softness of the induced superconducting gap to disorder at the semiconductor-superconductor interface \cite{Takei:PRL13} and to 
multiple subbands \cite{Pientka:PRL12,PhysRevB.90.085302}. 

Additional experimental observations of the existence of Majorana quasiparticles were subsequently reported by several independent groups, showing the celebrated 
zero-bias peak at non-zero magnetic field \cite{xu,Das:NP12,Finck:PRL13,Churchill:PRB13,Lee:13}. 
All these experiments point towards the clear evidence of MBSs in NWs, although it was shown that detection of sub-gap zero modes through zero-bias peaks 
in transport can be obscured, or even mimicked by other effects, such as owing to Kondo physics \cite{Lee:PRL12, Finck:PRL13,Churchill:PRB13,Zitkoetal}, 
disorder\cite{Pientka:PRL12,Bagrets:PRL12,Liu:PRL12,Sau:13}, smooth confinement \cite{Prada:PRB12,Kells:PRB12}, parity crossings of Andreev levels \cite{Lee:13,Zitkoetal}.

In all previous experiments a characteristic zero-bias tunnelling peak appearing at finite magnetic field have been reported, where in all cases a soft gap is also seen, 
indicated by sizable subgap conductance \cite{xu,Das:NP12,Finck:PRL13,Churchill:PRB13,Lee:13}.
As mentioned before, preliminary studies attribute it to disorder at the semiconductor-superconductor interface \cite{Takei:PRL13}, but also the 
emergence of quasiparticle states in the gap tends to destroy topological protection, since quasiparticles occupying such sub-gap states will inevitably participate 
affecting the quantum state when braiding the MBSs and thus inducing decoherence effects \cite{PhysRevB.85.165124,PhysRevB.85.174533}.
Their findings were partially in agreement with the existence of such zero energy Majorana bound states \cite{Sengupta:PRB01,Bolech:PRL07,Law:PRL09}, and
subsequent theoretical explanations showed that these features, as the non-closing of the gap, can be understood with realistic NS calculations 
\cite{Prada:PRB12, Stanescu:PRL12,Rainis:PRB13,StanescuModel13}.

\begin{figure}[!ht]
\centering
\includegraphics[width=.4\textwidth]{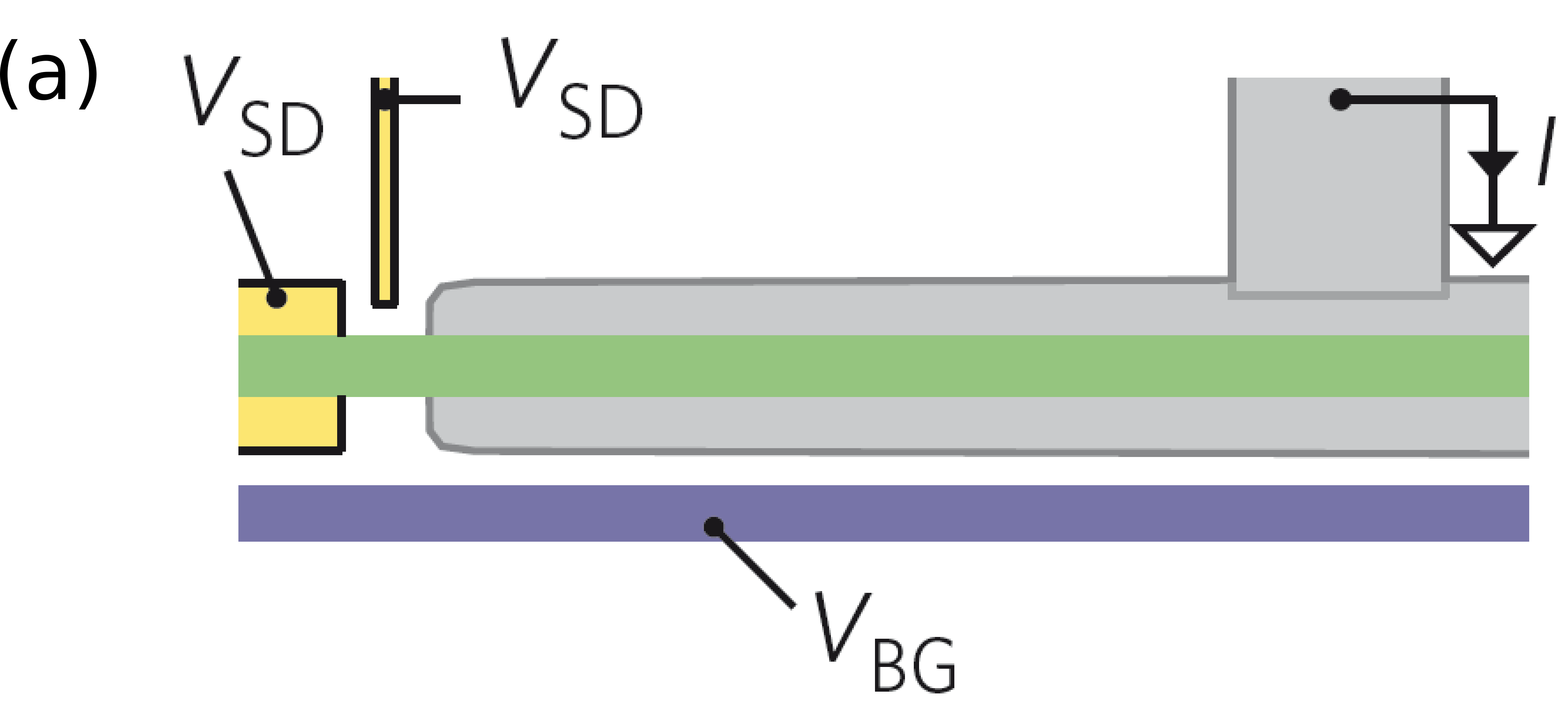} \includegraphics[width=.4\textwidth]{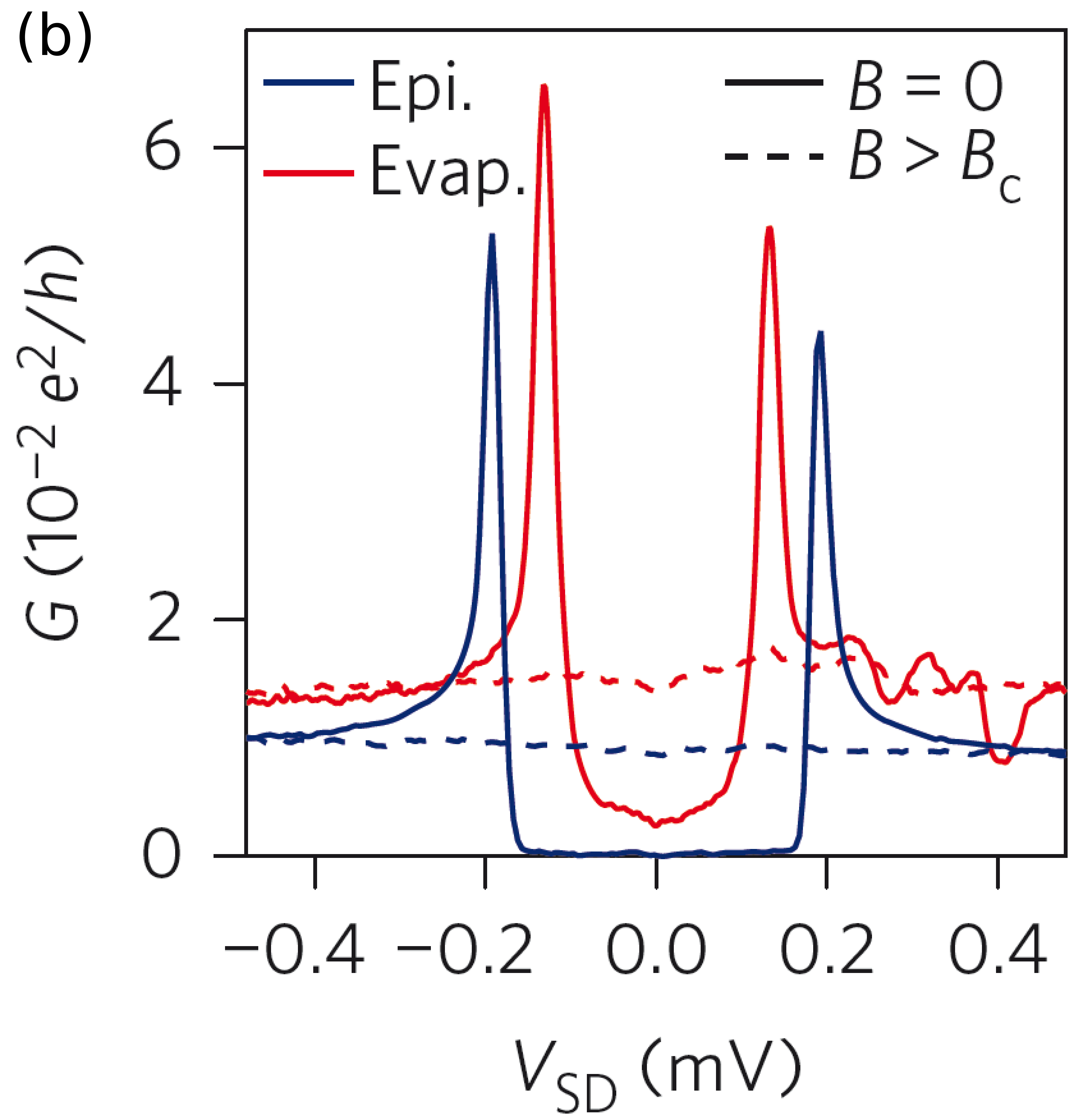} 
\caption[Hard gap experiments: Copenhagen ]{(Color online)
Hard gap experiments: Copenhagen group. (a) Measurement set-up, showing Ti/Au leads (yellow), InAs nanowire (green) and Al shell (grey).
 (b) Differential conductance as a function of source-drain voltage of an epitaxial full-shell device (blue) and
an evaporated control device (red) at $B = 0$ (solid line) and above a critical
field $B > B_{c}$ ($\sim75$\,mT for the epitaxial device, $\sim250$\,mT for the
control) (dashed line). 
Adapted from \cite{chang15}.}
\label{HardCop}
\end{figure}

New experiments have been also reported within the last year, where in nanowires the soft-gap problem was recently resolved by growing  Al superconductor epitaxially on 
InAs nanowires (see Fig.\,\ref{HardCop}), yielding greatly reduced sub-gap conductance \cite{chang15,Higginbotham,Krogstrup15} and later it was realised in a two dimensional 
semiconductor-superconductor heterostructure \cite{Kjaergaard}. Further studies on these materials showed improved Majorana signatures \cite{Albrecht16}.
The InAs nanowire still contain residual disorder, which shows up as unintentional quantum dots in transport measurements \cite{Lee:PRL12,Lee:13,chang15}.

\begin{figure}[!ht]
\centering
\includegraphics[width=.3\textwidth]{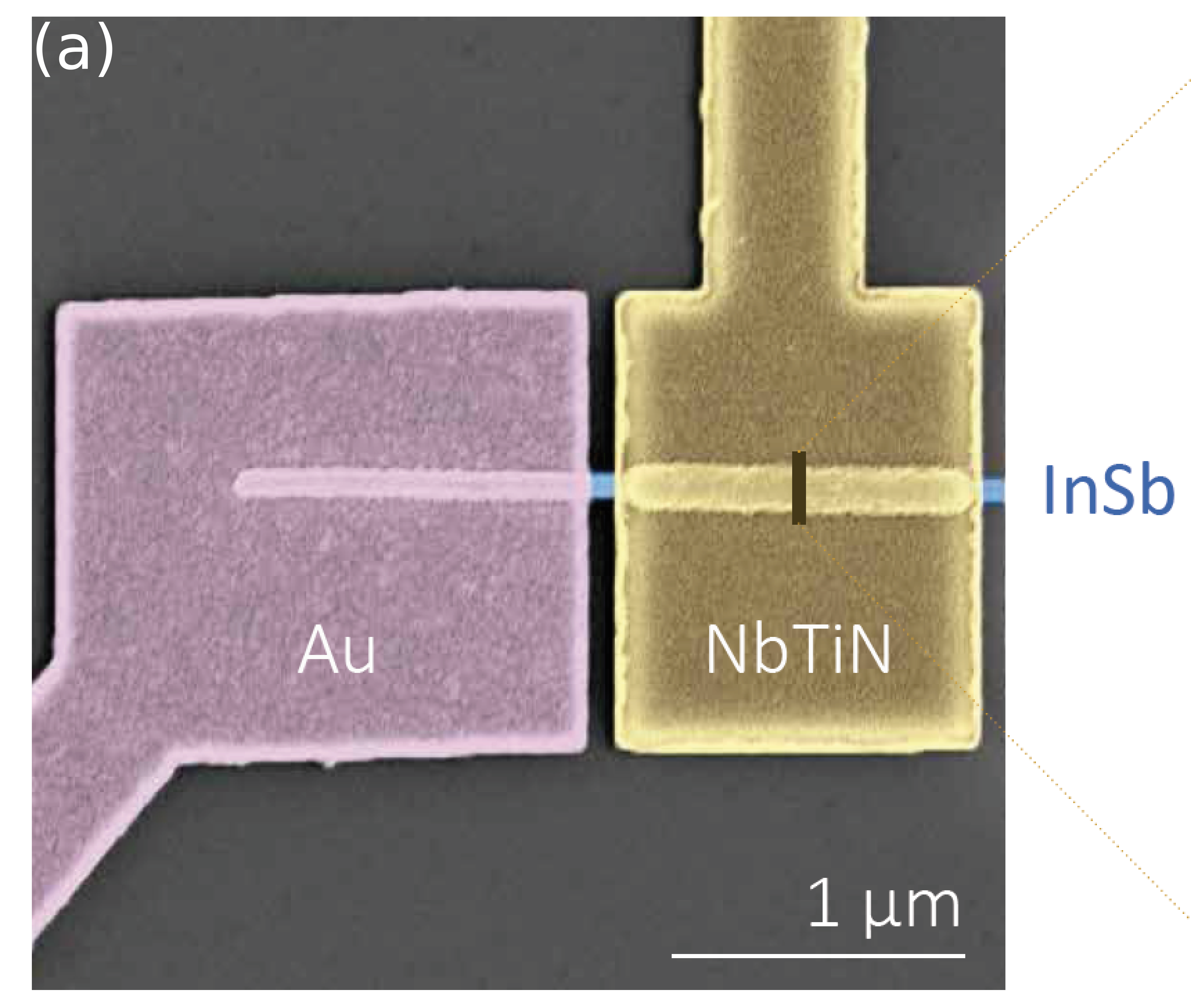} \includegraphics[width=.6\textwidth]{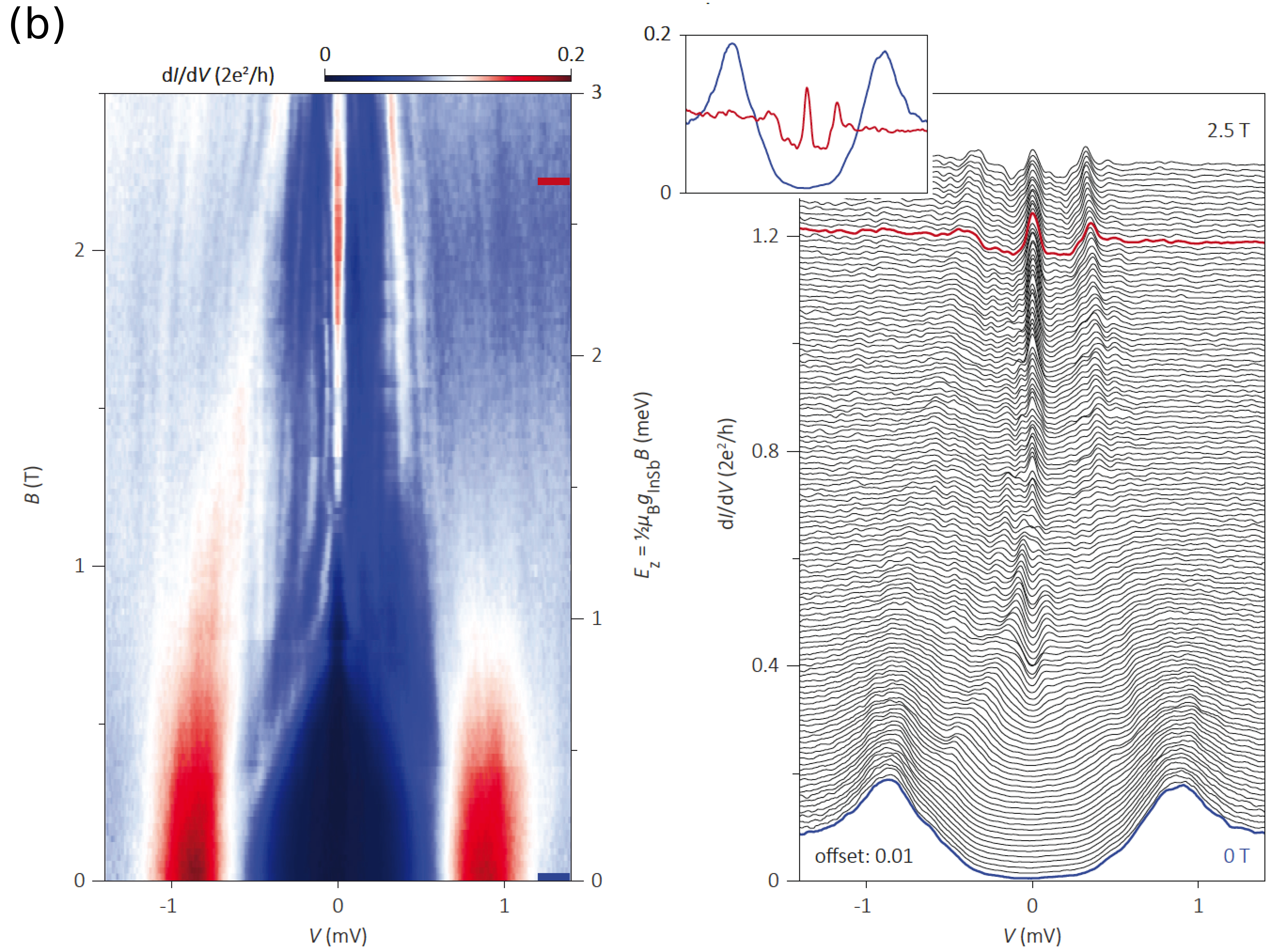} 
\caption[Hard gap experiments: Delft ]{(Color online)
Hard gap experiments: Delft group. (a) Top view of the measurement set-up, showing normal metal contact (Cr/Au, $10$\,nm/$125$\,nm) and superconducting contact (NbTi/NbTiN, 
$5$\,nm/$85$\,nm). Contact spacing is $\sim100$\,nm and InSb nanowire in light blue.
 (b) (left panel) An un-split zero-bias peak visible between $1.2$ and $2.5$\,T. (right panel) Line cuts from the data in left panel with vertical offsets $0.01\times 2e^{2}/h$.
Adapted from \cite{zhang16}.}
\label{Harddelft}
\end{figure}

Then, as an alternative material, the Delft group has recently reported new experiments showing significant improvements in reducing disorder in a high-quality 
interface between an InSb nanowire and NbTiN superconductor \cite{zhang16} (see Fig.\,\ref{Harddelft}). 
They demonstrate  ballistic transport and induced hard-gap with strongly reduced subgap density of states by gate tuning the device to a tunnel probe. 
In their experiment, the induced zero-bias peak is consistent with theory and exclude other explanations based on disorder effects. 
The choice for InSb nanowires is fully justified. Indeed, in general, InSb nanowires are cleaner (showing higher electron mobility \cite{0957-4484-26-21-215202,VanWeperen}) than InAs. Additionally InSb has a larger $g$-factor (about $5$ times larger), thus reducing the required external magnetic field needed to induce the topological phase transition.

According to previous discussion, recently there has been a remarkably progress in fabrication of junctions with hard gaps and good transparency in the single-channel 
limit. Therefore, it is natural to ask what can we learn from these good junctions beyond the zero-bias paradigm and investigate new geometries such as SNS junctions, which 
offer a number of advantages over NS junctions. These issues are investigated along this thesis.


The spin-texture needed for entering into the topological superconducting phase and generated by the SOC and Zeeman fields in previous description can be also mimicked  by a chain of magnetic atoms with an spatially modulated spin arrangement \cite{PhysRevB.88.020407,PhysRevB.88.155420,PhysRevLett.111.186805,PhysRevLett.111.147202,PhysRevLett.111.206802,PhysRevB.88.180503}.  Similar ideas allowed Nadj-Perge et al. \cite{Nadj-Perge31102014} to investigate an alternative scenario, where Fe atoms were placed on a conventional $s$-wave superconductor (Pb) with strong SOC. Here, the topological superconducting phase arises from the ferromagnetic interaction between Fe atoms and the strong SOC in the superconductor.
Nadj-Perge et al. \cite{Nadj-Perge31102014} reported preliminary evidence 
of zero bias peaks in the density of states, by Scanning tunnelling microscopy (STM), associated with the ends of the magnetic chains as possible Majorana signatures.
Still, it has triggered interesting discussions on the physical origin of these peaks due to the short values for the Majorana localization lengths. Although there are some possible theoretical explanations \cite{PhysRevLett.114.106801}, additional investigations are needed for supporting their findings.
More recently, Pawlak et al. \cite{lossjelena} investigated the spatial and electronic characteristics of topological superconducting chains of iron atoms on the surface of Pb(110) by combining STM and atomic force microscopy (AFM). They demonstrated that the Fe chains are mono-atomic and exhibit zero-bias conductance peaks at their ends, which are interpreted as signature for the Majorana bound state. It is also shown another strong fingerprint associated with the localization of the Majorana wavefunction, which exhibits an exponential decay. In this experiment the proximity gap is driven into the topological phase by a spin texture which gives rise to a helical field in a similar fashion as in \cite{Nadj-Perge31102014}.

\section{This thesis}
As discussed in previous sections, there has been an increasing theoretical and experimental interest 
towards real signatures for engineering one-dimensional topological superconductivity and therefore Majorana bound states. 
In spite of the efforts, conclusive experiments are urgently needed, and further theoretical analysis has to be done in more complex geometries such as 
Superconductor-Normal metal-Superconductor (SNS) junctions \cite{pablo,PhysRevLett.112.137001}. This geometry has a number of advantages including the possibility of 
studying supercurrents \cite{Deng:NL12,Doh:S05,Nishio:N11,Nilsson:NL12,Gunel:JOAP12}, or direct spectroscopy of Andreev bound states (ABS) 
\cite{PhysRevLett.110.217005, nphys1811, PhysRevLett.104.076805, PhysRevB.88.045101,PhysRevB.89.045422,PhysRevX.3.041034,Levy,Lee:13,PhysRevLett.85.170}.  
As we shall discuss along this thesis, this latter technique can be used, in principle, to directly monitor the detailed evolution from the trivial to the topological phase. 
Moreover, it contains information about the peculiar dependence of Majorana bound states hybridization with superconductor phase difference, despite not requiring any external 
control on it.
In the following we describe our main findings.
\subsection*{Chapter 2}
In this chapter we provide an introduction to the technical part of this thesis. Indeed, an important part of this thesis was devoted to investigate an appropriate scheme for modelling hybrid junctions.
For a complete description, we firstly introduce the classification of SNS junctions based on the comparison between the length of the normal N region, $L_{N}$, and the superconducting coherence length, $\xi$: short junctions for $L_{N}<\xi$, and long for $L_{N}>\xi$.
Then, we show in a simple example how the concepts of Andreev reflection, Andreev bound states and Multiple Andreev reflections emerge.
Later, we concentrate on giving the necessary details for modelling superconducting nanowires and hybrid one-dimensional NS and SNS junctions made of nanowires with Rashba SOC and Zeeman interaction. 
Afterwards, we make a detailed calculation of Andreev bound states and Josephson currents in SNS junctions made of Rashba nanowires with Zeeman interactions and induced $s$-wave superconductivity. Here, we show the detailed evolution of the Andreev bound states from the trivial phase into the topological phase with Majorana bound states. Moreover, we show that the Josephson and the critical current exhibit non-trivial signatures when Majorana bound states are present, which should be experimentally accessible with the outstanding advance in fabrication techniques. 

\subsection*{Chapter 3}
Motivated by the experiments based on nanowires hybrid junctions coupled to superconductors, we focused on the study of transport in voltage biased short SNS Josephson junctions made of NWs with strong spin-orbit coupling, as the system undergoes into the topological superconducting phase for increasing the Zeeman field \cite{PabloJorge}. In this work, by means of the Keldysh Green's function technique, we proposed the multiple Andreev reflection (MAR) and critical currents as an alternative and powerful tool to study the topological transition. This is possible by the direct effect that gap inversion, MBS formation and fermion-parity conservation have on the MAR current at various junction transparencies. On the other hand, we also showed that the critical current remains unexpectedly finite for all Zeeman fields due to a significant continuum contribution, and exhibits an anomaly at the topological transition that could be experimentally traced.

\subsection*{Chapter 4}
To support new experiments and provide further insights towards the origin of Majorana-like physics in hybrid superconductor-semiconductor-superconductor junctions, we made a detailed study on the role that confinement and helicity have on normal transport and on the sub-gap Andreev spectrum in short and long SNS junctions made of semiconducting nanowires (NWs) with strong Rashba spin-orbit coupling \cite{PhysRevB.91.024514}. We identified different normal transport regimes that lead to very interesting physics when superconducting leads are attached. Indeed, we found that a long junction with a helical normal section, but still in the topologically trivial regime, supports a low-energy sub-gap spectrum consisting of multiple-loop structures and parity crossings that smoothly evolve towards Majorana bound states as the Zeeman field exceeds its critical value. This suggests an interesting connection between sub-gap parity crossings in helical junctions in the trivial phase and Majorana bound states in the topological one.
\subsection*{Chapter 5}
Motivated and ispired by our previous work, where we found a connection between sub-gap parity crossings in helical junctions, in the trivial phase, and Majorana bound states, 
in the topological phase, we have investigated a novel approach to engineer Majorana bound states in non-topological superconducting wires \cite{JorgeEPs}.
The recipe of our scheme consists that instead of inducing a topological transition in a proximized Rashba NW, 
we propose to create a sufficiently transparent normal-superconductor junction on a Rashba wire, with a topologically trivial superconducting side and a helical normal side.
The strong coupling to the half-metallic environment forces a single long-lived resonance to emerge at precisely zero energy above a threshold
transparency, which evolves as the transparency is increased further into a stable state
localised at the junction. The robust zero energy pinning is protected by electron-hole symmetry, and it is not the result of any fine tuning, due to eigenstate bifurcation of the scattering matrix poles at exceptional points. We show that relevant transport and spectral properties associated to these zero energy states, here dubbed exceptional point Majorana bound states, are indistinguishable from those of conventional Majorana bound states.
\subsection*{Chapter 6}
In this Chapter we analyse screening properties  in superconducting one-dimensional nanowires with SOC and with Zeeman interaction. 
Our study is based on the linear response theory using the Bogoliubov De Gennes formulation of excitations in a superconductor and on the Randon Phase Approximation (RPA) approach.
Firstly, we calculate the density-density response function and then we apply it to obtain the RPA dielectric function and the screening properties of these wires. This calculation is relevant for experiments trying to measure 
Majorana bound states and their non-trivial overlap (which might change depending on the screening potential inside the wire). 

\chapter{\bf SNS junctions made of nanowires with spin-orbit coupling\footnote{The results of this chapter are being prepared for publication.}} 
\label{Chap2a}
\lhead{Chapter \ref{Chap2a}. \emph{SNS junctions in nanowires with spin-orbit coupling}} 

\begin{small}
In this chapter, we formally introduce hybrid Normal metal-Superconductor (NS) and Superconductor-Normal metal-Superconductor (SNS) junctions. 
For a better understanding, we firstly describe a situation where the spin-orbit coupling and the Zeeman interactions are neglected. Here, we discuss the basic phenomena in these systems such as Andreev reflection (AR), Andreev bound states (ABSs) and Multiple Andreev reflections (MARs) based mainly on scattering arguments, and then we classify SNS junctions depending on the length of the normal region with respect to the superconducting coherence length. 

Later, we provide a full tight-binding description for modelling superconducting nanowires, NS and SNS junctions made of one-dimensional semiconducting nanowires with SOC in presence of Zeeman interaction, where superconducting correlations are induced via proximity effect.
Afterwards, we investigate the ABSs formation and the emergence of Majorana bound states in short and long SNS junctions, as well as in superconducting nanowires and NS junctions.
In this part, we present the Andreev bound states as function of the superconducting phase difference for different values of the Zeeman field, spanning the trivial and topological phases.
Then, we concentrate on the Josephson current dependence on the superconducting phase difference, calculated from the Adreev spectrum in short and long SNS junctions. We show that the presence of MBSs in the topological phase is a distinguishable signature in both the Andreev spectrum and the Josephson current despite of being $2\pi$ periodic.
We also present the critical current as a function of the Zeeman field, which allows to make a detailed study of the topological transition and trace the behaviour of the topological gap, which after the band inversion gives rise to the topological phase with MBSs. 
We report that the topological transition point is a robust feature, which we expect to be distinguishable from other mechanisms in real experiments.

\end{small}

\newpage

\section{Introduction}
All physical proposals for investigating the emergence of Majorana bound states (MBSs) include $s$-wave superconductors.
In the previous chapter, we have introduced the platform based on nanowires with strong spin-orbit coupling (SOC) placed on a trivial $s$-wave superconductor.
Assuming good contact between the superconductor and the nanowire, superconducting correlations are then induced into the nanowire via proximity effect, giving rise to a superconducting nanowire.
Then, by applying an external Zeeman  field, $B$, the system reaches the topological phase for $B>B_{c}\equiv\sqrt{\Delta^{2}+\mu^{2}}$ with Majorana bound states (MBSs) at the end of the wire, one at each end. Here, $\mu$ is the chemical potential of the wire and $\Delta$ the induced superconducting $s$-wave pairing.

Performing measurements on single superconducting nanowires is a challenging task and instead it is usually required to fabricate junctions between normal and superconducting 
systems. This allows to tune different system parameters in the different regions of the system and therefore perform,  for instance, transport measurements.
 These systems are known as hybrid, since they combine the effects of regions in the superconducting state with others in the normal state.
For instance, a nanowire can be partially placed on a superconductor. Then, due to the proximity effect, the region in contact with the superconductor acquires superconductivity (denoted by S) leaving the other region (the non-proximitized region) in the normal state (denoted by N). This results in a NS junction. Additionally, one can place another superconductor bellow the left part of the nanowire. In this geometry, superconducting correlations are induced onto the regions that are in 
contact with the superconductors and therefore leaving a central region of the wire in the normal state. This system is refereed to as SNS junction.

Important effects take place in these hybrid systems which lead to very unusual physics and provide a powerful platform for investigating topological superconductivity and therefore Majorana bound states. In this chapter we first describe the basic phenomena in NS and SNS structures and then we focus on the low-energy Adreev spectrum in NS and SNS junctions made of semiconducting nanowires with Rashba SOC and Zeeman interaction. Additionally, we study phase-biased transport in SNS junctions and suggest it as a useful tool for the search of MBSs in nanowires. 
\section{Basic phenomena in NS and SNS junctions}
\label{introhybridNSSNS}
In this section, we introduce two important effects which take place in NS and SNS junctions: the Andreev reflection in NS junction and the formation of Andreev bound states (ABSs)  and multiple Andreev reflections (MAR) in SNS junctions.
\subsection{NS junctions: Andreev reflection}
Semiconducting nanowire junctions represent a very important platform for investigating scattering processes.
Consider, for instance, a junction formed out by two regions, where a right moving electron hits the interface between them. At this point, the electron experiments reflection or transmission according to scattering theory and depend on various properties of the system under investigation.
Within the reflection process, the electron is reflected as an electron, if the right part of the system is a non-superconducting region. The situation, however, changes when the right part is a superconductor.

 \begin{figure}[!ht]
\centering
\includegraphics[width=.5\textwidth]{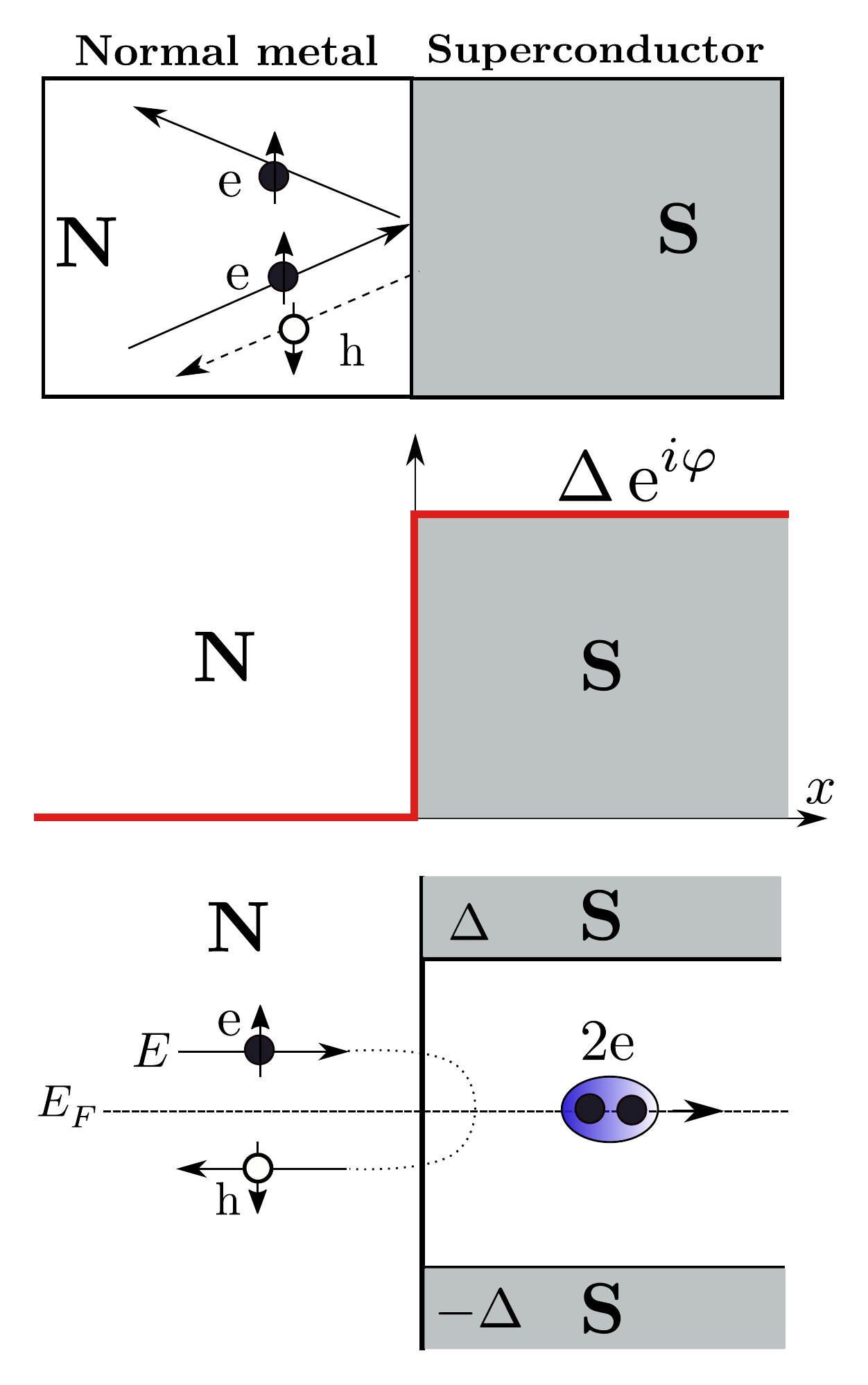} 
\caption[NS junction: profile of the order parameter and sketch of the Andreev reflection]{(Color online) 
 (Top) Schematic representation of a Normal metal (N)-Superconductor (S) junction. At the NS interface, a right-moving electron with energy $E$ can be reflected as a left-moving electron and also as a hole.
(middle) Profile of the superconducting pairing in the NS junction. The pairing potential is zero in the normal part, while finite in the superconducting one (red line), $\Delta {\rm e}^{i\varphi}$, where $\varphi$ is the superconducting phase. This picture is not completely true since the order parameter gets slightly suppressed in both sides close to the interface, however it represents a good platform for our interests. 
(bottom) Representation of the Andreev reflection process. At the NS interface, for $E<\Delta$ and within the Andreev approximation, an electron with energy $E$ is fully  reflected as a hole. As a result of this process a Cooper pair with charge $2e$ is transferred from the normal metal to the superconductor.}
\label{figchap21}
\end{figure}

Consider a NS junction, where neither SOC nor Zeeman effects are involved and superconductivity is allowed only in the S part, as sketched in Fig.\,\ref{figchap21}. In this case, the Fermi energy of the N side tends to be aligned with the Fermi energy of the S region. 
Now, suppose an electron is created in the normal region N with energy $E$, and travels towards the NS interface.
When such incident electron meets the NS interface it can be reflected or transmitted according to the scattering point of view. The analysis of this scattering problem is carried out by solving the BdG equations, given by Eq.\,(\ref{BdGEqs}), in the two regions. For determining the wave-functions and the energy spectrum, one follows the wave-matching technique taking into account the relation between the energy of the incident electron and the superconducting energy gap $\Delta$.
Indeed, when the energy of the incident electron is less than the superconducting gap, $E<\Delta$, there are no states in the S part available for transmission, and therefore the only possibility for the incident electron is to be fully reflected as a hole when there is no potential barrier at the interface. See Fig.\,\ref{figchap21}. 
This reflection process is known as \emph{Andreev reflection (AR)} \cite{andreev}. 
Because the charges of an electron and a hole are opposite, a charge of $2e$, a Cooper pair, is transferred from the normal metal into the superconductor.
Following wave-matching techniques, the Andreev reflection amplitude of an incoming electron and an outgoing hole can be written as
\begin{equation}
r_{eh}(\varphi)={\rm e}^{- i\varphi}\,{\rm e}^{-i{\rm arccos}\big(\frac{E}{\Delta}\big)}\,,
\end{equation}
where $\varphi$ is the superconducting phase. For the calculation of $r_{eh}$ it is also assumed Andreev approximation, meaning that $\mu\gg\Delta,E$, where $\mu$ is the chemical potential.
Following similar ideas and under similar physical conditions, an incident hole with energy $-E$ from the N region is  reflected as an electron with amplitude $r_{he}(\varphi)={\rm e}^{+ i\varphi}\,{\rm e}^{-i{\rm arccos}\big(\frac{E}{\Delta}\big)}\,
$. The Andreev reflection amplitude for a left going electron into a right going hole is $r_{eh}(-\varphi) $, while for a left going hole and a right going electron $r_{he}(-\varphi)$.
Notice that during this special reflection process the reflected quasiparticle acquires a phase from the superconductor, which leads to very interesting physics when a left superconducting contact is additionally attached to the NS junction (see next).
 \begin{figure}[!ht]
\centering
\includegraphics[width=.6\textwidth]{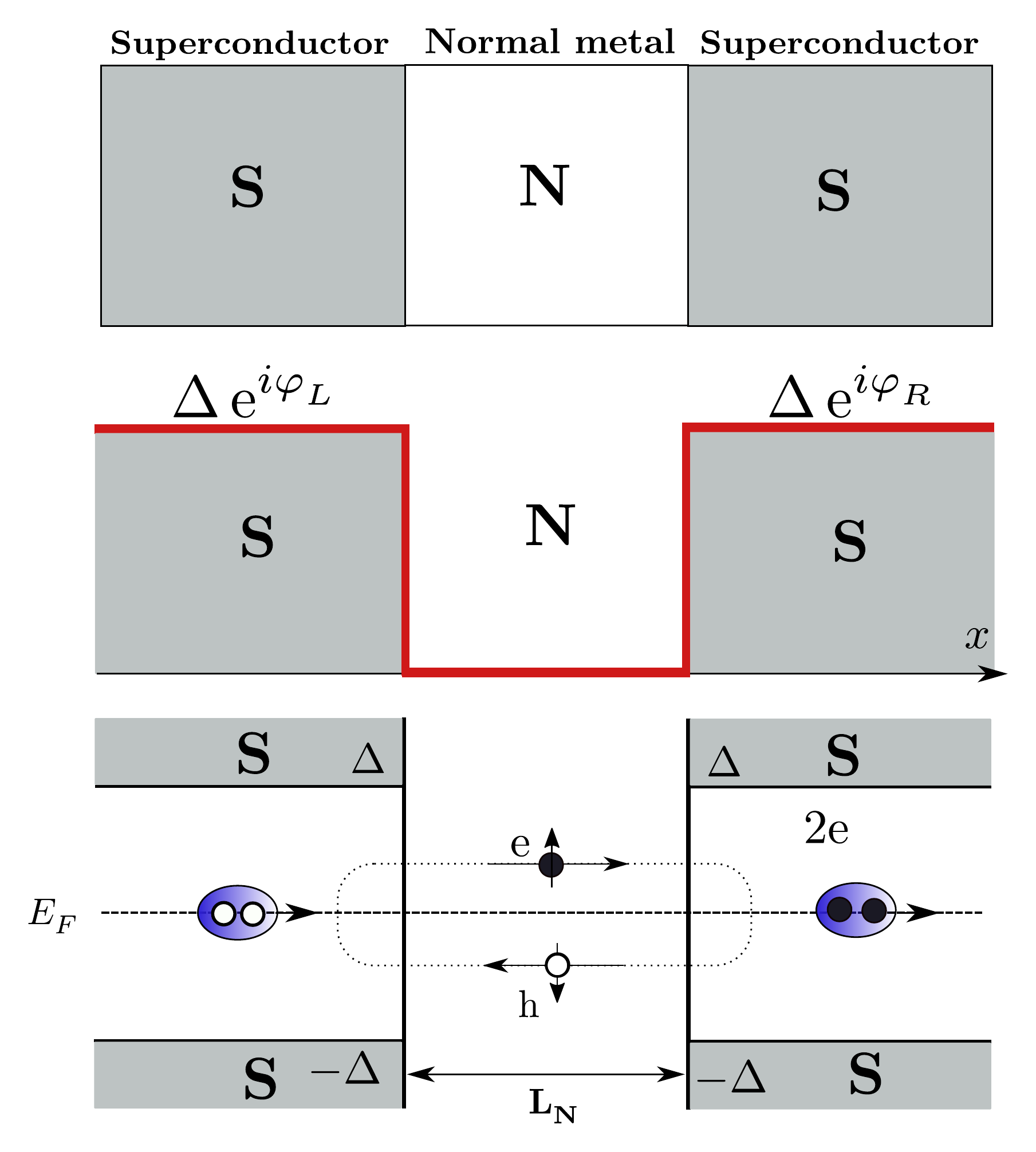} 
\caption[SNS junction: profile of the order parameter and sketch of ABSs]{(Color online) 
(Top) Schematic representation of a Superconductor (S)-Normal metal (N)-Superconductor (S) junction. At both interfaces, the Andreev reflection arises.
(middle) Profile of the pairing potential in a SNS junction. The pairing potential is zero in the normal part, while finite in the superconducting one (red line). 
(bottom) Representation of the emergence of an Andreev bound state. At the NS interface, for $E<\Delta$ and within the Andreev approximation, a right moving electron with energy $E$ is fully reflected into a left moving hole. The left moving hole is reflected back at the left SN interface into a right moving electron. As a result of this process a Cooper pair with charge $2e$ is transferred from the left to the right superconductor. The finite length of N and the finite phase difference $\varphi=\varphi_{L}-\varphi_{R}$ gives rise to a discrete energy spectrum also known as Andreev bound states.}
\label{figchap22}
\end{figure}
\subsection{SNS junctions: Andreev bound states}
Let us now attach another superconductor on the left side of our previous NS junction. This gives rise to the so-called SNS junction. Our system is now composed of a left superconductor (S) with pairing potential $\Delta_{L}=\Delta{\rm e}^{i\varphi_{L}}$, a central normal system (N) of length $L_{N}$ with zero superconducting pairing and a right superconductor (S) with $\Delta_{R}=\Delta{\rm e}^{i\varphi_{R}}$, as depicted in Fig.\,\ref{figchap22}. 
Notice that we focus on ballistic or perfectly transmitted SNS junctions. A brief discussion of finite transmission junction will be given at the end of this section.
Consider again a right-moving electron at low energy $E<\Delta$. According to the Andreev reflection process described in the previous section, such electron is reflected at the right NS interface into a left moving-hole, leaving a charge of $2e$ in the right superconductor.
 Then, the left-moving hole meets the left SN interface and is reflected back into a right-moving electron, taking a charge of $2e$ from the left superconductor. During one of these cycles a Cooper pair of charge $2e$ is transferred from the left to the right superconductor, thus creating a \emph{supercurrent} flow across the junction, as it is schematically shown in Fig.\,\ref{figchap22}.
 
Because of the finite length of  region N, $L_{N}$, discrete energy levels are formed. When the phase difference, $\varphi=\varphi_{L}-\varphi_{R}$, is non-zero, the SNS structure host standing bound waves with quantised energies. The corresponding states are referred to as \emph{Andreev bound states (ABSs)}
\cite{kuliksns}.
In general, the condition for the formation of bound states in one-dimension dictates that the total phase acquired during one cycle is a multiple of $2\pi$ \cite{zagoskin},
\begin{equation}
\label{SNSlevels}
-2{\rm arccos}\frac{E^{\pm}_{n}}{\Delta}\pm(\varphi_{L}^{}-\varphi_{R}^{})+[k_{+}(E^{\pm}_{n})-k_{-}(E^{\pm}_{n})]L_{N}=2\pi n\,, \,\, n=0,\pm 1,\ldots
\end{equation}
where $k_{\pm}(E)=k_{F}\sqrt{1\pm\frac{E}{\mu}}$, $k_{F}=\sqrt{2m\mu/\hbar^{2}}$, and the signs $\pm$ in front of $(\varphi_{L}^{}-\varphi_{R}^{})$ corresponds to a process starting with a right-moving electron (and its left-moving Andreev reflected hole), and to a process starting with a left-moving electron (and its right-moving Andreev reflected hole), respectively. 
Eq.\,(\ref{SNSlevels}) can be explicitly solved in two special limits: $L_{N}\rightarrow 0$ and $L_{N}\rightarrow \infty$. 
Within the Andreev approximation $E\ll \mu$, one can approximate $k_{+}(E^{\pm}_{n})-k_{-}(E^{\pm}_{n})\approx\frac{2E}{\hbar v_{F}}$, where $v_{F}=\hbar k_{F}/m$. Moreover, we consider states within the superconducting energy gap, $E<\Delta$.
Then, the two special limits give rise to the classification of such SNS junctions, which is based according to the relation between the length of the normal region $L_{N}$ 
and the superconducting coherence length $\xi=\frac{\hbar v_{F}}{\pi\Delta}$ \cite{RevModPhys.51.101}\footnote{We point out that the superconducting coherence length $\xi$
can be also defined without the factor $\pi$ in the denominator: $\xi=\hbar v_{F}/\Delta$.},
\begin{equation}
\begin{split}
L_{N}\ll\xi&\quad\Rightarrow\quad \textbf{short SNS junctions}\,,\\
L_{N}\gg\xi &\quad\Rightarrow \quad\textbf{long SNS junctions}\,.
\end{split}
\end{equation}
Such classification can be also given in terms of natural energy scales of the problem, the Thouless energy, $E_{T}=\hbar v_{F}/L_{N}$, 
and the induced superconducting pairing potential $\Delta$, being $v_{F}$ the Fermi velocity, and $L_{N}$ the length of the normal region. 
The above conditions, in terms of these energy scales, are 
\begin{equation}
\begin{split}
\Delta\ll E_{T}&\quad\Rightarrow\quad \textbf{short SNS junctions}\,,\\
\Delta\gg E_{T} &\quad\Rightarrow \quad\textbf{long SNS junctions}\,.
\end{split}
\end{equation}
Therefore, for a short junction the third term in Eq.\,(\ref{SNSlevels}) is neglected and we get two degenerated Andreev bound states
\begin{equation}
\label{shortABSs}
E_{\pm}(\varphi)=\pm\Delta\,{\rm cos}\Big(\frac{\varphi}{2}\Big)\,,
\end{equation}
where $\varphi=\varphi_{L}^{}-\varphi_{R}^{}$. The special case of $L_{N}\rightarrow0$, corresponds to the so-called \emph{weak link junctions}, where the only need for having ABSs is finite superconducting phase difference. This limit corresponds to the so-called \emph{superconducting point contact}.
For the long-junction regime, considering $E\ll\Delta$ and therefore setting ${\rm arccos}(E_{n}/\Delta)=\pi/2$, we have
\begin{equation}
\label{longABSs}
E_{\pm}^{n}(\varphi)=\frac{\hbar v_{F}}{2L_{N}}\,\Big[2\pi\Big(n+\frac{1}{2}\Big)\pm\varphi\Big]\,, \,\, n=0,\pm 1,\ldots\,.
\end{equation}
Notice that here we have a set of equally spaced energy levels until the levels approach $\Delta$, and the position of the levels is set by the superconducting phase difference $\varphi$.
 \begin{figure}[!ht]
\centering
\includegraphics[width=.9\textwidth]{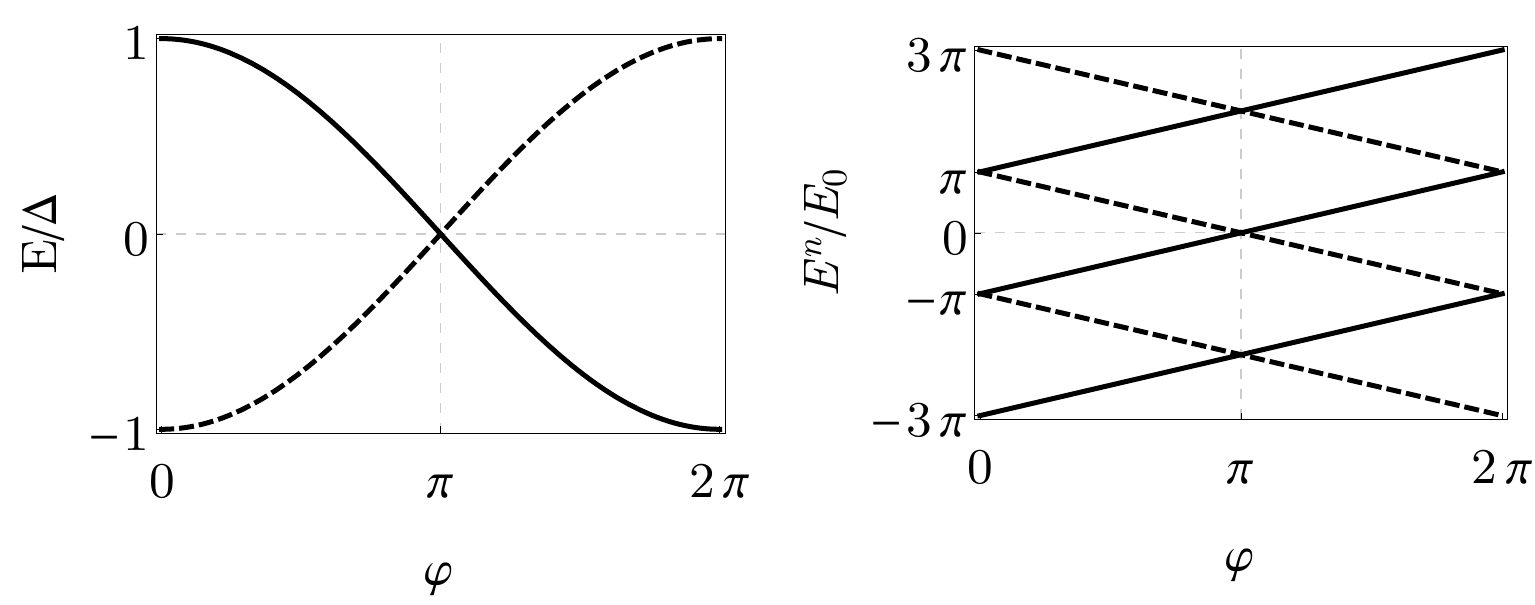} 
\caption[ABSs in ballistic short and long SNS junctions]{Energy levels in short (left), $L_{N}\ll\xi$ given by Eq.\,(\ref{shortABSs}), and long (right), $L_{N}\gg\xi$ given by Eq.\,(\ref{longABSs}), ballistic SNS junctions as function of the superconducting phase difference. In short junctions, two generate ABSs appear in the junction, while in the long junctions more levels fit within the junction. $E_{0}=\frac{\hbar v_{F}}{2L_{N}}$.}
\label{figchap23}
\end{figure}
In Fig.\,\ref{figchap23} we present the ABSs for short ($L_{N}\ll\xi$) and long ($L_{N}\gg\xi$) SNS junctions given by 
Eqs.\,(\ref{shortABSs}) and \,(\ref{longABSs}), respectively. 

\begin{figure}[!ht]
\centering
\includegraphics[width=.5\textwidth]{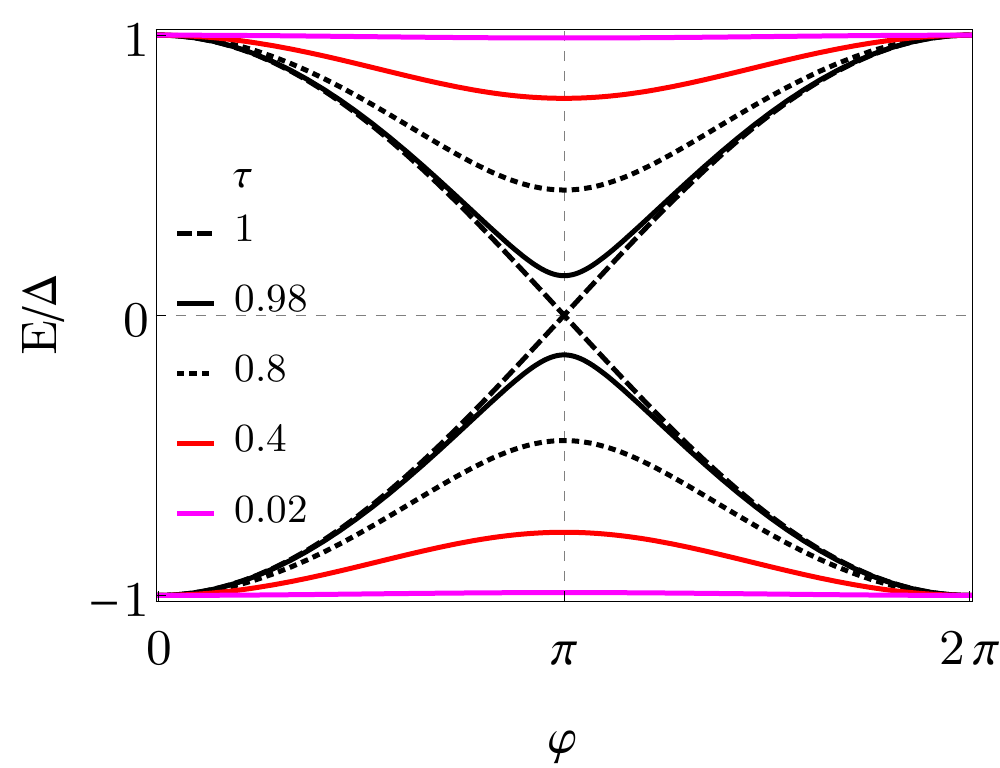} 
\caption[ABSs in a short SNS junction with finite transmission]{(Color online) Energy levels in a short SNS junction ($L_{N}\ll\xi$) with finite transmission as a function of the superconducting phase difference, given by Eq.\,(\ref{shorttauABSs}). A finite transmission $\tau$ leads to an avoided crossing at $\varphi=\pi$ between the two ABSs of a short SNS junction. For sufficiently small $\tau$, the ABSs become insensitive to the phase difference and tend to be at the edge of continuous quasiparticle spectrum in the superconductor.}
\label{figchap24}
\end{figure}

Notice that in the previous analysis we have considered ballistic SNS junctions, where elastic scattering is negligible. 
When the normal region N has a finite transmission $\tau$, there exists a finite backscattering probability in the normal region. Hence, in the case of a single channel short junction ($L_{N}\ll\xi$), a right-moving electron (or hole) can be reflected back into a left-moving electron (or hole), therefore giving rise to a coupling between the two ballistic ABSs, which have energies smaller than the superconducting gap $\Delta$ and are localised at the short junction over a distance of the order of $\xi$ \cite{kuliksns,Beenakker:92},
\begin{equation}
\label{shorttauABSs}
E_{\pm}(\varphi)=\pm\Delta\sqrt{1-\tau \,{\rm sin}^{2}\Big(\frac{\varphi}{2}\Big)}\,.
\end{equation}
The dependence of the energy levels on the superconducting phase difference $\varphi$, from previous expression, is plotted in Fig.\,\ref{figchap24} for different transmission values $\tau$. 
In the ballistic regime ($\tau=1$), two levels, corresponding to left-moving ($E_{+}$) and right-moving ($E_{-}$) electrons, cross at $\varphi=\pi$, while $\tau<1$ 
gives rise to an anticrossing at $\varphi=\pi$ of size $2\Delta\sqrt{1-\tau}$. Decreasing $\tau$, the opened gap increases and for sufficiently small transmission the ABSs lie at $\pm\Delta$. The transmission $\tau$ can be tuned electrostatically with gates for instance.

\subsection{The Josephson current}
As a result of the formation of Andreev bound states in SNS junctions a pair of correlated electrons is transferred from the left to the right superconductor, creating a supercurrent flow  across the junction. This phenomenon is known as the dc-Josephson effect and takes place as long as the superconducting phase difference is finite  \cite{RevModPhys.46.251}.
An important characteristic among the expressions given by Eqs.\,(\ref{shortABSs}), \,(\ref{longABSs}) and \,(\ref{shorttauABSs}) is the energy levels dependence on the superconducting phase difference between the two superconductors. This constitute the base of the Josephson effect \cite{RevModPhys.46.251}, which states that a finite current flows between the two superconductors when the superconducting phase difference is finite. In fact, the discovery of the Josephson effect was done in Superconductor-Insulator-Superconductor (SIS) tunneling junctions, where  Cooper pairs coherently tunnel through the insulating barrier between the superconductors and lead to a supercurrent flow which exhibits a dependence on the superconducting phase difference $I(\varphi)=I_{c}\,{\rm sin}(\varphi)$, being the critical current $I_{c}$ the maximum supercurrent that the junction can support \cite{RevModPhys.46.251}.

The Josephson current can be calculated from the thermodynamic potential, $F$, as a function of the superconducting phase difference between the two superconductors, $\varphi$,
\begin{equation}
\label{eqjc}
I(\varphi)=\frac{2e}{\hbar}\frac{dF}{d\varphi}\,,
\end{equation}
where $F$ thermodynamic potential (the grand canonical for instance) of the superconductor \cite{zagoskin, Beenakker:92}. The factor of $2$ in previous equation accounts for the Cooper pair charge and all the degeneracies are included in $F$.
Important to notice here is that previous formula is a general expression and remains valid for any kind of Josephson junction \cite{zagoskin}.
In order to find the current, one should be able to calculate $F$.  It can be done following the Green's function technique \cite{PhysRev.187.556} or directly from the BdG equations given by Eq.\,(\ref{BdGEqs}) \cite{zagoskin,Beenakker:92}. The latter approach leads to \cite{zagoskin,Beenakker:92}
\begin{equation}
\label{josehpcurrent}
I(\varphi)=-\frac{e}{\hbar}\sum_{p>0}{\rm tanh}\Big(\frac{E_{p}}{2\kappa_{B}^{}T} \Big)\frac{dE_{p}}{d\varphi}-
\frac{e}{\hbar}2\kappa_{B}^{}T\int_{\Delta}^{\infty}dE\,{\rm ln}\Big[2{\rm cosh}\Big(\frac{E}{2\kappa_{B}^{}T}\Big)\Big]\frac{d\rho_{c}^{}}{d\varphi}\,,
\end{equation}
where the first term represents the contributions from the discrete positive Andreev levels within the gap and the second term contains the contribution from the excited states in the continuum with energies above $\Delta$, whose density of states is $\rho_{c}$. Notice that in previous expression we have considered that our system is not spin degenerate, otherwise we would have obtained a factor of 2 multiplying the two terms on the right side.

In the short-junction limit ($L_{N}\ll\xi$), $\rho_{c}$ is the same as in a bulk superconductor and therefore is phase independent. Hence, in this regime, the energy levels from the continuous spectrum do not contribute to the Josephson current $I(\varphi)$ and Eq.\,(\ref{josehpcurrent}) at zero temperature, $T=0$, reduces to
\begin{equation}
\label{shortJosephcurrent}
I(\varphi)=-\frac{e}{\hbar}\sum_{p>0}\frac{dE_{p}}{d\varphi}\,.
\end{equation}

\begin{figure}[!ht]
\centering
\includegraphics[width=.9\textwidth]{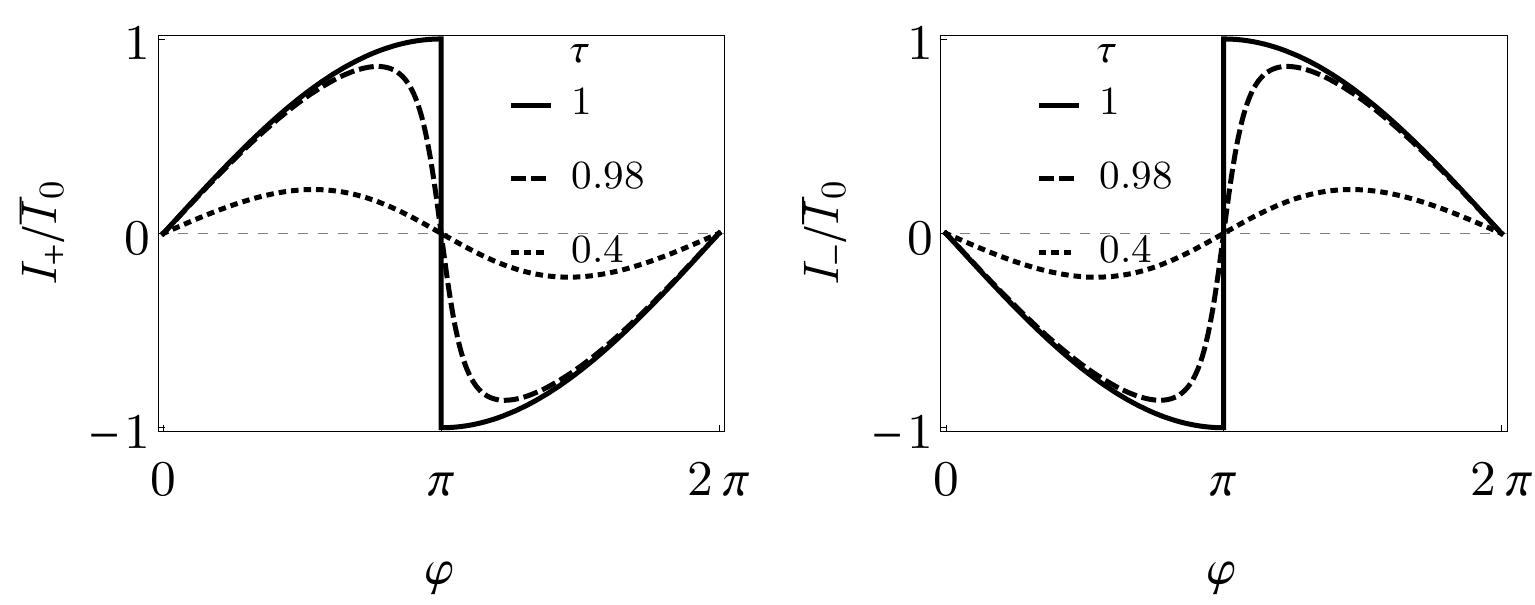} 
\caption[Josephson current in short SNS junction]{(Color online) Josephson current $I(\varphi)$ in a short SNS junction ($L_{N}\ll\xi$) carried by each ABS with finite transmission, given by Eq.\,(\ref{Ishort2}), where $\bar{I}_{0}=e\Delta/2\hbar$.}
\label{figchap25}
\end{figure}

For a short junction ($L_{N}\ll\xi$) we have obtained in  Eqs.\,(\ref{shorttauABSs}) two degenerate ABSs which are spin degenerate. Applying the previous approach for finding the Josephson current carried by the two ABSs, $I_{\pm}=-(2e/\hbar)dE_{\pm}/d\varphi$ (the factor of $2$ here counts for spin degeneracy). Then,
\begin{equation}
\label{Ishort2}
I_{\pm}(\varphi)=\pm\frac{e\Delta}{2\hbar}\frac{\tau\,{\rm sin}\varphi}{\sqrt{1-\tau\,{\rm sin}^{2}\big(\frac{\varphi}{2}\big)}}\,,
\end{equation} 
where each ABS carry supercurrent in opposite direction, which is plotted in Fig.\,\ref{figchap25}.
Notice that the supercurrent or Josephson current is an odd periodic function of $\varphi$ with period $2\pi$ and is zero at $\pi n$ for $n=0,1,\ldots$.
The current-phase characteristics strongly depends on the transmission $\tau$. 
For $\tau<<1$, the junction is in the tunnel regime and the current follows $I(\varphi)=I_{c}\,{\rm sin}\varphi$, where $I_{c}=(e\Delta/2\hbar)\tau$ is the critical current, 
which is the maximum supercurrent at $\varphi=\pi/2$. On the other hand, for $\tau=1$, one has  $I(\varphi)=I_{c}\,{\rm sin}(\varphi/2)$, where $I_{c}=(e\Delta/\hbar)\tau$ is the 
critical current, which is the maximum supercurrent at $\varphi=\pi$.

Notice that, in general, in a short junction with $N$ channels the number of ABSs within the gap is $4N$, where due to electron-hole symmetry of the Nambu description $2N$ are positive and 
$2N$ negative. 

In general, long junctions ($L_{N}\gg\xi$) host many energy levels within the gap $\Delta$ (see Fig.\,\ref{figchap23}), unlike the short-junction case.
We have shown that for $E\ll\Delta$ the energy levels are given by Eq.\,(\ref{longABSs}). For calculating the Josephson current, however, the levels above the gap may also contribute to the current as it is given by Eq.\,(\ref{josehpcurrent}).
Thus, we cannot just perform the derivative of the energy levels found in Eq.\,(\ref{longABSs}) with respect to $\varphi$, and then use Eq.\,(\ref{shortJosephcurrent}). 
We do not calculate the supercurrent in long junctions but rather discuss some differences with short junctions. 
It was shown that the Josephson current in a long junction at zero temperature is a $2\pi$ periodic function in the phase difference $\varphi$ and it has a linear relation with the phase gradient $\approx e v_{F}\varphi/L_{N}$ within $(-\pi,\pi)$, being zero at $\varphi=0$ and exhibiting a sawtooth profile at odd multiples of $\pi$ \cite{PhysRevB.5.72, ishisns}.

Along this thesis, however, we will not only use the approach presented in this part for calculating the Josephson current, but we also employ the Green's function formalism. The latter is described in Chapter\,\ref{Chap3}.

\subsubsection{Multiple Andreev reflections (MAR)}
\begin{figure}[!ht]
\centering
\includegraphics[width=.7\textwidth]{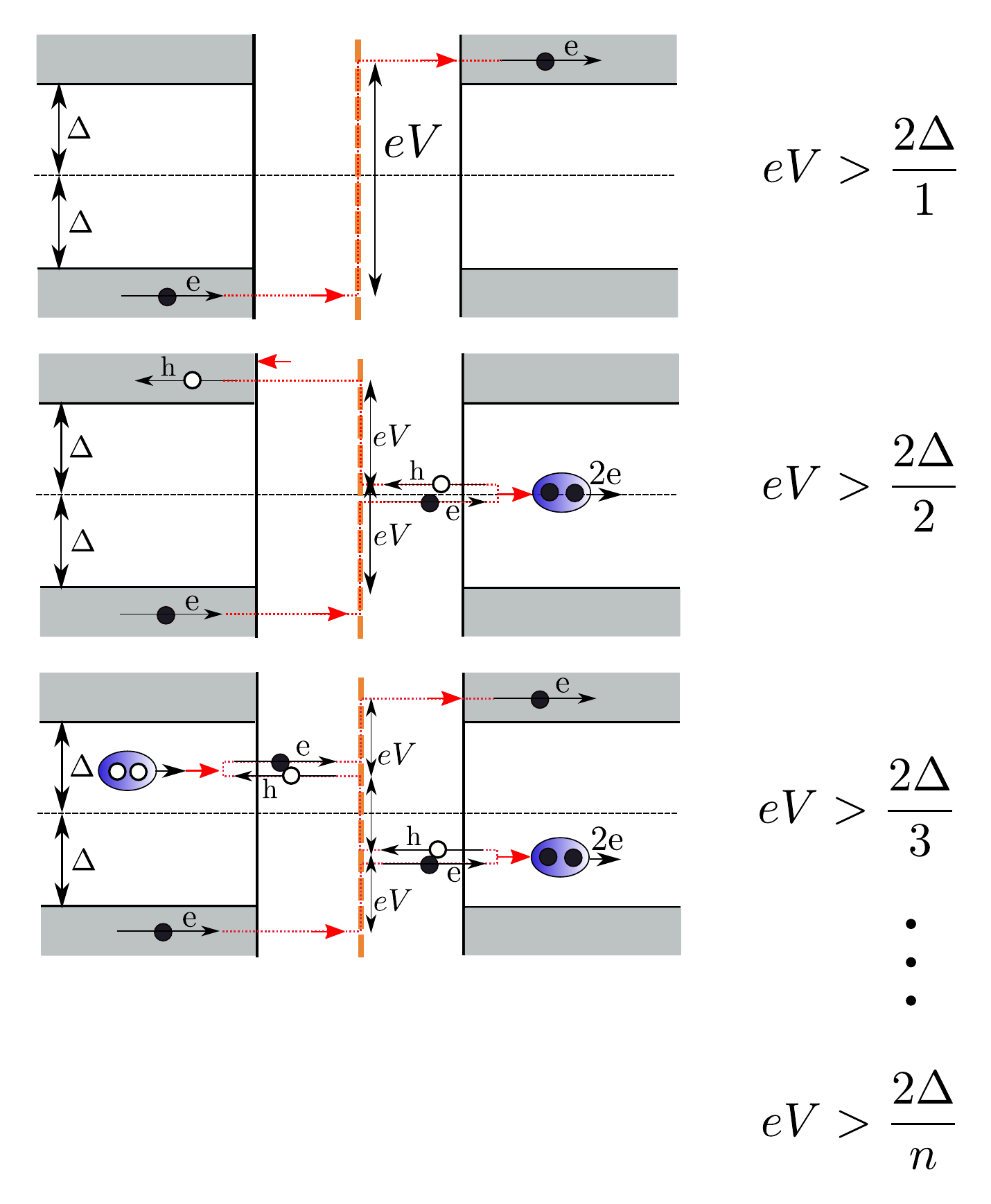} 
\caption[Multiple Andreev reflections (MAR)]{(Color online) Scattering processes in a voltage-biased SNS junction. Electrons (holes) gain energy $eV$ when crossing the dashed orange line (this denotes the voltage drop) from the left (right). Quasiparticles states are available in the continuum, shaded regions. Top panel: For $eV>2\Delta/1$, a single quasiparticle process is needed to transfer a charge of $1e$ from the left to the right superconductor. Middle panel: When $eV>2\Delta/2$ a single Andreev reflection takes places and a charge of $2e$ is transferred. For $eV<\Delta$, multiple Andreev reflections (MAR) take place until the energy of the incident particle exceeds the superconducting gap. The $n$ processes involves $n-1$ Andreev reflections, and starts with threshold voltages $eV_{n}>2\Delta/n$, which are reflected in the I-V curves at such threshold voltages.}
\label{figchapMAR}
\end{figure}

We have seen that in a SNS junction a right-moving electron is Andreev reflected as a hole at the right NS interface into a left-moving hole, which is also Andreev reflected into a right-moving hole at the left SN interface. Thus, Andreev reflections favour the formation of Andreev bound states. 
When a constant voltage $V$ is applied across the junction, the situation changes drastically. 
First, the superconducting phase difference becomes time-dependent and increases linearly with time $\varphi=(2eV/\hbar)t$. The Josephson current becomes oscillatory with $\omega_{J}=2eV/\hbar$, giving rise to the ac Josephson effect \cite{RevModPhys.46.251}.
We assume that the voltage drops over an arbitrary point of the N region, denoted by the orange dashed line in Fig.\,\ref{figchapMAR}
A right-moving electron (hole) increases (decreases) its energy each time it crosses the normal region N from left to right, 
while the Andreev reflected hole (electron) increases its energy passing through N from right to left.
Since energies are changing in the process of transmission, quasiparticle energies of both signs are considered and in left and right superconducting electrodes only quasiparticle states with $|E|>\Delta$ are available (Fig.\,\ref{figchapMAR}).
A right-moving electron from the left superconductor with energy below $-\Delta$ crosses the N region and arrives at the right superconductor only if $eV>2\Delta$ (top panel Fig.\,\ref{figchapMAR}). This represents a single quasiparticle process, where a charge of $1e$ is transferred from the left to the right superconductor.
For $eV$ slightly greater than $\Delta$, the right-moving electron arrives at the NS interface and performs an Andreev reflection, since it is the unique possibility as within $\Delta$ there are no available quasiparticle states. As a result, a left-moving hole emerges and when crossing the N region, it increases its energy by $eV$ arriving at the continuum of the left superconductor. During this process a charge of $2e$ is transferred from the left to the right superconductor.
For $eV<\Delta$, the right-moving electron from the left superconductor performs \emph{multiple Andreev reflections (MAR)} as needed to exit the superconducting gap. In each process, a charge of $2\Delta/eV$ is transferred, at least. The $n$ processes involves $n-1$ Andreev reflections, and starts with threshold voltages $eV_{n}>2\Delta/n$.
Important to remark is that a new transport channel arises whenever the full gap si an odd integer multiple of $eV$: $2\Delta/(2n+1)$, otherwise the right-moving electron ends up at the continuum of states of the left superconductor (second panel in Fig.\,\ref{figchapMAR}).
These processes give rise to a number of singularities in the I-V curves at the corresponding voltage.

\newpage\clearpage

\section{NS and SNS junctions made of Rashba nanowires: Details on modelling}
\label{SNSRashbaNWs}
In Chap.\,\ref{Chapter01} we have discussed the emergence of Majorana bound states (MBSs) in the Kitaev's toy model and then in a physical system based on Rashba nanowires. We have pointed out that MBSs emerge at interfaces of systems with different topology. 
Within the Rashba nanowire proposal what one needs is a nanowire with Rashba SOC placed on a $s$-wave superconductor and then apply a Zeeman field $B$, perpendicular to the SO axis.
The system exhibits two topologically different phases, which can be driven by varying $B$.
The topological transition is defined by the critical field $B_{c}\equiv\sqrt{\mu_{S}^{2}+\Delta^{2}}$, where $\mu_{S}$ is the chemical potential of the superconducting wire and $\Delta$ the induced pairing potential. For high fields, $B>B_{c}$, the superconducting nanowire is in the topological phase with Majorana bound states at its ends, one at each end. Here, the interfaces between two topologically different regions are between the topological superconducting nanowire and vacuum, which is in the trivial non-topological phase. 
We have also discussed that the wave-functions of the two Majorana bound states decay exponentially into the bulk of the superconducting wire \cite{DasSarma:PRB12,PhysRevB.86.085408}. 
The relevant decay length characterising this overlap is the Majorana localisation length $\ell_{M}$, which tells us how well are MBSs localised at the ends of the wire. For finite $L<2\ell_{M}$, being $L$ the length of the superconducting wire, the overlap between MBSs is significant and therefore they are no longer true zero modes. If there is a finite overlap between the MBSs wave functions, then the energy of the MBSs splits leaving behind their zero energy character. However, for long enough wires, they can still be considered as zero modes.

Junctions made of nanowires also represent experimental feasible platforms for the search of MBSs. 
Placing only the right region of a nanowire on a $s$-wave superconductor gives rise to NS junctions, while 
placing a nanowire on top of two $s$-wave superconductors, as making a bridge, lead to SNS junctions.
In this section we describe how to model superconducting nanowires, NS and SNS junctions based on the  semiconducting nanowires with SOC. Then, we perform a full analysis of the low-energy Andreev spectrum in such junctions and finally we focus on supercurrents and critical currents.

The calculations presented in this part are carried out considering experimental values 
for InSb nanowires: the electron's effective mass $m=0.015m_{e}$, the spin-orbit strength $\alpha_{R}=20$\,meVnm and typical induced pairing gap of $\Delta=0.25$\,meV \cite{Mourik:S12}. The other parameters such as the chemical potential is considered arbitrary since in principle it can be controlled by means of electric gates.
\newpage
\subsection{Tight-Binding discretization}
\label{TBM}
\begin{figure}[!hb]
\centering
\includegraphics[width=.8\textwidth]{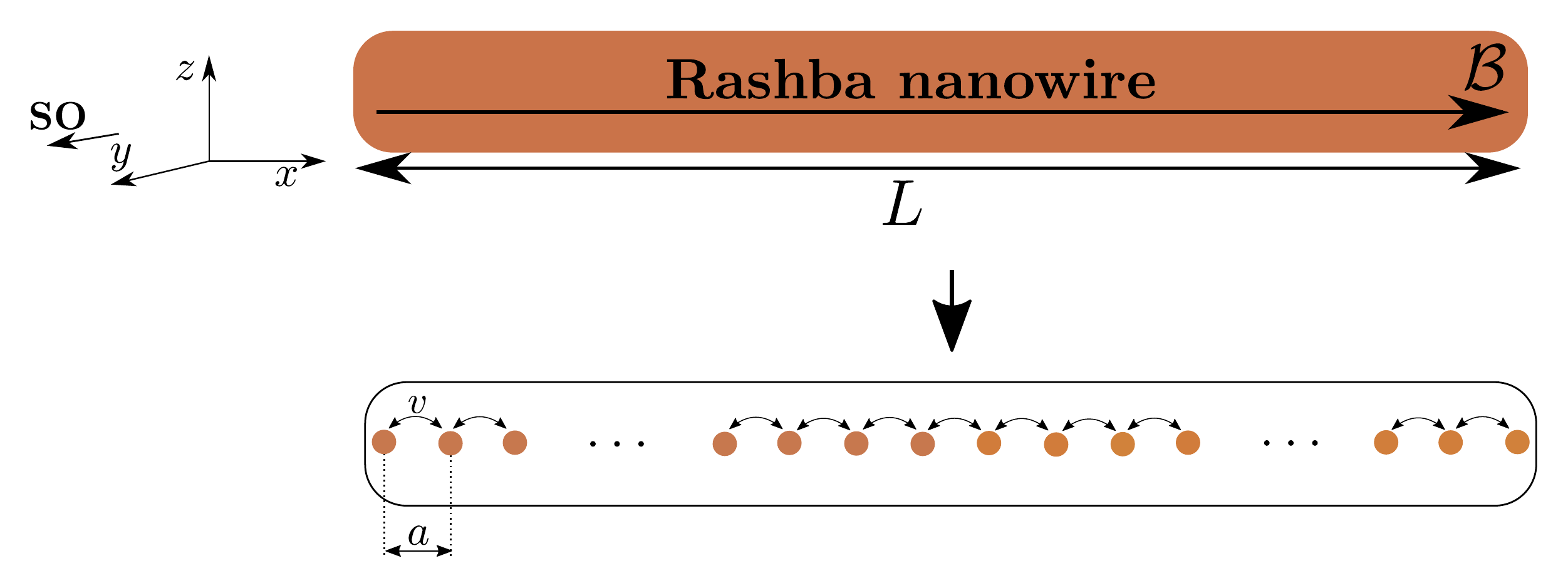} 
\caption[Schematic for tight-binding discretization]{A Rashba nanowire subjected to a magnetic field $\mathcal{B}$ is discretised into a tight-binding lattice with lattice spacing $a$.  The length of the system is $L=Na$, where $N$ is the number of sites and $a$ the lattice spacing. The applied magnetic field opens a gap of $2B$ in the energy spectrum as shown in Sec.\,\ref{Rashbawire}, where $B=g\mu_{B}\mathcal{B}/2$ is the associated Zeeman field in the wire, $g$ is the wire's $g$-factor and $\mu_{B}$ the Bohr magneton.}
\label{figchap27x}
\end{figure}

For computation purposes, we consider a discretisation of the 1D continuum model given by Eq.\,(\ref{H0Hamil}) for the Rashba nanowire into a tight-binding (TB) lattice with a small lattice spacing $a$, see Fig.\,\ref{figchap27x}.
The smaller is $a$, the better is the description and by letting such lattice constant tend to zero one recovers the usual continuum limit.
We choose a discrete lattice whose points are located at $x=a\,i$, where $i$ is an integer and $a$ the small lattice spacing. Notice that by doing so, we are trying to find a matrix representation of the 1D continuum model in site space.
Within the TB approach, Eq.\,(\ref{H0Hamil}) reads
\begin{equation}
\label{HamilTB}
H_{0}\,=\,\sum_{i}c_{i}^{\dagger}\,h\,c_{i}\,+\,\sum_{<ij>}c_{i}^{\dagger}\,v\,c_{j}\,+\,\text{h.c}\,,
\end{equation}
where the symbol $<ij>$ means that $v$ couples nearest-neighbor $i,j$ sites and 
\begin{equation}
\label{hopp}
h_{ii}\,\equiv\,h\,=\,
\begin{pmatrix}
2t\,-\,\mu&B\\
B& 2t\,-\,\mu
\end{pmatrix}\,,\,
h_{i+1,i}\,\equiv\,v\,=\,
\begin{pmatrix}
-t&t_{SO}\\
-t_{SO}& -t
\end{pmatrix}\,=\,h_{i,i+1}^{\dagger}\,,
\end{equation}
are matrices in spin space, $t=\hbar^{2}/2m^{*}a^{2}$ is the hopping parameter and $t_{SO}=\alpha_{R}/2a$ the SO hopping. The dimension of the matrix $H_{0}$ is set by the number of sites of the wire.

In the following subsections we describe how one can model superconducting wires, NS and SNS junctions based on the tight-binding description presented here.

\newpage
\subsection{The superconducting wire model}
\begin{figure}[!ht]
\centering
\includegraphics[width=.7\textwidth]{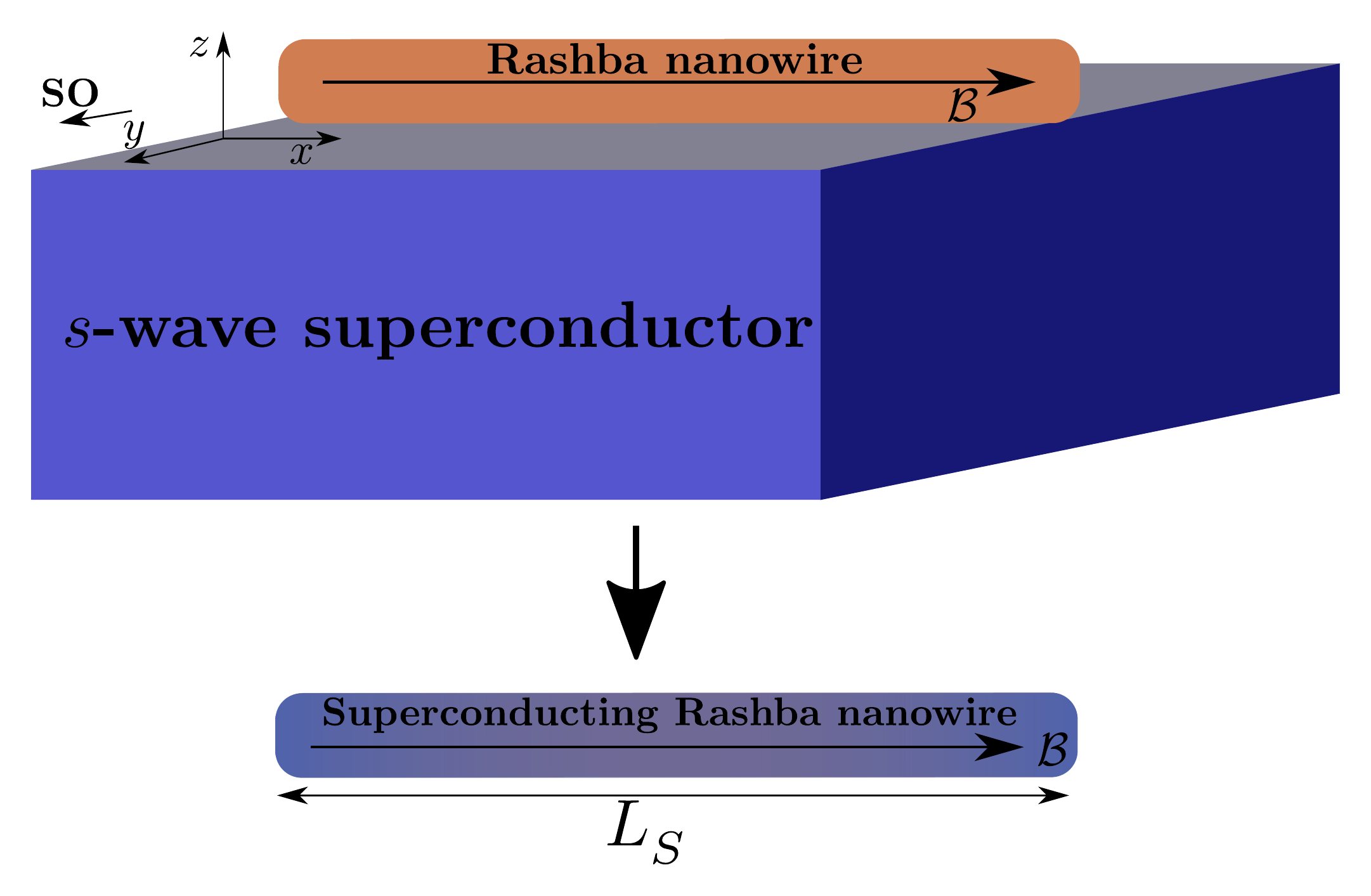} 
\caption[Superconducting Rashba nanowire]{(Color online) Top: A Rashba nanowire is placed on top of a $s$-wave superconductor and an external magnetic field $\mathcal{B}$, perpendicular to the spin-orbit axis, is applied. The applied magnetic field opens a gap of $2B$ in the energy spectrum as shown in Sec.\,\ref{Rashbawire}, where $B=g\mu_{B}\mathcal{B}/2$ is the associated Zeeman field in the wire, $g$ is the wire's $g$-factor and $\mu_{B}$ the Bohr magneton.
Bottom: Due to the proximity effect, this results in a superconducting Rashba nanowire of length $L_{S}=N_{S}a$, where $N_{S}$ is the number of sites of the superconducting wire and $a$ the lattice spacing.}
\label{figchap28}
\end{figure}

Consider a semiconducting nanowire placed on top of a superconductor, similar to the original proposal discussed in Sec.\,\ref{Rashbawire} of Chap.\,\ref{Chapter01}.
Assuming good contact between the nanowire and the superconductor, superconducting correlations are induced into the nanowire via proximity effect, resulting in a superconducting wire. This is schematically shown in Fig.\,\ref{figchap28}.

The nanowire is single mode and modelled with the Hamiltonian given by Eq.\,(\ref{H0Hamil}), which is discretised into a tight-binding lattice resulting in the Hamiltonian given by Eq.\,(\ref{HamilTB}).  The length of the superconducting wire is $L_{S}=N_{S}a$, where $N_{S}$ is the number of sites and $a$ the lattice spacing.
Superconducting correlations are of $s$-wave type as described in Sec.\,\ref{Rashbawire} of Chap.\,\ref{Chapter01} and the pairing potential is given by
\begin{equation}
\Delta(x)=i\sigma_{y}\,\Delta\,{\rm e}^{i\varphi}\,,
\end{equation}
where $\varphi$ is the superconducting phase and $\sigma_{y}$ the $y$ Pauli matrix.
The full Hamiltonian describing the superconductivity nanowire is written in Nambu space
\begin{equation}
\label{SCNW}
H=
\begin{pmatrix}
H_{0}&\Delta(x)\\
\Delta^{\dagger}(x)&-H_{0}^{*}
\end{pmatrix}\,.
\end{equation}
Previous Hamiltonian is diagonalised numerically and exactly solved with the dimensions given by the number of sites of the wire, which implies a dependence on the length of the wire.
Previous description accounts for finite length wires and therefore it is important upon investigating the overlap of MBSs discussed in in Sec.\,\ref{Rashbawire} of Chap.\,\ref{Chapter01}. The superconducting phase in the order parameter is assumed to be zero in our calculations of this part as it is only relevant when investigating Andreev bound states (see next sections).
\begin{figure}[!ht]
\centering
\includegraphics[width=.99\textwidth]{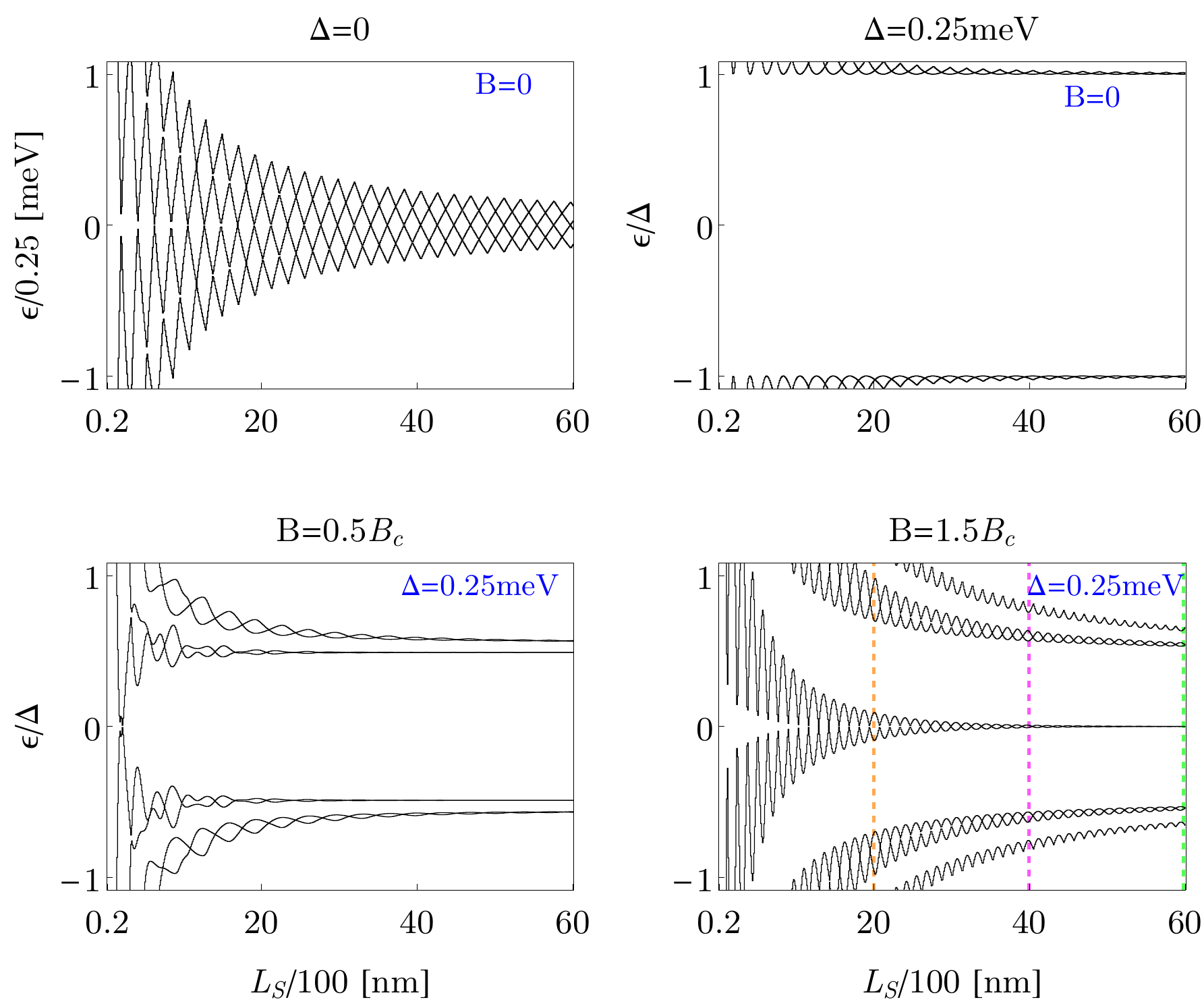} 
\caption[Energy levels as function of the wire's length for a Rashba nanowire]{(Color online) Energy levels as function of the wire's length $L_{S}$ for a superconducting Rashba nanowire at zero and finite Zeeman field, top and bottom rows, respectively.
Top row: at zero superconducting pairing, $\Delta=0$ (top left panel), the spectrum is gapless, while for $\Delta\neq0$ a clear induced gap is opened (top right panel).
Bottom row: as a Zeeman field is applied, the induced gap is reduced. While in the trivial phase, $B<B_{c}$, there are no levels within the reduced gap (bottom left panel), in the topological phase, $B>B_{c}$, two energy levels decay and develop and oscillatory behaviour as the length of the wire, $L_{S}$, is increased. For long enough wires, the the amplitude of the oscillations is reduced and such levels become zero energy.
Parameters: $\alpha_{R}=20$\,meVnm, $\mu=0.5$\,meV.}
\label{SC1}
\end{figure}

\begin{figure}[!ht]
\centering
\includegraphics[width=.99\textwidth]{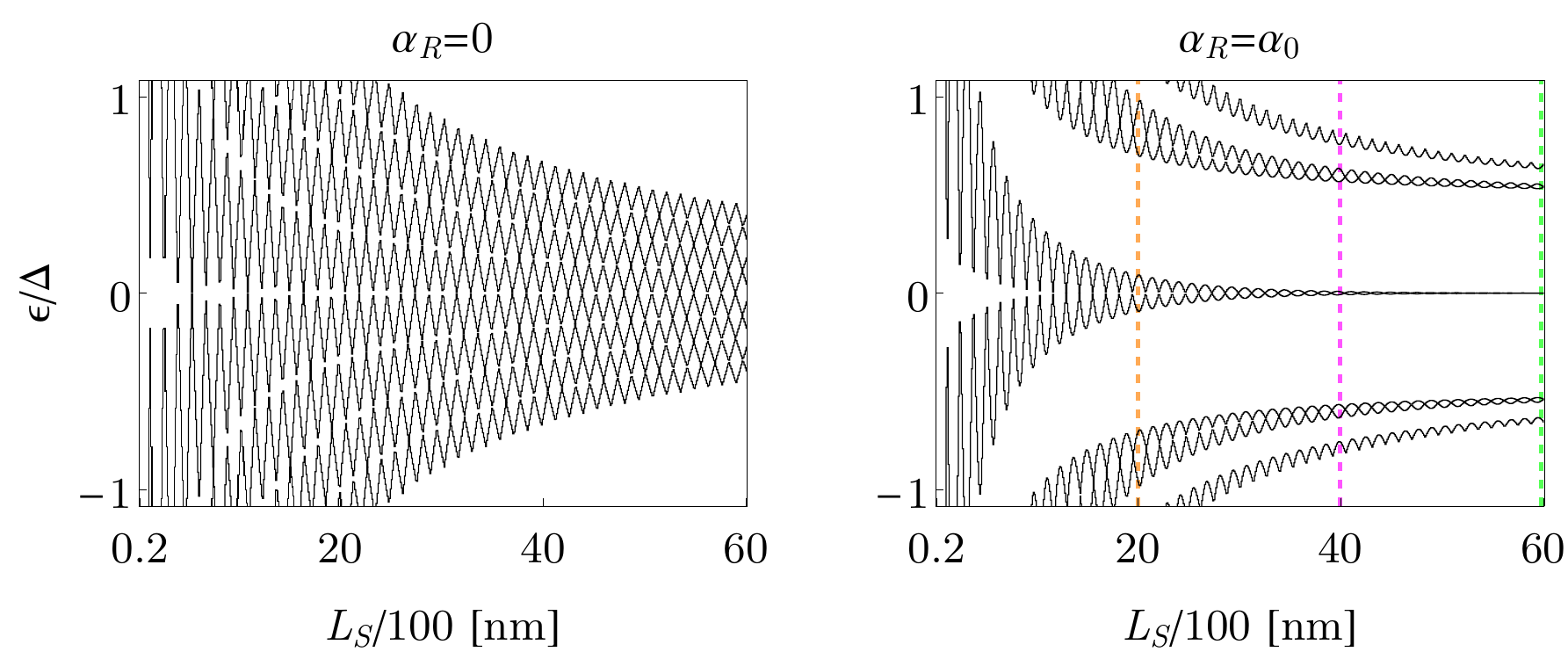} 
\caption[Energy levels as function of the wire's length for a Rashba nanowire in the topological phase]{(Color online) Energy levels as function of the wire's length $L_{S}$ for a superconducting Rashba nanowire in the topological phase, $B=1.5B_{c}$. Left panel show the case for zero SOC, while the right one for finite SOC. 
Notice that a finite SOC in the superconducting wire opens a gap in the energy spectrum and originates two decaying levels with oscillatory behaviour as the length of the wire increases (see orange dashed line). For long enough wires such levels become zero energy leading to MBSs (see magenta and green dashed lines).
Parameters: $\alpha_{0}=20$\,meVnm, $\mu=0.5$\,meV and $\Delta=0.25$\,meV.}
\label{SC2}
\end{figure}

Based on the Hamiltonian model given by Eq.\,(\ref{SCNW}), we first analyse what happens to the low-energy spectrum as a function of the wire's length $L_{S}$. 
This is shown in Figs.\,\ref{SC1} and \,\ref{SC2}.
Top row in Fig.\,\ref{SC1} shows the energy spectrum of a normal nanowire  at zero Zeeman field  for zero and finite superconducting pairing, respectively. 
While in the normal state (top left panel) the energy spectrum is gapless, by inducing superconductivity with a finite $\Delta$ (top right panel), a clear gap is opened in 
the energy spectrum, as expected, and remains finite as the length of the wire is increased. This is in accordance with the Anderson's theorem which prevents the existence of bound states inside the gap of an s-wave superconductor for non-magnetic impurities \cite{Anderson-theorem}.
For a finite Zeeman field (bottom row), but still in the trivial phase (left bottom panel), the induced gap, opened by superconductivity,
 gets reduced  as expected according to the description given in Sec.\,\ref{Rashbawire}. 
 While in the trivial phase, $B<B_{c}$, there are no levels within the induced gap (bottom left panel), in the topological phase, $B>B_{c}$, two energy levels  from energies around $\Delta$ develop an oscillatory decaying behaviour as the length of the wire, $L_{S}$, is increased. As such length is made really long, the amplitude of the oscillations is reduced and even totally negligible, where the levels become zero energy (see green dashed line). These lowest energy levels represent the oscillating MBSs, which for long wires become zero energy.
 This discussion is in accordance with description made in Sec.\,\ref{Rashbawire}. Indeed, the low momentum gap gets reduced as $B$ increases and closes at $B=B_{c}$ marking the topological phase transition point and for $B>B_{c}$ the system is in the topological superconducting phase with MBSs.

Before going further, in Fig.\,\ref{SC2} we demonstrate the importance of the SOC towards the emergence of MBSs in nanowires.
We examine the energy spectrum of a Rashba nanowire for  $B>B_{c}$ at zero and at finite SOC (left and right panels in Fig.\,\ref{SC2}).
While at zero SOC, left panel, the spectrum consists of a dense quasi-continuum as the length $L_{S}$ increases, a finite SOC removes all finite energy crossings while preserving the lowest two oscillatory decaying levels coming from energies around $\Delta$. The oscillatory pattern is visible for $L_{S}<2\ell_{M}$, while for long enough wires $L_{S}>2\ell_{M}$ the oscillations are considerable reduced and assumed to be zero, as marked by magenta and green dashed lines. 
Notice that $\ell_{M}$ represents the Majorana localisation length. We therefore conclude that, as expected for $B>B_{c}$, a topological superconducting nanowire host two MBSs (one at each end of the wire) which  exhibit an energy splitting when their overlap is finite,  while for long enough wires they become zero energy.

 \begin{figure}[!ht]
\centering
\includegraphics[width=.99\textwidth]{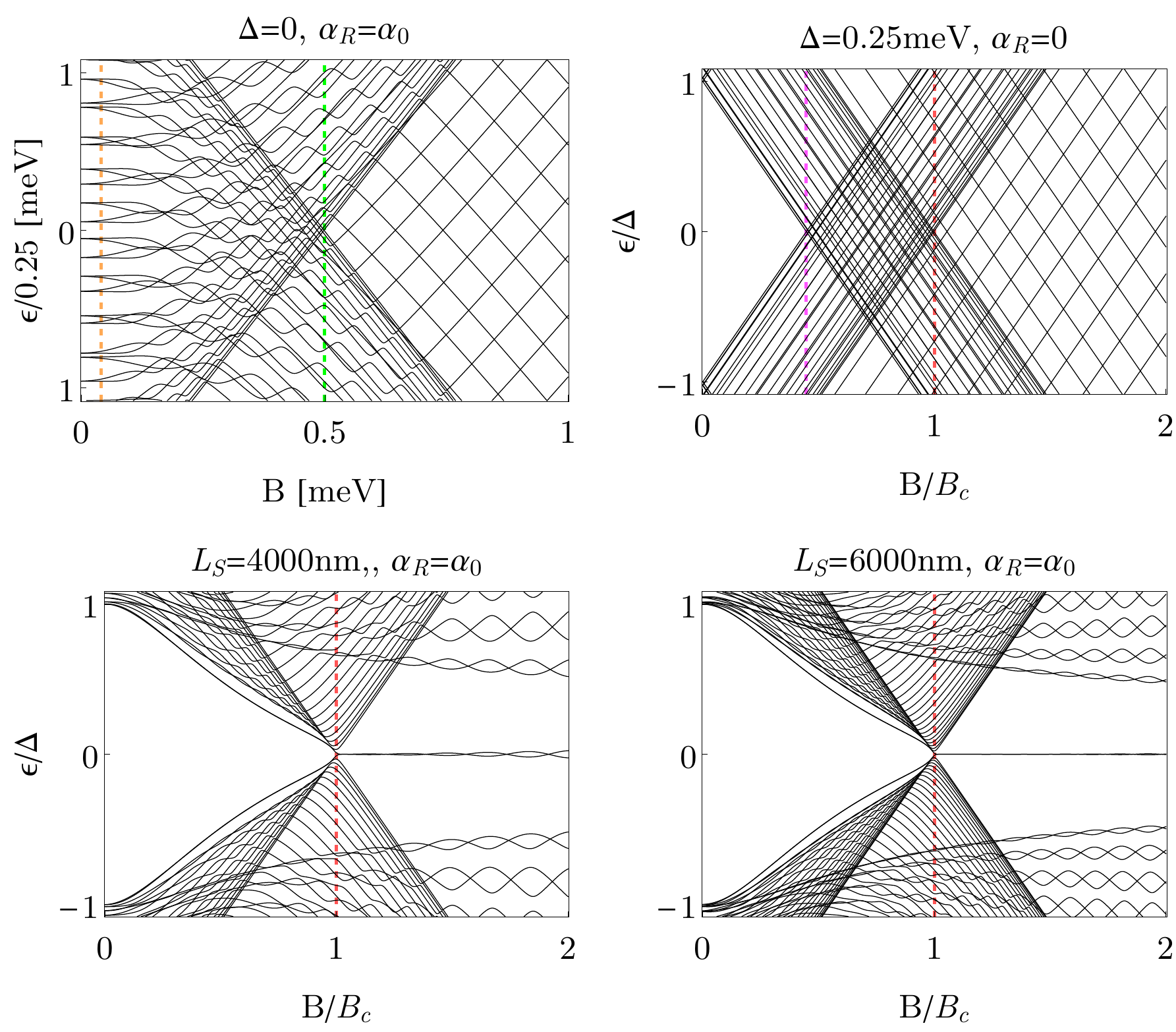} 
\caption[Energy levels as function of the Zeeman field]{(Color online) Energy levels as function of the Zeeman field for a Rashba nanowire at zero superconducting pairing 
with finite SOC (top left panel) and at finite superconducting pairing with zero SOC (top right panel). Bottom panels correspond to finite both 
pairing and SOC for $L_{S}=4000$\,nm and $L_{S}=6000$\,nm.
Notice that as the Zeeman field increases for $B>B_{c}$ two MBSs emerge and oscillate around zero energy.  
Parameters: $\alpha_{0}=20$\,meVnm, $\mu=0.5$\,meV and $\Delta=0.25$\,meV and $L_{S}=4000$\,nm (for top panels).}
\label{SC3}
\end{figure}
In order to clarify the previous discussion, we show in Fig.\,\ref{SC3} the low-energy spectrum as a function of the Zeeman field at fixed wire's length (see caption for more details).
Top left  panel represents the case of zero superconducting pairing, $\Delta=0$, and finite SOC, $\alpha_{R}\neq0$, while top right panel a situation of 
finite pairing, $\Delta\neq0$, but with zero SOC, $\alpha_{R}=0$.  Bottom panels show a calculation for finite both SOC and superconducting pairing for $L_{S}<2\ell_{M}$ and
$L_{S}>2\ell_{M}$, respectively.
In Fig.\,\ref{SC1} (top left panel) we have seen that at $B=0$ and $\Delta=0$ the energy spectrum consists of a dense quasi-continuum. 
The same is observed in Fig.\,\ref{SC3} (top left panel). Notice that as the Zeeman field takes finite values and increases, the energy levels split as shown in top 
left panel for $\Delta=0$ and $\alpha_{R}\neq0$. As the Zeeman field increases, within the weak Zeeman phase $B<\mu$, the energy levels contain both spin components and for 
certain values of the Zeeman field they may cross zero energy, while in the strong Zeeman phase $B>\mu$ one spin sector is completely removed giving rise to a spin-polarised 
spectrum as one can indeed observe in top left panel of Fig.\,\ref{SC3}. The transition point from weak to strong Zeeman phases is marked by the chemical potential $\mu$ 
(green dashed line), while the spin-orbit energy is marked by the orange dashed line. We point out that none of the zero energy crossings are protected in top left panel of Fig.\,\ref{SC3}.

Top right panel of Fig.\,\ref{SC3} shows the energy levels at finite superconducting pairing, $\Delta\neq0$, and zero SOC, $\alpha_{R}=0$.
The first thing one notices, in comparison to the top left panel of Fig.\,\ref{SC3}, is that at zero Zeeman field the superconducting pairing induces a gap with no levels 
within it according to the Anderson theorem as discussed before \cite{Anderson-theorem}. The magnetic field applied to a finite length superconducting nanowire tends to destroy 
the pairing of the electrons, inducing the so-called Zeeman depairing. Here, the spins of the electrons couple with the field, the up and down levels shift by $\pm B/2$, and roughly 
speaking, when $B$ exceeds the induced superconducting gap $\Delta$ the superconductivity ceases. This is indeed observed in top right panel of Fig.\,\ref{SC3}. As the Zeeman 
increases, the energy levels from energies around $\Delta$ get coupled to the Zeeman field and reduce their energies, thus reducing the induced gap. 
Different spin components of the energy levels cross zero energy at $B=\Delta$, signalling the closing of the induced superconducting gap, marked by the magenta dashed line. 
Further increasing of the Zeeman field $\Delta<B<B_{c}$ gives rise to a region (between red and magenta dashed lines) where the energy levels contain both spin components, 
which depend on the finite value of the chemical potential. Indeed, we have checked that for $\mu=0$ magenta and red dashed lines coincide (not shown).
On the other hand, for $B>B_{c}$, one spin sector is removed and the energy levels are spin polarised, showing a family of Zeeman crossings, which are not protected.

\begin{figure}[!ht]
\centering
\includegraphics[width=.99\textwidth]{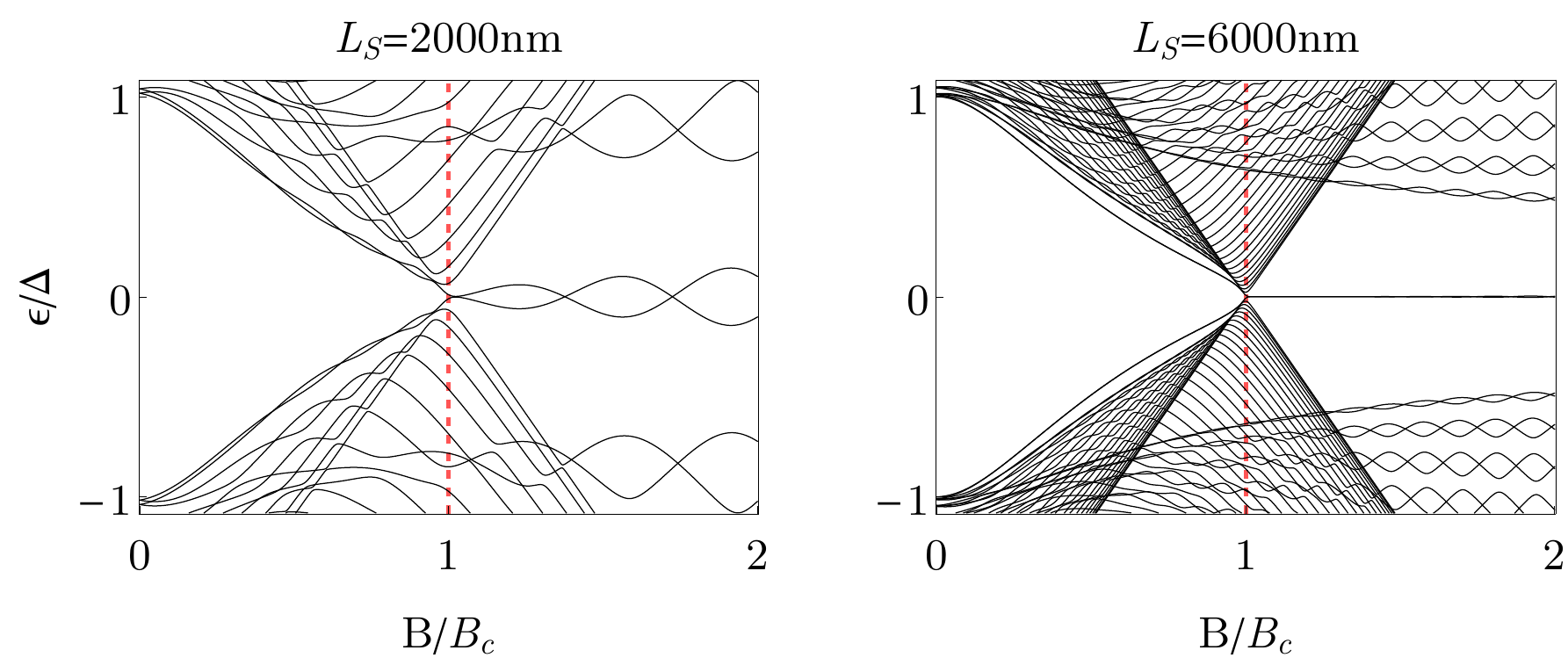} 
\caption[Energy levels with the Zeeman field: Rashba nanowire]{(Color online) Energy levels as function of the Zeeman field for a superconducting Rashba nanowire with length (left) $L_{S}=2000$\,nm and (right) $L=6000$\,nm. Notice that by increasing the Zeeman field, the gap closes at $B_{c}$ and for $B>B_{c}$ two MBSs emerge and their energy oscillates, as predicted by theory \cite{DasSarma:PRB12}. By increasing the length of the wire (see left panel, bottom panel in Fig.\,\ref{SC3} and right panel in this figure), the oscillations are considerable reduced and the MBSs can be considered as zero modes.
Parameters: $\alpha_{R}=20$\,meVnm, $\mu=0.5$\,meV and $\Delta=0.25$\,meV.}
\label{energysplitting}
\end{figure}

Remarkably, when the SOC is switched on we observe a number of changes in the low-energy spectrum (bottom panels).
First, the gap closing changes from $\Delta$, shown in top right panel for zero SOC, to $B_{c}=\sqrt{\Delta^{2}+\mu^{2}}$ (bottom panels). 
The critical field $B_{c}$ is the field which marks the topological transition point into the topological superconducting phase with MBSs.
Second, a clear closing at $B=B_{c}$ and reopening for $B>B_{c}$ of the induced gap is observed as the Zeeman field increases.
Third, the spin polarised energy spectrum shown in top right panel at zero SOC for $B>B_{c}$ is washed out, keeping only the crossings around zero energy of the two lowest levels. 
The two lowest energy levels for $B>B_{c}$ are the celebrated MBSs, which in this case exhibit the expected oscillatory behaviour due to their finite spatial overlap. 
For long enough wires, the amplitude of the oscillations is considerable reduced (even negligible) thus these levels acquire zero energy (see next discussion for Fig.\,\ref{energysplitting}). 
Fourth, the SOC introduces a finite energy separation (\emph{minigap}) between the lowest levels, crossings around zero, and the rest of the levels. The value of this minigap is related to the outer gap we have discussed in Sec.\,\ref{Rashbawire}, where we pointed out that such minigap is roughly constant for strong SOC.

In order to support the view on the lowest levels for $B>B_{c}$ and have a deeper insight about the topological transition, in Fig.\,\ref{energysplitting} we present the energy levels of a superconducting nanowire as function of the Zeeman field for $L_{S}<2\ell_{M}$ and $L_{S}>2\ell_{M}$. In both panels, the low-energy levels react to the increasing of the Zeeman field and trace the gap closing at $B=B_{c}$, as discussed previously. By further increasing of the Zeeman field, $B>B_{c}$, two energy levels remain around zero energy separated from the rest of the spectrum by a minigap, which is strongly dependent on the SOC.
The energy levels around zero oscillate when the Zeeman field $B$ increases giving rise to an energy splitting between them, as predicted by Eq.\,(\ref{esplitting}), \cite{DasSarma:PRB12}.  
The energy splitting can be small for long enough wires, see right panel in Figs.\,\ref{energysplitting} and \,\ref{SC2}, though strictly speaking it is non-zero for any length \cite{Lim:PRB12,Prada:PRB12,Rainis:PRB13,DasSarma:PRB12}. The MBSs become merely quasi-stationary states, with their energy splitting representing a Rabi frequency at which one oscillates into the other. It is also important to mention here that above $\Delta$ there is a considerable amount of discrete levels which become visible when the length of the wire is increased as observed in Fig.\,\ref{energysplitting}. This is due to the fact that since we are in a finite length nanowire the energy spectrum is discrete, where part of it is within the gap and the other above the gap forming the quasi-continuum. The discrete quasi-continuum will have important consequences in the calculation of the Josephson current across SNS junctions (see Sec.\,\ref{ABSsJIc}).

Thus, to conclude, we strongly point out that this part proves that only the combination of SOC, $s$-wave superconductivity and Zeeman interaction gives rise to MBSs at the end of the wire for $B>B_{c}$ in accordance with the original proposals \cite{PhysRevLett.105.177002,Lutchyn:PRL10}.
\subsection{The NS junction model }
\begin{figure}[!ht]
\centering
\includegraphics[width=.6\textwidth]{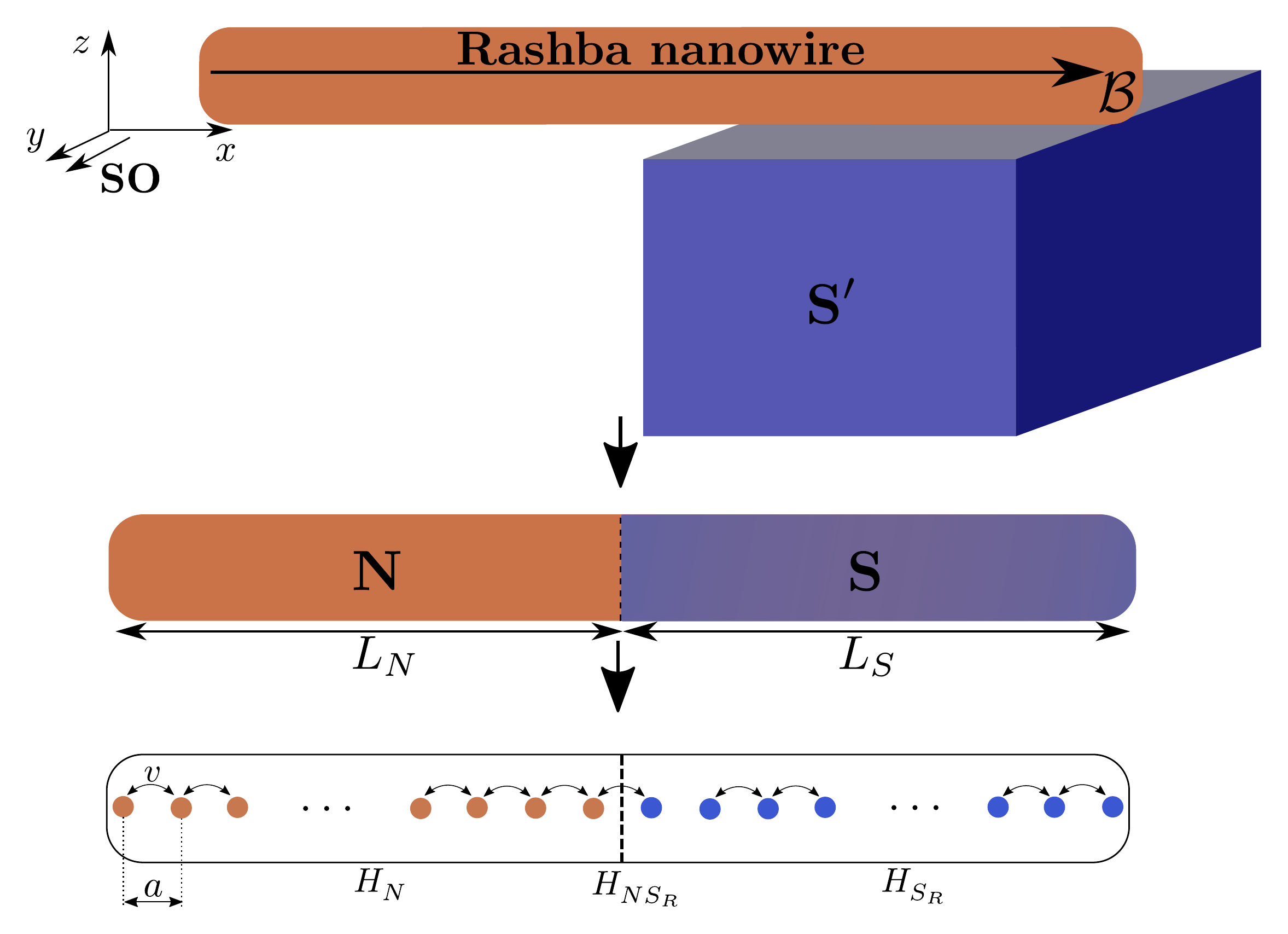} 
\caption[Sketch: NS junction]{(Color online) Top: A Rashba nanowire is partially placed on a $s$-wave superconductor and an external magnetic field $\mathcal{B}$ is applied along the wire such that it is perpendicular to the spin-orbit axis.  The applied magnetic field opens a gap of $2B$ in the energy spectrum as shown in Sec.\,\ref{Rashbawire}, where $B=g\mu_{B}\mathcal{B}/2$ is the associated Zeeman field in the wire, $g$ is the wire's $g$-factor and $\mu_{B}$ the Bohr magneton.
Due to the proximity effect the right part of the wire acquires superconductivity (denoted by S), leaving normal the left part of the nanowire (denoted N). This results in a NS junction of length $L_{S}+L_{N}$ (middle) which is discretized into a tight-binding lattice with spacing $a$ (bottom). $L_{S}=N_{S}a$ and $L_{N}=N_{N}a$ are the lengths of the N and S regions, being $N_{S}$ and $N_{N}$ the number of sites of S and N, respectively.}
\label{NSfig}
\end{figure}

Now, we consider a nanowire with Rashba SOC partially placed on top of a $s$-wave superconductor as schematically shown in Fig.\,\ref{NSfig}. The right part of the nanowire acquire superconducting correlations ($S_{R}$) and the left region remains in the normal state (N). This results in a NS junction.
Notice that an external magnetic field $\mathcal{B}$ is applied along the wire and perpendicular to the SO axis, which opens a gap of $2B$ in the energy spectrum, where $B=g\mu_{B}\mathcal{B}/2$ is the associated Zeeman field in the wire, $g$ is the wire's $g$-factor and $\mu_{B}$ the Bohr magneton.
 
For modelling the NS structure, we consider that the regions N and S are 
described by the tight-binding Hamiltonian $H_{0}$ given by Eq.\,(\ref{HamilTB}), with their respective chemical potential, Zeeman field, hopping parameters,
\begin{equation}
H_{N}=H_{0}(\mu_{N},B_{N},t_{N}, t_{SO,N})\,,\quad H_{S_{R}}=H_{0}(\mu_{S},B_{S},t_{S}, t_{SO,S})\,,
\end{equation}
where N and S denotes the normal and superconducting regions of the junction.
We will assume that only the chemical potentials are different in the N and S regions. Notice that the NS junction of length $L_{S}+L_{N}$ is discretized into a tight-binding lattice with spacing $a$ following Subsec.\,\ref{TBM}. Here, $L_{S}=N_{S}a$ and $L_{N}=N_{N}a$ are the lengths of the N and S regions, being $N_{S}$ and $N_{N}$ the number of sites of S and N, respectively.

The junction NS without superconducting correlations is therefore modelled by the Hamiltonian 
\begin{equation}
h_{NS}=
\begin{pmatrix}
H_{N}&H_{NS_{R}}\\
H_{NS_{R}}^{\dagger}&H_{S_{R}}
\end{pmatrix}\,,
\end{equation}
where $H_{N}$ and  $H_{S_{R}}$ is the Hamiltonian for the normal and superconducting regions given by Eq.\,(\ref{HamilTB}) and the Hamiltonian $H_{NS_{R}}$ couples the normal region to the superconducting part and contains non-zero elements for adjacent sites that lie at the interfaces of $N$ and $S_{R}$. 
One can control the transmission in these NS junctions by a hopping matrix $v_{0}=\tau v$, 
$\tau\in[0,1]$, which parametrises the coupling between the sites that define the interfaces of the NS junction. A tunnel junction can be modelled by considering $\tau\ll1$, while a fully transparent junction with $\tau=1$. We consider fully transparent junctions unless otherwise stablished.
The $s$-wave pairing potential must have the same structure as previous Hamiltonian. 
Therefore, 
\begin{equation}
\Delta(x)=
\begin{pmatrix}
\Delta_{N}&0\\
0&\Delta_{S_{R}}
\end{pmatrix}\,=\,
\begin{pmatrix}
0&0\\
0&\Delta_{S_{R}}
\end{pmatrix}\,,
\end{equation}
where $\Delta_{N}=0$ and $\Delta_{S_{R}}=i\sigma_{y}\,\Delta\,{\rm e}^{i\varphi}$
are the pairing potentials in the normal and in the superconducting regions, respectively. As for superconducting nanowires, the superconducting phase is assumed to be zero in our calculations for NS junctions.

Therefore, the full Hamiltonian describing the NS junction is given by
\begin{equation}
H_{NS}=
\begin{pmatrix}
h_{NS}&\Delta(x)\\
\Delta^{\dagger}(x)&-h_{NS}^{*}
\end{pmatrix}\,.
\end{equation}
In NS junctions described here (see  Fig.\,\ref{NSfig})  the S region becomes topological  for $B>B_{c}$ and therefore
Majorana bound states emerge at its ends in a similar fashion as in the topological superconducting wire with the sole difference that now the left part of the superconducting wire is in its normal state. We will see that this N part plays an important role as it introduces additional energy levels into the energy spectrum and therefore reduces the minigap which separates the MBSs from the rest of the levels in the topological phase.

In the discussion made for topological superconducting wires we have observed that in the trivial phase there are no levels within the induced gap when the length of the wire $L_{S}$ is increased (see bottom panels in Fig.\,\ref{SC1}). We have also seen that the low-energy spectrum of such superconducting wires in the topological phase develops decaying energy levels with an oscillatory behaviour as the length of the wire is increased. Here, the lowest two levels represent the Majorana bound states of the wire which oscillate due to the finite spatial overlap between their wavefunctions and they acquire zero energy when the length of the wire is long enough $L_{S}>>\ell_{M}$, being $\ell_{M}$ is the Majorana localization length. 
 \begin{figure}[!ht]
\centering
\includegraphics[width=.98\textwidth]{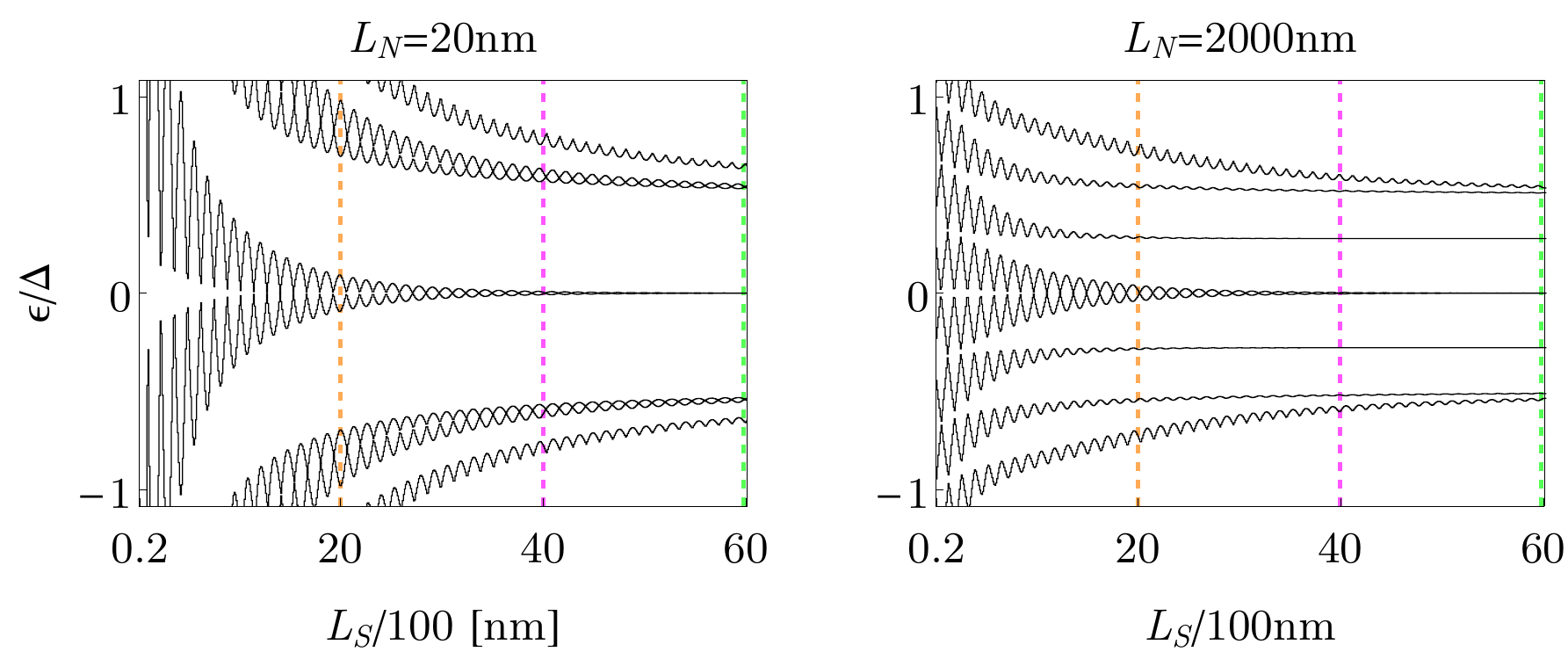} 
\caption[Energy levels in a NS junction as function of $L_{S}$]{(Color online) Energy levels in a NS junction as function of the length of the superconducting region $L_{S}$ both in the topological phase at $B=1.5B_{c}$: (left) for $L_{N}=20$\,nm and  (right) for $L_{N}=2000$\,nm. In both cases, by increasing $L_{S}$, two energy levels emerge from finite energy and become zero energy levels for long enough wires. In the latter, however, the mini-gap separating the zero modes from the rest of the levels is smaller than in the former. This suggests that a NS junction with a short N section is more practical in terms of Majorana modes protection. 
Parameters: $\alpha_{R}=20$\,meVnm, $\mu=0.5$\,meV and $\Delta=0.25$\,meV.}
\label{NS1}
\end{figure}

Now, we focus on the role of the length of the normal region $L_{N}$ in the low-energy spectrum of the NS junction.
This is reported in Fig.\,\ref{NS1}, where we present the energy levels as function of the  length of the superconducting region $L_{S}$ for a very short (left panel) and long (right panel) N regions. 
In the former, two levels emerge from the gap $\Delta$ and exponentially decay developing an oscillatory behaviour that is suppressed as the length of the superconducting region $L_{S}$ increases. This case is very similar to the one shown in right panel of Fig.\,\ref{SC2} for the topological superconducting wire. The latter, however, posses some differences. Indeed, when the length of the N region is increased, the number of energy levels of the system increases and even inside the gap one finds levels which coexist with the MBSs (see right panel in Fig.\,\ref{NS1}). This has a very important consequence as such levels affect the mini-gap by reducing it almost to a half of the one in case of a very short N region.
This apparent negative role of having long N regions is not completely true. In fact, the additional levels from N reduce the minigap but they also reduce the amplitude of the oscillations developed by the MBSs around zero energy. Compare for instance dashed magenta lines in both panels of Fig.\,\ref{NS1}. Therefore, the condition for having zero modes in NS junctions with long N region is reached before than in NS junctions with short N junctions. This is a very important conclusion and represents an advantage over NS junctions with short N regions towards practical implementation of MBSs for testing their braiding statistics. 

   \begin{figure}[!ht]
\centering
\includegraphics[width=.99\textwidth]{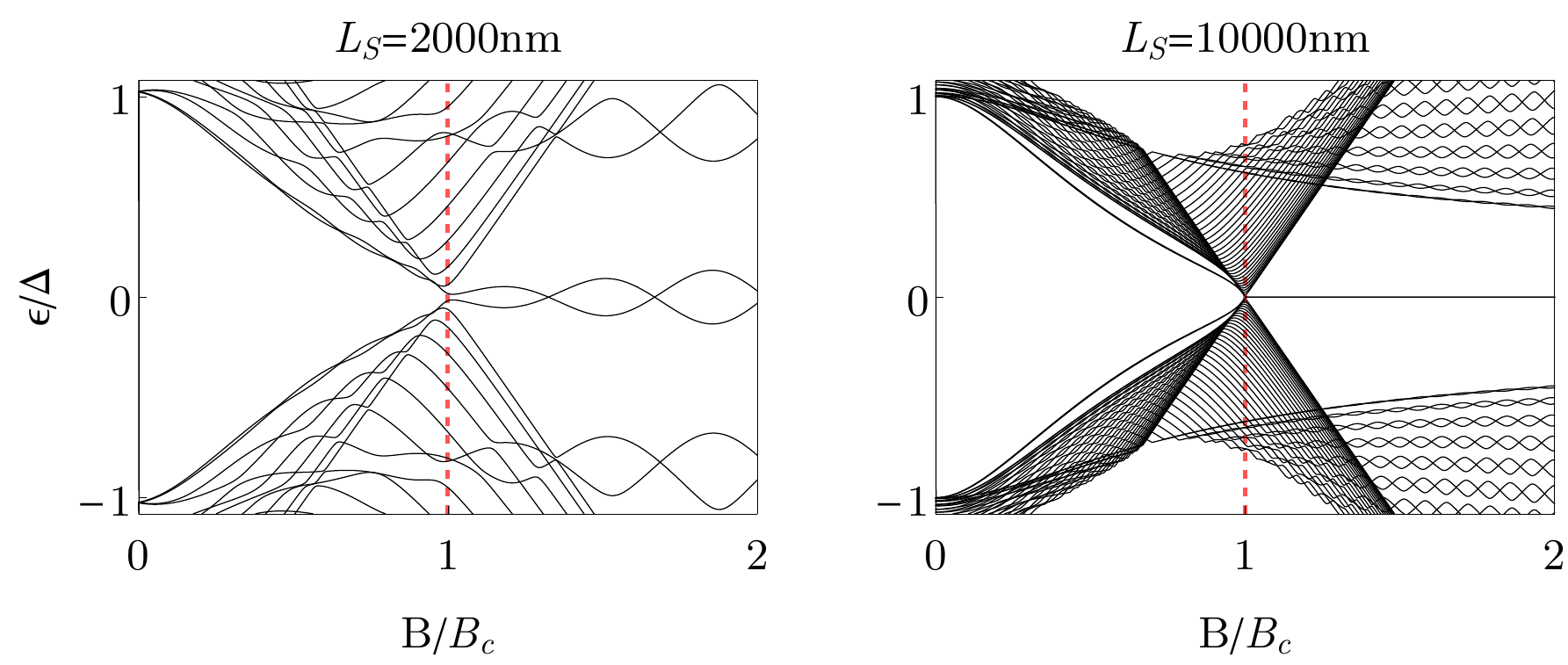} 
\caption[Energy levels in a NS junction as function of $B$: $L_{N}=20$\,nm]{(Color online) Energy levels in a NS junction as function of the Zeeman field for $L_{N}=20$\,nm: (left) $L_{S}=2000$\,nm and (right) $L_{S}=10000$\,nm.
 In both cases, by increasing $L_{S}$, two energy levels emerge from finite energy and become zero energy levels for long enough wires. In the latter, however, the mini-gap separating the zero modes from the rest of the levels is smaller than in the former. This suggests that a NS junction with a short N section is more practical in terms of Majorana modes protection. 
Parameters: $\alpha_{R}=20$\,meVnm, $\mu=0.5$\,meV and $\Delta=0.25$\,meV.}
\label{NS2}
\end{figure}

 \begin{figure}[!ht]
\centering
\includegraphics[width=0.99\textwidth]{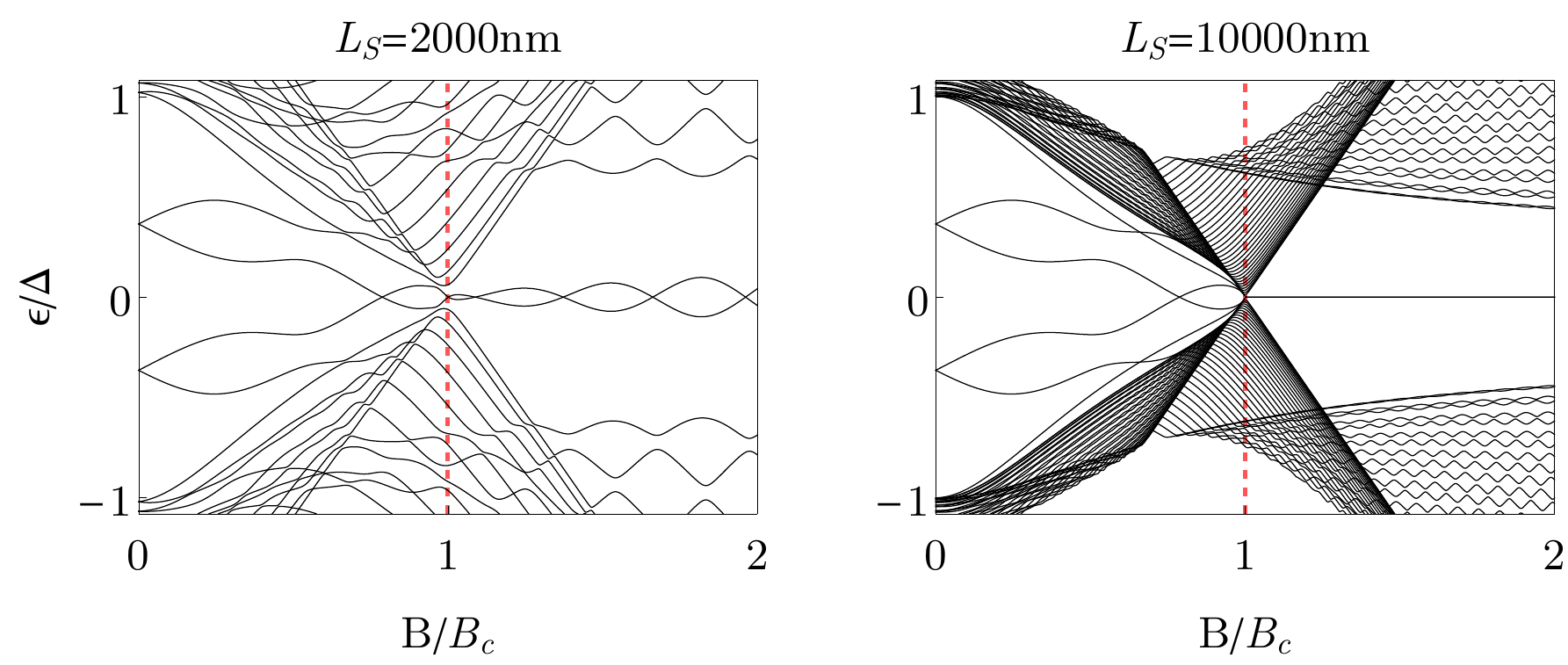} 
\caption[Energy levels in a NS junction as function of $B$: $L_{N}=400$\,nm]{(Color online) Energy levels in a NS junction as function of the Zeeman field for $L_{N}=400$\,nm: (left) $L_{S}=2000$\,nm and (right) $L_{S}=10000$\,nm.
 Unlike Fig.\,\ref{NS2} more levels enter from N inside the gap $\Delta$ and therefore the separation of the MBSs from the rest of the levels is reduced.
Parameters: $\alpha_{R}=20$\,meVnm, $\mu=0.5$\,meV and $\Delta=0.25$\,meV.}
\label{NS3}
\end{figure}

\begin{figure}[!ht]
\centering
\includegraphics[width=0.99\textwidth]{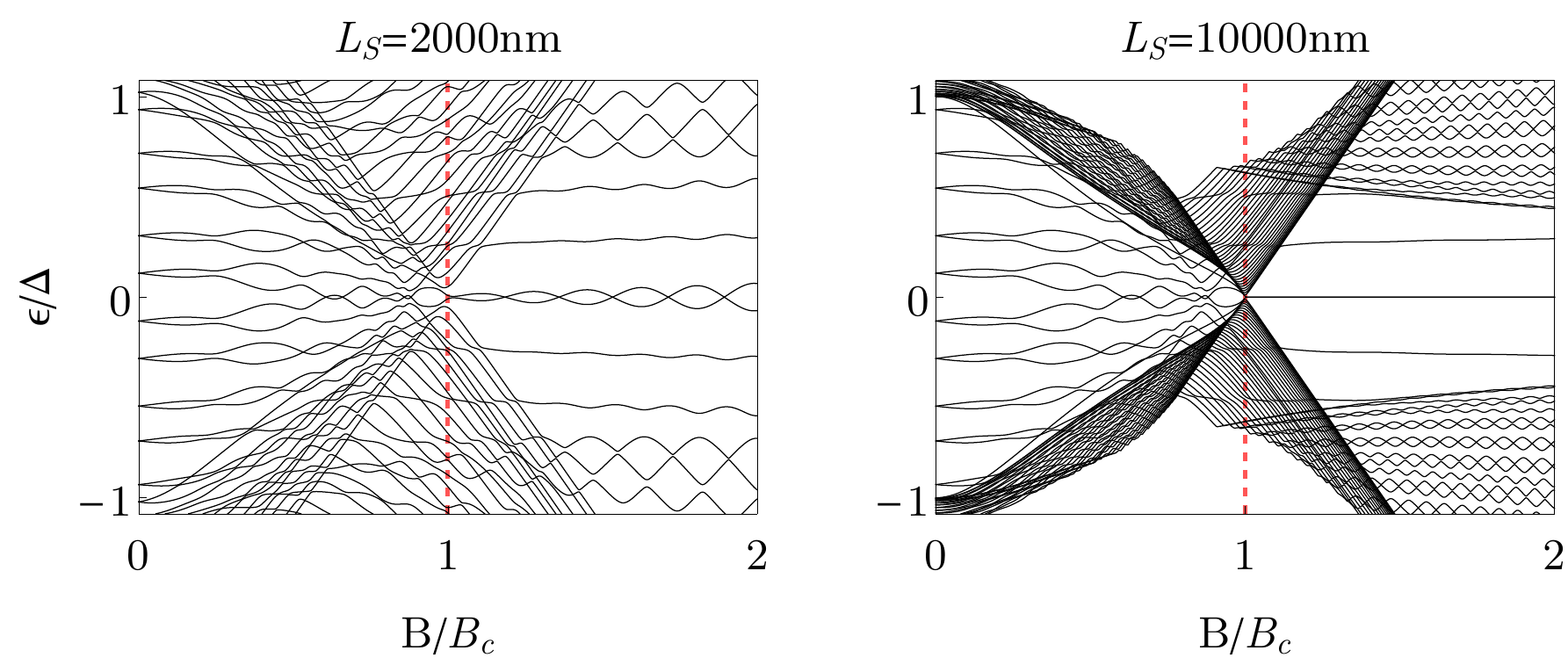} 
\caption[Energy levels in a NS junction as function of $B$: $L_{N}=2000$\,nm]{(Color online) Energy levels in a NS junction as function of the Zeeman field for $L_{N}=2000$\,nm: (left) $L_{S}=2000$\,nm and (right) $L_{S}=10000$\,nm.
More levels emerge inside $\Delta$ and the mini-gap is also reduced as expected when comparing with Figs.\,\ref{NS2} and \ref{NS3}.
Parameters: $\alpha_{R}=20$\,meVnm, $\mu=0.5$\,meV and $\Delta=0.25$\,meV.}
\label{NS4}
\end{figure}

In Figs.\,\ref{NS2}, \ref{NS3} and \ref{NS4}, we explore the dependence of the energy levels as function of the Zeeman field $B$ for very short, intermediate and long N regions of the NS junctions, respectively. In each figure we present two cases for (left panel) visible and (right panel) neglectible oscillations of the MBSs around zero energy. One observes that, as expected, the low energy spectrum traces the gap closing at $B=B_{c}$ and its respective reopening for $B>B_{c}$. We strongly point out that, at least in principle, spectroscopy of energy levels could experimentally test the emergence of MBSs in NS junctions. For very short N regions, shown in Fig.\,\ref{NS2}, the energy levels behave in a similar way as in the superconducting nanowire discussed in previous subsection (Fig.\,\ref{energysplitting}). The increase of $L_{N}$, presented in Figs.\,\ref{NS3} and \ref{NS4}, introduces additional levels (coming from N) into the low-energy spectrum inside $\Delta$ as can be indeed observed and tend to reduce the minigap for $B>B_{c}$. Another important feature to notice in this part is that although the energy levels follow the closing of the gap, they also develop a number of crossings around zero energy before the topological transition. Remarkably, this behaviour is not related to the topological phase but rather to a sign that the N region has reached its helical phase $B>\mu_{N}$. More about this discussion will be given in Chap.\,\ref{Chap2}. Moreover, we point out here that the oscillations around zero energy of the MBSs can be reduced by either increasing $L_{N}$ or $L_{S}$. The price to pay in the former is that by doing so one introduces additional levels and therefore the minigap separating the MBSs from the rest is also reduced, while the latter represents the standard picture for reducing such oscillations.

\newpage\clearpage
\subsection{The SNS junction model}
\label{SNSjunctionmodel}
\begin{figure}[!ht]
\centering
\includegraphics[width=.8\textwidth]{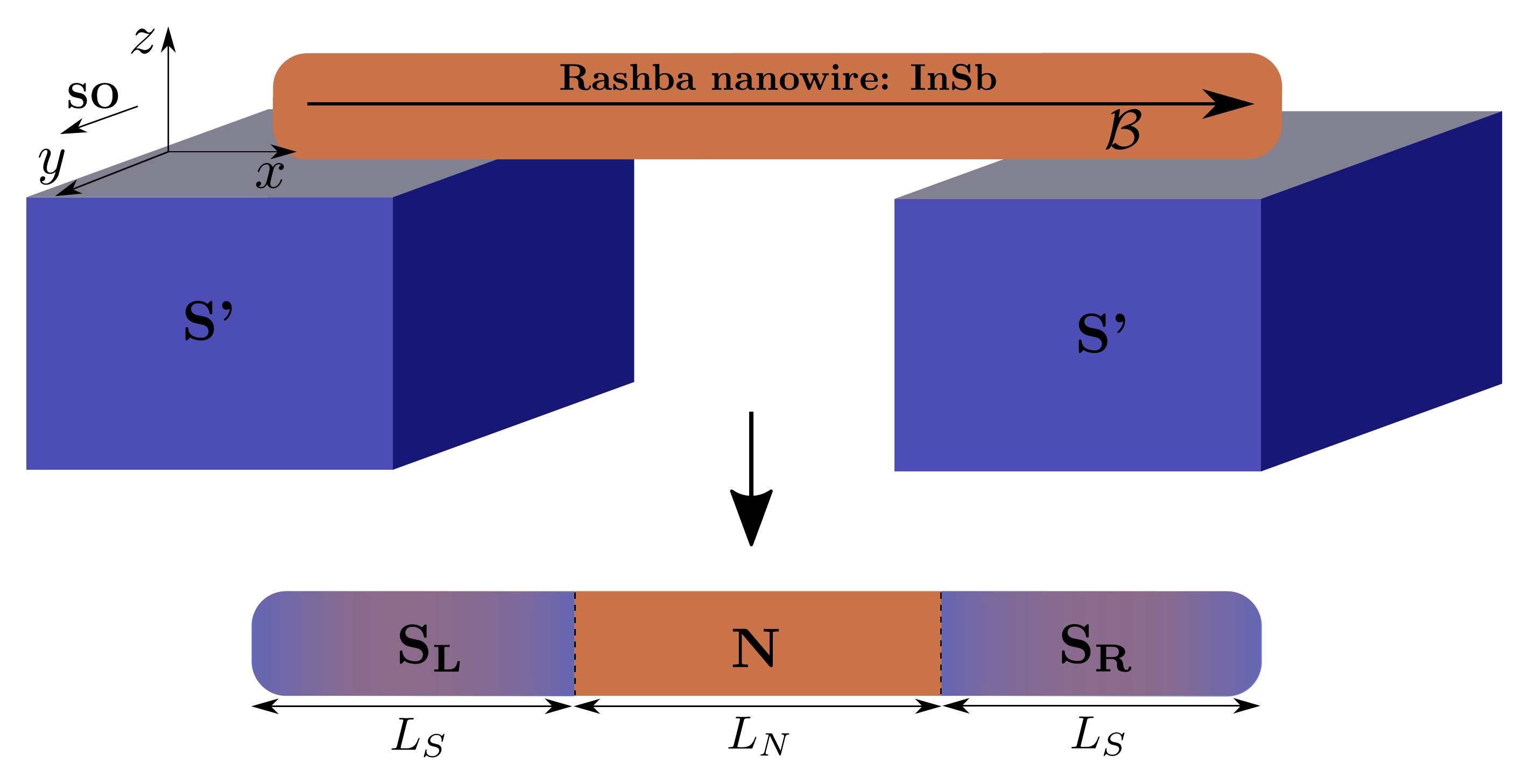} 
\caption[SNS junction based on Rashba nanowires]{(Color online) Top: A nanowire with Rashba SOC of length $L=L_{S}+L_{N}+L_{S}$ is placed on top of two $s$-wave superconductors (S') with pairing potentials $\Delta_{S'}$ and it is subjected to an external magnetic field $\mathcal{B}$ (denoted by the thick arrow). The applied magnetic field opens a gap of $2B$ in the energy spectrum as shown in Sec.\,\ref{Rashbawire}, where $B=g\mu_{B}\mathcal{B}/2$ is the associated Zeeman field in the wire, $g$ is the wire's $g$-factor and $\mu_{B}$ the Bohr magneton. Superconducting correlations are induced into the nanowire via proximity effect. Bottom: Left and right regions of the nanowire become superconducting, denoted by $S_{L}$ and $S_{R}$, with induced pairing potentials 
$\Delta_{S_{L(R)}}<\Delta_{S'}$ and chemical potentials $\mu_{S_{L(R)}}$, while the central region remains in the normal state with $\Delta_{N}=0$ and chemical potential $\mu_{N}$. This results in a Superconductor-Nanowire-Superconductor (SNS) junction.}
\label{figchap26}
\end{figure}

\begin{figure}[!ht]
\centering
\includegraphics[width=.7\textwidth]{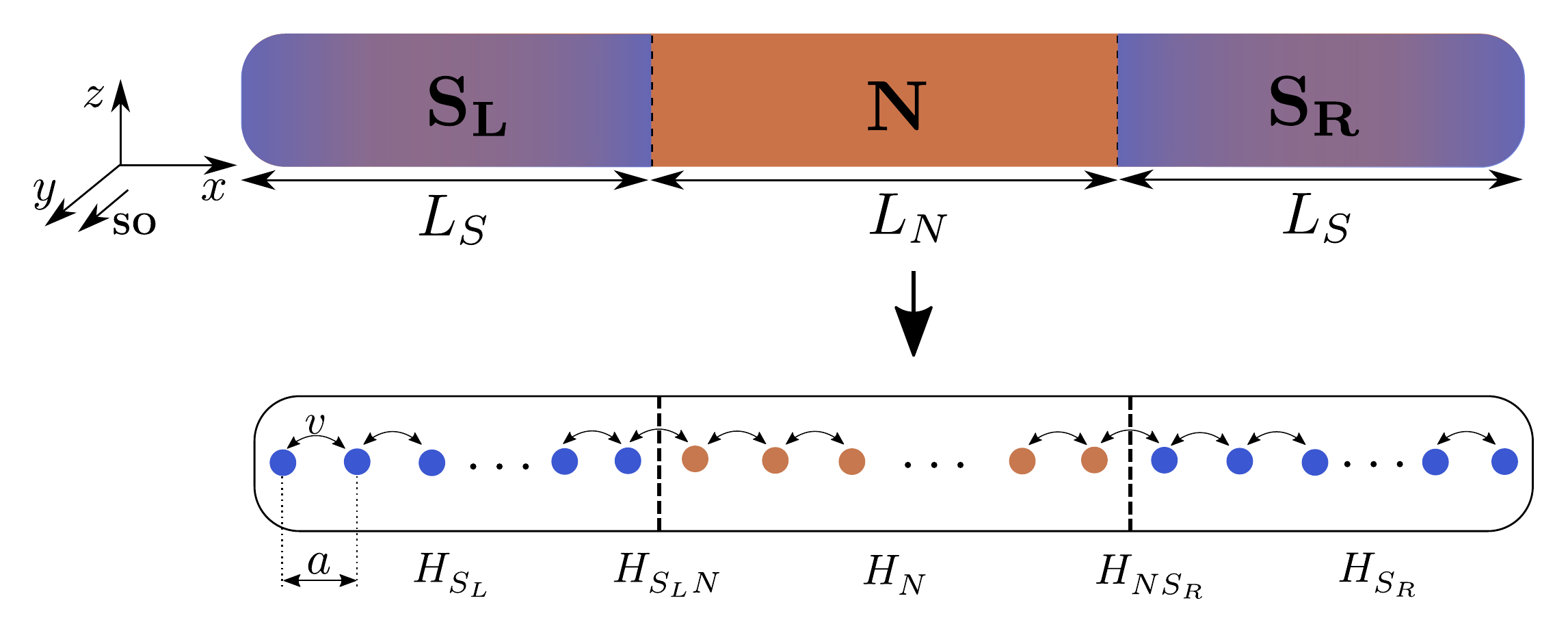} 
\caption[Tight-binding discretization: SNS junction]{(Color online) Schema of a SNS junction of length $L_{S}+L_{N}+L_{S}$ made of Rashba nanowires, which for computational purposes is discretised into a tight-binding lattice with lattice spacing $a$. Here, $L_{S}=N_{S}a$ and $L_{N}=N_{N}a$, where $N_{S}$ and $N_{N}$ are the number of sites in $S$ and $N$, respectively.}
\label{figchap10}
\end{figure}

 We consider a nanowire with Rashba SOC subjected to an external magnetic field $\mathcal{B}$ along the wire and perpendicular to the SO axis. 
 The applied magnetic field opens a gap of $2B$ in the energy spectrum, where $B=g\mu_{B}\mathcal{B}/2$ is the associated Zeeman field in the wire, $g$ is the wire's $g$-factor and $\mu_{B}$ the Bohr magneton.
 This nanowire is placed on top of two $s$-wave superconductors (S' with pairing potentials $\Delta_{S'}$) as it is schematically shown in Fig.\,\ref{figchap26}. 
We assume that there is a good contact interface between the nanowire and the superconductors (S'), so that superconducting correlations are induced by proximity effect into the regions of the nanowire that are in contact with the superconductors. This results in a nanowire containing left ($S_{L}$) and right ($S_{R}$) regions with superconducting properties leaving the central region in the normal state (N), as shown in Fig.\,\ref{figchap26}. 
These nanowires with left and right superconducting regions (denoted by S) and a central normal one (N) represent the hybrid Superconductor-Nanowire-Superconductor (SNS) junctions discussed here, as shown in Fig.\,\ref{figchap10}.  
It is common to consider that as a result of the proximity effect, the $S_{L(R)}$ parts have acquired an induced pairing $\Delta_{S_{L(R)}}<\Delta_{S'}$. Although rigorously speaking this assumption is incorrect, it provides a good description of the proximity effect for large gaps $\Delta_{S'}$.
The superconducting regions of the nanowire are considered to have chemical potential $\mu_{S_{L(R)}}$ and superconducting pairing potential 
$\Delta_{S_{L}}=\Delta\,{\rm e}^{i\varphi_{L}}$ and $\Delta_{S_{R}}=\Delta\,{\rm e}^{i\varphi_{R}}$, where $\Delta<\Delta_{S'}$. 
The central region of the nanowire is in the normal state without superconductivity, $\Delta_{N}=0$, and with chemical potential $\mu_{N}$.
For modelling SNS junctions made of Rashba nanowires, we proceed as in previous subsections, where the system is discretised into a tight-binding lattice with lattice spacing $a$. 
Here, the SNS junction is of length $L_{S}+L_{N}+L_{S}$, where $L_{S}=N_{S}a$ and $L_{N}=N_{N}a$, being $N_{S}$ and $N_{N}$ are the number of sites in $S$ and $N$, respectively.
The regions N and $S_{L(R)}$ are 
described by the tight-binding Hamiltonian $H_{0}$ given by Eq.\,(\ref{HamilTB}), see Fig.\,\ref{figchap10}, with their respective chemical potential, Zeeman field, hopping parameters,
\begin{equation}
\label{SSNSEQDS}
H_{N}=H_{0}(\mu_{N},B_{N},t_{N}, t_{SO,N})\,,\quad H_{S_{L(R)}}=H_{0}(\mu_{S_{L(R)}},B_{S_{L(R)}},t_{S_{L(R)}}, t_{SO,S_{L(R)}})\,
\end{equation}
where $N$ and $S_{L(R)}$ denotes the normal and left(right) superconducting regions of the SNS junction.

The Hamiltonian describing the SNS junction without superconductivity is then given by
\begin{equation}
\label{hsnszero}
h_{SNS}=
\begin{pmatrix}
H_{S_{L}}&H_{S_{L}N}&0\\
H^{\dagger}_{S_{L}N}&H_{N}&H_{NS_{R}}\\
0&H_{NS_{R}}^{\dagger}&H_{S_{R}}
\end{pmatrix}\,.
\end{equation}
where $H_{S_{i}}$ is the Hamiltonian of the superconducting region $i=L/R$ that we consider to be the same, $H_{S_{i}N}$ the Hamiltonian that couples the  superconducting region $S_{i}$ to the normal region $N$, while $H_{N S_{i}}$ the Hamiltonian that couples the normal region to $S_{i}$. 
The elements of these coupling matrices are non-zero for adjacent sites that lie at the interfaces of the superconducting regions and of the normal region, only.
This coupling is parametrized by a hopping matrix $v_{0}=\tau v$ between the sites that define the interfaces of the SNS junction, where $\tau\in[0,1]$. 
A tunnel junction can be modelled by considering $\tau\ll1$, while a full transparent junction with $\tau=1$.
All the elements in the diagonal of matrix Eq.\,(\ref{hsnszero}) have the structure of $H_{0}$ given by Eq. (\ref{HamilTB}) taking into account that the superconducting regions have a chemical potential $\mu_{S_{L(R)}}$, while this is $\mu_{N}$ for the normal region. This was stablished in Eq.\,(\ref{SSNSEQDS}). It is important to point out here that the matrix of Eq.\,(\ref{hsnszero}) is of finite size and the dimensions of the matrices are set by the number of sites of the respective region.
Now, we are left with considering superconducting correlations. Again, $s$-wave pairing is taken into account and the pairing potential for the full system must have the same structure as the SNS Hamiltonian, $h_{SNS}$. Therefore,
\begin{equation}
\label{deltasns}
\Delta(x)=
\begin{pmatrix}
\Delta_{S_{L}}&0&0\\
0&\Delta_{N}&0\\
0&0&\Delta_{S_{R}}
\end{pmatrix}
=
\begin{pmatrix}
\Delta_{0,S}\,{\rm e}^{{\rm i}\varphi_{L}}&0&0\\
0&0&0\\
0&0&\Delta_{0,S}\,{\rm e}^{{\rm i}\varphi_{R}}
\end{pmatrix}\,,
\end{equation} 
where $\Delta_{N}=0$ since in the normal region the superconducting correlations are absent, and $\Delta_{0,S}\,=\,{\rm i}\sigma_{y}\Delta\,$. 
In the introduction of this Chapter we have seen that a finite phase difference between $S_{L}$ and $S_{R}$ gives rise to the physics of Andreev bound states and therefore to the Josephson current. Thus, it is crucial to consider, and we indeed do, a finite phase difference between $S_{L}$ and $S_{R}$, $\varphi$: $\Delta_{S_{L}}=\Delta\,{\rm e}^{-i\varphi/2}$ and $\Delta_{S_{R}}=\Delta\,{\rm e}^{i\varphi/2}$. 

The full system Hamiltonian is then written in Nambu space 
\begin{equation}
\label{hsnss}
H_{SNS}=
\begin{pmatrix}
h_{SNS}&\Delta(x)\\
\Delta^{\dagger}(x)&-h_{SNS}^{*}
\end{pmatrix}
\end{equation}
 
Previous Hamiltonian is diagonalised numerically and for our calculations we consider realistic system parameters for InSb as described in the introduction of this Chapter. In the next section we properly investigate the low-energy spectrum and Josephson and critical currents in SNS junctions, whose model was described in this part. 
In particular, we focus on the formation of Andreev bound states, their evolution into Majorana bound states and show that the presence of MBSs can tested either by Andreev spectroscopy or by measuring the Josephson and the critical currents. 

\newpage
\section{Andreev levels and Josephson current in SNS junctions}
\label{ABSsJIc}
As we have discussed in the introduction of this work, Sec.\,\ref{KitaevModel}, a topological superconductor nanowire hosts two Majorana bound states (MBSs)
 at its ends, $\gamma_{1}$ and $\gamma_{2}$, one at each end. These two MBSs define a non-local fermion and can be fused to form a Dirac fermion $c^{\dagger}=(\gamma_{1}-i\gamma_{2})/2$. The fermion parity operator is defined as $2c^{\dagger}c-1=i\gamma_{1}\gamma_{2}$ and has eigenvalues $-1$ or $1$ for empty $\ket{0}$ and occupied $\ket{1}=c^{\dagger}\ket{0}$ states. 
A spatial overlap of the two MBSs, in the superconducting wire, hybridises them into eigenstates of opposite fermion parity. Indeed, in a topological superconducting wire, the spatial overlap induces an energy splitting given by Eq.\,(\ref{esplitting}) and it is proportional to ${\rm e}^{-2L_{S}/\xi_{eff}}$ \cite{DasSarma:PRB12}. 

\begin{figure}[!ht]
\centering
\includegraphics[width=.7\textwidth]{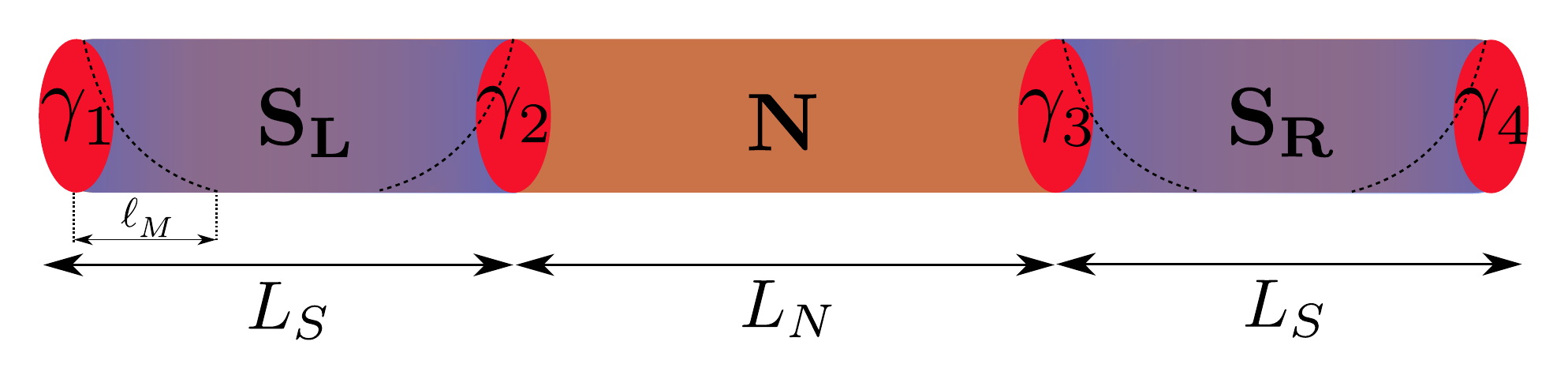} 
\caption[Sketch of SNS junction with four MBSs]{(Color online) SNS junction made of Rashba nanowires. When a Zeeman field is applied, perpendicular to the spin-orbit axis, the superconducting regions become topological. Due to finite length of the superconducting region, $L_{S}$, the junction host four Majorana bound states, $\gamma_{1},\gamma_{2},\gamma_{3},\gamma_{4}$, for a phase difference of $\pi$ with localisation length $\ell_{M}$.}
\label{figchap27}
\end{figure}

On the other hand, in the case of a SNS junctions discussed here, the superconducting regions of the nanowire, $S_{L}$ and $S_{R}$, are driven into the topological superconducting phase when $B>B_{c}\equiv\sqrt{\mu_{S}^{2}+\Delta^{2}}$, where $\mu_{S}=\mu_{S_{L(R)}}$. Therefore, due to the finite length $L_{S}$, we expect the formation of four MBSs in the SNS junction: two at the interfaces with vacuum, $\gamma_{1,4}$, (\emph{outer MBSs}), and two at the interfaces with the normal regions of the nanowire N, $\gamma_{2,3}$, (\emph{inner MBSs}), as schematically shown in Fig.\,\ref{figchap27}. Notice that when $L_{S}\rightarrow\infty$, the outer MBSs are at infinity and the problem involves only the two inner MBSs.
The four MBSs hybridise both through the region N or the finite region S. The two MBSs, $\gamma_{2,3}$, at either side of the junction with phase difference $\varphi$ and transparency $T$ will hybridise into states of opposite fermion parity with energies $\pm \Delta \sqrt{T}{\rm cos}(\varphi/2)$ \cite{Kwon:EPJB03, Fu:PRB09}, where $\Delta$ is the effective induced superconducting pairing.
It is important to point out here that a finite phase difference matters and the emergence of MBSs is not arbitrary but the system hosts four MBSs when the phase difference is $\pi$, while for zero phase difference only the outer MBSs are present,  $\gamma_{1,4}$. The wave-functions of the MBSs decay exponentially from both ends of the topological superconducting regions in the SNS junction into the bulk of S. For finite $L_{S}\ll2\ell_{M}$, being $L_{S}$ the length of the superconducting regions of the SNS junction and $\ell$ the Majorana localisation length, the overlap between MBSs is significant and therefore they are no longer true zero modes. However, if $L_{S}\gg2\ell_{M}$
the overlap is negligible and one can assume the MBSs to be true zero modes. 

Thus, we conclude that in these SNS junctions of finite length four MBSs emerge naturally. It was theoretically shown that the minimal platform for investigating the non-Abelian statistics  of MBSs consists of a system with four MBSs \cite{Ivanov:PRL01}. Therefore, research along these lines represents an important advance towards realistic implementations of MBSs and potential application in topological quantum computation \cite{RevModPhys.80.1083}.

In this part we present the low-energy Andreev spectrum, Josephson (supercurrent) and critical currents in ballistic short ($L_{N}\ll \xi$) and long ($L_{N}\gg\xi$)  SNS junctions based on the model described  in previous section, being $\xi$ the superconducting coherence length and $L_{N}$ the length of the normal region. In particular, we discuss the formation of Andreev bound states and their connection with the emergence of MBSs. Finally, we propose the Josephson and critical currents as a powerful tool for detecting MBSs in such hybrid systems.

\subsection{Andreev energy levels}
\begin{figure}[!ht]
\centering
\includegraphics[width=1\textwidth]{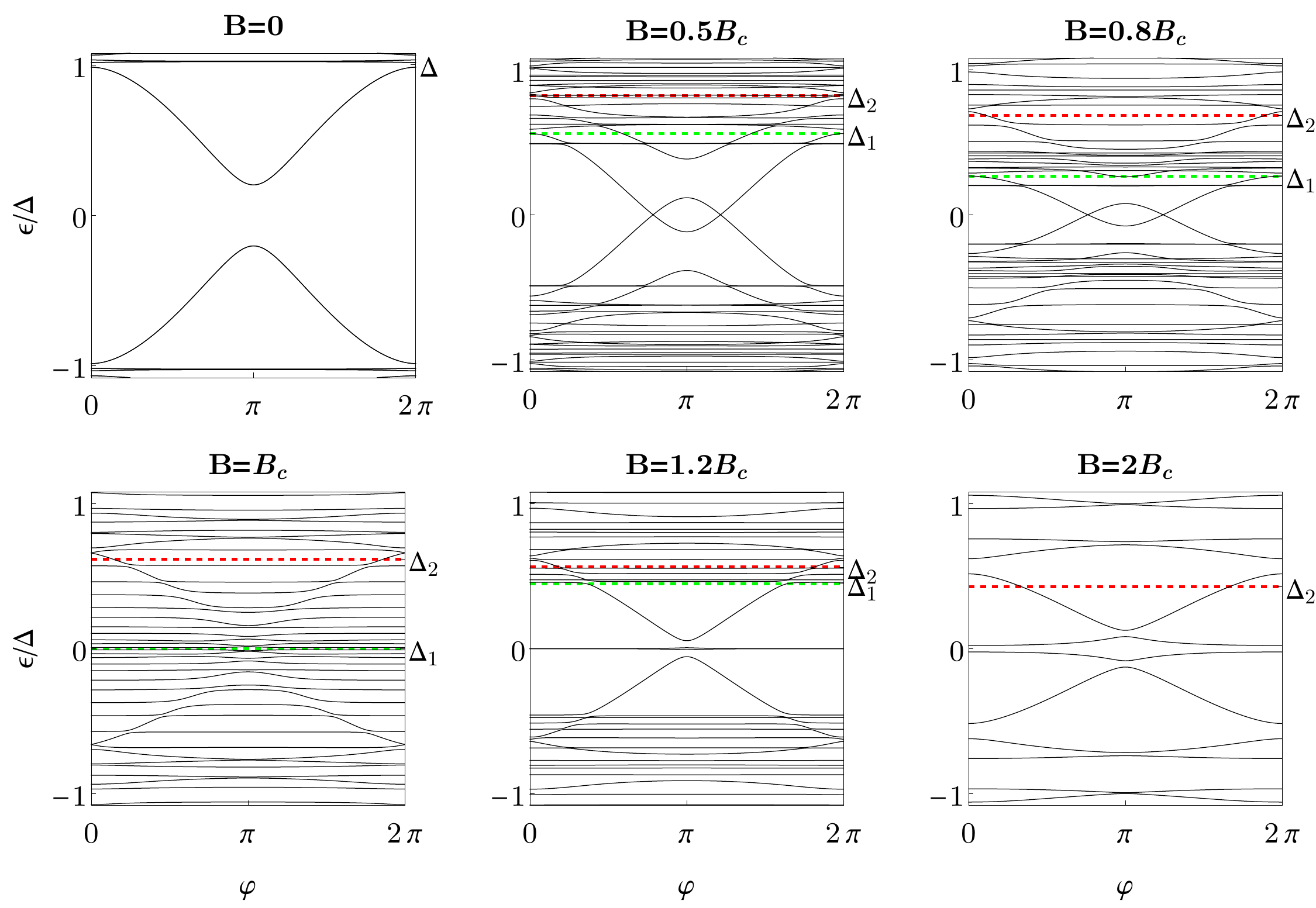} 
\caption[ABSs in a short SNS junction as function of $\varphi$: $L_{S}=2000$\,nm]{(Color online) Low-energy Andreev spectrum as a function of the superconducting phase difference in a short SNS junction for $L_{N}=20$\,nm and $L_{S}=2000$\,nm. Different panels show the evolution with the Zeeman field, from the trivial phase for $B<B_{c}$, at the topological transition $B=B_{c}$, and in the topological phase $B>B_{c}$. The energy spectrum traces the two different gaps which appear in the system for finite Zeeman field (marked by red and green dashed horizontal lines). Notice that after the gap inversion at $B=B_{c}$, two MBSs emerge at the ends of the junction as almost dispersions levels, while additionally two MBSs appear at $\varphi=\pi$. Parameters: $\alpha_{R}=20$\,meVnm, $\mu_{N}=\mu_{S}=0.5$\,meV and $\Delta=0.25$\,meV.}
\label{figchap29}
\end{figure}

\begin{figure}[!ht] 
\centering
\includegraphics[width=1\textwidth]{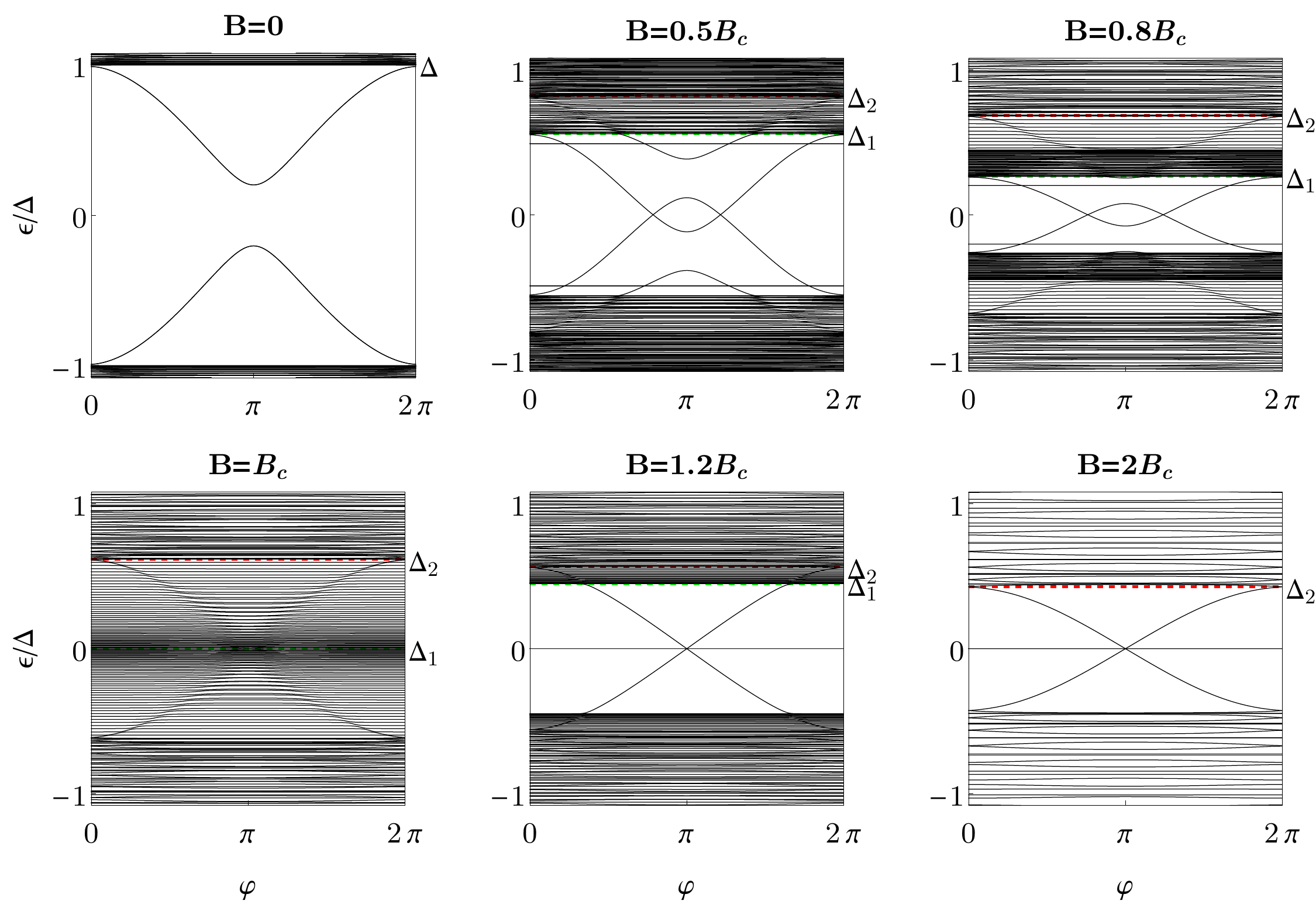} 
\caption[ABSs in a short SNS junction as function of $\varphi$: $L_{S}=10000$\,nm]{(Color online) Low-energy Andreev spectrum as a function of the superconducting phase difference in a short SNS junction for $L_{N}=20$\,nm and $L_{S}=10000$\,nm. Notice that in this case, the emergent outer MBSs are dispersionless with $\varphi$, while the inner touch zero at $\varphi=\pi$ acquiring Majorana character. 
Parameters: $\alpha_{R}=20$\,meVnm, $\mu_{N}=\mu_{S}=0.5$\,meV and $\Delta=0.25$\,meV.}
\label{figchap210}
\end{figure}

\begin{figure}[!ht]
\centering
\includegraphics[width=1\textwidth]{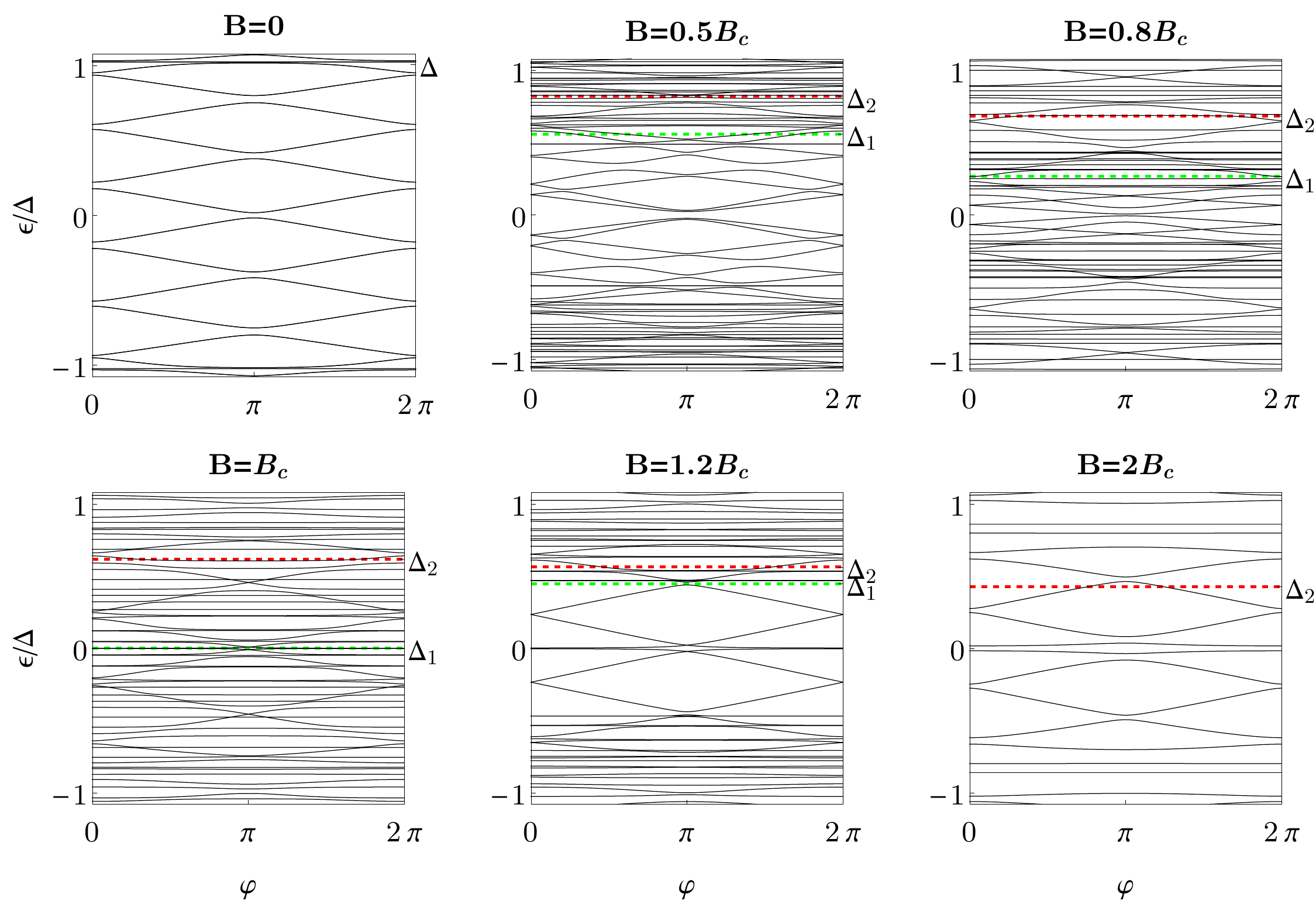} 
\caption[ABSs in a long SNS junction as function of $\varphi$: $L_{S}=2000$\,nm]{(Color online) Low-energy Andreev spectrum as a function of the superconducting phase difference in a long SNS junction for $L_{S}=2000$\,nm and $L_{N}=2000$\,nm. Different panels show the evolution with the Zeeman field. The low-energy levels linearly depend on the phase difference and the energy spectrum traces the two different gaps which appear in the system for finite Zeeman field (marked with red and green dashed horizontal lines).
For $B>B_{c}$ two dispersionless levels emerge and represent the outer MBSs, while the two levels tend to zero at $\varphi=\pi$, thus acquiring Majorana character and representing the inner MBSs.
Parameters: $\alpha_{R}=20$\,meVnm, $\mu_{N}=\mu_{S}=0.5$\,meV and $\Delta=0.25$\,meV.}
\label{figchap211}
\end{figure}

\begin{figure}[!ht]
\centering
\includegraphics[width=1\textwidth]{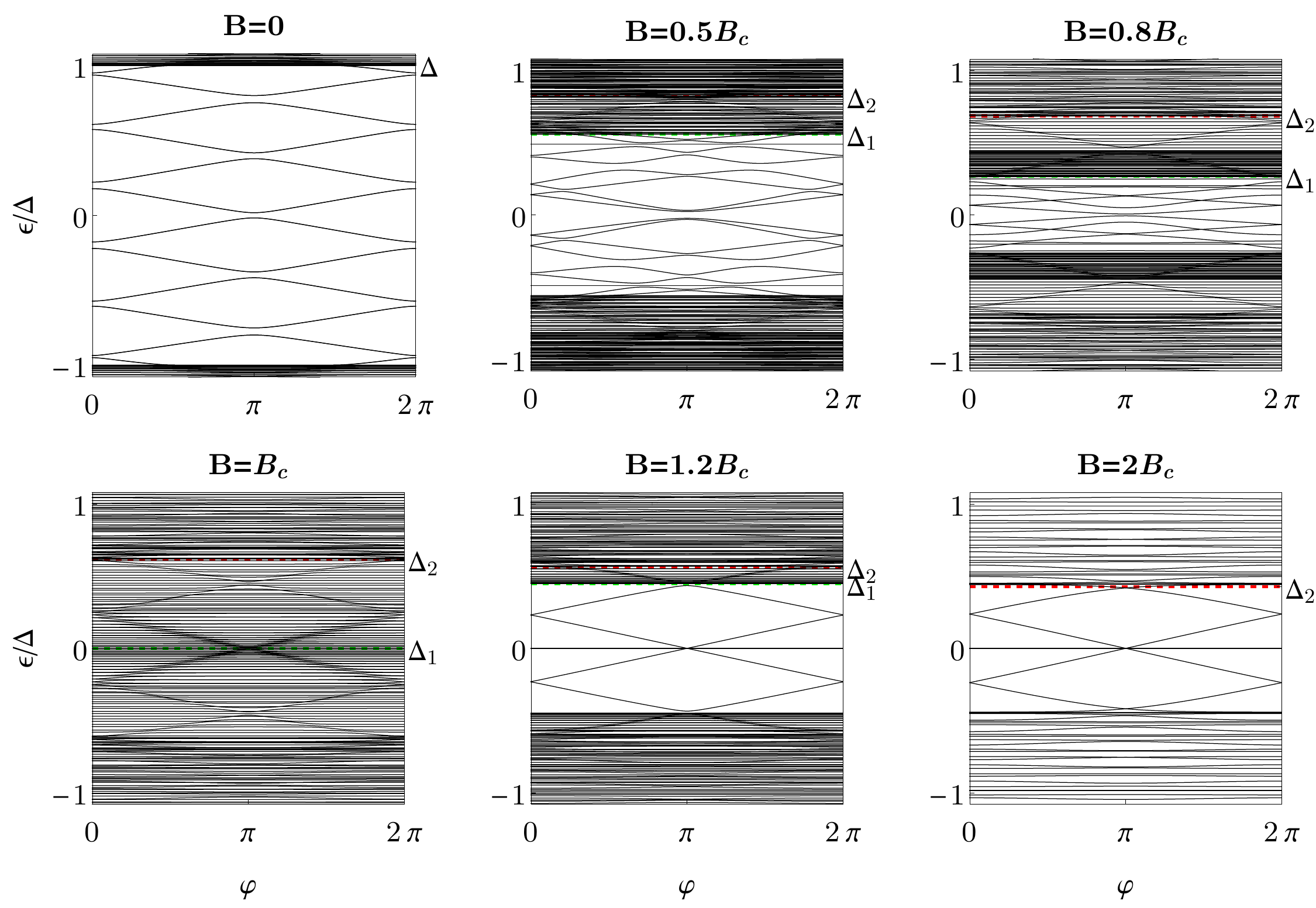} 
\caption[ABSs in a long SNS junction as function of $\varphi$: $L_{S}=10000$\,nm]{(Color online) Low-energy Andreev spectrum as a function of the superconducting phase difference in a long SNS junction for $L_{S}=10000$\,nm. Note that in this case, the outer MBSs lie at zero energy and the inner reach zero at $\varphi=\pi$ acquiring Majorana character. Parameters: $\alpha_{R}=20$\,meVnm, $\mu_{N}=\mu_{S}=0.5$\,meV and $\Delta=0.25$\,meV.}
\label{figchap212}
\end{figure}

In Figs.\ref{figchap29} and \ref{figchap210}, we show the Andreev levels as function of the superconducting phase difference and their evolution with the Zeeman field in a short SNS junction for $L_{S}\ll\ell_{M}$ and $L_{S}\gg\ell_{M}$, respectively. On the other hand, the dependence of the Andreev levels on the phase difference for long junctions is presented in Figs.\ref{figchap211} and \ref{figchap212}.

Let us first concentrate on the short-junction regime, Figs.\ref{figchap29} and \,\ref{figchap10} for $L_{S}\ll\ell_{M}$ and $L_{S}\gg\ell_{M}$, being $\ell_{M}$ the Majorana 
localisation length. At zero magnetic field (top left panel in Figs.\ref{figchap29} and \,\ref{figchap10}), two degenerate ABSs appear within $\Delta$ as solutions to the BdG 
equations. Note that even this non-topological case is anomalous as 
the ABS energies do not reach zero at $\varphi=\pi$, unlike predicted by the standard theory for a transparent channel  within the Andreev approximation $\mu_{S}\gg \Delta$ 
\cite{Beenakker:92}. The dense amount of levels above $|E|>\Delta$ represent the quasi-continuum of states, which consists on a discrete set of levels due to the finite length of the SNS 
junction. Indeed, all the calculations presented here correspond to systems of finite length and therefore their energy spectrum is discrete, where the levels in the energy spectrum 
corresponds to the full system, the N and S regions.
Within the gap $\Delta$ the junction host levels which are clearly discrete, however, above $\Delta$ one observes the formation of a discrete quasi-continuum being denser for junctions with longer $L_{S}$ as comparing Figs.\ref{figchap29} and \,\ref{figchap10}. This will have important consequences for the calculation of the Josephson current (see next subsection).
Moreover, the detachment of the quasi-continuum at $0$ and $2\pi$ is not clear but what is clear is that it is not zero. In what follows we will see that such detachment depends on the finite length of the S regions (see Fig.\,\ref{figchap210}).

For a non-zero Zeeman field the ABSs split and the two different gaps $\Delta_{1}$ and $\Delta_{2}$, discussed in Sec.\,\ref{Rashbawire}, emerge (top middle and right panels).
The former was defined at low momentum, while the latter at higher momentum. By increasing the Zeeman field the low momentum gap $\Delta_{1}$ gets reduced as expected from the band structure for the superconducting nanowire discussed in Sec.\,\ref{Rashbawire}, while the gap $\Delta_{2}$ remains roughly unchanged for strong SOC. At $B=B_{c}$, the energy spectrum exhibits the closing of the low momentum gap $\Delta_{1}$ (bottom left panel). This signals the topological transition point, where according to the adiabatic theorem two gapped topologically different phases are only connected through a gap closing.
By further increasing of the Zeeman field (bottom middle and right panels), $B>B_{c}$, the system enters into the topological phase and the superconducting regions denoted by $S_{L(R)}$ become topological, while the $N$ region remains in the normal state (see Fig.\,\ref{figchap10}).
In the topological superconducting nanowire, MBSs emerge at the end of the wire because it defines an interface between two topologically different regions: the vacuum, which is in the trivial phase and the topological superconducting wire.

In SNS junctions, on the other hand, MBSs are expected to appear for $B>B_{c}$ at the ends of the two topological superconducting sectors $S_{L(R)}$, since they define interfaces between topologically different regions: the normal region $N$ and vacuum. 
This is what we indeed observe for $B>B_{c}$ in Fig.\,\ref{figchap29}, where the low energy spectrum has Majorana properties.
For $B>B_{c}$, the topological phase is characterised by the emergence of two (almost) dispersionless levels with $\varphi$, which represent the outer MBSs $\gamma_{1,4}$ formed at the ends of the junction and schematically shown in Fig.\,\ref{figchap27}. Additionally, two energy levels tend to reach zero at $\varphi=\pi$ and represent the inner MBSs $\gamma_{2,3}$ formed inside the junction. 
 For sufficiently strong fields, $B=2B_{c}$, the lowest gap is $\Delta_{2}$, which in principle bound the MBSs.
Notice that the four MBSs exhibit a considerable splitting around $\varphi=\pi$. This is a result of the finite spatial overlap between the MBSs wave-functions, $L_{S}\ll2\ell_{M}$, being $\ell_{M}$  the Majorana localisation length.
The problem can be solved by either increasing the spin-orbit coupling strength and therefore reducing the Majorana localisation length, as discussed in Appendix \,\ref{Majorana-length}, or by increasing the length of the superconducting regions $L_{S}$, which indeed demands longer computational times. Since realistic experiments consider nanowires with an intrinsic spin-orbit coupling \cite{Mourik:S12}, which is a material property, for reducing the energy splitting at $\pi$ we 
consider SNS junctions with longer superconducting regions, $L_{S}\gg\ell_{M}$. As an example, we present in Fig.\,\ref{figchap210} the energy levels as function of the phase difference for $L_{S}\gg\ell_{M}$. First, we notice that the energy spectrum at $B=0$ for $|E|>\Delta$, exhibits a visible denser  spectrum than in Fig.\,\ref{figchap29} signalling the quasi-continuum of states. 
 The behaviour of the two different gaps remains as one increases $B$.
Notice that in the topological phase, $B>B_{c}$, the energy levels associated to two outer MBSs 
are really dispersionless with $\varphi$ and therefore they can be considered as truly zero modes.
Remarkably, the energy splitting at $\varphi=\pi$ is considerable reduced, however it will be always non-zero, though not visible by naked eye, due to the finite length and thus due to the presence of the outer MBSs. Observe also in Fig.\,\ref{figchap210} that an increase in the length of the superconducting regions favours the reduction of the detachment between the discrete spectrum and the quasi-continuum at $0$ and $2\pi$, as it should be for a ballistic junction \cite{Kwon:EPJB03, Fu:PRB09}.
 In the topological phase it can be associated to the finite spatial overlap between the wavefunctions of the MBSs at the end of the S regions.

A true crossing is reached by considering semi-infinite superconducting leads, $L_{S}\rightarrow\infty$, as it is done in Chap.\,\ref{Chap3}. In such geometry the outer MBSs are not present in the description of the system and the crossing at $\pi$ is protected by conservation of the total fermion parity.
Notice also that Fig.\,\ref{figchap210} shows that the inner MBSs, the ones giving rise to ABSs due to hybridization, are truly bound within $\Delta_{2}$, unlike Fig.\,\ref{figchap29}. 

The energy levels dependence on the superconducting phase difference in long SNS junctions is shown in Figs.\,\ref{figchap211} and \ref{figchap212} for $L_{S}\ll\ell_{M}$ and $L_{S}\gg\ell_{M}$, respectively.
In the introduction of this Chapter, Sec.\,\ref{introhybridNSSNS}, we have discussed the emergence of Andreev levels in short and long SNS junctions. At this levels, neither Zeeman nor spin-orbit coupling were considered. We have seen that a long SNS junction host more levels than a short one. This is because in a longer junction more levels fit within the region of length $L_{N}$. We have obtained the behaviour of the energy levels analytically taking into account a number of approximations that in general are not necessarily true.

A long junction, at $B=0$, contains more levels within the energy gap $\Delta$ than a short junction (see top left panels in Figs.\,\ref{figchap211} and \ref{figchap212}). This is expected as it is a situation similar to the one discussed in Sec.\,\ref{introhybridNSSNS}. When a finite Zeeman field is switched on, the two different gaps, discussed in \ref{Rashbawire}, emerge (see top middle and right panels in Figs.\,\ref{figchap211} and \ref{figchap212}).

As for short junctions, in long junctions, the Andreev energy spectrum traces the closing of the low momentum gap $\Delta_{1}$, which reaches zero at $B=B_{c}$ (see bottom left panels in Figs.\,\ref{figchap211} and \ref{figchap212}). This signals the topological phase transition point into the topological superconducting phase, where the $S_{L(R)}$ regions of the SNS junction become topological. For $B>B_{c}$, 
the four lowest levels represent the MBSs formed at the ends of the two topological superconducting regions $S_{L(R)}$. Notice that for strong Zeeman fields in the topological phase, the lowest gap is $\Delta_{2}$ and MBSs coexist with additional levels, which arise due to the long length of the N region (see bottom left and right panels in Figs.\,\ref{figchap211} and \ref{figchap212}).
When the length of the superconducting regions $L_{S}$ is increased, $L_{S}\gg\ell_{M}$, the two lowest energy levels do not disperse with $\varphi$ and the anti-crossing at $\varphi=\pi$ is negligible. Thus, this anti-crossing can be reduced as much as needed taking into account $L_{S}\gg\ell_{M}$.
In a junction made of two semi-infinity wires, only the inner MBSs are present and the crossing at $\varphi=\pi$ is also protected by conservation of the total fermion parity, as in the short junction regime.
 
 \begin{figure}[!ht]
\centering
\includegraphics[width=.99\textwidth]{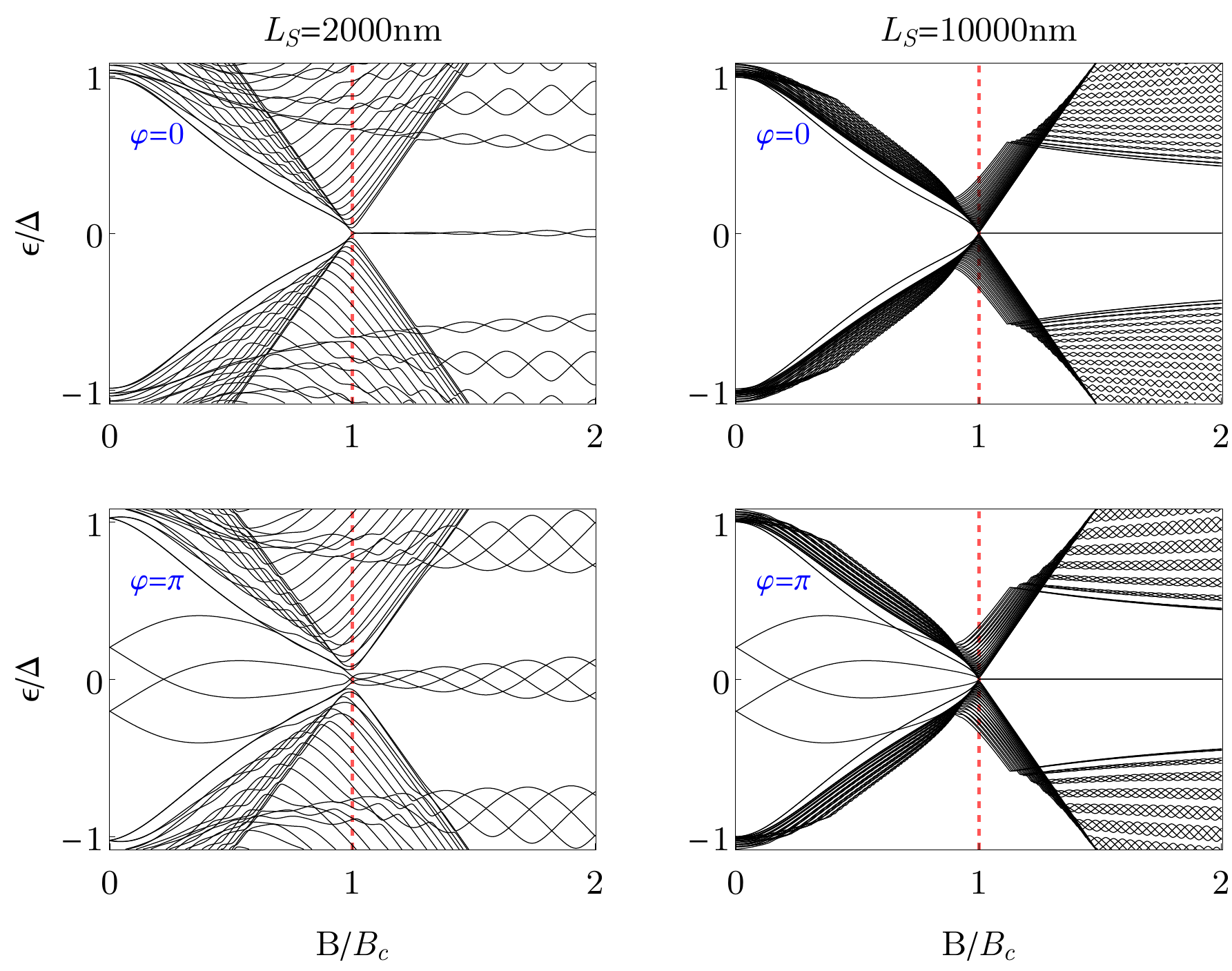} 
\caption[ABSs in a SNS junction as function of $B$ for $L_{N}=20$\,nm]{(Color online) Low-energy Andreev spectrum as a function of the Zeeman field in a short SNS junction: (left column) $L_{S}=2000$\,nm and (right column) $L_{S}=10000$\,nm. Top panels for $\varphi=0$, while bottom for $\varphi=\pi$. Notice that the low-energy spectrum traces the gap closing and reopening, while for $B<B_{c}$ one observes the emergence of MBSs, which oscillates with $B$ for $L_{S}<2\ell_{M}$ and do not for $L_{S}>2\ell_{M}$. Parameters: $L_{N}=20$\,nm, $\alpha_{R}=20$\,meVnm, $\mu=0.5$\,meV and $\Delta=0.25$\,meV.}
\label{SNSa}
\end{figure}
\begin{figure}[!ht]
\centering
\includegraphics[width=.99\textwidth]{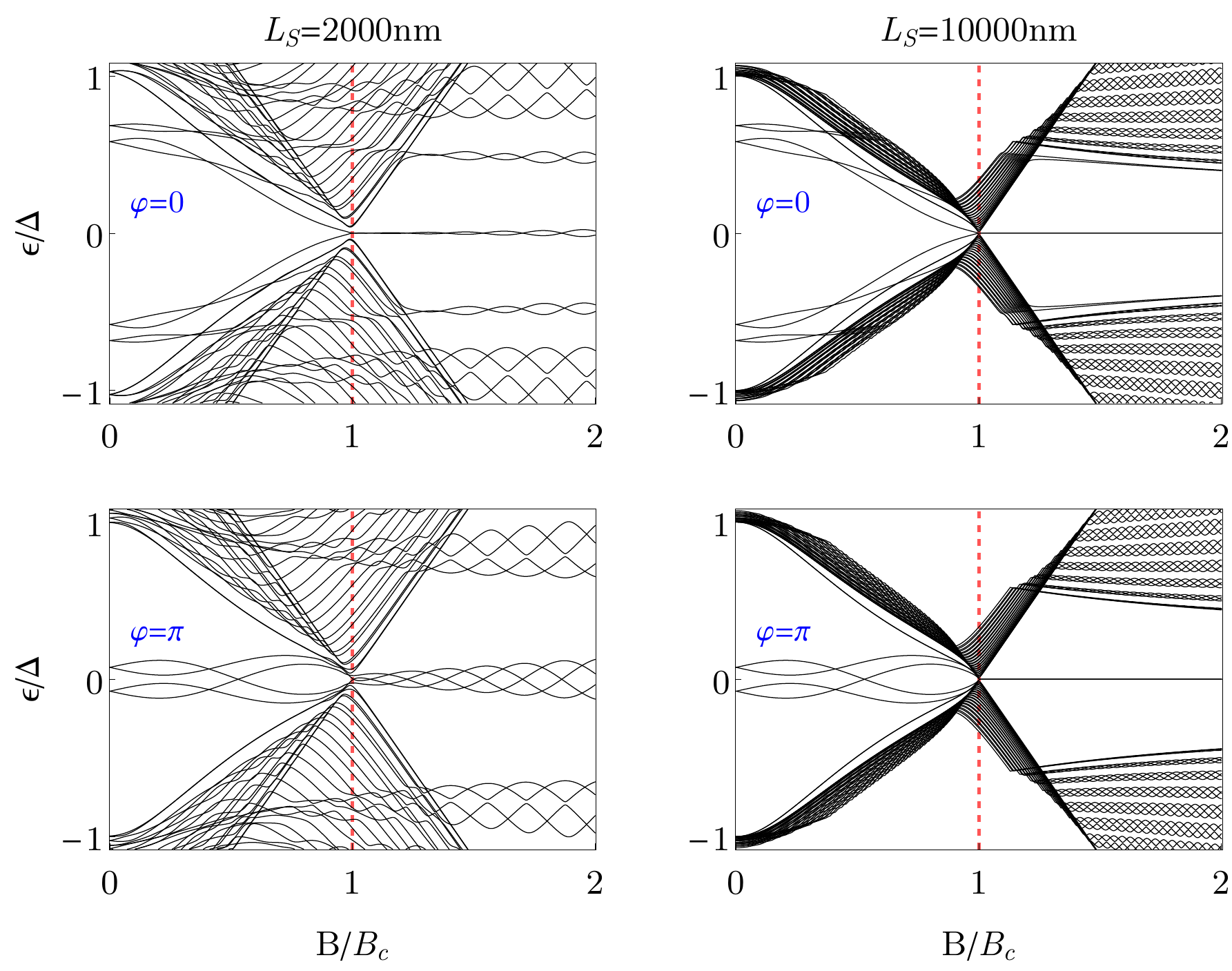} 
\caption[ABSs in a SNS junction as function of $B$ for $L_{N}=400$\,nm]{(Color online) Low-energy Andreev spectrum as a function of the Zeeman field in a intermediate SNS junction: (left column) $L_{S}=2000$\,nm and (right column) $L_{S}=10000$\,nm. Top panels for $\varphi=0$, while bottom for $\varphi=\pi$. In this case, the length of the normal regions is greater than in Fig.\,\ref{SNSa}. This increase introduces additional energy levels in the low-energy spectrum as seen in top panels.
Parameters: $L_{N}=400$\,nm, $\alpha_{R}=20$\,meVnm, $\mu=0.5$\,meV and $\Delta=0.25$\,meV.}
\label{SNSb}
\end{figure}
\begin{figure}[!ht]
\centering
\includegraphics[width=.99\textwidth]{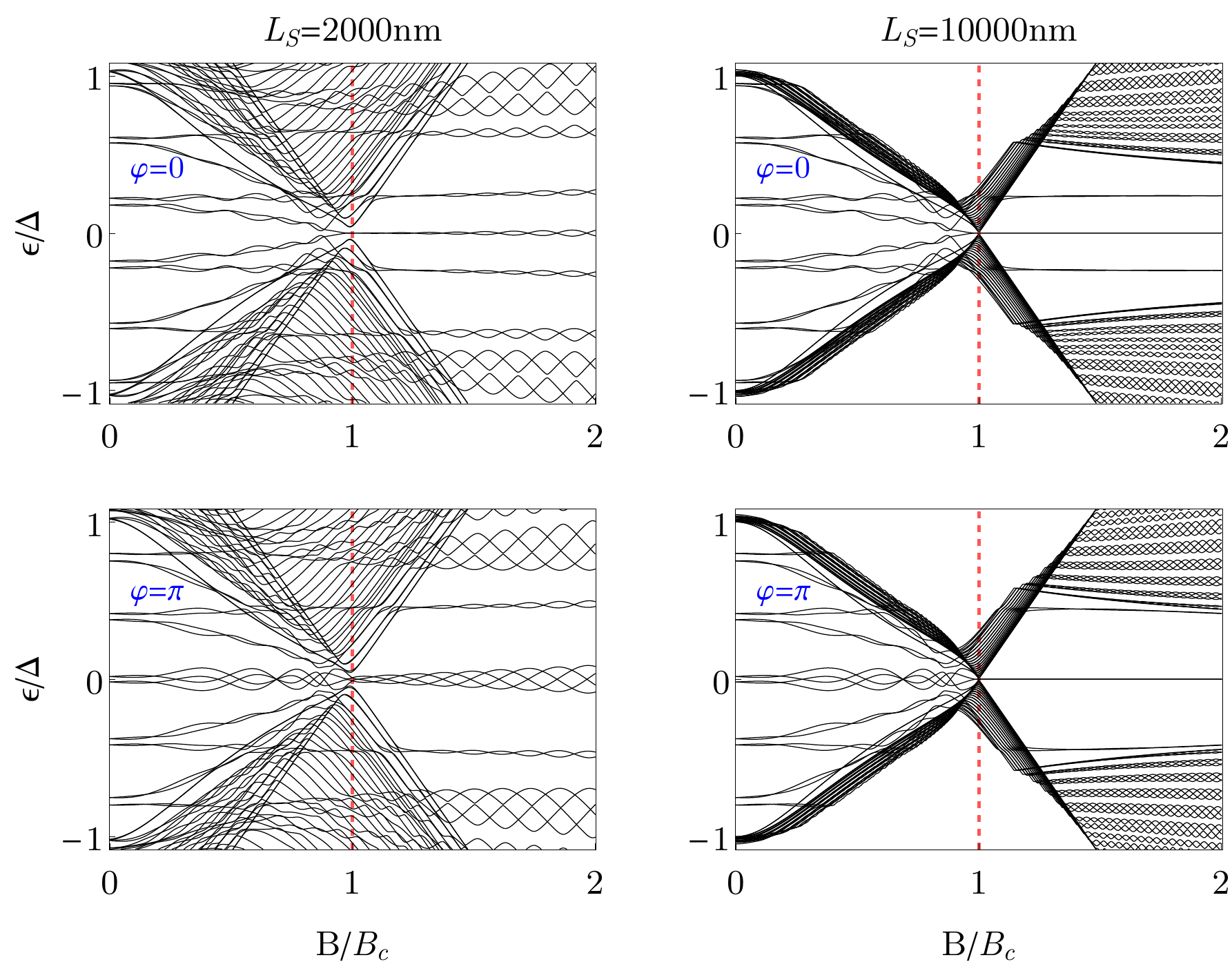} 
\caption[ABSs in a SNS junction as function of $B$ for $L_{N}=2000$\,nm]{(Color online) Low-energy Andreev spectrum as a function of the Zeeman field in a long SNS junction: (left column) $L_{S}=2000$\,nm and (right column) $L_{S}=10000$\,nm. Top panels for $\varphi=0$, while bottom for $\varphi=\pi$. In this case, the length of the normal regions is greater than in Figs.\,\ref{SNSa} and \,\ref{SNSb}. Again, notice that the low-energy spectrum now host more energy levels.
Parameters: $L_{N}=2000$\,nm, $\alpha_{R}=20$\,meVnm, $\mu=0.5$\,meV and $\Delta=0.25$\,meV.}
\label{SNSc}
\end{figure}

The discussion made before can be clarified by considering the dependence of the low-energy spectrum on the Zeeman field. This is shown in Figs.\,\ref{SNSa}, \ref{SNSb} and \ref{SNSc} for short, intermediate and long SNS junctions at (top row) $\varphi=0$ and (bottom row) $\varphi=\pi$, while left column for $L_{S}\ll2\ell_{M}$ and right column for $L_{S}\gg2\ell_{M}$.
As already discussed before for superconducting wires or NS junctions, the low-energy levels as function of the Zeeman field trace the gap inversion.
Here, either at $\varphi=0$ or $\varphi=\pi$ one clearly observes such prediction, with the sole difference that at $\varphi=\pi$ there are four lowest levels instead of two at $\varphi=0$, as can be indeed seen in the phase-dependence energy spectrum ( see for instance Fig.\,\ref{figchap29}). In the topological phase $B>B_{c}$ at  $\varphi=\pi$, 
the four lowest levels oscillate around zero energy for $L_{S}\ll\ell_{M}$, while such oscillations are remarkably suppressed by making the superconducting section longer $L_{S}\gg\ell_{M}$ (see bottom panels of Figs.\,\ref{SNSa}, \ref{SNSb} and \ref{SNSc}). On the other hand, in the topological phase $B>B_{c}$ at  $\varphi=0$, only the outer MBSs are present developing an oscillatory patter for $L_{S}\ll\ell_{M}$, which are suppressed for $L_{S}\gg\ell_{M}$ (see top panels of Figs.\,\ref{SNSa}, \ref{SNSb} and \ref{SNSc}).
An increase in the length of the normal section is reflected by increasing the amount of levels in the energy spectrum. Indeed, observe in Figs.\,\ref{SNSb} and \ref{SNSc}, for intermediate and long junctions, respectively, that additional levels fit in the junction and they appear inside the superconducting induced gap. As discussed before, in the topological phase, $B>B_{c}$, these levels tend to reduce the minigap (separation between the energy levels oscillating around zero and the rest of the spectrum). However, they also slightly reduce the amplitude of the oscillations of the energy levels around zero in the topological phase and therefore allowing such levels to reach their Majorana character in reasonable long superconducting regions (see Figs.\,\ref{SNSb} and \ref{SNSc}).
 
Notice that by comparing the low-energy spectrum of NS junctions, Figs.\,\ref{NS2}, \,\ref{NS3} and \,\ref{NS4} with those of SNS junctions at $\varphi=0$, in the latter additional levels are introduced and those arise from the finite left S region which is considered to model the SNS junction. In the topological phase these levels also slightly reduce the amplitude of the oscillations of the lowest levels around zero, introducing also a reduction of the mini-gap, which is bad for potential applications in topological quantum computation as it determines the degree of protection for the qubit. Although there are some limitations, these hybrid systems offer a direct way for monitoring the dependence if the MBSs on the superconducting phase difference, thus we believe that such hybrid structures constitute a powerful platform for the search of MBSs in one-dimensional systems.

To close this part, we emphasise that the energy spectrum of SNS nanowire junctions offers the possibility to directly monitor the Andreev bound states which trace the gap inversion 
and allows us to investigate their evolution into Majorana bound states in a powerful way. 
\newpage\clearpage
\subsection{Josephson and critical currents}
In Sec.\,\ref{introhybridNSSNS} we have introduced how to calculate the Josephson current based on the discrete Andreev spectrum of SNS junctions
\begin{equation}
\label{shortJosephcurrent2}
I(\varphi)=-\frac{e}{\hbar}\sum_{p>0}\frac{dE_{p}}{d\varphi}\,.
\end{equation}
Before going further we firstly discuss the validity of previous equation in short and long junctions. Notice that we are considering SNS junctions of finite length and therefore due to this 
fact their energy spectrum is discrete. Indeed, it consists of a discrete spectrum wihin the gap and of a quasi-continuum, which is also discrete due to the finite length of the
system, above the gap. Therefore, Eq.\,(\ref{shortJosephcurrent2}) allows us to include the discrete quasi-continuum contribution in the calculation of the Josephson 
current (supercurrent) $I(\varphi)$. Having established this we are in a position to use Eq.\,(\ref{shortJosephcurrent2}) for calculating the Josephson current in both short and long 
SNS junctions. 

Moreover, we define the critical current, $I_{c}$, which is the maximum supercurrent that flows across the junction.
 It is in general calculated by maximizing the Josephson current $I(\varphi)$ with respect to the superconducting phase difference $\varphi$, 
\begin{equation}
I_{c}={\rm max}[I(\varphi),\varphi]\,. 
\end{equation}
As before, here we present calculations for InSb nanowires, where the spin-orbit strength is $\alpha_{R}=20$\,meVnm and the typical 
induced superconducting gap $\Delta=0.25$\,meV \cite{Mourik:S12}. We expect that the Josepshon and the critical currents capture the presence of Majorana bound states for realistic 
parameters.

\begin{figure}[!ht]
\centering
\includegraphics[width=\textwidth]{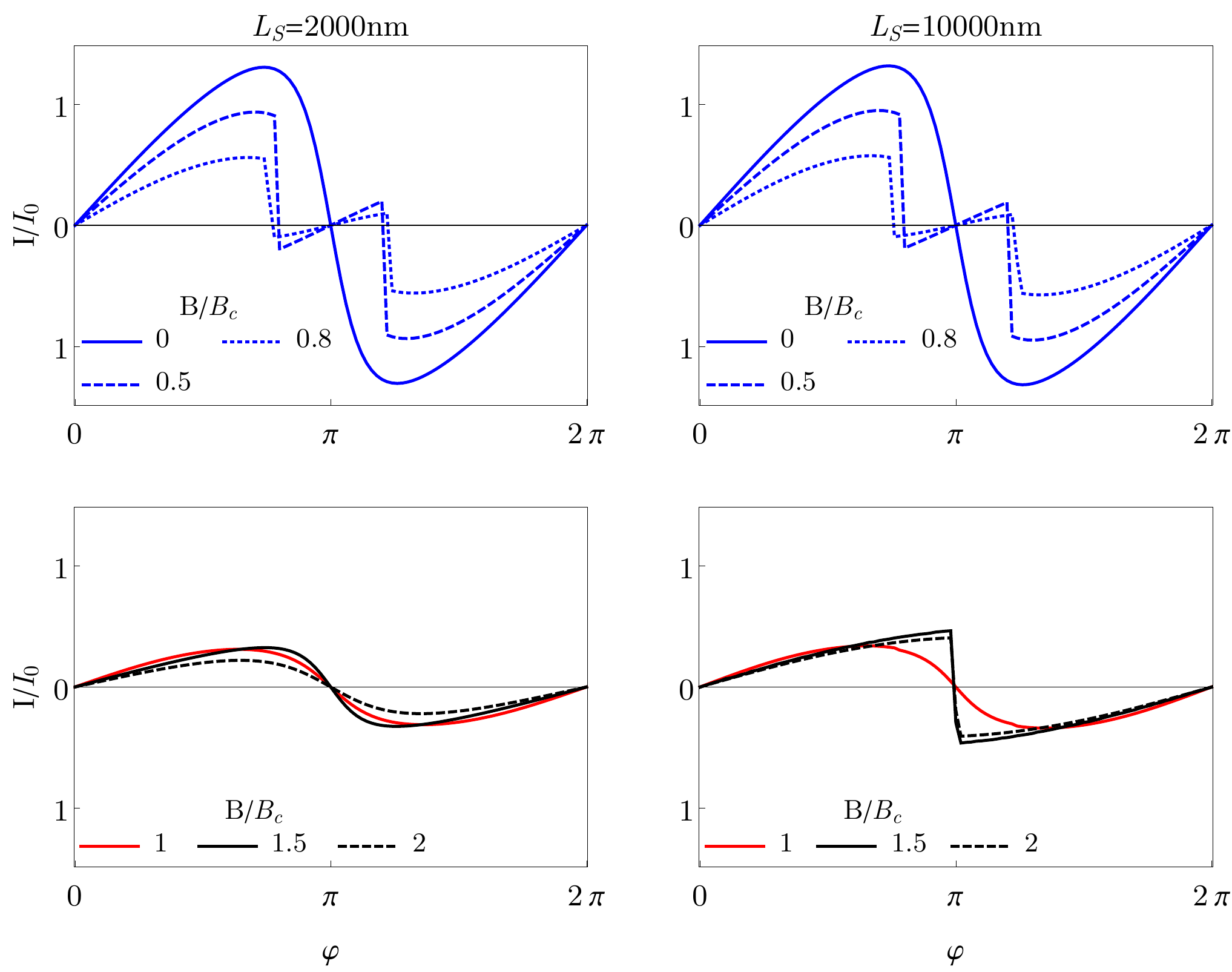} 
\caption[Josephson current in a short SNS junction]{(Color online) Josephson current in a short SNS junction as a function of the superconducting phase difference, 
$I(\varphi)$ for  (left column)  $L_{S}\ll2\ell_{M}$, $L_{S}=2000$\,nm, and (right column) $L_{S}\gg2\ell_{M}$, $L_{S}=10000$\,nm. Top row shows the Josephson current in the trivial phase for different values of 
the Zeeman field, $B<B_{c}$, while bottom for different values of the Zeeman field in the topological phase $B\geq B_{c}$. 
Notice the sawtooth feature at $\varphi=\pi$ for $L_{S}\gg2\ell_{M}$.
 Parameters: $L_{N}=2000$\,nm, $\alpha_{R}=20$\,meVnm, $\mu=0.5$\,meV, $\Delta=0.25$meV and $I_{0}=e\Delta/2\hbar$.}
\label{Iphishort1}
\end{figure}

\begin{figure}[!ht]
\centering
\includegraphics[width=\textwidth]{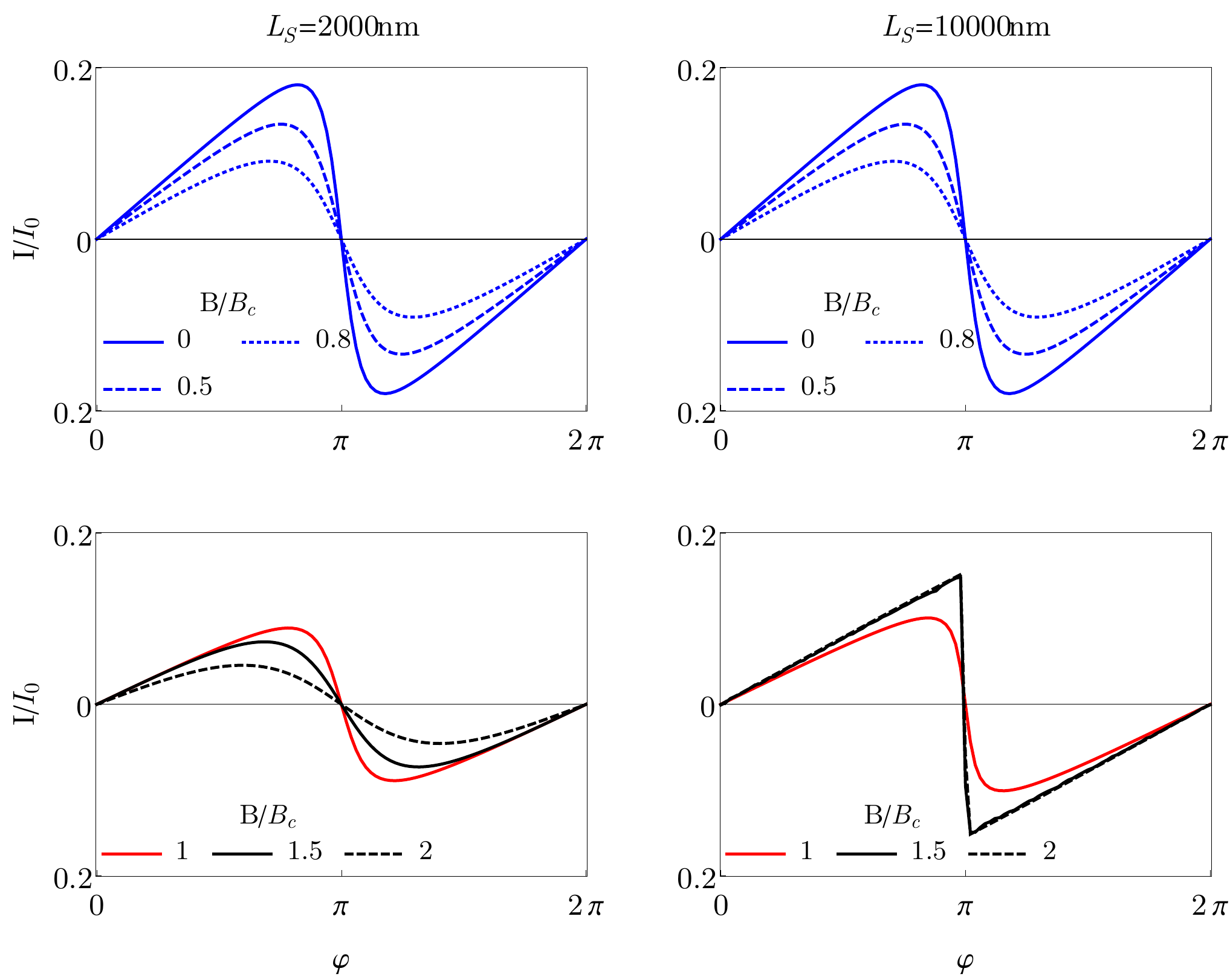} 
\caption[Josephson current in a long SNS junction]{(Color online) Josephson current in a long SNS junction as a function of the superconducting phase difference, 
$I(\varphi)$ for for  (left column)  $L_{S}\ll2\ell_{M}$, $L_{S}=2000$\,nm, and (right column) $L_{S}\gg2\ell_{M}$, $L_{S}=10000$\,nm. Top row shows the Josephson current in the trivial phase for different values 
of the Zeeman field, $B<B_{c}$, while bottom for different values of the Zeeman field in the topological phase $B\geq B_{c}$.
 Parameters: $L_{N}=2000$\,nm, $\alpha_{R}=20$\,meVnm, $\mu=0.5$\,meV, $\Delta=0.25$meV and $I_{0}=e\Delta/2\hbar$.}
\label{Iphilong1}
\end{figure}

Next, we present the Josephson current based on Eq.\,(\ref{shortJosephcurrent2}).
In Figs.\,\ref{Iphishort1} and \ref{Iphilong1} we present the evolution with the Zeeman field of the Josephson current as a function of the 
superconducting phase difference for finite length short and long SNS junctions, respectively.  Left row shows the supercurrent for $L_{S}\ll2\ell_{M}$, while right row for 
$L_{S}\gg2\ell_{M}$.
In short SNS junctions (Fig.\,\ref{Iphishort1}) at $B=0$ the Josephson current has the usual sine-like behaviour, which is reduced as the Zeeman field increases due to the reduction of the induced 
superconducting gap (see blue curves in top row). Notice that the supercurrent develops two kinks just before and after $\varphi=\pi$ for finite Zeeman field $B<B_{c}$ 
(see top panels). These correspond to the energy level crossings shown in Fig.\,\ref{figchap29}, which arise due to SOC.
Besides these features, the supercurrent curves for $B<B_{c}$, for both $L_{S}\ll2\ell_{M}$ and $L_{S}\gg2\ell_{M}$, exhibit a similar behaviour (see top row).

Interestingly, at the topological transition, $B=B_{c}$, the system is gapless but the current is not zero (see red curve in bottom row). 
For $B>B_{c}$, the S regions of the SNS junction become topological and Majorana bound states emerge at the ends of S. 
We have shown in Sec.\,\ref{Rashbawire} that the combination of SOC, Zeeman interaction and $s$-wave superconductivity generates two intraband $p$-wave pairings $\Delta_{--,++}$ 
and one interband $s$-wave $\Delta_{+-}$, where $+$ and $-$ denotes the energy bands. These two $\pm$ sectors have different gaps $\Delta_{1,2}$ at low and high momemtum, respectively,
that depend on the Zeeman field. The gap $\Delta_{1}$ is the topological gap which closes at $B=B_{c}$ and gives rise to the topological phase for $B>B_{c}$, while the high momentum gap 
remains finite and roughly constat for strong SOC. 
We have also seen that the topological phase is effectively reached due to the generation of an effective $p$-wave superconductor, which is the result of projecting the system Hamiltonian onto the lower band keeping 
only the intraband $p$-wave pairing $\Delta_{--}$. Deep in the topological phase the lowest gap is $\Delta_{2}$.
We have shown in Fig.\,\ref{figchap29} that the Andreev levels are associated with the gaps $\Delta_{1,2}$ and they trace the closing of the reopening of the gap. Remarkably, 
in the topological phase the lowest gap is $\Delta_{2}$ and therefore the number of levels within such gap is less than in the trivial phase.
This causes a reduction in the supercurrent as it is the direct result of the Andreev levels (see bottom row in Fig.\,\ref{Iphishort1}).
Interestingly, despite the presence of MBSs the current is $2\pi$ periodic.
For $L_{S}\ll2\ell_{M}$, there is a finite spatial Majorana wave function overlap and therefore the Josephson current in the topological phase tends to 
decrease (black curves in bottom left panel), while for $L_{S}\gg2\ell_{M}$ such overlap is negligible and the current do not decreases but it rather develops a 
clear sawtooth profile at $\varphi=\pi$ (see  black curves in bottom right panel). Notice that the supercurrent $I(\varphi)$ for $B<B_{c}$ shown in 
left ($L_{S}\ll2\ell_{M}$) and right ($L_{S}\gg2\ell_{M}$) top panels of Fig.\,\ref{Iphishort1} exhibit a similar behaviour even despite the length of the superconducting regions $L_{S}$ is different in both cases. 
Quite remarkably the supercurrent for $B>B_{c}$, shown in left and right bottom panels of Fig.\,\ref{Iphishort1}, exhibits a different dependence on $\varphi$ due to the 
emergence of MBSs.
While for $L_{S}\ll\ell_{M}$ (left bottom panel) the finite spatial overlap of Majorana wave-functions obscures the direct detection of MBSs, 
for  $L_{S}\gg\ell_{M}$ (right bottom panel) the sawtooth profile at $\varphi$ determines the presence of MBSs and this happens only for $B>B_{c}$.

As the length of the normal region is increased, one expects the Josephson current to be reduced. 
This is shown in Fig.\,\ref{Iphilong1} for a long junction considering $L_{S}\ll2\ell_{M}$ and $L_{S}\gg2\ell_{M}$ and spanning the trivial and topological phase. 
Notice that although the amplitude of the current is reduced in this case, the behaviour is similar to the short junction regime. 
Indeed, in the trivial phase (blue curves in top row), as one increases the Zeeman field, the current gets reduced. 
As the system enters into the topological phase, we distinguish two situations. For a finite Majorana wave function overlap $L_{S}\ll2\ell_{M}$, in the topological phase, 
the current decreases signalling the non-zero splitting at $\varphi=\pi$ in the energy spectrum. However, for $L_{S}\gg2\ell_{M}$, as the Zeeman field increases in the topological phase, the Josephson current acquires a constant value and it develops a clear sawtooth signature at $\varphi=\pi$ due fact that the splitting in the energy spectrum at $\varphi=\pi$ is zero.

Now, we present results for the critical current in short SNS junctions.
\begin{figure}[!ht]
\centering
\includegraphics[width=\textwidth]{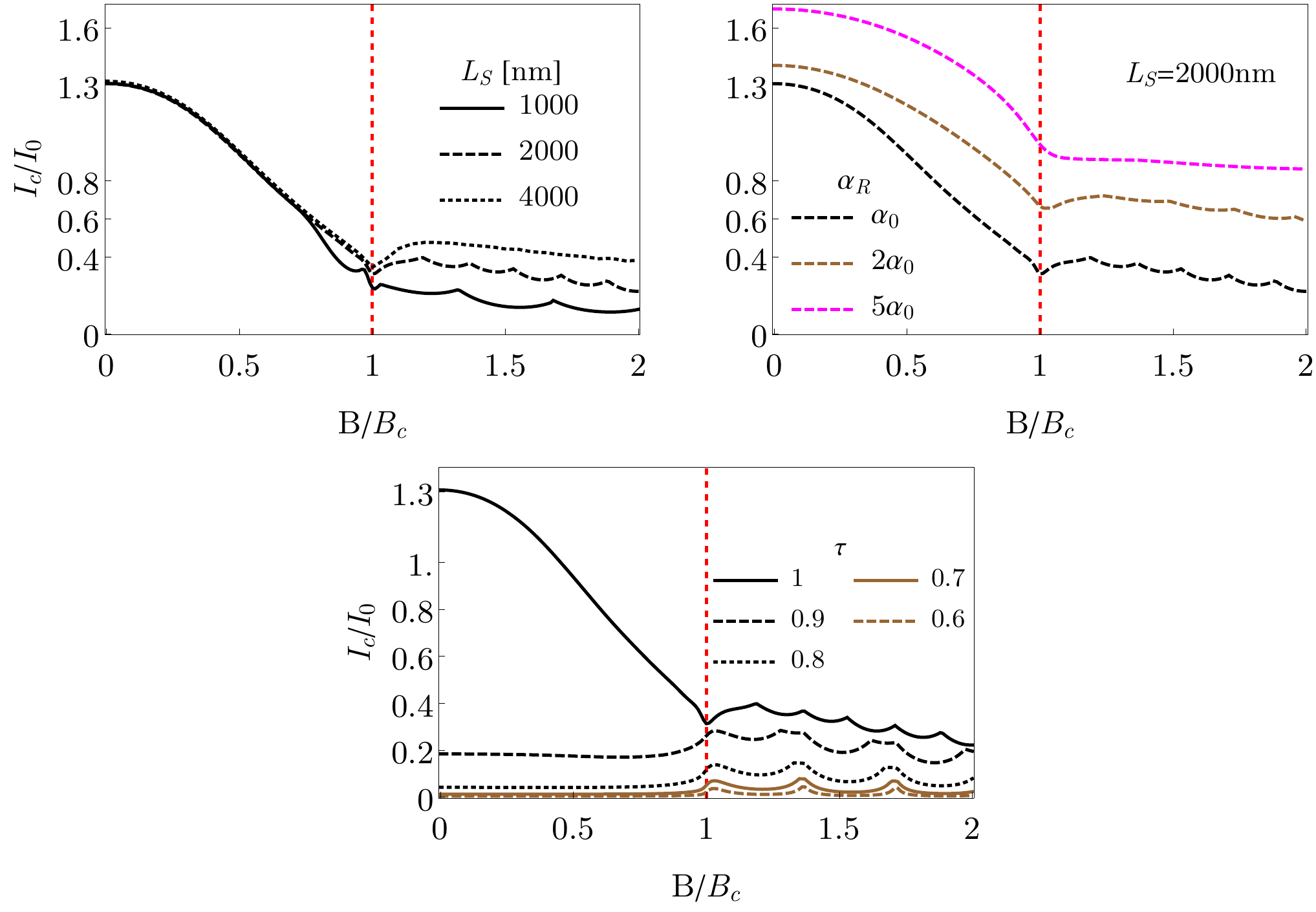} \\
\caption[Critical current in a short SNS junction]{(Color online) Critical current $I_{c}$ in a short SNS junction as a function of the Zeeman field. The red vertical dashed line marks the topological transition $B=B_{c}$. Top left panel: different curves represent different values of the superconducting region length $L_{S}$. 
Notice that the critical current is finite at $B=0$ and as $B$ increases it gets reduced and traces the gap closing and its reopening, exhibiting a non-trivial feature at the topological transition $B=B_{c}$, where, remarkably, it remains finite. For $B>B_{c}$, the critical current is finite and develop an oscillatory behaviour, which is related to the finite Majorana overlap, which is decreased by increasing the length $L_{S}$. Top right panel: Fixed $L_{S}=2000$\,nm and increasing the value of SOC $\alpha_{R}$. Notice that here the Majorana overlap is decreased by increasing the SOC and thus decreasing the Majorana localisation length $\ell_{M}$. 
 Bottom panel: fixed $L_{S}=2000$\,nm and decreasing the hopping between the superconducting and normal regions, $\tau$. In the trivial phase, $B<B_{c}$, the critical current is considerable reduced and for small values of $\tau$ it reaches zero. On the other hand, in the topological phase, $B>B_{c}$, although the current is reduced by decreasing $\tau$, it is not zero and rather exhibit oscillations due to Majorana overlap.
  Parameters: $L_{N}=20$\,nm, $\alpha_{0}=20$\,meVnm, $\mu=0.5$\,meV, $\Delta=0.25$meV and $I_{0}=e\Delta/2\hbar$.}
\label{Icshort1}
\end{figure}
In Fig.\,\ref{Icshort1} we present calculations of the critical current $I_{c}$ in short SNS junctions as the Zeeman field $B$ increases from the trivial ($B<B_{c}$) to the topological phase ($B>B_{c}$).
Different curves in top left panel correspond to different values of the superconducting region length $L_{S}$. Observe that the current is maximum at zero field and as the Zeeman 
field $B$ increases the current decreases almost linearly tracing the reduction of the low-momentum gap $\Delta_{1}$ discussed in Sec.\,\ref{Rashbawire}. We have shown that such gap closes at $B=B_{c}$ and marks the topological transition point into the topological superconducting phase with MBSs.

Remarkably, at the topological transition, $B=B_{c}$, the critical current remains finite and develops a robust non-trivial 
feature, which is the result of the closing of the gap in the energy spectrum. Roughly speaking half of its value at $B=0$ because the system is losing one channel 
after the topological transition. As the the Zeeman field is further increased for $B>B_{c}$, the critical current develop oscillations, 
which arise from the finite spatial overlap of the Majorana wave functions when $L_{S}\ll2\ell_{M}$. Notice that one way to reduce such overlap consists of increasing $L_{S}$, as it can be seen in top left panel. On the other hand, one can also reduce it by increasing the SOC and therefore decreasing the Majorana localisation length, as shown in top right panel. Although both ways are efficient, the former is more realistic including the variation of the length of the proximitized region, while the latter requires to increase a material property (the SOC) or to increase it by electrostatic gates, which is not an easy task at all. Notice that while in the former case, the effect of increasing $L_{S}$ only reduces the oscillations of the critical current in the topological phase without affecting its value in general, in the latter situation, by increasing the SOC, the critical current is increased.
Experimentally, however, is not very easy to fabricate SNS junctions with a perfect coupling between N and S regions, although recently there has been a considerable increase in fabrication techniques of such systems. 
In subsection \,\ref{SNSjunctionmodel}, we have shown that the coupling between the N and S regions can be parametrised by the parameter $\tau\in[0,1]$. In principle, by varying  such coupling, one can control the transparency, where $\tau=1$ determines a full transparent junction.
In bottom panel of Fig.\,\ref{Icshort1} we show the critical current as a function of the Zeeman field for different values of $\tau$ in a finite length SNS short junction.

Notice that as the control parameter $\tau$ decreases, the critical current also decreases. In the trivial phase, $B<B_{c}$, for small values of $\tau$, the critical current is considerable reduced and even zero. Remarkable, however, the critical current develops a robust feature at the closing of the topological gap and it exhibits a finite value for all $\tau$. 
In the topological phase, $B>B_{c}$, the critical current shows a non-monotonic behaviour for increasing $B$ with oscillations due to Majorana overlap. 

Thus, from changes in $I_{c}$ these features could be used to detect the topological transition and therefore the topological phase with Majorana bound states.
\newpage\clearpage

\section{Conclusions}
In this Chapter we have introduced the basic phenomena that takes place in hybrid NS and SNS junctions. Then, we have shown how to model such systems based on nanowires with spin-orbit coupling assuming good contact between nanowire and superconductor.

A detailed study of the Andreev bound states have been presented and the emergence of Majorana 
bound states have been investigated for short and long SNS junctions. In this part, we showed the Andreev bound states as function of the superconducting phase difference for different values of the Zeeman field, spanning the trivial and topological phases.
At this level, we point out that, in principle, Andreev levels spectroscopy
should be able to test the emergence of MBSs.

Afterwards we have focused on the Josephson current dependence on the superconducting phase difference, calculated from the Adreev spectrum in short and long SNS junctions. We have found that the presence of MBSs in the topological phase is a distinguishable signature in the Josephson current despite of being $2\pi$ periodic. Indeed, in the topological regime, the Josephson current is reduced since the number of ABSs is reduced. Additionally, when the spatial overlap is negligible between the MBSs, the Andreev spectrum is translated into the Josephson current in a clear sawtooth profile at $\varphi=\pi$, representing a clear manifestation of the MBSs.

We also present the critical current as a function of the Zeeman field. This quantity allows us to make a detailed study of the topological transition and trace the behaviour of the topological gap, which after the band inversion gives rise to the topological phase with MBSs. 
We report that the topological transition point is a robust feature, which we expect to be distinguishable from other mechanisms in real experiments.



\chapter{\bf Voltage-biased transport in SNS junctions\footnote{The results presented in this chapter have been published in \cite{PabloJorge}.}} 
\label{Chap3}
\lhead{Chapter \ref{Chap3}. \emph{Voltage-biased transport in SNS junctions}} 

\begin{small}
We study transport in a voltage biased superconductor-normal-superconductor (SNS) junction made of semiconducting nanowires with strong spin-orbit coupling, as it transitions into a topological superconducting phase for increasing Zeeman field. Despite the absence of a fractional steady-state ac Josephson current in the topological phase, the dissipative multiple Andreev reflection (MAR) current $I_{dc}$ at different junction transparencies is particularly revealing. It exhibits unique  features related to topology, such as the gap inversion, the formation of Majorana bound states, and fermion-parity conservation. Moreover, the critical current $I_c$, which remarkably does not vanish at the critical point where the system becomes gapless, provides direct evidence of the topological transition. These results demonstrate the feasibility to probe the formation of Majorana states without the need of phase sensitive or noise-related measurements. Therefore, for junctions with tuneable transparency, this collection of signatures constitutes an unmistakable fingerprint of Majorana bound states.
\end{small}

\newpage

\section{Introduction}
In Chapter \ref{Chapter01} we have shown that semiconducting nanowires (NWs) with a strong spin-orbit (SO) coupling in the proximity of s-wave superconductors and in 
the presence of an external Zeeman magnetic field $B$ are a promising platform to study Majorana physics. 
Theory predicts that above a critical field $B_c\equiv\sqrt{\mu^2+\Delta^2}$, defined in terms of the Fermi energy $\mu$ and the induced s-wave pairing $\Delta$, the 
wire undergoes a topological transition into a phase hosting zero energy Majorana bound states (MBSs) at the ends of the wire \cite{Lutchyn:PRL10,Oreg:PRL10}. 
We have discussed in Chapter \ref{Chapter01} that the emergence of a zero-bias peak in the differential conductance $dI/dV$ supports the existence of MBSs at normal-superconductor 
(NS) junctions.
Stronger evidence could be provided by the observation of non-Abelian interference (braiding) \cite{Nayak:RMP08}, or by transport in phase-sensitive 
superconductor-normal-superconductor (SNS) junctions. The latter approach, which typically involves the measurement of an anomalous ``fractional" $4\pi$-periodic ac 
Josephson effect \cite{Kitaev:P01,Fu:PRB09,Kwon:EPJB03}, is much less demanding than performing braiding. Realistically, however, the fractional effect, detected through, 
e.g., the absence of odd steps in Shapiro experiments \cite{Kwon:EPJB03,Jiang:PRL11,Dominguez:PRB12,Rokhinson:NP12}, may be difficult to measure (dissipation is expected to 
destroy it in the steady state), or may even develop without relation to topology \cite{Sau:12b}. 
Although it has been shown that the $4\pi$ periodicity survives in the dynamics, such as noise and transients \cite{Badiane:PRL11, San-Jose:PRL12a,Pikulin:PRB12}, simpler experimental probes of MBSs are extremely desirable.

Here we propose the multiple Andreev reflection (MAR) current in voltage-biased SNS junctions made of NWs \cite{Doh:S05,Nilsson:NL12}
as an alternative, remarkably powerful, yet simple tool to study the topological transition. 
This is made possible by the direct effect that gap inversion, MBS formation and fermion-parity conservation have on the MAR current $I_{dc}(V)$ at various junction transparencies $T_N$. 
For tunnel junctions, $I_{dc}(V)$ traces the closing and reopening of the superconducting gap at $B_c$, $\Delta_\mathrm{eff}\sim|B-B_c|$. This gap inversion can be shown to be a true topological transition by tuning the junction to perfect transparency $T_N=1$. In this regime, the limiting current $I_{dc}(V\to 0)$ shows signatures of the parity conservation effects that are responsible for the fractional Josephson current in the presence of MBSs, but which, in contrast to the latter, survive in the steady state limit. Moreover, the detailed dependence of MAR as a function of $T_N$ has the fundamental advantage over NS junctions in that it contains information about the peculiar dependence of MBS hybridization with superconductor phase difference $\phi$, despite not requiring any external control on it. Similarly, we show that another important phase-insensitive quantity, the critical current $I_c$, remains unexpectedly finite for all $B$ due to a significant continuum contribution, and exhibits an anomaly at the topological transition.

This chapter is organized as follows. In section \ref{Rashba NW}, we review the model for Rashba nanowires in the presence of both s-wave superconducting pairing and an external Zeeman field and describe how a
spinless p-wave superconductor regime can be achieved. In particular, we discuss how the problem can be understood in terms of two independent p-wave superconductors, originated from the Rashba helical bands, and weakly coupled by an interband pairing term. This two-band description is very useful in order to understand the main results of this paper. Such results are discussed in section \ref{SNS} which is divided in two parts. The first part (subsection  \ref{ABS-SNS}) is devoted to the ABSs which are confined in the junction. The detailed evolution of these ABSs as the system undergoes a topological transition has not been discussed in the literature, to the best of our knowledge, and becomes essential in order to gain a deep understanding of transport across the junction, which is discussed in subsection \ref{ac Josephson}. In this part we study the ac Josephson effect in nanowire SNS junctions, with focus on how the MAR currents reflect the topological transition in the nanowires as the Zeeman field increases. In particular, we present a thorough analysis of MAR transport in topological SNS junctions for arbitrary transparency of the normal part. Finally, we discuss in section \ref{Critical current} how the critical current $I_c$ does not vanish at the critical point where the system becomes gapless and, importantly, how $I_c$ provides direct evidence of the topological transition. 
%

\section{Rashba nanowire model and effective p-wave pairing}\label{Rashba NW}
\begin{figure} 
   \includegraphics[width=1 \columnwidth]{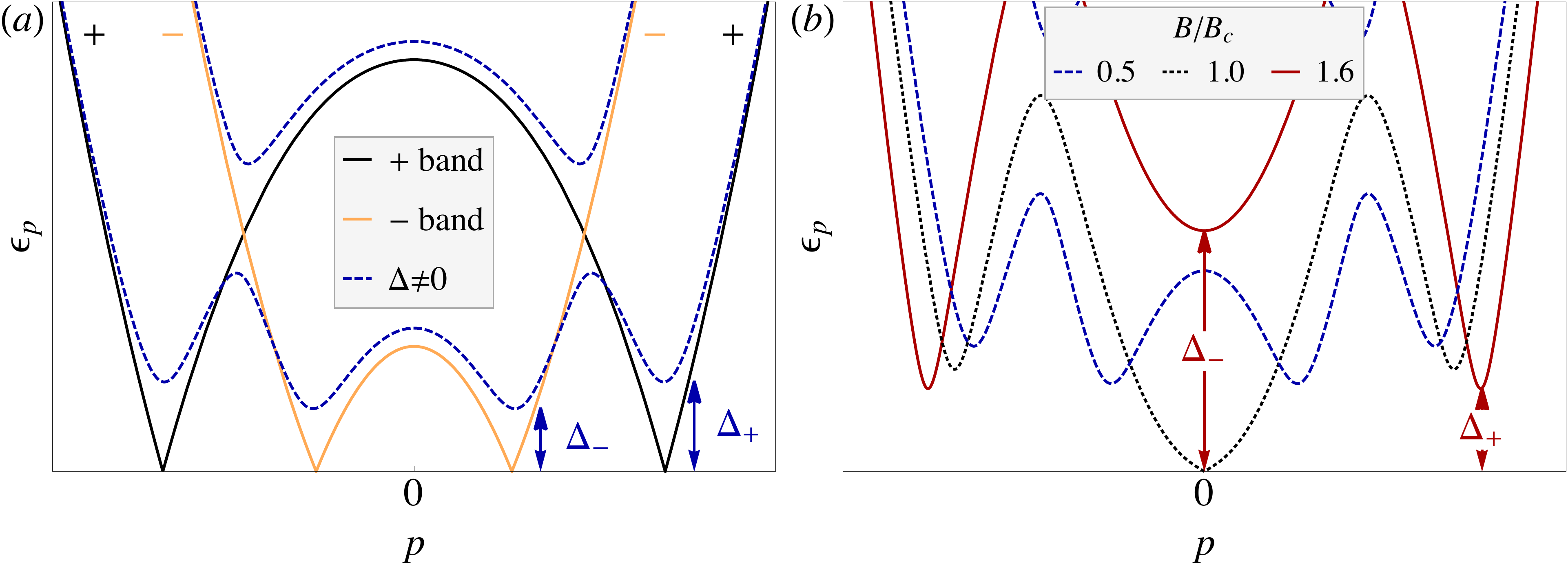} 
     \caption[Lowest bands of a nanowire]{(Color online) (a) Lowest bands of a $B=0.5B_c$ nanowire, with (dashed) and without (solid) pairing $\Delta$. 
     (b) Evolution of bands with Zeeman field $B$. Gap $\Delta_{1}=\Delta_-$ closes at $B=B_c$, while $\Delta_{2}=\Delta_+$ does not.
}
   \label{fig:bands}
\end{figure}

Before going further we firstly review the possibility for engineering one-dimensional topological superconductivity based based on
nanowires with SOC in proximity to a $s$-wave superconductor (see Chapter \ref{Chapter01} for more details).
A single one-dimensional NW in the normal state is described by the Hamiltonian  
\begin{equation}
\label{modelX0}
H_0=\frac{p^2}{2m^*}-\mu-\frac{\alpha_\mathrm{R}}{\hbar}\sigma_y p+B\sigma_x,
\end{equation}
where $m^*$ is the effective mass, $\alpha_\mathrm{R}$ the SO strength, $\mu$ the chemical potential and $\sigma_i$ the spin Pauli matrices. An external 
magnetic field $\mathcal{B}$ along the wire produces a Zeeman splitting $B=\frac{1}{2}g\mu_B \mathcal{B}$, where $\mu_B$ is the Bohr
magneton and $g$ is the wire g-factor. 
The Nambu Hamiltonian
\begin{equation}\label{modelX1}
H=
\begin{pmatrix}
H_0 & -i\Delta\sigma_y\\
i\Delta^*\sigma_y & -H^*_0\\
\end{pmatrix}\,,\end{equation}
models the NW in the presence of an induced s-wave superconducting pairing $\Delta$ (here assumed real without loss of generality).
The essential ingredient for a topological superconductor is an effective p-wave pairing acting on a single (``spinless") fermionic species 
\cite{Kitaev:P01}. SO coupling splits NW states into two subbands of opposite helicity at $B=0$. At finite $B$, these two subbands, 
which we label $+$ and $-$ [black and orange lines in Fig. \ref{fig:bands}(a)], have spins canted away from the SO axis.
Notice that here we have a notation opposite to the one given in Chap.\,\ref{Chapter01}.
The s-wave pairing $\Delta$, 
expressed in the $\pm$ basis, takes the form of an intraband p-wave $\Delta^{++/--}_{p}(p)= \pm i p\Delta\alpha_\mathrm{R}/\sqrt{B^2+(\alpha_\mathrm{R} p)^2}$, 
plus an interband s-wave pairing $\Delta^{+-}_s(p)=\Delta B/\sqrt{B^2+(\alpha_\mathrm{so} p)^2}$ \cite{Alicea:RPP12}. 

Without the latter, the problem decouples into two independent p-wave superconductors, while $\Delta^{+-}_s$ acts as a weak coupling between them. 
Each quasi-independent $\pm$ sector has a different ($B$-dependent) gap, which we call $\Delta_{1}=\Delta_-$ (at small $p$) and $\Delta_{2}=\Delta_+$ (large $p$), 
see Fig. \ref{fig:bands}(a,b). While $\Delta_+$ remains roughly constant with $B$ (for strong SO coupling  \cite{Prada:PRB12,Rainis:PRB13}), $\Delta_-$ vanishes linearly as $B$ approaches the critical field, $\Delta_-\approx |B-B_c|$ \footnote{Note that, in general, $\Delta_-$ is at a small but finite momentum. However, as $B$ approaches $B_c$, $\Delta_-$ becomes centered at $p=0$ and is approximately equal to $|E_0|$, where $E_0$ is the zero momentum energy of the lowest subband, $E_0=B-B_c$, and is related to the topological charge of the lowest superconducting band. \cite{Ghosh:PRB10}}. This closing and reopening (gap inversion) signals a topological transition, induced by the effective removal of the $-$ sector away from the low-energy problem. Below $B_c$ the NW is composed of two 
spinless p-wave superconductors, and is therefore topologically trivial. Above $B_c$, $\Delta_-$ is no longer a p-wave gap, but rather a normal (Zeeman) spectral gap already present in the normal state, transforming the wire into a single-species p-wave superconductor with non-trivial topology. This phase contains MBSs, protected by the effective gap $\Delta_\mathrm{eff}=\mathrm{Min}(\Delta_+,\Delta_-)$, at the wire ends. Above a certain field $B_c^{(2)}$, the gap $\Delta_\mathrm{eff}$ saturates at $\Delta_+$ and the physics of superconducting helical edge states in spin-Hall insulators is recovered \cite{Fu:PRB09,Badiane:PRL11}. 

\section{Nanowire SNS junctions}\label{SNS}

\begin{figure}[t] 
   \centering
   \includegraphics[width=0.5\columnwidth]{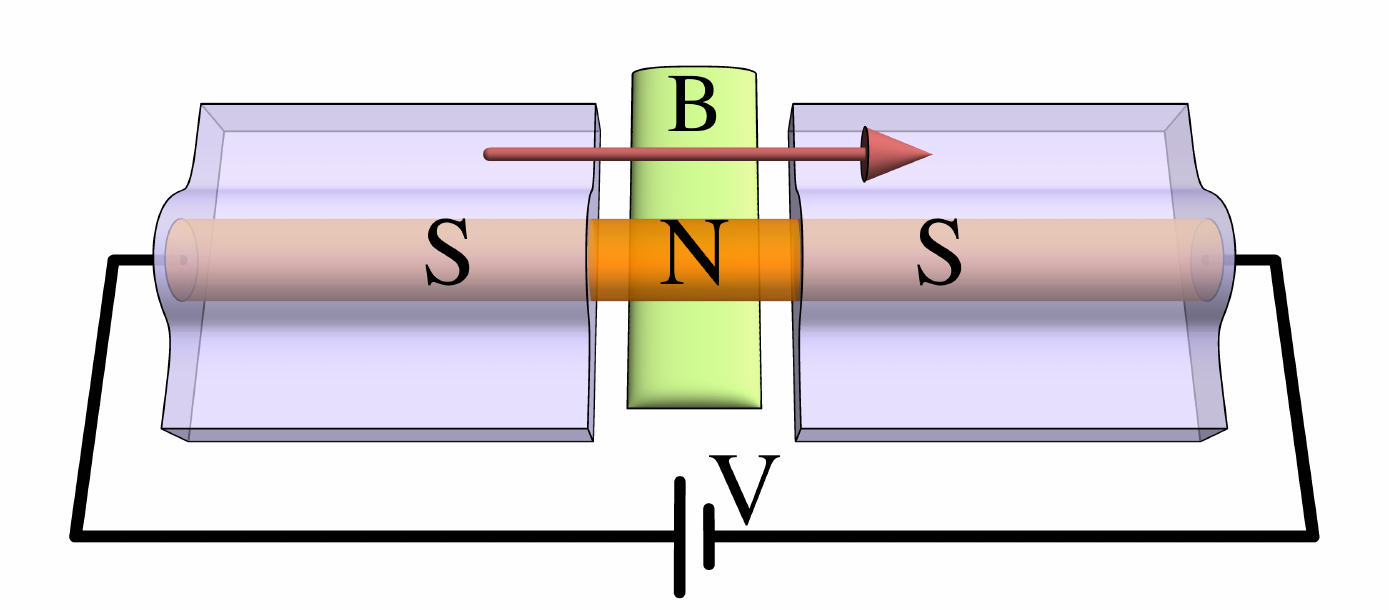}
     \caption[Sketch of a short SNS junction]{(Color online) Short SNS junction fabricated by covering a semiconductor nanowire with two S-wave superconductors. A bias $V$ and a longitudinal Zeeman field $B$ can be applied to the wire. The central normal region has tuneable transparency via a depletion bottom gate.
}
   \label{fig:sys}
\end{figure}
In the previous section we described how a semiconducting nanowire with a strong SO coupling in the proximity of an s-wave superconductor and in the presence of an external 
Zeeman magnetic field $B$ behaves as a topological superconductor above a critical field $B_c$. See also Chapter \ref{Chapter01} for further details.

Here we are concerned with the effects of this topological transition on the MAR current $I_{dc}(V)$ across junctions formed with such nanowires. 
In particular, we consider SNS junctions of different normal transparencies $T_N$. 
Experimentally, such geometry can be fabricated by partially covering a 
single NW with two superconducting leads and leaving an uncovered normal region in the middle. The coupling of the normal part of the NW to the superconducting leads 
can be tuned by local control of the electron density with a gated constriction. This can be realized by using, e.g., bottom-gates forming a quantum point contact, see 
Fig. \ref{fig:sys}. Such geometry has been successfully implemented experimentally in Ref.  \cite{Churchill:PRB13} for NS junctions, where control of the coupling between 
the superconducting and normal sections from near pinch-off (tunneling limit) to the multichannel regime is demonstrated. We point out that the SNS junctions taken into account here 
differ from the ones discussed in Chapter \ref{Chap2a} in that 
 the superconducting leads are semi-infinite unlike the finite length ones in Chapter \ref{Chap2a}.
\begin{figure}[t] 
   \centering
   \includegraphics[width=\columnwidth]{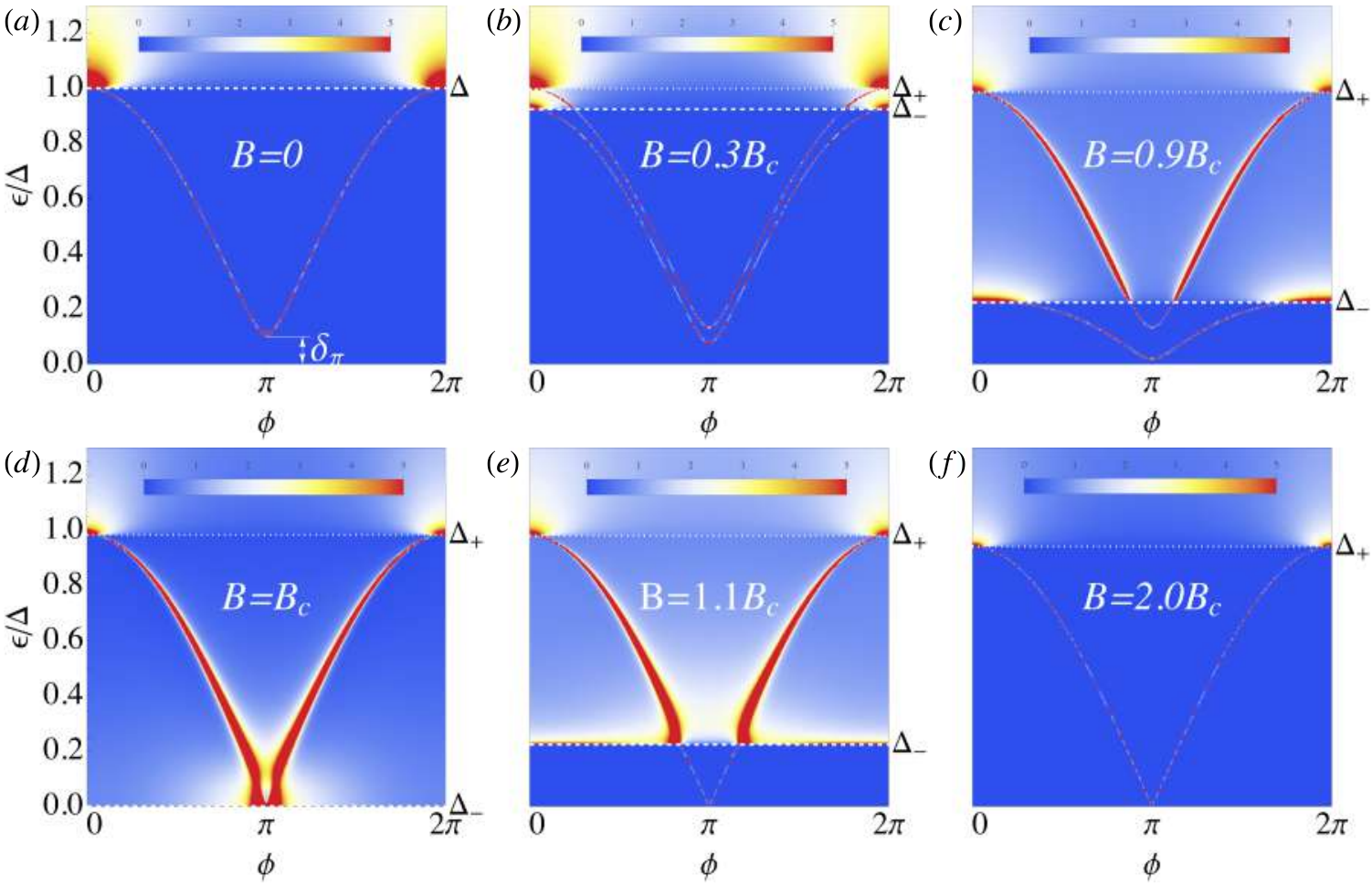} 
     \caption[Andreev (quasi)bound states $\epsilon(\phi)$]{(Color online) Local density of states at the junction for perfect normal transparency $T_N=1$, which is peaked at the energy $\epsilon_\pm(\phi)$ of Andreev (quasi)bound states. Different panels show how the Andreev states evolve as the system undergoes the topological transition.
}
   \label{fig:ABS}
\end{figure}
For simplicity, here we focus on short SNS junctions\footnote{Results for the long junction limit are discussed in Chap.\,\ref{Chap2a}} with single channel nanowires. 
For computation purposes, we consider a discretisation of the continuum model Eq. (\ref{modelX0}) for the Rashba nanowire into a tight-binding lattice with a small lattice spacing 
$a$, as describes in Chapter \ref{Chap2a}. This transforms terms containing the momentum operator $p$ into nearest-neighbour hopping matrices $v$. 
Namely $H_0=\sum_{i}c^+_ihc_i+\sum_{\langle ij\rangle}c^+_i vc_j + \mathrm{h.c.}$,  with
\[
h=\left(\begin{array}{cc}2t-\mu & B \\B & 2t-\mu\end{array}\right),\hspace{0.5cm} v=\left(\begin{array}{cc}-t & \frac{\hbar}{2a}\alpha_\mathrm{so} \\-\frac{\hbar}{2a}\alpha_\mathrm{so} & -t\end{array}\right),
\] 
are matrices in spin space, and $t=\hbar^2/2m^*a^2$. The pairing is incorporated like in Eq. (\ref{modelX1}). A short SNS junction is modelled by suppressing the hopping matrix $v_0=\nu v$ between two sites in the middle of the wire, which represent the junction. The  dimensionless factor $\nu\in[0,1]$ controls the junction's normal transparency at $B=0$, which we denote $T_N(\nu)$. A phase difference $\phi$ across the junction is implemented by multiplying $\Delta$ to the left and right of the junction by $e^{\mp i\phi/2}$, respectively. Despite the simplicity of this description, it contains the relevant physics of a short SNS junction. As it has been shown for standard junctions \cite{Cuevas:PRB96}, such physics essentially depend on the contact normal transmission as well as the voltage drop across it \footnote{Note that, for the sake of simplicity, we do not include the possibility of junctions containing resonant levels or quantum dots. A study of such junctions, including Coulomb blockade effects, is beyond the scope of this paper but might be useful in order to analyse the possibility of Majorana physics arising in experiments with short SNS junctions containing quantum dot nanowires, such as the ones reported in Ref. \cite{Deng:NL12}}. Thus, we expect that a more detailed modeling, including e.g. a spatially-dependent voltage drop, would only modify the effective transmission $T_N(\nu)$ which defines the different regimes we shall explain in the following.

\subsection{Andreev bound states}\label{ABS-SNS}
In Chapter \ref{Chap2a} finite length SNS junctions and therefore we discussed the system with four MBSs, two inner at the junction and two outer at the ends of the superconducting 
regions. Here, however, take into account semi-infinite superconducting leads so that the outer MBSs are not involved in the description.

In such short SNS junction, an ABS should form for each of the two p-wave sectors described in section \ref{Rashba NW} for $B<B_c$, 
while only one, associated to $\Delta_+$, should remain for $B>B_c$. To support this picture, we present calculations of the local density of states (LDOS) 
at the junction in the transparent limit ($T_N=1$). This LDOS is peaked at the energy $\epsilon_\pm$ of the ABS, which is a function of the phase difference $\phi$ across the 
junction. For $B=0$ (Fig. \ref{fig:ABS}a) the two ABSs are degenerate and confined within the gap $\Delta$ \footnote{Note that even this non-topological case is anomalous as 
the ABS energies do not reach zero at $\phi=\pi$, unlike predicted by the standard theory for a transparent channel $T_N=1$ within the Andreev approximation $\mu\gg \Delta$ 
\cite{Beenakker:92}. We have checked that the energy minimum $\delta_\pi$ does indeed vanish as $\mu/\Delta$ grows, see Fig. \ref{fig:deltapi} in \ref{AppendixB3}}. 
As the Zeeman field increases, Fig. \ref{fig:ABS}b, the two ABS split and the system develops the two distinct gaps $\Delta_+$ and $\Delta_-$ described in Section 
\ref{Rashba NW}. Note that both ABSs are truly bound at energies below the lowest gap $\Delta_-$, but only  \emph{quasibound}  in the energy window $\Delta_-<\epsilon<\Delta_+$. 
This is clearly seen in the plots as broadening of the ABS resonances (see, e.g.  Fig. \ref{fig:ABS}c). As $B$ approaches the critical field $B_c$, $\Delta_-$ gets reduced, 
and becomes exactly zero at $B=B_c$. Note that at this point the upper ABS has reached zero energy at  $\phi=\pi$ and is quasibound for all energies, Fig. \ref{fig:ABS}d. 
Upon entering the topological phase ($B\geq B_c$), $\Delta_-$ reopens but one of the ABSs of the problem \emph{has disappeared} (Fig. \ref{fig:ABS}e). 
The surviving ABS associated to $\Delta_+$ arises as the hybridisation of the two emerging MBSs across the junction. Global fermion-parity conservation protects 
the $\phi=\pi$ level crossing. Due to the residual $\Delta^{+-}_s$ coupling between the two sectors, the $\Delta_+$ Andreev state is once more quasibound in the energy 
window $\Delta_-<\epsilon<\Delta_+$. At high enough magnetic fields, $\Delta_+$ is the smallest gap of the problem and hence the Majorana ABS is truly bound, 
Fig. \ref{fig:ABS}f. In long junctions, more ABSs can be confined in the junction as discussed in Chap.\,\ref{Chap2a} \cite{PhysRevB.91.024514,Chevallier:PRB12}. 
These extra ABSs coexist with the ones described here and may, for example, anti-cross with the Majorana-like $\Delta_+$ Andreev level, affecting its character near zero energy 
\cite{Prada:PRB12,PhysRevB.91.024514}.
\subsection{ac Josephson effect and MAR currents}\label{ac Josephson}
\begin{figure}[t] 
   \centering
   \includegraphics[width=\linewidth,clip]{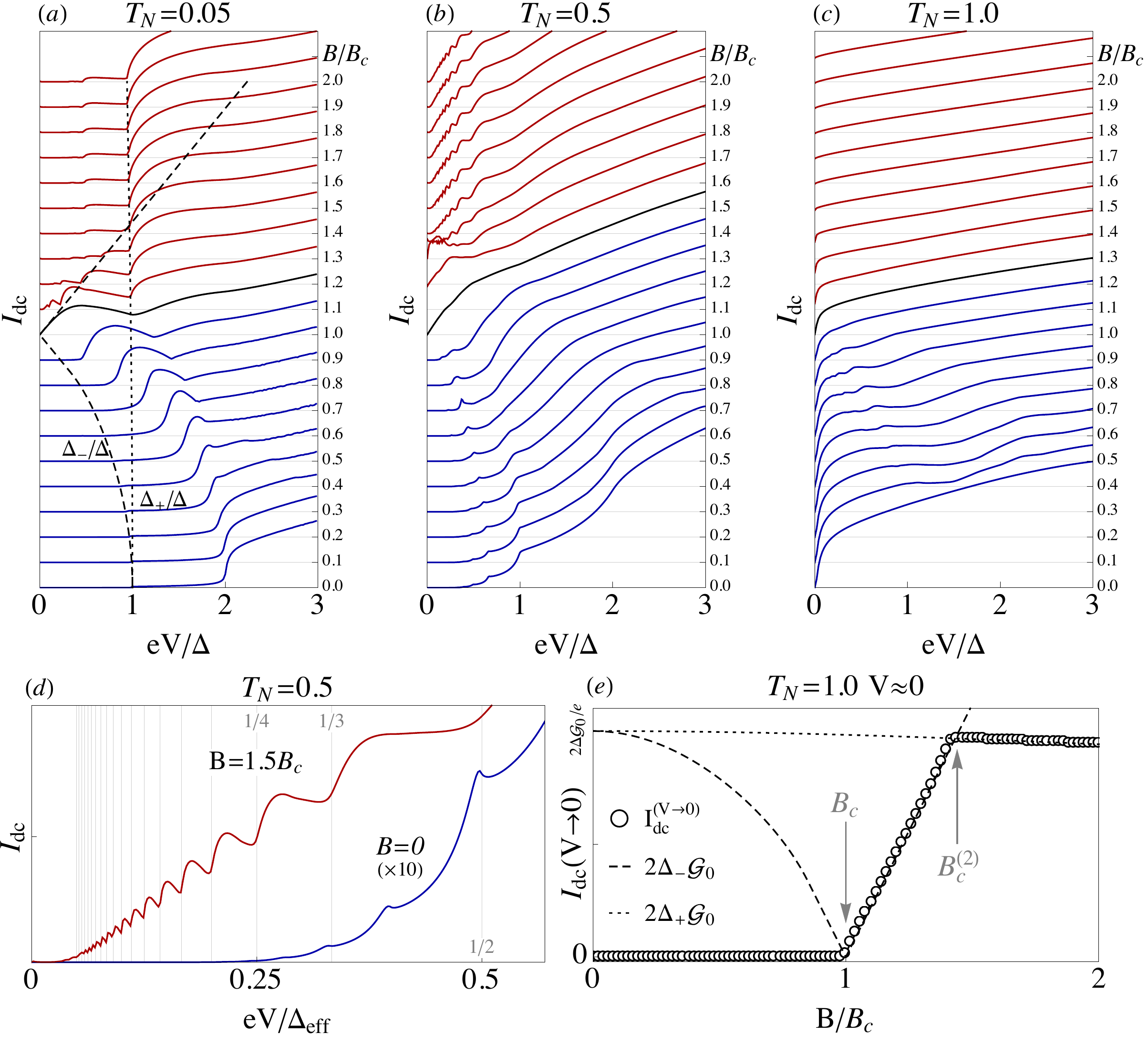} 
     \caption[MAR current for tunnelling, intermediate and full transparency.]{(Color online) Time-averaged Josephson current $I_{dc}$ as a function of bias $V$ for increasing Zeeman field $B$. Curves are offset by a constant $2\Delta \mathcal{G}_0/e$, with $\mathcal{G}_0=e^2/h$. Blue and red curves correspond to the non-topological ($B<B_c$) and topological ($B>B_c$) phases respectively. Panels (a) to (c) show the cases of tunnelling, intermediate and full transparency.  Panel (d) is a blowup of the low bias MAR subharmonics at intermediate transparency. Panel (e) shows the asymptotic $I_{dc}(V\to 0)$ at full transparency (circles), along with the dependence of the quantities $2\Delta_- \mathcal{G}_0$ and $2\Delta_+\mathcal{G}_0$ with $B$ across the topological transition  [dashed/dotted lines, evolution also shown in panel (a)].
}
   \label{fig:I0}
\end{figure}

Under a constant voltage bias $V$, the pairings $\Delta$ to the left and right of the junction acquire an opposite and time-dependent phase difference, 
\begin{equation}
\phi(t)=2 eVt/\hbar\,.
\end{equation}
This induces Landau-Zener transitions between the ABSs and into the continuum, thereby developing a time dependent Josephson current with both $I_{dc}$ and $I_{ac}$ components. 
Such is the point of view in e.g. Refs. \cite{Averin:PRL95,San-Jose:PRL12a}. 
Alternatively, $\phi(t)$ can be gauged away into the hopping across the junction, 
$v_0(t)=\nu e^{-i\frac{eV}{\hbar} t\tau_z}\sum_{\sigma\sigma'} c^+_{r\sigma}v_{\sigma\sigma'}c_{l\sigma'} + \mathrm{h.c.}$, where $\tau_z$ is the $z$-Pauli matrix in Nambu space. 
By employing Keldysh-Floquet theory \cite{Cuevas:PRB96, Sun:PRB02}, we obtain the stationary-state time-dependent ac Josephson current 
\begin{equation}
I(t)=\sum_n e^{in\frac{eV}{\hbar} t}I_n
\end{equation}
(note that only even harmonics survive, see \ref{AppendixB1} for full details).  
Here, we concentrate on the dc-current $I_{dc}=I_0$. The results for $I_{dc}(V)$ at small, intermediate and full transparency are summarised in Fig. \ref{fig:I0}(a-c) 
for increasing values of $B$ spanning the topological transition.

\subsubsection{Tunneling regime.} 
For non-topological tunnel junctions, dc-transport vanishes below an abrupt threshold voltage (Fig. \ref{fig:I0}(a), blue curves)
\begin{equation}
V_t=2\Delta_\mathrm{eff}/e=2\Delta_-/e\,.
\end{equation}
This well known result follows from the fact that there are no quasiparticle excitations in the decoupled wires for energy 
$\epsilon\in(-\Delta_\mathrm{eff},\Delta_\mathrm{eff})$ if $B<B_c$. Indeed, to second order in perturbation theory in $\nu$, 
the MAR current takes the form of a convolution between $A_0(\omega)$ and $A_0(\omega\pm eV/\hbar)$, where $A_0$ is the decoupled ($\nu=0$) 
spectral density at each side of the junction (\ref{AppendixB2}). [The trace of $A_0(\epsilon)$, proportional to the LDOS, is shown in Fig. \ref{fig:Ic}(a)]. 
Hence, as $B$ increases, the tunnelling current threshold follows the closing of the gap in the LDOS, until $V_t$ vanishes and $I_{dc}$ becomes linear in small 
$V$ at $B_c$ (black curve). As $B>B_c$, the gap reopens, but the threshold is now halved to (Fig. \ref{fig:I0}(a), red curves) 
\begin{equation}
V_t=\Delta_\mathrm{eff}/e\,.
\end{equation}
Note that the small step visible at $eV=\Delta_\mathrm{eff}/2$ is the second-order MAR, whose relative height vanishes as $T_N\to 0$. 
The change, easily detectable as a halving of the slope of the threshold $dV_t/dB$ across $B_c$, is due to the emergence of an intra-gap zero-energy MBS in the topological phase 
[see zero energy peak in Fig. \ref{fig:Ic}(a)], which opens a tunnelling transport channel from or into the new zero energy state. 
Moreover, when $B=B_c^{(2)}$, $\Delta_-$ surpasses $\Delta_+$, and $\Delta_\mathrm{eff}$ saturates at $\Delta_+$. This is directly visible in $V_t(B)$ as a kink at $B_c^{(2)}$ 
[see dashed and dotted lines in Fig. \ref{fig:I0}(a)]  \footnote{Similar considerations may apply to recent experiments with lead nanoconstrictions formed in an STM tip, 
see Ref. \cite{Rodrigo:PRL12}.}. 

\subsubsection{Intermediate transparency regime.} As transparency increases, subharmonic MAR steps develop at voltages
\begin{equation}
V_t/n=2\Delta_\mathrm{eff}/en\,,\quad (n=2,3,4,\dots)\,,
\end{equation}
as shown Fig. \ref{fig:I0}(b)
The specific profile of each step with $V$ still contains information on the LDOS of the junction at energies around $\Delta_\mathrm{eff}$. At $B=0$, the power-law LDOS for $|\epsilon|>\Delta$ results in a staircase-like curve $I_{dc}(V)$ 
[blue line in Fig. \ref{fig:I0}(d)]. This shape is roughly preserved up to $B=B_c$. For $B>B_c$ the MAR profile changes qualitatively, however. The subharmonic threshold voltages 
\begin{equation}
V_t/n
\end{equation}
are halved (since $V_t$ is halved), and the MAR current profile becomes oscillatory instead of step-like. A blowup of the oscillations is presented in Fig. \ref{fig:I0}(d) (red curve), together with guidelines for the corresponding $V_t/n$ in gray. 


The emergence of oscillatory MAR steps, which here is connected to the formation of  zero energy peaks in the LDOS owing to the localized MBSs, is well known in Josephson junctions containing a resonant level \cite{Johansson:PRB99,Yeyati:PRL03,Jonckheere:PRB09}. Note, however, that the oscillations in a topologically trivial system,  such as for instance a quantum dot between two superconductors, arise at odd fractions of $2\Delta_\mathrm{eff}$, i.e. at voltages $2\Delta_\mathrm{eff}/(2n-1)e$, instead of the $\Delta_\mathrm{eff}/en$ of the Majorana case. Interestingly, this difference is ultimately due to the fact that a resonant level spatially localised within the junction cannot carry current directly into the reservoirs, while a zero energy MBS (essentially half a non local fermion) can. This same situation arises in $d$-wave Josephson junctions, which also exhibit oscillatory $\Delta_\mathrm{eff}/en$ MAR subharmonics owing to the presence of mid gap states \cite{Cuevas:PRB01}.

\subsubsection{Transparent limit.} In the limit $T_N\to 1$, ABS energies $\epsilon_\pm(\phi)$ [Fig. \ref{fig:ABS}(c)] touch the continuum at $\phi=0$. This has an important consequence. From the Landau-Zener point of view of the ac Josephson effect \cite{Averin:PRL95}, the time dependence of $\phi(t)=2eV t/\hbar$ for an arbitrarily small $V$ will induce the escape of any quasiparticle occupying an ABS into the continuum after a single $\phi(t)$ cycle. A given ABS becomes occupied with high probability in each cycle around $\phi=\pi$ if the rate $\hbar\,d\phi(t)/dt=2eV$ exceeds its energy minimum $\epsilon(\pi)\equiv \delta_\pi$. (Recall this energy is finite, since the Andreev approximation does not apply, see \ref{AppendixB3}.) 
One quasiparticle is then injected into the continuum per cycle, and a finite $I_{dc}(V\gtrsim \delta_\pi/e)$ arises. Below such voltage, however, the ABS remains empty, so that if $\delta_\pi$ is finite, as is the case of a realistic non-topological junction [see Fig. \ref{fig:ABS}(a-c)], one obtains $I_{dc}(V\to 0)=0$ (valid for any transparency at $B<B_c$). This is in contrast to the conventional $B=0$, $T_N=1$ result $I(V\to 0)=4\Delta\mathcal{G}_0/e$, predicted within the Andreev approximation ($\mathcal{G}_0=e^2/h$).


After the topological transition, this picture changes dramatically. The two MBSs at each side of the junction hybridise for a given $\phi$ into a \emph{single} ABS. This seemingly innocent change has a notable consequence. Since fermion parity in the superconducting wires is globally preserved, an anticrossing at $\phi=\pi$, which would represent a mixing of a state with one and zero fermions in the lone ABS, is forbidden. Parity conservation therefore imposes $\delta_\pi=0$ in the presence of MBSs, irrespective of $T_N$ or $\mu/\Delta$ \footnote{Note that residual splitting may survive in the topological phase for finite length nanowires, for which a finite (albeit exponentially small) coupling between four MBSs exist.}. This is a true topologically protected property of the junction, and gives rise to a finite $I_{dc}(V\to 0)=2\Delta_\mathrm{eff} \mathcal{G}_0/e$, i.e.  half the value expected for the  non-topological junction in the Andreev approximation. This abrupt change is shown in Fig. \ref{fig:I0}(c,e).
 The $I_{dc}(V\to 0)$ MAR current in transparent junctions, therefore, directly probes the emergence of parity protection.

\section{Critical current}\label{Critical current} 
\begin{figure}[t] 
   \centering
   \includegraphics[width=0.75\linewidth,clip]{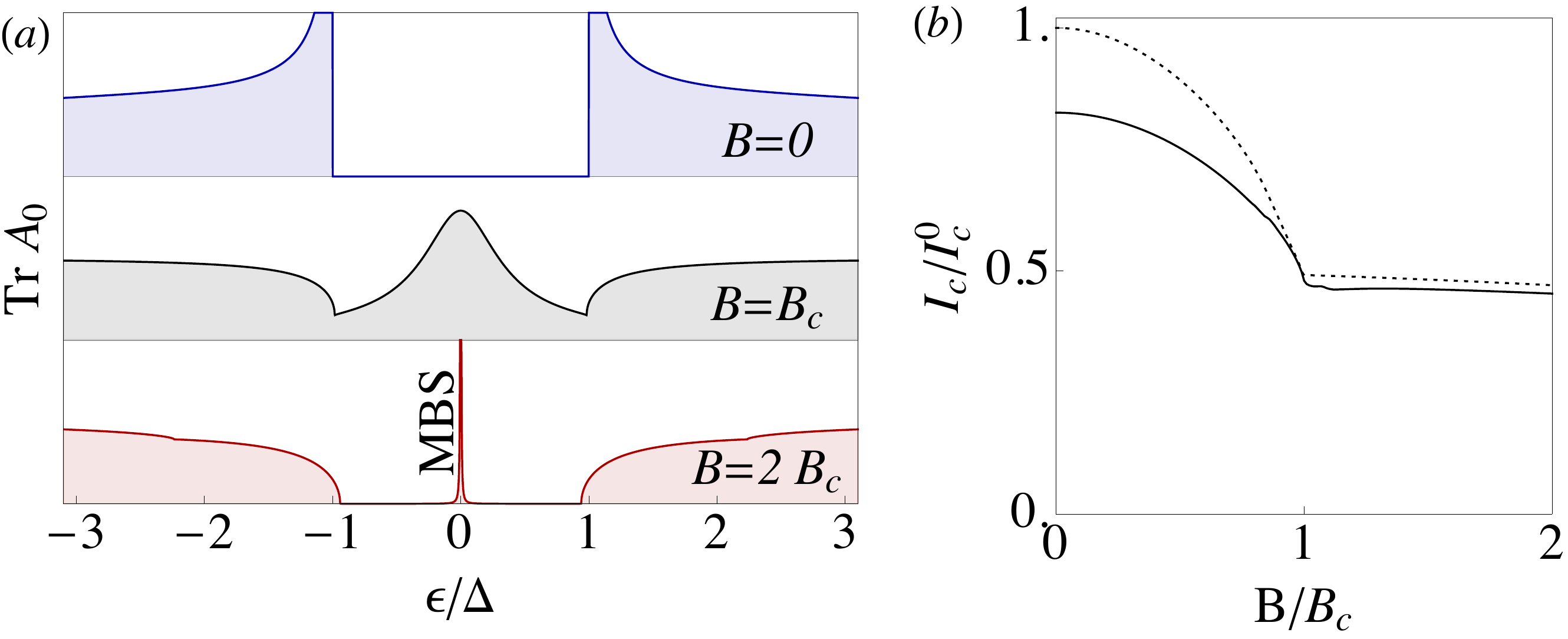} 
     \caption[LDOS at the end of a NW and critical current $I_{c}$]{(Color online) (a) Local density of states at the end of a single nanowire in the non-topological (top), critical (middle) and topological phase (bottom). A zero-energy Majorana peak appears in the latter case. (b) The critical current $I_c(B)$ for $T_N=1$ across the the topological transition in units of $I_c^0=e\Delta/\hbar$. 
     The dotted line corresponds to $\frac{1}{2}(\Delta_+ + \Delta_-)/\Delta$ for $B<B_c$, and $\frac{1}{2}\Delta_+/\Delta$ for $B>B_c$. 
}
   \label{fig:Ic}
\end{figure}

In the transparent limit, a supercurrent peak  \cite{Doh:S05,Nilsson:NL12, Chauvin:PRL07} may hinder the experimental identification of the $I_{dc}(V\to 0)$ limit, but itself holds valuable information about the transition.  
The critical current $I_c$ may be computed in general by maximizing the $V=0$ (time-independent) current $I(\phi)$ respect to $\phi$ (including the contribution from the continuum) 
as shown in the last part of Chapter \ref{Chap2a}. 
For a short transparent junction at $B=0$, $I_c$ is maximum, and equal to $I_c^0\equiv e\Delta/\hbar$ in the Andreev approximation.
 Fig. \ref{fig:Ic}(b) shows $I_c$ for increasing values of $B$. Naively, one may expect that a junction without a superconducting gap should not 
 carry a finite supercurrent, but this is not the case here. At $B=B_c$, $I_c$ is finite \footnote{Note also that in junctions with trivial superconductors, 
 $I_c\rightarrow 0$ as the nanowire becomes helical at $B=\mu<B_c$ \cite{Cheng:PRB12}. Our result is therefore another nontrivial consequence of topology in the junction.}, 
 while $\Delta_\mathrm{eff}=0$ [the junction LDOS at criticality is also gapless, see Fig. \ref{fig:Ic}(a)]. 
 This gapless supercurrent comes from the $\epsilon_+(\phi)$ \emph{quasi-bound} Andreev state in the continuum, which contributes almost as if it were a subgap ABS. 
 It is thus a reasonable approximation to write $I_c$ as the sum of the critical current from each ABS. For $B<B_c$, $I_c\approx \frac{1}{2}I_c^0(\Delta_++\Delta_-)/\Delta$. 
 The $\Delta_-$ contribution, however, should not be included for $B>B_c$, leading to a discontinuity in $\partial I_c/\partial B$. 
 This simple model gives a qualitative fit [dotted line in Fig. \ref{fig:Ic}(b)] to the exact numerics (solid line), with deviations coming from corrections to the Andreev 
 approximation, and contributions above $\Delta_+$. Additional deviations in experiments, coming e.g. from the finite impedance of the electromagnetic environment, are not 
 expected to alter the discontinuity in $\partial I_c/\partial B$, which remains a signature of the  topological transition. 

Note that we also present results for the critical current in short SNS junctions in Chap.\,\ref{Chap2a}. 
%
 
\section{Conclusions}
In conclusion, we have shown that the dc-current in voltage biased Josephson junctions is a flexible experimental probe into the various aspects of the topological superconducting transition in semiconducting nanowires. Tuning the junction transparency one may obtain evidence of MBS formation as conclusive as a fractional Josephson effect, without requiring control of the junction phase. Moreover, we have found that the critical current in the wire does not vanish at the transition due to above-gap contributions, although its derivative with $B$ exhibits a discontinuity as a result of the disappearance of one ABS. This behavior of $I_c$ provides a direct evidence of the topological transition. MAR spectroscopy and critical current measurements in nanowires similar to the ones studied here have already been reported \cite{Doh:S05,Nilsson:NL12}. 

Although we have focused here on the simplest case (single-band, short junction limit) we expect the main features of the topological transition to remain robust under more general conditions. Preliminary results in the quasi-one dimensional multiband case show that $I_c$ is a non-monotonic function for increasing magnetic fields. For weak interband SO mixing \cite{Lim:PRB12}, the behavior discussed in Fig. \ref{fig:Ic}(b) can be generalized and $I_c$  presents a series of minima at different fields corresponding to the topological transition of each subband.

Importantly, the alternative physical scenarios, such as, e. g., disorder \cite{Pientka:PRL12,Bagrets:PRL12,Liu:PRL12,Sau:13} or Andreev bound states  \cite{Lee:13}, that produce ZBAs in NS junctions (and thus mimic Majorana physics), cannot give the distinct features associated to global parity that were discussed here for SNS junctions. We therefore believe that experiments along the lines discussed in this paper could provide the first unambiguous report of a topological transition in nanowires, and the emergence of Majorana bound states.


\chapter{\bf SNS junctions in nanowires with spin-orbit coupling: role of confinement and helicity on the sub-gap spectrum\footnote{The results presented in this chapter have been published in \cite{PhysRevB.91.024514}.}} 
\label{Chap2}
\lhead{Chapter \ref{Chap2}. \emph{SNS junctions in nanowires with SOC: confinement and helicity}} 





\begin{small}
We study normal transport and the sub-gap spectrum of superconductor-normal-superconductor (SNS) junctions made of semiconducting nanowires with strong Rashba spin-orbit coupling. We focus, in particular, on the role of confinement effects in long ballistic junctions. In the normal regime, scattering at the two contacts gives rise to two distinct features in conductance, Fabry-Perot resonances and Fano dips. The latter arise in the presence of a strong Zeeman field $B$ that removes a spin sector in the leads (\emph{helical} leads), but not in the central region. Conversely, a helical central region between non-helical leads exhibits helical gaps of half-quantum conductance, with superimposed helical Fabry-Perot oscillations. These normal features translate into distinct subgap states when the leads become superconducting. In particular, Fabry-Perot resonances within the helical gap become parity-protected zero-energy states (parity crossings), well below the critical field $B_c$ at which the superconducting leads become topological.  As a function of Zeeman field or Fermi energy, these zero-modes oscillate around zero energy, forming characteristic loops, which evolve continuously into Majorana bound states as $B$ exceeds $B_c$. The relation with the physics of parity crossings of Yu-Shiba-Rusinov bound states is discussed.
\end{small}

\newpage

\section{Introduction}
 Majorana fermions, particles that are their own antiparticles, have been the subject of intense research over the past 
decades in the context of particle physics and cosmology \cite{majorana, Avignone:RMP08}. 
In Chapter \ref{Chapter01} we have discussed that during last few years this interest extended to the condensed matter arena where Majorana fermions are intensely studied 
nowadays \cite{wilcek,RevModPhys.87.137}. This state of affairs has been driven by the key observation that emergent quasiparticles in superconductors can be described as 
Majorana fermions \cite{Alicea:RPP12,Leijnse:SSAT12,Beenakker:11,StanescuModel13,RevModPhys.87.137}. This, together with the recent advances in the field of topological 
materials \cite{Hasan:RMP10,Qi:RMP11}, has  spurred an intense search for condensed matter realizations of Majorana fermions. Most of these realizations focus on 
zero-energy modes inside the gap of topological superconductors. Thus, instead of Majorana fermions, they are now more precisely referred to as Majorana bound states (MBSs) or 
Majorana zero modes. \footnote{It has been argued that Bogoliubov quasiparticles in conventional superconductors are true Majorana fermions \cite{PhysRevB.81.224515}). 
Their Majorana fermion nature can be revealed by annihilation processes, see \cite{PhysRevLett.112.070604}.}.


We have seen in Chapter \ref{Chapter01} that MBSs can emerge in exotic superconductors, such as p-wave, since they realize topological phases that support edge excitations 
with Majorana fermion character \cite{PhysRevB.44.9667, volovik, PhysRevB.61.9690, PhysRevB.61.10267, kitaev, PhysRevB.73.220502}. 
and a physical promising platform is based on the proximity effect between a conventional s-wave superconductor and a semiconductor nanowire  (NW) with strong spin-orbit (SO) 
coupling \cite{Sato:PRL09,Sau:PRL10,Alicea:PRB10,Lutchyn:PRL10,Oreg:PRL10}. 
It has been shown \cite{Lutchyn:PRL10,Oreg:PRL10} that if an external Zeeman field $B$, orthogonal to the SO axis, exceeds a critical value 
$B_{c}\equiv\sqrt{\mu^2+\Delta^2}$, where $\mu$ is the Fermi energy and  $\Delta$ the induced s-wave pairing, zero energy MBSs emerge at the nanowire ends signaling a 
topologically non-trivial phase.

Unfortunately, the outcome of the simplest detection protocol for MBSs in NW devices \cite{Sengupta:PRB01,Bolech:PRL07,Law:PRL09}, detection of subgap zero modes 
through zero-bias anomalies in transport  \cite{Mourik:S12,Deng:NL12,Das:NP12,Finck:PRL13,Churchill:PRB13,Lee:13}, can be obscured, or even mimicked, by other effects as discussed in 
Chapter \ref{Chapter01} \cite{Lee:PRL12,Pientka:PRL12,Bagrets:PRL12,Liu:PRL12,Sau:13,Prada:PRB12,Kells:PRB12,Zitkoetal}.  
In this Chapter we go beyond zero-bias anomaly experiments and study more complex geometries such as Superconductor-Normal-Superconductor (SNS) junctions 
\cite{pablo,PhysRevLett.112.137001,Flensberg-Xu}. This geometry has a number of advantages including the possibility of studying supercurrents 
\cite{Deng:NL12,Doh:S05,Nishio:N11,Nilsson:NL12,Gunel:JOAP12}, or direct spectroscopy of Andreev bound states (ABS) 
\cite{PhysRevLett.110.217005, nphys1811, PhysRevLett.104.076805, PhysRevB.88.045101,PhysRevB.89.045422,PhysRevX.3.041034,Levy,Lee:13,PhysRevLett.85.170} as 
shown in Chapter \ref{Chap2a}.  As we shall discuss, this latter technique can be used, in principle, to directly monitor the detailed evolution from the trivial 
to the nontrivial regime (see also \ref{Chap2a}). Previous papers have mostly focused on short 
junctions \cite{pablo,Bena1,gilbertini2012,Bena2,stefano2014} and detailed studies of ABS in long and intermediate-length junctions has been carried out in Chap.\,\ref{Chap2a}, 
while the study of other effects (such as confinement and helicity) remain largely unaddressed. In particular, the role of Fabry-Perot resonances occurring in normal transport as the middle NW finite-lenght section of the junction is depleted has never been studied to the best of our knowledge. In this work we fill this void and present detailed calculations of the normal conductance and Andreev spectra in such geometries. 
We emphasize here that all nanowire experiments should ideally belong to the category studied here, as confinement effects should be present when a ballistic quasi-one dimensional conductor is contacted between leads, especially when the normal part of the NW (in our geometry, the region of length $L_{nw}$ not directly in contact with leads) is gated. This electrical gating naturally creates quantum wells (or barriers) with their associated confined quantum levels in the middle region of the NW.

In the first half of this work, we discuss normal transport across a finite length ballistic NW. We show how bandstructure details in the presence of strong Rashba SO coupling and Zeeman fields may dominate transport, and give rise to distinct features associated to helical phases (defined by singly-degenerate subbands at the Fermi level with spin locked to momentum) known as helical gaps (Fig.\,\ref{fig1}). Likewise, finite contact resistance induces confinement resonances in conductance as quasibound states develop in the NW. In the simplest case of non-interacting electrons \footnote{Coulomb blockade effects will be discussed elsewhere.}, we find that confinement generates two types of resonances: Fabry-Perot resonances and helical Fano dips.  Fabry-Perot resonances for a spinful mode \cite{Kretinin:NL10} will give conductance oscillations with a ceiling of $2e^2/h$, unless the central NW is depleted into its helical regime, in which case one may observe \emph{helical} Fabry-Perot resonances with a half-quantum $e^2/h$ ceiling.  For long enough junctions, many helical Fabry-Perot resonances may occur. We discuss that, while confinement effects may mask the helical gap, the characteristic reentrance of helical Fabry-Perot resonances with Zeeman field or gate voltage contains valuable information about non-trivial helical transport through the NW. The second kind of resonances are sharp Fano dips when the central section of the NW is gated to form a quantum well (non-helical) and the NW sections below the contacts (the leads) are helical. Therefore, both types of resonant features in normal transport may signal the helical regime in different sections (central or below the contacts) of the NW.  In the presence of superconducting leads, the two lead to distinct effects.


In the second half of this work we consider the connection of this phenomenology to transport in the superconducting regime. Each helical Fano dip in the normal phase translates, in an SNS geometry, into a single subgap state that crosses zero energy as a function of external parameters (Fermi energy or Zeeman field). Such a crossing is often known as a \emph{parity crossing}, since it is protected by conservation of number parity in the junction. As we discuss, these parity crossings are made possible by the nontrivial topology in the underlying effective p-wave superconductor for $B>B_c$. Similar bound states originated from nonmagnetic impurities in topological superconductors and superfluids have been recently discussed  in Refs.  \cite{Sau-Demler,Huetal} and can be considered the p-wave counterparts of Yu-Shiba-Rusinov bound states \cite{subgap1,subgap2,subgap3,subgap4,subgap5} in standard s-wave superconductors with magnetic impurities.
A more direct analogy with standard Yu-Shiba-Rusinov magnetic bound states actually applies in the non-topological phase $B<B_c$. In this situation, helical Fabry-Perot resonances in normal conductance translate, in the superconducting regime, into loops around zero energy in the ABS spectrum as a function of external parameters. For long junctions, many of these loops are visible, each separated by a parity crossing at zero energy. As a result, the $B<B_c$ subgap spectrum contains near-zero energy subgap states that oscillate as a function of Fermi energy or Zeeman field when the N region of the junction is helical. Interestingly, we find that these oscillating near-zero subgap states in the trivial regime are smoothly connected to MBS when Zeeman is increased beyond $B_c$.


This paper is organized as follows. In section \,\ref{sec1} we describe the Hamiltonian model employed in our work. 
Section \ref{normCon} focuses on the normal conductance and how the two types of resonances, helical Fabry-Perot and helical Fano dips, appear in the system. The rest of the paper is devoted to analysing the consequences of these resonant levels in the sub-gap spectrum in the superconducting regime.
After a brief discussion on how the SNS junction is modeled, as well as a discussion about the relevant length scales of the problem, section \ref{ElevelsSNS} presents a systematic study of the subgap spectrum of SNS junctions, including its dependence on the superconducting phase difference across the junction, $\varphi$.
We discuss in detail how the presence of confined levels within the central region affect the ABS and lead to parity crossings in the topological phase.  
The dependence of the ABS on phase difference, Fermi energy of the normal region and Zeeman field is discussed for both short and long junctions in subsections 
\ref{short} and \ref{long}, respectively. 
Our conclusions are presented in Section \ref{concl}. Here we employ the model described in detail in Section \ref{SNSjunctionmodel} of Chapter \ref{Chap2a} for 
SNS junctions by using a tight-binding version of the continuum model Hamiltonian.  Appendix \ref{tightc} discusses an effective model that fully explains the phenomenology behind helical Fano resonances. 
\section{Nanowire model}
\label{sec1}
We present the model for a nanowire with Rashba SOC and in the presence of an external Zeeman, whose description was fully studied in Chapter \ref{Chapter01}. 
We restrict ourselves to the strictly one dimensional (single-mode) case for simplicity. Generalisations to multimode nanowires are relatively straightforward. The model Hamiltonian reads
\begin{equation}
\label{Leq1}
H_{0}\,=\,\frac{p^{2}}{2m^{*}}\,-\,\mu\,-\,\frac{\alpha_{R}}{\hbar}\,\sigma_{y}\,p\,+\,B\,\sigma_{x}\,,
\end{equation}
where $p$ is the momentum operator, $m^{*}$ is the effective electron mass, $\alpha_{R}$ the Rashba SOC strength, $\mu$ the Fermi energy and $\sigma_{i}$ the spin Pauli matrices. An external magnetic field $\mathcal{B}$ along the wire produces a Zeeman splitting $B=g\mu_{B}\mathcal{B}/2$, where $\mu_{B}$ is the Bohr magneton and $g$ the wire $g$-factor.  The Rashba coupling defines a typical length, the spin-orbit length $l_{SO}\equiv\hbar/\sqrt{2m^*E_{SO}}$, with the spin-orbit energy defined as $E_{SO}=\frac{1}{2}\alpha_{R}^2m^{*}/\hbar^2$. For typical InSb values
$m^{*}=0.015\,m_{e}$,  with $m_{e}$ the electron mass and $\alpha_{R}=0.2$ eV \AA, the spin-orbit energy is $E_{SO}\approx 50 \mathrm{\mu eV}$ which gives SO lengths of the order of $l_{SO}\approx 200$nm. 

Note that the Rashba and Zeeman fields in Eq.\,(\ref{Leq1}) are perpendicular. As a result, the two spinful bands (shifted by SO) become mixed by the Zeeman term and the zero-field crossing point at zero momentum becomes an anticrossing of size $2B$. When the chemical potential lies within this anti crossing gap, the system has two Fermi points, as opposed to four Fermi points for $\mu$ above or below this gap. This window is a \emph{helical} gap, since the two fermi points correspond to counter propagating states with different spins (the spin projection is locked to momentum) \cite{helical}, see right panel in Fig.\,\ref{fig1}. 


\section{The normal conductance}
\label{normCon}
  \begin{figure}[!ht]
\centering
\includegraphics[width=.99\textwidth]{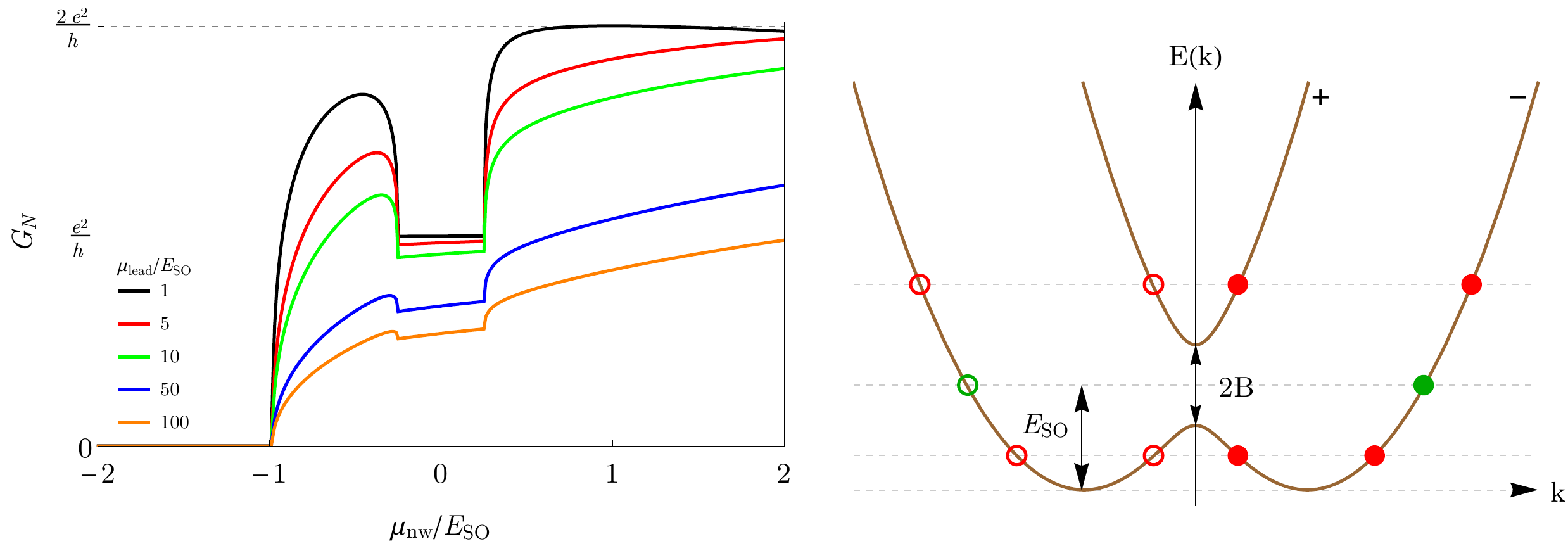} 
\caption[$G_{N}(\mu_{\mathrm{nw}})$ for a semi-infinite N-NW junction]{(Color online) (left) Normal conductance $G_{N}$ as a function of the Fermi energy $\mu_{\mathrm{nw}}$ in the left lead for a semi-infinite N-NW junction. Parameters: $\alpha_{R}=20$\,meV\,nm (which corresponds to $E_{SO}=0.05$\,meV) and $B=0.0125$\,meV. Different curves show how $G_{N}(\mu_{\mathrm{nw}})$ evolves for increasing Fermi energy $\mu_{\mathrm{lead}}$ in the right lead. (right) Dispersion relation for a Rashba NW in the presence of a transverse $B$ field. Within the gap there is only one right mover per energy (green filled circle), while outside the gap there are two (red filled circles). This gives rise to the reentrant behavior of conductance, from $\sim 2e^2/h$ to $e^2/h$ and back to $2e^2/h$, as a function of Fermi energy in the main panel. The spin of the counter propagating states (open circles) is opposite to the propagating ones (filled circles), hence the name helical.}
\label{fig1}
\end{figure}
Before discussing the sub-gap Andreev spectrum of a NW coupled to superconducting leads, we characterize the normal regime in the presence of a Zeeman field. We are interested in particular in the normal conductance $G_{N}$ as the Fermi energy ($\mu_{\mathrm{nw}}$) in the middle section of the NW (length $L_{nw}$) varies with respect to the one in the left and right leads $\mu_{\mathrm{leads}}$. Such situation models a NW contacted between normal electrodes and with a Fermi energy tuned by a central gate, see e. g. Ref. \cite{Churchill:PRB13}. For simplicity in the discussion, we model the gate-induced electrostatic potential with an abrupt profile (the role of smooth gate potentials has been recently discussed in Ref. \cite{PhysRevB.90.235415}).

For computations purposes we discretize Eq.\,(\ref{Leq1}) into a tight-binding lattice.
The momentum operator introduces hopping elements $v$ between nearest-neighbor sites. 
The transparency of the left and right contacts is parameterised by a factor $\tau\in[0,1]$, introduced in the hopping matrix $v_{0}=\tau v$ across the two interfaces, see Section \ref{SNSjunctionmodel}.
$G_{N}$ is calculated by means of the Greens function technique \cite{Caroli,Haug:07}, 
\begin{equation}
\begin{split}
 G_{N}&=4\frac{e^{2}}{h}\,{\rm Tr}[\Gamma_{L}\,G^{r}\,\Gamma_{R}\,G^{a}]\\
   \end{split}
 \end{equation}
 where 
 \begin{equation}
 G^{r}=g_{0}^{r}+g_{0}^{r}\,\Sigma^{r}\,G^{r}=(G^{a})^{\dagger}
 \end{equation}
 is the full retarded Green's function. The bare Green's function of the normal region without the presence of the leads is given by 
  \begin{equation}
 g_{0}^{r}=[\omega-h_{0}+i 0^{+}]^{-1}\,.
  \end{equation}
 The hamiltonian $h_{0}$ corresponds to $H_{0}$ in Eq.\,\ref{Leq1} with $\mu=\mu_{\mathrm{nw}}$. 
 The leads are taken into account through the self-energies 
  \begin{equation}
 \Sigma_{L(R)}^{r}=v\,g^{r}_{L(R)}v^{\dagger}\,,
  \end{equation}
 where $g_{L(R)}^{r}=[\omega-h_{L(R)}+i 0^{+}]^{-1}$ stands for the left/right lead's propagator, when decoupled from the system. 
 In this case, $h_{L(R)}$ corresponds to $H_{0}$ in Eq.\,(\ref{Leq1}) with $\mu=\mu_{\mathrm{leads}}$.  
 Finally, 
  \begin{equation}
 \Gamma_{L(R)}=\frac{\Sigma_{L/R}^{r}-\Sigma_{L/R}^{a}}{2i}\,. 
 \end{equation}
 In practice, $G_{N}$ is computed recursively with the boundary conditions imposed by the leads. 
 \begin{figure}[!ht]
\centering
\includegraphics[width=.8\textwidth]{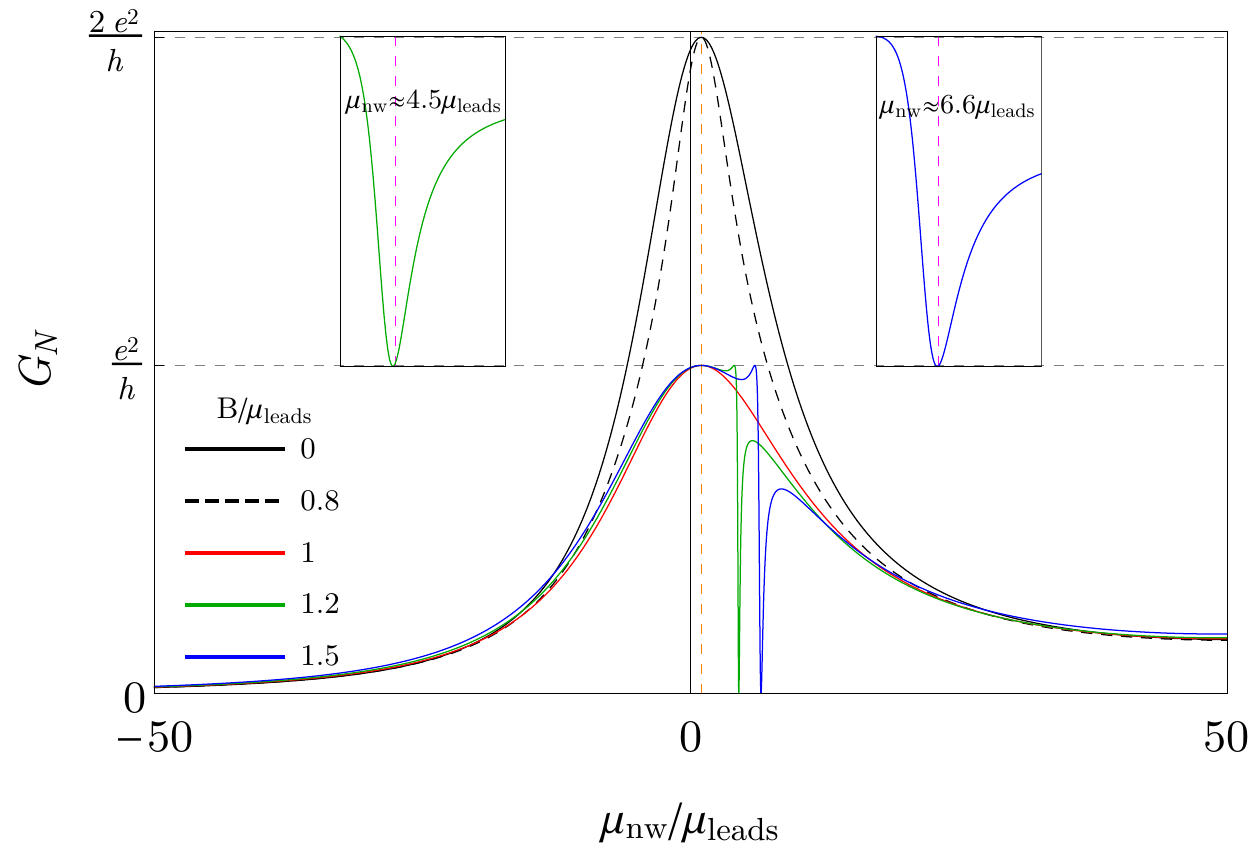} 
\caption[$G_{N}(\mu_{\mathrm{nw}})$ for a short N-NW-N junction]{(Color online) Normal conductance $G_{N}$ as a function of the Fermi energy $\mu_{\mathrm{nw}}$ for a short N-NW-N junction, $L_{\mathrm{nw}}=20$nm. Different curves show how $G_{N}(\mu_{\mathrm{nw}})$ evolves with the Zeeman field $B$. Confinement induces two types of resonances: Fabry-Perot (confinement in the central region) and Fano dips (leads have become helical).
Rest of parameters $E_{SO}=0.05$\,meV and $\mu_{\mathrm{leads}}=10E_{SO}$.
The insets show a blow-up of $G_{N}(\mu_{\mathrm{nw}})$ around the Fano dip for two different $B$.}
\label{fig2}
\end{figure}

To set the stage, we first consider an N-NW junction between a good metal and a semi infinite nanowire, which will allow us to discuss deviations when we consider confinement effects. 
Fig.\,\ref{fig1} shows the expected conductance profile as a function of the NW Fermi energy $\mu_{\mathrm{nw}}$, for different values $\mu_{\mathrm{lead}}$ of the Fermi energy in the metal. At finite magnetic fields, the normal conductance exhibits a gap (with $G_N\approx e^2/h$) of size $\Delta\mu_{\mathrm{nw}}=2B$. As we explained, this gap is a direct consequence of the combined action of Zeeman effect and strong SO coupling and reflects the presence of helical transport, namely spin-polarized counter propagating states \cite{helical}. As discussed in Ref. \cite{PhysRevB.90.235415}, the visibility of this helical gap depends on various factors which, importantly, include the actual value of the SO energy. Indeed, as the ratio $\mu_{\mathrm{lead}}/E_{SO}$ is made larger, the visibility of the gap in $G_N$ is rapidly degraded (see lower curves in Fig.\,\ref{fig1}).

We now consider the confined N-NW-N junction geometry. Due to the confinement of the central NW section, Fabry-Perot resonances are expected. Fig.\,\ref{fig2} shows the extreme case of a very short central region with only one resonant quasibound state in the junction. As expected, the conductance without external Zeeman field (solid curve) has a Lorentzian shape and reaches its maximum value $G_{N}=2e^2/h$ when $\mu_{\mathrm{nw}}=\mu_{\mathrm{leads}}$ (vertical dashed guideline). Similar results are found for small Zeeman fields $B<\mu_{\mathrm{leads}}$ (dashed). When $B=\mu_{\mathrm{leads}}$, however, the leads becomes spin-polarized (or helical, to be precise) and hence the maximum conductance is halved to $G_{N}=e^2/h$ (red curve). 

\begin{figure}[!ht]
\centering
\includegraphics[width=.99\textwidth]{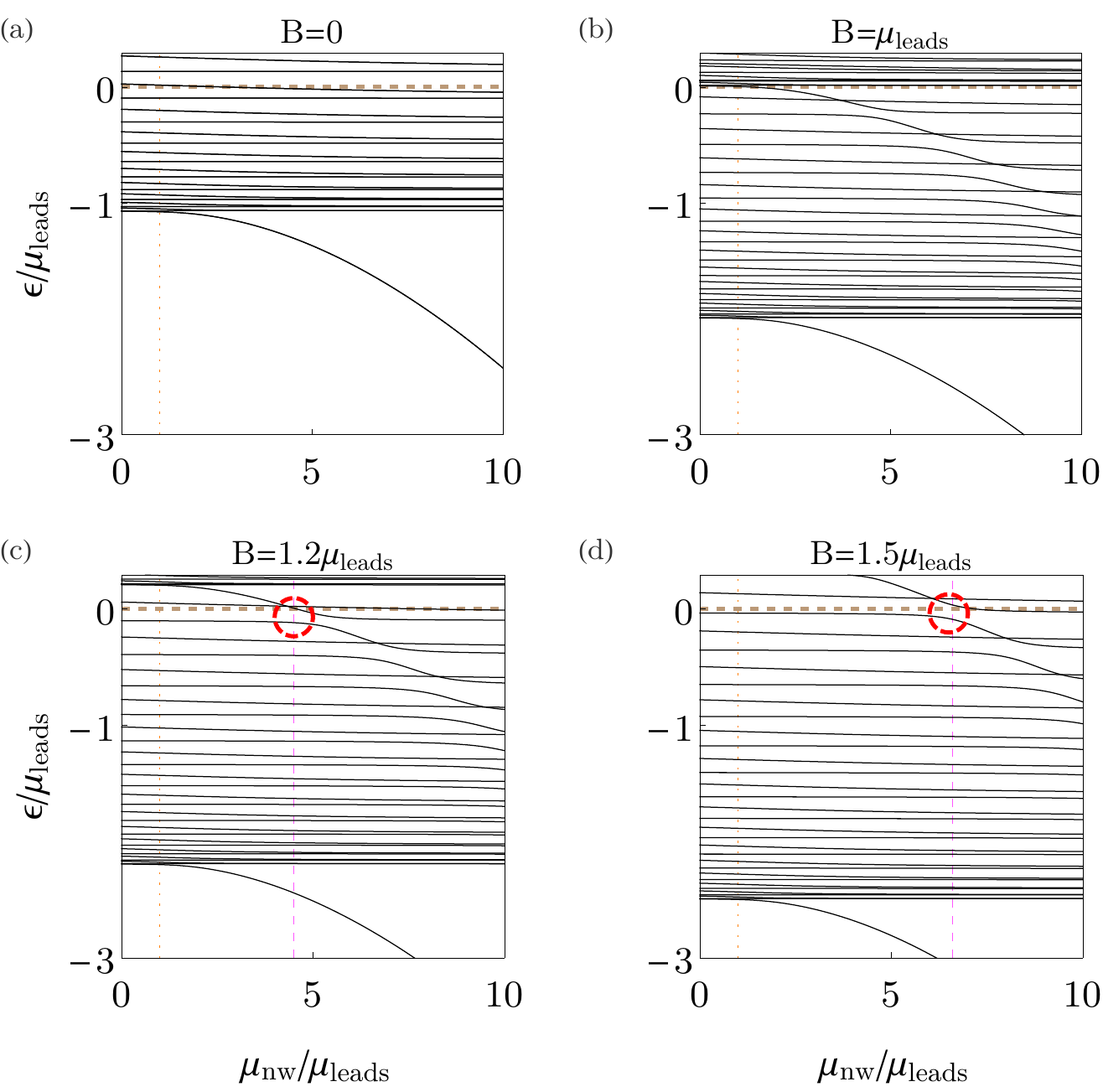} 
\caption[Energy levels as a function of the Fermi energy $\mu_{\mathrm{nw}}$]{(Color online) Energy levels as a function of the Fermi energy $\mu_{\mathrm{nw}}$ for the same system as in Fig.\,\ref{fig2}, short N-NW-N junction. Different panels show how the levels evolves with the Zeeman field $B$. The red dashed circle shows the value of $\mu_{\mathrm{nw}}$ for which one of the projections of the Zeeman-split bound state resonates with carriers at the Fermi level (horizontal dashed line), leading to a Fano resonance in conductance. Fano originates from spin-split quantum-well levels interacting with helical leads at $B>\mu_{\mathrm{leads}}$: one spin-projection is strongly coupled to the continuum of states in the leads and the other is weakly coupled (owing to Rashba canting). This mimics the physics of a Fano resonance.}
\label{fig3}
\end{figure}

We consider first the situation with $B>\mu_{\mathrm{leads}}$. This regime is characterised by strong Fano dips that appear when $\mu_{\mathrm{nw}}$ is positive, namely when the junction is gated to create a quantum dot instead of a barrier, see Eq. (\ref{Leq1}). At these Fano dips destructive interference is maximum and $G_{N}=0$. Moreover, the position of these Fano resonances moves to higher $\mu_{\mathrm{nw}}$ as $B$ increases (Fig.\,\ref{fig2}, insets).
The Fano dips can be understood by noticing that the system develops a truly bound state at an energy below $\mu_{\mathrm{leads}}$ as $\mu_{\mathrm{nw}}$ increases (Fig.\,\ref{fig3}a). 

While for $B\ll\mu_{\mathrm{leads}}$ this level lies far below the chemical potential of the leads and cannot significantly affect $G_N$, in the case $B>\mu_{\mathrm{leads}}$ at hand, the situation is markedly different. At such high fields, one spin sector in the leads is removed away from the chemical potential, and the leads become helical. Similarly, the bound state below $\mu_{\mathrm{leads}}$ is Zeeman-split, such that the component corresponding to the removed spin sector may then cross the Fermi level at a given $\mu_{\mathrm{nw}}$ (Fig.\,\ref{fig3}b-d).  
This results in one spin projection strongly coupled to the continuum (the sector that is not removed), while the other spin projection remains weakly coupled to this helical continuum through the split bound state (dashed circles), owing to the small spin canting induced by SO. This configuration mimics the physics of a Fano resonance, as we explicitly demonstrate in Appendix  \ref{tightc} with an effective model.  Note that SO is essential to mimic the physics of the Fano effect (two channels with very different coupling to the continuum). Indeed, we have checked that for $\alpha_R=0$ (namely a fully spin-polarized system without canting) the effect disappears (not shown). The general behaviour is related to the so-called Fano-Rashba effect in systems with inhomogeneous Rashba couplings \cite{PhysRevB.74.153313,Fano-Rashba-B} although in our case the bound states originate from the Fermi energy inhomogeneity, which is probably more realistic for NWs with gates. For intermediate lengths, the system can accommodate many of the above resonances but the helical gap is not visible (not shown). 

 \begin{figure}[!ht]
\centering
\includegraphics[width=.8\textwidth]{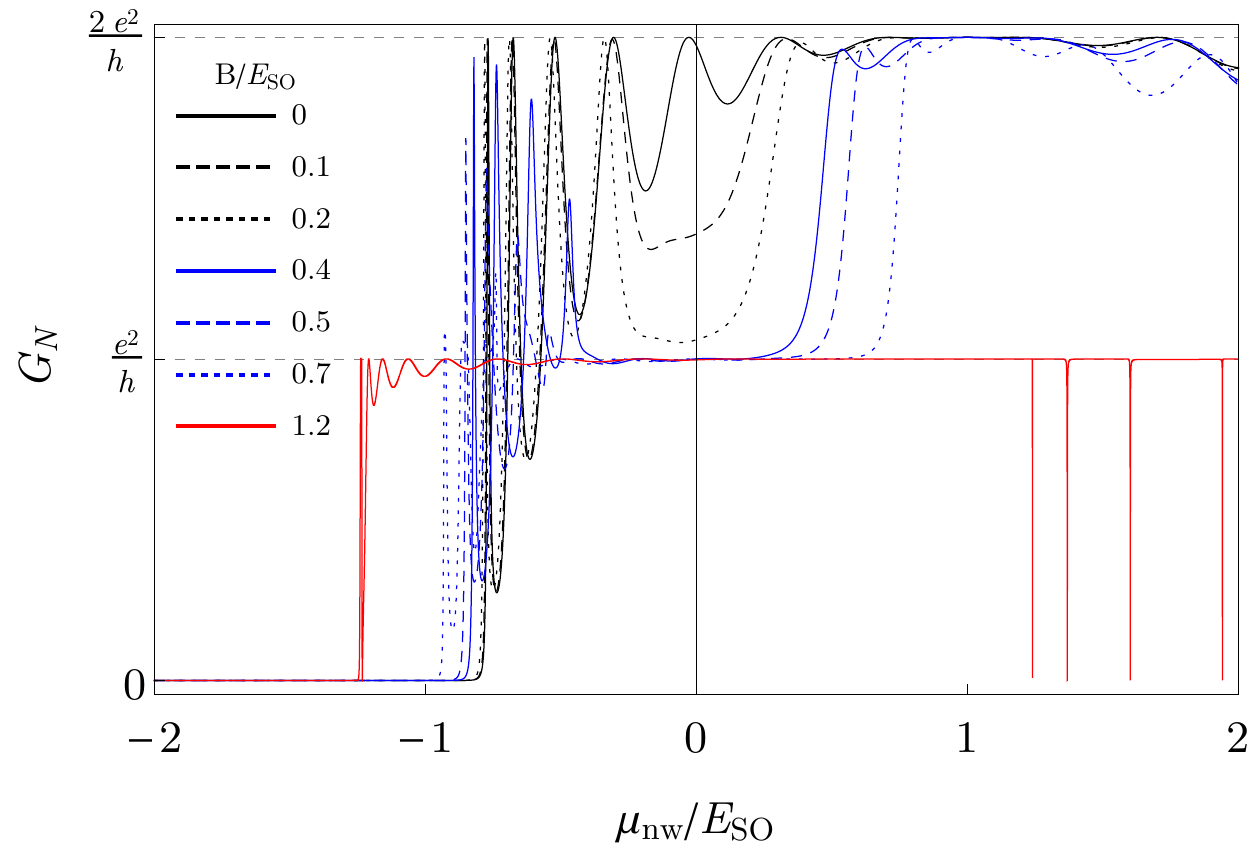} 
\caption[$G_{N}( \mu_{\mathrm{nw}})$ for a long junction for different $B$]{(Color online) Normal conductance $G_{N}$ as a function of the Fermi energy $\mu_{\mathrm{nw}}$ for a long junction with $L_{nw}=4\mu m$,  $E_{SO}=0.05$\,meV and $\mu_{\mathrm{leads}}=E_{SO}$. For intermediate magnetic fields, $B\leq E_{SO}$ the conductance develops a clear helical gap inside the Fabry-Perot resonant structure. This gap signals the region where the middle section of the NW becomes helical. When $B\geq \mu_{\mathrm{leads}}$, the contacts become helical too and the conductance shows helical Fano dips (red curve).}
\label{fig4}
\end{figure}

Consider now the $B<\mu_{\mathrm{leads}}$ regime complementary to the preceding discussion. In this situation, there exist two propagating channels in the leads, and conductance may reach $2e^2/h$ at Fabry-Perot maxima, as long as the central NW is likewise non-helical ($B>|\mu_{\mathrm{nw}}|$). Otherwise, for long enough junctions ($L_{\mathrm{nw}}\geq 4\mathrm{\mu m}$ for the realistic NW parameters in our simulation) a helical gap develops in conductance, such that $G_N\lesssim e^2/h$. As central $\mu_{\mathrm{nw}}$ is tuned into and out of the helical regime, conductance exhibits a reentrant behavior, switching from $\sim 2e^2/h$ to $e^2/h$ and back to $2e^2/h$. This reentrance can be resolved across multiple resonant helical Fabry-Perot oscillations.
This is illustrated in Fig.\,\ref{fig4} where we plot the conductance for a 4$\mathrm{\mu m}$-long nanowire as a function of the central Fermi energy $\mu_{\mathrm{nw}}$. Note the reentrant conductance, and the helical Fabry-Perot resonances with an $e^2/h$ ceiling, signalling helical transport in the junction.
The visibility of the conductance reentrance and the helical gap is lost for fields $B>E_{SO}$, see bandstructure, right panel in Fig. \,\ref{fig1}. At such fields, the helical gap becomes an extended $G_N\sim e^2/h$ half-plateau (potentially with superimposed Fano resonances if $B$ also exceeds $\mu_{\mathrm{leads}}$) that emerges directly from pinch-off $G_N=0$.
Note that the regime with helical Fano dips in the normal conductance is quite relevant towards reaching topological superconductivity: the NW under the contacts can become a non-trivial topological superconductor in the presence of pairing as long as it can be depleted and made helical in the normal phase. Hence our prediction of helical Fano dips superimposed on a half-plateau of $G_N\sim e^2/h$ constitutes a strong signature of helical behaviour as precursor of non-trivial superconductivity. 
 \begin{figure}[!ht]
\centering
\includegraphics[width=.8\textwidth]{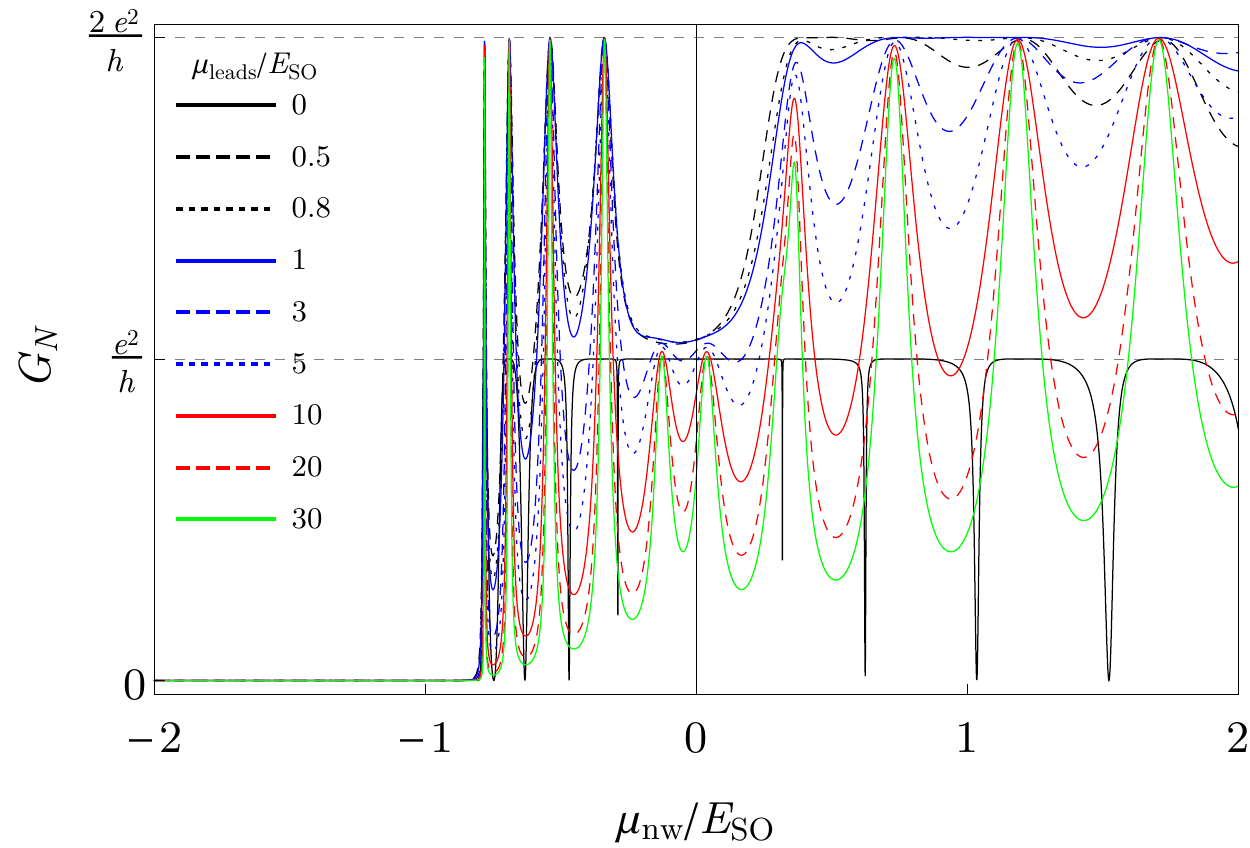} 
\caption[$G_{N}( \mu_{\mathrm{nw}})$ for a long junction for fixed $B$ and different $\mu_{\mathrm{leads}}$]{(Color online) Same as Fig.\,\ref{fig4} for fixed magnetic field $B=0.2E_{SO}$ and increasing $\mu_{\mathrm{leads}}$: Normal conductance $G_{N}$ as a function of the Fermi energy $\mu_{\mathrm{nw}}$ for a long junction with $L_{nw}=4\mu m$. Notice that the helical Fano dips are only seen when the leads become helical, for $\mu_{\mathrm{leads}}<B$ (solid black curve).}
\label{fig5}
\end{figure}

\begin{figure}[!ht]
\centering
\includegraphics[width=.99\textwidth]{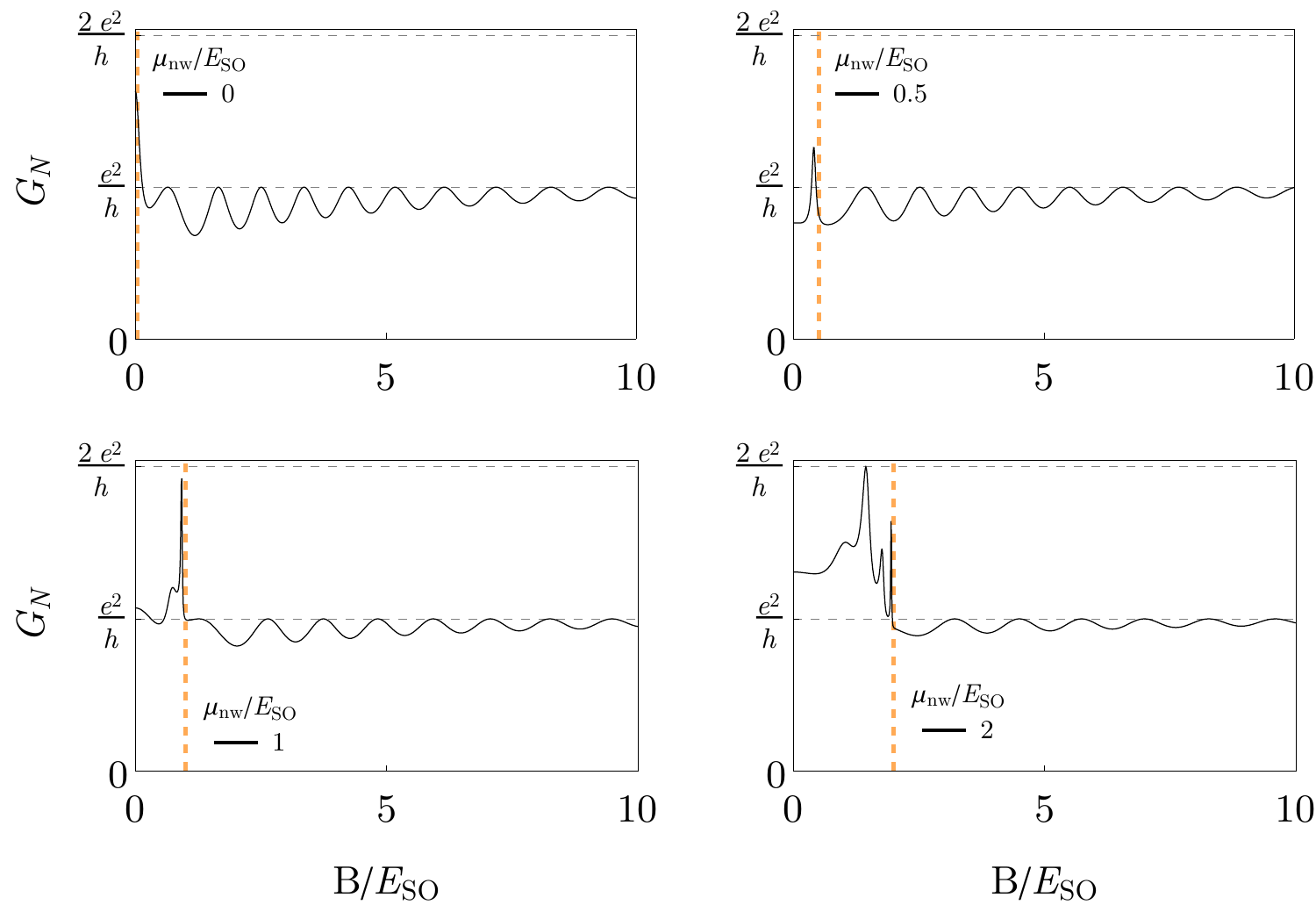} 
\caption[ $G_{N}(B)$ for different $\mu_{\mathrm{nw}}$ ]{(Color online) Normal conductance $G_{N}$ as a function of magnetic field (same parameters as in Fig.\,\ref{fig4}, except $\mu_{\mathrm{leads}}=10E_{SO}$). Different panels show the evolution for different values of the Fermi energy $\mu_{\mathrm{nw}}$. The oscillatory behavior when $B>\mu_{\mathrm{nw}}$ reflects the transition to the helical regime in the normal side (see orange dashed line).}
\label{fig6}
\end{figure}

Similar phenomenology is obtained for conductance at fixed magnetic fields and increasing $\mu_{\mathrm{leads}}$ (Fig.\,\ref{fig5}). As expected, the Fano dips disappear as soon as $\mu_{\mathrm{leads}}>B$ while the gap coming from helicity in the central section in the NW is much more robust. Increasing $\mu_{\mathrm{leads}}$ results in well defined Fabry-Perot resonances in the helical gap region. The normal conductance as a function of magnetic field is shown in Fig.\,\ref{fig6}. Here, a change from irregular behavior to regular $e^2/h$ oscillations as a function of magnetic field signals the helical regime when $B\geq\mu_{\mathrm{nw}}$ \cite{PhysRevB.90.235415}. 

Having in mind that there exists no conclusive experimental evidence of the helical regime in nanowires in the literature\cite{nphys610,VanWeperen}, the nontrivial resonant effects in finite-length junctions that we have described, both helical Fabry-Perot resonances and helical Fano dips, could be used as an interesting option for detecting such helical transport regime in long junctions. Even more significant, these helical resonant features give rise to a non-trivial subgap spectrum when the leads become superconducting, as we discuss in what follows.
\clearpage

\section{Subgap levels in SNS junctions}
\label{ElevelsSNS}
In this section we investigate the role of the effects we have studied in previous section on the sub-gap spectrum.
First, we discuss how we model a SNS junction and refer to Section \ref{SNSjunctionmodel} for more details.
Then, we make an important distinction of the different length scales of the problem. At the end we present our results and point out the relevance of our calculations towards experimental detection of Majorana bound states.

\subsection{SNS junction model and relevant length scales}
\begin{figure}[!ht]
\centering
\includegraphics[width=.7\textwidth,height=.5\textwidth]{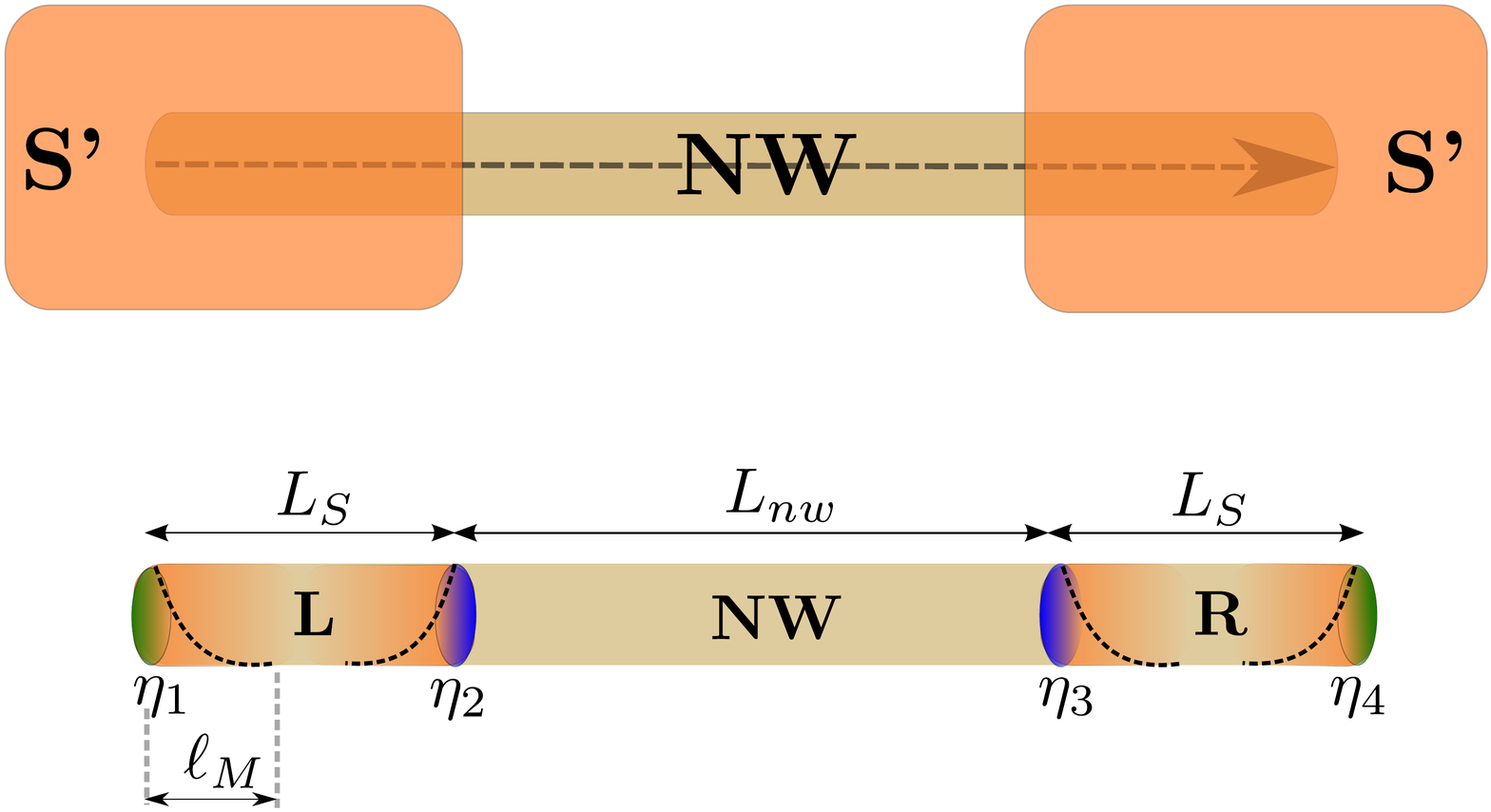} 
\caption[Sketch of the SNS junction]{(Color online) Two s-wave superconducting contacts (S', with gaps $\Delta_{S'}$) deposited on top of a Rashba nanowire (NW) of length $L=L_{S}+L_{\mathrm{nw}}+L_{S}$. The superconductors induce superconducting correlations into some regions of the nanowire via proximity effect, giving rise to regions which we refer to as superconducting leads (left L and right R) with gaps $\Delta<\Delta_{S'}$ and Fermi energies $\mu_{\mathrm{leads}}$, and a central region in the normal state with $\mu_{\mathrm{nw}}$. The dashed arrow in the first figure denotes the applied Zeeman field along the NW. Due to the finite length $L_S$, the junction in the topological phase hosts four Majorana bound states, $\eta_{1}, \eta_{2}, \eta_{3}, \eta_{4}$, for a phase difference of $\pi$ between the superconductors, with localisation length  $\ell_{M}$.}
\label{fig7}
\end{figure}

A full description for modeling finite length SNS nanowire junctions is provided in Chapter \ref{Chap2a}.  
To model a SNS junction we assume that the outer parts of the wire are coupled to an s-wave superconductor (with bulk values $\mu_{S'}$ and pairing $\Delta_{S'}$), while the central is not (see Fig.\,\ref{fig7}).
Superconducting correlations are induced by proximity effect into the nanowire. For good enough contact between the NW and the superconductor, the Cooper pair amplitude is finite inside the NW regions below the superconductor. In most papers in the literature, this situation is modeled by including by hand a pairing term, $\Delta<\Delta_{S'}$, in the hamiltonian of such NW regions.  While, rigorously speaking, this is incorrect (the superconducting coupling constant is zero inside the NW), it is well known that it provides a good description of the proximity effect for large enough gaps (in such cases, the parameter $\Delta$ is essentially the low frequency limit of a tunneling self-energy and is given by the tunnel coupling between the normal and superconducting parts, see e.g. \cite{Bena2}). Therefore, we adopt this approximation here for simplicity (we have checked that all our conclusions remain unaltered irrespective of whether we use this simplified model or a full NW + SC coupling model, see appendix \ref{inducspe}). In cases where the interface transparencies are small, extra Fabry-Perot resonances coming from insulating layers could complicate our analysis, see Ref. \cite{Galaktionov-Zaikin}. 

In particular, we model the regions of the nanowire below the superconducting contacts as regions with Fermi energy $\mu_{leads}$ and pairing potential on the left (L) and right (R) contact given by $\Delta_{L}=\Delta\,{\rm e}^{- i \varphi/2}$ and $\Delta_{R}=\Delta\,{\rm e}^{ i \varphi/2}$, with $\Delta<\Delta_{S'}$.
The region in the middle of the nanowire without superconducting correlations is the normal region (N) with Fermi energy denoted by $\mu_{\mathrm{nw}}$ as before. At high enough magnetic fields, the regions of the NW below the superconductors (S regions of the junction) can be driven into a topological superconducting phase when $B>B_{c}\equiv\sqrt{\mu_S^2+\Delta^2}$. Owing to the finite length $L_S$, this results in a SNS junction with four Majorana bound states for a phase difference of $\pi$ between the
superconductors: two inner Majorana bound states, labeled $\eta_{2,3}$, that form inside the junction, and two outer Majorana bound states, $\eta_{1,4}$, see Fig.\,\ref{fig7}. On the other hand, for a zero phase difference, only the outer MBSs are present.

SNS Josephson junctions are classified in two types, depending on the relationship between the length of the normal region $L_{\mathrm{nw}}$ 
(i.e. distance between the superconducting contacts) and the coherence length $\xi=\hbar v_{F}/(\pi\Delta)$, where $v_{F}$ is the Fermi velocity. 
Short junctions are characterized by $L_{\mathrm{nw}}\ll\xi$, whereas $L_{\mathrm{nw}}\gg\xi$ in long junctions. See Chapter \ref{Chap2a} for additional details.
Such classification can be also given in terms of natural energy scales of the problem, the Thouless energy, $E_{T}=\hbar v_{F}/L_{\mathrm{nw}}$, and the induced superconducting 
pairing potential $\Delta$, being $v_{F}$ the Fermi velocity, and $L_{\mathrm{nw}}$ the length of the normal region. 
The above conditions, in terms of these energy scales, are $\Delta\ll E_{T}$ for short junctions and $\Delta\gg E_{T}$ for long ones. The significance of this classification is related to the typical number $\sim \Delta/E_{T}$ of Andreev subgap states of the junction, in addition to the MBSs at zero energy.

The MBSs wave functions decay from both ends of the topological superconducting leads.  The inner and outer MBSs may feel their mutual presence if their wave functions exhibit a non zero spacial overlap. The relevant decay distance characterizing this overlap is the Majorana localization length $\ell_{M}$ (appendix \ref{Majorana-length}). 
For finite $L_{S}\ll2\ell_{M}$ the overlap between MBSs is significant and therefore they are no longer true zero modes.

In what follows, we discuss the subgap spectrum of short SNS junctions in the topological regime $B>B_c$ as well as the subgap spectrum of long SNS junctions as one goes from the \emph{helical} junction regime to the topological one. The helical junction regime is defined by a central region depleted into the helical regime, while the S regions remain non-topological, namely by $\mu_{leads}>\mu_{\mathrm{nw}}$, and $\mu_{\mathrm{nw}}<B<B_c$.
\begin{figure}[!ht]
\centering
\includegraphics[width=.9\textwidth]{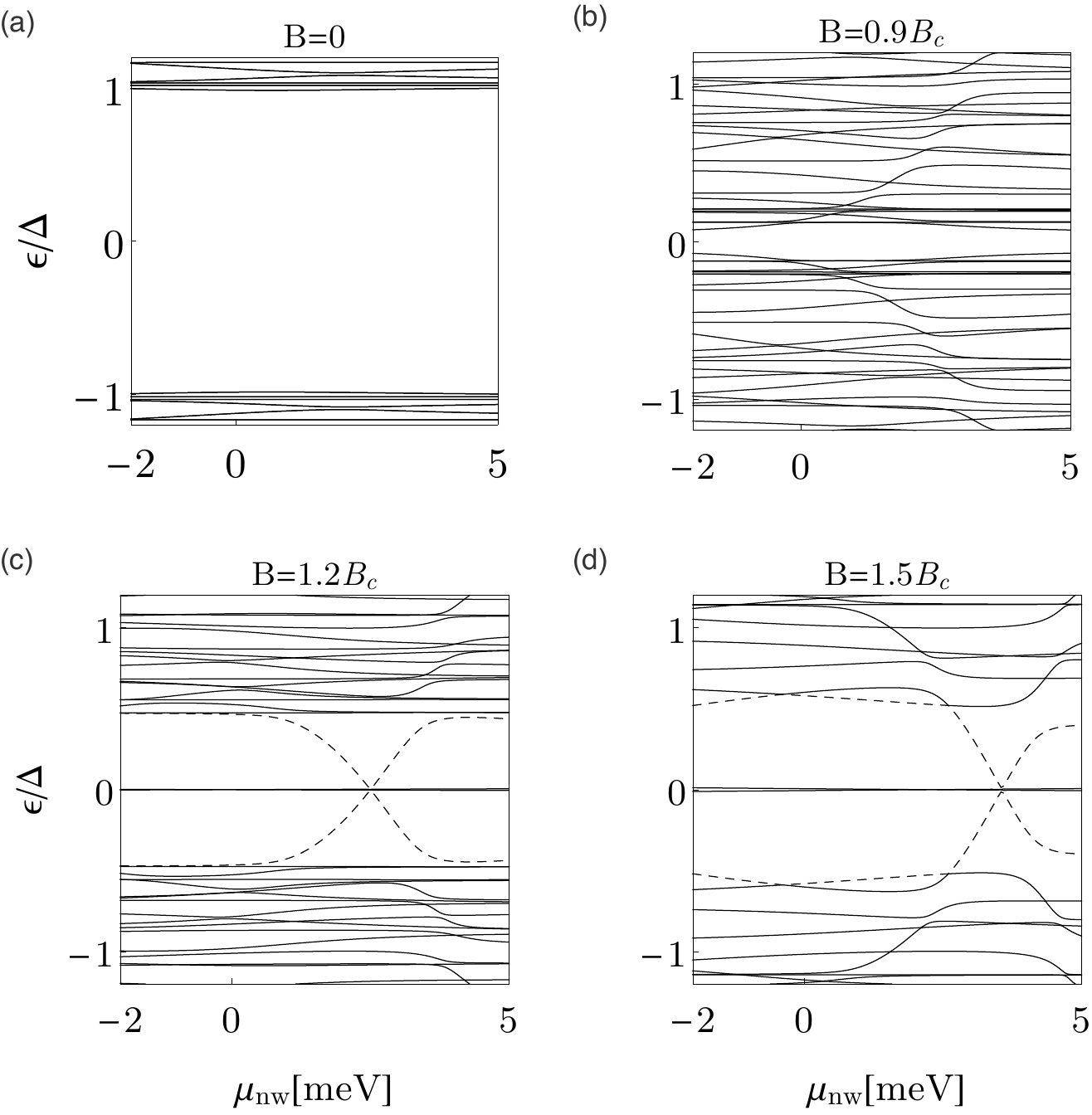} 
\caption[$\varepsilon(\mu_{\mathrm{nw}})$ for different $B$]{Andreev levels at $\varphi=0$ as function of the Fermi energy $\mu_{\mathrm{nw}}$ for a short junction, $L_{\mathrm{nw}}=20$\,nm. Different panels show the evolution of the spectrum for increasing magnetic fields. No sub-gap states arise for trivial superconducting leads, (a) and (b). For $B>B_{c}$, (c) and (d), the outer Majorana bound states coexists with a Shiba-like resonance. The Fano resonance induces sub-gap states owing to p-wave nature of the topological leads. Parameters: $E_{SO}=0.05$\,meV, $\mu_{\mathrm{leads}}=10E_{SO}$\,meV, $L_{S}=2\mu$m, $\Delta=0.25$\,meV.}
\label{fig8}
\end{figure}
\subsection{Short junctions}
\label{short}
For very short junctions, the ABS spectrum at $B<B_c$ and $\varphi=0$ does not contain sub-gap states (Figs. \ref{fig8}a and b). This is expected for a short junction with 
$\xi\gg L_{nw}$. 
The $B>B_c$ spectrum (Figs. \ref{fig8}c and d), on the other hand, is much more interesting. It contains the expected subgap state near zero energy for all 
$\mu_{\mathrm{nw}}$ (coming from the weakly coupled outer Majoranas for $L_S\gg2\ell_{M}$, the inner MBS at $\varphi=0$ are strongly hybridized and form standard 
ABS at energy $\sim\Delta$). Notably, this MBS coexists  with a bound state that crosses zero energy for a given $\mu_{\mathrm{nw}}>0$ (dashed line).

This bound state originates from the single resonance that the junction accommodates for increasing $\mu_{\mathrm{nw}}>0$ (see Fig. \ref{fig3}), which we discussed in connection to Fano resonances. If we interpret this resonant state as an impurity level, our results for $B<B_c$ are consistent with Anderson's theorem which prevents the existence of bound states inside the gap of an s-wave superconductor for non-magnetic impurities \cite{Anderson-theorem}. The reason is that the zero-enery crossing appears for $B>B_c$, such that the superconductor is effectively p-wave. Therefore, the emergence of these subgap states crossing zero energy should be understood as a direct consequence of nontrivial topology in the junction \cite{Sau-Demler,Huetal}. The precise condition for the level crossing coincides with the condition for having a Fano dip. As we discussed in section \ref{normCon}, this is the condition in the normal regime for having a single resonant state which interferes destructively with a helical contact; the latter condition is here fulfilled because $\mu_{\mathrm{leads}}<B_c<B$. These subgap states and zero-energy crossings should be understood as the p-wave counterparts of so-called Yu-Shiba-Rusinov sub-gap states \cite{subgap1,subgap2,subgap3,subgap4} and their corresponding parity crossings  \cite{subgap5} in s-wave superconductors with magnetic impurities. 

\begin{figure}[!ht]
\centering
\includegraphics[width=\textwidth]{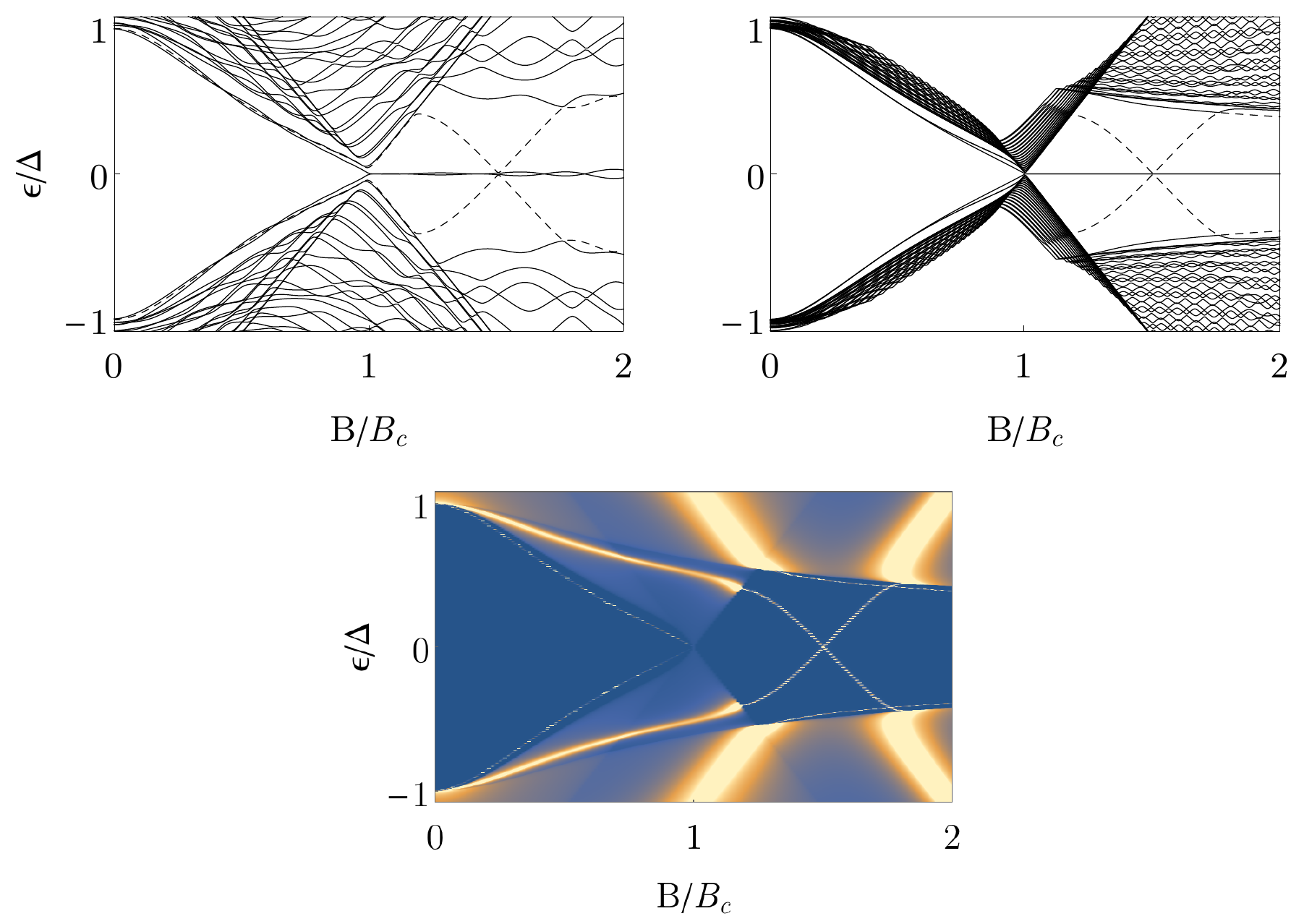} 
\caption[$\epsilon(B)$ for $\mu_{\mathrm{nw}}=3.57$\,meV ]{Andreev levels at $\varphi=0$ as function of the Zeeman field at $\mu_{\mathrm{nw}}=3.57$\,meV. Left and middle panels show the coexistence of the zero energy crossing (dashed line) with the two lowest zero energy levels (outer Majorana bound states) for: (top left) $L_{S}=2$\,$\mu$m and (top right) $L_S=10$\,$\mu$m. In the former we have a situation of overlapping Majorana bound states, while in the latter such overlapping is negligible.
Bottom panel shows the situation for $L_S=\infty$, where the zero energy states are not involved anymore. In this case the bound state develops a perfect crossing at zero energy. The rest of parameters are the same as in Fig. \ref{fig8}.}
\label{fig9}
\end{figure}

 \begin{figure}[!ht]
\centering
\includegraphics[width=.9\textwidth]{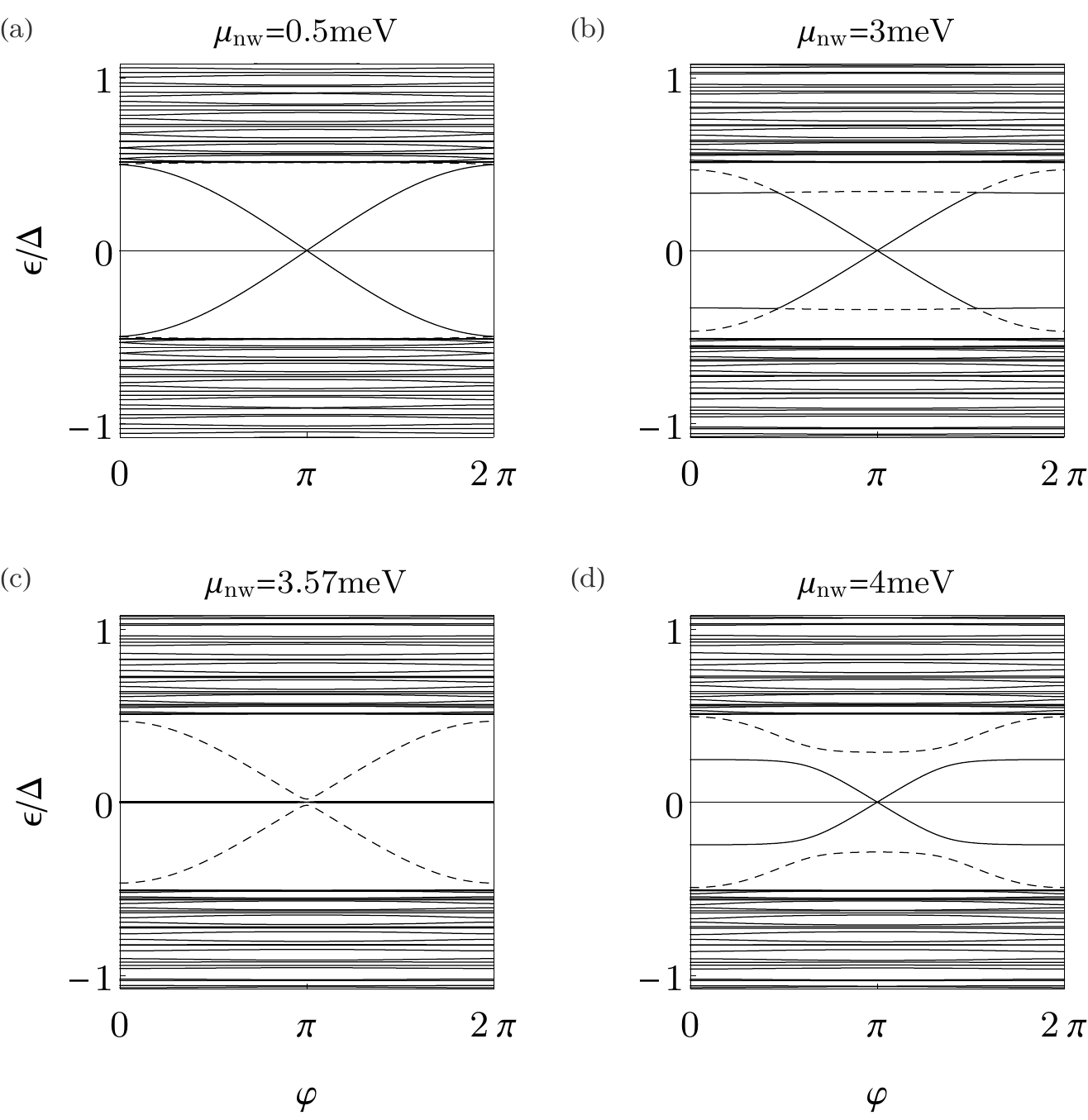} 
\caption[$\epsilon(\varphi)$ for a short-junction]{Andreev levels as function of the superconducting phase difference in the short-junction regime, $L_{\mathrm{nw}}=20$\,nm, at $B=1.5B_{c}$. Notice the emergence of a bound state (dashed line) coming from the continuum. Parameters: $\alpha_{R}=20$\,meV\,nm for InSb nano wires, $\mu_{leads}=0.5$\,meV, $L_{S}=10\mu m$, and $\Delta=0.25$\,meV. Different panels show the Andreev levels around $\mu_{\mathrm{nw}}=3.57$\,meV near the zero-energy crossing in Fig.\,\ref{fig8}d.}
\label{fig10}
\end{figure}

Further insight comes from the magnetic field dependence at fixed $\mu_{\mathrm{nw}}$ (Fig. \ref{fig9}), where we show three different situations: (top left) $L_{S} \ll 2\ell_M$, $L_S\gg 2\ell_M$ (top right) and $L\rightarrow\infty$ (bottom).
 After the closing of the gap across the topological phase transition at $B=B_c$, the spectrum of the junction exhibits  a perfect zero-energy state (left and middle panels) accompanied by a zero-energy crossing (dashed line in top left and right panels) similar to the one discussed in Fig. \ref{fig8}. 
On the other hand, when $L\rightarrow\infty$, the zero energy states (outer Majorana bound states) are not present anymore and a truly crossing at zero energy is observed.

Note here that despite the finite length of the central NW (in the top right panel) the zero energy state for $B>B_c$ does not oscillate as a function of Zeeman field, unlike what is typical of overlapping MBSs (top left panel)\cite{Lim:PRB12,Prada:PRB12,Rainis:PRB13,DasSarma:PRB12}. This can be easily understood as this state comes from the \emph{outer} MBSs which at $\varphi=0$ are effectively decoupled across the junction, since we assume $L_S\gg 2\ell_M$ for the top right panel. 

We now analyse in more detail the full phase dependence in the topological phase for different values of $\mu_{\mathrm{nw}}$. The low-energy sector is characteristic of a short junction: two almost $\varphi$-independent levels near zero energy coming from outer MBSs and two dispersive levels coming from hybridization of inner MBSs across the junction. The anti crossings near $\varphi=\pi$ are only visible for finite $L_S/(2\ell_{M})$. For $L_S\gg 2\ell_{M}$ (Fig.  \ref{fig10}a), the zero-energy levels are flat and the anti crossing at $\varphi=\pi$ becomes negligible \footnote{In $L_S\rightarrow\infty$ limit, the outer Majoranas are no longer involved in transport while the levels at $\varphi=\pi$ exactly cross (not shown) giving rise to anomalous $4\pi$-periodic spectrum and Josephson currents if fermionic parity is conserved}. In the following, we refer to the dispersive ABS with almost perfect crossings at $\varphi=\pi$ as Majorana ABSs. As $\mu_{\mathrm{nw}}$ increases, an extra bound state emerges from the continuum as an almost dispersionless subgap state and interacts very weakly with the Majorana ABSs (Fig.  \ref{fig10}b). Importantly, after crossing zero energy (Fig.  \ref{fig10}c) and reemerging at finite energy (Fig.  \ref{fig10}d), the anti crossing with the Majorana ABS is considerably larger, indicating that the bound state has changed its parity character.

\begin{figure}[!ht]
\centering
\includegraphics[width=.9\textwidth]{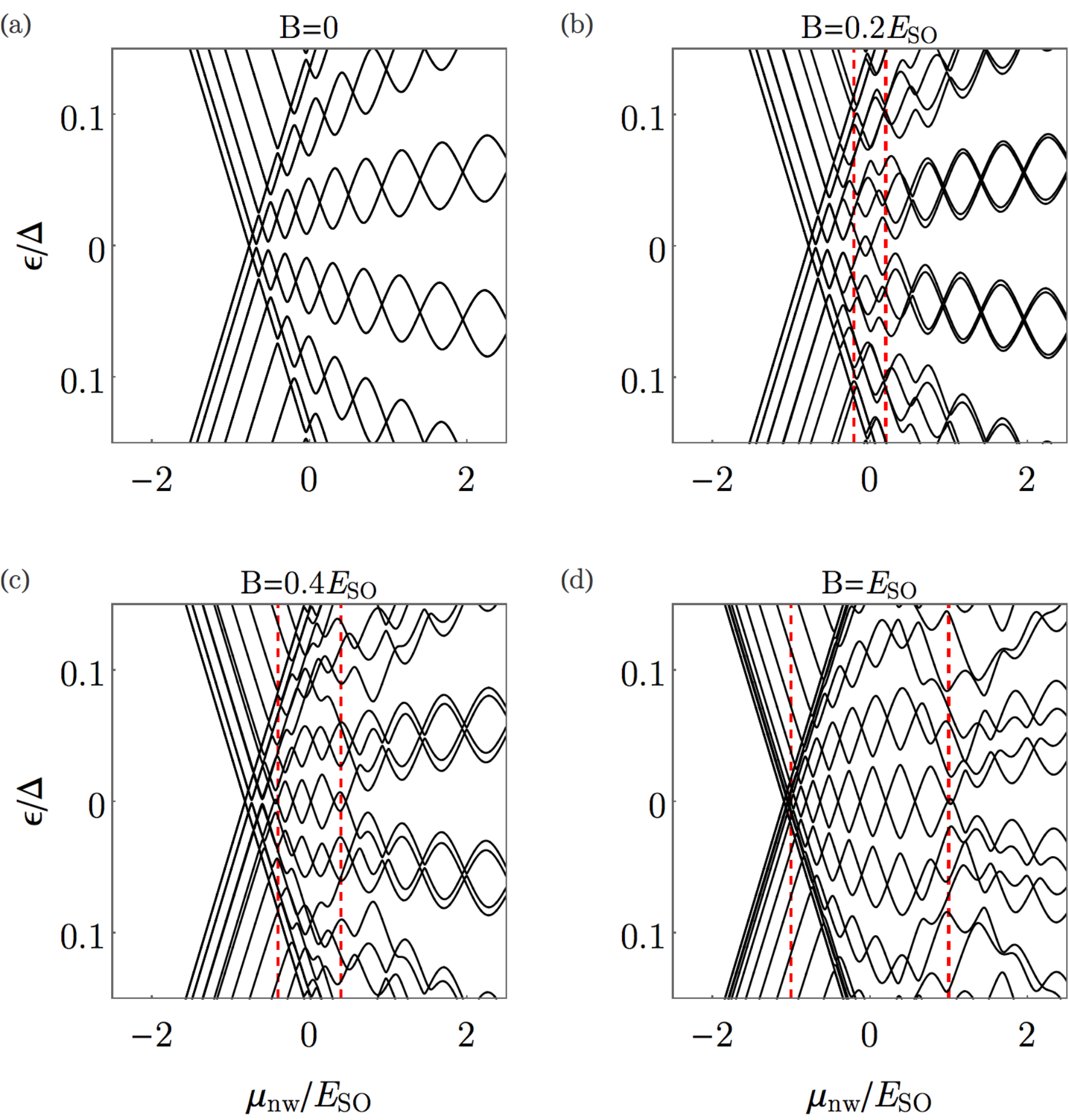} 
\caption[$\epsilon(\mu_{\mathrm{nw}})$ for a long junction]{Andreev levels at $\varphi=0$ as function of $\mu_{\mathrm{nw}}$ for a long junction, $L_{\mathrm{nw}}=4\mu$m and various magnetic fields. At finite B, the ABS spectrum shows a loop structure around zero energy in the region where the normal section becomes helical (marked by dashed lines) for fields that can be well bellow the topological transition ($B_c\approx 11.2 E_{SO}$). These helical loops resemble overlapping Majoranas but note that the superconducting section is deep in the trivial regime. Parameters: $E_{SO}=0.05$\,meV, $\mu_{leads}=10E_{SO}$, $L_{S}=2\mu$m, $\Delta=0.25$\,meV.}
\label{fig11}
\end{figure}  
\subsection{Long junctions}
\label{long}
The ABS spectrum of long junctions at small magnetic fields $B<B_c$ differs considerably from the one of short junctions. Even for $B=0$ (Fig.  \ref{fig11}a), the spectrum is very sensitive to the sharp increase of conductance at small negative $\mu_{\mathrm{nw}}$, when the junction goes rapidly from pinch-off to fully transmitting (solid black line in Fig.\,\ref{fig4}). This is reflected in a feature that resembles the closing and reopening of a gap (but, of course, is  related to the central region becoming metallic, rather than with a gap closing). The emergence of Fabry-Perot resonances in the normal phase is translated into the appearance of level pairs at finite energies, or loops, that oscillate with system parameters in the superconducting phase. A distinct change in the loop structure takes place as $B$ is increased within a window $|\mu_{\mathrm{nw}}|<B$. This, recall, corresponds to the helical regime of the normal region, characterised in normal transport by a helical gap and helical Fabry-Perot oscillations. The loops inside said window reconnect, and give rise to new loops around zero-energy, separated by parity crossings (Fig.  \ref{fig11}b). Each of these crossings corresponds to a helical Fabry-Perot resonance in the normal regime.
For larger Zeeman energies, supporting many helical Fabry-Perot resonances within the helical gap, correspondingly many consecutive zero-energy loops become visible in the superconducting regime. As soon as the normal side ceases to be helical ($|\mu_{\mathrm{nw}}|>B$), the spectrum does no longer show loops around zero energy. Since depleting the normal section of the NW junction should be much easier than gating the proximized region, we expect that said near-zero loops and parity crossings should be ubiquitous for finite size junctions near depletion \footnote{Intermediate $L_{\mathrm{nw}}$ junctions also show the same behaviour, not shown.} and constitute yet another alternative scheme to detect the helical regime. 
\begin{figure}[!ht]
\centering
\includegraphics[width=0.9\textwidth]{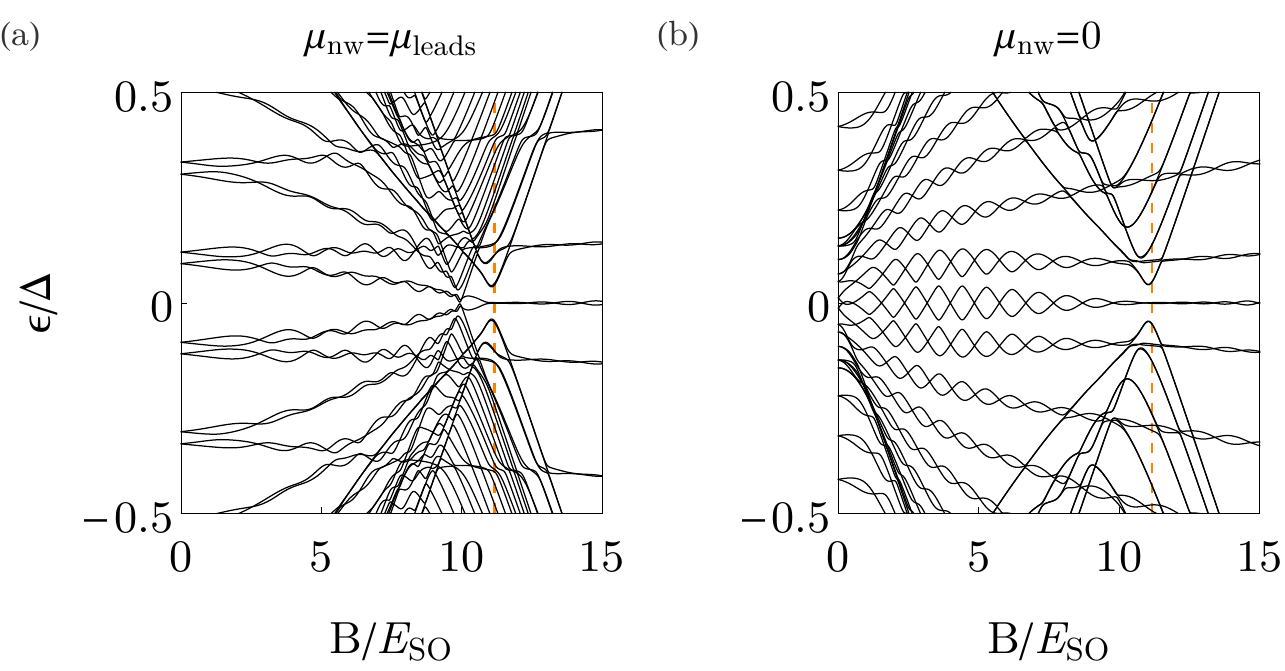} 
\caption[Andreev levels as function of $B$]{Andreev levels at $\varphi=0$ as function of the Zeeman field $B$. 
In the left panel we present the standard picture at high fields: the normal region never becomes helical before the topological transition.
In the right panel, an alternative escenario is shown when N is depleted: Majorana zero modes continuously emerge from helical loops. Indeed, the loop structure appear as soon as the normal section becomes helical for fields that can be well below the topological transition. Notice that the N region becomes helical as we decrease $\mu_{\mathrm{nw}}$, at fixed $\mu_{\mathrm{leads}}$. Important to remark here is that these loops and the ones that oscillate around zero in the left panel of Fig.\,\ref{fig9} share the same origin: they are Zeeman-induced crossings protected by spin-orbit.
Same parameters as in Fig.\,\ref{fig11}. The critical field $B_c\approx11.2E_{SO}$ is marked by vertical dashed line.}
\label{fig12}
\end{figure}

Each loop in the helical regime (see e.g. Fig.  \ref{fig11}b) is similar to the ones expected for magnetic impurities \cite{subgap1,subgap2,subgap3,subgap4}, or quantum dots in the Coulomb blockade regime \cite{Lee:13,PhysRevLett.110.217005} coupled to superconductors (we emphasize here that our junction is noninteracting). This result again suggests an interesting analogy with the physics of Yu-Shiba-Rusinov states in superconductors with magnetic impurities. Here, the combined action of Zeeman-induced spin-polarization \emph{and depletion} is crucial. 
Consecutive loops around zero energy, resemble the oscillatory behavior expected from overlapping MBSs in finite length NWs. However, since the helical gap condition $|\mu_{\mathrm{nw}}|<B$ does not involve $\mu_S$, which may be large, the zero-energy loops may exist while the proximized S regions are still in the topologically trivial regime $B<B_c$ (Fig. \ref{fig11} c and d). Remarkably, there exists a profound connection between zero-loops and MBSs. We find that the former actually evolve continuously into outer MBSs as $B$ is increased beyond $B_c$.  

To illustrate this key idea, we compare in  Fig.  \ref{fig12}, a situation without near-zero energy loops at low B fields ($\mu_{\mathrm{nw}}=\mu_{leads}$, panel a) with another with loops at very low B coming from a helical normal region ($\mu_{\mathrm{nw}}=0$,  panel b). While the MBSs in the first configuration emerge from a situation without zero energy states/crossings at low fields, the ones corresponding to the second configuration are clearly evolving from the low B-field loops around zero energy. We emphasize here that both configurations correspond to the same physical nanowire junction with the sole difference of a depletion in the \emph{normal part} of the junction in the second case. Fig.  \ref{fig12} nicely illustrates two of our main results: 1) long loops with parity crossings in the ABS spectrum can be used to identify the helical regime in a Rashba NW and 2) such helical loops, coming from depletion in the \emph{normal} side of the junction, continuously evolve into MBS for large enough magnetic fields.
\begin{figure}[!ht]
\centering
\includegraphics[width=.9\textwidth]{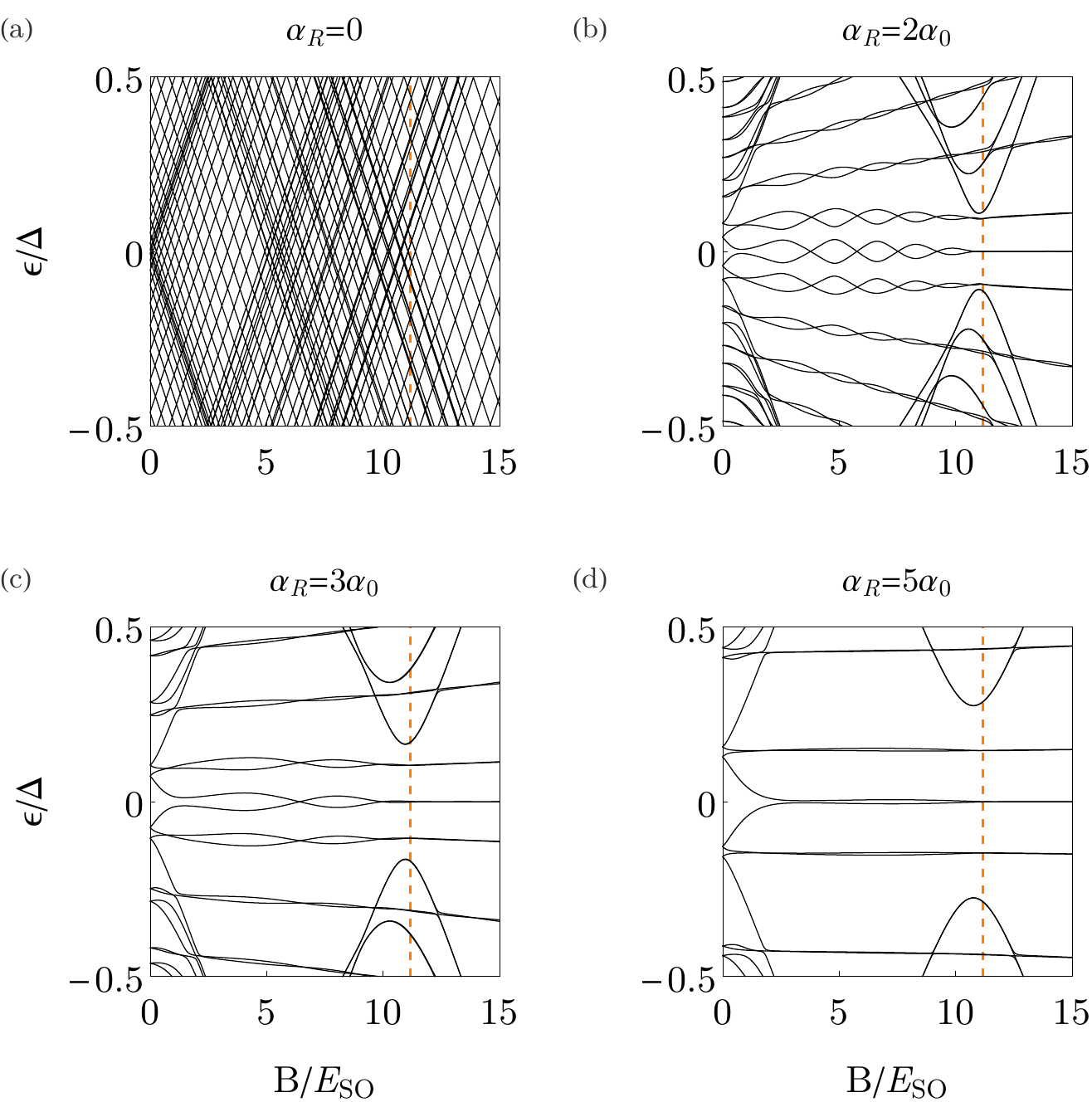} 
\caption[Andreev levels for different $\alpha_{R}$]{Same as Fig.\,\ref{fig12} for $\mu_\mathrm{nw}=0$ and increasing values of the SO coupling $\alpha_R$. The critical field $B_c$ is marked by a vertical orange dashed line.}
\label{fig13}
\end{figure}
To obtain more precise information about the nature of this interesting connection between $B<B_c$ near-zero loops and MBS states, we study their evolution for increasing SO coupling (Fig.  \ref{fig13}). For $\alpha_R=0$ (Fig.  \ref{fig13}a), Zeeman-induced depairing closes the superconducting gap and the spectrum becomes a dense quasi-continuum (the full junction is in the normal regime), as expected. Any $\alpha_R\neq 0$ removes all finite energy crossings while preserving the parity-protected crossings at zero energy. As a result, the spectrum is still gapped after the first parity crossing (the Zeeman field is no longer fully depairing) and \emph{many parity crossings} are possible. This important observation is illustrated in Fig.  \ref{fig13}(b,c,d)  (see also Fig.  \ref{fig12}b). For finite $\alpha_R$, the low-energy spectrum remains gapped after the first crossing and also after subsequent crossings. 
\begin{figure}[!ht]
\centering
\includegraphics[width=.9\textwidth]{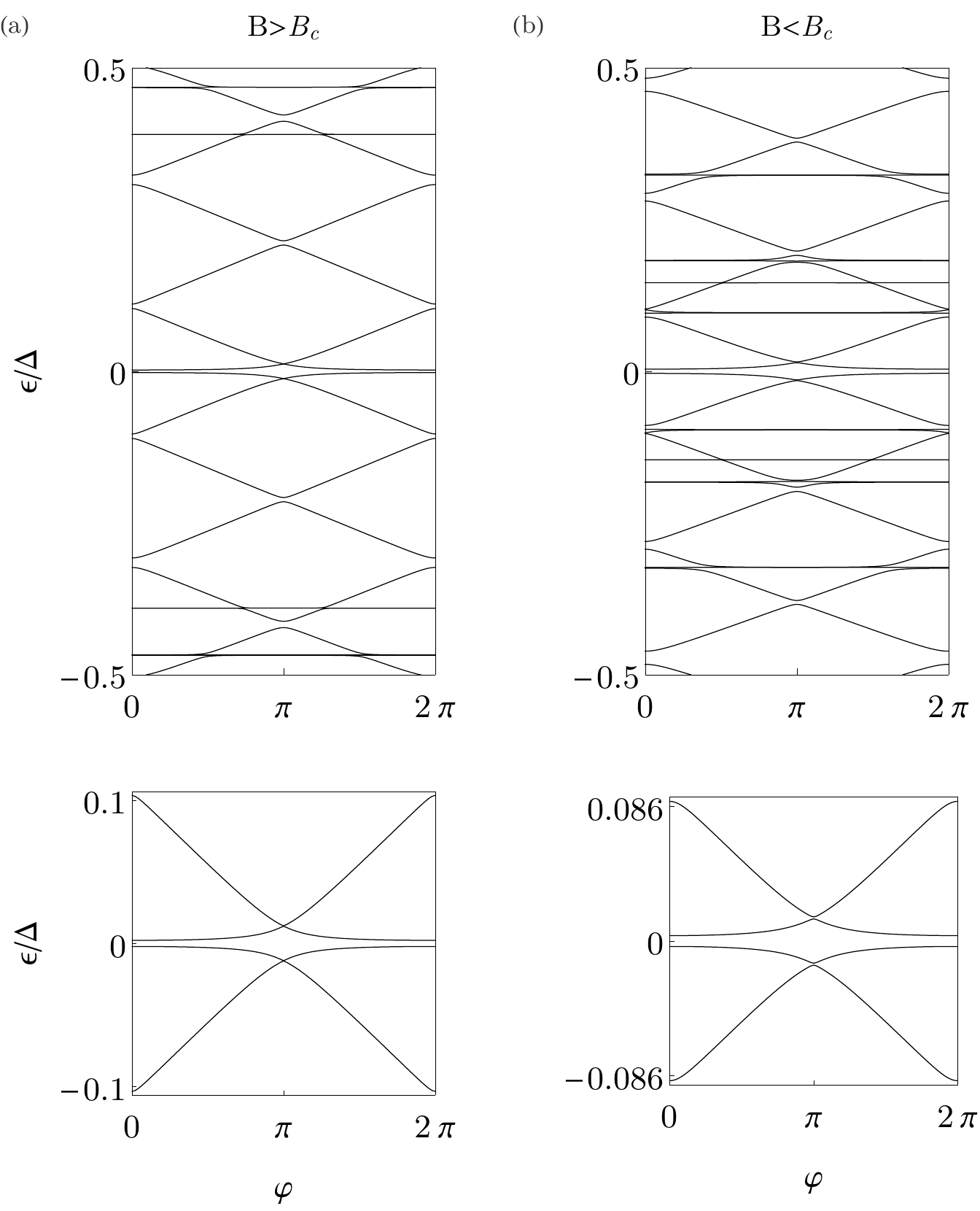} 
\caption[Andreev levels $\varepsilon(\varphi)$ for different Zeeman fields]{Andreev levels as function of the superconducting phase difference $\varphi$ for two values of the Zeeman field (a) $B=13E_{SO}$ and (b) $B=10E_{SO}$. Bottom panels show the lowest four levels in each situation. Rest of parameters same as in Fig. \ref{fig12}b. Notice that the resemblance between helical loops and Majorana modes is also clear in the phase dispersion of the spectrum: same low energy properties with small anti-crossings near $\varphi=\pi$.} 
\label{fig14}
\end{figure}
Another interesting conclusion that we can draw from our results is that a clear distinction between the near-zero states in the $B<B_c$ and $B>B_c$ regions can no longer be made. The only difference is quantitative, in that the amplitude of MBS oscillations in the topological regime become smaller for increasing $\alpha_R$, unlike for $B<B_c$. (The SO length becomes much shorter and, hence $L_S\gg\ell_{M}$). However, other spectral properties, such as the mini gap separating the near-zero modes from the first excited states, is roughly the same in both the trivial $B<B_c$ and non-trivial $B>B_c$ phases.

To finish, we consider the phase dependence of the subgap spectrum in Fig.\,\ref{fig14}. While topological SNS junctions with $L_S\rightarrow\infty$ are $4\pi$-periodic as a function of phase difference $\varphi$ due to the characteristic parity-protected crossing at $\varphi=\pi$ (see e.g. Fig.\,\ref{fig10}a), in finite $L_S$ junctions  (Fig. \,\ref{fig14}a), said crossing is avoided, and splits by a small energy due to the hybridization of MBSs at the junction (inner) and MBSs at the far ends of each S region (outer), which leads to a more conventional $2\pi$-periodicity  \cite{San-Jose:PRL12a}. Interestingly, the subgap spectrum at $B<B_c$ (Fig. \,\ref{fig14}b) shows essentially the same phase-dependence, shown in bottom panels of Fig. \,\ref{fig14}, which further confirms the deep connection between the $B<B_c$ and $B>B_c$ parity crossings. Note that the resulting Josephson current \cite{Cheng:PRB12}, which only depends on the Andreev spectrum, would be effectively the same (not shown).

\section{Conclusions}
\label{concl}
We have studied the normal transport and the sub-gap spectrum of SNS junctions based on semiconducting nanowires with strong Rashba spin-orbit coupling. In particular, we have focused on the role of confinement effects in ballistic finite-length junctions and analyzed the distinct properties of the ABS for short and long junctions as different sections of the underlying NW (N or S or both) become helical. For $B>B_c$, confined levels in the normal section give rise to bound subgap states, as expected from the effective p-wave nature of the topological superconductor. In normal transport, such bound states give rise to helical Fano dips. Perhaps more strikingly, we have found that a long junction with a helical normal section, but still in the topologically trivial regime with $\mu_{\mathrm{nw}}<B<B_c$, supports a low-energy subgap spectrum consisting of multiple-loop structures and parity crossings. Such states are derived from helical Fabry-Perot resonances in the normal regime. We have argued that such multiple loop structure in the ABS spectrum could be used to unambiguously identify the helical regime in NWs. Interestingly, these multiple loops smoothly evolve towards Majorana bound states as the Zeeman field exceeds the critical value. This suggests an interesting connection between subgap parity crossings in helical junctions with $B<B_c$ and Majorana bound states in topological ones with $B>B_c$. A recent study of fully open helical-N/trivial-S contacts\cite{JorgeEPs} further confirms the profound connection between subgap states in the helical regime and Majorana physics. 

\chapter{\bf Majorana bound states from exceptional points in non-topological superconductors\footnote{This Chapter is part of a work published in \cite{JorgeEPs}}} 
\label{ChapEPs}
\lhead{Chapter \ref{ChapEPs}. \emph{MBSs from exceptional points in non-topological superconductors}} 

\begin{small}
Recent experimental efforts towards the detection of Majorana bound states have focused on creating the conditions for topological superconductivity. Here we demonstrate an alternative route, which achieves fully localised zero-energy Majorana bound states when a topologically trivial superconductor is strongly coupled to a helical normal region. Such a junction can be experimentally realised by e.g. proximitizing a finite section of a nanowire with spin-orbit coupling, and combining electrostatic depletion and a Zeeman field to drive the non-proximitized (normal) portion into a helical phase. Majorana zero modes emerge in such an open system without fine-tuning as a result of charge-conjugation symmetry, and can be ultimately linked to the existence of `exceptional points' (EPs) in parameter space, where two quasibound Andreev levels bifurcate into two quasibound Majorana zero modes.  After the EP, one of the latter becomes non-decaying as the junction approaches perfect Andreev reflection, thus resulting in a Majorana dark state (MDS) localised at the NS junction. We show that MDSs exhibit the properties associated to conventional closed-system Majorana bound states, while not requiring topological superconductivity.
\end{small}

\newpage

\section{Introduction}
The emergence of topologically protected Majorana zero modes in topological superconductors has recently entered the spotlight of condensed matter research   \cite{Alicea:RPP12,Leijnse:SSAT12,Beenakker:11,StanescuModel13,RevModPhys.87.137}
One of the main reasons is the remarkable prediction that such Majorana bound states (MBSs), also known as Majorana zero modes, should obey non-Abelian braiding statistics \cite{Kitaev:P01,Ivanov:PRL01}, much like the 5/2 states in the fractional Hall effect, without requiring many-body correlations. 
It has been argued that the successful generation, detection and manipulation of MBSs would open the possibility of practical topologically protected quantum computation \cite{Nayak:RMP08,Sarma:16}. 
Despite impressive experimental progress  \cite{Mourik:S12,Deng:NL12,Das:NP12,Rokhinson:NP12,Finck:PRL13,Churchill:PRB13,Lee:13,Hart:NP14,Pribiag:NN15,Nadj-Perge31102014}, such ambitious goals have still not been conclusively achieved.

A number of practical proposals have been put forward aiming to generate the conditions for the spontaneous emergence of robust MBSs in real devices. 
Some of the most studied ones are based on proximitising topological insulators \cite{Fu:PRL08} or semiconductor nanowires \cite{Lutchyn:PRL10,Oreg:PRL10}. 
The core challenge in all these proposals has been to artificially synthesise a topologically non-trivial superconductor with a well-defined and robust topological 
gap \cite{Takei:PRL13}. The bulk-boundary correspondence principle dictates that the superconductor surface is then host to topologically protected MBSs. Creating a topological gap is arguably the main practical difficulty of such proposals, particularly since topological superconductors are rather sensitive to disorder.

In this Chapter we present an alternative scheme for the creation of MBSs that does not require topological superconductivity at all. 
The possibility of engineering Majoranas in topologically trivial setups has been studied in other contexts before. It has been shown, for example, that topological excitations, 
and MBSs in particular, may arise in trivial superconductors under adequate external driving \cite{Jiang:PRL11,Kundu:PRL13}, similarly to the mechanism behind Floquet topological 
insulators \cite{Lindner:NP11}. Also, cold-atom systems with specifically engineered dissipation \cite{Diehl:NP11,PhysRevLett.109.130402} may relax into a topologically non-trivial 
steady state that are host to dark states at zero energy with Majorana properties.  
Our approach is implemented in a solid state setup and is based on proximitized semiconductor nanowires. In its topologically trivial regime, such a wire will not generate 
MBSs when terminated with vacuum (i.e. at a closed boundary). Its spectrum is instead a set of Bogoliubov quasiparticles that can be seen as pairs of Majoranas hybridized to 
finite energy. By creating a sufficiently transparent normal-superconductor (NS) junction at one end of the wire, we create a different kind of \emph{open} boundary, to which 
the bulk-boundary correspondence principle does not apply. Such a high-transparency junction can be fabricated by proximitizing only one half of a pristine semiconducting 
nanowire (Fig.\ref{fig:WireSketch}).
We demonstrate that, as one tunes the normal side into a helical (half-metallic) regime via a parallel Zeeman field, one Majorana pair becomes decoupled into two zero energy 
resonances. One of which is subsequently removed into the reservoir, leaving behind a stable Majorana `dark state' (MDS) at the NS junction without requiring a non-trivial 
superconductivity. A dark state here is defined as a bound state that despite having an energy embedded in a continuum of delocalized excitations is orthogonal to them and, 
therefore, non-decaying.

\subsection{General considerations on NS junctions}
The emergence of these MDSs cannot be described using the conventional band topology language, but rather needs to be understood in the context of open quantum systems. 
Unlike in closed systems, eigenstates in open systems decay with time, as the state leaks into the reservoir. Hence, their energies
 are no longer real but have a negative imaginary part $\Gamma_p$, $\epsilon_p=E_p-i\Gamma_p$. Such a complex spectrum is sometimes modelled by a non-Hermitian Hamiltonian. A more precise and general description is  obtained by considering the analytic continuation of the scattering matrix $S(\omega)$, where the energy $\omega$ of incoming states from the reservoir  is allowed to extend into the lower complex half-plane. The analogous to the real eigenvalues of the closed system then becomes the poles of $S(\omega)$ for the open system. 
\begin{figure}[!h]
   \centering
   \includegraphics[width=.6\columnwidth]{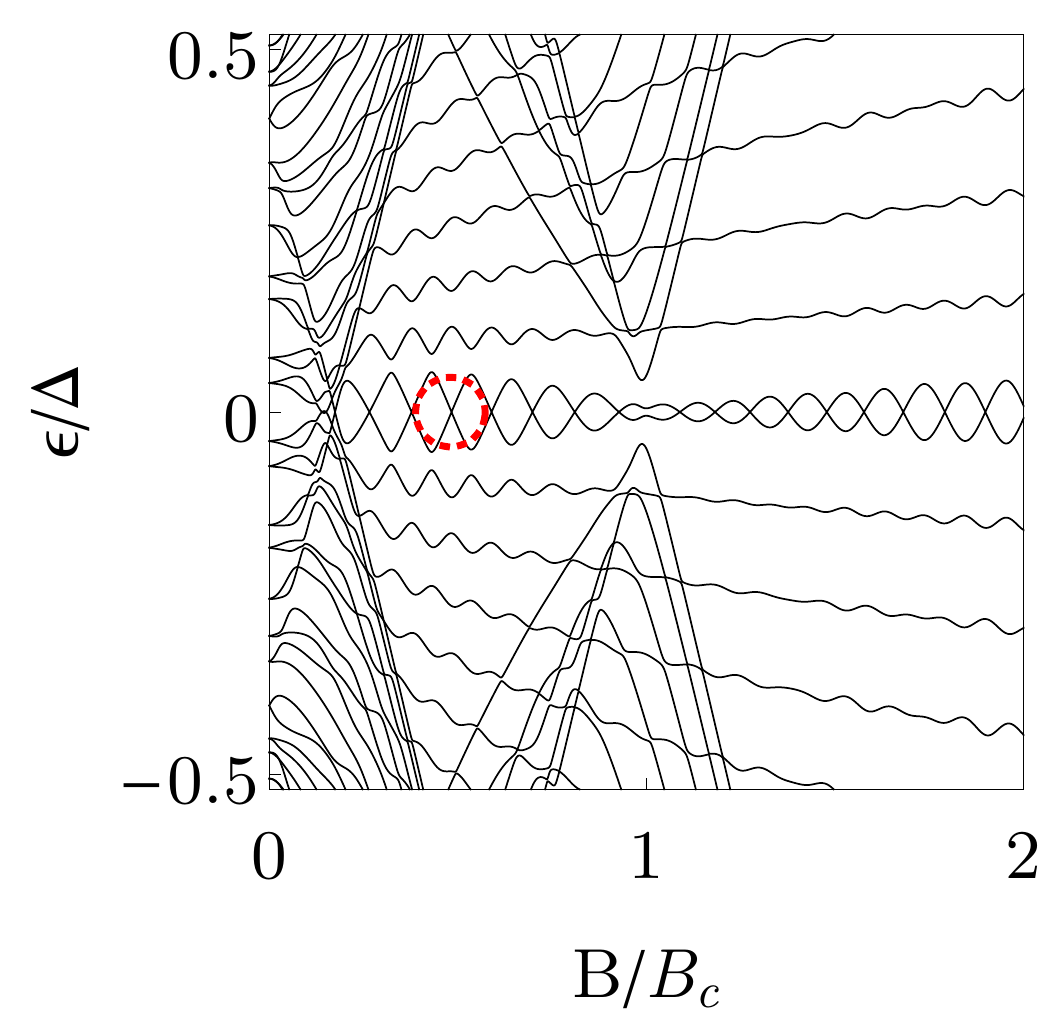}
   \caption[Low-energy spectrum in a closed NS junction as a function of $B$]{(Color online) Low-energy spectrum in a closed NS junction as a function of the Zeeman field. The red dashed circle marks the BdG parity crossing which is analysed in a open NS system.}
   \label{fig:NSlevelsEps} 
\end{figure}

Scattering processes at the NS interface in the stationary regime are determined by the scattering matrix ($S$-matrix), whose general form is given by
\begin{equation}
\label{Smatrix}
S(E)=1-2\pi iW^{\dagger} (E-H_{eff})^{-1}W\,, 
\end{equation}
where $H_{eff}=H-i\pi WW^{\dagger}$ is the effective non-Hermitian Hamiltonian that describes this \emph{open system}, $W$ is the coupling matrix to the normal reservoir N and $H$ 
the isolated system's time-independent Hermitian Hamiltonian.

As an example we consider a minimal description that consists of a low-energy energy BdG parity crossing, a pair of Andreev levels near zero-energy crossing 
as the ones discussed in Chap.\,\ref{Chap2} (see for instance Fig.\,\ref{fig:NSlevelsEps}) and then we analyse its properties in an open NS system, when coupled to a single pair of electron-hole modes in the normal reservoir N. Here, we follow the discussion made in \cite{PhysRevB.92.144306}.
The low-energy problem restricts the dimension of matrices $H$ and $W$ to be $2\times2$. Therefore, they can be in general written as
\begin{equation}
H=
\begin{pmatrix}
a&b\\
c&d
\end{pmatrix}\,,\quad
W=\begin{pmatrix}
e&f\\
g&h
\end{pmatrix}
\end{equation}
where the entries are in general complex numbers.
Since we are dealing with a superconducting system, electron-hole symmetry requires $H=-\tau_{x}H^{*}\tau_{x}$ and $W=\tau_{x}W^{*}\tau_{x}$ \cite{RevModPhys.87.1037}. Thus, taking into account Hermicity of $H$, and electron-hole symmetry, we arrive at
\begin{equation}
H=
\begin{pmatrix}
E_{0}&0\\
0&-E_{0}
\end{pmatrix}\,,\quad
W={\rm e}^{i\varphi \tau_{z}}\begin{pmatrix}
\lambda_{+}&\lambda_{-}\\
\lambda_{-}&\lambda_{+}
\end{pmatrix}{\rm e}^{i\varphi' \tau_{z}}\,,
\end{equation}
where $\lambda_{\pm}$ are defined from the eigenvalues of the coupling matrix $WW^{\dagger}$, 
\begin{equation}
\gamma_{1}=(\lambda_{+}+\lambda_{-})^{2}\,,\quad \gamma_{2}=(\lambda_{+}-\lambda_{-})^{2}\,,
\end{equation}
and $\varphi,\varphi'$ being real coefficients. Therefore, the effective Hamiltonian can be written as
\begin{equation}
\label{fulleffH}
H_{eff}=
\begin{pmatrix}
E_{0}&0\\
0&-E_{0}
\end{pmatrix}
-\frac{i\pi}{2}
\begin{pmatrix}
\gamma_{1}+\gamma_{2}&\gamma_{1}-\gamma_{2}\\
\gamma_{1}-\gamma_{2}&\gamma_{1}+\gamma_{2}
\end{pmatrix}\,.
\end{equation}
In order to understand the meaning of $\gamma_{1,2}$, we write down the effective Hamiltonian in the Majorana representation, which is defined by 
the following rotation matrix $\Omega$ \cite{RevModPhys.87.1037},
\begin{equation}
\Omega=\begin{pmatrix}
1&1\\
i&-i
\end{pmatrix}\,.
\end{equation}
In this basis, the rotation of the effective Hamiltonian, $ H_{eff}\rightarrow \Omega H_{eff}\Omega^{\dagger}$, is written as
\begin{equation}
\label{MajRepHee}
H_{eff}=
\begin{pmatrix}
-i\pi\gamma_{1}&-iE_{0}\\
iE_{0}&-i\pi\gamma_{2}
\end{pmatrix}\,,
\end{equation}
and exhibits a clear physical meaning. Indeed, note that in Eq.\,(\ref{MajRepHee}) each Majorana contributing to a BdG excitation may have a different lifetime $\gamma_{1,2}$, where $E_{0}$ represents the Majorana overlap. 

Further information is acquired from analysing the poles of the $S$-matrix given by Eq.\,(\ref{Smatrix}). Such poles are the eigenvalues of the effective Hamiltonian $H_{eff}$. Then, from Eq.\,(\ref{fulleffH}), we get the eigenvalues and eigenfunctions
\begin{equation}
\label{EPSEp}
E_{\pm}=-i\frac{\pi(\gamma_{1}+\gamma_{2})}{2}\pm\sqrt{E_{0}^{2}-\bigg[\frac{\pi(\gamma_{1}-\gamma_{2})}{2}\bigg]^{2}}\,,\quad
\psi_{\pm}=\begin{pmatrix}
i\frac{E_{0}\pm\sqrt{E_{0}^{2}-\big[\frac{\pi(\gamma_{1}-\gamma_{2})}{2}\big]^{2}}}{\frac{\pi(\gamma_{1}-\gamma_{2})}{2}}\\
1
\end{pmatrix}\,.
\end{equation}
The NS junction is now an open system with quasi bound Andreev states that acquire a finite imaginary part, which gives rise to a finite lifetime. 

The discussion made before can be generalised to a multichannel case \cite{PhysRevB.92.144306}, and therefore we can conclude that in NS junctions, all such poles come in pairs 
$\pm E_p-i\Gamma_p$ for all poles with non-zero real part $E_p$, due to the charge-conjugation symmetry of the Nambu representation \cite{Pikulin:JL12,Pikulin:PRB13}. 
Here, we denote the real part as $E_{p}$, while $\Gamma_{p}$ as the imaginary part of eigenvalues given in Eq.\,(\ref{EPSEp}). In this sense, poles with zero real part $E_p=0$  
are special, as they do not come in pairs. We denote their total number by $Z$, which excludes any pole with zero real \emph{and} imaginary part.

\begin{figure}
   \centering
   \includegraphics[width=\columnwidth]{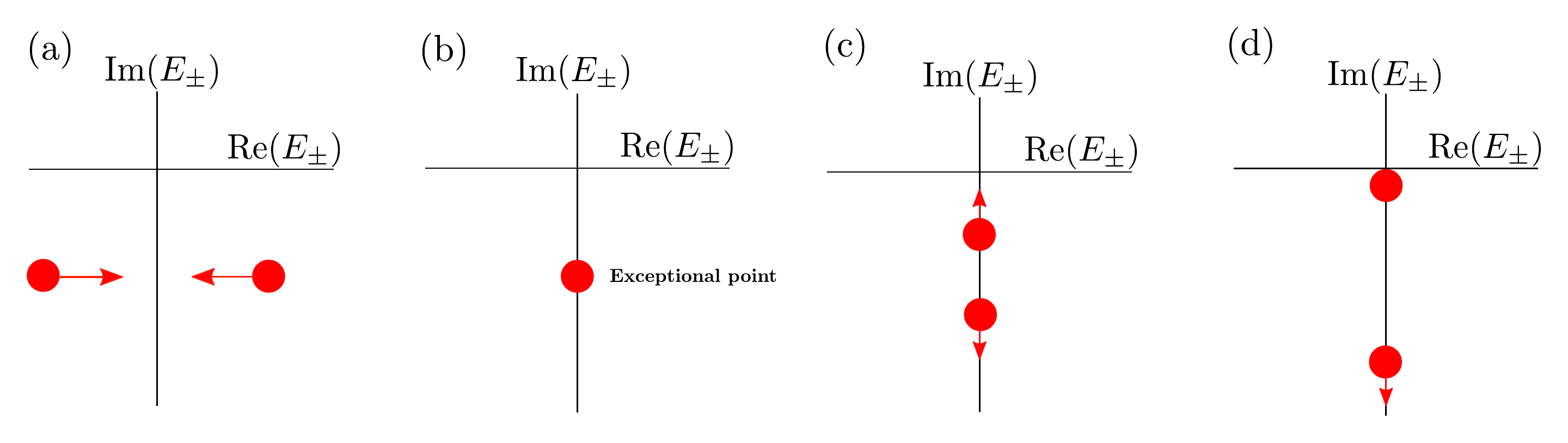}
   \caption[General considerations on NS junctions]{(Color online) Sketch of the evolution of the two S-matrix lowest poles of an open NS junction. 
   (a) The S-matrix poles are the eigenvalues of the respective effective Hamiltonian and they acquire a finite imaginary part, emerging in the lower complex half-energy plane.
 (b) As a given parameter of the system is varied, the lowest poles first approach each other and become degenerate at the imaginary axis, giving rise to an exceptional point, where eigenvalues and eigenfunctions coalesce. (c) After the exceptional point, the S-matrix has two purely imaginary poles, which are modes with zero real part but with different lifetimes (different couplings to the reservoir N). (d) When one of the couplings  evolves towards zero, it is said that the corresponding pole is \emph{buried}, implying the existence of a zero-energy non-decaying (dark) state somewhere in the system with Majorana properties.}
   \label{PoleEvolution} 
\end{figure}

From Eq.\,(\ref{EPSEp}), one notices that for $E_{0}>\pi(\gamma_{1}-\gamma_{2})/2$, there are two kind of poles in the low complex plane with standard BdG degeneracy 
$E_{+}=-E_{-}^{*}$ (see Fig.\,\ref{PoleEvolution}a). 
As a given parameter of the system is varied, a pair of poles $\epsilon_p=\pm E_p-i\Gamma_p$ first approach each other (see Fig.\,\ref{PoleEvolution}a)
and then become degenerate at the imaginary axis  when 
$E_{0}=\pi(\gamma_{1}-\gamma_{2})/2$ (see Fig.\,\ref{PoleEvolution}b), being the real part of the eigenvalues is zero, $E_p=0$. 
The S-matrix has a single degenerate pole and the eigenvalues and eigenvectors of the effective Hamiltonian, which are poles of the S-matrix, have the special form
\begin{equation}
\label{eoseos}
E_{\pm}=i\pi\frac{\gamma_{1}+\gamma_{2}}{2}\,,\quad \psi_{\pm}=
\begin{pmatrix}
i\\
1
\end{pmatrix}\,\quad\quad \rightarrow\quad\quad E_{+}=E_{-}\,,\quad \psi_{+}=\psi_{-}\,.
\end{equation}
The fusion of the two poles of the $S$-matrix is known as a \emph{pole transition}, which is a particular instance of a more general phenomenon
 in open quantum systems and known as \emph{exceptional point} (EP) \cite{Kato:95,Moiseyev:11,Berry:CJOP04,Heiss:JPA12,Heiss:PRE00}.
It differs from a closed-system degeneracy in that the corresponding eigenstates do not remain orthogonal, but rather coalesce into one as they pass through a branch point singularity, as one can indeed check from Eq.\,(\ref{eoseos}),
\begin{equation}
 E_{+}=E_{-}\,,\quad \psi_{+}=\psi_{-}\,.
\end{equation}
It turns out that this phenomenon is very well known in the open quantum systems, where complex eigenvalues and eigenfunctions coalesce at an exceptional point in the spectrum owing to branch point singularities. 
Indeed, exceptional points have been extensively studied in photonics  where they have been shown to give rise to novel phenomena unique to open systems \cite{Bender:PRL98,Klaiman:PRL08,Regensburger:N12,Guo:PRL09,Liertzer:PRL12,Brandstetter:NC14,Peng:S14,Poli:NC15}. Their implications in electronic systems, however, have been seldom discussed \cite{Mandal:EEL15,Malzard:15}.

After the exceptional point, in our example for $E_{0}<\pi(\gamma_{1}-\gamma_{2})/2$ (see Fig.\,\ref{PoleEvolution}c), the $S$-matrix has two purely imaginary poles
\begin{equation}
 E_{\pm}=-i\frac{\pi(\gamma_{1}+\gamma_{2})}{2}\pm i\sqrt{\bigg[\frac{\pi(\gamma_{1}-\gamma_{2})}{2}\bigg]^{2}-E_{0}^{2}}\,,
\end{equation}
representing two modes with zero real energy but with different lifetime, which arises from the imaginary part.
The two degenerate poles branch along the imaginary axis and their decay rates bifurcate into different values as one can indeed observe in previous equation.
A maximum width bifurcation is reached when $E_{0}=0$ (see Fig.\,\ref{PoleEvolution}d), that is at the parity crossing (see red circle in Fig.\,\ref{fig:NSlevelsEps}), then the eigenvalues read
\begin{equation}
E_{+}=-i\pi\gamma_{2}\,,\quad E_{-}=-i\pi\gamma_{1}\,.
\end{equation}
Observe that they acquire different lifetime which depends on the couplings to the reservoir N, $\gamma_{1,2}$, as they are eigenvalues of the coupling matrix $W$.

The total number $Z$ (which excludes any pole with zero real \emph{and} imaginary part) has a very important meaning in open NS 
junctions, and defines the analogue of band topology of a closed quantum system. Indeed, it has been shown that the topology of the scattering matrix in quasi-1D NS junctions is 
classified by the invariant $\nu=Z\,\mathrm{mod}\,2$,  i.e. the parity of the number of poles with zero real energy, with $\nu=1$ signalling an  open system with non-trivial 
topology from the point of view of scattering \cite{Pikulin:JL12,Pikulin:PRB13}.
In terms of its S-matrix poles, the topological transitions of an open NS junction follow a characteristic pattern.
At the beginning, $Z=0$, since there are no poles with zero real energy, and the S-matrix  is trivial, $\nu=0$. 
The EP transition discussed above is the open-system counterpart of a band inversion in a closed system \cite{Pikulin:JL12,Pikulin:PRB13}. 
We have seen that after the exceptional point, the $S$-matrix has two purely imaginary poles and their decay rates bifurcates into different values denoted now by
$\Gamma_{0}<\Gamma_{1}$. The exceptional point thus involves a change of $Z$ by 2, but the topology of the $S$-matrix remains trivial, $\nu=0$.
If $\Gamma_{0}$ evolves towards zero (or close enough to zero for all practical purposes), it is said that the corresponding pole is 
\emph{buried}, and it is excluded from the $N$ count, effectively signalling a change of topology $\nu=1$.  Crucially, the existence of a buried pole implies 
the existence of a zero-energy non-decaying (dark) state somewhere in the system with Majorana properties. In this sense, S-matrix topology is a true generalization of the band-structure topology of closed systems, and has the same implications in terms of topologically protected excitations, albeit in the context of open systems. It is also closely linked to the existence of an exceptional point in the system that occurs before the pole burying, in the trivial $\nu=0$ phase.

\begin{figure}
   \centering 
   \includegraphics[width=0.85\columnwidth]{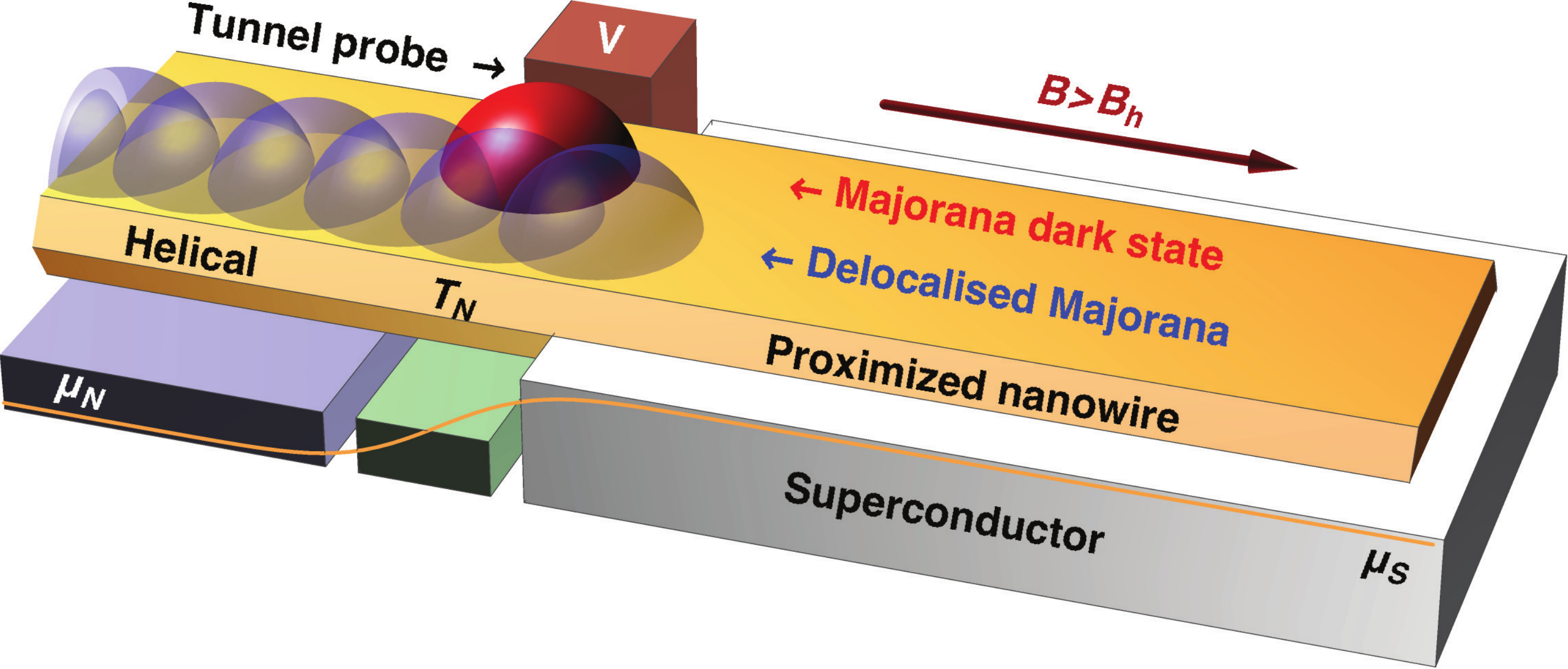}
   \caption[Sketch of a semiconductor nanowire, partially proximitized with a $s$-wave superconductor]{(Color online) A sketch of a semiconductor nanowire, partially proximitized with a conventional superconductor on the right side. The normal side may be depleted (small Fermi energy $\mu_N$), and may become helical under a Zeeman field, $B>B_h\equiv\mu_N$, while the superconducting side remains topologically trivial at small fields. For sufficiently transparent junctions in the Andreev limit ($\Delta\ll\mu_S$), this results in Majorana dark state bound to the junction, in red.} 
   \label{fig:WireSketch} 
\end{figure}

In this Chapter we show that, while at weak couplings between the normal environment and the superconductor a non-trivial S-matrix implies that the superconductor is also non-trivial in 
isolation, this is not the case at strong couplings. In specific but experimentally relevant conditions, a sufficiently transparent junction between a normal metal and a 
\emph{trivial} superconductor has a non-trivial S-matrix with $\nu=1$, and is thus host to a Majorana dark state. The required conditions are: (1) the system should have a 
finite spin-orbit coupling at the contact, (2) the normal part of the junction should be sufficiently depleted (small Fermi energy $\mu_N$) and polarised by a Zeeman field $B$ 
into a helical half-metallic phase $B>\mu_N$, (3) the normal transmission of the junction should be close to one, and (4) the trivial superconductor should be in the Andreev 
limit $\Delta\ll \mu_S$, where $\Delta$ is the superconducting gap and $\mu_S$ is its Fermi energy. The rationale of these conditions is to achieve good Andreev reflection of 
helical carriers from the normal side, which generates a MDS strongly localized at the junction. 

The intuitive mechanism behind the process is as follows. 
When isolated, the trivial superconducting wire is host to a Majorana \emph{pair} at each end, which is strongly hybridized into a fermionic Bogoliubov state, with real energies 
$\pm E_p$. As the contact is opened onto a helical wire (which has a single decay channel), one (and only one) of the two Majoranas escapes into the reservoir (blue state in 
Fig. \ref{fig:WireSketch}), leaving behind the orthogonal Majorana as a dark state (red), pinned at zero energy since it no longer overlaps with the escaped Majorana. 
This process takes place formally by crossing an exceptional point bifurcation of the $\pm E_p$ poles into poles $i\Gamma_{0,1}$ on the imaginary axis.
As the conditions above are fulfilled, the zero energy dark state becomes truly non-decaying $\Gamma_0\to 0$. Deviations from these conditions result in a residual decay rate 
$\Gamma_0$. The dark state is then a sharp Majorana resonance centered at zero, but with Majorana properties surviving at times shorter than $\tau_0=1/\Gamma_0$.
We consider a realistic model of an NS contact in a proximitized semiconducting wire.
We show that the residual decay rate of the MDS in non-ideal conditions depend exponentially with junction lengthscales, just 
like the Majorana splitting in closed topological superconductors.

\section{Majorana dark states in a proximitized Rashba wire}
In Chapter \ref{Chapter01} we have discussed that in recent years experimental progress has been reported towards the detection of MBS in Rashba nanowires  
\cite{Mourik:S12,Deng:NL12,Das:NP12,Rokhinson:NP12,Finck:PRL13,Churchill:PRB13,Lee:NN14}. 
These efforts were in large part stimulated by the prediction by Lutchyn \textit{et al.}\cite{Lutchyn:PRL10} and Oreg \textit{et al.} \cite{Oreg:PRL10} 
that these type of systems would undergo a topological transition into an effective p-wave superconducting phase when a Zeeman field $B$ parallel to the wire exceeds a critical 
value $B_c$.  See Chap.\,\ref{Chapter01} for a full description about 
engineering 1D topological superconductors based on nanowires with Rashba spin-orbit coupling. We now consider models relevant to single-mode Rashba wires, and demonstrate the formation of MDSs for $B<B_c$.

Following Refs. \cite{Lutchyn:PRL10,Oreg:PRL10}, we model a thin proximitized Rashba nanowire under a Zeeman field by a spinful 1D tight-binding chain, 
\begin{equation}
\label{model}
\begin{split}
H_\textrm{S}&=(2t-\mu_S)\sum_{\sigma n}c_{\sigma n}^\dagger c^{\phantom{\dagger}}_{\sigma n} 
-\mathop{\sum_{ \langle n,n'\rangle}}_{\sigma}t\, c_{\sigma n'}^\dagger c^{\phantom{\dagger}}_{\sigma n} 
-i\mathop{\sum_{\langle n,n'\rangle}}_{\sigma,\sigma'}t^{\mathrm{SO}\phantom{\dagger}}_{n'-n}c_{\sigma' n'}^\dagger \sigma^{y}_{\sigma'\sigma}c^{\phantom{\dagger}}_{\sigma n}\\
&+\sum_{\sigma,\sigma' n}B\,c_{\sigma' n}^\dagger \sigma^{x}_{\sigma'\sigma}c^{\phantom{\dagger}}_{\sigma n}+
\sum_{\sigma n}\Delta\, c_{\sigma n}^\dagger c^\dagger_{\bar\sigma n}+\mathrm{H.c}\,.
\end{split}
\end{equation} 
Notice that previous Hamiltonian is the same given by Eq.\,(\ref{SCNW}) in Chap.\,\ref{Chap2a}.
The parameters of the model are the chemical potential measured from depletion $\mu_S$, the hopping $t=\hbar^2/(2m^*a_0^2)$, where $m^*$ is the effective mass and $a_0$ 
is the lattice spacing, the induced pairing $\Delta$, the SO hopping 
$t^\mathrm{SO}_{\pm 1}=\pm \frac{1}{2}\alpha_\mathrm{SO}/a_0$, where $\alpha_\mathrm{SO}=\hbar^2/(m^*\lambda_\mathrm{SO})$ is the SO coupling and $\lambda_\mathrm{SO}$ is 
the SO length, and the Zeeman field $B=\frac{1}{2}g\mu_B \mathcal{B}_x$, where $g$ is the g-factor, and $\mathcal{B}_x$ is the magnetic field along the wire. 
In what follows we present simulations with parameters corresponding to an InSb proximitized wire \cite{Mourik:S12} ($\Delta=0.25$ meV, $\alpha_\mathrm{SO}=20$ meV nm, 
$m^*=0.015 m_e$, $g=40$). We will consider both an isolated proximitized wire of finite length, and an open NS contact between proximitized and non-proximitized sections of a 
nanowire, Fig. \ref{fig:WireSketch}. The latter is assumed infinite (see the supplemental information for finite length effects), and is modelled by the same Hamiltonian, 
albeit with $\Delta=0$ and a $\mu_N$ in place of $\mu_S$. The normal-state average transparency per mode $T_N$ of the contact is physically controlled by a electrostatic gating in 
an actual device, and is modelled here either by a hopping $t'\leq t$ across the contact, or by a spatial interpolation between $\mu_N$ and $\mu_S$ {across a certain contact length 
$L_C$ determined by the distance of the wire to the depletion gate ($\Delta$ is always abrupt, see Supplementary Material). Note that if the density of defects in the wire is small, 
$T_N\sim 1-\exp(-L_C/\lambda)$, with $\lambda$ a lengthscale of the order of the average Fermi wavelength.}
 \begin{figure}
   \centering
   \includegraphics[width=\columnwidth]{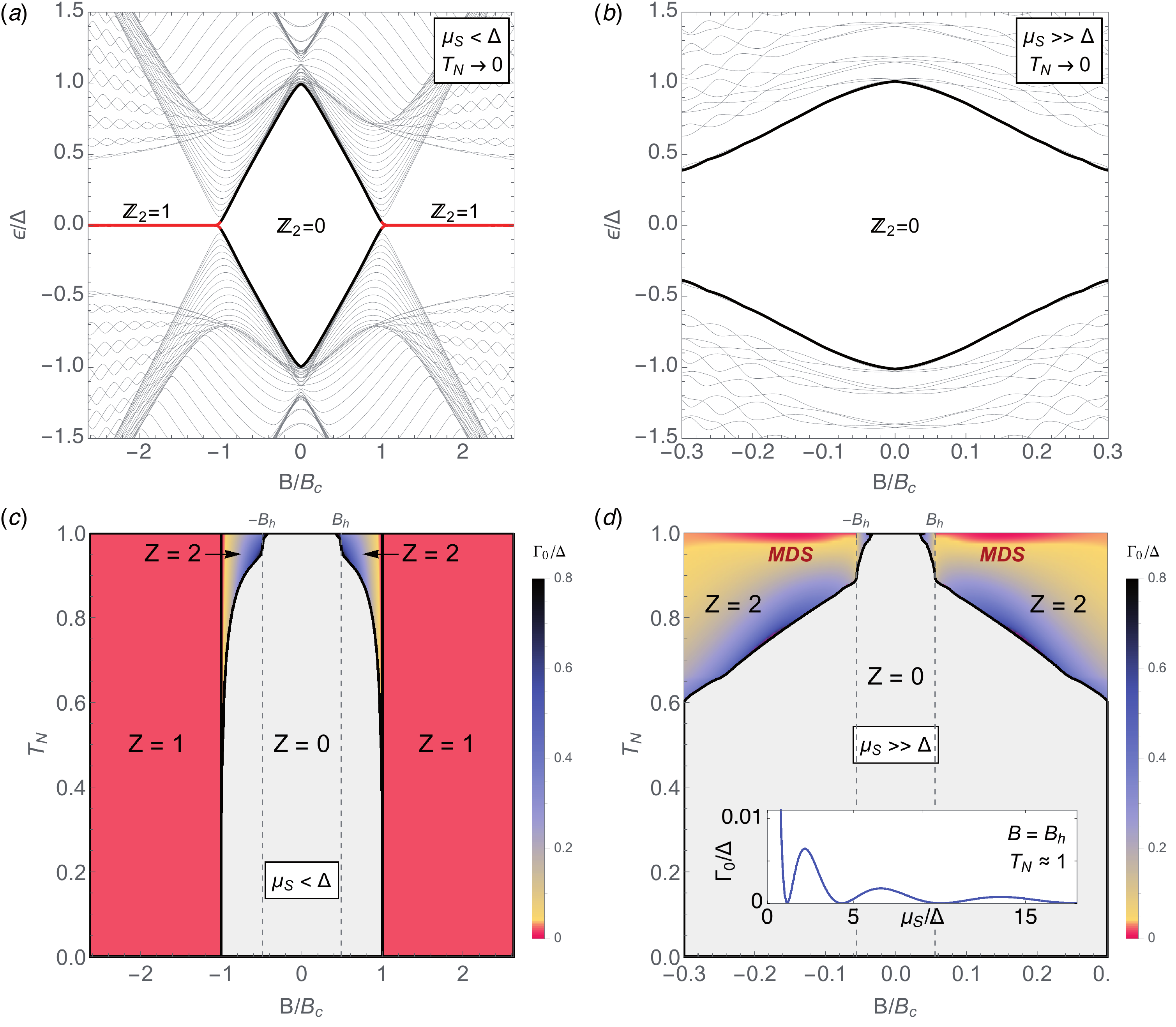}
   \caption[Exceptional points in a proximitized Rashba nanowire]{(Color online) Exceptional points in a proximitized Rashba nanowire. (a,b) The spectrum for an InSb proximitized 
wire ($\Delta=0.25$ meV, $\alpha_\mathrm{SO}=20$ meV nm, $m^*=0.015 m_e$, $g=40$) at vanishing NS transparency both for $\mu_S/\Delta\approx 0.5$ (a) and $\mu_S/\Delta\approx 10$ (b), for the same range of  Zeeman field $B$. (c,d) The corresponding phase diagram of the NS junction in the $B-T_N$ plane. Colors denote the residual decay rate $\Gamma_0$ after the exceptional point (thick black line). Panel (d) in the Andreev limit shows the formation of Majorana dark states (MDSs) in the trivial phase $B<B_c$ at high transparency $T_N$ [red regions]. Inset: the residual decay rate of the MDSs vanishes as $\mu_S$ is pushed into the Andreev limit $\mu_S\gg \Delta$. For the realistic parameters of the simulations, this residual decay rate corresponds to a lifetime of the MDSs of $\sim$0.2 microseconds.}
   \label{fig:OpenWire} 
\end{figure}

Topologically, the isolated nanowire belongs, for finite $B$, to the same one-dimensional D-class as the multimode Kitaev wire. For $|B|$ smaller than a critical $B_c=(\mu_S^2+\Delta^2)^{1/2}$, the nanowire is trivial ($\nu=Z=0$). 
As $|B|$ exceeds $B_c$, the topological invariant becomes non-trivial ($\nu=1$) through a band inversion, see Fig. \ref{fig:OpenWire}a, with the peculiarity that the two hybridized Majoranas in the trivial $|B|<B_c$ phase are not deep inside the gap, but at the band edge. As $\mu_S$ grows, $B_c$ quickly becomes unrealistically large, and the nanowire remains trivial for all reasonable fields, Fig. \ref{fig:OpenWire}b. {(A large $\mu_S$, incidentally, is the natural experimental regime, since the superconductor will typically transfer charge to the proximitized section of the wire that is difficult to deplete due to screening.)}

The nanowire contains a non-proximitized normal section, Fig. \ref{fig:WireSketch}, and we define the effective Hamiltonian as $H_{S}+\Sigma_N(\omega=0)$, with the exact self energy from the normal portion of the wire evaluated at zero frequency $\Sigma_N(\omega=0)$ (see Appendix\,\ref{AppChapEPs}). As the coupling to the reservoirs increases (the junction transparency $T_N$ grows), two eigenvalues drop out from the band edge into the lower complex plane, and merge at the imaginary axis at an exceptional point. 
This exceptional point is only reached if the Zeeman field exceeds a certain value $|B|\gtrsim B_h\equiv \mu_N$, see dashed lines in \ref{fig:OpenWire}(c,d). 
This $B_h$ is the field required for the normal nanowire to become \emph{helical}. For $|B|>B_h$, the normal nanowire hosts a single propagating mode, with the other spin sector 
completely depleted by the Zeeman field, and behaves as a single-mode reservoir like the one discussed for the Kitaev wire in \cite{JorgeEPs}. Note that this helical regime should 
be achievable using electrostatic gating, since it only requires a sufficient depletion of the non-proximitized section of the semiconductor nanowire, 
{unscreened by the superconductor}.

After crossing the exceptional point (region above thick black curves in \ref{fig:OpenWire}[c,d]), the scattering matrix $S(\omega)$ at the trivial NS contact acquires $Z=2$ purely imaginary poles. One of the two moves towards the origin. The asymptotic decay rate $\Gamma_0$ in the limit $T_N\to 1$ is not vanishing in general, so that the corresponding states should be denoted as Majorana resonances \cite{Sau:PRL12}. However, in the experimentally relevant Andreev limit $\mu_S\gg\Delta$, Fig. \ref{fig:OpenWire}d, the asymptotic $\Gamma_0$ vanishes exponentially with $\mu_S/\Delta$ (see inset). When the wire is tuned into the regime $\mu_S\gg\Delta$ and the contact is made sufficiently transparent, the Majorana resonances are stabilised into proper non-decaying MDSs. For example, for the realistic parameters of the simulations in Fig. \ref{fig:OpenWire}d the lifetime for the minimum widths (see inset) corresponds to $\sim$0.2 microseconds. While this is already a rather long time, this is not an upper bound since even longer lifetimes can in principle be obtained by increasing $\mu_S$.

The interpretation of this mechanism is as follows. While two MBSs at opposite ends of an isolated topological nanowire can be considered exact zero modes up to exponentially small corrections (in the wire length) coming from their mutual overlap, an MDSs (red in Fig. \ref{fig:WireSketch}) also becomes an exact Majorana zero mode without any fine tuning by a suppression of its overlap with its sibling Majorana (blue in Fig. \ref{fig:WireSketch}), which escapes into the helical reservoir. It is important to note that, in contrast to isolated topological wires, any residual overlap that remains after the exceptional point does not translate into a finite energy splitting, but rather into a residual decay rate $\Gamma_0$.

\section{Physical properties of Majorana dark states}

Having established the emergence of zero energy dark states at a transparent helical metal-trivial superconductor junction in the Andreev limit, we now turn to the analysis of the physical properties of said states, and compare them to conventional MBSs. We will  study their signatures in transport, their wavefunction locality, particle-hole conjugation, their charge neutrality, uniform charge oscillations and finally and the low energy spectrum properties in SNS geometries.

\subsection*{Signatures in dI/dV}
 \begin{figure*}
   \centering
   \includegraphics[width=0.99\textwidth]{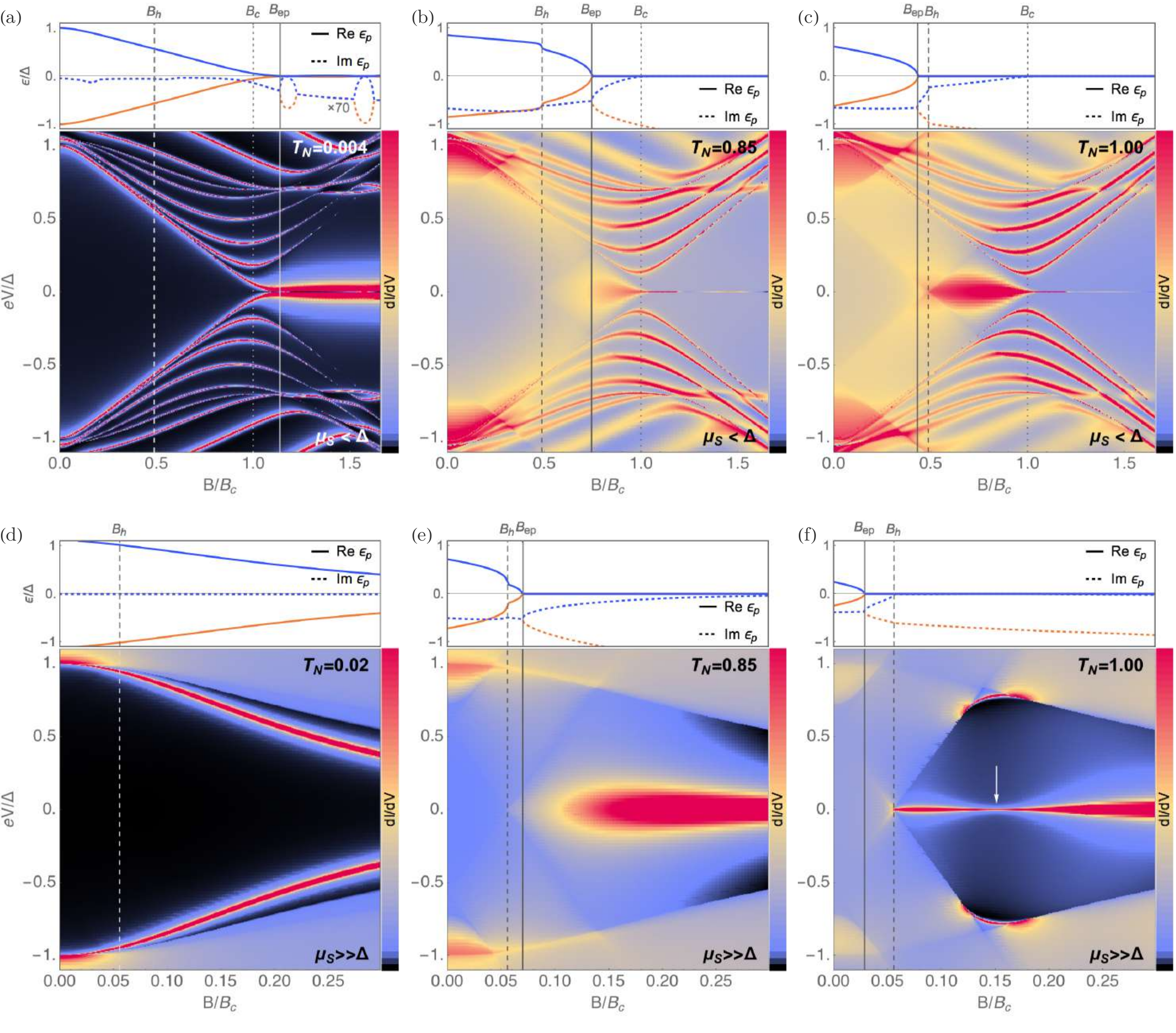}   
   \caption[Signatures in $dI/dV$]{(Color online) Signatures in $dI/dV$. (a-f) Tunnelling differential conductance $dI/dV$ through a third probe contacted at the junction (brown in Fig. \ref{fig:OpenWire}a), for different junction transparencies $T_N$ and for the same range of Zeeman fields $B$ and bias voltages $V$. First row corresponds to $\mu_S=0.5\Delta$, second row to $\mu_S=10\Delta$, as in Fig. \ref{fig:OpenWire}. The length of the proximitized wire is 1.5 $\mu$m. Evolution of lowest poles across a $B=B_\mathrm{ep}$ exceptional point is shown atop each panel {(higher poles not plotted for clarity)}. Note the sharp zero bias anomaly (ZBA) in the transparent, Andreev limit of panel (f), signalling the presence of MDSs in the topologically trivial regime $B_h<B<B_c$. (g,h) Differential conductance across the junction, for the same parameters as panels (c,f). The ZBA becomes a sharp dip on a constant $2G_0=2e^2/h$ background.}
   \label{fig:dIdV} 
\end{figure*}

We start by analysing the differential conductance $dI/dV$ through a normal tunnelling probe weakly coupled to the neighbourhood of the junction, brown in Fig. \ref{fig:WireSketch}. 
In the tunneling limit, this is proportional to the local density of states at the junction at energy $\epsilon=eV$, where $V$ is the bias voltage. 
(Note that this is different from the differential conductance across the NS contact, which is not necessarily in the tunnelling regime, see below). 
We computed the tunnelling $dI/dV$ versus $B$ and $V$ using standard quantum transport techniques\cite{Datta:97} (see Appendix\,\ref{AppChapEPs}), 
and we present the results in Fig. \ref{fig:dIdV}(a-f) for several transparencies $T_N$, both far from the Andreev limit (top row), and deep into the Andreev limit (bottom row). 
Atop each panel, the evolution of the lowest complex eigenvalues with $B$ is shown, with an exceptional point bifurcation at $B=B_\mathrm{ep}$. 
 Panel (a) corresponds to the tunnelling limit $T_N\ll 1$, which exhibits a sharp peak at zero-energy due to conventional MBSs for $B\gtrsim B_c$, and slightly split in two due 
 to their hybridization across $L_S$. As $T_N$ is increased, panels (b,c), the $B\gtrsim B_c$ zero-energy peak in the LDOS is washed away, since the Majorana at the contact becomes 
 a finite-lifetime Majorana resonance, increasingly delocalized into the reservoir. Interestingly, however, as $T_N$ is increased the onset of the Majorana resonance is shifted to fields $B<B_c$, for which the band topology of the superconducting nanowire is trivial. The Majorana resonance emerges upon crossing an exceptional point at a certain $B_{ep}$ (see bifurcation in Fig. \ref{fig:dIdV}(b,c) [top plots]. As $T_N$ is increased from 0 to 1, $B_{ep}$ decreases from $B_{ep}\approx  B_c$ (solid guidelines) to $B\approx B_h$ (dashed). $B_h\equiv \mu_N$ is the field for which the N side of the junction becomes helical (spin becomes locked to the value of momentum), as is characterised by the loss of one propagating channel.

Two important conclusions can be drawn from these results. On the one hand, we see that in the particular $T_N\to 0$ regime, $L_S\to\infty$ limit, the topological transition and the exceptional point coincide $B_{ep}=B_c$.\cite{Pikulin:PRB13} For wires sufficiently open to reservoirs, however, the exceptional point and its associated Majorana resonance will occur at lower fields (around $B_h$) than the topological crossover, or pole burying (at $B_c$), and might therefore be easier to reach experimentally. 
Secondly, the difference between a Majorana resonance at $B<B_c$ and a conventional $B>B_c$ MBSs in \emph{finite} wires seems to be merely qualitative, i.e. differing simply in their lifetime, since both correspond to the same pole of the scattering matrix at different positions in the complex plane. 
Conventional MBSs above $B_c$ have a finite (real) energy splitting due to the finite length, while Majorana resonances below $B_c$ have a finite width $\Gamma$. However, while the former can be decreased by increasing the length of the proximitised wire (making the MBSs effectively eigenstates), the inverse lifetime $\Gamma$ of Majorana resonances is always sizeable in the present regime ($\mu_S\lesssim\Delta$); it decreases with Zeeman field, and exhibits a kink for high $T_N$ at the helical transition $B=B_h$ of the normal side (see top  panel of Fig.\,\ref{fig:dIdV}c, where the imaginary part of the poles develops a kink at $B_{h}$), but it remains relatively large all the way till $B_c$. Next, we will show that in the experimentally relevant regime $\mu_S\gg\Delta$ this is no longer the case. The kink at $B_h$ evolves, at high $T_N$, into an S-matrix topological crossover far below $B_c$, which gives rise to the formation of a new type of exceptional MBSs localised at the junction, with $\Gamma\rightarrow 0$.

So far, we have shown that Majorana resonances for $B>B_{ep}$ evolve continuously into conventional MBSs as $B$ crosses $B_c$. A natural question arises as to why a zero-energy Majorana resonance may appear in a topologically trivial wire in the first place. This may be intuitively understood by considering that a topologically trivial $B<B_c$ isolated nanowire actually hosts a \emph{pair} of  MBSs pairs at each end (one for each of two opposite p-wave sectors in the trivial wire \cite{Alicea:RPP12}), that are strongly hybridized away from zero energy to an energy close to $\Delta$, due to their large overlap. However, when one end is fully opened into a helical wire with a single propagating mode ($B>B_h>B_{ep}$), one (and only one) of the two MBSs escapes into the reservoir, and decouples from the other MBS, which remains well localized at the contact and returns to zero energy. This mechanism does not apply for $B<B_h$ since in this case both MBSs may immediately escape into the reservoir. 
The origin of the finite lifetime of a Majorana resonance is connected to the probability for one of the two hybridized Majoranas to completely escape into the helical wire and out into the reservoir. The higher this probability, the longer-lived the remaining Majorana will be, since its overlap with its delocalised sibling will be suppressed. {For perfect escape probability, the Majorana becomes a zero-energy, non-decaying dark state at the junction}. One might guess that this escape probability should be $T_N$ itself,  but we showed in Fig. \ref{fig:dIdV}(a,b,c) that a $T_N=1$ contact with $\mu_S\lesssim \Delta$ has a Majorana resonance of finite lifetime. The reason is that a perfect \emph{normal} transparency $T_N=1$ only implies escape probability one across the NS junction in the limit in which the induced pairing $\Delta$ is a small perturbation respect to the normal phase, i.e. in the Andreev limit $\mu_S\gg \Delta$. This is, incidentally, the realistic regime of experimental samples, since electrostatically depleting a proximitised wire to have $\mu_S\lesssim \Delta$ is much more difficult than depleting an exposed section, whose $\mu_N$ can typically be tuned all the way to zero by a gate.

To demonstrate this scenario, we show in Fig. \ref{fig:dIdV}(d,e,f) the LDOS and pole evolution of the NS junction in the Andreev limit, $\mu_S=10\Delta$. This moves the topological crossover $B_c$ to higher fields ($B_c=2.2$ T, up from $0.25$ T), while the helical field $B_h=\mu_N$ (dashed guideline) remains the same. The range of physical $B$ fields is the same as in the top row of Fig. \ref{fig:dIdV}, but now $B_c$ falls outside of this range. 
As $T_N$ is increased (by making the spatial transition between $\mu_N$ and $\mu_S$ at the contact smooth \cite{PhysRevB.90.235415}), the width of the Majorana resonance becomes greatly reduced, {i.e. the resonance becomes gradually decoupled from the reservoir}. Note the similarity between the sharp zero energy peak in the LDOS emerging in Fig. \ref{fig:dIdV}f for $B>B_h$ and the one in Fig. \ref{fig:dIdV}a for $B>B_c$. The width of the latter, a conventional MBSs associated to a topologically non-trivial superconducting bulk, vanishes as $T_N\rightarrow 0$. The width of the former, in contrast, vanishes as $T_N\rightarrow 1$ and $\mu_S/\Delta$ is increased. This is demonstrated in the inset of Fig. \ref{fig:OpenWire}d, which shows $\Gamma_0=-\mathrm{Im}\,\epsilon_p$ at the helical transition $B_h$ for $T_N\approx 1$ as a function of $\mu_S/\Delta$. We see that $\Gamma_0$ is quickly suppressed as soon as we approach the Andreev limit $\mu_S>\Delta$, and becomes arbitrarily small as $\mu_S$ grows. This supports the intuitive picture of the preceding paragraphs. 

Panel (f), with $T_N\to 1$, corresponds to a cut along the top of Fig. \ref{fig:OpenWire}(d), for which MDSs are fully developed. Their presence gives rise to sharp zero bias anomaly (ZBA) in transport at fields $B>B_h$, with a sharpness that increases exponentially with $\mu_S/\Delta$. This type of ZBA was the first signature of MBSs explored experimentally \cite{Mourik:S12}, though in the present context they arise far from the topological regime $B\ll B_c$. The ZBA is not preceded by signatures of a gap closing. Note also that away from the ideal conditions $T_N\to 1$, $\mu_S\gg \Delta$, wide Majorana resonances are also visible in the topologically trivial regime, panels (b,c,e), albeit of finite lifetime.
 Exceptional MBSs are therefore zero-energy dark state that arise at a sufficiently transparent junction between a conventional superconductor and a metal with a single propagating helical channel. By the S-matrix definition, such junction becomes effectively non-trivial topologically, while by the band topology definition, both normal and superconducting bulks remain trivial. In spite of the system being arbitrarily far from a band topological transition, the more general S-matrix point of view shows that the zero energy state associated to the buried pole is a genuine MBS. In contrast to a $B>B_c$ MBS, the exceptional MBS is located \emph{at} the junction (red sphere in Fig. \ref{fig:WireSketch}), and its residual energy scale can be tuned all the way to zero by increasing $T_N$ instead of $L_S$. In the following we will demonstrate its Majorana character, and analyse its associated phenomenology.

\subsection*{Spatial localization and Majorana character}
\begin{figure}  
   \centering
   \includegraphics[width=0.99\columnwidth]{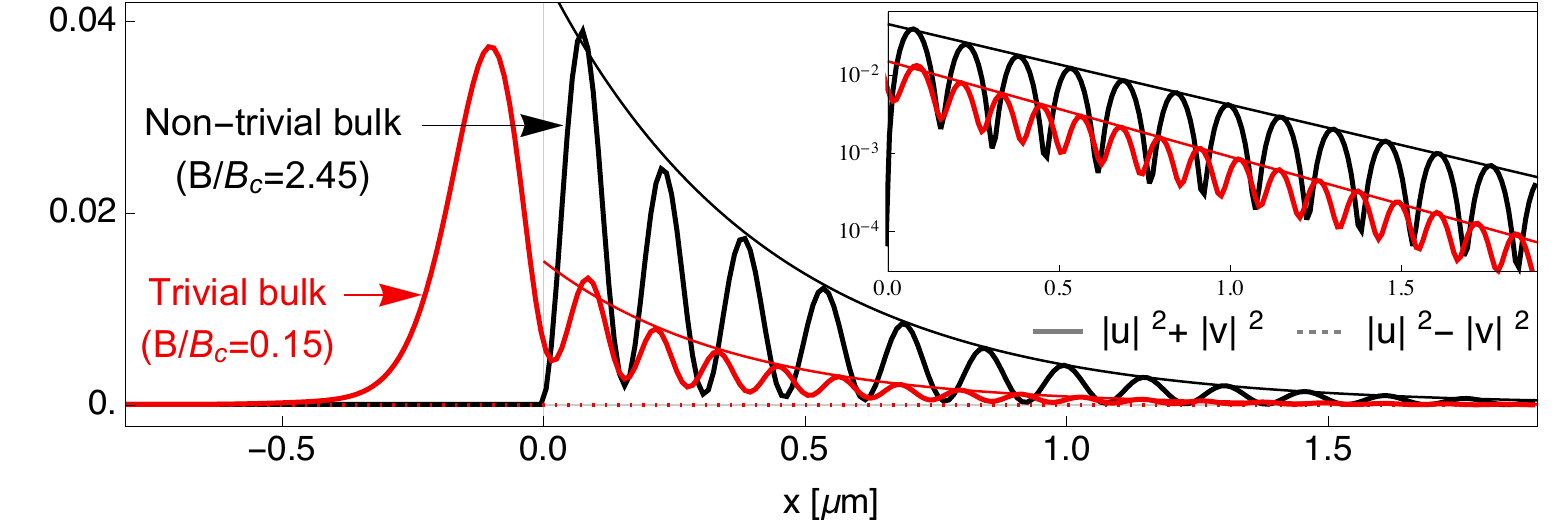}
   \caption[Spatial localization, Majorana character]{(Color online) Spatial localization, Majorana character. Spatial quasiparticle density $|u|^2+|v|^2$ for Majorana bound states in an NS junction, located at $x=0$. The solid red curve corresponds to an MDS between trivial bulks ($T_N\approx 1$, $B_h<B<B_c$, see white arrow in Fig. \ref{fig:dIdV}f), while the solid black curve corresponds to a conventional MBS in a closed topological wire ($T_N=0$, $B>B_c$). Both decay exponentially $\sim e^{-x/\xi}$ with a Majorana localization length $\xi=\hbar v_F/\Delta(B)$ \cite{Klinovaja:PRB12}  (see inset). Dotted lines are the corresponding charge densities  $|u|^2-|v|^2$, zero everywhere, revealing the Majorana character of both states. }
   \label{fig:SpatLoc}  
\end{figure}

\begin{figure}  
   \centering
   \includegraphics[width=0.99\columnwidth]{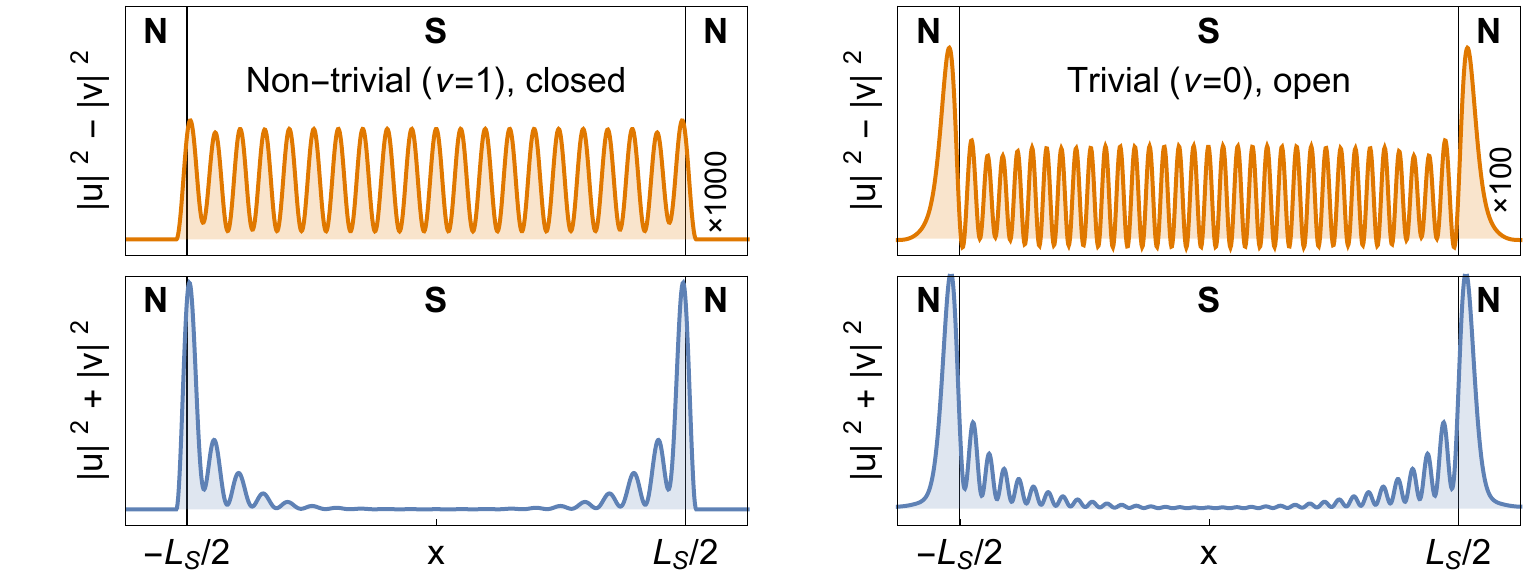}
   \caption[Spatial localization, Majorana character]{(Color online) Spatial localization, Majorana character. Same as in Fig.\,\ref{fig:SpatLoc}, albeit in a NSN geometry with finite $L_S$, hosting two overlapping Majoranas. The charge densities $|u|^2-|v|^2$ exhibit spatially uniform oscillations due to the overlap.}
   \label{fig:Neutral}  
\end{figure}

The spatial locality and Majorana self-conjugation $\gamma=\gamma^\dagger$ are assessed next, by analysing the wavefunction of the MDSs. Figure \ref{fig:SpatLoc} shows, in red, the quasiparticle density $|\psi(x)|^2=|u|^2+|v|^2$ (solid lines) and charge density $|\rho(x)|^2=|u|^2-|v|^2$ (dashed lines) of the MDS marked by the white arrow in Fig \ref{fig:dIdV}f ($u$ and $v$ are particle and hole components of its wavefunction, respectively). As discussed above, the MDS represents a non-decaying state at zero energy. The figure shows that it is furthermore well localised at the junction, decaying exponentially as $\sim e^{-x/\xi}$ with a Majorana localization length $\xi=\hbar v_F/\Delta(B)$ \cite{Klinovaja:PRB12} into the superconductor (see envelopes and inset in \ref{fig:SpatLoc}).  For comparison we also show in black the spatial probability $|\psi(x)|^2$ of a conventional $B>B_c$ MBS for a topological bulk at {zero transparency (isolated topological wire)}. For both states, the charge density $\rho(x)$ is zero, as implied by the Majorana relation $\gamma=\gamma^\dagger$. 
We now examine the charge density patterns that arise from the weak overlap of two MDSs. It was shown \cite{Lin:PRB12,Ben-Shach:PRB15} that the charge density $\rho(x)=|u|^2-|v|^2$, which is zero everywhere for an isolated MBS (see Fig. \ref{fig:SpatLoc}), develops a spatial oscillatory pattern that is uniform throughout space when two MBSs approach each other in a 1D superconductor, irrespective of their particular positions. This is a very specific and non-trivial signature of MBSs that probes the state wavefunction itself, and was proposed as a way to detect MBSs through charge sensing. A transparent NSN junction (with N portions coupled to reservoirs) provides a convenient geometry to study this effect in the topologically trivial regime. Two localized MDSs appear for $|B|>B_h$ at the two ends of the S section. They weakly overlap, and should thus be expected to exhibit uniform spatial charge oscillation throughout the superconductor in analogy to conventional MBSs. Figs. \ref{fig:Neutral} compare the charge density $\rho(x)$ for $B>B_c$ MBSs in the tunnelling limit and $B_h<B<B_c$ in the transparent limit. Once more, we see the strong similarity between the two cases, which points to an essential equivalence between the two types of states.
\subsection{Low-energy phase dependence}
\begin{figure} 
   \centering
   \includegraphics[width=0.99\columnwidth]{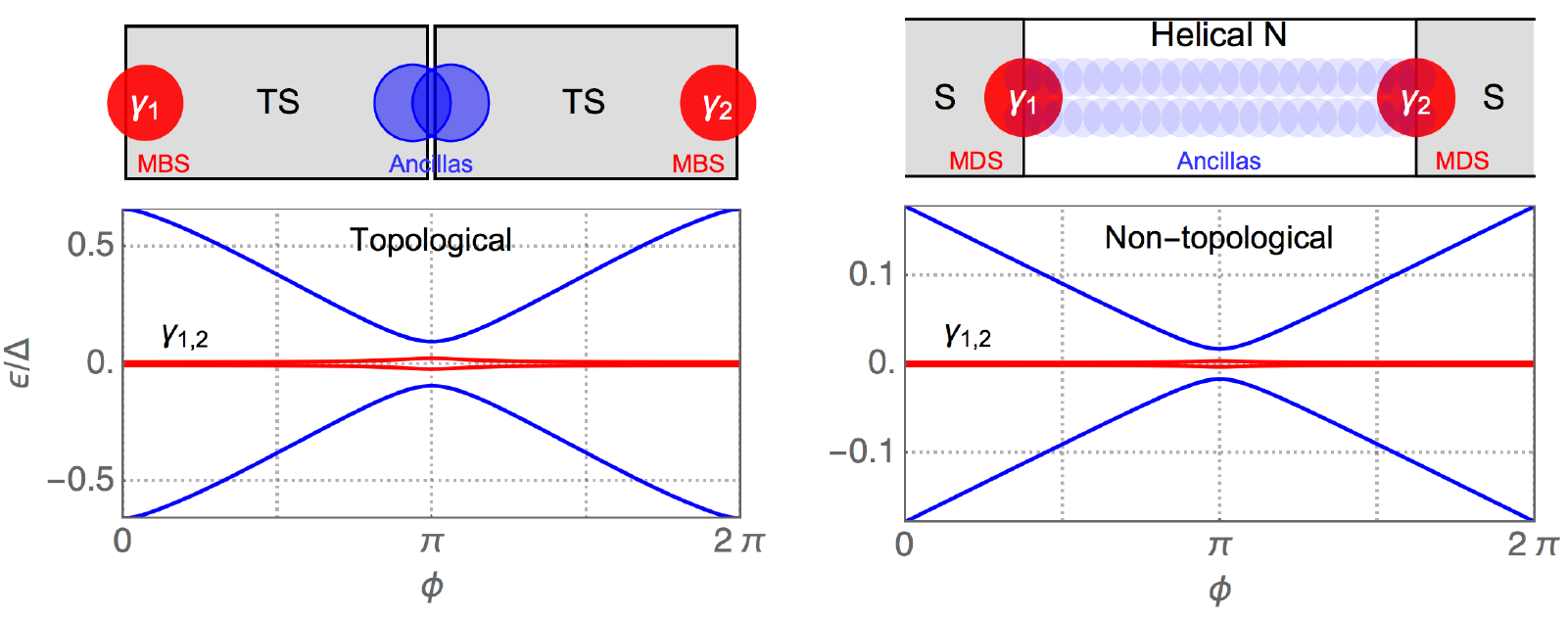}
   \caption[Low-energy phase dependence]{(Color online) Low-energy phase dependence. Topological (a) and non-topological (b) four-Majorana Josephson junction (zero energy $\gamma_{1,2}$ Majoranas in red, ancilla Majoranas in blue), and the corresponding low energy Andreev spectrum as a function of superconducting phase difference $\phi$.}
   \label{fig:phasedep}  
\end{figure}

In Fig.\,\ref{fig:phasedep} we present the low-energy phase dependence in a topological (left) and non-topological (right) four-Majorana Josephson junction.
MBSs at the outer (non-contacted) ends of the wire will form a zero energy fermion for long enough wires (in red), while the inner MBSs (the ancilla pair, in blue) will form a fermion with $\phi$-dependent energy. 
The ancilla Majoranas in this situation correspond to the two Majoranas delocalised over the helical region (blue in Fig. \ref{fig:phasedep}, while the outer Majoranas are the MDSs localised at each contact (red). Regardless of their different positions in space, the low energy Andreev spectrum in this system is essentially the same as for a four-MBS topological Josephson junction \cite{PhysRevB.91.024514} (compare left and right panels in Fig. \ref{fig:phasedep}). 

\section{Conclusions} 
We have presented a novel approach to engineer Majorana bound states in non-topological superconducting wires. 
{Instead of inducing a topological transition in a proximitized Rashba wire, we consider a sufficiently transparent normal-superconductor junction created on a Rashba wire, with a topologically trivial superconducting side and a helical normal side. The strong coupling to the helical environment forces a single long-lived resonance to emerge from an exceptional point at precisely zero energy above a threshold transparency. This resonance evolves as the transparency is increased further into a stable dark state localised at the junction.} 
We have demonstrated this phenomenon in a realistic model for a proximitized semiconducting nanowire. 

The zero-energy state emerges as the junction traverses an exceptional point at the threshold transparency, and becomes robustly pinned to zero energy without fine tuning by virtue of charge-conjugation symmetry. Moreover, its residual decay rate at perfect transparency is exponentially suppressed in the experimentally relevant Andreev limit.
Finally, we have shown that relevant transport and spectral properties associated to these zero energy states, here dubbed {Majorana dark states}, are indistinguishable from those of conventional Majorana bound states. 

{Thus, our proposal offers a new promising strategy towards generating and detecting Majorana bound states in the lab, with potential advantages over more conventional approaches in situations where manipulating the metallic environment and contact properties proves to be simpler than engineering a topological superconducting transition.
Most importantly, the condition for reaching a helical phase in the normal side, while the proximitized region of the nanowire remains in the large $\mu_S$ Andreev limit, is expected to be accessible experimentally. All the necessary ingredients for our proposal are already available in the lab: dramatic advances in fabrication of thin semiconducting nanowires, proximitized with conventional s-wave superconductors, were recently reported \cite{chang15,Krogstrup15}. Highly transparent, single-channel, NS contacts and a high-quality proximity effect in quantitative agreement with theory were demonstrated. Reaching the helical regime in such high-quality and fully tunable devices should be within reach, so we expect that our proposal for MDSs will be soon tested.}

The general connection demonstrated here between Majorana states and the bifurcation of zero energy complex eigenvalues at exceptional points in open systems offers a new perspective on the mechanisms that may give rise to Majorana states in condensed matter systems. {In this sense it extends conventional strategies based on topological superconductors. It moreover expands on the extensive studies on exceptional point physics in optics, where it has been shown that state coalescence has far-reaching physical consequences, such as e.g. non-Abelian geometric phases \cite{Heiss:EPJD99,Dembowski:PRL01,Kim:FDP13}. To date, most studies of this kind have been concerned with open photonic systems under parity-time (PT) symmetry \cite{Bender:PRL98,Klaiman:PRL08,Regensburger:N12} and with its spontaneous breakdown through exceptional point bifurcations. This leads to intriguing physical phenomena, such as unidirectional transmission or reflection \cite{Regensburger:N12}, loss-induced transparency \cite{Guo:PRL09}, lasers with reversed pump dependence and other exotic properties \cite{Liertzer:PRL12,Brandstetter:NC14,Peng:S14}. Such striking optical phenomena are seemingly unrelated to the physics described in this work, but interesting connections are being made. These include open photonic systems with charge-conjugation symmetry \cite{Malzard:15} (as opposed to PT symmetry), and which show spectral transitions analogous to the zero mode bifurcation discussed here. Also, radiation-induced non-hermicity has been demonstrated as a way to convert Dirac cones into exceptional points \cite{Zhen:N15}, a phenomenon completely analogous to the conversion of zero energy crossings of Bogoliubov-de Gennes excitations in Josephson junctions into double exceptional point structures. Further research should extend the understanding of these interesting connections.  More importantly, our study and others \cite{Malzard:15,Zhen:N15} raise relevant questions, such as whether there is a deeper connection between topological transitions in closed systems and spectral bifurcations in non-hermitian systems. Answering such questions would help in advancing our understanding of the meaning of non-trivial topology in open systems.}


\chapter{\bf Density response function in Rashba nanowires: a linear response approach\footnote{The results of this chapter are being prepared for publication.}} 
\label{ChapDensity}
\lhead{Chapter \ref{ChapDensity}. \emph{Density response function in Rashba nanowires: a linear response approach}} 

\begin{small}
In this Chapter we investigate the density-density response function in nanowires with Rashba spin-orbit coupling, where a Zeeman field is applied perpendicular to the spin-orbit axis. Then, we extend the analysis and consider induced $s$-superconductivity into the nanowire, which is solve by taking zero interband pairing and then the strong Zeeman field limit. This analysis is carried out based on the linear response theory. 
 Then, we use these results in order to investigate electron-electron interactions in the longitudinal direction of the wire within the Random Phase Approximation approach (RPA). In this part, we have calculated the dielectric function, and then the charge density-density response as well as the screened potential which allows us to describe screening effects in such nanowires. 

\end{small}

\newpage

\section{Introduction}
Nanowires with Rashba spin-orbit coupling (SOC) are a solid platform for investigating Majorana bound states (MBSs). 
In fact, it was shown that when a Zeeman field $B$ is applied, perpendicular to the spin-orbit axis, and $s$-wave superconductivity is induced, the nanowire becomes a topological superconducting nanowire for fields exceeding $B_{c}=\sqrt{\Delta^{2}+\mu^{2}}$, which is defined in terms of the induced superconducting pairing $\Delta$ and the wire's chemical potential $\mu$. See Chapter \ref{Chapter01} for details.
A significative number of measurements on electronic systems can be investigated by applying a small external field, as a probe of a certain type, to the system at an initial time and then focus on how the system responds.
When the field is small enough, the response is proportional (a linear function) to the external perturbation, and the coefficient of proportionality is the correlation function of the system in the equilibrium ensemble without the perturbation. Therefore, the description of the response is based on linear response theory \cite{vignale}. Within this approach, the correlations functions are also known as linear response functions. The change in the density of a system due to a local change in its density (the response of the expectation value of the number density operator) is an example of such linear response functions and it is described by the density-density response function (also known as Lindhard function) \cite{vignale}.
The density-density response function provides crucial information upon understanding static screening, and, to the best of our knowledge, in nanowires with SOC and Zeeman interaction it was not explored so far. The chemical potential in the nanowire model for MBSs plays an important role, and in real experiments it is tuned by means of electrostatic gates. Electron-electron interactions can affect the real value of $\mu$ and and therefore changing the topological transition point. 
A first step for investigating electron-electron interactions in such nanowires can be described within the Random Phase Approximation (RPA) and follows from the knowledge of the density-density response function \cite{vignale}. Indeed, in a non-interacting electron gas, the (linear response) RPA approach is the approximation in which the proper response function is replaced by the density-density response function (also known as the Lindhard function) \cite{vignale}. 
We are interested in the electron-electron interactions along the longitudinal direction of the wire as we believe it is the most realistic situation.
The calculation of the Lindhard function is trivial for free electrons \cite{vignale}, however, it becomes complex when the system involves SOC, Zeeman fields or superconductivity. 
In this Chapter we firstly provide a brief introduction and then calculate the density-density response function in one-dimensional nanowires with SOC and Zeeman interaction. Afterwards, we consider induced $s$-wave superconductivity into Rashba nanowires, where our main interest is to calculate the density-density response function in the regime of strong Zeeman field. 
The results of the density-density response function allow us to investigate the dielectric function, the charge density and screened potential within the RPA approximation.
 \newpage
\subsection{Model}
We consider a one-dimensional one mode wire with Rashba SOC and Zeeman field, perpendicular to the spin-orbit axis, modelled by the following Hamiltonian
\begin{equation}
\mathcal{H}_{0}=\mathcal{H}_{kin}+\mathcal{H}_{so}+\mathcal{H}_{Z}\,,
\end{equation}
where the first, second and third terms are the kinetic, Rashba and Zeeman Hamiltonians
\begin{equation}
\begin{split}
\mathcal{H}_{kin}&=\sum_{\sigma}\int\,dx\,\psi_{\sigma}^{\dagger}(x)\left[ -\frac{\hbar^{2}\partial_{x}^{2}}{2m}-\mu\right]\psi_{\sigma}(x)\,,\\
\mathcal{H}_{SO}&=\sum_{\sigma \sigma^{\prime}}\int dx\,\psi_{\sigma}^{\dagger}(x)\left[ \boldsymbol{\alpha}\cdot\boldsymbol{\sigma}\right]_{\sigma\sigma^{\prime}}\,\left[ -i\hbar\partial_{x}\right]\psi_{\sigma^{\prime}}(x)\,,\\
\mathcal{H}_{Z}&=B\sum_{\sigma\sigma^{\prime}}\int dx\,\psi^{\dagger}_{\sigma}(x)\left[\sigma_{x}\right]_{\sigma\sigma^{\prime}}\psi_{\sigma^{\prime}}(x)\,.
\end{split}
\end{equation}
The spin direction is such that $\boldsymbol{\alpha}\cdot\boldsymbol{\sigma}=-\alpha_{R}\sigma_{y}/\hbar$.
$\sigma=\uparrow,\downarrow$ denotes the spin direction along the $y$-axis. 
The Hamiltonian $\mathcal{H}_{0}$ in the basis $\psi(x)=(\psi_{\uparrow}(x),\psi_{\downarrow}(x))^{T}$, $T$ being the transpose operation, can be written as
\begin{equation}
\mathcal{H}_{0}=\int dx\,
\psi^{\dagger}(x)\,H_{0}
\psi(x)\,
\end{equation}
where the Hamiltonian density, $H_{0}$, reads
\begin{equation}
H_{0}=\frac{p^{2}}{2m}\,-\,\frac{\alpha_{R}}{\hbar}\sigma_{y}p+B\sigma_{x}\,,
\end{equation}
where $p=-i\hbar\partial_{x}$ is the momentum operator, $m$ the electron's effective mass in the nanowire, $\alpha_{R}$ the spin-orbit coupling strength, and $B$ the Zeeman splitting due to the magnetic field $\mathcal{B}$.

\begin{figure} 
   \centering
   \includegraphics[width=0.8\columnwidth]{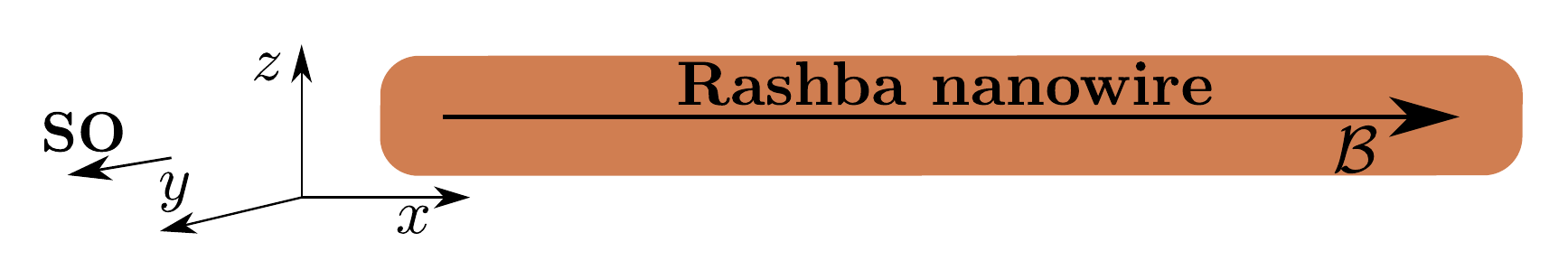}
   \caption[Sketch of a nanowire with Rashba spin-orbit coupling]{Sketch of a nanowire with Rashba spin-orbit coupling, where a Zeeman field $\mathcal{B}$ is applied along the wire and perpendicular to the spin-orbit axis.}
   \label{fig:wire}  
\end{figure}
The eigenvalues and eigenvectors of the previous Hamiltonian are given by,
\begin{equation}
\label{eq2}
\varepsilon_{k\sigma}=\xi_{k}+\sigma\sqrt{B^{2}+(\alpha_{R}k)^{2}}\,,\quad 
\psi_{k\sigma}(r)= \eta_{k\sigma}\phi_{k}(r)\,,
\end{equation}
where $\phi_{k}(r)=\frac{1}{\sqrt{L}}{\rm e}^{ikr}$ and $\eta_{k\sigma}=\frac{1}{\sqrt{2}}
\begin{pmatrix}
 \sigma\gamma_{k}\\
1
 \end{pmatrix}\,$, 
 $\xi_{k}=\hbar^{2} k^{2}/2m-\mu$ is the dispersion relation for free electrons, $\sigma=\pm$ labels the two bands, $\mu$ the chemical potential measured from the bottom of the band, $\gamma_{k}=\frac{v_{k}}{|v_{k}|}$, with $v_{k}=B+i\alpha_{R}k$,  and $L$ is the length of the wire. Notice that we have renamed the functions $\phi_{k,\sigma}$ in Eq.\,(\ref{H0vEnVec}) by $\eta_{k,\sigma}$ in Eq.\,(\ref{eq2}).

\subsection{General concepts}
\label{generalconcepts}
When a system is perturbed by a external potential $\phi_{ext}$, its density deviates from the equilibrium value, and the change induces an additional potential $\phi_{ind}$. The total potential in the system is the sum of the external and the induced potentials
\begin{equation}
\label{phitotal}
\phi_{scr}=\phi_{ind}+\phi_{ext}\,,
\end{equation}
and it is also known as the screened potential.
If the perturbation is weak, there is a linear relationship between the induced density and the external potential 
\begin{equation}
\label{ninduced}
n_{ind}(q,\omega)=\chi(q,\omega)\phi_{ext}(q,\omega)\,,
\end{equation}
where $\chi$ is the density-density response function. By using Poisson and Eq.\,(\ref{phitotal}) and $\phi_{ext}=V(q)$, the dielectric function is given by
 \begin{equation}
 \label{funcdielectric}
\epsilon(q,\omega)=\frac{1}{1+V(q)\,\chi(q,\omega)}\,.
 \end{equation}
 In general, previous equation is the exact structure of the dielectric response where we need to know $\chi(q,\omega)$. 
 
 The screened effective potential is different from the bare external potential, and therefore it is useful to define a proper density-density response function, which gives the response of the induced density to the screened potential $n_{ind}(q,\omega)=\tilde{\chi}(q,\omega)\phi_{scr}(q,\omega)$.
 Therefore, the dielectric function and the density response become,
 \begin{equation}
 \epsilon(q,\omega)=1-V(q)\,\tilde{\chi}(q,\omega)\,,
 \end{equation}
 and
 \begin{equation}
 \chi(q,\omega)=\frac{\tilde{\chi}(q,\omega)}{ \epsilon(q,\omega)}
 \end{equation}
 
\begin{figure} 
   \centering
   \includegraphics[width=0.9\columnwidth]{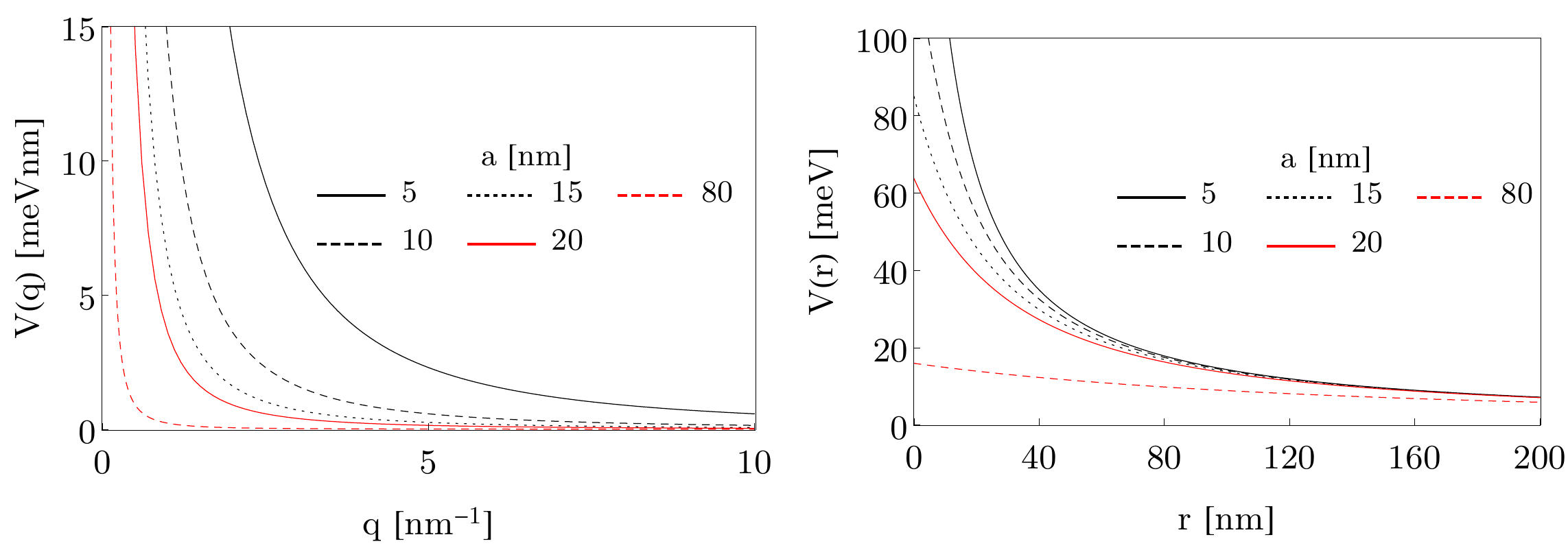}
   \caption[Coulomb potential in 1D]{(Color online) Coulomb potential in an infinite cylindrical wire of radius $a$.}
   \label{vqr}  
\end{figure}

  The Randon Phase Approximation replaces the induced charge without interactions with the screened charge. Thus, in a non-interaction electron gas, the (linear response) RPA approach is the approximation in which the proper response function is replaced by the Lindhard function\cite{vignale}. The Lindhard function is the density-density response function defined in Eq.\,(\ref{densityresponsefunction}).
 \begin{equation}
 \tilde{\chi}(q,\omega)=\chi_{nn}^{}(q,\omega)\,.
 \end{equation}
 So that the RPA dielectric function is given by
  \begin{equation}
  \label{dielectricEPRPA}
 \epsilon^{RPA}(q,\omega)=1-V(q)\,\chi_{nn}^{}(q,\omega)\,,
 \end{equation}
 and the RPA charge density by,
  \begin{equation}
 \label{chiRPA}
 \chi^{RPA}(q,\omega)=\frac{\chi_{nn}^{}(q,\omega)}{1-V(q)\chi_{nn}^{}(q,\omega)}\,.
 \end{equation}
 Notice that $V(q)$ represents the Coulomb interaction which depends on the dimensionality of the problem. In our case, it corresponds to a one-dimensional wire. To derive such potential, one examine the representation of the Yukawa interaction in an infinite cylindrical wire of radius $a$, where the transverse wave-function is assumed to be gaussian. Thus, arriving at \cite{vignale},
\begin{equation}
\label{pot1d}
V(q)=-\frac{e^{2}}{4\pi\epsilon_{0}}\,{\rm e}^{q^{2}a^{2}}{\rm Ei}[-q^{2}a^{2}]\,,
\end{equation}
where ${\rm Ei}$ is the exponential-integral function
\begin{equation}
{\rm Ei}(x)=-\int_{-x}^{\infty}\frac{{\rm e}^{-u}}{u}du\,,
\end{equation}
being $a$ the radius of the wire and $\epsilon_{0}$ the vacuum permittivity. Notice that ${\rm Ei}(x)$ diverges logarithmically for $x\rightarrow0$, imposing that the use of finite $a$ in one-dimensional wires is both a physically realistic characteristic and a mathematical requirement \cite{vignale}. See for instance right panel in Fig.\,\ref{vqr}.
The Fourier transform of the Coulomb potential is given by, see left panel in Fig.\,\ref{vqr}, 
\begin{equation}
\label{pot1d2}
V(r)=\int dq \, V(q)\,{\rm e}^{iqr}\,.
\end{equation}

The dielectric function is a property of the material being investigated, and usually assumed constant.
We will see that, in general, it depends on various parameters of the system and considering it as a constant implies somehow a naive approximation.

The screened potential for a point charge with $V(q)$, given by Eq.\,(\ref{pot1d}), can be calculated as follows\cite{vignale}
\begin{equation}
\label{phirpa1}
\phi^{RPA}_{scr}(q,\omega)=\frac{V(q)}{\epsilon^{RPA}(q,\omega)}\,,
\end{equation}
or its Fourier transform as, in the static limit $\omega=0$,
\begin{equation}
\label{phirpa2}
\phi^{RPA}_{scr}(r-r')=\int \frac{dq}{2\pi}\,\phi_{scr}^{RPA}(q)\,{\rm e}^{iq(r-r')}=\int \frac{dq}{2\pi}\,\frac{V(q)}{\epsilon^{RPA}(q)}\,{\rm e}^{iq(r-r')}\,,
\end{equation}
where the integral includes the whole $q$-space, $q\in[-\infty,\infty]$.
Previous equation represents the screened potential of the point charge potential $V(q)$ within the RPA approximation, since the dielectric function $\epsilon^{RPA}(q)$ is calculated in such approximation. 
Likewise, the induced density from Eq.\,(\ref{ninduced}) can be written in the RPA limit as
\begin{equation}
\label{ninduced2}
n_{ind}^{RPA}(q)=\chi(q)\phi^{RPA}_{scr}(q)\,,
\end{equation}
or its Fourier transform
\begin{equation}
\label{ninduced3}
n_{ind}^{RPA}(r-r')=\int \frac{dq}{2\pi}\,n_{ind}^{RPA}(q)\,{\rm e}^{iq(r-r')}\,.
\end{equation}

We have seen in this subsection that, within the Random Phase Approximation (RPA), the Lindhard function (density response function) $\chi_{nn}$ provides a simple but powerful tool in order to investigate the dielectric function, charge density and screened potential. This, of course, represents a first step in order to understand electron-electron interactions in Rashba nanowires.

\subsubsection{Finite length considerations}
In order to take into account the length of the wire, one can define a new potential $\bar{\phi}_{scr}^{RPA}(r)$, 
\begin{equation}
\begin{split}
\bar{\phi}_{scr}^{RPA}(r)&=\int_{-\infty}^{\infty}\,d\,r'\,n(r,r',B,\mu,\alpha_{R})\phi_{scr}^{RPA}(r-r')\,,
\end{split}
\end{equation}
where we have introduced $n(r,r',B,\mu,\alpha_{R})$ to count for the one dimensional electron density which in general depends on $r,r'$ and on the parameters of the system under investigation. In principle, the electron density should be calculated self-consistently using the Poisson and BdGs equations, however, for simplicity we assume that the electron density remains roughly constant within $[0,L]$, $n(r,r',B,\mu,\alpha_{R})\approx n_{0}\equiv1/L$, where $L$ represents the wire's length. Thus, the integral over $r'$ in previous equation is finite for $r'\in[0,L]$, then
for the screened potential, in the wire of length $L$, we get
\begin{equation}
\label{phirftxx2}
\begin{split}
\bar{\phi}_{scr}^{RPA}(r)
&=\frac{n_{0}}{2\pi}\int_{-\infty}^{\infty}dq\,{\rm e}^{iqr}\frac{V(q)}{\epsilon^{RPA}(q)}\,\frac{i}{q}\Big[{\rm e}^{-iqL}-1\Big]\,,
\end{split}
\end{equation}
where we have assumed constant density in the wire. 
Similarly, for the RPA induced density in the finite length wire we get from Eq.\,(\ref{ninduced3})
\begin{equation}
\label{ninduced4}
\bar{n}_{ind}^{RPA}(r)=\frac{n_{0}}{2\pi}\int_{-\infty}^{\infty} dq\,n_{ind}^{RPA}(q)\,{\rm e}^{iq(r-r')}\frac{i}{q}\Big[{\rm e}^{-iqL}-1\Big]\,.
\end{equation}

Instead of using a self-consistent solution of the Poisson-BdGs equations in $n_{0}$,  for simplicity we just consider $n_{0}=1/L$.
Notice that $\bar{n}_{ind}^{RPA}(r)$ is the perturbation to the electron density of the infinite wire $n$, which can be calculated as follows.
Indeed, from Eq.\,(\ref{eq2}) we can solve for $k$ and get $\pm k_{\pm}$, where
\begin{equation}
k_{\pm}(\mu,B,\alpha_{R})=\sqrt{k_{\mu}^{2}+2k_{so}^{2}\pm\sqrt{(k_{\mu}^{2}+2k_{so}^{2})^{2}+k_{Z}^{4}-k_{\mu}^{4}}}
\end{equation}
where $k_{\mu}=\sqrt{2m\mu/\hbar^{2}}$, $k_{Z}=\sqrt{2mB/\hbar^{2}}$ and $k_{so}=m\alpha_{R}/\hbar^{2}$. The density of states can be calculated from previous equation as $D(E)=(1/\pi)dk/dE$. Thus, the electron density $n(\mu,B,\alpha_{R})$ is calculated by integrating the density of states up to the Fermi level, which is defined by the chemical potential $\mu$.
Then, for $\mu>B$, both Rashba bands are occupied and therefore 
\begin{equation}
n(\mu,B,\alpha_{R})=\frac{1}{\pi}\big[ k_{+}(\mu,B,\alpha_{R})+ k_{+}(\mu,B,\alpha_{R})\big]\,.
\end{equation}
When the chemical potential is within the Zeeman gap, $-B<\mu<B$, only the lower band is occupied and therefore
\begin{equation}
\label{nBhigh}
n(\mu,B,\alpha_{R})=\frac{1}{\pi} k_{+}(\mu,B,\alpha_{R})\,.
\end{equation}
And for $\mu<-B$
\begin{equation}
n(\mu,B,\alpha_{R})=\frac{1}{\pi}\big[ k_{+}(\mu,B,\alpha_{R})- k_{+}(\mu,B,\alpha_{R})\big]\,.
\end{equation}
Previous formulae provide analytics for the electron density for a Rashba nanowire. 


\subsection{Screening of the Coulomb potential with constant dielectric function}
In our previous discussion, we have shown that the dielectric function can be calculated from the density-density response function following RPA and linear response theory.
On the other hand, it is usually considered that the static dielectric function is a parameter of the system and assumed to be a known quantity, being approximately $\epsilon_{r}\approx16$ for InSb. However, later in this Chapter we will see that it strongly depends on the wave-vector $q$, but before going further, we first discuss the situation with constant dielectric function, $\epsilon^{RPA}(q)\approx \epsilon_{r}$. Then, we can write Eq.\,(\ref{phirpa2}) as
\begin{equation}
\label{phirpa0a}
\phi_{scr}(r-r')=\frac{1}{\epsilon_{r}}\int  \frac{dq}{2\pi}V(q){\rm e}^{iq(r-r')}\,,
\end{equation}
where we have suppressed the notation $RPA$ in $\phi_{scr}$ for obvious reasons and $V(q)$ is the 1D Coulomb potential \cite{vignale} given by Eq.\,(\ref{pot1d}).
\begin{figure} 
   \centering
   \includegraphics[width=0.9\columnwidth]{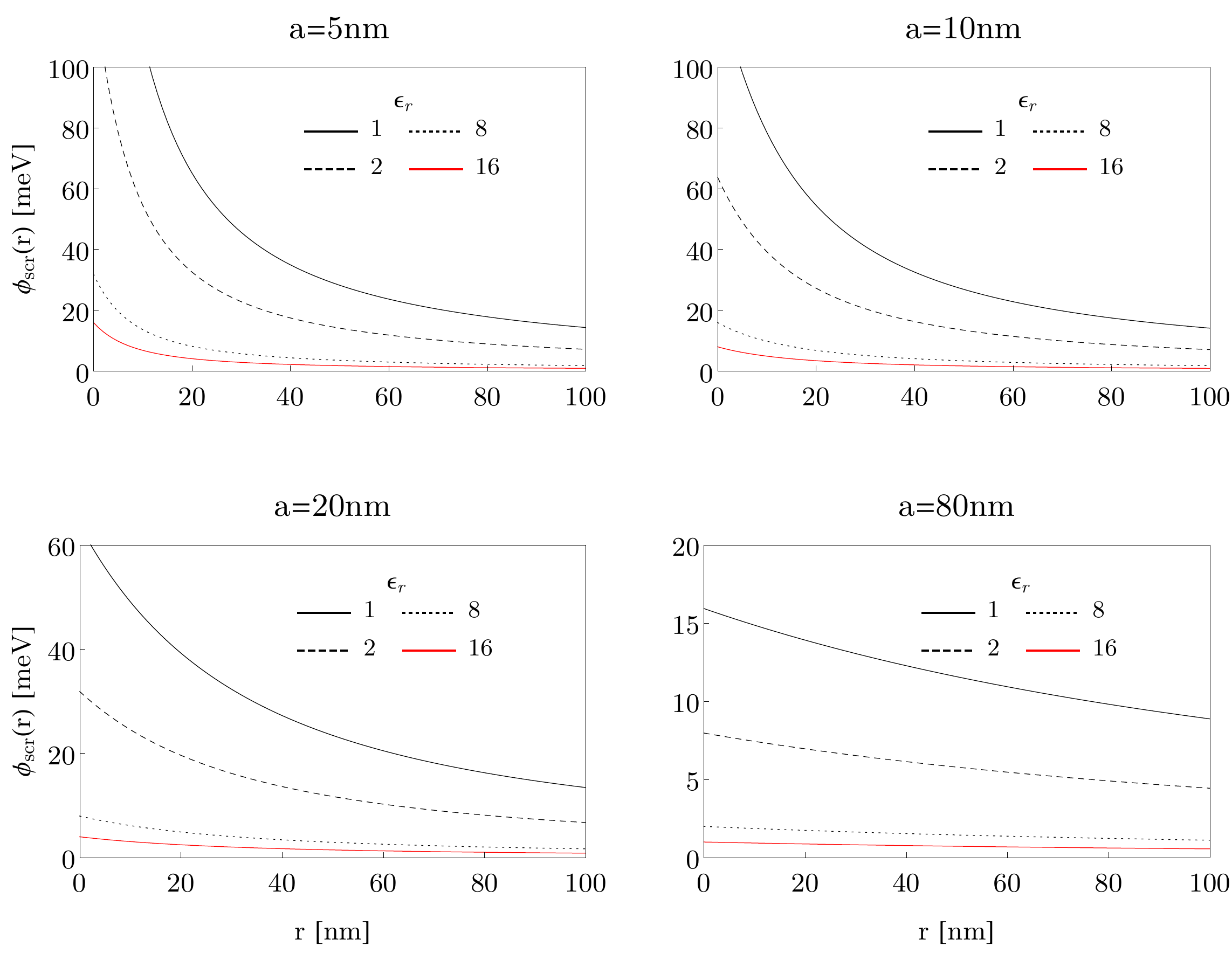}
   \caption[Screened potential with constant dielectric function in an infinite wire]{(Color online) Screened Coulomb potential in an infinite cylindrical wire of radius $a$ as a function of $r$ for different values of the dielectric constant, from Eq.\,(\ref{phirpa0a}). Notice that an increase in the dielectric constant reduces the screened potential but preserving the $1/r$ decay, while by increasing the wire's radius $a$ the potential is also reduced and it tends to decrease linearly for large $a$. }
   \label{vqscr}  
\end{figure}
 As before, a finite length wire $L$ is taken into account by integrating over $r'$,  with $n(\mu,B,\alpha_{R})$  the electron density,
\begin{equation}
\label{phirpa0b}
\begin{split}
\bar{\phi}_{scr}(r)
&=\frac{n_{0}}{\epsilon_{r}}\int \frac{dq}{2\pi}\,{\rm e}^{iqr}\,V(q)\,\frac{i}{q}\Big[{\rm e}^{-iqL}-1\Big]\,.
\end{split}
\end{equation}
In Fig.\,\ref{vqscr} we plot the screened Coulomb potential as a function of $r$, given by Eq.\,(\ref{phirpa0a}), for different values of the dielectric constant $\epsilon_{r}$ and wire's radius $a$. For small $a$ (top left panel), the behaviour of the screened potential $\phi_{scr}(r)$ follows the typical $1/r$ decay, however, as $\epsilon_{r}$ is increased, $\phi_{scr}(r)$ experiments a considerable reduction for all $r$ and roughly  preserving the $1/r$ shape for small $r$.
This implies that the screened lengths are reduced by increasing $\epsilon_{r}$.
Notice also that by increasing the wire's radius $a$ (top right and bottom panels), the values of the screened potential at small $r$ considerable decrease. For huge values of $a$ (bottom right panel), the  $1/r$ decay is practically lost and the potential tends to decrease linearly. 
\begin{figure} 
   \centering
   \includegraphics[width=0.8\columnwidth]{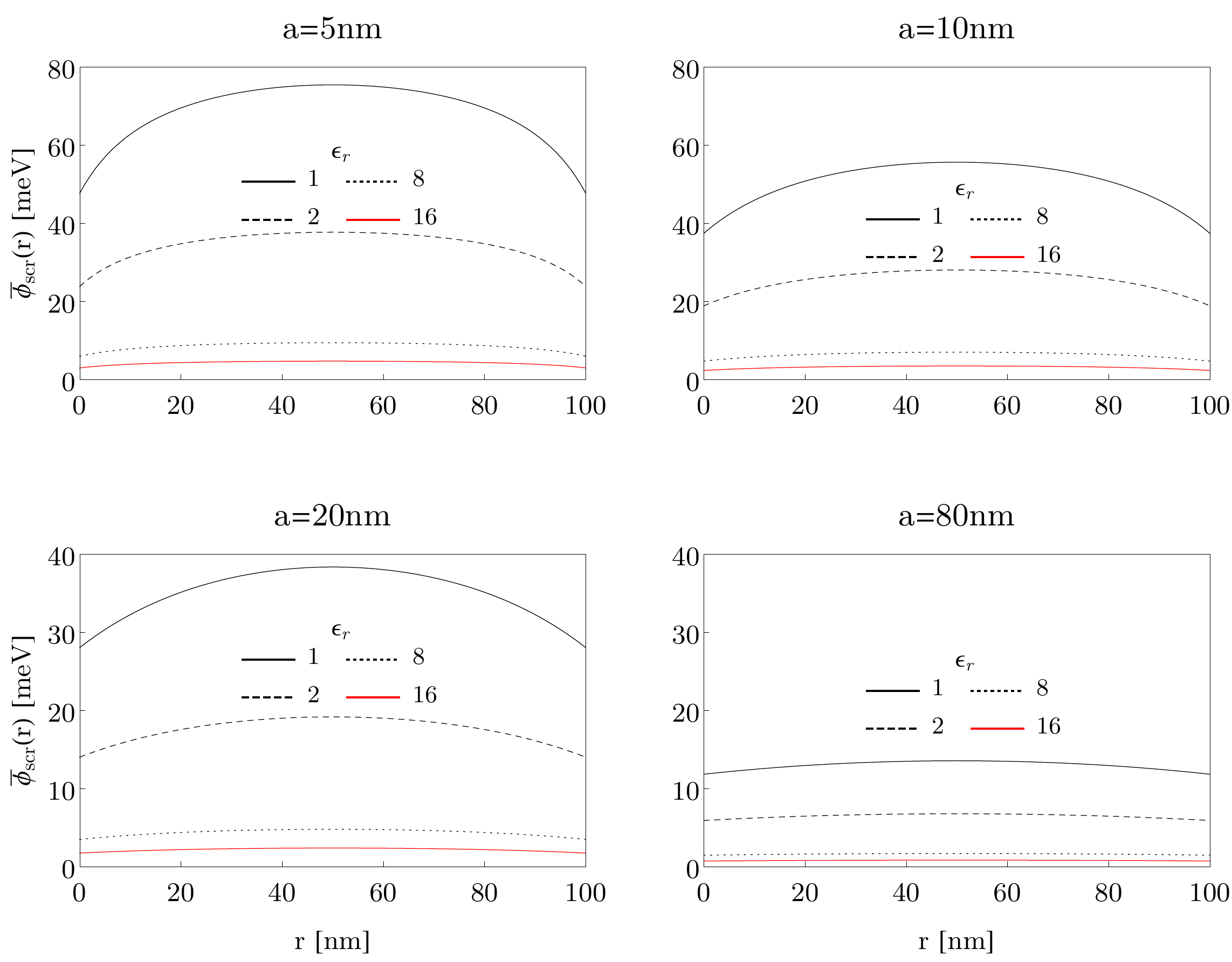}
   \caption[Screened potential with constant dielectric function in a finite wire]{(Color online) Screened Coulomb potential in an finite cylindrical wire of radius $a$ and length $L=100$\,nm as a function of $r$ for different values of the dielectric constant, $\epsilon_{r}$, from Eq.\,(\ref{phirpa0b}). Parameters: $B=2$\,meV, $\alpha_{R}=20$\,meVnm and $\mu=1$\,meV.}
   \label{vqscrL1}  
\end{figure}
\begin{figure} 
   \centering
   \includegraphics[width=0.8\columnwidth]{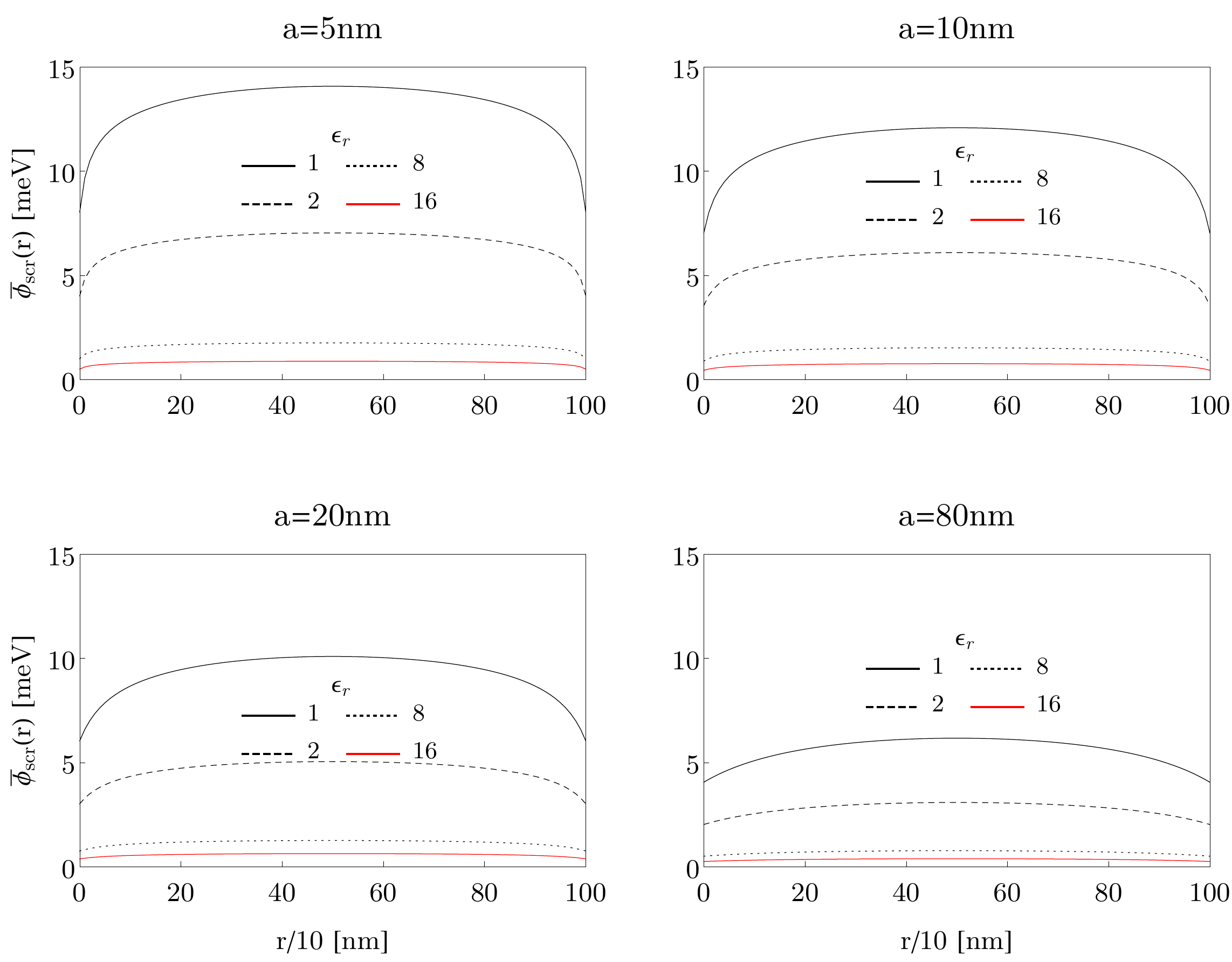}
   \caption[Screened potential with constant dielectric function in a finite wire]{(Color online) Screened Coulomb potential in an finite cylindrical wire of radius $a$ and length $L=1000$\,nm as a function of $r$ for different values of the dielectric constant, $\epsilon_{r}$, from Eq.\,(\ref{phirpa0b}). Parameters: $B=2$\,meV, $\alpha_{R}=20$\,meVnm and $\mu=1$\,meV.}
   \label{vqscrL2}  
\end{figure}


In Figs.\,\ref{vqscrL1} and \ref{vqscrL2} we plot the screened potential in a wire of finite length $L$ as a function of $r$, $\bar{\phi}_{scr}(r)$, given by Eq.\,(\ref{phirpa0b}), for different values of the dielectric constant and the wire's radius.
In both cases, the screened potential tends to be reduced by approaching the edges of the wire, while acquiring a flat shape in the bulk.
Notice that by increasing the dielectric constant, $\bar{\phi}_{scr}(r)$ is considerable reduced (see for instance different curves in left top panel of Fig.\,\ref{vqscrL1}), and  an increase in the wire's radius also introduces a reduction in the screened potential (compare top panels), as also observed in  Fig.\,\ref{vqscr}. 
On the other hand, as one increases the length of the wire $L$, the values of the screened potential are enhanced and the flatness behaviour in the bulk is more robust. 
The lowest curves (red curves) in previous figures correspond to $\epsilon_{r}=16$, typical value of the dielectric constant for InSb wires. Observe that for long finite wires with large radius $a$  the values of the screened potential $\bar{\phi}_{scr}(r)$ tend to be around $\mu$.

In this part we have learned that depending on the values of the dielectric constant $\epsilon_{r}$, the screened potential 
may acquire different values. Small (big) values of $\epsilon_{r}$ lead to big (small) values of the screened potential. Additionally, we have seen that increasing the wire's radius also tend to reduce the screened potential. These effects remain in both infinite and finite wires. The introduction of the finite wire's length also affect the screened potential at the wire edges. 

Next we will see how previous results change when the dielectric function is calculated from the density response function within the RPA and linear response approaches.
\section{Density response function in Rashba nanowires}
In this part we calculate the density response function $\chi$ in nanowires with spin-orbit coupling and subjected to an external Zeeman field, perpendicular to the spin-orbit axis, following the linear response approach \cite{vignale}.
From previous section we know that $\chi$ allows us for studying screening effects in these one-dimensional systems.
Although a full exact analysis should be done by solving the Poisson-BdGs equations, we believe that the treatment we do here represents a first step towards the understanding of electron-electron interactions in such nanowires.
\subsection{Density response function with SOC and Zeeman interaction}
We calculate the density-density response function for a homogeneous non-interacting 1D nanowire in the presence of Rashba spin-orbit and Zeeman interactions following the linear response theory \cite{vignale}.

The response function in momentum and frequency space is given by \cite{vignale}
\begin{equation}
\label{densityresponsefunction}
\chi(q,\omega)=\int d(r-r')\int d(t-t')\,{\rm e}^{-iq(r-r'))}\,{\rm e}^{i(\omega+i\eta)(t-t')}\,\chi(1,1')\,,
\end{equation}
where we have denoted $\chi(r,t,r',t')\equiv\chi(1,1')$ and
\begin{equation}
\label{shortdefnn}
\chi(1,1')=-\frac{i}{\hbar}\big<[\hat{n}(1),\hat{n}(1')]\big>\,.
\end{equation}
Here, $\hat{n}$ represents the electron density operator.
In second quantisation, $\hat{n}$ can be written in terms of electron field operators.
The relation between the annihilation field operator $\Psi(r,t)$, at position $r$ and time $t$, and any other set of annihilation operators $c$ is given by $\Psi_{\sigma}(r,t)=\sum_{\alpha}\psi_{\alpha}(r,t)c_{\alpha}(t)$, where $\psi_{\alpha}$ represent the one-electron wave functions and $\alpha$ labels spin, momentum, etc.
Therefore, it is natural to use the eigenstates $\psi_{k,\sigma}$, given by Eq.\,(\ref{eq2}), to construct such field operators. The electron density operator is then written as \cite{vignale}
\begin{equation}
\hat{n}(r,t)=\sum_{\alpha}\Psi^{\dagger}_{\alpha}(r,t)\Psi_{\alpha}(r,t)\,.
\end{equation}
Then, we can write
\begin{equation}
\begin{split}
\hat{n}(1)&=\mathop{\sum_{k_{1}\sigma_{1}}}_{k_{2}\sigma_{2}}\psi_{k_{1}\sigma_{1}}^{\dagger}(r)\psi_{k_{2},\sigma_{2}}(r)\,c^{\dagger}_{k_{1}\sigma_{1}}(t)c^{\dagger}_{k_{2}\sigma_{2}}(t)\,,\\
\hat{n}(1')&=\mathop{\sum_{k_{3}\sigma_{3}}}_{k_{4}\sigma_{4}}\psi_{k_{3}\sigma_{3}}^{\dagger}(r')\psi_{k_{4},\sigma_{4}}(r')\,c^{\dagger}_{k_{3}\sigma_{3}}(t')c^{\dagger}_{k_{4}\sigma_{4}}(t')\,,
\end{split}
\end{equation}
where $\hat{n}(1)=\hat{n}(r,t,r',t')$.
Now, we plug the wave functions given in Eq.\,(\ref{eq2}) into previous equations, and get
\begin{equation}
\label{densityoperators}
\begin{split}
\hat{n}(1)&=\mathop{\sum_{k_{1}\sigma_{1}}}_{k_{2}\sigma_{2}}
\eta_{k_{1}\sigma_{1}}^{\dagger}\eta_{k_{2}\sigma_{2}}\,\phi_{k_{1}}^{\dagger}(r)\phi_{k_{2}}(r)\,c^{\dagger}_{k_{1}\sigma_{1}}(t)c_{k_{2}\sigma_{2}}(t)\,,\\
\hat{n}(1')&=\mathop{\sum_{k_{3}\sigma_{3}}}_{k_{4}\sigma_{4}}
\eta_{k_{3}\sigma_{3}}^{\dagger}\eta_{k_{4}\sigma_{4}}\,\phi_{k_{3}}^{\dagger}(r')\phi_{k_{4}}(r')\,c^{\dagger}_{k_{3}\sigma_{3}}(t')c_{k_{4}\sigma_{4}}(t')\,,
\end{split}
\end{equation}
where $\sigma_{i}=\pm$, and 
$
 c_{k\sigma}^{\dagger}(t)=c_{k\sigma}^{\dagger}\,{\rm e}^{i\varepsilon_{k\sigma}t/\hbar}\,,\quad
  c_{k\sigma}(t)=c_{k\sigma}\,{\rm e}^{-i\varepsilon_{k\sigma}t/\hbar}\,.
$
 Then, we introduce the equations for the electron density operator, given by Eqs.\,(\ref{densityoperators}), into the average of the commutator given by Eq.\,(\ref{shortdefnn}). 
 After some algebra we can get the static limit of the density-density response function, see Eq.\,(\ref{chinnstatic}) in Appendix \ref{AppChapDensity} for more details on the derivation,
 \begin{equation}
  \label{MainTextchi0}
  \chi(q)=\frac{1}{2L}\sum_{k\sigma,\sigma'}\bigg[1+\sigma\sigma'\frac{B^{2}+\alpha_{R}^{2}k(k+q)}{|v_{k}||v_{k+q}|} \bigg]\frac{n_{k\sigma}-n_{k+q\sigma'}}{\varepsilon_{k\sigma}-\varepsilon_{k+q\sigma'}}\,.
 \end{equation}
At zero temperature $T=0$, the Fermi distribution function $n_{k,\sigma}$ becomes a step function
$n_{k,\sigma}=\theta(-\varepsilon_{k,\sigma})$, which is $1$ for $-\varepsilon_{k,\sigma}>0$ or  $0$, otherwise. The function $\theta(-\varepsilon_{k,\sigma})$ restricts the $k$ integration to interval $[-k_{F,\sigma},+k_{F,\sigma}]$, with $k_{F}$ defined from Eq.\,(\ref{eq2}), $\varepsilon_{k_{F},\sigma}=0$, where 
\begin{equation}
\label{fermik}
k_{F,\sigma}=\pm\sqrt{k_{\mu}^{2}+2k_{so}^{2}-\sigma\sqrt{(k_{\mu}^{2}+2k_{so}^{2})^{2}-k_{\mu}^{4}+k_{z}^{4}}}\,,
\end{equation}
with $k_{\mu}=\sqrt{2m\mu/\hbar^{2}}$, $k_{so}=m\alpha_{R}/\hbar^{2}$ and $k_{Z}=\sqrt{2mB/\hbar^{2}}$.

\begin{figure} 
   \centering
   \includegraphics[width=0.9\columnwidth]{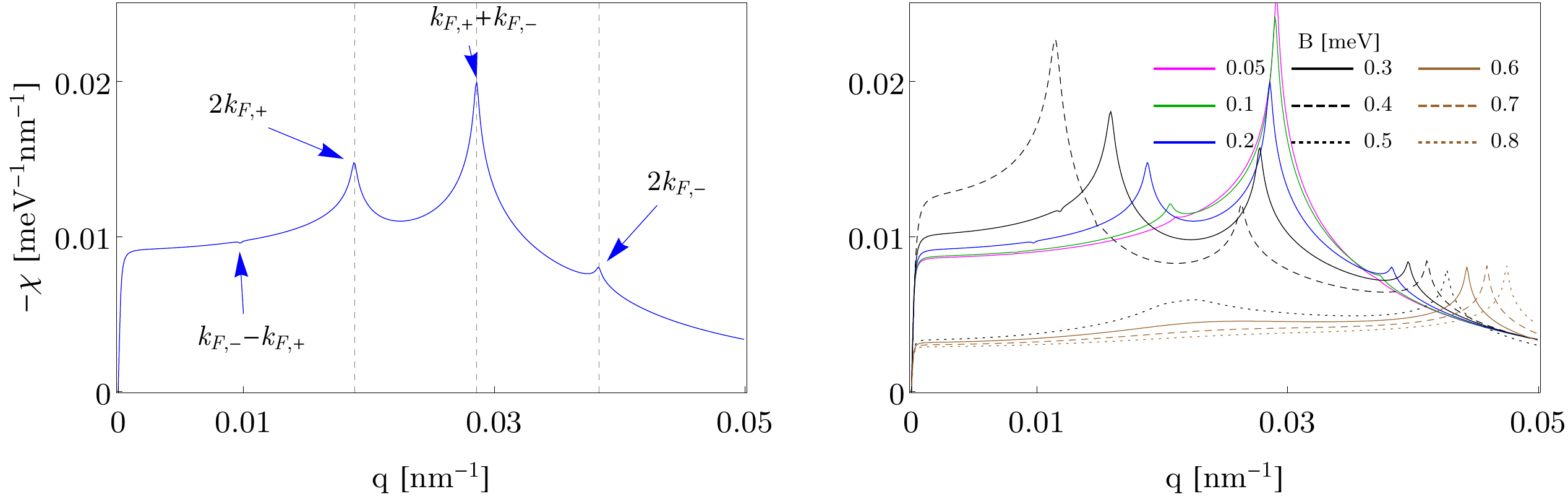}
   \caption[Density response function in Rashba nanowires]{(Color online) Density response function as a function of momentum $q$ in a nanowire with Rashba SOC and Zeeman interaction. Left panel shows a case for fixed Zeeman interaction, $B=0.2$\,meV, while in right panel different curves correspond to different values of the Zeeman field $B$. Notice that the response function exhibits resonance peaks at $2k_{F,\pm}$ and at $k_{F,+}+k_{F,-}$, where such wave vectors are defined by Eq.\,(\ref{fermik}). The almost invisible kink close to $q\approx0.01$ corresponds to $k_{F,-}-k_{F,+}$. Parameters: $\alpha_{R}=20$\,meVnm, $\mu=0.5$\,meV}
   \label{chi0}  
\end{figure}
In Fig.\,\ref{chi0} we present the static density-density response function, given by Eq.\,(\ref{MainTextchi0}), as a function of the momentum $q$ for different values of the Zeeman field. 
The $q\rightarrow0$ limit of the static density response function is a measure of the number of excited states available to the system for vanishing excitation energy.
Therefore, unlike the free electron case \cite{vignale}, the density response function, given by Eq.\,(\ref{MainTextchi0}), is zero at $q=0$, and that is because the external magnetic field introduces a gap into the system \cite{vignale}. 
Another important feature of the density response function given by Eq.\,(\ref{MainTextchi0}) is that it develops three clear resonance peaks at $2k_{F,+}$, $2k_{F,-}$ and $k_{F,+}+k_{F,-}$, and also an almost invisible kink for $B<\mu$ close to $q\approx0.01$  corresponds to $k_{F,-}-k_{F,+}$.
This is different from to the free electron case, where only one peak at $2k_{\mu}$ \cite{vignale}. 
The positions of these peaks are shown in left panel of Fig.\,\ref{chi0}, while in the right panel one observes its evolution as the Zeeman field increases.
The peaks are visible as long as they are real. 
Indeed, in Fig.\,\ref{chi0}, for experimentally reasonable values of $B$, $\mu$ and $\alpha_{R}$, one observes that for $B<\mu$, the density response $\chi(q)$ exhibits the three clear resonant peaks at the corresponding $q$ and an almost invisible kink as explained above. 
On the other hand, when $B>\mu$, $k_{F,+}$, defined by Eq.\,(\ref{fermik}), becomes completely imaginary and therefore the system involves only one Fermi momentum leaving only the resonance at $2k_{F,-}$. Notice that the almost invisible kink close to $q\approx0.01$ happens at $k_{F,-}-k_{F,+}$ as long as such wave vectors are real, which means for $B<\mu$, otherwise such resonance peak is the same as the one at  $k_{F,-}+k_{F,+}$.
Important, such peaks can be interpreted as evidence of static charge density waves.

\begin{figure} 
   \centering
   \includegraphics[width=0.9\columnwidth]{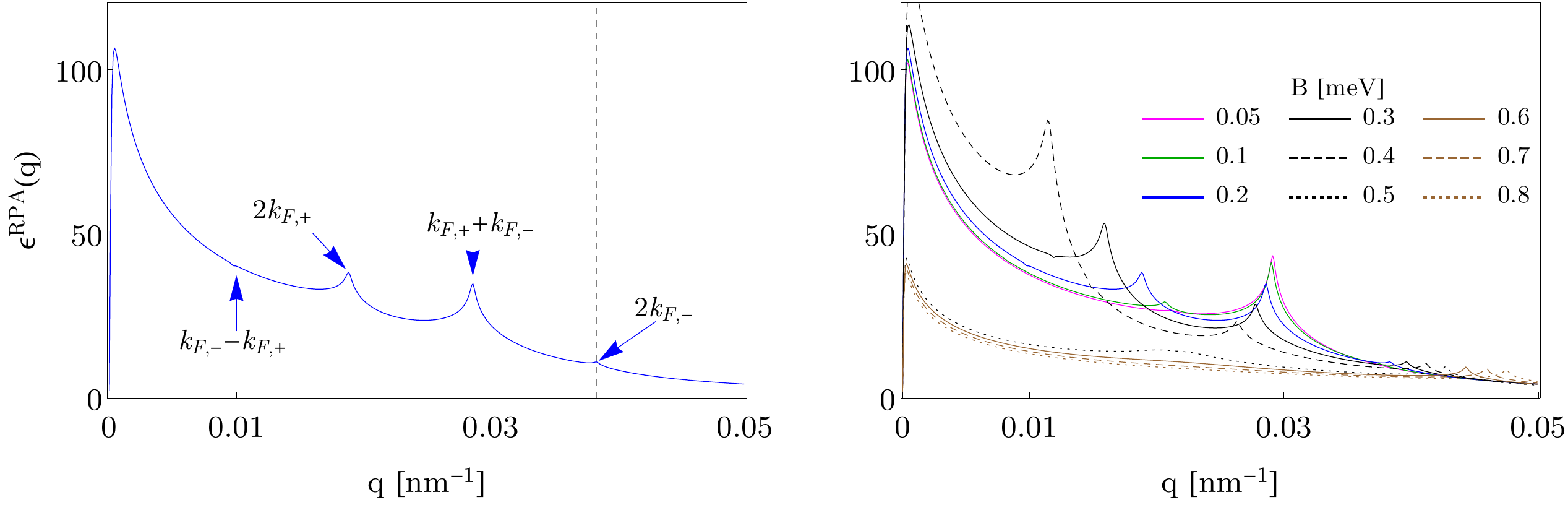}
   \caption[RPA Dielectric function in Rashba nanowires]{(Color online) Dielectric function in the RPA approach as a function of momentum $q$ in a nanowire with Rashba SOC for different values of the Zeeman field and fixed wire's radius $a$. Left panel shows a case for fixed Zeeman interaction, $B=0.2$\,meV, while in right panel different curves correspond to different values of the Zeeman field $B$. Notice that the resonance peaks from the density response, at $2k_{F,\pm}$ and at $k_{F,+}+k_{F,-}$, are also present in the dielectric function, where such wave vectors are defined by Eq.\,(\ref{fermik}). 
There is also an almost invisible kink close to $q\approx0.01$, which corresponds to $k_{F,-}-k_{F,+}$.
   Parameters: $\alpha_{R}=20$\,meVnm, $\mu=0.5$\,meV, $a=20$nm.}
   \label{ep0}  
\end{figure}

In  Subsec. \,\ref{generalconcepts} we have introduced the RPA approach in order to calculate the dielectric function. We have pointed out that in this approach the proper response function is the density-density response function ( or Lindhard function), see Eq.\,(\ref{dielectricEPRPA}). Thus, in order to calculate the dielectric function for Rashba nanowires, we use the density-density response given by Eq.\,(\ref{MainTextchi0}), and then plug into Eq.\,(\ref{dielectricEPRPA}).
The results are shown in Fig.\,\ref{ep0}, where we present the dielectric function as a function of momentum $q$ in the RPA approach. In this part we have also the Coulomb potential 
in 1D derived in \cite{vignale}, see Eq.\,(\ref{pot1d}).
In Fig.\,\ref{ep0} the nanowire's radius $a$ is fixed.

\begin{figure} 
   \centering
   \includegraphics[width=0.9\columnwidth]{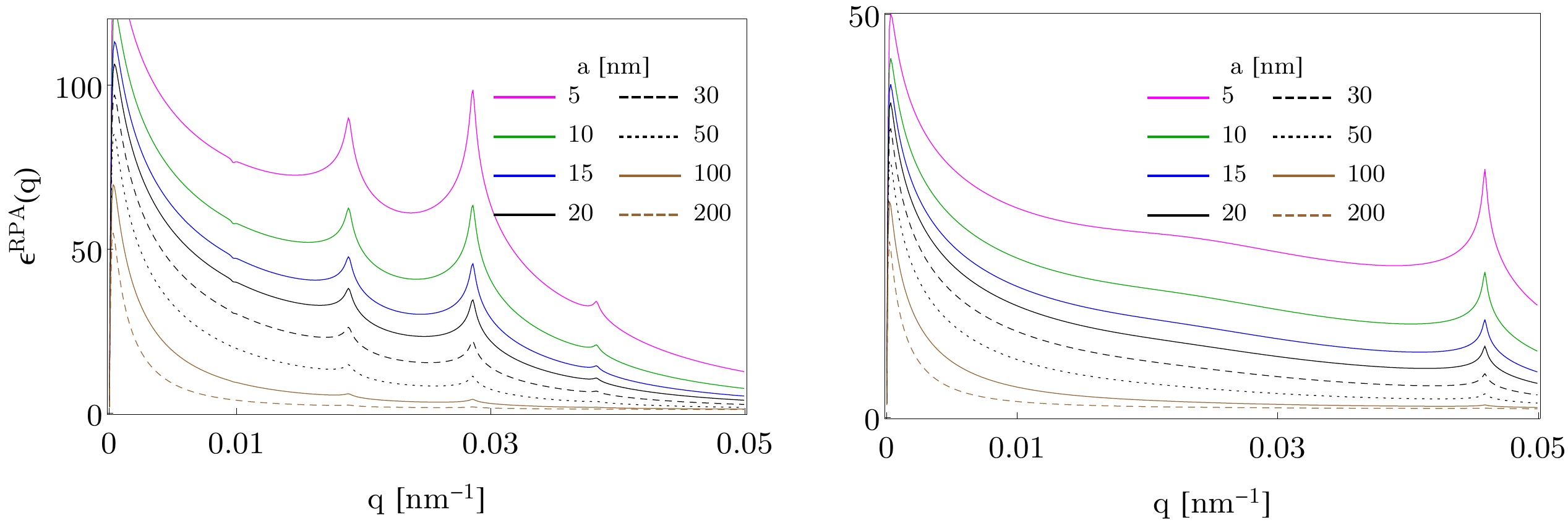}
   \caption[RPA Dielectric function in Rashba nanowires]{(Color online) Dielectric function in the RPA approach as a function of momentum $q$ in a nanowire with Rashba SOC for different values of the wire's radius $a$.
   In left panel, $\mu>B=0.2$\,meV, while right panel for $\mu<B=0.7$\,meV. Notice that in both cases the resonance peaks are washed out as $a$ is increased. 
The small kink close to $q\approx0.01$, which corresponds to $k_{F,-}-k_{F,+}$, is also visible and suffers from the same effect.
   Parameters: $\alpha_{R}=20$\,meVnm, $\mu=0.5$\,meV.}
   \label{ep1}  
\end{figure}
Since the dielectric function is calculated from the density response function, one expects that such function still contains information about the resonance peaks, as one can indeed observe in Fig.\,\ref{ep0}. We also notice that the resonance peaks in the dielectric function $\epsilon^{RPA}(q)$ are not of the same high as in the density response function $\chi(q)$, see Fig.\,(\ref{chi0}).
\begin{figure} 
   \centering
   \includegraphics[width=0.9\columnwidth]{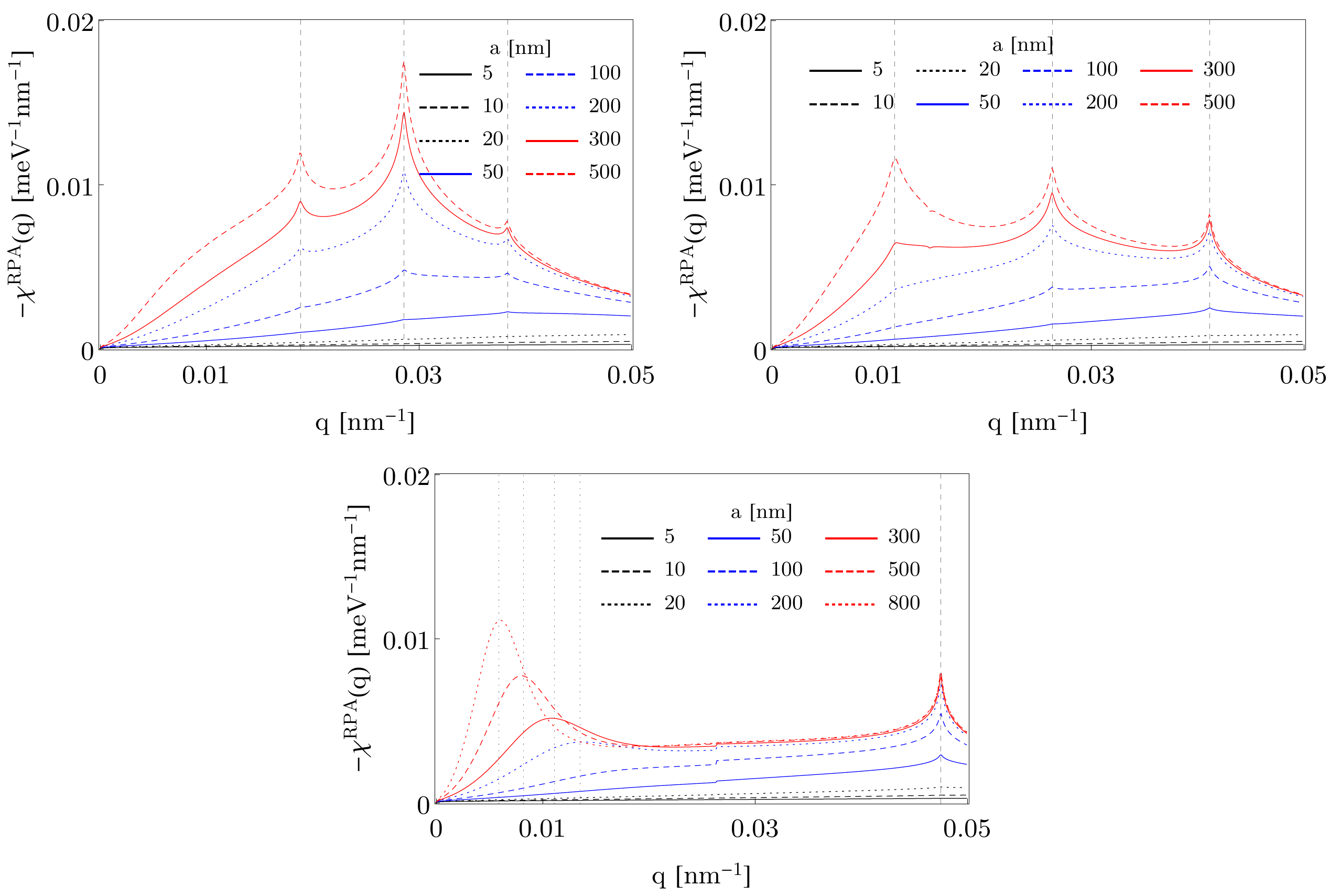}
   \caption[RPA Charge density in Rashba nanowires]{(Color online) Charge density in the RPA approach as a function of momentum $q$ in a nanowire with Rashba SOC for different values of the wire's radius $a$.
 Left and right top panels for $B=0.2$\,meV$<\mu$ and $B=0.4$\,meV$<\mu$, respectively, while bottom panel for $B=0.8$\,meV$>\mu$. 
 Notice that the visibility of the resonance peaks at the Fermi wave vectors takes place at large values of $a$. While for $B<\mu$ (top panels) one distinguish three peaks associated with the wave vectors $2k_{F,+}$, $2k_{F,-}$, $k_{F,+}+k_{F,-}$, for $B>\mu$ (bottom panel) only the one at $2k_{F,-}$ survives around $0.05$\,nm$^{-1}$. Additionally, a resonance at low $q$ emerges, being visible for $a\geq 200$\,nm and follows the rule $q_{res}=2\pi/(a+\lambda_{F,-})$, where $\lambda_{F,-}$ is the Fermi length $\lambda_{F,-}=2\pi/k_{F,-}$. 
   Parameters: $\alpha_{R}=20$\,meVnm, $\mu=0.5$\,meV.}
   \label{chirpa1}  
\end{figure}

The visibility of the resonance peaks also depends on the new length scale introduced with the wire's radius. Indeed, observe the evolution of the dielectric function $\epsilon^{RPA}(q)$ at fixed Zeeman field for $B<\mu$ and $B>\mu$ in left and right panels of Fig.\,(\ref{ep1}). In the former case, the two Fermi wave vectors $k_{F,\pm}$ are real and therefore the dielectric function develops the three resonance peaks at $2k_{F,+}$, $2k_{F,-}$, $k_{F,+}+k_{F,-}$, and a small kink at $k_{F,+}-k_{F,-}$.
While in the latter, $k_{F,+}$ is imaginary and only one resonance peak at $2k_{F,-}$ is observed.
In both cases, as the wire's radius increases, the resonance peaks are reduced and even completely washed out when $a$ is long enough. It seems that the resonance peaks are visible as long as $a<{\rm min}(2\pi/q_{F})$, where $q_{F}=\{2k_{F,+},2k_{F,-},k_{F,+}+k_{F,-},k_{F,-}-k_{F,+}\}$.

In Fig.\,\ref{chirpa1}, we present the charge density response function in the RPA approach  as a function of the momentum, $\chi^{RPA}(q)$, for different values of the wire's radius, calculated from Eq.\,(\ref{chiRPA}). Top panels correspond for $B<\mu$, while bottom for $B>\mu$. Notice that, in the former case, the expected resonance peaks, which appeared in the density response function in Fig.\,\ref{chi0} at $2k_{F,+}$, $2k_{F,-}$, $k_{F,+}+k_{F,-}$ and a small kink at $k_{F,+}-k_{F,-}$, are only visible when the radius of the wire is made very large (see dashed lines in top panels). On the other hand, in the latter, for $B>\mu$, $k_{F,+}$ is imaginary and only one resonance peak is the one associated to $2k_{F,-}$, as one indeed observes (see dashed line at high momentum).
Interestingly, at small momentum, the charge density, $\chi_{RPA}(q)$, develops a resonance, which is visible for $a\geq 200$\,nm (for parameters given in caption of Fig.\,\ref{chirpa1}) and follows the rule $q_{res}=2\pi/(a+\lambda_{F,-})$, where $\lambda_{F,-}$ is the Fermi length $\lambda_{F,-}=2\pi/k_{F,-}$. 

\section{Density response function in superconducting nanowires}
The aim of this part is to calculate the response function in strong Zeeman fields, as it is the one that corresponds to the topological superconducting phase with MBSs.

In this part we consider the same Rashba nanowire as in the previous section, and place it in proximity to a $s$-wave superconductor with Hamiltonian
 \begin{equation}
\mathcal{H}_{sc}=\int dx\Big[\Delta\psi_{\uparrow}^{\dagger}(k)\psi_{\downarrow}^{\dagger}(-k)+\Delta^{\dagger}\psi_{\downarrow}(-k)\psi_{\uparrow}(k)\Big]\,.
\end{equation}
For introducing superconductivity, we first write down the full system Hamiltonian, $\mathcal{H}_{0}+\mathcal{H}_{sc}$, in the so-called helical basis given by Eqs.\,(\ref{basis1}), whose description was given in Sections \ref{Rashbawire} and \ref{appRashba}.
In such basis, the full Hamiltonian is written as 
\begin{equation}
\label{HhelicalDen}
\begin{split}
\mathcal{H}&=
\int \frac{dk}{2\pi}
\big[
\varepsilon_{k,+}\psi_{+}^{\dagger}(k)\psi_{+}(k)
+
\varepsilon_{k,-}\psi_{-}^{\dagger}(k)\psi_{-}(k)
\big]\\
&+
\bigg[\frac{\Delta_{--}(k)}{2}\psi_{-}^{\dagger}(k)\psi_{-}^{\dagger}(-k)
+\frac{\Delta_{++}(k)}{2}\psi_{+}^{\dagger}(k)\psi_{+}^{\dagger}(-k)
+\Delta_{+-}(k)\psi_{+}^{\dagger}(k)\psi_{-}^{\dagger}(-k)+h.c
\bigg],
\end{split}
\end{equation}
where 
\begin{equation}
\label{pairingshelical}
\Delta_{--,++}(k)=\frac{\pm i\alpha_{R} k \Delta}{\sqrt{B^{2}+\alpha_{R}^{2}k^{2}}}\equiv \pm i \Delta_{p}(k)\,,\quad \Delta_{+-}(k)=\frac{B \Delta}{\sqrt{B^{2}+\alpha_{R}^{2}k^{2}}}
\equiv\Delta_{s}(k)\,,
\end{equation}
represent the different pairing functions that arise in our nanowire due to the interplay of Rashba SOC and Zeeman interaction when placed on a $s$-wave superconductor. 
The first line of Eq.\,(\ref{HhelicalDen}) is just the normal Rashba nanowire Hamiltonian, while in the second line, the first and second terms associated to $\Delta_{--,++}(k)$, describe pairing between states of the same $\mp$ band, while the third term associated to $\Delta_{+-}(k)$, represent pairing between states of different band. 

Eq.\,(\ref{HhelicalDen}) can be written in Nambu space,
 \begin{equation}
 \mathcal{H}=\frac{1}{2}\int \frac{dk}{2\pi}\Psi^{\dagger}(k)\,H_{BdG}\,\Psi(k)\,, \quad \Psi(k)=\begin{pmatrix}
\psi_{+}^{\dagger}(k),
\psi_{-}^{\dagger}(k),
\psi_{+}(-k),
\psi_{-}(-k)
\end{pmatrix}^{\dagger}
 \end{equation}
 where the Bogoliubov-de Gennes Hamiltonian reads
 \begin{equation}
 H_{BdG}=
 \begin{pmatrix}
 \varepsilon_{k,+}&0&\Delta_{++}(k)&\Delta_{+-}(k)\\
 0&\varepsilon_{k,-}&-\Delta_{+-}(k)&\Delta_{--}(k)\\
 \Delta_{++}^{\dagger}(k)&-\Delta_{+-}^{\dagger}(k)&-\varepsilon_{-k,+}&0\\
 \Delta_{+-}^{\dagger}(k)&\Delta_{--}^{\dagger}(k)&0&-\varepsilon_{-k,-}
 \end{pmatrix}\,,
 \end{equation}
whose eigenvalues read, see Eq.\,(\ref{SpectrumFull}),
 \begin{equation}
 \label{SpectrumFullhelical}
E_{\pm}^{2}(k)=|\Delta_{++}(k)|^{2}+\Delta_{+-}^{2}(k)+\frac{\varepsilon_{k,+}^{2}+\varepsilon_{k,-}^{2}}{2}
\pm|\varepsilon_{k,+}-\varepsilon_{k,-}|\sqrt{\Delta_{+-}^{2}(k)+\Big[\frac{\varepsilon_{k,+}+\varepsilon_{k,-}}{2}\Big]^{2}}\,,
\end{equation}
where $\varepsilon_{\pm}=\xi_{k}\pm\sqrt{B^{2}+\alpha^{2}k^{2}}$, $\xi_{k}=\hbar^{2}k^{2}/2m-\mu$, and $\Delta_{++,--,+-}$ by Eqs.\,(\ref{pairingshelical}).
 At $k=0$ the lower band, $-$, develops a gap, as shown in Sec.\,\ref{Rashbawire}, and follows
\begin{equation}
E_{-}^{2}(k=0)=(B-\sqrt{\Delta^{2}+\mu^{2}})^{2}\,\quad\quad\Rightarrow \quad\quad E_{-}(k=0)=|B-B_{c}|\,,
\end{equation}
which closes at $B_{c}=\sqrt{\Delta^{2}+\mu^{2}}$, determining the topological phase transition point into a topological superconducting phase with Majorana bound states at the end of the wire.

On the other hand, for zero interband pairing, $\Delta_{+-}=0$, we are left with pairing potentials between states of the same band, only.
This could mimic the physics of two p-wave superconductors with pairing potentials $\Delta_{++}$ and $\Delta_{--}$.
In this regime, the BdG Hamiltonian reads,
\begin{equation}
H_{BdG}=\begin{pmatrix}
 \varepsilon_{+}(k)&0&\Delta_{++}(k)&0\\
 0&\varepsilon_{-}(k)&0&\Delta_{--}(k)\\
 \Delta_{++}^{\dagger}(k)&0&-\varepsilon_{+}(-k)&0\\
0&\Delta_{--}^{\dagger}(k)&0&-\varepsilon_{-}(-k)
 \end{pmatrix}
\,,
 \end{equation}
 where the energy spectrum reads,
\begin{equation}
E_{\pm}^{2}(k)=|\Delta_{++}(k)|^{2}+\varepsilon_{\pm}^{2}(k)\,,
\end{equation}
with $\varepsilon_{\pm}(k)=\frac{\hbar^{2}k^{2}}{2m}-\mu\pm\sqrt{B^{2}+\alpha_{R}^{2}k^{2}}$ and the superconducting pairing potentials are of $p$-wave nature
\begin{equation}
\Delta_{--,++}(k)=\frac{\pm i\alpha_{R} k\Delta}{\sqrt{B^{2}+\alpha_{R}^{2}k^{2}}}\equiv \pm i\Delta_{p}(k)\,,.
\end{equation}
At $k=0$,
\begin{equation}
E_{\pm}^{2}(k)=(-\mu\pm B)^{2}\,,
\end{equation}
so that there is a closing of the gap at low momentum when $\mu=\pm B$. This resembles to the closing of the gap at $B=\sqrt{\Delta^{2}+\mu^{2}}$ when the interband pairing, $\Delta_{+-}$, is not zero.
\subsection{Density response function: zero interband pairing, two band model}
For simplicity, we calculate the density response function when the pairing of states of different band is zero.
Effectively, it is similar to a situation with two p-wave superconductors.
We follow \cite{vignale}. We start but calculating the density response with SO and Zeeman field, and then use the bogoliubov transformation found in previous chapter in order to write the helical basis in the quasiparticle one.

The density-density response function is given by
\begin{equation}
\label{definitionchiq}
\chi(q,\omega)\int\,d(r-r')\,\int\,d(t-t')\,{\rm e}^{-iq(r-r')}\,{\rm e}^{i(\omega+i\eta)(t-t')}\chi(r,r',t,t')\,,
\end{equation}
where
\begin{equation}
\label{chichichi}
\chi(r,r',t,t')\equiv=\chi(1,1')=-\frac{i}{\hbar}\big< \big[ \hat{n}(1),\hat{n}(1')\big]\big>\,.
\end{equation}
Here, $\hat{n}$ is the electron density operator.
We have seen in previous section that, in second quantisation, the electron density operator can be written in terms of field operators
\begin{equation}
\label{densityscsc}
\begin{split}
\hat{n}(1)&=\mathop{\sum_{k_{1}\sigma_{1}}}_{k_{2}\sigma_{2}}\psi_{k_{1}\sigma_{1}}^{\dagger}(r)\psi_{k_{2},\sigma_{2}}(r)\,c^{\dagger}_{k_{1}\sigma_{1}}(t)c^{\dagger}_{k_{2}\sigma_{2}}(t)\,,\\
\hat{n}(1')&=\mathop{\sum_{k_{3}\sigma_{3}}}_{k_{4}\sigma_{4}}\psi_{k_{3}\sigma_{3}}^{\dagger}(r')\psi_{k_{4},\sigma_{4}}(r')\,c^{\dagger}_{k_{3}\sigma_{3}}(t')c^{\dagger}_{k_{4}\sigma_{4}}(t')\,,
\end{split}
\end{equation}
where $\psi_{k\sigma}$ are wave functions of the Rashba-Zeeman problem
\begin{equation}
\psi_{k\sigma}(r)=\frac{1}{\sqrt{L}}{\rm e}^{ikr}\frac{1}{\sqrt{2}}
\begin{pmatrix}
\sigma\gamma_{k}\\
1
\end{pmatrix}\,,\quad\quad \gamma_{k}=\frac{B+i\alpha_{R} k}{\sqrt{B^{2}+\alpha_{R}^{2}k^{2}}}\,,\quad \sigma=\pm\,.
\end{equation}
We then write
\begin{equation}
\label{eqswave}
\psi_{k\sigma}(r)=\eta_{k\sigma}\phi_{k}(r)
\end{equation}
where
\begin{equation}
\eta_{k\sigma}=\frac{1}{\sqrt{2}}
\begin{pmatrix}
\sigma\gamma_{k}\\
1
\end{pmatrix}\,,
\quad\quad\phi_{k}(r)=\frac{1}{\sqrt{L}}{\rm e}^{ikr}\,.
\end{equation}
Therefore,
\begin{equation}
\begin{split}
\hat{n}(1)&=\mathop{\sum_{k_{1}\sigma_{1}}}_{k_{2}\sigma_{2}}
\eta_{k_{1}\sigma_{1}}^{\dagger}\eta_{k_{2}\sigma_{2}}\,\phi_{k_{1}}^{\dagger}(r)\phi_{k_{2}}(r)\,c^{\dagger}_{k_{1}\sigma_{1}}(t)c_{k_{2}\sigma_{2}}(t)\,,\\
\hat{n}(1')&=\mathop{\sum_{k_{3}\sigma_{3}}}_{k_{4}\sigma_{4}}
\eta_{k_{3}\sigma_{3}}^{\dagger}\eta_{k_{4}\sigma_{4}}\,\phi_{k_{3}}^{\dagger}(r)\phi_{k_{4}}(r)\,c^{\dagger}_{k_{3}\sigma_{3}}(t)c_{k_{4}\sigma_{4}}(t)\,.
\end{split}
\end{equation}
Now, we insert previous equations into Eq.\,(\ref{chichichi}), and then sum over $\sigma=\pm$. Moreover, since we are dealing with a superconducting system, operators $c$ are transformed into new ones according to the Bogoliubov transformation described in detail in Appendix\,\ref{BOGOGOs}.
We are interested on the zero temperature and zero frequency limit, and after some algebra, see Appendix \ref{appdensity} for details, we get
\begin{equation}
\label{twobandensity}
\begin{split}
\chi(r,r')&=
-\frac{1}{L}\sum_{k}
\Bigg\{
\frac{A_{k,k+q}^{+}}{E_{k,+}+E_{k+q,+}}
\big(u_{k,+}v_{k+q,+} -v_{k,+}u_{k+q,+}\big)^{2}\\
&\quad\quad\quad\quad\quad\quad\quad+
\frac{A_{k,k+q}^{+}}{E_{k,-}+E_{k+q,-}}
\big(u_{k,-}v_{k+q,-} -v_{k,-}u_{k+q,-}\big)^{2}\\
&\quad\quad\quad\quad\quad\quad\quad+
\frac{A_{k,k+q}^{-}}{E_{k,-}+E_{k+q,+}}
\big(u_{k,-}v_{k+q,+} -v_{k,-}u_{k+q,+}\big)^{2}\\
&\quad\quad\quad\quad\quad\quad\quad+
\frac{A_{k,k+q}^{-}}{E_{k,+}+E_{k+q,-}}
\big(u_{k,+}v_{k+q,-} -v_{k,+}u_{k+,-}\big)^{2}
\Bigg\}\,,
\end{split}
\end{equation}
with the superconducting coherence factors given by
\begin{equation}
\begin{split}
\big(u_{k,+}v_{k+q,+} -v_{k,+}u_{k+q,+}\big)^{2}&=\frac{1}{2}\bigg[1-\frac{\varepsilon_{k,+}\varepsilon_{k+q,+}+\Delta_{p}(k)\Delta_{p}(k+q)}{E_{k,+}E_{k+q,+}} \bigg]\\
\big(u_{k,-}v_{k+q,-} -v_{k,-}u_{k+q,-}\big)^{2}&=\frac{1}{2}\bigg[1-\frac{\varepsilon_{k,-}\varepsilon_{k+q,-}+\Delta_{p}(k)\Delta_{p}(k+q)}{E_{k,-}E_{k+q,-}} \bigg]\\
\big(u_{k,-}v_{k+q,+} -v_{k,-}u_{k+q,+}\big)^{2}&=\frac{1}{2}\bigg[1-\frac{\varepsilon_{k,-}\varepsilon_{k+q,+}+\Delta_{p}(k)\Delta_{p}(k+q)}{E_{k,-}E_{k+q,+}} \bigg]\\
\big(u_{k,+}v_{k+q,-} -v_{k,+}u_{k+,-}\big)^{2}&=\frac{1}{2}\bigg[1-\frac{\varepsilon_{k,+}\varepsilon_{k+q,-}+\Delta_{p}(k)\Delta_{p}(k+q)}{E_{k,+}E_{k+q,-}} \bigg]\,.
\end{split}
\end{equation}
where
\begin{equation}
\Delta_{p}(k)=\frac{\alpha_{R} k\Delta }{\sqrt{B^{2}+\alpha_{R}^{2}k^{2}}}
\end{equation}
represent the $p$-wave nature of the superconducting pairing potential, and
\begin{equation}
A_{k_{1},k_{2}}^{\sigma}=\frac{1}{2}\bigg[1+\sigma\frac{B^{2}+\alpha_{R}^{2}k_{1}k_{2}}{\sqrt{B^{2}+\alpha_{R}^{2}k_{1}^{2}}\sqrt{B^{2}+\alpha_{R}^{2}k_{2}^{2}}} \bigg]\,,\quad \sigma=\pm\,.
\end{equation}
is the coefficient arising from the interplay of the Rashba SOC and Zeeman interaction.
Notice that the summation in $k$ in Eq.\,(\ref{twobandensity}) is replaced by an integration over $k$: $(1/L)\sum_{k}\rightarrow(1/2\pi)\int dk$, and the limits of integration includes all $k$, i.e. $k\in[-\infty,\infty]$. We do not discuss the results obtained in this part because what interests us is the 
strong Zeeman regime, which is discussed next.
\subsection{Density response function in strong Zeeman field: one band model}
In this part we address the situation of high Zeeman field $B\gg\mu, \Delta$. In this case, only the lowest band, $-$, is occupied. 
In a situation of high magnetic field, the system is in the topological phase and one can project the full Hamiltonian onto the lower band $\sigma=-$,  then
\begin{equation}
\mathcal{H}=
\int \frac{dk}{2\pi}
\varepsilon_{-}(k)\psi^{\dagger}_{-}(k)\psi_{-}(k)+\int \frac{dk}{2\pi} 
\Big[\frac{\Delta_{--}(k)}{2}\psi_{-}^{\dagger}(k)\psi_{-}^{\dagger}(-k)
+h.c\Big]\,.
\end{equation}
Previous Hamiltonian can be written in the BdG form in the $(\psi_{-}(k),\psi_{-}^{\dagger}(-k))$ basis, thus
\begin{equation}
H_{BdG,-}(k)= \begin{pmatrix}\,
\varepsilon_{-}(k)&\Delta_{--}(k)\\
\Delta_{--}^{\dagger}(k)&-\varepsilon_{-}(-k)
 \end{pmatrix}\,,
\end{equation}
with energy dispersion given by
\begin{equation}
\label{energies}
\tilde{E}_{-}(k)=\pm\sqrt{|\Delta_{--}|^{2}+\varepsilon_{-}^{2}(k)}\equiv\pm E_{k,-}\,,
\end{equation}
and $p$-wave superconducting pairing potential
\begin{equation}
\Delta_{--}(k)=\frac{i\alpha_{R} k \Delta}{\sqrt{B^{2}+\alpha_{R}^{2}k^{2}}}\equiv i\Delta_{p}(k)\,.
\end{equation}

Now, we calculate the density-density response function following the linear response theory described in \cite{vignale}, as we have done in previous subsection.
The response function in momentum and frequency space is given by
\begin{equation}
\chi(q,\omega)=\int d(r-r')\int d(t-t')\,{\rm e}^{-iq(r-r'))}\,{\rm e}^{i(\omega+i\eta)(t-t')}\,\chi(r,r',t,t')\,,
\end{equation}
where, $\chi(r,r',t,t')\equiv\chi(1,1')$,
\begin{equation}
\chi(1,1')=-\frac{i}{\hbar}\big<[\hat{n}(1),\hat{n}(1')]\big>\,,
\end{equation}
where $\hat{n}$ is the electron density operator.
In second quantisation , $\hat{n}$ can be written in terms of the helical field operators described in the previous section
\begin{equation}
\label{elecdensity}
\begin{split}
\hat{n}(1)&=\mathop{\sum_{k_{1},\sigma_{1}}}_{k_{2},\sigma_{2}}\psi^{\dagger}_{k_{1},\sigma_{1}}(r)\psi_{k_{2},\sigma_{2}}(r)\,c^{\dagger}_{k_{1},\sigma_{1}}(t)c_{k_{2},\sigma_{2}}(t)\,,\\
\hat{n}(1')&=\mathop{\sum_{k_{3},\sigma_{3}}}_{k_{4},\sigma_{4}}\psi^{\dagger}_{k_{3},\sigma_{3}}(r')\psi_{k_{4},\sigma_{4}}(r')\,c^{\dagger}_{k_{3},\sigma_{3}}(t')c_{k_{4},\sigma_{4}}(t')\,,
\end{split}
\end{equation}
where 
\begin{equation}
\psi_{k,\sigma}(r)=
\frac{1}{\sqrt{L}}\,{\rm e}^{ikr}
\frac{1}{\sqrt{2}}\begin{pmatrix}
\sigma\gamma_{k}\\
1
\end{pmatrix}=\phi_{k}(r)\eta_{k,\sigma}
\end{equation}
are the wave functions of the Rashba-Zeeman problem, and $\gamma_{k}=\frac{B+i\alpha_{R} k}{\sqrt{B^{2}+\alpha_{R}^{2}k^{2}}}$.
Hence, Eq.\,(\ref{elecdensity}) can be written as
\begin{equation}
\label{elecdensity2}
\begin{split}
\hat{n}(1)&=\mathop{\sum_{k_{1},\sigma_{1}}}_{k_{2},\sigma_{2}}
\phi^{\dagger}_{k_{1}}(r)\phi_{k_{2}}(r)
\eta^{\dagger}_{k_{1},\sigma_{1}}\eta_{k_{2},\sigma_{2}}\,
c^{\dagger}_{k_{1},\sigma_{1}}(t)c_{k_{2},\sigma_{2}}(t)\,,\\
\hat{n}(1')&=\mathop{\sum_{k_{3},\sigma_{3}}}_{k_{4},\sigma_{4}}
\phi^{\dagger}_{k_{3}}(r')\phi_{k_{4}}(r')
\eta^{\dagger}_{k_{3},\sigma_{3}}\eta_{k_{4},\sigma_{4}}\,
c^{\dagger}_{k_{3},\sigma_{3}}(t')c_{k_{4},\sigma_{4}}(t')\,.
\end{split}
\end{equation}
By inserting previous relations into the equation for the density response we get
\begin{equation}
\label{densityfull}
\begin{split}
\chi(1,1')&=-\frac{i}{\hbar}\frac{1}{L^{2}}\mathop{\sum_{k_{1}\cdots k_{4}}}_{\sigma_{1}\cdots\sigma_{4}}
\eta^{\dagger}_{k_{1},\sigma_{1}}\eta_{k_{2},\sigma_{2}}
\eta^{\dagger}_{k_{3},\sigma_{3}}\eta_{k_{4},\sigma_{4}}\,
{\rm e}^{i(k_{2}-k_{1})r+i(k_{4}-k_{3})r'}\\
&\times\big<\big[c^{\dagger}_{k_{1},\sigma_{1}}(t)c_{k_{2},\sigma_{2}}(t),
c^{\dagger}_{k_{3},\sigma_{3}}(t')c_{k_{4},\sigma_{4}}(t')\big]\big>\,.
\end{split}
\end{equation}
Now, we assume the situation of high Zeeman field, where only the lowest band, $\sigma=-$, is occupied. Therefore,
in Eq.\,(\ref{densityfull}) the unique combination that we need is 
\begin{equation}
\label{densityfull2}
\begin{split}
\chi(1,1')&=-\frac{i}{\hbar}\frac{1}{L^{2}}\sum_{k_{1}\cdots k_{4}}
\eta^{\dagger}_{k_{1},-}\eta_{k_{2},-}
\eta^{\dagger}_{k_{3},-}\eta_{k_{4},-}\,
{\rm e}^{i(k_{2}-k_{1})r+i(k_{4}-k_{3})r'}\\
&\times\big<\big[c^{\dagger}_{k_{1},-}(t)c_{k_{2},-}(t),
c^{\dagger}_{k_{3},-}(t')c_{k_{4},-}(t')\big]\big>\,.
\end{split}
\end{equation}
In order to describe the system with superconducting effects, we need to transform our $c$ operators in Eq.\,(\ref{densityfull2}) into the new ones $\alpha$ according to the Bogoliubov transformation we have found in Eq.\,(\ref{bogogog1}) or \,(\ref{bogogog2}).
In momentum space, the static response function at zero temperature, see Appendix \ref{StringBdensity} , then we get
 \begin{equation}
 \label{chisca}
 \chi(q)
 =-\frac{1}{L}\sum_{k}
A_{k,k+q}^{+}
\frac{\big(u_{k}v_{k+q}-u_{k+q}v_{k}
\big)^{2}}{E_{k,-}+E_{k+q,-}}\,,
\end{equation}
which is valid only for high Zeeman fields $B\gg\mu,\Delta$. Previous equation is the density-density response function also known as the Lidnhard function.
Previous equation represents the density-density response function in a 1D superconducting wire with Rashba and Zeeman interaction with only the lowest band occupied, also known as the Lindhard function. The coherence factors coming from the spin-orbit and Zeeman effects is given by
\begin{equation}
\begin{split}
A_{k,k+q}&=\frac{1}{2}\Bigg[1+\frac{B^{2}+\alpha_{R}^{2} k(k+q)}{\sqrt{B^{2}+\alpha_{R}^{2}k^{2}}\sqrt{B^{2}+\alpha_{R}^{2}(k+q)^{2}}}\Bigg]\,,
\end{split}
\end{equation}
while the one from the superconductor
\begin{equation}
\begin{split}
\big(u_{k}v_{k+q}-u_{k+q}v_{k}
\big)^{2}&=\frac{1}{2}\bigg[1-\frac{\varepsilon_{k,-}\varepsilon_{k+q,-}+\Delta_{p}(k)\Delta_{p}(k+q)}{E_{k,-}E_{k+q,-}} \bigg]\,,
\end{split}
\end{equation}
and 
\begin{equation}
\begin{split}
\Delta_{p}(k)&=\frac{\alpha_{R} k \Delta}{\sqrt{B^{2}+\alpha_{R}^{2}k^{2}}}\,,\\
\varepsilon_{k,-}&=\xi_{k}-\sqrt{B^{2}+\alpha_{R}^{2}k^{2}}\,,\\
 E_{k,-}&=\sqrt{\Delta^{2}_{k}+\varepsilon_{k,-}^{2}}\,.
\end{split}
\end{equation}

 \begin{figure}[!ht]
\centering
\includegraphics[width=.9\textwidth]{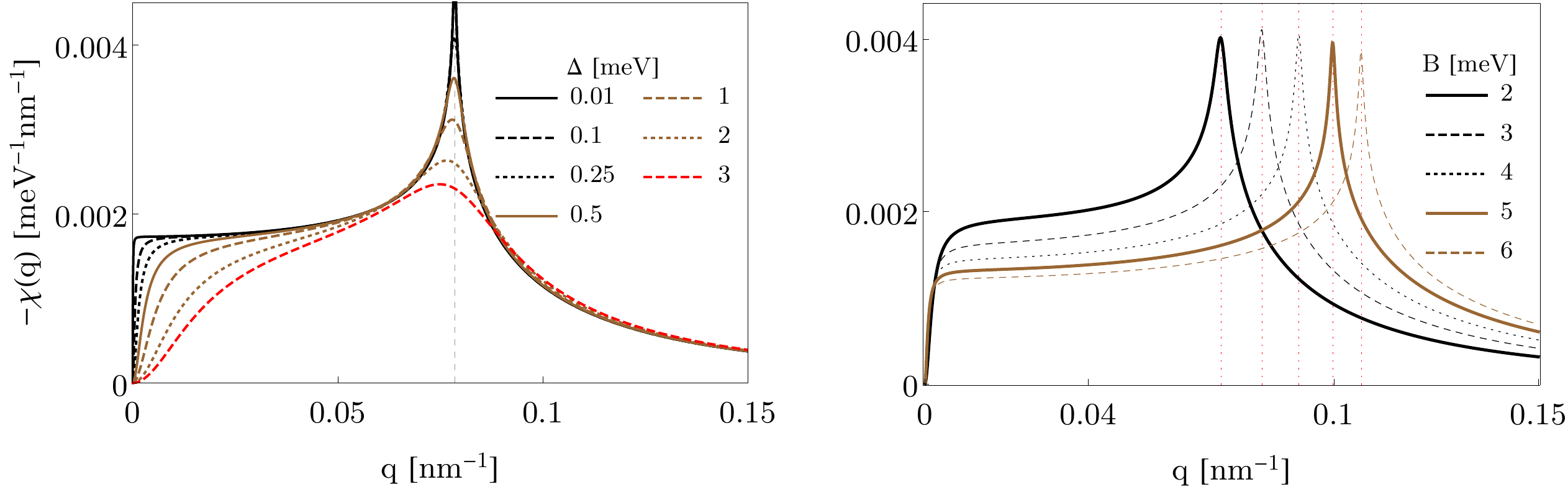} 
\caption[Density response function in a superconducting Rashba wire at high Zeeman]{(Color online) (Left) Density-density response function in a superconducting Rashba nanowire as a function of the momentum $q$ for different values of the superconducting pairing $\Delta$ at $B=2.6$\,meV (left panel), and different values of the Zeeman field (right panel).
In the former, observe the clear resonance at $q=2k_{F}$, which is washed out when $\Delta$ increases, as we indeed expected, while in the latter such resonance move to higher values of momentum as the Zeeman field increases.
Parameters here are: $\alpha_{R}=30$\,meVnm, $\mu=1$\,meV and $\Delta=0.25$\,meV.
}
\label{figchiq}
\end{figure}
In Fig.\,\ref{figchiq} the density response function, given by Eq.\,(\ref{chisca}), is plotted. The expected resonance at $q=2k_{F}$ is observed, which consequently is suppressed by increasing the superconducting pairing. Notice that the small momentum limit is really sensible to any change in $\Delta$, while higher momentum is roughly constant.
This resonance at $2k_{F}$ gives rise to the so-called Friedel oscillations, which can be observed by Fourier transformed Eq.\,(\ref{chisc}) into $r$-space. The Fermi momentum is
 $k_{F}=k_{+}=\sqrt{k_{\mu}^{2}+2k_{SO}^{2}+\sqrt{(k_{\mu}^{2}+2k_{SO}^{2})^{2}-k_{\mu}^{4}+k_{Z}^{4}}}$, where $k_{\mu}=\sqrt{2m\mu/\hbar^{2}}$, $k_{Z}=\sqrt{2mB/\hbar^{2}}$ and $k_{SO}=m\alpha_{R}/\hbar^{2}$.
 
 \begin{figure}[!ht]
\centering
 \includegraphics[width=.9\textwidth]{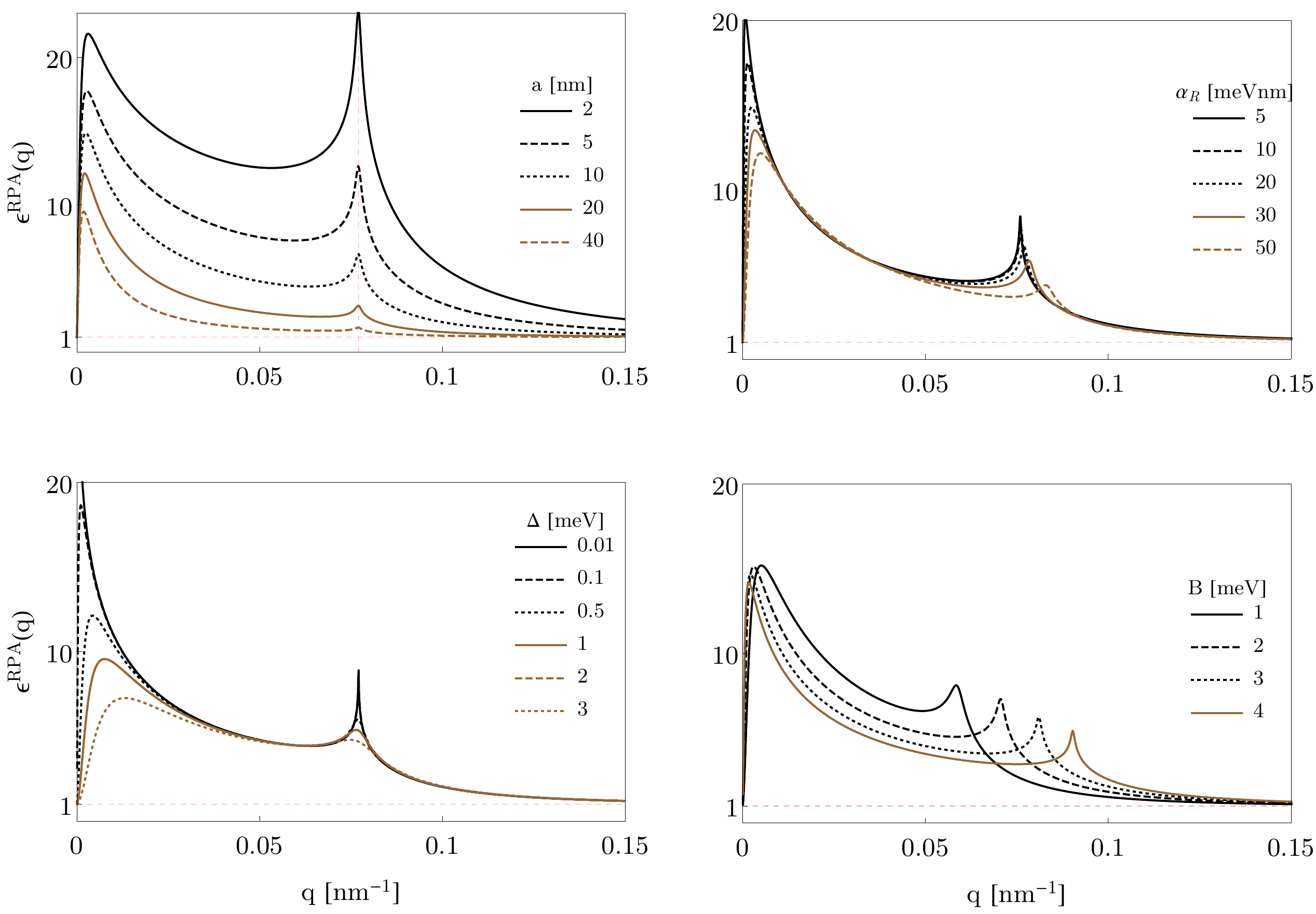} 
\caption[RPA dielectric function in a superconducting Rashba wire at high Zeeman]{(Color online) Dielectric function in the RPA limit in a superconducting Rashba nanowire at high Zeeman field as a function of the momentum $q$. Top left: for different values of the wire's radius. Top right: for different values of the SOC. Bottom left: for different values of the superconducting pairing. Bottom right: for different values of the Zeeman field.
 Important to notice here is that an enhancement in the Zeeman field, moves the resonances to higher momentum as expected, and changes the low momentum behaviour of the dielectric function. 
Parameters here are: $\alpha_{R}=30$\,meVnm, $\mu=1$\,meV and $\Delta=0.25$\,meV.
}
\label{figepsc}
\end{figure}

In Fig.\,\ref{figepsc} the dielectric function in the RPA limit, given by Eq.\,(\ref{dielectricEPRPA}), is plotted. 
The resonance at $q=2k_{F}$ is still visible for a certain range of the wire's radius $a$, but
it is completely lost when $k_{F}a\gg1$ (see top right panel). A feature that is really sensible to any change in $a$ is the small $q$ region with a complex dependence on $q$, while the high momentum one exhibit a decay behaviour converging to $1$.
The dependence of the dielectric function on the spin-orbit coupling,  superconductivity, Zeeman and superconductivity, is given in top right, bottom left and right panels, respectively.  An increase in the SOC, the resonance peak is slightly moved towards higher momenta, as indeed expected indeed, and the high of $\epsilon^{RPA}$ at small $q$ is decreased, but its width is constant. 
The effect of the superconducting pairing is observed in the bottom left panel. While it washes out the resonance when increased but maintaining fixed its the position, the hight of the dielectric function is considerably reduced.
In the bottom right panel, the Zeeman field is increased. The position of the resonance is then also moved towards higher momenta as in the case when the SOC is increased. An important feature of the small momentum window is that the width of resonance-like behaviour is varied, unlike the situation with different SOC and superconductivity.

Now, we are in a position to investigate the screened potential within the RPA approach $\phi_{scr}^{RPA}(r)$ \cite{vignale}.
In Subsection\,\ref{generalconcepts} we described the screened potential $\phi_{scr}^{RPA}(r)$ for a point charge with electron-electron interaction in one dimension $V(q)$ given by Eq.\,(\ref{pot1d}). Such screened potential is described by Eq.\,(\ref{phirpa1}) in momentum space or by Eq.\,(\ref{phirpa2}) in real space, for $\omega=0$,
\begin{equation}
\label{phirpa1xx}
\phi^{RPA}_{scr}(q)=\frac{V(q)}{\epsilon^{RPA}(q)}\,,\quad \rightarrow\quad  \phi^{RPA}_{scr}(r-r')=\int \frac{dq}{2\pi}\,\phi_{scr}^{RPA}(q)\,{\rm e}^{iq(r-r')}\,.
\end{equation}
\begin{figure}[!ht]
\centering
\includegraphics[width=.8\textwidth]{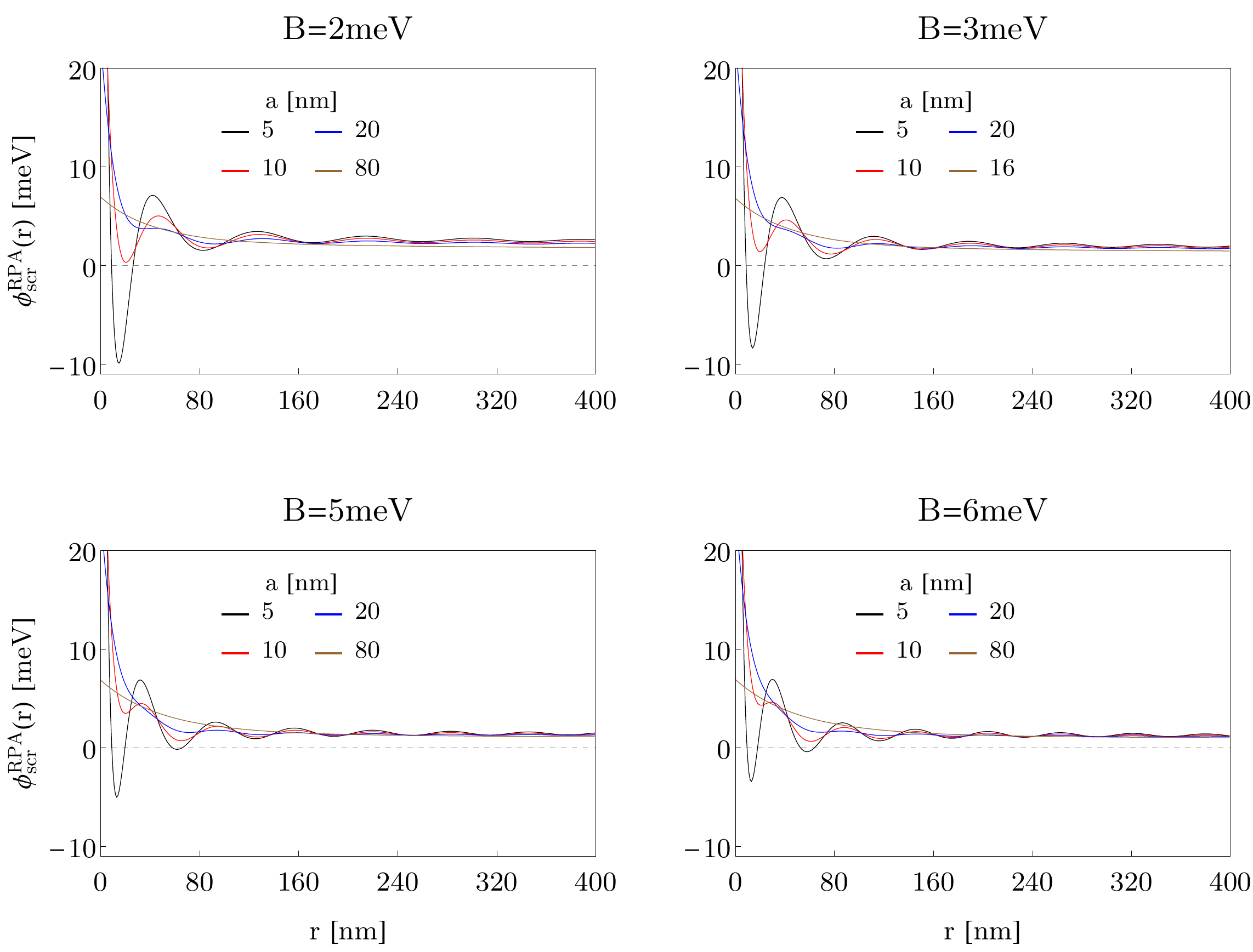} 
\caption[RPA screened potential in an infinite superconducting Rashba wire as a function of $r$]{(Color online) RPA screened potential in an infinite superconducting Rashba nanowire as a function of $r$, from Eq.\,(\ref{phirpa1xx}) for different values of the Zeeman field.
Parameters here are: $\alpha_{R}=30$\,meVnm, $\mu=1$\,meV, and $\Delta=0.25$\,meV.}
\label{phirLZERO}
\end{figure}
\begin{figure}[!ht]
\centering
\includegraphics[width=.8\textwidth]{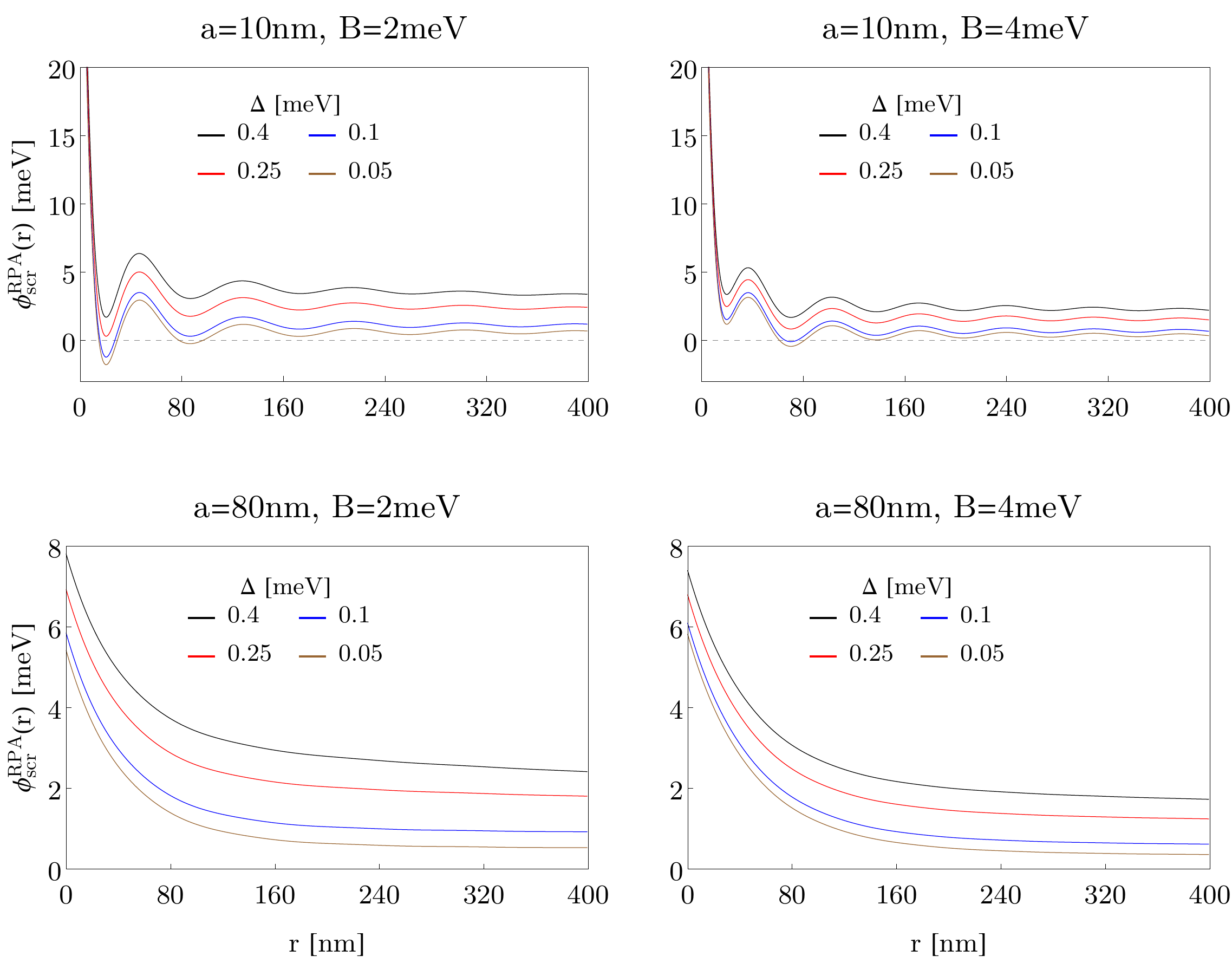} 
\caption[RPA screened potential in an infinite superconducting Rashba wire as a function of $r$ for different $\Delta$]{(Color online) RPA screened potential in an infinite superconducting Rashba nanowire as a function of $r$, from Eq.\,(\ref{phirpa1xx}) for different values of the pairing potential $\Delta$. Different panel correspond to different Zeeman fields.
Parameters here are: $\alpha_{R}=30$\,meVnm, $\mu=1$\,meV, and $\Delta=0.25$\,meV.}
\label{phirLZERO2}
\end{figure}
Now,  we discuss the screened potential $\phi^{RPA}_{scr}(r)$ for an infinite nanowire with Rashba SOC, Zeeman interaction and superconductivity calculated from Eq.\,(\ref{phirpa1xx}) as a function of the distance $r$. 
Fig.\,\ref{phirLZERO} show such situation for different values of the Zeeman field $B$  and different values of the wire's radius $a$. The potential exhibits a clear decay for small $r$ with the so-called Friedel oscillations whose amplitude is reduced as $r$ increases. Note that the oscillations are  reduced by increasing the wire's radius and completely washed out for realistic experimental values (see brown curve). Notice that for small values of $a$ and $r$ the screened potential takes huge values and diverges for $\rightarrow0$ due to the logarithmic divergence of the 1D Coulomb potential $V(r)$.
The oscillatory behaviour depends on the Fermi wave vector $k_{F}$ defined in Eq.\,(\ref{fermik}) and therefore on the system parameters ($B,\alpha_{R},\mu$). 
For increasing such parameters the oscillations become more extended so that they are not observed (not shown).
The decay of the screened potential is slow and the finite value of the potential for long distances $r$ strongly depends on the superconducting pairing $\Delta$ as shown in Fig.\,\ref{phirLZERO2}, where we present the screened potential as a function of $r$ for different values of the superconducting pairing $\Delta$ with oscillations (small $a$, top row) and without them (large $a$, bottom panel).
From such plots one can conclude that the long $r$ limit do not decay, however, we have checked that $\phi^{RPA}_{scr}(r)$ in fact exhibits a really slow decay.
First, the superconducting pairing does not influence on the shape of the oscillations for small $r$. As increasing $r$ the potential for different $\Delta$ acquire different values but still preserving the oscillatory behaviour. For long $r$ and small $\Delta$ the values of the screened potential are reduced. This effect is observed for different values of the wire's radius $a$, as shown in Fig.\,\ref{phirLZERO2}.

On the other hand the induced density was defined in momentum and real space by Eqs.\,(\ref{ninduced2}), 
\begin{equation}
\label{ninduced2x}
n_{ind}^{RPA}(q)=\chi(q)\phi^{RPA}_{ext}(q)\,,
\end{equation}
and in real space by
\begin{equation}
\label{ninduced2x2}
n_{ind}^{RPA}(r-r')=\int \frac{dq}{2\pi}\,n_{ind}^{RPA}(q)\,{\rm e}^{iq(r-r')}\,,
\end{equation}
where $\chi$ is the density-density response function.

\begin{figure}[!ht]
\centering
\includegraphics[width=.6\textwidth]{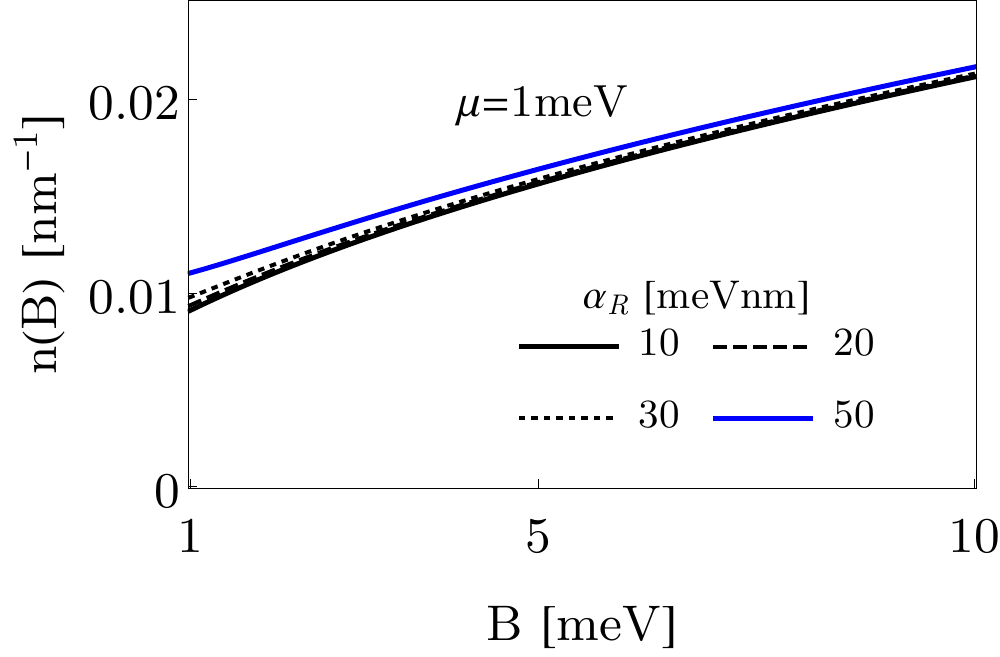} 
\caption[Electron density for an infinite wire as a function of the Zeeman field]{(Color online) Electron density for an infinite wire as a function of the Zeeman field.
Parameters here are: $\alpha_{R}=30$\,meVnm, $\mu=1$\,meV.}
\label{n0}
\end{figure}

The induced density $n_{ind}^{RPA}$ has to be understood as a correction for the electron density $n$ of the infinite wire given by Eq.\,(\ref{nBhigh} for high Zeeman fields), resulting in $n+n_{ind}^{RPA}$. In Fig.\,\ref{n0} we plot the electron density given by Eq.\,(\ref{nBhigh}) for an infinite Rashba wire. Notice the dependence on the system's parameters and it increased as one increases the Zeeman field.

\begin{figure}[!ht]
\centering
\includegraphics[width=.8\textwidth]{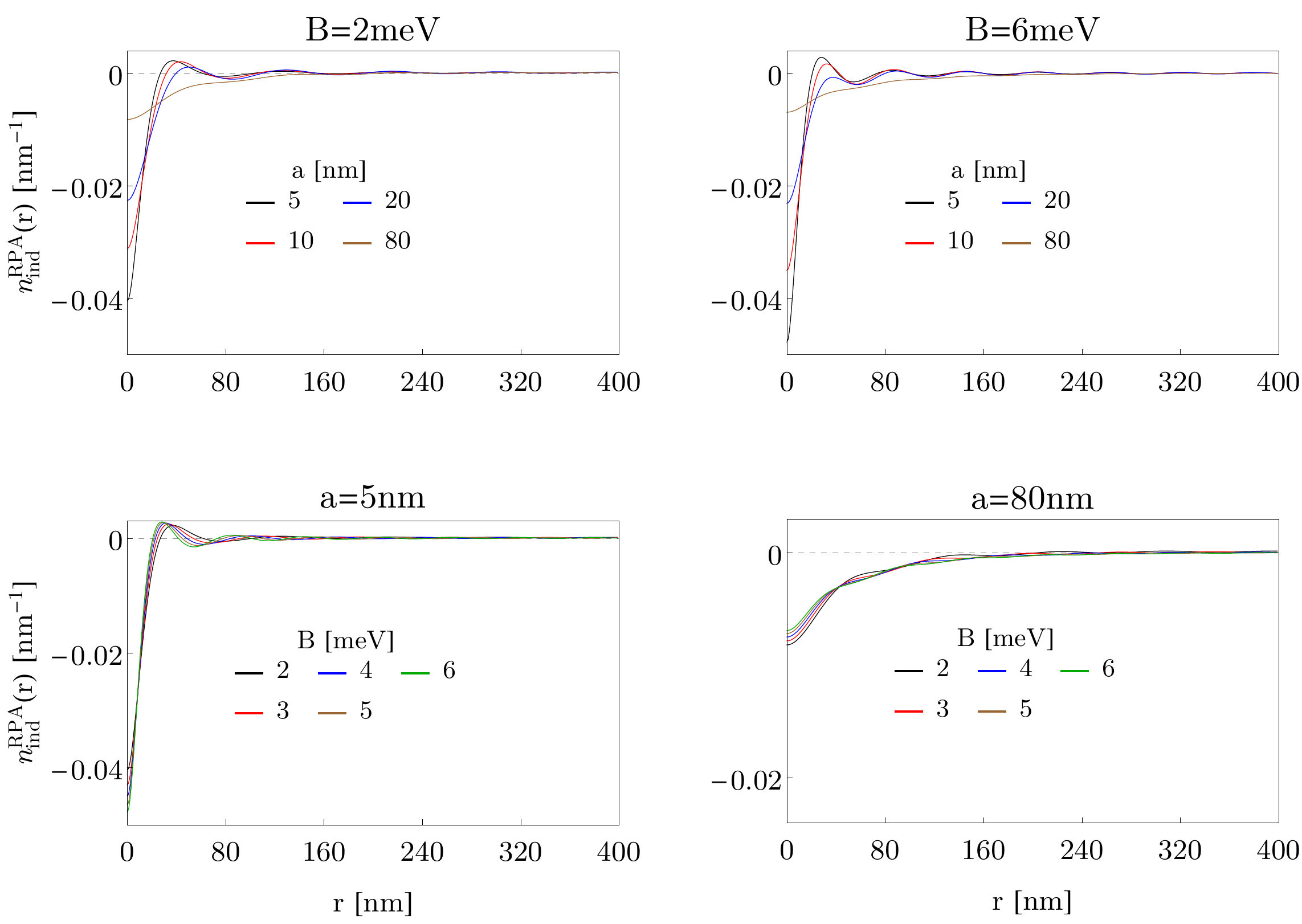} 
\caption[RPA induced density in an infinite superconducting Rashba nanowire as a function of $r$]{(Color online) RPA induced density in an infinite superconducting Rashba nanowire as a function of $r$, from Eq.\,(\ref{ninduced2x}). Different panels correspond for different values of the Zeeman field $B$ and in each panel we considered different values of the wire's radius $a$.
Parameters here are: $\alpha_{R}=30$\,meVnm, $\mu=1$\,meV, and $\Delta=0.25$\,meV.}
\label{ninduced1}
\end{figure}

In Fig.\,\ref{ninduced1} is plotted the induced density  $n_{ind}^{RPA}$ given by Eq.\,(\ref{ninduced2x2}) as a function of the distance $r$  (top row) for  two values of the Zeeman field and different radius of the wire $a$ and (bottom row)  for two values of the wire's radius and different Zeeman fields. 
The induced density  $n_{ind}^{RPA}$ at small $r$ tends to be negative which arises from the logarithmic divergence (small $a$) of the 1D Coulomb potential $V(q)$ and it is suppressed when the wire's radius $a$ takes realistic values $a\approx80$\,nm. As we have explained before, the negative value the induced density $n_{ind}^{RPA}$ means that the total electron density $n+n_{ind}^{RPA}$ is reduced at small $r$ due to the electron-electron interaction, since $n_{ind}^{RPA}$ corresponds to deviations from the electron density in equilibrium without perturbation ($n$ for an infinity wire given by Eq.\,(\ref{nBhigh})). As $r$ increases  the induced density captures the oscillatory pattern of the density-density response function $\chi$, which are suppressed at long distances $r$.

In order to take into account the finite length of the wire, $L$, we have defined a new potential $\bar{\phi}_{scr}^{RPA}(r)$, where $L$ represents the wire's length,
\begin{equation}
\label{newphiL2}
\begin{split}
\bar{\phi}_{scr}^{RPA}(r)
&=\frac{n_{0}}{2\pi}\int dq\,{\rm e}^{iqr}\,\frac{V(q)}{\epsilon^{RPA}(q)}\,\frac{i}{q}\Big[{\rm e}^{-iqL}-1\Big]\,,
\end{split}
\end{equation}
and for the induced density from Eq.\,(\ref{ninduced4}), we get
\begin{equation}
\label{ninduced4x}
\bar{n}_{ind}^{RPA}(r)=\frac{n_{0}}{2\pi}\int_{-\infty}^{\infty} dq\,n_{ind}^{RPA}(q)\,{\rm e}^{iq(r-r')}\frac{i}{q}\Big[{\rm e}^{-iqL}-1\Big]\,.
\end{equation}
Notice that for simplicity we have defined the electron density in the wire as $n_{0}=1/L$, being $L$ the length of the wire, and do not follow more complicated schemes as the Poisson approach.
By doing such assumption for $n_{0}$ we do not have any dependence on the system parameters such as $B,\alpha_{R},\mu$ that in principle $n_{0}$ has.
A full Poisson-BdGs solution should take into account these details.
\begin{figure}[!ht]
\centering
\includegraphics[width=.9\textwidth]{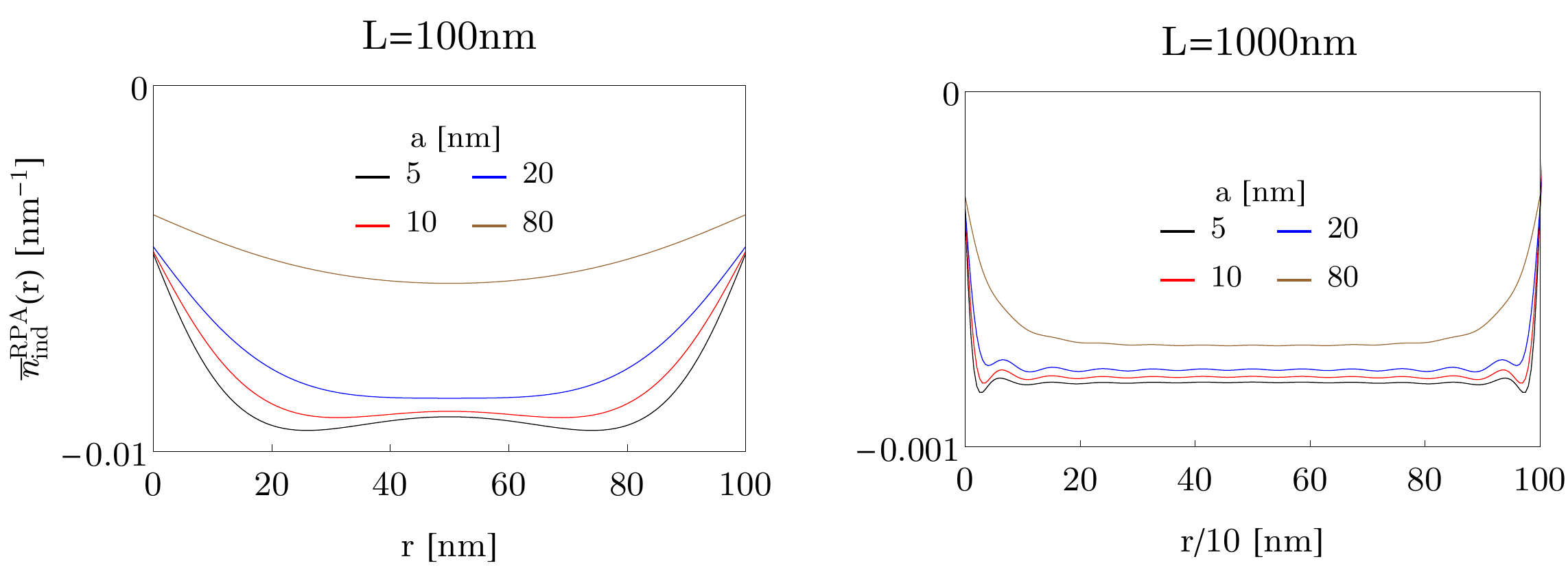} 
\caption[RPA induced density in a finite superconducting Rashba nanowire as a function of $r$]{(Color online) RPA induced density in a finite superconducting Rashba nanowire as a function of $r$, from Eq.\,(\ref{ninduced4x}). Different panels correspond for different values of the Zeeman field $B$ and in each panel we considered different values of the wire's radius $a$.
Parameters here are: $L=100$\,nm, $\alpha_{R}=30$\,meVnm, $\mu=1$\,meV, and $\Delta=0.25$\,meV.}
\label{ninduced1L}
\end{figure}

\begin{figure}[!ht]
\centering
\includegraphics[width=.9\textwidth]{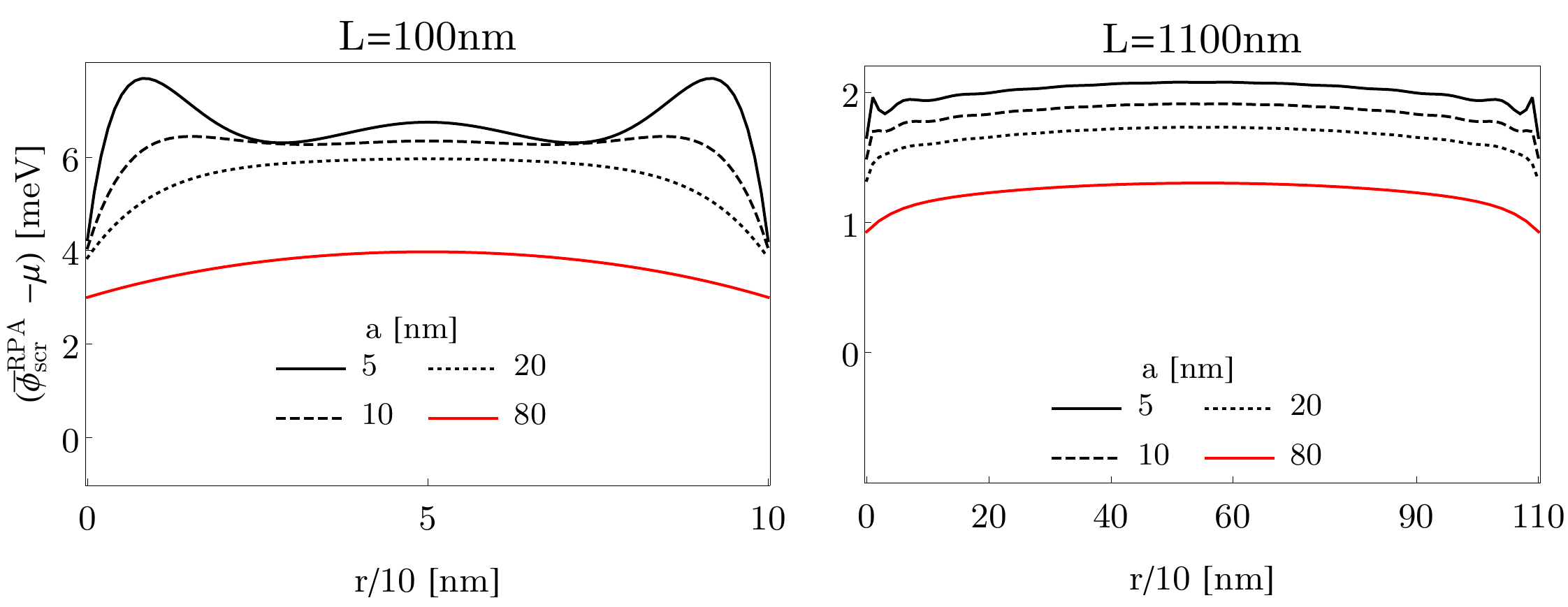} 
\caption[RPA screened potential in a finite length superconducting Rashba nanowire as a function of $r$]{(Color online) Screened potential in the RPA limit in a superconducting Rashba nanowire as a function of $r$, from Eq.\,(\ref{newphiL2}), in a short (left panel) and long wire (right).
Parameters here are: $\alpha_{R}=30$\,meVnm, $\mu=1$\,meV, $B=2$\,meV and $\Delta=0.25$\,meV.}
\label{screr}
\end{figure}

 In Fig.\,\ref{screr} we present the screened potential as a function of the position calculated from Eq.\,(\ref{newphiL2}) for a short and long wire, respectively, for different values of the wire's radius. The denomination short and long is made by comparing the wire's length with $1/k_{+}$.
 The screened potential develops the expected oscillatory behaviour in the wire, which represent the well-known Friedel oscillations. In a long wire it acquires a higher amplitude, although such oscillations are more visible in a short wire than in a long one as one can indeed see. Moreover, one notices that by increasing the wire's radius  the oscillatory behaviour of the screened potential is reduced, and even washed out. For a reasonable long wire with a realistic radius, the potential is constant in the bulk, while it is reduced at the edges.
The screened potential $\phi_{scr}^{RPA}$ introduces a change in the chemical potential $\mu$ so that the new chemical potential is $\mu^{new}=\mu-\phi_{scr}^{RPA}$. First, one notices that due to this dependence, the topological transition point determined by $B_{c}=\sqrt{\Delta^{2}+\mu^{2}}$ is affected and 
becomes  $B_{c}^{new}=\sqrt{\Delta^{2}+(\mu^{new})^{2}}$. A full analysis requires to include the full band calculation in the density response function and therefore be able to map the topological transition by increasing the Zeeman field, which is planned for a future study.
Based on the results obtained in the strong Zeeman regime, this new chemical $\mu^{new}$ allows us to investigate the Majorana wave function  amplitude as a function of the position in the wire and compare it to the situation with $\mu$. Additionally, we can also calculate the low-energy Andreev spectrum and focus on its evolution of $B$, being valid only for strong $B$. All these calculations form part of a future study in order to provide a full RPA analysis of the Majorana problem in nanowires.

\section{Conclusions}
 In this Chapter we have calculated the density-density response function in one-dimensional superconducting nanowires with spin-orbit coupling and Zeeman fields. Firstly, we have investigated the density-density response function in isolated wires. Then, we have considered a proximitized nanowire, where the density-density response was calculated taking into account zero-interband pairing and later in a strong Zeeman regime, being the latter of great relevance as such limit corresponds to a situation with MBSs. The analysis was done within the linear response theory framework. 
 Then, we used these results in order to investigate electron-electron interactions in the longitudinal direction of the wire within the Random Phase Approximation approach. In this part, we have calculated the dielectric function, and then the charge density-density response as well as the screened potential for the infinite wire as well as for a finite length wire. 
 



\chapter{{\bf Conclusions}} 
\label{Chapter7}
\lhead{Chapter 7. \emph{Conclusions}} 

In this thesis we have investigated hybrid superconductor-semiconductor junctions made of semiconducting nanowires with Rashba spin-orbit coupling.  
Along this thesis, we emphasise the importance of employing hybrids superconductor-semiconductor nanowire junctions 
towards the unambiguously detection of Majorana bound states (MBSs) beyond zero-bias anomalies. We have analysed the low energy Andreev spectrum, phase and voltage-biased transport, have proposed a new scheme for engineering MBSs in trivial NS junctions and at the end have performed an analysis of screening properties in Rashba nanowires based on linear response theory and Random Phase Approximation.

The main results of the thesis can be summarised as follows:
\begin{itemize}

\item In chapter~\ref{Chap2a} we have presented in detail how hybrid NS and SNS junctions are modelled based on semiconductor nanowires with Rashba spin-orbit coupling. In this part, we have showed that emergence of Majorana bound states is distinguishable in the Andreev spectrum (see for instance Fig.\,\ref{figchap29}). Despite of being $2\pi$ periodic, the Josephson current exhibits a clear  reduction in the topological phase and when, additionally, the Majorana overlap is negligible (see for instance Fig.\,\ref{Iphishort1}) it develops a distinguishable sawtooth profile at $\varphi=\pi$.
 Remarkably,  the critical current traces the closing and reopening of the topological gap, exhibiting a robust and non-trivial feature at the gap closing, remaining finite in the topological phase and revealing the Majorana oscillations (see Fig.\,\ref{Icshort1}). 

\item In chapter~\ref{Chap3} we have studied transport in a voltage-biased short superconductor-normal-superconductor (SNS) junction made of semiconducting nanowires with Rashba spin-orbit coupling,  as the applied Zeeman field drives the system into the topological superconducting phase. We have shown that the dissipative multiple Andreev reflection (MAR) current at different junction transparencies exhibits unique features related to topology such as gap inversion, the formation of MBSs and fermion-parity conservation. 
We predict the halving of MAR steps owing to the presence of Majorana bound states in the junction (see Fig.\,\ref{fig:I0}(a,d)) as well as the possibility of observing the transition into the topological superconducting phase directly in the critical current of the junction (see Fig.\,\ref{fig:Ic}).

\item In chapter~\ref{Chap2} we have extended traditional conceptions and made a detailed study on the role that confinement and helicity have on normal transport and on the sub-gap Andreev spectrum in short and long SNS junctions made of semiconducting nanowires (NWs) with strong Rashba spin-orbit coupling. We found that a long junction with a helical normal section but still in the topologically trivial regime, supports a low-energy sub-gap spectrum consisting of multiple zero-energy crossings 
that smoothly evolve towards MBSs as the junction becomes topological (see for instance Fig.\,\ref{fig12}).  
This indeed suggests an interesting connection between subgap parity crossings in helical junctions with $B<B_c$ and Majorana bound states in topological ones with $B>B_c$.

\item In chapter~\ref{ChapEPs} we have proposed an alternative route for engineering MBSs with non-topological superconducting wires, which consists of creating a sufficiently transparent normal-superconductor (NS) junction on a Rashba wire, with a topologically trivial superconducting side and a helical normal side. The zero-energy state emerges as the junction traverses an exceptional point at the threshold transparency and becomes robustly pinned to zero energy without fine tuning by virtue of charge-conjugation symmetry.  Finally, we have shown that relevant transport and spectral properties associated to these zero energy states, here dubbed are indistinguishable from those of conventional MBSs (see for instance Fig.\,\ref{fig:phasedep}).

\item In chapter~\ref{ChapDensity} we have investigated screening properties of proximitized nanowires with SOC and Zeeman fields within RPA and linear response theory. This, indeed, represents a relevant step for experiments trying to measure Majorana bound states and their non-trivial overlap. 
\end{itemize}

With the recent advance in fabrication techniques and induced hard-gaps technology \cite{chang15,Krogstrup15}, we expect that our research reported here will support new experiments towards the unambiguously detection of MBSs.


\chapter{\bf Resumen y conclusiones }
\label{Chapter7spanish}
\lhead{Resumen y conclusiones }


En 1937, Ettore Majorana postul\'{o} que la equation de Dirac se puede escribir en una base donde sus soluciones son campos reales \cite{majorana}.
Esto permite dividir la ecuaci\'{o}n de Dirac en dos  sistemas de ecuaciones independientes, cada una con dos ecuaciones acopladas. 
La solucion a uno de estos representa un campo real, $\psi$, y por ende la part\'{i}cula asociada a dicho campo es su propia 
antipart\'{i}cula, $\psi=\psi^{\dagger}$.
Esta part\'{i}cula es fermi\'onica, esp\'{i}n $1/2$, neutral, y se denomina fermi\'{o}n de Majorana.
El origen de los fermiones de Majorana est\'{a} en f\'{i}sica de part\'{i}culas, donde aun no se ha determinado completamente 
si los neutrinos obedecen dicha fenomenolog\'{i}a. 

Por otro lado, en materia condensada ha despertado mucho inter\'{e}s la idea de tener part\'{i}culas que sean sus propias antipart\'{i}culas. 
El origen de este inter\'{e}s est\'{a} en varias predicciones te\'{o}ricas en torno al a\~{n}o 2010 que demostraron que los superconductores topol\'{o}gicos, la variante superconductora de los aislantes topol\'{o}gicos, poseen excitaciones de cuasipart\'{i}cula de tipo Majorana \cite{kitaev,Fu:PRL08}.  La conexi\'{o}n formal con la f\'{i}sica de altas energ\'{i}as se encuentra en la simetr\'{i}a de conjugaci\'{o}n de carga: la simetr\'{i}a electr\'{o}n-hueco, en lenguaje de f\'{i}sica de la materia condensada, que todo superconductor posee. 
Efectivamente, las excitaciones sobre el estado fundamental BCS en un superconductor son superposiciones cu\'{a}nticas de electrones y huecos que se denominan cuasipart\'{i}culas de Bogoliubov. Una cuasipart\'{i}cula de Bogoliubov a energ\'{i}a cero es perfectamente neutra (mitad electr\'{o}n-mitad hueco) y por tanto, igual a su propia antipart\'{i}cula. El \'{u}ltimo ingrediente para que una excitaci\'{o}n de Bogoliubov sea de tipo Majorana es que no tenga degeneraci\'{o}n de esp\'{i}n y, por tanto,  todas las propuestas para generar superconductividad topol\'{o}gica se basan en superconductores de onda $p$, debido a que estos poseen fases topol\'{o}gicas.
T\'{e}ngase en cuenta que, en materia condensada, lo que ha llamado mucho la atenci\'{o}n son estas cuasipart\'{i}culas con car·cter Major\'{a}nico pero a energ\'{i}a cero.
 Estas no tienen 
analog\'{i}a en f\'{i}sica de part\'{i}culas, pero a\'{u}n obedecen $\gamma=\gamma^{\dagger}$ (donde $\gamma$ viene a ser un operador Major\'{a}nico). Las fases con estados de Majorana se pueden distinguir de otras 
por medio de invariantes topol\'{o}gicos, que en este caso es el n\'{u}mero Major\'{a}nico. Los estados de Majorana a energ\'{i}a cero aparecen como estados ligados 
a interfaces con diferente topolog\'{i}a y es por eso que tambi\'{e}n se les denomina estados ligados de Majorana (MBSs).
Aparte de su inter\'{e}s fundamental, estas excitaciones en un superconductor topol\'{o}gico poseen estad\'{i}stica cu\'{a}ntica no-Abeliana, sin an\'{a}logos en el modelo est\'{a}ndar de la f\'{i}sica de part\'{i}culas, que da lugar a una forma de computaci\'{o}n cu·ntica muy robusta \cite{RevModPhys.80.1083}. 

Hasta hace poco, la necesidad de trabajar con superconductores de onda $p$ se consideraba una limitaci\'{o}n poco menos que insalvable, dada su escasez en la naturaleza y su sensibilidad al desorden. Esta situaci\'{o}n ha cambiado en los \'{u}ltimos 
a\~{n}os con varias propuestas te\'{o}ricas que demuestran que es posible generar de manera efectiva superconductividad de onda $p$ combinando materiales con fuerte acoplo esp\'{i}n-\'{o}rbita (aislantes topol\'{o}gicos o semiconductores con acoplo Rashba) a superconductores usuales con simetr\'{i}a de onda s \cite{Alicea:PRB10,PhysRevLett.105.177002,Lutchyn:PRL10}. 
Una de las plataformas mas prometedoras consiste en nanohilos semiconductores  con fuerte acoplamiento de esp\'{i}n \'{o}rbita \cite{PhysRevLett.105.177002,Lutchyn:PRL10}.
Cuando en el nanohilo se induce superconductividad de onda $s$, el nanohilo entra en la fase topol\'{o}gica cuando $B>B_{c}\equiv B_{c}$, donde 
$B_{c}=\sqrt{\mu^{2}+\Delta^{2}}$ es el campo cr\'{i}tico y se\~{n}ala la transition topol\'{o}gica dando origen a la inversi\'{o}n del gap, $\mu$ es el potencial qu\'{i}mico del nanohilo y $\Delta$ el pairing superconductor inducido en el nanohilo.

La posibilidad de obtener superconductividad topol\'{o}gica combinando estos materiales m\'{a}s comunes gener\'{o} altas expectativas y muchos laboratorios del mundo se lanzaron a la carrera para detectar por primera vez una part\'{i}cula de Majorana. 
El experimento pionero corresponde al grupo de Delft \cite{Mourik:S12}, donde miden la  emergencia de un pico a voltaje cero en la conductancia diferencial  de una uni\'{o}n h\'{i}brida semiconductor-superconductor como caracter\'{i}stica de la emergencia de un estado ligado de Majorana. 
Experimentos posteriores \cite{xu,Das:NP12,Finck:PRL13,Churchill:PRB13,Lee:13} tambi\'{e}n  muestran datos consistentes con cuasipart\'{i}culas de Majorana  sin embargo no es posible descartar otras explicaciones alternativas del origen de los picos a voltaje cero, tales como desorden, efecto Kondo, etc. 

Muchas de las preguntas han sido parcialmente resultas \'{u}ltimamente, sin embargo, a\'{u}n se requiere estudios adicionales con nuevas propuestas. Adicionalmente, recientemente ha habido un gran avance en la fabricaci\'{o}n de uniones h\'{i}bridas, donde se ha logrado sintetizar sistemas con hard-gaps inducidos \cite{chang15,Krogstrup15,zhang16}.

El objetivo de esta tesis es proponer y analizar posibles protocolos de detecci\'{o}n m\'{a}s all\'{a} de esta primera generaci\'{o}n experimental.
Por lo antes expuesto, esta tesis est\'{a} dedicada a investigar propiedades de transporte en sistemas h\'{i}bridos NS y SNS basaos en nanohilos semiconductores, as\'{i} como tambi\'{e}n estudiar la emergencia de los estados de Majorana en dichas estructuras.
Una de las ventajas de las uniones SNS es, por ejemplo, que permite estudiar el monitoreo detallado de algunos observables con respecto a la diferencia 
de fase superconductora.

En el cap\'itulo~\ref{Chapter01} se da una introducci\'{o}n general a la superconductividad topol\'{o}gica en una dimension basados en el modelo de Kitaev para superconductividad de onda $p$, y adem\'{a}s mostramos la emergencia de los estados de Majorana a energ\'{i}a cero al final del nanohilo. 
Luego se describe el modelo f\'{i}sico basado en nanohilos semiconductores con esp\'{i}n \'{o}rbita tipo Rashba, donde un campo magn\'{e}tico Zeeman $B$ es aplicado y superconductividad de onda $s$ es inducida en el nanohilo. Se asume que existe buen contacto entre el nanohilo y el superconductor de tal manera que correlaciones superconductoras sean inducidas al 
nanohilo por medio del efecto proximidad. Esto da origen a un nanohilo superconductor que cuando el el campo magn\'{e}tico aplicado es mayor que $B_{c}$ se convierte en un 
nanohilo superconductor topol\'{o}gico con estados de Majorana al final del nanohilo, uno en cada lado.
Aqu\'{i}, resaltamos que los estados de Majorana a energ\'{i}a cero aparecen en la fase topol\'{o}gica ligados a interfaces de diferente topolog\'{i}a, con su funci\'{o}n 
de onda decayendo hacia el bulk.

En el cap\'itulo~\ref{Chap2a}, en primer lugar, se introducen los conceptos b\'{a}sicos en uniones h\'{i}bridas Normal-Superconductor (NS) y 
Superconductor-Normal-Superconductor (SNS): reflexi\'{o}n de Andreev (AR) y estado ligado de Andreev (ABS). Aqu\'{i} tambi\'{e}n clasificamos las uniones SNS en cortas ($L_{N}<\xi$) 
o largas ($L_{N}>\xi$), donde $L_{N}$ es la longitud de la regi\'{o}n normal y $\xi$ es la longitud de coherencia superconductora.
Luego nos centramos en describir detalladamente como modelamos 
estas uniones h\'{i}bridas a partir de nanohilos semiconductores con esp\'{i}-\'{o}rbita tipo Rashba, sujetos a un campo magn\'{e}tico y donde superconductividad de onda $s$ es 
inducidad en el nanohilo. Para esto, discretizamos el Hamiltoniano para el nanohilo en una red tight-binding, para luego escribir el Hamiltoniano del sistema en espacio Nambu.
A partir de esto calculamos los niveles de energ\'{i}a ya sea para un nanohilo superconductor, uni\'{o}n NS o uni\'{o}n SNS. 
En el caso de uniones SNS, mostramos la evoluci\'{o}n detallada de los niveles como funci\'{o}n de la diferencia de fase desde la fase trivial hacia la fase topol\'{o}gica con 
estados de Majorana. Debido a la longitud finita de S, en este caso el sistema posee cuatro estados de Majorana cuando la diferencia de fase es $\pi$.
Resaltamos en esta parte que, debido a que \'{u}ltimamente se viene investigando experimentalmente espectroscop\'{i}a de niveles, es posible realizar un estudio experimental, similar 
al que hacemos, para detectar la presencia de los estados de Majorana. 
Adem\'{a}s mostramos que tanto la corriente Josephson como la corriente cr\'{i}tica, calculadas a partir de los niveles, poseen informaci\'{o}n relevante sobre los estados de Majorana.
Sorprendentemente, la corriente cr\'{i}tica como funci\'{o}n del campo Zeeman traza perfectamente la evolucion hacia la fase topol\'{o}gica. Esta adquiere un valor no nulo 
en la transici\'{o}n topol\'{o}gica, aqu\'{i} el sistema no tiene gap, y se mantiene finito a medida que el campo aumenta. Sin lugar a dudas, esto da origen a 
caracter\'{i}sticas no usuales y solamente relacionadas con la emergencia de los estados de Majorana. Por eso creemos que estas caracter\'{i}sticas 
pueden ser distinguibles experimentalmente en uniones con nanohilos.

En el cap\'{i}tulo~\ref{Chap3} se propone una forma potente y sencilla para detectar estados ligados de Majorana (MBSs) en una uni\'{o}n h\'{i}brida SNS, 
donde un voltage constante es aplicado.
Aqu\'{i}, se estudia transporte Josephson bajo voltage constante en una uni\'{o}n SNS hecha de nanohilos con esp\'{i}n-\'{o}rbita fuerte, a medida 
que el sistema experimenta una transici\'{o}n hacia la fase topol\'{o}gica cuando aumentamos el campo Zeeman. 
La uni\'{o}n SNS es formada por dos leads superconductores semi-infinitos, y por lo tanto solo dos estados de Majorana emergen en la uni\'{o}n. 
Usando la t\'{e}cnica de las funciones de Green de Keldysh, proponemos que las corrientes debido a las m\'{u}ltiples reflecciones de Andreev (MAR) as\'{i}como tambi\'{e}n
la corriente cr\'{i}tica sirven como una forma alternativa y potente para estudiar la transici\'{o}n topol\'{o}gica. Esto se hace posible debido al efecto directo 
que la inversi\'{o}n del gap, la formaci\'{o}n de MBSs y la conservaci\'{o}n de la paridad fermi\'{o}nica tienen en la corriente MAR para diferentes transparencias de la uni\'{o}n. 
Por otro lado, tambi\'{e}n demostramos que la corriente cr\'{i}tica es  inesperadamente finita para todo campo Zeeman debido a la contribuci\'{o}n del continuo, y exhibe una 
anomal\'{i}a en la transici\'{o}n topol\'{o}gica que podr\'{i}a ser experimentalmente localizada y distinguida, como tambi\'{e}n se predice en el cap\'itulo~\ref{Chap2a}.

En el cap\'{i}tulo~\ref{Chap2} estudiamos transporte normal y el espectro subgap en uniones superconductoras SNS hechas de nanohilos semiconductores 
con un fuerte acoplamiento esp\'{i}n-\'{o}rbita tipo Rashba. Nos centramos, en particular, en el papel que los efectos de confinamiento tienen en uniones bal\'{i}sticas largas. 
En el r\'{e}gimen normal, scattering en los dos contactos da lugar a dos caracter\'{i}sticas distintas de la conductancia: resonancias Fabry-Perot y dips Fano. 
Estas \'{u}ltimas se producen en presencia de un fuerte campo Zeeman $B$ que elimina un sector de esp\'{i}n en los leads (leads helicales), pero no en la regi\'{o}n 
central. Inversamente, una regi\'{o}n central helical entre los conductores no helicales presenta gaps helicales de la mitad de la cu\'{a}ntica conductancia, 
con oscilaciones helicales Fabry-Perot superpuestas. 
Estas caracter\'{i}sticas normales se traducen en distintos estados subgap cuando los leads se convierten en superconductores. 
En particular, resonancias Fabry-Perot dentro del gap helical se convierten en cruces de estados a energ\'{i}a cero protegidos por la paridad (cruces de paridad), muy por debajo del 
campo cr\'{i}tico $B_{c}$ en el que los leads superconductores se convierten en topol\'{o}gicos.
Como funci\'{o}n del campo Zeeman o energ\'{i}a de Fermi, estos modos oscilan alrededor de energ\'{i}a cero, formando loops caracter\'{i}sticos, que evolucionan continuamente 
hacia estados ligados de Majorana cuando $B$ excede $B_{c}$. En esta parte tambi\'{e}n se discute la relaci\'{o}n con la 
f\'{i}sica de los cruces de paridad de estados ligados de Yu-Shiba-Rusinov.

Sintetizar superconductores en su fase topol\'{o}gica con un gap robusto y bien definido es la mayor dificultad pr\'{a}ctica que presenta la propuesta original.
En el cap\'{i}tulo~\ref{ChapEPs} hemos investigado una plataforma alternativa para crear estados ligados de Majorana en nanohilos superconductors no topol\'{o}gicos.
Nuestra propuesta consiste en que en lugar de inducir la transici\'{o}n topol\'{o}gica en un nanohilo en proximidad a un superconductor de onda $s$, 
se debe conectar el nanohilo superconductor trivial a un metal normal helical por medio de un contacto transparente. 
Esta nueva propuesta representa un gran avance y tiene muchas ventajas sobre los esquemas convencionales, 
ya que los superconductores topol\'{o}gicos son sensitivos al desorden y llegar a la fase topol\'{o}gica implica un gran desaf\'{i}o experimental en muchos de ellos.
 Dicha uni\'{o}n puede ser realizad experimentalmente, por ejemplo, poniendo en proximidad una secci\'{o}n finita del nanohilo con acoplamiento esp\'{i}n-\'{o}rbita, y combinando 
 gates electrost\'{a}ticos y un campo Zeeman para llevar la secci\'{o}n que no est\'{a} en proximidad al superconductor hacia su fase helical.
En nuestra propuesta, los estados de Majorana a energ\'{i}a cero emergen en dicho sistema 
abierto sin la necesidad de fine-tuning como resultado de la simetr\'{i}a de conjugaci\'{o}n de carga, y puede ser ligado a la existencia de puntos excepcionales (EPs) en el espacio 
de par\'{a}metros, donde dos estados cuasi ligados de Andreev se bifurcan en dos estados ligados de Majorana a energ\'{i}a cero.
Despu\'{e}s del EP, uno de los estados cuasi ligados de Andreev se vuelve no-decaying a medida que la uni\'{o}n se aproxima a reflexi\'{o}n Andreev perfecta, resultando asi en un estado
de Majorana dark (MDS) localizado en la uni\'{o}n NS.
Aqu\'{i} se muestra que los MDSs poseen propiedades asociadas a los estados ligados de Majorana convencionales en sistemas cerrados, sin la necesidad de requerir superconductividad 
topol\'{o}gica.

Por otro lado, en el cap\'{i}tulo~\ref{ChapDensity} usamos teor\'{i}a de respuesta lineal para calcular la funci\'{o}n de respuesta de densidad en nanohilos con acoplamiento esp\'{i}n-\'{o}rbita y 
campo Zeeman aplicado. Mostramos tambi\'{e}n que en esto puede generalizarse cuando el sistema se vuelve superconductor. Para ello nos centramos en dos l\'{i}mites: el primero, donde el sistema 
posee dos bands pero el pairing entre bandas es cero y por ende posee una transici\'{o}n topol\'{o}gica; mientras que en el segundo nos centramos en el regimen de campos altos cuando el sistema
est\'{a} en la fase topol\'{o}gica. Luego, dentro de la Random Phase Approximation (RPA), calculamos la funci\'{o}n diel\'{e}ctrica, luego la densidad de carga y el potencial 
de apantallamiento. Debido a la simpleza del enfoque, esperamos que los resultados sean cualitativamente consistentes con m\'{e}todos auto-consistentes.

Esperamos que el trabajo presentado en esta tesis abra nuevos horizontes hacia la detecci\'{o}n de estados ligados de Majorana en materia condensada. 


\addtocontents{toc}{\vspace{1em}} 

\begin{appendices}
\chapter{\bf Appendix for Chapter \ref{Chapter01}}
\label{AppA0} 
\lhead{Appendix \ref{AppA0}. \emph{For Chapter 1}}
In this chapter we provide additional details for a complete description of the introduction, thus making this thesis self-contained.
\section{Details of the Bogoliubov-de Gennes Hamiltonian}
\label{BCS}
The reduced pairing Hamiltonian, given in the main text by Eq.\,(\ref{eq2}), emerges as a result of the electron-electron interaction Hamiltonian \cite{zagoskin},
\begin{equation} 
\label{eq1}
H=\sum_{k,\sigma}\xi_{k}\,c^{\dagger}_{k\sigma}c_{k\sigma}+\sum_{k,k',q}\sum_{\sigma,\sigma'}V(q)\,c^{\dagger}_{k+q,\sigma}c^{\dagger}_{k'-q,\sigma'}c_{k',\sigma'}c_{k,\sigma}\,,
\end{equation}
where the first term describes free electrons, while the second one the interactions with strength $V(q)$.
Regarded to Eq.\,(\ref{eq2}), we write down the two parts of such Hamiltonian, $H=H_{kin}+H_{int}$,
\begin{equation}
H_{kin}=\sum_{k,\sigma}\xi_{k}\,c^{\dagger}_{k\sigma}c_{k\sigma}\,,\quad
H_{int}=-\sum_{k,k'}V_{k,k'}\,c^{\dagger}_{k,\uparrow}c^{\dagger}_{-k,\downarrow}
c_{-k',\downarrow}c_{k',\uparrow}\,.
\end{equation}
Because fluctuations are small with respect to non-zero expectation values of operators 
$c_{-k,\downarrow}c_{k,\uparrow}$, it is appropriate to write down \cite{tinkham}
\begin{equation}
\begin{split}
c_{-k',\downarrow}c_{k',\uparrow}&=\average{c_{-k',\downarrow}c_{k',\uparrow}}+\big(c_{-k',\downarrow}c_{k',\uparrow}-\average{c_{-k',\downarrow}c_{k',\uparrow}}\big)\\
c_{k,\uparrow}^{\dagger}c_{-k,\downarrow}^{\dagger}&=\big<c_{k,\uparrow}^{\dagger}c_{-k,\downarrow}^{\dagger}\big>+\big(c_{k,\uparrow}^{\dagger}c_{-k,\downarrow}^{\dagger}-
\big<c_{k,\uparrow}^{\dagger}c_{-k,\downarrow}^{\dagger}\big>\big)\,,
\end{split}
\end{equation}
where the second terms in the right hand side, in previous two equations, represents the fluctuation around which the averages of the operators, the first terms, fluctuate. Then, the interaction part of the pairing Hamiltonian can be written as,
\begin{equation}
\begin{split}
H_{int}&=\sum_{k,k'}V_{k,k'}
\Big[\big<c_{k,\uparrow}^{\dagger}c_{-k,\downarrow}^{\dagger}\big>\average{c_{-k',\downarrow}c_{k',\uparrow}}
+
\big<c_{k,\uparrow}^{\dagger}c_{-k,\downarrow}^{\dagger}\big>(c_{-k',\downarrow}c_{k',\uparrow}-\average{c_{-k',\downarrow}c_{k',\uparrow}})\\
&+
(c_{k,\uparrow}^{\dagger}c_{-k,\downarrow}^{\dagger}-\big<c_{k,\uparrow}^{\dagger}c_{-k,\downarrow}^{\dagger}\big>)\average{c_{-k',\downarrow}c_{k',\uparrow}}
+
(c_{k,\uparrow}^{\dagger}c_{-k,\downarrow}^{\dagger}-\big<c_{k,\uparrow}^{\dagger}c_{-k,\downarrow}^{\dagger}\big>)(c_{-k',\downarrow}c_{k',\uparrow}-\average{c_{-k',\downarrow}c_{k',\uparrow}})
\Big]\,\nonumber.
\end{split}
\end{equation}
In the mean-field approach, the terms $\big<\big>\big<\big>$ in previous expression are 
neglected, since fluctuations are assumed to be small. Therefore, the mean-field interaction Hamiltonian can be written as
\begin{equation}
\sum_{k,k'}V_{k,k'}
\Big[
\big<c_{k,\uparrow}^{\dagger}c_{-k,\downarrow}^{\dagger}\big>c_{-k',\downarrow}c_{k',\uparrow}
+
c_{k,\uparrow}^{\dagger}c_{-k,\downarrow}^{\dagger}\average{c_{-k',\downarrow}c_{k',\uparrow}}
-
\big<c_{k,\uparrow}^{\dagger}c_{-k,\downarrow}^{\dagger}\big>\average{c_{-k',\downarrow}c_{k',\uparrow}}
\Big]\,\nonumber.
\end{equation}
Now, we are in position to define fundamental the averages of two creation or annihilation operators as the superconducting pairing potential 
$\Delta_{k}=-\sum_{k'}V_{k,k'}\average{c_{-k',\downarrow}c_{k',\uparrow}}$, leading to macroscopic quantum coherence of the system. This represents the fundamental characteristic 
of the superconducting state. A superconducting system is in a ordered state, and the pairing potential is also known as the order parameter \cite{DeGennes,tinkham,zagoskin}.
Thus, the mean-field pairing Hamiltonian reads,
\begin{equation}
\label{mfaH0A}
H_{MFA}=\sum_{k,\sigma}\xi_{k}\,c^{\dagger}_{k\sigma}c_{k\sigma}
+\sum_{k}\Big[\Delta_{k}^{*}\,c_{-k,\downarrow}c_{k,\uparrow}+\Delta_{k}\,c_{k,\uparrow}^{\dagger}c_{-k,\downarrow}^{\dagger}\Big]-\sum_{k}\Delta_{k}\average{c_{-k,\downarrow}c_{k,\uparrow}}\,,
\end{equation}
which is the one given in Eq.\,(\ref{mfaH0}).

Superconducting systems require to treat electrons and holes on the same footing. 
This is why one follows the Bogoliubov-de Gennes formalism \cite{zagoskin}, where one introduces a redundant description that will be clear in the following.
Let us first write the kinetic term of Eq.\,(\ref{mfaH0A}),
\begin{equation}
 \begin{split}
 \sum_{k,\sigma}\xi_{k}\,c^{\dagger}_{k\sigma}c_{k\sigma}&=\frac{1}{2} 
	 \sum_{k,\sigma}\Big[\xi_{k}\,c^{\dagger}_{k\sigma}c_{k\sigma}+\xi_{k}\,c^{\dagger}_{k\sigma}c_{k\sigma}\Big]=
	 \frac{1}{2} 
	 \sum_{k,\sigma}\Big[\xi_{k}\,c^{\dagger}_{k\sigma}c_{k\sigma}+\xi_{k}-\xi_{k}\,c_{k\sigma}c^{\dagger}_{k\sigma}\Big]\,,\\
	&=	 \frac{1}{2} 
	 \sum_{k,\sigma}\Big[\xi_{k}\,c^{\dagger}_{k\sigma}c_{k\sigma}-\xi_{-k}\,c_{-k\sigma}c^{\dagger}_{-k\sigma}\Big]+\frac{1}{2}\sum_{k}\xi_{k}\,,
 \end{split}
 \end{equation}
 where in the second equality in the first line, we have used usual anti-commutation relations for fermionic operators $\{c^{\dagger}_{k\sigma},c_{k'\sigma'}\}=c^{\dagger}_{k\sigma}c_{k'\sigma'}+c_{k'\sigma'}c^{\dagger}_{k\sigma}=\delta_{k,k'}\delta_{\sigma\sigma'}$, and in the second line we have just relabelled $k$ to $-k$. Now, we assume $\Delta_{k}=\Delta$ to be independent of $k$, and proceed with the second term in Eq.\,(\ref{mfaH0A}) similarly as for the kinetic term,
 \begin{equation}
 \begin{split}
 \Delta\sum_{k}c_{-k,\downarrow}c_{k,\uparrow}&=\frac{ \Delta}{2} \sum_{k}\big[c_{-k,\downarrow}c_{k,\uparrow}+c_{-k,\downarrow}c_{k,\uparrow}\big]=
 \frac{ \Delta}{2}\sum_{k}\big[c_{-k,\downarrow}c_{k,\uparrow}-c_{k,\uparrow}c_{-k,\downarrow}\big]\,,\\
  &=\frac{ \Delta}{2}\sum_{k}\big[c_{-k,\downarrow}c_{k,\uparrow}-c_{-k,\uparrow}c_{k,\downarrow}\big]\,,
 \end{split}
 \end{equation}
 and 
  \begin{equation}
 \begin{split}
 \Delta^{*}\sum_{k}c_{k,\uparrow}^{\dagger}c_{-k,\downarrow}^{\dagger}&=
 \frac{\Delta^{*}}{2}\sum_{k}\Big[c_{k,\uparrow}^{\dagger}c_{-k,\downarrow}^{\dagger}+c_{k,\uparrow}^{\dagger}c_{-k,\downarrow}^{\dagger}\Big]=
  \frac{\Delta^{*}}{2}\sum_{k}\Big[c_{k,\uparrow}^{\dagger}c_{-k,\downarrow}^{\dagger}-c_{-k,\downarrow}^{\dagger}c_{k,\uparrow}^{\dagger}\Big]
 \\
 &=
  \frac{\Delta^{*}}{2}\sum_{k}\Big[c_{k,\uparrow}^{\dagger}c_{-k,\downarrow}^{\dagger}-c_{k,\downarrow}^{\dagger}c_{-k,\uparrow}^{\dagger}\Big]
\,.
  \end{split}
 \end{equation}
 In previous two equations we have used $\{c_{k\sigma},c_{k'\sigma'}\}=c_{k\sigma}c_{k'\sigma'}+c_{k'\sigma'}c_{k\sigma}=0$, and $\{c^{\dagger}_{k\sigma},c^{\dagger}_{k'\sigma'}\}=c^{\dagger}_{k\sigma}c^{\dagger}_{k'\sigma'}+c^{\dagger}_{k'\sigma'}c^{\dagger}_{k\sigma}=0$. 
 Then, Eq.\,(\ref{mfaH0A}) reads
 \begin{equation}
 \begin{split}
 H_{MFA}&=\frac{1}{2} 
	 \sum_{k}\Big[\xi_{k}\,c^{\dagger}_{k,\uparrow}c_{k,\uparrow}+\xi_{k}\,c^{\dagger}_{k,\downarrow}c_{k,\downarrow}-\xi_{-k}\,c_{-k,\uparrow}c^{\dagger}_{-k,\uparrow}-\xi_{-k}\,c_{-k,\downarrow}c^{\dagger}_{-k,\downarrow}\Big]\,\\
	&+
	\frac{ \Delta}{2}\sum_{k}\big[c_{-k,\downarrow}c_{k,\uparrow}-c_{-k,\uparrow}c_{k,\downarrow}\big]
	  \frac{\Delta^{*}}{2}\sum_{k}\Big[c_{k,\uparrow}^{\dagger}c_{-k,\downarrow}^{\dagger}-c_{k,\downarrow}^{\dagger}c_{-k,\uparrow}^{\dagger}\Big]\,\\
	  &
	 +\frac{1}{2}\sum_{k}\xi_{k}-\sum_{k}\Delta_{k}\average{c_{-k,\downarrow}c_{k,\uparrow}}\,.
	 \end{split}
 \end{equation}
 Previous equation can be rewritten in a matrix form, then
 \begin{equation}
 \label{eqMFA}
 \begin{split}
 H_{MFA}&=\frac{1}{2}
 \sum_{k}
\begin{pmatrix}
c_{k,\uparrow},
c_{k,\downarrow},
c_{-k,\uparrow}^{\dagger},
c_{-k,\downarrow}^{\dagger}
 \end{pmatrix}^{\dagger}
 \begin{pmatrix}
\xi_{k}&0&0&\Delta\\
0&\xi_{k}&-\Delta&0\\
0&-\Delta^{*}&-\xi_{-k}&0\\
\Delta^{*}&0&0&-\xi_{-k}\\
\end{pmatrix}
\begin{pmatrix}
c_{k,\uparrow}\\
c_{k,\downarrow}\\
c_{-k,\uparrow}^{\dagger}\\
c_{-k,\downarrow}^{\dagger}
 \end{pmatrix}\,\\
 &	 +\frac{1}{2}\sum_{k}\xi_{k}-\sum_{k}\Delta_{k}\average{c_{-k,\downarrow}c_{k,\uparrow}}\,.
 \end{split}
 \end{equation}
 We notice that in this elegant form, we have introduced a new spinor basis containing creation and annihilation operators. These operators define the so-called Nambu representation \cite{zagoskin}, which emerges naturally in superconducting systems and we define them as,
\begin{equation}
\label{spinorNambu}
\Psi_{k}=\begin{pmatrix}
c_{k,\uparrow},
c_{k,\downarrow},
c_{-k,\uparrow}^{\dagger},
c_{-k,\downarrow}^{\dagger}
 \end{pmatrix}^{T}\,.
\end{equation}
We point out that such representation is not unique but the physics remains the same.
The matrix in Eq.\,(\ref{eqMFA}) is the so-called Bogoliubov-de Gennes Hamiltonian \cite{DeGennes,tinkham,zagoskin}
\begin{equation}
H_{BdG}
\,=\,
\begin{pmatrix}
\xi_{k}&0&0&\Delta\\
0&\xi_{k}&-\Delta&0\\
0&-\Delta^{*}&-\xi_{-k}&0\\
\Delta^{*}&0&0&-\xi_{-k}\\
\end{pmatrix}\,.
\end{equation}
The last two terms in Eq.\,(\ref{eqMFA}) are constant and usually dropped off in the literature. 
Thus, the spectrum of Eq.\,(\ref{eqMFA}) is found after solving the eigenvalue problem $H_{BdG}\Psi_{k}=E_{k}\Psi_{k}$, and given by Eq.\,(\ref{BdGSC}), 
where the eigenvalues form the diagonal elements of $H_{BdG}$ in the new basis and represent the quasiparticle excitations in a superconductor.
\subsection*{Bogoliubov transformation}
\label{bogotrans}
The ground state of a superconductor is formed by Cooper pairs, while the quasiparticle excitations in a superconductor are separated from such ground state by an energy gap, 
as seen in Eq.\,(\ref{BdGSC}), and can emerge by breaking a Cooper pairs.
Therefore, it is natural to look for a relation between operators $c$ and the new ones denoted with $\alpha$ next, where the $H_{MFA}$ is diagonal, with the energies 
given by Eq.\,(\ref{BdGSC}).
The new basis is found by projecting the spinor $\Psi_{k}^{\dagger}$, given in Eq.\,(\ref{spinorNambu}), onto the eigenvectors of $H_{BdG}$
\begin{equation}
\label{bogotrans1}
 \alpha^{\dagger}_{k}=\Psi_{k}^{\dagger}U_{k}\,\quad\rightarrow\quad
 \begin{pmatrix}
\alpha_{k,\uparrow},
\alpha_{k,\downarrow},
\alpha_{-k,\uparrow}^{\dagger},
\alpha_{-k,\downarrow}^{\dagger}
 \end{pmatrix}
 =
 \begin{pmatrix}
c_{k,\uparrow},
c_{k,\downarrow},
c_{-k,\uparrow}^{\dagger},
c_{-k,\downarrow}^{\dagger}
 \end{pmatrix}U_{k}
\end{equation}
where $\alpha_{k}$ is the spinor representing the new basis, $U_{k}$ is an unitary operator formed by the eigenvectors of $H_{BdG}$, which connects these two basis,
\begin{equation}
\label{Uk}
U_{k}=\begin{pmatrix}
\Psi_{k}(E_{k,+}),\Psi_{k}(E_{k,+}),\Psi_{k}(E_{k,-}),\Psi_{k}(E_{k,-})\,,
\end{pmatrix}
\end{equation}
being a $4\times4$ matrix due to spin and electron-hole symmetries. Notice that the columns of $U_{k}$ are the eigenvectors of $H_{BdG}$. 
We will see that the unitary matrix $U_{k}$ can be also constructed by placing the eigenvectors as rows, instead of columns, in $U_{k}$, giving rise to $U_{k}^{T}$, 
which is also unitary matrix and also diagonalizes $H_{BdG}$. 
Taking the Hermitian conjugate of Eq.\,(\ref{bogotrans1})
one has $ \alpha_{k}=U_{k}^{\dagger}\Psi_{k}$.
Then, because $U_{k}U_{k}^{\dagger}=1$, one has the inverse transformation
\begin{equation}
\label{bogotrans2}
\Psi_{k}=U_{k}\alpha_{k}\,\quad\rightarrow\quad 
\begin{pmatrix}
c_{k,\uparrow}\\
c_{k,\downarrow}\\
c_{-k,\uparrow}^{\dagger}\\
c_{-k,\downarrow}^{\dagger}
 \end{pmatrix}=U_{k}\begin{pmatrix}
\alpha_{k,\uparrow}\\
\alpha_{k,\downarrow}\\
\alpha_{-k,\uparrow}^{\dagger}\\
\alpha_{-k,\downarrow}^{\dagger}
 \end{pmatrix}
\end{equation}
The transformation given by Eq.\,(\ref{bogotrans1}), or the inverse given by Eq.\,(\ref{bogotrans2}), are known in the literature as the Bogoliubov-Valatin transformation, but we will refer to it as Bogoliubov transformation, only. 
We have pointed out that the unitary matrix, which connects the two basis is formed out by the eigenvectors of $H_{BdG}$, becoming our first task in the following part. 
\subsubsection*{The eigenvectors}
We start by solving the eigenvalue problem for $H_{BdG}$,
\begin{equation}
\label{appeq1}
H_{BdG}\Psi_{k}=E_{k}\Psi_{k}\quad\rightarrow\quad |H_{BdG}-IE_{k}|=0\,,\quad
\text{where}\,,
\Psi_{k}=A{\rm e}^{ikx}\begin{pmatrix}
u_{k,\uparrow},
u_{k,\downarrow},
v_{k,\uparrow},
v_{k,\downarrow}
\end{pmatrix}^{T}\,,
\end{equation}
where $I$ is the $4\times4$ identity matrix. The eigenvalues then read $E_{k,\pm}=\pm\sqrt{\xi_{k}^{2}+|\Delta|^{2}}$. Now, we go back to the first equation of 
Eqs.\,(\ref{appeq1}) and using the known eigenvalues and the form of the eigenvectors, we write
\begin{equation}
\begin{split}
(\xi_{k}-E)u_{k,\uparrow}+\Delta v_{k,\downarrow}&=0\,,\\
(\xi_{k}-E)u_{k,\downarrow}-\Delta v_{k,\uparrow}&=0\,,\\
-\Delta^{*} u_{k,\downarrow}+(-\xi_{k}-E)v_{k,\uparrow}&=0\,,\\
\Delta^{*} u_{k,\uparrow}+(-\xi_{k}-E)v_{k,\downarrow}&=0\,.
\end{split}
\end{equation}
From previous system of equations, one observes that they can be written
in two groups, 
\begin{equation}
\label{group1A}
\begin{split}
(\xi_{k}-E)u_{k,\uparrow}+\Delta v_{k,\downarrow}&=0\,,\\
\Delta^{*} u_{k,\uparrow}+(-\xi_{k}-E)v_{k,\downarrow}&=0\,.
\end{split}
\end{equation}
and 
\begin{equation}
\label{group1B}
\begin{split}
(\xi_{k}-E)u_{k,\downarrow}-\Delta v_{k,\uparrow}&=0\,,\\
-\Delta^{*} u_{k,\downarrow}+(-\xi_{k}-E)v_{k,\uparrow}&=0\,.
\end{split}
\end{equation}
which can be solved separately because spin up and down electrons are not coupled, but rather electrons and holes through the pairing potential $\Delta$. 
Additionally, one indeed notices that in these two groups only one equation matters, since the two equations are coupled and therefore contain redundant information. The fact that there are two groups corresponding to up and down spins, which can be solved separately, allows us to define eigenvectors associated to the first and second group, respectively,
\begin{equation}
\label{thetwogroups}
\Psi_{k,1}=
\begin{pmatrix}
u_{k,\uparrow}\\
0\\
0\\
v_{k,\downarrow}
\end{pmatrix}
=\begin{pmatrix}
u_{k,\uparrow}\\
v_{k,\downarrow}
\end{pmatrix}\,,\quad \quad 
\Psi_{k,2}=
\begin{pmatrix}
0\\
u_{k,\downarrow}\\
v_{k,\uparrow}\\
0
\end{pmatrix}=
\begin{pmatrix}
u_{k,\downarrow}\\
v_{k,\uparrow}
\end{pmatrix}\,,
\end{equation}
where the zeroes in $\Psi_{k,1}$ and $\Psi_{k,2}$ correspond to elements that are not present in Eqs.\,(\ref{group1A}) and Eqs.\,(\ref{group1B}), respectively.
The first component in the second equality of $\Psi_{k,1(2)}$ corresponds electron part with spin up (down), while the second to the respective hole part with spin down (up) (because a hole is a time-reversed electron).
 The other equation that helps us to solve such systems is the normalisation condition associated to $\Psi_{k,1(2)}$: $|u_{k,\uparrow}|^{2}+|v_{k,\downarrow}|^{2}=1$ and $|u_{k,\downarrow}|^{2}+|v_{k,\uparrow}|^{2}=1$

From Eqs.\,(\ref{group1A}), for $E=E_{k,+}$, one has
\begin{equation}
\label{eqAp}
u_{k,\uparrow,+}=\frac{\xi_{k}+E_{k,+}}{\Delta^{*}}v_{k,\downarrow,+}\quad\rightarrow\quad 
\Psi_{k,1,+}=
\begin{pmatrix}
u_{k,\uparrow,+}\\
v_{k,\downarrow,+}
\end{pmatrix}=
\begin{pmatrix}
 \frac{\xi_{k}+E_{k,+}}{\Delta^{*}}\\
 1
\end{pmatrix}v_{k,\downarrow,+}\,,
\end{equation}
and normalization condition dictates
\begin{equation}
  \bigg(\frac{\xi_{k}+E_{k,+}}{\Delta^{*}}\bigg)^{2}|v_{k,\downarrow,+}|^{2}+|v_{k,\downarrow,+}|^{2}=1\rightarrow
\quad |v_{k,\downarrow,+}|^{2}=\frac{1}{2}\bigg[1-\frac{\xi_{k}}{E_{k,+}} \bigg]\,.
\end{equation}
For finding $u$, one takes the positive value of $v$ in previous equation and then plug it into first equation of Eqs.\,(\ref{eqAp}), then
\begin{equation}
u_{k,\uparrow,+}=\sqrt{\frac{1}{2}\bigg[1+\frac{\xi_{k}}{E_{k,+}} \bigg]}\,.
\end{equation}
Then, 
\begin{equation}
\Psi_{k,1,+}=
\begin{pmatrix}
u_{k}\\
v_{k}
\end{pmatrix}=\begin{pmatrix}
u_{k}\\
0\\
0\\
v_{k}
\end{pmatrix}
=\Psi_{k,1}(E_{k,+})\,,
\end{equation}
where we have rewritten it as a four component vector according to Eqs.\,(\ref{thetwogroups}), and we have defined, up to a phase, 
\begin{equation}
\label{defuv}
 \begin{split}
  u_{k}=\sqrt{\frac{1}{2}\Bigg[1+\frac{\xi_{k}}{E_{k,+}}\Bigg]}\,,\quad &
  v_{k}=\sqrt{\frac{1}{2}\Bigg[1-\frac{\xi_{k}}{E_{k,+}}\Bigg]}\,.
  \end{split}
\end{equation}
In the same way, for $E=E_{k,-}=-E_{k,+}$, we have
\begin{equation}
\label{eqAp2}
u_{k,\uparrow,-}=\frac{\xi_{k}-E_{k,+}}{\Delta^{*}}v_{k,\downarrow,-}\quad\rightarrow\quad 
\Psi_{k,1,-}=
\begin{pmatrix}
u_{k,\uparrow,-}\\
v_{k,\downarrow,-}
\end{pmatrix}=
\begin{pmatrix}
 \frac{\xi_{k}-E_{k,+}}{\Delta^{*}}\\
 1
\end{pmatrix}v_{k,\downarrow,-}\,,
\end{equation}
and normalization condition dictates
\begin{equation}
 \begin{split}
  \bigg(\frac{\xi_{k}-E_{k,+}}{\Delta^{*}}\bigg)^{2}|v_{k,\downarrow,-}|^{2}+|v_{k,\downarrow,-}|^{2}=1\rightarrow
\quad |v_{k,\downarrow,-}|^{2}=\frac{1}{2}\bigg[1+\frac{\xi_{k}}{E_{k,+}} \bigg]\,,
\end{split}
\end{equation}
then, by taking the positive value of $v_{k}$, up to a phase, and plugging it back into the first equation of Eqs.\,(\ref{eqAp2})
one finds 
\begin{equation}
u_{k,\uparrow,-}=-\sqrt{\frac{1}{2}\bigg[1-\frac{\xi_{k}}{E_{k,+}} \bigg]
}\,.
\end{equation}
We notice that, according to definitions given by Eqs.\,(\ref{defuv}), for $E_{k,-}$, we have obtained $u_{k,\uparrow}=-v_{k}$ and $v_{k,\uparrow}=u_{k}$. Then, 
\begin{equation}
\Psi_{k,1,-}=
\begin{pmatrix}
-v_{k}\\
u_{k}
\end{pmatrix}=
\begin{pmatrix}
-v_{k}\\
0\\
0\\
u_{k}
\end{pmatrix}
=\Psi_{k,1}(E_{k,-})\,,
\end{equation}
where according to Eqs.\,(\ref{thetwogroups}) we have written as a four component vector.
Therefore, the results we have obtained are, as four component vectors,
\begin{equation}
\Psi_{k,1,+}\equiv\Psi_{k,1}(E_{k,+})=
\begin{pmatrix}
u_{k}\\
0\\
0\\
v_{k}
\end{pmatrix}\,,\quad
\,\quad
\Psi_{k,1,-}\equiv\Psi_{k,1}(E_{k,-})=
\begin{pmatrix}
-v_{k}\\
0\\
0\\
u_{k}
\end{pmatrix}\,,\quad
\end{equation}
In a similar fashion one can proceed for Eqs.\,(\ref{group1B}), and after some algebra obtains 
\begin{equation}
\Psi_{k,2,+}\equiv\Psi_{k,2}(E_{k,+})=
\begin{pmatrix}
0\\
u_{k}\\
v_{k}\\
0
\end{pmatrix}\,,\quad
\Psi_{k,2,-}\equiv\Psi_{k,2}(E_{k,-})=
\begin{pmatrix}
0\\
-v_{k}\\
u_{k}\\
0
\end{pmatrix}\,.\quad
\end{equation}

\subsubsection*{The transformation}
Now, going back to the unitary transformation, it is important to remember that the problem was defined in two parts and therefore the four equations were 
divided in two groups of two equations each and solved separately. The unitary matrix, given by Eq.\,(\ref{Uk}), is then carefully rewritten as
\begin{equation}
\begin{split}
U_{k}&=
\begin{pmatrix}
\Psi_{k}(E_{k,1,+}),\Psi_{k}(E_{k,2,+}),\Psi_{k}(E_{k,2,-}),\Psi_{k}(E_{k,1,-})\,,
\end{pmatrix}\,.
\end{split}
\end{equation}
Therefore, the unitary operators are given by 
\begin{equation}
U_{k}=
\begin{pmatrix}
u_{k}&0&0&-v_{k}\\
0&u_{k}&-v_{k}&0\\
0&v_{k}&u_{k}&0\\
v_{k}&0&0&u_{k}
\end{pmatrix}\,,\quad 
U_{k}^{\dagger}=
\begin{pmatrix}
u_{k}&0&0&v_{k}\\
0&u_{k}&v_{k}&0\\
0&-v_{k}&u_{k}&0\\
-v_{k}&0&0&u_{k}
\end{pmatrix}\,.
\end{equation}
Then, according Eq.\,(\ref{bogotrans1}), $\alpha_{k}=U_{k}^{\dagger}\Psi_{k}$,
\begin{equation}
\begin{pmatrix}
\alpha_{k,\uparrow}\\
\alpha_{k,\downarrow}\\
\alpha_{-k,\uparrow}^{\dagger}\\
\alpha_{-k,\downarrow}^{\dagger}
 \end{pmatrix}
 =\begin{pmatrix}
u_{k}&0&0&v_{k}\\
0&u_{k}&v_{k}&0\\
0&-v_{k}&u_{k}&0\\
-v_{k}&0&0&u_{k}
\end{pmatrix}
 \begin{pmatrix}
c_{k,\uparrow}\\
c_{k,\downarrow}\\
c_{-k,\uparrow}^{\dagger}\\
c_{-k,\downarrow}^{\dagger}
 \end{pmatrix}\,,
\end{equation}
one arrives at
\begin{equation}
\label{transBGD1}
\begin{split}
\alpha_{k,\uparrow}&=u_{k}c_{k,\uparrow}+v_{k}c_{-k,\downarrow}^{\dagger} \,,\\
\alpha_{k,\downarrow}&= u_{k}c_{k,\downarrow}+v_{k}c_{-k,\uparrow}^{\dagger} \,,\\
\alpha_{-k,\uparrow}^{\dagger}&=-v_{k}c_{k,\downarrow}+u_{k}c_{-k,\uparrow}^{\dagger} \,,\\
\alpha_{-k,\downarrow}^{\dagger}&= -v_{k}c_{k,\uparrow}+u_{k}c_{-k,\downarrow}^{\dagger}\,,
\end{split}
\end{equation}
 or the inverse, $\Psi_{k}=U_{k}\alpha_{k}$, by Eq.\,(\ref{bogotrans2}),
 \begin{equation}
\begin{pmatrix}
c_{k,\uparrow}\\
c_{k,\downarrow}\\
c_{-k,\uparrow}^{\dagger}\\
c_{-k,\downarrow}^{\dagger}
 \end{pmatrix}=\begin{pmatrix}
u_{k}&0&0&-v_{k}\\
0&u_{k}&-v_{k}&0\\
0&v_{k}&u_{k}&0\\
v_{k}&0&0&u_{k}
\end{pmatrix}
\begin{pmatrix}
\alpha_{k,\uparrow}\\
\alpha_{k,\downarrow}\\
\alpha_{-k,\uparrow}^{\dagger}\\
\alpha_{-k,\downarrow}^{\dagger}
 \end{pmatrix}\,,
\end{equation}
 leading to
\begin{equation}
\label{transBGD2}
\begin{split}
c_{k,\uparrow}&=u_{k}\,\alpha_{k,\uparrow}-v_{k}\,\alpha^{\dagger}_{-k,\downarrow}\\
c_{k,\downarrow}&=u_{k}\,\alpha_{k,\downarrow}-v_{k}\,\alpha^{\dagger}_{-k,\uparrow}\\
c_{-k,\uparrow}^{\dagger}&=v_{k}\,\alpha_{k,\downarrow}+u_{k}\,\alpha^{\dagger}_{-k,\uparrow}\\
c_{-k,\downarrow}^{\dagger}&=v_{k}\,\alpha_{k,\uparrow}+u_{k}\,\alpha^{\dagger}_{-k,\downarrow}\,.
\end{split}
\end{equation}
 
To finish this part, we point out that the other unitary matrix that diagonalises $H_{BdG}$ is $U^{T}$, being $T$ the transpose operation. Then, according Eq.\,(\ref{bogotrans1}), $\alpha_{k}=(U_{k}^{T})^{\dagger}\Psi_{k}$, one has
\begin{equation}
\label{transBGD3}
\begin{split}
\alpha_{k,\uparrow}&=u_{k}c_{k,\uparrow}-v_{k}c_{-k,\downarrow}^{\dagger} \,,\\
\alpha_{k,\downarrow}&= u_{k}c_{k,\downarrow}-v_{k}c_{-k,\uparrow}^{\dagger} \,,\\
\alpha_{-k,\uparrow}^{\dagger}&=v_{k}c_{k,\downarrow}+u_{k}c_{-k,\uparrow}^{\dagger} \,,\\
\alpha_{-k,\downarrow}^{\dagger}&= v_{k}c_{k,\uparrow}+u_{k}c_{-k,\downarrow}^{\dagger}\,,\\
\end{split}
\end{equation}
 or the inverse, $\Psi_{k}=U_{k}^{T}\alpha_{k}$, following Eq.\,(\ref{bogotrans2}),
\begin{equation}
\label{transBGD4}
\begin{split}
c_{k,\uparrow}&=u_{k}\,\alpha_{k,\uparrow}+v_{k}\,\alpha^{\dagger}_{-k,\downarrow}\\
c_{k,\downarrow}&=u_{k}\,\alpha_{k,\downarrow}+v_{k}\,\alpha^{\dagger}_{-k,\uparrow}\\
c_{-k,\uparrow}^{\dagger}&=-v_{k}\,\alpha_{k,\downarrow}+u_{k}\,\alpha^{\dagger}_{-k,\uparrow}\\
c_{-k,\downarrow}^{\dagger}&=-v_{k}\,\alpha_{k,\uparrow}+u_{k}\,\alpha^{\dagger}_{-k,\downarrow}\,.
\end{split}
\end{equation}

Equations \ref{transBGD1} and \ref{transBGD2} or \ref{transBGD3} and \ref{transBGD4} constitute the so-called Bogoliubov-Valatin transformations.

Notice that, for instance, from Eqs.\,(\ref{transBGD1}) one can write
\begin{equation}
\begin{split}
\alpha_{k,\uparrow(\downarrow)}&=
 \begin{pmatrix}
c_{k,\uparrow(\downarrow)},c_{-k,\downarrow(\uparrow)}^{\dagger} \end{pmatrix}
\begin{pmatrix}
u_{k}\\
v_{k} \end{pmatrix}\,,\quad \Psi_{k}(E_{k,+})=\begin{pmatrix}
u_{k}\\
v_{k} \end{pmatrix}\,,\quad\text{for}\,,\quad E_{k,+}\,,\\
\alpha_{-k,\downarrow(\uparrow)}^{\dagger}&=
 \begin{pmatrix}
c_{k,\uparrow(\downarrow)},c_{-k,\downarrow(\uparrow)}^{\dagger} \end{pmatrix}
\begin{pmatrix}
-v_{k}\\
u_{k} \end{pmatrix}\,,\quad \Psi_{k}(E_{k,-})=\begin{pmatrix}
-v_{k}\\
u_{k} \end{pmatrix}\,,\quad\text{for}\,,\quad E_{k,-}\,,
\end{split}
\end{equation}
and therefore conclude that $\alpha_{k,\uparrow(\downarrow),E_{k,+}}$ is connected to $\alpha_{-k,\downarrow(\uparrow),E_{k,-}}^{\dagger}$ through the so-called electron-hole symmetry of $H_{BdG}$ discussed in the main text.

\subsection*{Comment on singlet and triplet superconductors}
\label{pairings}
Pauli inclusion principle imposes that the pairing function must be antisymmetric, then 
\begin{equation}
\Delta_{k,\alpha,\beta}\propto\average{c_{-k,\alpha}c_{k,\beta}}=-\Delta_{-k,\beta,\alpha}\,,
\end{equation} 
where the $\alpha, \beta$ denotes the spin indices.
We can separate the spin part from the orbital part, then
 \begin{equation}
 \Delta_{k,\alpha,\beta}=f_{\alpha,\beta}\Delta_{k}\,.
 \end{equation}
Therefore, in the case of singlet pairing, the spin part is odd, while the orbital part is even, the orbital part fulfils 
\begin{equation}
\Delta_{k}=\Delta_{-k}\,,
\end{equation}
and  thus we have the pairing symmetry of the so-called $s$-wave superconductor. 

On the other hand, in the case of triplet pairing, the spin part is even, while the orbital part is odd, then for the orbital part 
\begin{equation}
\Delta_{k}=-\Delta_{-k}\,,
\end{equation}
and we therefore have the pairing symmetry of the so-called $p$-wave superconductor.

Of course that our discussion is oversimplified and does not consider other pairing symmetries such as $d,f,$ etc. The aim of this part is  just  to point out the difference between $s$ and $p$-wave superconductors, since these two types are discussed in this thesis.
\section{Kitaev's model: open and closed periodic boundary conditions}
In this section we provide additional details, which complete the description of the Kitaev's model.
\subsection*{Open chain}
Here, we concentrate on the chain with $N$ fermionic sites, where the first and the last are not connected. We refer to this approach as to the open chain case.
The Kitaev model, $H=H_{\mu}+H_{t}+H_{\Delta}$, given by the Hamiltonian in Eq.\,(\ref{kiatev0}) with elements
\begin{equation}
\label{hamilS0}
\begin{split}
H_{\mu}&=-\mu\sum_{j=1}^{N}\Big(c^{\dagger}_{j}c_{j}-\frac{1}{2}\Big)\,,\\
H_{t}&=-\,\sum_{i=1}^{N-1}\Big[ t\,\big(c^{\dagger}_{j}c_{j+1}+c^{\dagger}_{j+1}c_{j}\big)\Big]\,,\\
H_{t}&=\,\sum_{i=1}^{N-1}\Big[\,\Delta\,c_{j}c_{j+1}\,+\,\Delta^{*}\,c^{\dagger}_{j+1}c^{\dagger}_{j}\Big]\,,
\end{split}
\end{equation}
was written as Eq.\,(\ref{kitaev0a}) by transforming fermionic operators $c^{\dagger}_{j}(c_{j})$ into Majorana ones according to Eqs.\,(\ref{MajTrans}). 
Here, we complete the derivation by expressing each Hamiltonian term into the new basis, assuming real pairing potential $\Delta$.
Therefore, in terms of the Majorana operators, we have
\begin{equation}
\begin{split}
c^{\dagger}_{j}c_{j}&=\frac{1}{2}\big(\gamma^{A}_{j}-i\gamma^{B}_{j}\big)\frac{1}{2}\big(\gamma^{A}_{j}+i\gamma^{B}_{j}\big)\,,\\
c^{\dagger}_{j}c_{j+1}&=\frac{1}{2}\big(\gamma^{A}_{j}-i\gamma^{B}_{j}\big)\frac{1}{2}\big(\gamma^{A}_{j+1}+i\gamma^{B}_{j+1}\big)\,,\\
c^{\dagger}_{j+1}c_{j}&=\frac{1}{2}\big(\gamma^{A}_{j+1}-i\gamma^{B}_{j+1}\big)\frac{1}{2}\big(\gamma^{A}_{j}+i\gamma^{B}_{j}\big)\,,\\
c_{j}c_{j+1}&=\frac{1}{2}\big(\gamma^{A}_{j}+i\gamma^{B}_{j}\big)\frac{1}{2}\big(\gamma^{A}_{j+1}+i\gamma^{B}_{j+1}\big)\,,\\
c^{\dagger}_{j+1}c^{\dagger}_{j}&=\frac{1}{2}\big(\gamma^{A}_{j+1}-i\gamma^{B}_{j+1}\big)\frac{1}{2}\big(\gamma^{A}_{j}-i\gamma^{B}_{j}\big)\,.
\end{split}
\end{equation}
Previous equations can be rewritten as
\begin{equation}
\begin{split}
c^{\dagger}_{j}c_{j}&=\frac{1}{4}\Big[\gamma^{A}_{j}\gamma^{A}_{j}+i\gamma^{A}_{j}\gamma^{B}_{j}-i\gamma^{B}_{j}\gamma^{A}_{j}+\gamma^{B}_{j}\gamma^{B}_{j}\Big]\,,\\
c^{\dagger}_{j}c_{j+1}&=\frac{1}{4}\Big[\gamma^{A}_{j}\gamma^{A}_{j+1}+i\gamma^{A}_{j}\gamma^{B}_{j+1}
-i\gamma^{B}_{j}\gamma^{A}_{j+1}+\gamma^{B}_{j}\gamma^{B}_{j+1}\Big]\,,\\
c^{\dagger}_{j+1}c_{j}&=\frac{1}{4}\Big[\gamma^{A}_{j+1}\gamma^{A}_{j}+i\gamma^{A}_{j+1}\gamma^{B}_{j}
-i\gamma^{B}_{j+1}\gamma^{A}_{j}+\gamma^{B}_{j+1}\gamma^{B}_{j}\Big]\,,\\
c_{j}c_{j+1}&=\frac{1}{4}\Big[\gamma^{A}_{j}\gamma^{A}_{j+1}+i\gamma^{A}_{j}\gamma^{B}_{j+1}
+i\gamma^{B}_{j}\gamma^{A}_{j+1}-\gamma^{B}_{j}\gamma^{B}_{j+1}\Big]\,,\\
c^{\dagger}_{j+1}c^{\dagger}_{j}&=\frac{1}{4}\Big[\gamma^{A}_{j+1}\gamma^{A}_{j}-i\gamma^{A}_{j+1}\gamma^{B}_{j}
-i\gamma^{B}_{j+1}\gamma^{A}_{j}-\gamma^{B}_{j+1}\gamma^{B}_{j}\Big]\,.
\end{split}
\end{equation}
In the following, we make use of the algebra for Majorana operators given by 
Eqs.\,(\ref{MajAlgebra}), where
\begin{equation}
\gamma^{A(B)}_{j}\gamma^{A(B)}_{j}=\gamma^{A(B)2}_{j}=1\,,\quad
\gamma_{j}^{A(B)}\gamma_{j+1}^{A(B)}+\gamma_{j+1}^{A(B)}\gamma_{j}^{A(B)}=0\,.
\end{equation}
Thus, according to Eqs.\,(\ref{hamilS0}), the necessary expressions are given by
\begin{equation}
\begin{split}
c^{\dagger}_{j}c_{j}-\frac{1}{2}&=\frac{i}{2}\gamma_{j}^{A}\gamma_{j}^{B}\,\\
c^{\dagger}_{j}c_{j+1}+c^{\dagger}_{j+1}c_{j}&=\frac{i}{2}\Big[\gamma_{j}^{A}\gamma_{j+1}^{B}
-\gamma_{j}^{B}\gamma_{j+1}^{A}\Big]\,,\\
c_{j}c_{j+1}+c^{\dagger}_{j+1}c^{\dagger}_{j}&=\frac{i}{2}\Big[\gamma_{j}^{A}\gamma_{j+1}^{B}
+\gamma_{j}^{B}\gamma_{j+1}^{A}\Big]\,.
\end{split}
\end{equation}
Therefore, the Hamiltonian elements given by Eqs.\,(\ref{hamilS0}) in the Majorana basis read
 \begin{equation}
\begin{split}
H_{\mu}&=-\frac{i\mu}{2}\sum_{j=1}^{N}\gamma_{j}^{A}\gamma_{j}^{B}\,\\
H_{t}&=-\frac{it}{2}\sum_{j=1}^{N-1}\Big[\gamma_{j}^{A}\gamma_{j+1}^{B}
-\gamma_{j}^{B}\gamma_{j+1}^{A}\Big]\,,\\
H_{\Delta}&=\frac{i\Delta}{2}\sum_{j=1}^{N-1}\Big[\gamma_{j}^{A}\gamma_{j+1}^{B}
+\gamma_{j}^{B}\gamma_{j+1}^{A}\Big]\,.
\end{split}
\end{equation}
The full Hamiltonian, $H=H_{\mu}+H_{t}+H_{\Delta}$, is then written as
\begin{equation}
H\,=\,-\frac{i\mu}{2}\sum_{j=1}^{N}\gamma_{j}^{A}\gamma_{j}^{B}+\frac{i}{2}\sum_{j=1}^{N-1}
\Big[(\Delta+t)\gamma_{j}^{B}\gamma_{j+1}^{A}
+(\Delta-t)\gamma_{j}^{A}\gamma_{j+1}^{B}\Big]\,,
\end{equation}
which is the one given in the main text by Eq.\,(\ref{kitaev0a}), with $\omega_{-}=\Delta-t$ and $\omega_{+}=\Delta+t$.

We have discussed in the main text the emergence of zero energy Majorana modes at the end of the chain, one at each end, in the topological phase. 
At this point, Majorana operators $\gamma_{1}^{A}$ and $\gamma_{N}^{B}$, at the end of the chain, do not enter in the Hamiltonian given by Eq.\,(\ref{kitaev0c}), and therefore 
the following relations hold
\begin{equation}
\label{commuzero}
[\gamma_{1}^{A},H]=0\,,\quad [\gamma_{N}^{B},H]=0\,.
\end{equation} 
Thus, if $E_{0}$ is the energy of the ground state $\ket{GS}$, then $H\ket{GS}=E_{0}\ket{GS}$. 
Now, by employing Eqs.\,(\ref{commuzero}), we write
\begin{equation}
\begin{split}
\gamma_{1}^{A}H-H\gamma_{1}^{A}=0\,\quad \rightarrow\,\quad \gamma_{1}^{A}H=H\gamma_{1}^{A}\,,\\
\gamma_{N}^{B}H-H\gamma_{N}^{B}=0\,\quad \rightarrow\,\quad \gamma_{N}^{B}H=H\gamma_{N}^{B}\,.
\end{split}
\end{equation}
Then, previous discussions imply the following relations 
\begin{equation}
\label{appmaj}
H\gamma_{1}^{A}\ket{GS}=E_{0}\gamma_{1}^{A}\ket{GS}\,,\quad
 H\gamma_{N}^{B}\ket{GS}=E_{0}\gamma_{N}^{B}\ket{GS}\,.
\end{equation}
Previous two equations manifest the fact that we can create a pair of Majorana states at the end of the chain, represented by operators $\gamma_{1}^{A}$ and $\gamma_{N}^{B}$, with
one Majorana at each end. Remarkably, after we create this pair, the new state is still the ground state with the same energy $E_{0}$ as $\ket{GS}$, 
indicating that there is no cost of energy to create a pair of Majorana states at the ends of a superconducting chain.

By fusing the two unpaired Majorana operators $\gamma_{1}^{A}$ and $\gamma_{N}^{B}$, we have defined a zero energy fermion represented by new operators $f$ and $f^{\dagger}$, 
see Eq.\,(\ref{newopMaj}). Hence, from Eq.\,(\ref{appmaj}), one can show that
\begin{equation}
Hf\ket{GS}=E_{0}f\ket{GS}\,,\quad
 Hf^{\dagger}\ket{GS}=E_{0}f^{\dagger}\ket{GS}\,.
\end{equation}
Then, the ground states of $H$ can be spanned by operators $f$ and $f^{\dagger}$. The occupation number operator is defined as $n=f^{\dagger}f$, while the fermion parity 
operator as $P=(-1)^{n}=-i\gamma_{1}^{A}\gamma_{N}^{B}$.
Let us now calculate
\begin{equation}
\begin{split}
\bra{GS} -i\gamma_{1}^{A}\gamma_{N}^{B}\ket{GS}&\,=\,
-\bra{GS}\Big[2f^{\dagger}f-1 \Big]\ket{GS}\\
&\,=\,
\begin{cases}
    -1\,,  & f^{\dagger}f=1\,, \text{Occupied}, \\
      +1\,,& f^{\dagger}f=0\,, \text{empty}.
\end{cases}
\end{split}
\end{equation}
Therefore, previous equations indicate that we can use the occupation number $n$ in order to label the ground states. The two ground states are then the empty state $\ket{0}$ and the occupied one $\ket{1}=f^{\dagger}\ket{0}$ satisfying $f\ket{1}=0$ and  $f^{\dagger}\ket{1}=0$.

\subsection*{Closed periodic boundaries}
The assumption of a chain forming a closed loop with periodic boundary conditions, requires to add an extra term given by Eq.\,(\ref{kiatev0}) to the Hamiltonian, arriving at the Hamiltonian given by Eq.\,(\ref{kiatev0x}). 
The goal is to write down  the Hamiltonian given by Eq.\,(\ref{kiatev0x}), which is in site space, in momentum space. 
Thus, we perform Fourier transformations of the fermion operators $c_{j}$ as follows
\begin{equation}
c_{j}=\frac{1}{\sqrt{N}}\sum_{k}\,{\rm e}^{-i kx_{j}}c_{k}\,,\quad 
c_{j}^{\dagger}=\frac{1}{\sqrt{N}}\sum_{k}\,{\rm e}^{i kx_{j}}c_{k}^{\dagger}\,.
\end{equation}
An implication of the periodicity of the lattice is that wave-vectors $k$ that differ by a reciprocal lattice vector are the same, so that the sum over $k$ is limited to the first Brillouin zone: 
\begin{equation}
k=\frac{2\pi}{a}\frac{n}{N}\,,\quad -\frac{N}{2}\leqslant n\leqslant \frac{N}{2}\,,\quad N\text{: even}\,.
\end{equation}
In the following we will make use of the discrete representation of the Kronecker delta function
\begin{equation}
\delta_{k,k'}=\frac{1}{N}\sum_{j=1}^{N}{\rm e}^{i(k-k')x_{j}}\,.
\end{equation}
According to Eq.\,(\ref{kiatev0x}), the fermion operators we need are the followings,
\begin{equation}
\begin{split}
\sum_{j=1}^{N}c_{j}^{\dagger}c_{j}&\,=\,\sum_{j=1}^{N}\frac{1}{\sqrt{N}}\sum_{k}{\rm e}^{ikx_{j}}\,c_{k}^{\dagger}\,\frac{1}{\sqrt{N}}\sum_{k'}{\rm e}^{-ik'x_{j}}\,c_{k'}\,=\,\frac{1}{N}\sum_{j=1}^{N}\sum_{kk'}{\rm e}^{i(k-k')x_{j}}c_{k}^{\dagger}c_{k'}\,,\\
&\,=\,\sum_{kk'}c_{k}^{\dagger}c_{k'} \frac{1}{N}\sum_{j=1}^{N}{\rm e}^{i(k-k')x_{j}}\,=\,\sum_{kk'}c_{k}^{\dagger}c_{k'} \delta_{k,k'}\,,\\
&\,=\,\sum_{k}c_{k}^{\dagger}c_{k} \,,
\end{split}
\end{equation}
\begin{equation}
\begin{split}
\sum_{j=1}^{N}c_{j}^{\dagger}c_{j+1}&\,=\,\sum_{j=1}^{N}\frac{1}{\sqrt{N}}\sum_{k}{\rm e}^{ikx_{j}}\,c_{k}^{\dagger}
\,\frac{1}{\sqrt{N}}\sum_{k'}{\rm e}^{-ik'x_{j+1}}\,c_{k'}\,=\,\frac{1}{N}\sum_{j=1}^{N}\sum_{kk'}{\rm e}^{-ik'a}\,{\rm e}^{i(k-k')x_{j}}c_{k}^{\dagger}c_{k'}\,,\\
&\,=\,\sum_{kk'}c_{k}^{\dagger}c_{k'}\,{\rm e}^{-ik'a} \frac{1}{N}\sum_{j=1}^{N}{\rm e}^{i(k-k')x_{j}}\,=\,\sum_{kk'}c_{k}^{\dagger}c_{k'} \,{\rm e}^{-ik'a}\delta_{k,k'}\,,\\
&\,=\,\sum_{k}c_{k}^{\dagger}c_{k}\,{\rm e}^{-ika} \,.
\end{split}
\end{equation}
where we have used that $x_{j+1}=x_{j}+a$. This is also used next,
\begin{equation}
\begin{split}
\sum_{j=1}^{N}c_{j+1}^{\dagger}c_{j}&\,=\,\sum_{j=1}^{N}\frac{1}{\sqrt{N}}\sum_{k}{\rm e}^{ikx_{j+1}}\,c_{k}^{\dagger}
\,\frac{1}{\sqrt{N}}\sum_{k'}{\rm e}^{-ik'x_{j}}\,c_{k'}\,=\,\frac{1}{N}\sum_{j=1}^{N}\sum_{kk'}{\rm e}^{ika}\,{\rm e}^{i(k-k')x_{j}}c_{k}^{\dagger}c_{k'}\,,\\
&\,=\,\sum_{kk'}c_{k}^{\dagger}c_{k'}\,{\rm e}^{ika} \frac{1}{N}\sum_{j=1}^{N}{\rm e}^{i(k-k')x_{j}}
\,=\,\sum_{kk'}c_{k}^{\dagger}c_{k'} \,{\rm e}^{ika}\delta_{k,k'}\,,\\
&\,=\,\sum_{k}c_{k}^{\dagger}c_{k}\,{\rm e}^{ika} \,.
\end{split}
\end{equation}
\begin{equation}
\begin{split}
\sum_{j=1}^{N}c_{j}c_{j+1}
&\,=\,\sum_{j=1}^{N}\frac{1}{\sqrt{N}}\sum_{k}{\rm e}^{-ikx_{j}}\,c_{k}
\,\frac{1}{\sqrt{N}}\sum_{k'}{\rm e}^{-ik'x_{j+1}}\,c_{k'}
\,=\,\frac{1}{N}\sum_{j=1}^{N}\sum_{kk'}{\rm e}^{-ik'a}\,{\rm e}^{-i(k+k')x_{j}}c_{k}c_{k'}\,,\\
&\,=\,\sum_{kk'}c_{k}c_{k'}\,{\rm e}^{-ik'a} \frac{1}{N}\sum_{j=1}^{N}{\rm e}^{-i(k+k')x_{j}}
\,=\,\sum_{kk'}c_{k}c_{k'} \,{\rm e}^{-ik'a}\delta_{k,-k'}\,,\\
&\,=\,\sum_{k}c_{k}c_{-k}\,{\rm e}^{ika}\,=\,\sum_{k}c_{-k}c_{k}\,{\rm e}^{-ika} \,.
\end{split}
\end{equation}

\begin{equation}
\begin{split}
\sum_{j=1}^{N}c_{j+1}^{\dagger}c_{j}^{\dagger}
&\,=\,\sum_{j=1}^{N}\frac{1}{\sqrt{N}}\sum_{k}{\rm e}^{ikx_{j+1}}\,c_{k}^{\dagger}
\,\frac{1}{\sqrt{N}}\sum_{k'}{\rm e}^{ik'x_{j}}\,c_{k'}^{\dagger}
\,=\,\frac{1}{N}\sum_{j=1}^{N}\sum_{kk'}{\rm e}^{ika}\,{\rm e}^{i(k+k')x_{j}}c_{k}^{\dagger}c_{k'}^{\dagger}\,,\\
&\,=\,\sum_{kk'}c_{k}^{\dagger}c_{k'}^{\dagger}\,{\rm e}^{ika} \frac{1}{N}\sum_{j=1}^{N}{\rm e}^{i(k+k')x_{j}}
\,=\,\sum_{kk'}c_{k}^{\dagger}c_{k'}^{\dagger} \,{\rm e}^{ika}\delta_{k,-k'}\,,\\
&\,=\,\sum_{k}c_{k}^{\dagger}c_{-k}^{\dagger}\,{\rm e}^{ika} \,.
\end{split}
\end{equation}

Therefore, the Hamiltonian given by Eq.\,(\ref{kiatev0x}) in momentum space reads
\begin{equation}
\label{eqintro}
\begin{split}
H
&\,=\,\sum_{k}\,\xi_{k}\,\Big(c_{k}^{\dagger}c_{k}-\frac{1}{2} \Big)
+\frac{1}{2}\sum_{k}( -2t\cos ka)\,+\,\sum_{k}\Delta\Big(c_{-k}c_{k}\,{\rm e}^{-ika}+c_{k}^{\dagger}c_{-k}^{\dagger}{\rm e}^{ika}\Big)\Big]\,,
\end{split}
\end{equation}
where $\xi_{k}=-\mu-2t\cos ka$ is the normal energy dispersion.
In what follows, we make use of the redundant behaviour to treat electrons and hole at the same footing, introduced when we discussed the Bogoliubov formalism in Sec.\,\ref{bdgformalism}. 
Thus, formally, we can rewrite the fermion operators involved in previous Hamiltonian as
\begin{equation}
\begin{split}
\sum_{k}\xi_{k}\Big(c_{k}^{\dagger}c_{k}-\frac{1}{2}\Big)
&=\frac{1}{2}\sum_{k}\xi_{k}\Big(2 c^{\dagger}_{k}c_{k}-1\Big)\,,
\,=\,\frac{1}{2}\sum_{k}\big(\xi_{k}c_{k}^{\dagger}c_{k}-\xi_{k}c_{k}c_{k}^{\dagger} \big)\\
&=\frac{1}{2}\sum_{k}\big(\xi_{k}c_{k}^{\dagger}c_{k}-\xi_{-k}c_{-k}c_{-k}^{\dagger} \big)\,,\\
\sum_{k}c_{-k}c_{k}\,{\rm e}^{-ika}&=
\frac{1}{2}\sum_{k}\Big[c_{-k}c_{k}\,{\rm e}^{-ika}+c_{-k}c_{k}\,{\rm e}^{-ika} \Big]
=\frac{1}{2}\sum_{k}\Big[c_{-k}c_{k}\,{\rm e}^{-ika}-c_{k}c_{-k}\,{\rm e}^{-ika} \Big]\,,\\
&\,=\,\frac{1}{2}\sum_{k}\Big[c_{-k}c_{k}\,{\rm e}^{-ika}-c_{-k}c_{k}\,{\rm e}^{ika} \Big]=\frac{1}{2}\sum_{k}c_{-k}c_{k}\Big[{\rm e}^{-ika}-\,{\rm e}^{ika} \Big]\,,\\
&\,=\,\frac{1}{2}\sum_{k}(-2i\sin ka)\,,\\
\sum_{k}c_{k}^{\dagger}c_{-k}\,{\rm e}^{ika}
\,&=\,\frac{1}{2}\sum_{k}\Big[c_{k}^{\dagger}c_{-k}^{\dagger}\,{\rm e}^{ika}+c_{k}^{\dagger}c_{-k}^{\dagger}\,{\rm e}^{ika} \Big]
=\frac{1}{2}\sum_{k}\Big[c_{k}^{\dagger}c_{-k}^{\dagger}\,{\rm e}^{ika}-c_{-k}^{\dagger}c_{k}^{\dagger}\,{\rm e}^{ika} \Big]\,,\\
&\,=\,\frac{1}{2}\sum_{k}\Big[c_{k}^{\dagger}c_{-k}^{\dagger}\,{\rm e}^{ika}-c_{k}^{\dagger}c_{-k}^{\dagger}\,{\rm e}^{-ika} \Big]=\frac{1}{2}\sum_{k}c_{k}^{\dagger}c_{-k}^{\dagger}\Big[{\rm e}^{ika}-\,{\rm e}^{-ika} \Big]\,,\\
&\,=\,\frac{1}{2}\sum_{k}(2i\sin ka)\,.
\end{split}
\end{equation}
Therefore, Eq.\,(\ref{eqintro}) reads,
\begin{equation}
\label{introhk}
\begin{split}
H&\,=\,-\frac{1}{2}\sum_{k}\big(\xi_{k}c_{k}^{\dagger}c_{k}-\xi_{-k}c_{-k}c_{-k}^{\dagger} \big)\,+\,\frac{1}{2}\sum_{k}(-2t\cos ka)\\
&\,+\,\frac{1}{2}\sum_{k}\big[(2i\Delta\sin ka)c^{\dagger}_{k}c^{\dagger}_{-k}+(-2i\Delta\sin ka)c_{-k}c_{k} \big]\,,
\end{split}
\end{equation}
where the first term represents the kinetic part for electrons and holes, respectively, the second term is a constant usually neglected. 
The third term in previous equation is the pairing part of the Hamiltonian, where fermions with momentum $k$ and $-k$ are paired together. Notice that there are two special points in $k$-space with zero contribution to the pairing Hamiltonian because $\sin0=\sin\pi=0$.  
Eq.\,(\ref{eqintro}) can be written as
\begin{equation}
\label{introk3}
\begin{split}
H&\,=\,\frac{1}{2}\sum_{k}\xi_{k}
\begin{pmatrix}
c_{k}^{\dagger}& c_{-k}
 \end{pmatrix}
 \begin{pmatrix}
1& 0\\
0&-1
 \end{pmatrix}
 \begin{pmatrix}
c_{k}\\
 c_{-k}^{\dagger}
 \end{pmatrix}\\
&\,+\,
 \frac{1}{2}\sum_{k}(-2\Delta\sin ka)
\begin{pmatrix}
c_{k}^{\dagger}& c_{-k}
 \end{pmatrix}
 \begin{pmatrix}
0& -i\\
i&0
 \end{pmatrix}
 \begin{pmatrix}
c_{k}\\
 c_{-k}^{\dagger}
 \end{pmatrix}
\,+\,\frac{1}{2}\sum_{k}(-2t\cos ka)
\end{split}
\end{equation}
We have seen in Sec.\,\ref{bdgformalism} that when dealing with electrons and holes, in superconducting systems for instance, it is natural to define Nambu spinors
\begin{equation}
\psi_{k}\,=\,\begin{pmatrix}
c_{k}\\
c_{-k}^{\dagger}
 \end{pmatrix}\,.
\end{equation}  
Then, Eq.\,(\ref{introk3}) can be written as
\begin{equation}
\label{introk4}
\begin{split}
H&\,=\,\frac{1}{2}\sum_{k}\xi_{k}
\psi^{\dagger}_{k}
\tau_{z}
\psi_{k}\,+\,
 \frac{1}{2}\sum_{k}\Delta_{k}
\psi^{\dagger}_{k}
\tau_{y}
\psi_{k}
\,+\,\frac{1}{2}\sum_{k}(-2t\cos ka)\,,\\
&\,=\,\frac{1}{2}\sum_{k}
\psi^{\dagger}_{k}
H_{BdG}\psi_{k}\,+\,\frac{1}{2}\sum_{k}(-2t\cos ka)\,,\\
\end{split}
\end{equation}
where we have used that $\xi_{k}=\xi_{-k}$ , since $\xi_{k}$ is an even function of $k$, $\Delta_{k}=-2\Delta\sin ka$, and $H_{BdG}=\xi_{k}\tau_{z}+\Delta_{k}\tau_{y}$ 
being the Bogoliubov-de Gennes Hamiltonian. Notice that $\Delta_{k}$ is odd in $k$, thus exhibiting its $p$-wave nature.
\section{Rashba nanowire model}
\label{appRashba}
In this section we provide additional details that we find useful in order to have a complete view of what we present in Sec.\,\ref{Rashbawire} of the main text.
\subsection{Eigenvalues and eigenfunctions for the Rashba normal system}
In this part we calculate the eigen vectors for the Rashba system subjected to a perpendicular Zeeman magnetic field. 
The Hamiltonian for the normal system $H_{0}$ is given by Eq.\,(\ref{H0Hamil}). 
The eigenvalues and eigenvectors are found by solving the eigenvalue equation $H_{0}\Psi=\varepsilon\Psi$, where $\varepsilon$ are the eigen-values and $\Psi$ 
the eigenvectors that due to spin are two dimensional spinors.
The problem can be solved by supposing that solutions are given in terms of plane waves of the form
$\Psi={\rm e}^{ikr}(a,b)^{T}$, where $T$ stands for the transpose operation, and then one introduces them back into Eq.\,(\ref{H0Hamil}). For the eigevalues, ${\rm det}(H_{0}-\varepsilon)=0$,
 \begin{equation}
{\rm det}\begin{pmatrix}
\xi_{k}-\varepsilon&i\alpha k+B\\
-i\alpha k+B& \xi_{k}-\varepsilon
 \end{pmatrix}=
 \begin{pmatrix}
0\\
0
 \end{pmatrix}\,\rightarrow
 \varepsilon_{k,\pm}=\xi_{k}\pm\sqrt{B^{2}+\alpha_{R}^{2}k^{2}}\,,
 \end{equation}
 where $\xi_{k}=\hbar^{2}k^{2}/2m-\mu$.
  Now, by using the calculated eigenvalues and the form of the wavefunctions, one is able to find the values of $a$ and $b$. Of course, we should not 
  forget the normalization condition $|\Psi|^{2}=1$. Then, by writing down the eigenvalue equation, 
 \begin{equation}
\begin{pmatrix}
\xi_{k}-\varepsilon_{k,\pm}&i\alpha k+B\\
-i\alpha k+B& \xi_{k}-\varepsilon_{k,\pm}
 \end{pmatrix}
\begin{pmatrix}
a\\
b
 \end{pmatrix}=
 \begin{pmatrix}
0\\
0
 \end{pmatrix}\,,
 \end{equation}
we can write
\begin{equation}
 \label{syseqation}
 \begin{split}
 (\xi_{k}-\varepsilon_{k,\pm})a+(i\alpha k+B)b&=0\\
 (-i\alpha k+B)a+( \xi_{k}-\varepsilon_{k,\pm})b&=0\,,
 \end{split}
 \end{equation}
which are two coupled equations. 
  Only one gives us relevant information: one can express $a$ in terms of $b$ or the inverse, the result won't change.
 Now, we use the first equation and express $a$ in terms of $b$
 \begin{equation}
a=-\frac{(i\alpha k+B)}{  (\xi_{k}-\varepsilon_{k,\pm})}b\,,
 \end{equation}
 and the wavefunctions hence reads,
 \begin{equation}
 \Psi_{k,\pm}(r)=\begin{pmatrix}
a\\
b
 \end{pmatrix}{\rm e}^{ikr}=
  \begin{pmatrix}
-\frac{(i\alpha k+B)}{  (\xi_{k}-\varepsilon_{k,\pm})}b\\
b
 \end{pmatrix}{\rm e}^{ikr}=b
   \begin{pmatrix}
-\frac{(i\alpha k+B)}{ (\xi_{k}-\varepsilon_{k,\pm})}\\
1
 \end{pmatrix}{\rm e}^{ikr}\,,
 \end{equation}
 where $\xi_{k}-\varepsilon_{k,\pm}=\mp\sqrt{B^{2}+\alpha^{2}k^{2}}$. 
 Hence,
 \begin{equation}
 \Psi_{\pm}(k)=b  \begin{pmatrix}
\pm\frac{(i\alpha k+B)}{\sqrt{B^{2}+\alpha^{2}k^{2}}}\\
1
 \end{pmatrix}{\rm e}^{ikr}\,,
 \end{equation}
where $b$ can be calculated from the normalization condition  
 \begin{equation}
 \begin{split}
 |\Psi_{k,\pm}(r)|^{2}&=b^{*}\bigg(\pm\frac{(-i\alpha k+B)}{\sqrt{B^{2}+\alpha^{2}k^{2}}}, 1\bigg)
 b \begin{pmatrix}
\pm\frac{(i\alpha k+B)}{\sqrt{B^{2}+\alpha^{2}k^{2}}}\\
1
 \end{pmatrix}=1\,,\\
 &=|b|^{2}2=1
 \end{split}
  \end{equation}
 therefore $|b|=1/\sqrt{2}$ up to a phase that we do not consider for making the discussion easier. Then
  \begin{equation}
   \label{soeq1}
 \Psi_{k,\pm}(r)=\frac{1}{\sqrt{2}}\begin{pmatrix}
\pm\frac{(i\alpha k+B)}{\sqrt{B^{2}+\alpha^{2}k^{2}}}\\
1
 \end{pmatrix}{\rm e}^{ikr}\,.
 \end{equation}
On the other hand, we could also consider the second equation in \ref{syseqation}, then 
\begin{equation}
b=-\frac{ (-i\alpha k+B)}{ \xi_{k}-\varepsilon_{k,\pm}}a\,,
\end{equation}
and proceeding in the same way as before we find 
 \begin{equation}
 \label{soeq2}
 \Psi_{k,\pm}(r)=\frac{1}{\sqrt{2}}\begin{pmatrix}
1\\
\pm\frac{(B-i\alpha k)}{\sqrt{B^{2}+\alpha^{2}k^{2}}}
 \end{pmatrix}{\rm e}^{ikr}\,.
 \end{equation}
 Both solutions, Eq.\,(\ref{soeq1}) and Eq.\,(\ref{soeq2}) are equivalent.
In the main text, we consider the wavefunctions in the form given by Eq.\,(\ref{soeq1}). We comment here that the eigenvectors shown in Eq.\,(\ref{H0vEnVec}) are normalised to a finite region of length $L$, that is why in such equation we find the factor $1/\sqrt{L}$.
 
   \begin{figure} 
   \centering\includegraphics[width=.7\columnwidth]{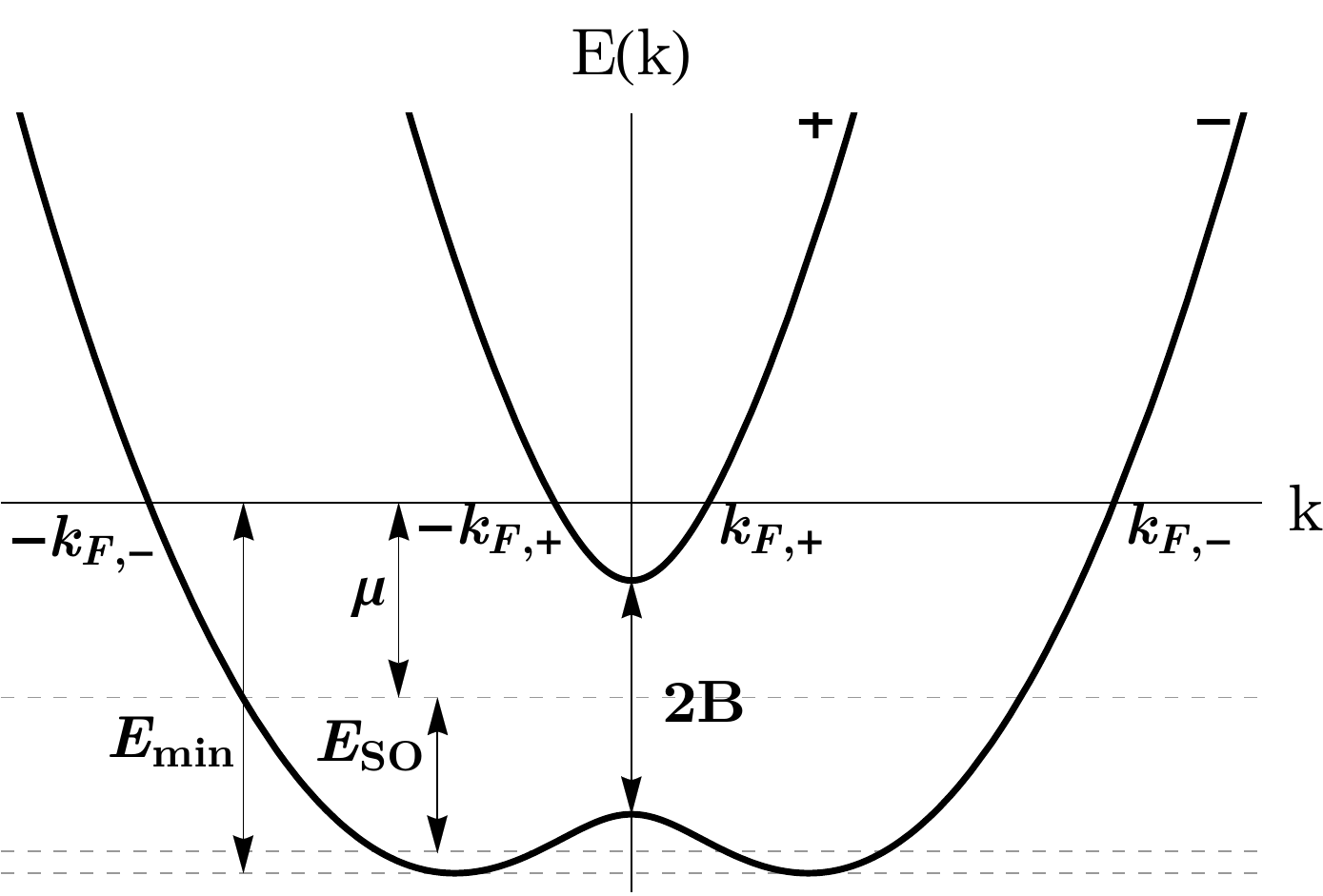} 
     \caption[Energy dispersion of a Rashba Nanowire]{(Color online) (left) 
     Energy dispersion for a Rashba Nanowire given by Eq.\,(\ref{H0vEnVec}) showing the points where the energy cross the zero energy line.
}
   \label{fig:RashbaNWx}
\end{figure}
The momenta where the energy intersects the $k$-axis are determined from $\varepsilon_{k,\pm}=0$, then,
\begin{equation}
\label{fermpoints}
\begin{split}
k_{1}&=-\sqrt{k_{\mu}^{2}+2k_{so}^{2}-\sqrt{\left( k_{\mu}^{2}+2k_{so}^{2}\right)^{2}-k_{\mu}^{4}+k_{Z}^{4}}}\,,\\
k_{2}&=+\sqrt{k_{\mu}^{2}+2k_{so}^{2}-\sqrt{\left( k_{\mu}^{2}+2k_{so}^{2}\right)^{2}-k_{\mu}^{4}+k_{Z}^{4}}}\,,\\
k_{3}&=-\sqrt{k_{\mu}^{2}+2k_{so}^{2}+\sqrt{\left( k_{\mu}^{2}+2k_{so}^{2}\right)^{2}-k_{\mu}^{4}+k_{Z}^{4}}}\,,\\
k_{4}&=+\sqrt{k_{\mu}^{2}+2k_{so}^{2}+\sqrt{\left( k_{\mu}^{2}+2k_{so}^{2}\right)^{2}-k_{\mu}^{4}+k_{Z}^{4}}}\,,
\end{split}
\end{equation}
where $k_{2}=k_{-}-k_{1}$ and $k_{4}=-k_{3}$ and
\begin{equation}
 k_{\mu}=\sqrt{2m\mu}/\hbar\,,\quad k_{so}=\alpha_{so} m/\hbar^{2}\,,\quad k_{Z}=\sqrt{2mB}/\hbar\,.
\end{equation}
We identify that $k_{2}=k_{+}$ and $k_{4}=k_{-}$, the wavevectors associate to the eigenvalues $\varepsilon_{k,\pm}$, as shown in Fig.\,\ref{fig:RashbaNWx}
The spectrum in this case has extremes at $k=0$ and at $\pm k_{min}=\pm\sqrt{k_{so}^{2}-k_{Z}^{4}/4k_{so}^{2}}$, and the corresponding energies are $-E_{F}\pm B$ and $E_{min}=-\mu-E_{so} -\Delta_{Z}^{2}/4E_{so}$, respectively.

\subsection{Role of the Zeeman field parallel to the spin-orbit axis }
\label{updownbasis}
In this subsection we discuss the role of a Zeeman magnetic field parallel to the spin-orbit axis,
which is considered to be 
\begin{equation}
\label{perpB}
\mathcal{H}_{Z\myparallel}=B_{\myparallel}\,\sum_{\sigma\sigma^{\prime}}\int\,dx\,
\psi^{\dagger}_{\sigma}(x)\left[\sigma_{y}\right]_{\sigma\sigma^{\prime}}\psi_{\sigma^{\prime}}(x)\,,
\end{equation}
where $B_{\myparallel}$ is the magnitude of such field.
The full Hamiltonian is then given by $\mathcal{H}=\mathcal{H}_{kin}+\mathcal{H}_{SO}+\mathcal{H}_{Z}+\mathcal{H}_{sc}$, and
the Zeeman field contains two components $H_{Z}=\mathcal{H}_{Z,\myparallel}+\mathcal{H}_{Z,\perp}$. 
The perpendicular component is given by Eq.\,(\ref{B1Hamil}) and denoted simple by $B$ in the main text. This term is 
the usual term considered in the literature and it plays a fundamental part towards the physical implementation for the search of MBSs in Rashba nanowires.
On the other hand, we point out that experimentally is not so easy to control the magnetic field along a certain direction, giving rise to 
a non-zero components in both perpendicular and parallel directions. This motivates us in order to discuss the role of the component parallel to the spin
orbit axis. 
Notice that now the Zeeman field is a vector with two components and can be parametrized as 
\begin{equation}
\label{decomposition}
B_{\perp}=B\sin\theta\cos\phi \,,\quad B_{\myparallel}=B\sin\theta\sin\phi\,.
\end{equation}

The full Hamiltonian, $\mathcal{H}$, can be then written in Nambu basis,
$\Psi=(\psi_{\uparrow}, \psi_{\downarrow}, \psi_{\uparrow}^{\dagger},\psi_{\downarrow}^{\dagger})^{\dagger}$, so that 
\begin{equation}
\mathcal{H}\,=\,\frac{1}{2}\int dx\,\Psi^{\dagger}H_{BdG}\,\Psi\,,
\end{equation}
where, 
\begin{equation}
\begin{split}
H_{BdG}&\,=\,\left(-\hbar^{2}\partial^{2}_{x}/2m-\mu\right)\tau_{z}\sigma_{0}\,-\,i\alpha_{R}\tau_{z}\sigma_{y}\,\partial_{x}+\,B_{\perp}\tau_{z}\sigma_{x}\,
+\,B_{\myparallel}\tau_{0}\sigma_{y}\,+\,\Delta\tau_{y}\sigma_{y}\,,
\end{split}
\end{equation}
is the Bogoliubov de Gennes Hamiltonian. The Pauli matrices $\sigma$ and $\tau$  act in spin and electron-hole subspaces, respectively.  
The spectrum is then given after solving the eigenvalue problem $H_{BdG}\Psi=E\Psi$, from where one gets the polynomial 
\begin{equation}
\label{quartic}
\begin{split}
P(E)\,&\equiv E^{4}+aE^{2}+bE+c\,=\,0\,,\\
a\,&=\,-2\left[B_{\perp}^{2}+B_{\myparallel}^{2}+(\alpha k)^{2}+\Delta^{2}+E_{k}^{2}\right]\\
b\,&=\,-8\left[B_{\myparallel}\,\alpha k\,E_{k}\right] \\
c\,&=\,E_{k}^{2}\left[ E_{k}^{2}-2\left(B_{\perp}^{2}+B_{\myparallel}^{2}+(\alpha k)^{2}-\Delta^{2}\right)\right]
+\left[ (\alpha k)^{2}-B_{\myparallel}^{2}+\Delta^{2}\right]^{2}\\
\,&\,\quad+\,B_{1}^{2}\left[B_{\perp}^{2}+2\left(B_{\myparallel}^{2}+(\alpha k)^{2}-\Delta^{2}\right)\right]\,\\
\xi_{k}\,&=\,\hbar^{2}k^{2}/2m-\mu\,.
\end{split}
\end{equation}
For $B_{\myparallel}=0$, the spectrum reads
\begin{equation}
E^{2}_{\pm}=\xi_{k}^{2}+(\alpha_{R} k)^{2}+B^{2}_{\perp}+\Delta^{2}\pm2\sqrt{B_{\perp}^{2}\Delta^{2}+\left[B_{\perp}^{2}+(\alpha_{R}k)^{2}\right]\xi_{k}^{2}}\,,
\end{equation}
and its evolution with the Zeeman field is depicted in Fig.\,\ref{fig:fullRashbaNambu}. Notice that  in the main text we denoted $B_{\perp}=B$.
We have seen that proximity induced superconductivity modifies the gap at $k=0$, originally opened by the perpendicular Zeeman field, and opens a gap of 
$\Delta_{2}=E_{-}(k_{F,-})$, at finite momentum. 
The topological transition is marked by vanishing the gap at small momenta $\Delta_{1}\approx|B-B_{c}|$. It happens at the critical field defined as $B_{c}=\sqrt{\Delta^{2}+\mu^{2}}$, where for $B>B_{c}$ the system host two Majorana bound states at the end of the wire.
While the gap at small momenta gets reduces as one increases $B_{\perp}$ and it closes at the critical field $B_{c}$, the outer gap $\Delta_{2}$ remains roughly constant in strong SOC at $k\approx k_{F,-}$ for any $B_{\perp}$. See App.\,\ref{appendixgapsapp}.

Hence, it is natural to ask whether the condition, $\Delta_{1}=0$ at $B_{c}$, holds or not for a non-zero paralell component $B_{\myparallel}\neq0$.
 Further insight can be extracted from the energy spectrum, but first let us check some situations.
We solve $P(E)=0$ and get
\begin{equation}
\label{solutions}
\begin{split}
E_{1}\,&=\,\frac{d_{5}}{2}-\frac{1}{2}\sqrt{-\frac{4a}{3}-d_{4}-\frac{2b}{d_{5}}}\,,\\
E_{2}\,&=\,\frac{d_{5}}{2}+\frac{1}{2}\sqrt{-\frac{4a}{3}-d_{4}-\frac{2b}{d_{5}}}\,,\\
E_{3}\,&=\,-\frac{d_{5}}{2}-\frac{1}{2}\sqrt{-\frac{4a}{3}-d_{4}+\frac{2b}{d_{5}}}\,,\\
E_{4}\,&=\,-\frac{d_{5}}{2}+\frac{1}{2}\sqrt{-\frac{4a}{3}-d_{4}+\frac{2b}{d_{5}}}\,,
\end{split}
\end{equation}
where
\begin{equation}
\begin{split}
d_{5}\,&=\,\sqrt{-\frac{2a}{3}+d_{4}}\,,\nonumber\\
d_{4}\,&=\,\frac{d_{1}}{3}\left( \frac{2}{d_{3}}\right)^{1/3}+\frac{1}{3}\left(\frac{d_{3}}{2}\right)^{1/3}\,,\nonumber\\
d_{3}\,&=\,d_{2}+\sqrt{-4d_{1}^{3}+d_{2}^{2}}\,,\quad
d_{2}\,=\,2a^{3}+27b^{2}-72ac\,,\quad
d_{1}\,=\,a^{2}+12c\,,\nonumber
\end{split}
\end{equation}
where $a,b,c$ and $E_{k}$ are defined in Eq.\,(\ref{quartic}).

The inner gap at zero momentum, $\Delta_{1}=E_{1,4}(k=0)\,$, is calculated from Eqs\,(\ref{solutions}), and reads
\begin{equation}
\Delta_{1}=\,\left| \sqrt{B_{\perp}^{2}+B_{\myparallel}^{2}}-\sqrt{\Delta^{2}+\mu^{2}}\right|\,.
\end{equation}
If one now introduces Eqs.\,(\ref{decomposition}) into previous equation, we get  $\Delta_{-}=|B-B_{c}|$. Therefore, we conclude that at this point the condition of the closing of the gap for entering into the topological phase does not change.

\begin{figure} 
 \centering \includegraphics[width=0.99\columnwidth]{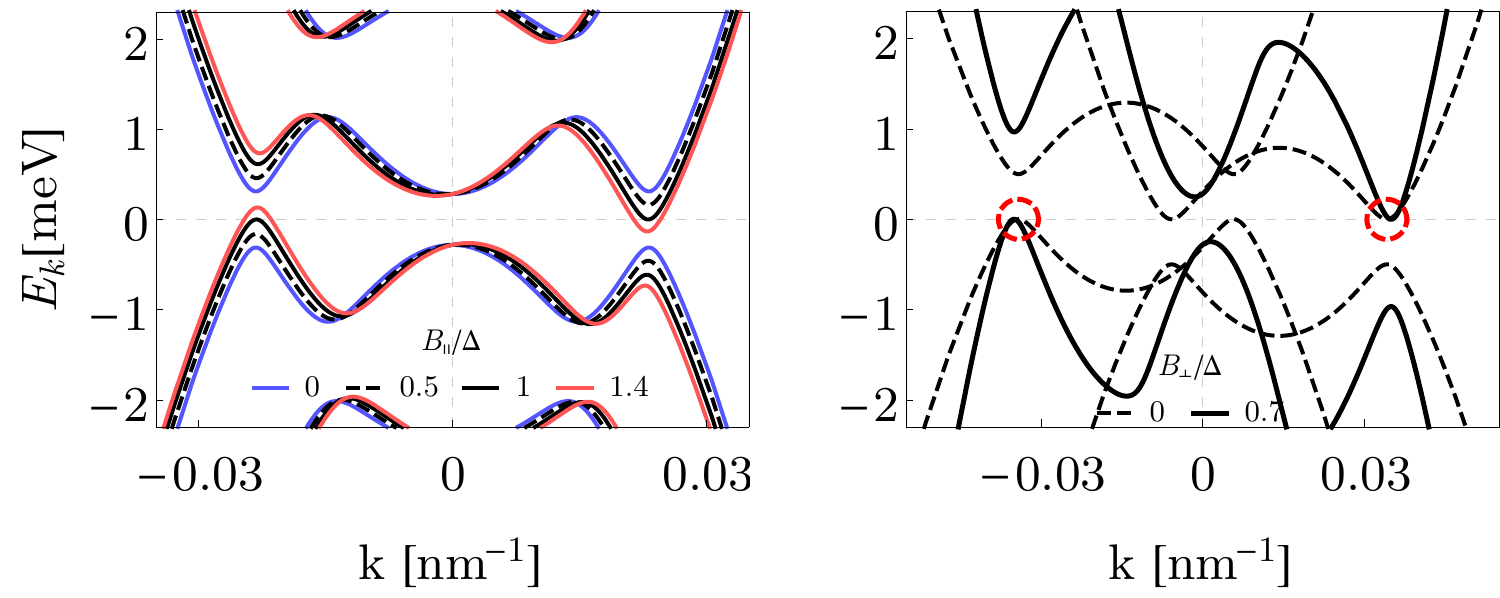}
     \caption[Band spectrum with both components of the Zeeman field]{(Color online) (Left) Band spectrum given by Eqs.\,(\ref{solutions}) in the topological phase $B>B_{c}$. Different curves represent different values of $\theta=90$ and $\phi=0,10,21,30$. Parameters are $\alpha_{R}=20$meVnm, $\mu=0.5meV$. 
     Right panel shows solutions given by Eqs.\,(\ref{solutions}) at $B_{\perp}=0$ and $B_{\perp}\neq0$. Both curves are for $\alpha_{R}=70$meV and $\mu=0.5$meV. 
     Notice the good agreement at $k=k_{F,-}$.}
   \label{fig1perp}
\end{figure}

On the other hand, the outer gap at high momentum, $\Delta_{2}=E_{1,4}(k=\pm k_{1})$, is not constant anymore and exhibits a different behavior as it was for zero parallel field, shown in App.\,\ref{appendixgapsapp}.
To have an intuition of what is happening, in Fig.\,\ref{fig1perp} we plot the solutions given by Eqs.\,(\ref{solutions}) for different values of $B_{\myparallel}$. 
One observes that the outer gap, for $E_{1}(k)$ at $-k_{1}$ and $E_{4}$ at $+k_{1}$, decreases as one increases $B_{\myparallel}$. 
This behavior is important as it is the constantness of such gap the one that offers topological protection towards physical implementation
of quantum computation free of decoherence.
Thus, we are interested in knowning when such gap reaches zero.
We solve $E_{1,4}=0$ at $\mp k_{F,-}$ for $B_{\myparallel}$ and get the condition when such bands reach zero energy,
\begin{equation}
 B_{\myparallel}\approx |\Delta|\,.
\end{equation}
This can be understood in the following oversimplified view. Indeed, despite of being in the topological phase or not, the perpendicular component does not play a transcendental role at the outer branch in case of strong SOC and the momentum 
 is almost constant $k\approx k_{F,-}$.
Therefore, the problem can be analyzed in a situation of zero perpendicular $B_{\perp}=0$. 
Thus from $P(E,B_{\perp}=0)=0$ one gets
\begin{equation}
\label{solutionb10}
\begin{split}
E_{1}^{0}\,&=\,-B_{\myparallel}+\sqrt{(\xi_{k}-\alpha_{R}k)^{2}+\Delta^{2}}\,,\\
E_{2}^{0}\,&=\,B_{\myparallel}+\sqrt{(\xi_{k}+\alpha_{R}k)^{2}+\Delta^{2}}\,,\\
E_{3}^{0}\,&=\,-B_{\myparallel}-\sqrt{(\xi_{k}-\alpha_{R}k)^{2}+\Delta^{2}}\,,\\
E_{4}^{0}\,&=\,B_{\myparallel}-\sqrt{(\xi_{k}+\alpha_{R}k)^{2}+\Delta^{2}}\,.
\end{split}
\end{equation}
In right panel of Fig.\,\ref{fig1perp} we plot the band spectrum in zero and finite perpendicular field, corresponding to Eqs.\,(\ref{solutionb10}) and (\ref{solutions}), respectively. One indeed notices that for $k_{F,-}\approx \pm 2k_{so}$ the agreement is very good, as we expected. 
Without loss of generality, one can tune the chemical potential to zero $\mu=0$. Then from Eqs.\,(\ref{solutionb10}) $E_{1,4}^{0} $at $k=\pm2k_{so}$ one has 
the condition for the closing of the outer gap: $B_{\myparallel}\approx|\Delta|$ \cite{PhysRevB.89.245405}.
{\bf Inverse iteration}:
To support our finding about the condition of the closing of the outer gap, inverse iteration method can be employed \cite{numrecipes}.
Here, the wave-function corresponding to $E_{1}(k_{1})=0$ can be used to find $E_{1}(k)$ around $k_{1}$ and 
therefore calculate the outer gap by means of inverse iteration \cite{numrecipes}. 
First, we solve $H_{BdG}\Psi_{0}=0$ at $k=k_{1}$, from where we get $\Psi_{0}$ at $k_{1}$. 
A general solution thus reads $\Psi_{k}(k)={\rm e}^{ikx}(U_{1},U_{2},U_{3},U_{4})^{\dagger}$, where $|U_{1}|^{2}+|U_{2}|^{2}+|U_{3}|^{2}+|U_{4}|^{2}=1$ is the normalization condition.
 \begin{figure} 
 \centering \includegraphics[width=.9\columnwidth]{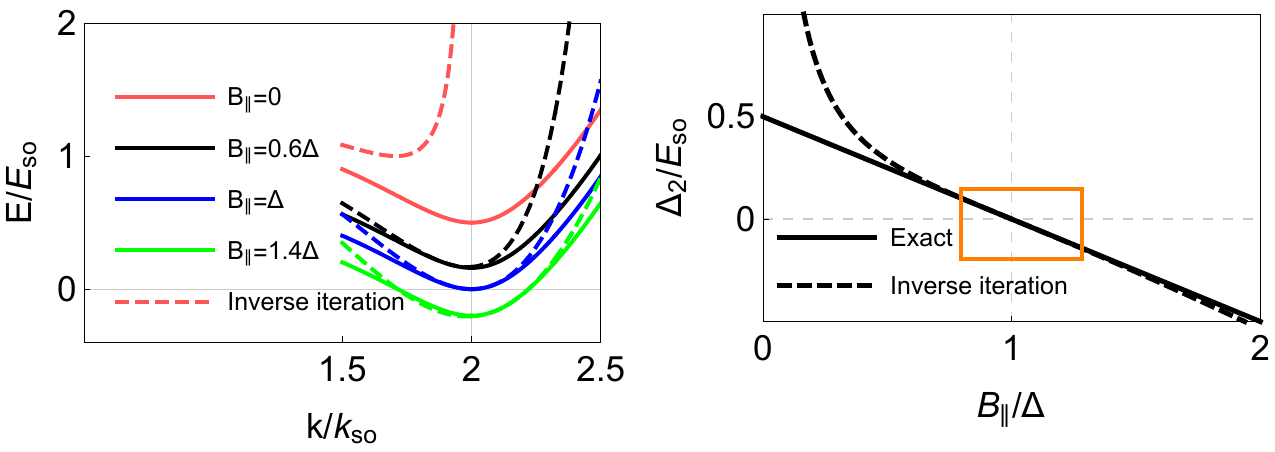}
      \caption[Exact and inverse iteration approaches for the gap at high momentum]{(Color online) (left) Evolution of $E_{1}(k)$ around $k_{1}$ for different $B_{\parallel}$, considering exact solution (solid) and inverse iteration (dashed). 
      (right) Outer gap as function of $B_{\parallel}$ for the exact solution (solid) and by inverse iteration (dashed). Notice in both panels the good agreement when $B_{\parallel}\approx\Delta$}
 \label{figx}
\end{figure}
For strong SOC, the wave-function reads $\Psi_{0}(x)\,\approx\,{\rm e}^{ik_{1}x}\left(B_{\myparallel},-iB_{\myparallel},-i\Delta,-\Delta\right)^{\dagger}/N$, where $N={ \sqrt{2}\sqrt{ B_{\myparallel}^{2}+\Delta^{2}}}$.
The improved eigenvector $\Psi_{N}(k)$ for any $k$ around $k_{1}$ is then calculated from $H_{BdG}(k)\Psi_{N}(k)=\Psi_{0}(k_{1})$. 
Since we already know the wave function at $k_{1}$ for $E_{1}=0$, we can improve such eigenvalue by inverse iteration using $\Psi_{N}$ for any $k$ around $k_{1}$.
Indeed, the improved eigenvalue correction reads, $E_{1}(k_{1})\approx E_{1}(k=k1,B_{\myparallel}=\Delta)+|\Psi_{0}|^{2}/(\Psi_{N}^{\dagger}\Psi_{0})$, 
where the first term is zero at $k=k_{1}$ and we only have a contribution from the second term. Here $\Psi_{0}$ and $\Psi_{N}$ are already known from previous discussion, therefore 
the improved eigenvalue then reads,
\begin{equation}
\label{sol}
E_{1}(k)=-\frac{(4B_{\myparallel}^{2}-4\Delta^{2}-k^{2}(k-2\alpha))(B_{\myparallel}^{2}+\Delta^{2})}{D_{1}}\,,
\end{equation}
where $D_{1}=2B_{\myparallel}^{2}(2B_{\myparallel}-k^{2}(k-2\alpha_{R}))+2(6B_{\myparallel}-k^{2}(k-2\alpha))\Delta^{2}\neq0$. Eq.\,(\ref{sol}) can be then expanded around $2\alpha_{R}$, $k\rightarrow2\alpha_{R}+\delta$ for small $\delta$, where we find
\begin{equation}
E_{1}(\delta)=\gamma(\delta-\delta_{0})^{2}-\delta_{0}+\eta\,,
\end{equation}
where
\begin{equation}
\begin{split}
\gamma=\frac{(B_{\myparallel}^{2}+\Delta^{2})D_{2}}{2B_{\myparallel}^{3}(B_{\myparallel}^{2}+3\Delta^{2})^{3}},\quad&
\beta=\frac{\alpha_{R}(B_{\myparallel}^{2}-\Delta^{2})^{2}(B_{\myparallel}^{2}+\Delta^{2})}{B_{\myparallel}^{2}(B_{\myparallel}^{2}+3\Delta^{2})^{2}}\,,\quad
\eta=\frac{-B_{\myparallel}^{4}+\Delta^{4}}{B_{\myparallel}(B_{\myparallel}^{2}+3\Delta^{2})},\quad
\delta_{0}=-\frac{\beta}{2\gamma}\,\nonumber,
\end{split}
\end{equation}
and $D_{2}=B_{\myparallel}^{4}[B_{\myparallel}^{3}+(B_{\myparallel}+18\alpha_{R}^{2})\Delta^{2}]+[B_{\myparallel}^{2}(-5B_{\myparallel}+12\alpha_{R}^{2})+(3B_{\myparallel}+2\alpha^{2})\Delta^{2}]\Delta^{4}$.
At $\delta= \delta_{0}$, the outer gap $E_{1}(\delta=\delta_{0})\approx\Delta_{2}$,
\begin{equation}
\Delta_{2}\approx\frac{-B_{\myparallel}^{4}+\Delta^{4}}{B_{\myparallel}(B_{\myparallel}^{2}+3\Delta^{2})}
\end{equation}
In Fig.\,\ref{figx} we plot previous equation as function $B_{\myparallel}$ and the exact value calculated from Eqs.\,(\ref{solutions}). 
Our result by inverse iteration is in a good agreement only when the parallel field $B_{\myparallel}$ is close to $\Delta$, as we have indeed expected.

\subsection{Full Rashba Hamiltonian in the helical basis}
In this part we show how the Hamiltonian for the Rashba nanowire in proximity to a $s$-wave superconductor, $\mathcal{H}_{0}+\mathcal{H}_{sc}$ given by Eq.\,(\ref{fullHamiltonian}), 
is written in the so-called helical basis \cite{Alicea:PRB10,Alicea:RPP12},
\begin{equation}
 \label{eq111}
 \psi(k)=\phi_{-}(k)\psi_{-}(k)+\phi_{+}(k)\psi_{+}(k)\,,
 \end{equation}
 where $\psi_{\pm}$ are operators that annihilates states in the upper/lower bands and $\phi_{\pm}$ the respective normalized wavefunctions given by Eq.\,(\ref{H0vEnVec}) that we decompose into the two spinor components
  \begin{equation}
   \label{soeq5}
 \phi_{+}(k)=\begin{pmatrix}
\phi^{\uparrow}_{+}(k)\\
\phi^{\downarrow}_{+}(k)
 \end{pmatrix}=
 \frac{1}{\sqrt{2}}\begin{pmatrix}
+\gamma_{k}\\
1
 \end{pmatrix}\,,
 \end{equation}
 and
   \begin{equation}
   \label{soeq6}
 \phi_{-}(k)=\begin{pmatrix}
\phi^{\uparrow}_{-}(k)\\
\phi^{\downarrow}_{-}(k)
 \end{pmatrix}=
 \frac{1}{\sqrt{2}}\begin{pmatrix}
-\gamma_{k}\\
1
 \end{pmatrix}\,,
 \end{equation}
where $\gamma_{k}=\frac{(i\alpha k+B)}{\sqrt{B^{2}+\alpha^{2}k^{2}}}$. 
According to previous discussion, it is more appropriate to write down Eq.\,(\ref{eq111}) as follows 
  \begin{equation} 
 \label{helicalbasis}
 \begin{split}
  \psi_{\uparrow}(k)&=\phi_{-}^{\uparrow}(k)\psi_{-}(k)+\phi_{+}^{\uparrow}(k)\psi_{+}(k)\,,\\
  \psi_{\downarrow}(k)&=\phi^{\downarrow}_{-}(k)\psi_{-}(k)+\phi^{\downarrow}_{+}(k)\psi_{+}(k)\,,
 \end{split}
 \end{equation}
 which using Eqs.\,(\ref{soeq5}) and (\ref{soeq5}) become
 \begin{equation}
  \label{helicalbasis2}
 \begin{split}
   \psi_{\uparrow}(k)&=\frac{1}{\sqrt{2}}\left[-\gamma_{k}\psi_{-}(k)+\gamma_{k}\psi_{+}(k)\right]\,,\\
   \psi_{\downarrow}(k)&=\frac{1}{\sqrt{2}}\left[\psi_{-}(k)+\psi_{+}(k)\right]\,,\\
 \end{split}
 \end{equation}
 In what follows, we proceed to express the terms of the full Hamiltonian given by Eq.\,(\ref{fullHamiltonian}) in the basis given by Eqs.\,(\ref{helicalbasis}), resp. by 
 Eqs.\,(\ref{helicalbasis2}).
 
 \subsubsection*{Kinetic term}
 The kinetic Hamiltonian is given by,
 \begin{equation}
 \label{kinhamil}
 \mathcal{H}_{kin}\,=\,\int dx\,\big[\psi_{\uparrow}^{\dagger}(x)\left[ \xi_{k}\right] \psi_{\uparrow}\,+\,\psi_{\downarrow}^{\dagger}(x)\left[ \xi_{k}\right] \psi_{\downarrow}\big]\,,
 \end{equation}
where $\xi_{k}=\hbar^{2}k^{2}/2m-\mu$\,. Now, we use Eqs.\,(\ref{helicalbasis2}), and then write the elements of previous Hamiltonian in the so called helical basis,
 \begin{equation}
 \begin{split}
 \psi_{\uparrow}^{\dagger}\psi_{\uparrow}
 &=\frac{1}{2}\left[ \psi_{-}^{\dagger}(k)\psi_{-}(k)- \psi_{+}^{\dagger}(k)\psi_{-}(k)-\psi_{-}^{\dagger}(k)\psi_{+}(k)+\psi_{+}^{\dagger}(k)\psi_{+}(k)\right]\,,\\
  \psi_{\downarrow}^{\dagger}\psi_{\downarrow}
 &=\frac{1}{2}\left[ \psi_{-}^{\dagger}(k)\psi_{-}(k)+ \psi_{+}^{\dagger}(k)\psi_{-}(k)+\psi_{-}^{\dagger}(k)\psi_{+}(k)+\psi_{+}^{\dagger}(k)\psi_{+}(k)\right]\,,
 \end{split}
 \end{equation}
 where we have used that $\gamma^{\dagger}_{k}\gamma_{k}=1$. 
 Now, we add up previous two expressions for $\psi_{\uparrow}^{\dagger}\psi_{\uparrow}$ and $ \psi_{\downarrow}^{\dagger}\psi_{\downarrow}$, and 
 plug them into the kinetic Hamiltonian given by Eq.\,(\ref{kinhamil}),
  \begin{equation}
 \label{kinhamil2}
 \mathcal{H}_{kin}\,=\,\int dx\,\xi_{k}\Big[\psi_{-}^{\dagger}(k) \psi_{-}(k)\,+\,\psi_{+}^{\dagger}(k) \psi_{+}(k)\Big]\,.
 \end{equation}
 \subsubsection*{Spin-orbit and Zeeman terms}
 The spin-orbit Hamiltonian from Eq.\,(\ref{soH}) reads
 \begin{equation}
\label{soH2}
\begin{split}
\mathcal{H}_{soc}&\,=\,(i\alpha k)\int\,dx\left[\psi_{\uparrow}^{\dagger}\,\psi_{\downarrow}\,-\,\psi_{\downarrow}^{\dagger}\,\psi_{\uparrow} \right]\,.
\end{split}
\end{equation}
Then we express the terms of previous Hamiltonian using Eqs.\,(\ref{helicalbasis2}),
\begin{equation}
\begin{split}
\psi^{\dagger}_{\uparrow}\psi_{\downarrow}
&= \frac{1}{2}\left[-\psi_{-}^{\dagger}(k)\psi_{-}(k)\gamma^{\dagger}_{k}+\psi_{+}^{\dagger}(k)\psi_{-}(k)\gamma_{k}^{\dagger}
-\psi_{-}^{\dagger}(k)\psi_{+}(k)\gamma^{\dagger}_{k}+\psi_{+}^{\dagger}(k)\psi_{+}(k)\gamma_{k}^{\dagger} \right]\,,\\
\psi^{\dagger}_{\downarrow}\psi_{\uparrow}&=
\frac{1}{2}
\left[-\psi_{-}^{\dagger}(k)\psi_{-}(k)\gamma_{k}-\psi_{+}^{\dagger}(k)\psi_{-}(k)\gamma_{k}
+
\psi_{-}^{\dagger}(k)\psi_{+}(k)\gamma_{k}+\psi_{+}^{\dagger}(k)\psi_{+}(k)\gamma_{k}
\right]\,.
\end{split}
\end{equation}
Therefore, subtracting previous two equations just to get the spin-orbit Hamiltonian we have
 \begin{equation}
\label{soH3}
\begin{split}
\mathcal{H}_{SO}
&=\frac{i\alpha k}{2}\int\,dx
\Big[
\psi_{-}^{\dagger}(k)(\gamma_{k}-\gamma^{\dagger}_{k})\psi_{-}(k)
+\psi_{+}^{\dagger}(k)(\gamma_{k}^{\dagger}-\gamma_{k})\psi_{+}(k)\\
&\quad\quad\quad \quad \quad \quad +\psi_{+}^{\dagger}(k)(\gamma_{k}^{\dagger}+\gamma_{k})\psi_{-}(k)
-\psi_{-}^{\dagger}(k)(\gamma_{k}^{\dagger}+\gamma_{k})\psi_{+}(k)
\Big]\,.
\end{split}
\end{equation}
The Zeeman Hamiltonian from Eq.\,(\ref{B1Hamil}) reads,
 \begin{equation}
\label{Hzemanz}
\begin{split}
\mathcal{H}_{z}&\,=\,B\int\,dx\left[\psi_{\uparrow}^{\dagger}\,\psi_{\downarrow}\,+\,\psi_{\downarrow}^{\dagger}\,\psi_{\uparrow} \right]\,.
\end{split}
\end{equation}
Therefore, using the same terms that helped us to express the spin-orbit Hamiltonian, we arrive at
 \begin{equation}
\label{Hzemanz2}
\begin{split}
\mathcal{H}_{Z}
&=\frac{B}{2}\int\,dx
\Big[
-\psi_{-}^{\dagger}(k)(\gamma_{k}+\gamma^{\dagger}_{k})\psi_{-}(k)
+\psi_{+}^{\dagger}(k)(\gamma_{k}^{\dagger}+\gamma_{k})\psi_{+}(k)\\
&\quad\quad\quad \quad \quad \quad 
+\psi_{+}^{\dagger}(k)(\gamma_{k}^{\dagger}-\gamma_{k})\psi_{-}(k)
-\psi_{-}^{\dagger}(k)(\gamma_{k}^{\dagger}-\gamma_{k})\psi_{+}(k)
\Big]\,.
\end{split}
\end{equation}
Now, we add up the spin-orbit and Zeeman terms derived above 
\begin{equation}
\label{eq110}
\begin{split}
\mathcal{H}_{SO}+\mathcal{H}_{Z}&=\int\frac{dx}{2}
\Big\{
\psi_{-}^{\dagger}(k)\big[i\alpha k(\gamma_{k}-\gamma^{\dagger}_{k})-B(\gamma_{k}+\gamma^{\dagger}_{k})\big]\psi_{-}(k)
+
\psi_{+}^{\dagger}(k)\big[i\alpha k(\gamma_{k}^{\dagger}-\gamma_{k})+B(\gamma_{k}^{\dagger}+\gamma_{k})\big]\psi_{+}(k)\\
&
+\psi_{+}^{\dagger}(k)\big[i\alpha k(\gamma_{k}^{\dagger}+\gamma_{k})+B(\gamma_{k}^{\dagger}-\gamma_{k})\big]\psi_{-}(k)
+\psi_{-}^{\dagger}(k)\big[-i\alpha k(\gamma_{k}^{\dagger}+\gamma_{k})-B(\gamma_{k}^{\dagger}-\gamma_{k})\big]\psi_{+}(k)
\Big\}\,.
\end{split}
\end{equation}
Next, we write down some useful relations, to be used later, derived from the definition of $\gamma_{k}=\frac{B+i\alpha k}{\sqrt{B^{2}+\alpha^{2}k^{2}}}$,
\begin{equation}
\begin{split}
\gamma_{k}^{\dagger}-\gamma_{k}&=\frac{B-i\alpha k}{\sqrt{B^{2}+\alpha^{2}k^{2}}}-\frac{(B+i\alpha k)}{\sqrt{B^{2}+\alpha^{2}k^{2}}}=
\frac{-2i\alpha k}{\sqrt{B^{2}+\alpha^{2}k^{2}}}\\
\gamma_{k}^{\dagger}+\gamma_{k}&=\frac{B-i\alpha k}{\sqrt{B^{2}+\alpha^{2}k^{2}}}+\frac{B+i\alpha k}{\sqrt{B^{2}+\alpha^{2}k^{2}}}=
\frac{2B}{\sqrt{B^{2}+\alpha^{2}k^{2}}}\,,
\end{split}
\end{equation}
Then, we work out the elements of Eq.\,(\ref{eq110}, and get
\begin{equation}
 \mathcal{H}_{SO}+\mathcal{H}_{Z}
=\int dx
\bigg\{
\psi_{-}^{\dagger}(k)\bigg[-\sqrt{B^{2}+\alpha^{2}k^{2}}\bigg]\psi_{-}(k)
+
\psi_{+}^{\dagger}(k)\bigg[\sqrt{B^{2}+\alpha^{2}k^{2}}\bigg]\psi_{+}(k)
\bigg\}\,.
\end{equation}

\subsubsection*{Kinetic, spin-orbit and Zeeman terms}
Now we add up the three terms we have calculated in previous subsections,
Therefore,
we can write
\begin{equation}
\begin{split}
\mathcal{H}_{kin}+\mathcal{H}_{SO}+\mathcal{H}_{Z}&=
\int dx
\big[
\varepsilon_{+}(k)\psi_{+}^{\dagger}(k)\psi_{+}(k)
+
\varepsilon_{-}(k)\psi_{-}^{\dagger}(k)\psi_{-}(k)
\big]\,,
\end{split}
\end{equation}
where $\varepsilon_{k,\pm}=\xi\pm\sqrt{B^{2}+\alpha^{2}k^{2}}$ are the eigenvalues of the Rashba problem with Zeeman field given by Eq.\,(\ref{H0vEnVec}).
\subsubsection*{The superconducting term}
\label{helicalsup}
The superconducting Hamiltonian given by Eq.\,(\ref{hsc}) is written as
\begin{equation}
\mathcal{H}_{sc}=\frac{1}{2}\int dx\Big\{\Delta\Big[\psi_{\uparrow}^{\dagger}(k)\psi_{\downarrow}^{\dagger}(-k)-\psi_{\downarrow}^{\dagger}(-k)\psi_{\uparrow}^{\dagger}(k)\Big]+h.c\Big\}\,,
\end{equation}
where we have used fermion anti-commutations relations 
\begin{equation}
\big\{\psi^{\dagger}_{\uparrow}(k),\psi^{\dagger}_{\downarrow}(-k)\big\}=\psi^{\dagger}_{\uparrow}(k)\psi^{\dagger}_{\downarrow}(-k)
+\psi^{\dagger}_{\downarrow}(-k)\psi^{\dagger}_{\uparrow}(k)=0\,.
\end{equation}
Now, we express the elements of $\mathcal{H}_{sc}$ in terms of the helical basis, given by Eqs.\,(\ref{helicalbasis2}),
\begin{equation}
\begin{split}
\psi^{\dagger}_{\uparrow}(k)\psi^{\dagger}_{\downarrow}(-k)&=
\frac{1}{2}\Big[-\psi_{-}^{\dagger}(k)\psi_{-}^{\dagger}(-k)\gamma_{k}^{\dagger}+\psi_{+}^{\dagger}(k)\psi_{-}^{\dagger}(-k)\gamma_{k}^{\dagger}
-\psi_{-}^{\dagger}(k)\psi^{\dagger}_{+}(-k)\gamma_{k}^{\dagger}+\psi_{+}^{\dagger}(k)\psi^{\dagger}_{+}(-k)\gamma_{k}^{\dagger}
 \Big]\,,\\
 \psi^{\dagger}_{\downarrow}(-k)\psi^{\dagger}_{\uparrow}(k)&=
 \frac{1}{2}\Big[
 -\psi_{-}^{\dagger}(-k)\psi_{-}^{\dagger}(k)\gamma_{k}^{\dagger}-\psi^{\dagger}_{+}(-k)\psi_{-}^{\dagger}(k)\gamma_{k}^{\dagger}
 +\psi_{-}^{\dagger}(-k)\psi_{+}^{\dagger}(k)\gamma_{k}^{\dagger}+\psi^{\dagger}_{+}(-k)\psi_{+}^{\dagger}(k)\gamma_{k}^{\dagger}\Big]
\end{split}
\end{equation}
Then, we further subtract previous two expressions in order to obtain $\mathcal{H}_{sc}$,
\begin{equation}
\begin{split}
&\psi^{\dagger}_{\uparrow}(k)\psi^{\dagger}_{\downarrow}(-k)-\psi^{\dagger}_{\downarrow}(-k)\psi^{\dagger}_{\uparrow}(k)=\\
&\frac{1}{2}\Big[-\psi_{-}^{\dagger}(k)\psi_{-}^{\dagger}(-k)\gamma_{k}^{\dagger}+\psi_{+}^{\dagger}(k)\psi_{-}^{\dagger}(-k)\gamma_{k}^{\dagger}
-\psi_{-}^{\dagger}(k)\psi^{\dagger}_{+}(-k)\gamma_{k}^{\dagger}+\psi_{+}^{\dagger}(k)\psi^{\dagger}_{+}(-k)\gamma_{k}^{\dagger}
 \\
& 
+ \psi_{-}^{\dagger}(-k)\psi_{-}^{\dagger}(k)\gamma_{k}^{\dagger}+\psi^{\dagger}_{+}(-k)\psi_{-}^{\dagger}(k)\gamma_{k}^{\dagger}
 -\psi_{-}^{\dagger}(-k)\psi_{+}^{\dagger}(k)\gamma_{k}^{\dagger}-\psi^{\dagger}_{+}(-k)\psi_{+}^{\dagger}(k)\gamma_{k}^{\dagger}\Big]\,.
\end{split}
\end{equation}
In the previous equation we identify four combinations of the operators $\psi_{\pm}$. In the following we write them separately as $[\cdots]_{\sigma\sigma'}$,
where the subscript $\sigma\sigma'$ stands for the combination $\psi_{\sigma}\psi_{\sigma'}$.
The $[\cdots]_{--}$ term then reads,
\begin{equation}
\begin{split}
\Big[\psi^{\dagger}_{\uparrow}(k)\psi^{\dagger}_{\downarrow}(-k)-\psi^{\dagger}_{\downarrow}(-k)\psi^{\dagger}_{\uparrow}(k)\Big]_{--}
&=
\frac{1}{2}\Big[-\psi_{-}^{\dagger}(k)\psi_{-}^{\dagger}(-k)\gamma_{k}^{\dagger}
+ \psi_{-}^{\dagger}(-k)\psi_{-}^{\dagger}(k)\gamma_{k}^{\dagger}\Big]\,,\\
&=\frac{1}{2}\Big[-\psi_{-}^{\dagger}(k)\psi_{-}^{\dagger}(-k)\gamma_{k}^{\dagger}
+ \psi_{-}^{\dagger}(k)\psi_{-}^{\dagger}(-k)\gamma_{-k}^{\dagger}\Big]\,,\\
&=\frac{1}{2}
\psi_{-}^{\dagger}(k)\psi_{-}^{\dagger}(-k)
\Big[
-\gamma_{k}^{\dagger}
+ \gamma_{-k}^{\dagger}
\Big]\,,\\
\end{split}
\end{equation}
where in the second equality we have made the substitution $k\rightarrow-k$ in the second term. The $[\cdots]_{++}$ term reads
\begin{equation}
\begin{split}
\Big[\psi^{\dagger}_{\uparrow}(k)\psi^{\dagger}_{\downarrow}(-k)-\psi^{\dagger}_{\downarrow}(-k)\psi^{\dagger}_{\uparrow}(k)\Big]_{++}&=
\frac{1}{2}\Big[
\psi_{+}^{\dagger}(k)\psi^{\dagger}_{+}(-k)\gamma_{k}^{\dagger}
-\psi^{\dagger}_{+}(-k)\psi_{+}^{\dagger}(k)\gamma_{k}^{\dagger}
\Big]\,,\\
&=\frac{1}{2}\Big[
\psi_{+}^{\dagger}(k)\psi^{\dagger}_{+}(-k)\gamma_{k}^{\dagger}
-\psi^{\dagger}_{+}(k)\psi_{+}^{\dagger}(-k)\gamma_{-k}^{\dagger}
\Big]\,,\\
&=\frac{1}{2}\psi_{+}^{\dagger}(k)\psi^{\dagger}_{+}(-k)
\Big[
\gamma_{k}^{\dagger}
-\gamma_{-k}^{\dagger}
\Big]\,,
\end{split}
\end{equation}
where in the second equality we have made the substitution $k\rightarrow-k$ in the second term. 
The terms $[]_{+-,-+}$ can be grouped in one 
\begin{equation}
\begin{split}
&\Big[\psi^{\dagger}_{\uparrow}(k)\psi^{\dagger}_{\downarrow}(-k)-\psi^{\dagger}_{\downarrow}(-k)\psi^{\dagger}_{\uparrow}(k)\Big]_{+-,-+}
\\
&=\frac{1}{2}
\Big[
\psi_{+}^{\dagger}(k)\psi_{-}^{\dagger}(-k)\gamma_{k}^{\dagger}
+\psi^{\dagger}_{+}(-k)\psi_{-}^{\dagger}(k)\gamma_{k}^{\dagger}
-\psi_{-}^{\dagger}(k)\psi^{\dagger}_{+}(-k)\gamma_{k}^{\dagger}
-\psi_{-}^{\dagger}(-k)\psi_{+}^{\dagger}(k)\gamma_{k}^{\dagger}
\Big]\,,\\
&=\frac{1}{2}
\Big[
\psi_{+}^{\dagger}(k)\psi_{-}^{\dagger}(-k)\gamma_{k}^{\dagger}
-\psi_{-}^{\dagger}(-k)\psi_{+}^{\dagger}(k)\gamma_{k}^{\dagger}
+\psi^{\dagger}_{+}(-k)\psi_{-}^{\dagger}(k)\gamma_{k}^{\dagger}
-\psi_{-}^{\dagger}(k)\psi^{\dagger}_{+}(-k)\gamma_{k}^{\dagger}
\Big]\\
&=\frac{1}{2}
\Big[
2\psi_{+}^{\dagger}(k)\psi_{-}^{\dagger}(-k)\gamma_{k}^{\dagger}
+2\psi^{\dagger}_{+}(-k)\psi_{-}^{\dagger}(k)\gamma_{k}^{\dagger}
\Big]\,\\
&=
\Big[
\psi_{+}^{\dagger}(k)\psi_{-}^{\dagger}(-k)\gamma_{k}^{\dagger}
+\psi^{\dagger}_{+}(k)\psi_{-}^{\dagger}(-k)\gamma_{-k}^{\dagger}
\Big]\,,\\
&=
\psi^{\dagger}_{+}(k)\psi_{-}^{\dagger}(-k)\Big[
\gamma_{k}^{\dagger}
+\gamma_{-k}^{\dagger}
\Big]\,,\\
\end{split}
\end{equation}
where in the second equality we have reordered the operator pairs just to make visible the use of the fermionic 
commutation relation, $\{\psi^{\dagger}_{\alpha},\psi^{\dagger}_{\beta}\}=0$, and the third equality we have used such relation, and in  fourth equality we have made the substitution $k\rightarrow-k$ in the second term. 

We still need to evaluate some expressions with combinations of $\gamma_{k}$
\begin{equation}
\gamma_{k}=\frac{B+i\alpha k}{\sqrt{B^{2}+\alpha^{2}k^{2}}}\,,\quad
\gamma_{k}^{\dagger}=\frac{B-i\alpha k}{\sqrt{B^{2}+\alpha^{2}k^{2}}}\,,\quad
\gamma_{-k}^{\dagger}=\frac{B+i\alpha k}{\sqrt{B^{2}+\alpha^{2}k^{2}}}=\gamma_{k}\,,
\end{equation}
then
\begin{equation}
\begin{split}
f_{--}&=-\gamma_{k}^{\dagger}
+ \gamma_{-k}^{\dagger}
=
\frac{2i\alpha k}{\sqrt{B^{2}+\alpha^{2}k^{2}}}
\,,\\
f_{++}&=\gamma_{k}^{\dagger}
-\gamma_{-k}^{\dagger}=-\frac{2i\alpha k}{\sqrt{B^{2}+\alpha^{2}k^{2}}}\,,\\
f_{+-}&=\gamma_{k}^{\dagger}
+\gamma_{-k}^{\dagger}=\frac{2B}{\sqrt{B^{2}+\alpha^{2}k^{2}}}\,.
\end{split}
\end{equation}

Therefore, the superconducting Hamiltonian reads,
\begin{equation}
\begin{split}
\mathcal{H}_{sc}&=\frac{1}{2}\int dx 
\Bigg\{
\frac{1}{2}\bigg[\frac{2i\alpha k \Delta}{\sqrt{B^{2}+\alpha^{2}k^{2}}}
 \bigg]\psi_{-}^{\dagger}(k)\psi_{-}^{\dagger}(-k)\\
&\quad\quad\quad\quad\quad+
\frac{1}{2}\bigg[\frac{-2i\alpha k \Delta}{\sqrt{B^{2}+\alpha^{2}k^{2}}}
 \bigg]\psi_{+}^{\dagger}(k)\psi_{+}^{\dagger}(-k)\\
&\quad\quad\quad\quad\quad+
\bigg[\frac{2B\Delta}{\sqrt{B^{2}+\alpha^{2}k^{2}}}
 \bigg]\psi_{+}^{\dagger}(k)\psi_{-}^{\dagger}(-k)+
 h.c
\Bigg\}\,.
\end{split}
\end{equation}
Notice in previous equation, the coefficients of the operators $\psi^{\dagger}_{\sigma}\psi^{\dagger}_{\sigma'}$. Such terms represent
the pairing between bands $\sigma$ and $\sigma'$. We therefore write 
\begin{equation}
\begin{split}
\mathcal{H}_{sc}&=\int dx 
\Big[\frac{\Delta_{--}(k)}{2}\psi_{-}^{\dagger}(k)\psi_{-}^{\dagger}(-k)
+\frac{\Delta_{++}(k)}{2}\psi_{+}^{\dagger}(k)\psi_{+}^{\dagger}(-k)
+\Delta_{+-}(k)\psi_{+}^{\dagger}(k)\psi_{-}^{\dagger}(-k)+h.c
\Big]\,.
\end{split}
\end{equation}
where 
\begin{equation}
\label{pairingsapp}
\Delta_{--}(k)=\frac{i\alpha k \Delta}{\sqrt{B^{2}+\alpha^{2}k^{2}}}\,,\quad
\Delta_{++}(k)=\frac{-i\alpha k \Delta}{\sqrt{B^{2}+\alpha^{2}k^{2}}}\,,\quad
\Delta_{+-}(k)=\frac{B \Delta}{\sqrt{B^{2}+\alpha^{2}k^{2}}}\,.
\end{equation}
\subsubsection*{Full Hamiltonian}
\label{fullhSC0}
The Hamiltonian describing a Rashba nanowire subjected to a Zeeman interaction and superconducting correlations is given by
\begin{equation}
\mathcal{H}=\mathcal{H}_{0}+\mathcal{H}_{sc}\,,
\end{equation} 
where
\begin{equation}
\begin{split}
\mathcal{H}_{0}&=
\int \frac{dk}{2\pi}
\Big[
\varepsilon_{+}(k)\psi^{\dagger}_{+}(k)\psi_{+}(k)
+
\varepsilon_{-}(k)\psi^{\dagger}_{-}(k)\psi_{-}(k)
\Big]\,,\\
&=
\int \frac{dk}{2\pi}
\frac{1}{2}
\Big[
\varepsilon_{+}(k)\psi^{\dagger}_{+}(k)\psi_{+}(k)
-
\varepsilon_{+}(-k)\psi_{+}(-k)\psi^{\dagger}_{+}(-k)
+
\varepsilon_{-}(k)\psi^{\dagger}_{-}(k)\psi_{-}(k)
-
\varepsilon_{-}(-k)\psi_{-}(-k)\psi^{\dagger}_{-}(-k)\\
&\quad\quad\quad\quad+
\varepsilon_{+}(k)
+\varepsilon_{-}(k)
\Big]\,,\\
&=\int \frac{dk}{2\pi}
\frac{1}{2}\,\Psi_{k}^{\dagger}
\begin{pmatrix}
 \varepsilon_{+}(k)&0&0&0\\
 0&\varepsilon_{-}(k)&0&0\\
 0&0&-\varepsilon_{+}(-k)&0\\
 0&0&0&-\varepsilon_{-}(-k)
 \end{pmatrix}
 \Psi_{k}+
\text{constant}\,.
\end{split}
\end{equation}
where $\Psi_{k}=\begin{pmatrix}
\psi_{+}^{\dagger}(k)&\psi_{-}^{\dagger}(k)&\psi_{+}(-k)&\psi_{-}(-k)
\end{pmatrix}^{T}$ is the Nambu spinor. 

For the superconducting part we proceed in a similar way
\begin{equation}
\begin{split}
\mathcal{H}_{sc}&=\int \frac{dk}{2\pi} 
\Big[\frac{\Delta_{--}(k)}{2}\psi_{-}^{\dagger}(k)\psi_{-}^{\dagger}(-k)
+\frac{\Delta_{++}(k)}{2}\psi_{+}^{\dagger}(k)\psi_{+}^{\dagger}(-k)
+\Delta_{+-}(k)\psi_{+}^{\dagger}(k)\psi_{-}^{\dagger}(-k)\\
&\quad\quad\quad\quad+
\frac{\Delta_{--}^{\dagger}(k)}{2}\psi_{-}(-k)\psi_{-}(k)
+\frac{\Delta^{\dagger}_{++}(k)}{2}\psi_{+}(-k)\psi_{+}(k)
+\Delta_{+-}^{\dagger}(k)\psi_{-}(-k)\psi_{+}(k)
\Big]\,,\\
&=\int \frac{dk}{2\pi} 
\Big[\frac{\Delta_{--}(k)}{2}\psi_{-}^{\dagger}(k)\psi_{-}^{\dagger}(-k)
+\frac{\Delta_{++}(k)}{2}\psi_{+}^{\dagger}(k)\psi_{+}^{\dagger}(-k)\\
&\quad\quad\quad\quad
+\frac{\Delta_{+-}(k)}{2}\Big(\psi_{+}^{\dagger}(k)\psi_{-}^{\dagger}(-k)-\psi_{-}^{\dagger}(-k)\psi_{+}^{\dagger}(k)\Big)\\
&\quad\quad\quad\quad+
\frac{\Delta_{--}^{\dagger}(k)}{2}\psi_{-}(-k)\psi_{-}(k)
+\frac{\Delta^{\dagger}_{++}(k)}{2}\psi_{+}(-k)\psi_{+}(k)\\
&\quad\quad\quad\quad+
\frac{\Delta_{+-}^{\dagger}(k)}{2}\Big(\psi_{-}(-k)\psi_{+}(k)-\psi_{+}(k)\psi_{-}(-k)\Big)
\Big]\,,\\
\end{split}
\end{equation}
then we can write previous Hamiltonian as
\begin{equation}
\begin{split}
\mathcal{H}_{sc}
&=\int \frac{dk}{2\pi} 
\frac{1}{2}
\Big[\Delta_{--}(k)\psi_{-}^{\dagger}(k)\psi_{-}^{\dagger}(-k)
+\Delta_{++}(k)\psi_{+}^{\dagger}(k)\psi_{+}^{\dagger}(-k)\\
&\quad\quad\quad\quad
+\Delta_{+-}(k)\psi_{+}^{\dagger}(k)\psi_{-}^{\dagger}(-k)-\Delta_{+-}(k)\psi_{-}^{\dagger}(-k)\psi_{+}^{\dagger}(k)\\
&\quad\quad\quad\quad+
\Delta_{--}^{\dagger}(k)\psi_{-}(-k)\psi_{-}(k)
+\Delta^{\dagger}_{++}(k)\psi_{+}(-k)\psi_{+}(k)\\
&\quad\quad\quad\quad+
\Delta_{+-}^{\dagger}(k)\psi_{-}(-k)\psi_{+}(k)-\Delta_{+-}^{\dagger}(k)\psi_{+}(k)\psi_{-}(-k)
\Big]\,,\\
&=\int \frac{dk}{2\pi}
\frac{1}{2}
\,\Psi_{k}^{\dagger}
\begin{pmatrix}
 0&0&\Delta_{++}(k)&\Delta_{+-}(k)\\
 0&0&-\Delta_{+-}(k)&\Delta_{--}(k)\\
 \Delta_{++}^{\dagger}(k)&-\Delta_{+-}^{\dagger}(k)&0&0\\
 \Delta_{+-}^{\dagger}(k)&\Delta_{--}^{\dagger}(k)&0&0
 \end{pmatrix}\Psi_{k}\,.
 \end{split}
\end{equation}
where $\Psi_{k}=\begin{pmatrix}
\psi_{+}^{\dagger}(k)&\psi_{-}^{\dagger}(k)&\psi_{+}(-k)&\psi_{-}(-k)
\end{pmatrix}$ is the Nambu spinor.
Therefore, the full Hamiltonian reads,
\begin{equation}
\mathcal{H}=\int \frac{dk}{2\pi}
\frac{1}{2}\,\Psi_{k}^{\dagger}
\begin{pmatrix}
 \varepsilon_{+}(k)&0&\Delta_{++}(k)&\Delta_{+-}(k)\\
 0&\varepsilon_{-}(k)&-\Delta_{+-}(k)&\Delta_{--}(k)\\
 \Delta_{++}^{\dagger}(k)&-\Delta_{+-}^{\dagger}(k)&-\varepsilon_{+}(-k)&0\\
 \Delta_{+-}^{\dagger}(k)&\Delta_{--}^{\dagger}(k)&0&-\varepsilon_{-}(-k)
 \end{pmatrix}\Psi_{k}\,,
\end{equation}
where the pairing potentials are given in the previous subsection by Eqs.\,(\ref{pairingsapp}). The matrix under the integral is the so-called BdG Hamiltonian.

\chapter{\bf Appendix for Chapter \ref{Chap2a}}
\label{App0} 
\lhead{Appendix \ref{App0}. \emph{For Chapter 2}}
In this Appendix we provide further details on NS and SNS junctions.
\section{Tight-binding discretisation of the Rashba nanowire}
For computation purposes, we consider a discretisation of the 1D continuum model given by Eq.\,(\ref{Leq1}) for the Rashba nanowire into a tight-binding (TB) lattice with a small lattice spacing $a$.
The smaller is $a$, the better is the description, and by letting such lattice constant tend to zero one recovers the usual continuum limit. 
We choose a discrete lattice whose points are located at $x=a\,i$, where $i$ is an integer and $a$ the small lattice spacing. Notice that by do so, we are trying to find a matrix representation of our 1D continuum model in site space.
\subsection*{The kinetic term}
The first term in our Hamiltonian is the kinetic energy operator $K$,
\begin{equation}
K=\frac{p^{2}}{2m}\,-\,\mu(x)\,,
\end{equation}
where $p=\frac{\hbar}{i}\frac{\partial}{\partial x}$ is the momentum operator, and $\mu$ is the chemical potential. In principle, $\mu$ depends on the spatial coordinate $x$, however, in most of our calculations we consider it as independent.
Therefore, in the TB representation, one is allow to write
\begin{equation}
\big[K\big]_{x=ai}\,=\,\bigg[\frac{p^{2}}{2m}\bigg]_{x=ai}\,=\,\bigg[-\frac{\hbar^{2}}{2m}\frac{\partial^{2}}{\partial x^{2}} \bigg]_{x=ai}\,-\,\mu(x=ai)\,.
\end{equation}
Now, the derivatives are expressed by using the method of finite differences and in terms of creations and annihilation operators. Thus the first derivative can be approximated by
\begin{equation}
\label{1stTB}
\bigg[\frac{\partial}{\partial x}\bigg]_{x=(i+1/2)a}\,=\,\frac{\big( c_{i+1}^{\dagger}-c_{i}^{\dagger}\big)c_{i}}{a}
\end{equation}
Therefore, taking into account previous equation, the second derivative can be approximated by
\begin{equation}
\begin{split}
\bigg[\pdd{}{x}\bigg]_{x=i\,a}\,&=\,\frac{1}{a}\bigg\{ \bigg[ \pd{}{x}\bigg]_{x=(i+1/2)a}\,-\,\bigg[ \pd{}{x}
\bigg]_{x=(i-1/2)a}\bigg\}\,,\\
&=\,\frac{1}{a^{2}}\big\{ \big(c_{i+1}^{\dagger}\,-\,c_{i}^{\dagger} \big)\,c_{i}\,-\,\big(c_{i}^{\dagger}\,-\,c_{i-1}^{\dagger} \big)\,c_{i}\big\}\,,\\
&=\,\frac{1}{a^{2}}\,\big(c_{i+1}^{\dagger}\,-\,2\,c_{i}^{\dagger}\,+\,c_{i-1}^{\dagger}\big)\,c_{i}\,.
\end{split}
\end{equation}
On the other hand, the chemical potential $\mu_{x=ai}=\mu_{i}$ is on-site dependent, unlike the momentum operator that gives rise to hopping elements between sites. Thus, we can write
\begin{equation}
\mu_{x=ai}=\mu_{i}\,c^{\dagger}_{i}c_{i}\,.
\end{equation}
Hence, the Kinetic energy operator in the tight-binding approach can be expressed as
\begin{equation}
\label{0stTB}
\left[\frac{p^{2}}{2m}\right]_{x=a\,i}\,=\,-t\,c^{\dagger}_{i+1}\,c_{i}\,+\,(2\,t\,-\,\mu_{i})c_{i}^{\dagger}\,c_{i}\,-\,t\,c_{i-1}^{\dagger}\,c_{i}\,,
\end{equation}
where $t=\hbar^{2}/2ma^{2}$ is the hopping energy parameter between nearest-neighbor sites. 
Since we also consider spin, Eq.\,(\ref{0stTB}) has to be written as
\begin{equation}
\label{0stTB1}
\left[\frac{p^{2}}{2m}\right]_{x=a\,i}\,=\,\sigma_{0}\big[-t\,c^{\dagger}_{i+1}\,c_{i}\,+\,(2\,t\,-\,\mu_{i})c_{i}^{\dagger}\,c_{i}\,-\,t\,c_{i-1}^{\dagger}\,c_{i}\big]\,,
\end{equation}
where $\sigma_{0}$ is the $2\times2$ identity matrix.
\subsection*{The spin-orbit term}
The spin-orbit term from Eq.\,(\ref{H0Hamil}) is
\begin{equation}
H_{SO}=-\frac{\alpha_{R}}{\hbar}\sigma_{y}p_{x}\,,
\end{equation}
where $p_{x}=\frac{\hbar}{i}\frac{\partial}{\partial x}$ is the momentum operator.
Again, as before
\begin{equation}
[H_{SO}]_{x=ai}=\Big[-\frac{\alpha_{R}}{\hbar}\sigma_{y}p_{x}\Big]_{x=ai}=
\bigg[-\frac{\alpha_{R}}{\hbar}\sigma_{y}\frac{\hbar}{i}\frac{\partial}{\partial x}\bigg]_{x=ai}
=
i\alpha_{R}\sigma_{y}\bigg[\frac{\partial}{\partial x}\bigg]_{x=ai}\,.
\end{equation}
For the first derivative we use
\begin{equation}
\bigg[\frac{\partial}{\partial x}\bigg]_{x=ai}=\frac{\big( c_{i+1}^{\dagger}-c_{i-1}^{\dagger}\big)c_{i}}{2a}\,,
\end{equation}
which can be understood as a symmetrized case of Eq.\,(\ref{1stTB}), since
\begin{equation}
\frac{\big( c_{i+1}^{\dagger}-c_{i}^{\dagger}\big)c_{i}}{a}=\frac{\big( c_{i+1}^{\dagger}-c_{i-1}^{\dagger}\big)c_{i}}{2a}\,.
\end{equation}
Then, the SO term reads
\begin{equation}
\begin{split}
[H_{SO}]_{x=ai}
&=
i\,t_{SO}\sigma_{y}
\big( c_{i+1}^{\dagger}c_{i}-c_{i-1}^{\dagger}c_{i}\big)\,,
\end{split}
\end{equation}
where $t_{SO}=\alpha_{R}/2a$ is the SO hopping.
\subsection*{The Zeeman term}
The Zeeman term from Eq.\,(\ref{H0Hamil}) is
\begin{equation}
H_{Z}=B\sigma_{x}\,,
\end{equation}
which is independent of the site index. Thus, it leads to an on-site term in the tight-binding
Hamiltonian in a similar way as the chemical potential in the kinetic term.
\begin{equation}
[H_{Z}]_{x=ai}=[B\sigma_{x}]_{x=ai}c^{\dagger}_{i}c_{i}=B\sigma_{x}c^{\dagger}_{i}c_{i}\,,
\end{equation}
\subsection*{The full Hamiltonian }
Therefore $H_{0}$ reads,
\begin{equation}
H_{0}\,=\,\sum_{i}c_{i}^{\dagger}\,h\,c_{i}\,+\,\sum_{<ij>}c_{i}^{\dagger}\,v\,c_{j}\,+\,\text{h.c}\,.\,
\end{equation}
where the symbol $<ij>$ means that $v$ couples nearest-neighbor $i,j$ sites and 
\begin{equation}
\label{hopp}
\begin{split}
h_{ii}&\,\equiv\,h\,=\,
\begin{pmatrix}
2t\,-\,\mu&B\\
B& 2t\,-\,\mu
\end{pmatrix}\,,\\
h_{i+1,i}&\,\equiv\,v\,=\,
\begin{pmatrix}
-t&t_{SO}\\
-t_{SO}& -t
\end{pmatrix}\,=\,h_{i,i+1}^{\dagger},
\end{split}
\end{equation}
are matrices in spin space, $t=\hbar^{2}/2m^{*}a^{2}$ and $t_{SO}=\alpha_{R}/2a$. 
\chapter{\bf Appendix for Chapter \ref{Chap3}}
\label{AppB} 
\lhead{Appendix \ref{AppB}. \emph{For Chapter 3}}
\section{General form of the current}
The current from contact $\alpha$ flowing through lead $\alpha$ to the central region can be 
calculated from the evolution of the total number operator of fermions in such lead, and it is defined as \cite{Haug:07}
\begin{equation}
\label{Idefinition}
J_{\alpha}(t)=-e\big<\dot{N}_{\alpha}(t)\big>,
\end{equation}
where $N_{\alpha}=\sum_{k}c^{\dagger}_{k\alpha}c_{k\alpha}$ is the number operator for lead $\alpha$, and $e$ is the electron charge, $\big<\big>$ denotes the average over the ground state and $\dot{N}(t)$ represents the derivative of $N$
 over time.
The time evolution of the occupation number operator $N_{\alpha}(t)$ is described by the Heisenberg picture, 
where the equation of motion for operators reads,
\begin{equation}
i\hbar\frac{\partial }{\partial t}N_{\alpha}(t)=\big[N_{\alpha}(t),H(t)\big]\,,
\end{equation}
and $H(t)$ is the Hamiltonian of the entire system, which in principle is time-dependent. The commutator of two operators reads $[A,B]=AB-BA$, then
 $[A,B]=-[B,A]$. Thus, Eq.\,(\ref{Idefinition}) can be rewritten as
 \begin{equation}
\label{Idefinition2}
J_{\alpha}(t)=-\frac{ie}{\hbar}\big<[H(t),N_{\alpha}(t)]\big>\,.
\end{equation}
The Hamiltonian $H$ usually has three terms: the first that corresponds to the leads, the second to the central 
region and the third one that take into account electron transitions between the leads and the central region.

In our case the SNS system is described by the Hamiltonian
\begin{equation}
\label{SNSHamilMAR}
H=\frac{1}{2}\Bigg[\sum_{\alpha=1,2}
\begin{pmatrix}
c_{\alpha}&c_{\alpha}^{\dagger}
\end{pmatrix}^{\dagger}
\check{h}_{\alpha}
\begin{pmatrix}
c_{\alpha}\\
c_{\alpha}^{\dagger}
\end{pmatrix}
+
\begin{pmatrix}
d&d^{\dagger}
\end{pmatrix}^{\dagger}
\check{h}_{0}
\begin{pmatrix}
d\\
d^{\dagger}
\end{pmatrix}
+
\sum_{\alpha=1,2}
\begin{pmatrix}
c_{\alpha}&c_{\alpha}^{\dagger}
\end{pmatrix}^{\dagger}
\check{t}_{\alpha}
\begin{pmatrix}
d\\
d^{\dagger}
\end{pmatrix}
\Bigg]\,,
\end{equation}
where $\check{h}_{\alpha}$, $\alpha=L,R$, corresponds to Hamiltonians for leads left L and right R, respectively, 
while $\check{h}_{0}$ is the central system Hamiltonian, which is not superconducting, and the last term represents
the tunnelling Hamiltonian, which counts for electron transitions between the leads and the central region . The matrices $\check{h}_{\alpha}$
and $\check{t}_{\alpha}$ have Nambu structure
\begin{equation}
\check{h}_{\alpha}=
\begin{pmatrix}
h_{\alpha}&\Delta_{\alpha}\\
\Delta_{\alpha}^{\dagger}&-h_{\alpha}^{*}
\end{pmatrix}\,,\quad 
\check{h}_{0}=
\begin{pmatrix}
h_{0}&0\\
0&-h_{0}^{*}
\end{pmatrix}\,,\quad
\check{t}_{\alpha}=
\begin{pmatrix}
t_{\alpha}&0\\
0&-t_{\alpha}^{*}
\end{pmatrix}\,,
\end{equation}
with known sub-matrices $h_{\alpha},h_{0},t_{\alpha},\Delta_{\alpha}$, which are in spin space.
The particle number operator $N_{\alpha}$ commutes both with the corresponding 
Hamiltonian of the isolated central system and with the Hamiltonian of the isolated leads. The unique term  changing
particle numbers in each separate lead is the tunnelling term, last term in Eq.\,(\ref{SNSHamilMAR}), whose Hamiltonian is
\begin{equation}
H_{T}=\sum_{\alpha}H_{\alpha}\,,\quad H_{\alpha}=c^{\dagger}_{\alpha}t_{\alpha}d-c_{\alpha}t^{*}_{\alpha}d^{\dagger}\,.
\end{equation}
Notice that for a correct derivation, one needs to consider also the hermitian conjugate of $H_{\alpha}$.
Now we define the mixed lesser Green's functions
\begin{equation}
\label{greenslab}
G^{<}_{\alpha}(t,t')\equiv i\big< c^{\dagger}_{\alpha}(t')d(t)\big>\,.
\end{equation}
Then, the current can be written as
\begin{equation}
J_{\alpha}(t)=\frac{2e}{\hbar}{\rm Re}{\rm Tr}\Big[G^{<}_{\alpha}(t,t)\check{t}_{\alpha}\tau_{z}\Big]\,,
\end{equation}
where $\tau_{z}$ counts for electrons and holes.
The current is given by the time diagonal components of the Green's functions defined in Eq.\,(\ref{greenslab}), $G^{<}_{\alpha}(t,t')=G^{<}_{\alpha}(t,t)$. The problem is to find the mixed lesser Green's function. One can proceed with the Keldysh technique, where one employs the mixed-contour-ordered Green's function, $G_{\alpha}=-i\big<T_{c}d(\tau)c^{\dagger}_{\alpha}(\tau')\big>$, and then perform analytic continuation to the real time to obtain the mixed lesser function $G^{<}_{\alpha}(t,t')$ \cite{Haug:07}
\begin{equation}
G^{<}_{\alpha}(t,t')=\int d\,t_{1}\Big[G^{r}(t,t_{1})\check{t}_{\alpha}^{\dagger}(t')g^{<}_{\alpha}(t_{1},t')+G^{<}(t,t_{1})\check{t}_{\alpha}^{\dagger}(t')g^{a}_{\alpha}(t_{1},t')\Big]\,,
\end{equation}
where $G^{r}$ and $G^{<}$ are the retarded and lesser full system Green's function, respectively. $g^{a}$ and $g^{<}$ are the isolated leads Green's functions defined as
\begin{equation}
\begin{split}
g^{<}_{\alpha}(t,t')&=i\big< c^{\dagger}_{\alpha}(t')c_{\alpha}(t)\big>\,,\\
g^{a}_{\alpha}(t,t')&=-\theta(t-t')i\big< \{c_{\alpha}(t),c^{\dagger}_{\alpha}(t')\}\big>\,,
\end{split}
\end{equation}
where $\theta$ is the step function being $1$ for $t>t'$, and $0$ otherwise. Then, the current acquires the form
\begin{equation}
J_{\alpha}(t)=\frac{2e}{\hbar}\int d\,t'\,{\rm Re}{\rm Tr}\big\{[G^{r}(t,t')\check{t}_{\alpha}^{\dagger}(t')g^{<}_{\alpha}(t',t)\check{t}_{\alpha}(t)+G^{<}(t,t')\check{t}_{\alpha}^{\dagger}(t')g^{a}_{\alpha}(t',t)\check{t}_{\alpha}(t)]\tau_{z}\big\}\,.
\end{equation}
Now, we can introduce the lead's self energies defined as
\begin{equation}
\Sigma_{\alpha}^{a(r)}=\check{t}^{\dagger}_{\alpha}g_{\alpha}^{a(r)}\check{t}_{\alpha}\,,\quad \Sigma_{\alpha}^{<}=\check{t}^{\dagger}_{\alpha}g_{\alpha}^{<}\check{t}_{\alpha}\,,
\end{equation}
therefore the current reads
\begin{equation}
J_{\alpha}(t)=\frac{2e}{\hbar}\int d\,t'\,{\rm Re}{\rm Tr}\big\{[G^{r}(t,t')\Sigma^{<}_{\alpha}(t',t)+G^{<}(t,t')\Sigma^{a}_{\alpha}(t',t)]\tau_{z}\big\}\,.
\end{equation}
\section{Floquet-Keldysh formalism}\label{AppendixB1}
Consider a mesoscopic system composed of two semi-infinite leads (labeled $L$ and $R$), each in thermal equilibrium at the same temperature $T$ and with the same chemical potential $\mu=0$. Each lead has a finite s-wave superconducting pairing $\Delta_\alpha$, where $\alpha=L,R$. A central system ($\alpha=S$), which may or may not be superconducting, is coupled to both leads through operator $v$. In its Nambu form, the Hamiltonian of the system reads
\[
\hat H=\frac{1}{2}\sum_{ij}\left(\begin{array}{c|c}
c_j&c^+_j
\end{array}\right)
H_{ij}\left(\begin{array}{c}
c_j\\\hline c^+_j
\end{array}\right),
\]
where the Nambu Hamiltonian matrix takes the general form
\[
H=\left(\begin{array}{ccc|ccc}h_L & v^+ & 0 & \Delta_L & 0 & 0 \\ v& h_S & v^+ & 0 & \Delta_S & 0 \\ 0 & v & h_R & 0 & 0 & \Delta _R\\\hline \Delta_L^+ & 0 & 0 & -h_L^* & (-v^+)^* & 0 \\ 0 & \Delta_S^+ & 0 & -v^* & -h_S^* & (-v^+)^* \\ 0 & 0 & \Delta_R^+ & 0 & -v^* & -h_R^*\end{array}\right).
\]
Here $h_\alpha$ is the normal Hamiltonian for each section of the system. The blocks delimited by lines denote the Nambu particle, hole and pairing sectors.

If we apply a left-right voltage bias $V$ through the junction, the Bardeen-Cooper-Schrieffer (BCS) pairing of the leads will become time dependent, $\Delta_{L/R} \to e^{\pm i Vt}\Delta_{L/R}$, while $h_{L/R}\to h_{L/R}\pm V/2$ (we take $e=\hbar=1$). Both these changes can be gauged away from the leads and into the system by properly redefining $c^+_i\to c^+_i(t)= e^{\pm iVt/2} c^+_i$. This transformation is done also inside the system $S$, thereby effectively dividing it into two, the portion with an $e^{iVt/2}$ phase (denoted $S_L$), and the portion with the opposite phase (denoted $S_R$). This restores $H$ to its unbiased form, save for a new time dependence in $h_S\to h_S(Vt)$, which is constrained to the coupling between the $S_L$ and $S_R$,
\[
h_S(Vt)=\left(\begin{array}{cc}h_{S_L} & e^{-iVt}v^+_0\\ e^{iVt}v_0 & h_{S_R} \end{array}\right).
\]  
It is important to note that $H(t)$ is periodic, with angular frequency $\omega_0=V$. In the steady state limit (at long times $t$ after switching on the potential $V$) all response functions and observables will exhibit the same time periodicity (all transient effects are assumed to be completely damped away). In particular, the steady state current $I(t)=I(t+2\pi/\omega_0)$, so that
\[
I(t)=\sum_n e^{in\omega_0 t}I_n\,,
\]
for some harmonic amplitudes $I_n$, in general complex, that satisfy $I_n=I_{-n}^*$ since $I(t)$ is real.
This current can be computed using the Keldysh Green's function formalism \cite{Haug:07}.
 The standard expression for $I(t)$ is computed starting from the definition of $I(t)=\partial_t N_L$, where $N_L$ is the total number of fermions in the left lead. By using Heisenberg equation and the Keldysh-Dyson equation, one arrives at, see previous section for details on the derivation, 
\[
I(t)=\mathrm{Re}[J(t)],
\]
where
\[
J(t)=\frac{2e}{\hbar} \int dt'~\mathrm{Tr}\left\{\left[
G^r(t,t')\Sigma_L^<(t',t) + G^<(t,t')\Sigma_L^a(t',t)\right]\tau_z\right\}.
\]
The z-Pauli matrix $\tau_z$ above acts on the Nambu particle-hole sector,
\[
\tau_z=\left(\begin{array}{c|c}\mathbbm{1} &0\\\hline 0 & -\mathbbm{1} \end{array}\right).
\] 
The self energy from the left lead is defined as $\Sigma_L^{a,<}(t',t)=v\,g_L^{a,<}(t',t) v^+$, where $g_L(t',t)=g_L(t'-t)$ stands for the left lead's propagator, when decoupled from the system (this propagator depends only on the time difference since the decoupled lead is time independent in this gauge).
We define the Fourier transform of $g$ as
\[
g(\omega)=\int_{-\infty}^\infty dt e^{i\omega t} g(t).
\]
The retarded propagator in Fourier space is 
\[
g_L^r(\omega)=\frac{1}{\omega-h_L+i\eta}\,,
\]
while the advanced $g^a_L(\omega)=\left[g^r_L(\omega)\right]^+$. One can compute $g_L^<(\omega)=i f(\omega) A_L(\omega)$, where  $f(\omega)=1/(e^{\omega/k_BT}+1)$ is the Fermi distribution in the leads, and $A_L(\omega)=i(g_L^r(\omega)-g_L^a(\omega))$ is the Nambu spectral function. The $g^r_{L/R}$ [and in particular $A_{L/R}(\omega)$] is assumed known, or at least easily obtainable from $h_{L/R}$ and $v$. 
Finally, the Green functions $G^r(t',t)$ and $G^<(t',t)$ correspond to the propagator for the full system, including the coupling to the leads. (Note that, in practice, since $G$ is inside a trace in $J(t)$, only matrix elements of $G$ \emph{inside} the $S$ portion of the full system are needed).
The retarded $G^r$ satisfies the equation of motion
\[
\left[i\partial_{t'}-H(t')\right]G(t',t)=\delta(t'-t),
\]
while $G^<$ (when projected onto the finite-dimensional system $S$) satisfies the Keldysh relation
\begin{eqnarray*}
G^<(t',t)&=&\int dt_1dt_2 G^r(t',t_1)\\
&&\times\left[\Sigma_L^<(t_1-t_2)+\Sigma_R^<(t_1-t_2)\right]G^a(t_2,t).
\end{eqnarray*}
Since $h_S$ in $H$ is time dependent, $G$ propagators depend on two times; unlike $\Sigma_{L/R}$ or $g_{L/R}$ they are not Fourier diagonal. Instead, we can exploit the steady-state condition, which reads
\[
G(t',t)=G(t'+\frac{2\pi}{\omega_0},t+\frac{2\pi}{\omega_0}),
\]
to expand the system's $G$ as a Fourier transform in $t'-t$ and a Fourier \emph{series} in $t$. We define
\[
G(t',t)=\sum_n e^{-in\omega_0 t}\int_{-\infty}^\infty \frac{d\epsilon}{2\pi}e^{-i \epsilon(t'-t)}G_{n}(\epsilon).
\]
The natural question is how the equation of motion is expressed in terms of the harmonics $G_n(\epsilon)$. It takes the most convenient form if we redefine $G_n(\epsilon)$ (where $\epsilon$ is unbounded) in terms of the quasienergy $\tilde{\epsilon}\in [0,\hbar\omega_0]$, i.e. $\epsilon=\tilde\epsilon+m\omega_0$
\[
G_{mn}(\tilde\epsilon)=G_{m-n}(\tilde\epsilon+m\omega_0).
\]
This has the advantage that the equation of motion translates to a matrix equation analogous to that of a static system in Fourier space
\[
\sum_m (\tilde \epsilon+n'\omega_0-H_{n'm})G^r_{mn}(\tilde\epsilon)=\delta_{n'n},
\]
where 
\[
H_{n'n}=\int dt e^{i(n'-n)t} H(t).
\] 
This is known as the Floquet description of the steady state dynamics in terms of sidebands, which appear formally as a new quantum number $n$. Time dependent portions of $H(t)$ act as a coupling between different sidebands. The effective Hamiltonian for the $n$-th sideband is the static portion of $H(t)$, shifted by $-n\omega_0$. One therefore sometimes defines the Floquet ``Hamiltonian'' of the system as 
\[
{\bm{h_S}}_{nm}={h_S}_{nm}-n\omega_0\delta_{nm},
\]
where, as before, ${h_S}_{n'n}=\int dt e^{i(n'-n)t} {h_S}(t)$. Likewise, one may define the Floquet self-energies as 
\[
{\bm{\Sigma_L}}_{nm}(\tilde\epsilon)=\delta_{nm}\Sigma_{L/R}(\tilde\epsilon+n\omega_0),
\]
(since the leads are static, $\bm{\Sigma}$ is sideband-diagonal).

The Floquet equation of motion for $G^r_{nm}(\tilde\epsilon)$ can be solved like in the case of a static system. Within the $S$ portion of the system, we have
\[
\bm{G}^r(\tilde\epsilon)=\left[\tilde\epsilon-\bm{h_S}-\bm{\Sigma_L}^r(\tilde \epsilon)-\bm{\Sigma_R}^r(\tilde \epsilon)\right]^{-1}.
\]
Boldface denotes the sideband structure implicit in all the above matrices.
Similarly, the Keldysh relation takes the simple form
\[
\bm{G^<}(\tilde \epsilon)=\bm{G^r}(\tilde \epsilon)\left[\bm{\Sigma_L}^<(\tilde \epsilon)+\bm{\Sigma_R}^<(\tilde \epsilon)\right]\bm{G^a}(\tilde \epsilon).
\]
Finally, the time averaged current $I_{dc}\equiv I_0$ takes the form
\begin{equation}
\label{KeldyshIdc}
I_{dc}=\frac{2e}{h} \int_0^{\hbar\omega_0} d\tilde\epsilon~\mathrm{Re}\mathrm{Tr}\left\{\left[\bm{G^r}(\tilde\epsilon)\bm{\Sigma_L^<}(\tilde\epsilon)+\bm{G^<}(\tilde\epsilon)\bm{\Sigma_L^a}(\tilde\epsilon)\right]\tau_z\right\},
\end{equation}
where the trace includes the sideband index. 
In a practical computation, the number of sidebands that must be employed is finite, and depends on the applied voltage bias $V$ (the typical number scales as $n_\mathrm{max}\sim v_0/V$). We employ an adaptive scheme that increases the number of sidebands recursively until convergence for each value of $V$.

\section{Derivation of the current in the tunneling limit}
\label{AppendixB2}
In this part we aim at calculating the current in the tunneling limit. 
The Hamiltonian for our system is given by Eq.\,(\ref{SNSHamilMAR}). 

According to Floquet's theory, a periodic time-dependent system is described in terms of sidebands, whose Hamiltonian is shifted by $n\omega_{0}$.
In this part we consider a subspace of one sideband, $n=-1,0,+1$, where the central system is composed of two sites, left and right. The role of the leads are considered through their self energies.
The full system Hamiltonian is written as
\begin{equation}
H(t)=H+V\,{\rm e}^{i\omega_{0}t}+V^{\dagger}\,{\rm e}^{-i\omega_{0}t}
\end{equation}
where $H$ is given by Eq.\,(\ref{SNSHamilMAR}) and
 \begin{equation}
  \label{couplingsidebands}
  \begin{split}
   V\,=\,
  \begin{pmatrix}
 0& 0&v_{0}^{ \dagger}& 0\\
0& 0& 0&0\\
 0& 0&0& 0\\
 0&-\,v_{0}^{*}& 0& 0 \\
  \end{pmatrix}\,,&\,\quad
V^{\dagger}\,=\, 
  \begin{pmatrix}
    0&0& 0& 0 \\
   0& 0& 0 &-\,v_{0}^{T} \\
v_{0}& 0& 0& 0\\
      0& 0& 0& 0\\
  \end{pmatrix}\,.
  \end{split}
  \end{equation}
  where $v_{0}$ represents that hopping between the two central sites $l$ and $r$, and it is given by
\begin{equation}
\label{transplink}
v_{0}\,=\, \eta\,v\,=\, \eta\left(
  \begin{array}{ c c }
 -\,t & t_{SO}  \\
  -t_{SO}& -\,t\\
  \end{array} \right)\,,
  \end{equation}  
  where $\eta$ is an small parameter that describes the nature of the junction that links the left and right superconductors, $t=\hbar^{2}/(2ma^{2})$ and $t_{SO}=\hbar\alpha_{R}/2a$.
As described in Floquet's theory, the Hamiltonians for the isolated central system sidebands are
\begin{equation}
 H_{11}=H_{S}-\omega_{0}\,,\quad
  H_{00}=H_{S}\,,\quad
 H_{11}=H_{S}+\omega_{0}\,,
 \end{equation}
 where 
 \begin{equation}
 H_{S}
=
\begin{pmatrix}
h_{l}&0\\
0&h_{r}
\end{pmatrix} \,,\quad
h_{l(r)}=
\begin{pmatrix}
h_{0}&\Delta\\
\Delta^{\dagger}&-h_{0}^{*}
\end{pmatrix}\,,
 \end{equation}
 and $h_{0}$ and $\Delta$ are matrices in spin space. 
Notice that the Hamiltonian for the left lead, considering a tight-binding model and Nambu in each site, is infinite.
 We assume that the left and right leads are attached at the left and right edges of the central system, so that their self-energy terms ($\Sigma_{L/R}$) only influence sites of the first and last column of the central system. That means that since we deal with a central system with sites $l=0$ and $r=1$, the self-energy terms $\Sigma_{L/R}$ will influence sites $l=0$ and $r=1$, respectively. Since a nearest-neighbor tight-binding model is used, the hopping matrices between leads and central system have nonzero elements only between sites on the surface of the lead and their neighboring sites in the central system. This means that only the surface Green's function $(g_{L/R})_{00}$ (big letters L and R label hte left and right leads, respectively) is needed when one calculates the lead's self-energies.
In the Floquet decomposition, one describes the central system in terms of sidebands. These sidebands are coupled by the matrices ${\bf V}$ and ${\bf V}^{\dagger}$, and they are identical copies of the original system with the difference that their energies are shifted (due to the voltage bias): for a $n$ sideband subspace we have $h+n\omega_{0}$, $h$ and $h-n\omega_{0}$, where the original system is $h$. $\omega_{0}=2eV/\hbar^{2}$. 
As we mentioned, in this part, we consider one-sideband subspace, with three sites, $n=-1$, $n=0$ and $n=1$. 
 
 The general expression for the current is given by Eq.\,(\ref{KeldyshIdc}), for $\alpha=L$,
 \begin{equation}
\label{currentjosephzerothqx}
I_{dc}\,=\,\frac{2\,e}{\hbar}\int_{0}^{\hbar\omega_{0}}\frac{d\,\omega}{2\pi}{\rm Tr}\left\{ \left[{\bf G}^{r}(\omega)\, {\bf \Sigma}^{<}_{L}(\omega)\,+\,{\bf G}^{<}(\omega)\,{\bf \Sigma}_{L}^{a}(\omega)\right]\tau_{z}\right\}\,,
 \end{equation}
 where all the elements in previous equation are matrices in Floquet, Nambu, sideband and central system subspaces. $\mathbf{G}$ is the full system Green's function, while $\mathbf{\Sigma}$ the self energy of both leads. 
 In this part we are interested in the tunnelling regime, meaning small transparency across the junction. As we have discussed, our system consists of two semiinfinite leads which are coupled through a central region with two sites $l=0$ and $r=1$. The hopping between these two sites is reduced and parametrised by $v_{0}=\eta v$, where $\eta\in[0,1]$ see Eq.\,(\ref{transplink}). For $\eta\ll1$ one models a tunnel junction. Then, we consider second order terms in $\eta$ in Eq.\,(\ref{currentjosephzerothqx}). Notice that the self-energies are proportional to $\Sigma\approx V^{\dagger} g V$, while the coupling matrix $V$ is given by Eq.\,(\ref{couplingsidebands}) and it is proportional to $\eta$. Hence, the self energies are of second order in $V$ and therefore in $\eta$. 
Thus, a zeroth order expansion of the Dyson's equation is enough $G\approx g$.
 
 Therefore, in one sideband subspace ($n=0,\pm1$) reads,
\begin{equation}
\label{Gr}
\begin{split}
{\bf G}^{r}\,&=\,
\begin{pmatrix}
G^{r}_{-1-1}&G^{r}_{-10}&G^{r}_{-11}\\
G^{r}_{0-1}&G^{r}_{00}&G^{r}_{01}\\
G^{r}_{1-1}&G^{r}_{10}&G^{r}_{11}\\
\end{pmatrix}\,,
\quad\quad
 {\bf G}^{a}(\omega)
 \,=\,
 \left({\bf G}^{r}(\omega)\right)^{\dagger}\,,\\
 {\bf G^{<}}\,&=\,
\begin{pmatrix}
G^{<}_{-1-1}&G^{<}_{-10}&G^{<}_{-11}\\
G^{<}_{0-1}&G^{<}_{00}&G^{<}_{01}\\
G^{<}_{1-1}&G^{<}_{10}&G^{<}_{11}\\
\end{pmatrix}\,.
 \end{split}
\end{equation}   
In the space of sidebands, the self-energy matrices are diagonal. Thus, the lesser and advanced self-energies ${\bf \Sigma}^{<(a)}$ corresponding to lead $\alpha$ are,
\begin{equation}
{\bf \Sigma}^{<}_{L}\,=\,
\begin{pmatrix}
\Sigma^{<}_{L,-1-1}&{\bf 0}&{\bf 0}\\
{\bf 0}&\Sigma^{<}_{L,00}&{\bf 0}\\\
{\bf 0}&{\bf 0}&\Sigma^{<}_{L,11}\\
\end{pmatrix}\,,\quad
{\bf \Sigma}^{a}_{L}\,=\,
\begin{pmatrix}
\Sigma^{a}_{L,-1-1}&{\bf 0}&{\bf 0}\\
{\bf 0}&\Sigma^{a}_{L,00}&{\bf 0}\\\
{\bf 0}&{\bf 0}&\Sigma^{a}_{L,11}\\
\end{pmatrix}\,,
\end{equation}
where each matrix element of previous two matrices is a $2\times2$ matrix in the central system subspace,
\begin{equation}
\Sigma^{<}_{L,-1-1}\,=\,
\begin{pmatrix}
\Sigma^{\prime,<}_{L,-1-1}&{\bf 0}\\
{\bf 0}&{\bf 0}
\end{pmatrix}\,,\quad
\Sigma^{<}_{L,00}\,=\,
\begin{pmatrix}
\Sigma^{\prime,<}_{L,00}&{\bf 0}\\
{\bf 0}&{\bf 0}
\end{pmatrix}\,,\quad
\Sigma^{<}_{L,-1-1}\,=\,
\begin{pmatrix}
\Sigma^{\prime,<}_{L,11}&{\bf 0}\\
{\bf 0}&{\bf 0}
\end{pmatrix}\,.
\end{equation}
The roll of the self-energies labelled with prime symbol will be understood in the following.

  The Nambu Pauli matrix $\tau_{z}$ in Eq.\,(\ref{currentjosephzerothqx}) is in the space of sidebands ($n=-1,0,1$), in the space of left and right sites of the central system, in spin space and electron hole, thus one can write
 \begin{equation}
 \label{tauaua}
 \tau_{z}=
 \begin{pmatrix}
 \tau_{z,-1-1}& 0& 0\\
 0& \tau_{z,00}& 0\\
 0& 0& \tau_{z,11}\\
 \end{pmatrix}=
  \begin{pmatrix}
 \tau_{-1-1}& 0& 0\\
 0& \tau_{00}& 0\\
 0& 0& \tau_{11}\\
 \end{pmatrix}
 \end{equation}
 where each element is in the space of left and right sites of the central system, in spin space and electron hole 
 $
 \sigma_{0,{\rm sites}}\otimes\sigma_{0,{\rm spin}}\otimes\tau_{z,{\rm Nambu}}$.
  
 


Multiplying the matrices in Eq.\,(\ref{currentjosephzerothqx}) in the sideband subspace, we get
 \begin{equation}
 \label{squadre}
\begin{pmatrix}
G^{r}_{-1-1}\Sigma^{<}_{L-1-1}+G^{<}_{-1-1}\Sigma^{a}_{L-1-1}&G^{r}_{-10} \Sigma^{<}_{L00}+G^{<}_{-10}\Sigma^{a}_{L00}&G^{r}_{-11}\Sigma^{<}_{11}+G^{<}_{-11}\Sigma^{a}_{11}\\
G^{r}_{0-1}\Sigma^{<}_{L-1-1}+G^{<}_{0-1}\Sigma^{a}_{L-1-1}&G^{r}_{00} \Sigma^{<}_{L00}+G^{<}_{00}\Sigma^{a}_{L00}&G^{r}_{01}\Sigma^{<}_{11}+G^{<}_{01}\Sigma^{a}_{11}\\
G^{r}_{1-1}\Sigma^{<}_{L-1-1}+G^{<}_{1-1}\Sigma^{a}_{L-1-1}&G^{r}_{10} \Sigma^{<}_{L00}+G^{<}_{10}\Sigma^{a}_{L00}&G^{r}_{11}\Sigma^{<}_{L11}+G^{<}_{11}\Sigma^{a}_{L11}\,,
\end{pmatrix}
 \end{equation} 
where all the elements are matrices in spin  and the central system (left and right) subspace.
Tracing previous matrix over sidebands, we get 
\begin{equation}
\begin{split}
&[G^{r}_{-1-1}\,\Sigma^{<}_{L,-1-1}\,+\,G^{<}_{-1-1}\,\Sigma^{a}_{L,-1-1}]\,\tau_{-1-1}\,\\
&+\,[G^{r}_{00}\,\Sigma^{<}_{L,00}\,+\,G^{<}_{00}\,\Sigma^{a}_{L,-1-1}]\,\tau_{00}
+\,[G^{r}_{11}\,\Sigma^{<}_{L,11}\,+\,G^{<}_{11}\,\Sigma^{a}_{L,11}]\,\tau_{11}\,.
\end{split}
\end{equation}
As discussed previously, we consider up to second order in $\eta$, and therefore a zeroth order expansion of the Dyson's equation is enough ($G\approx g$). 
Therefore, previous equation reads
\begin{equation}
\begin{split}
&[g^{r}_{-1-1}\,\Sigma^{<}_{L,-1-1}\,+\,g^{<}_{-1-1}\,\Sigma^{a}_{L,-1-1}]\,\tau_{-1-1}\,\\
&+\,[g^{r}_{00}\,\Sigma^{<}_{L,00}\,+\,g^{<}_{00}\,\Sigma^{a}_{L,-1-1}]\,\tau_{00}+[g^{r}_{11}\,\Sigma^{<}_{L,11}\,+\,g^{<}_{11}\,\Sigma^{a}_{L,11}]\,\tau_{11}\,,
\end{split}
\end{equation}
where $g_{ii}$ is the Green's function for the disconnected sideband $i$, whose matrix structure contains the influence of the left and right leads propagators on the central system. 
Thus, in the space of left and right leads, one has
 \begin{equation}
 \label{disconnectedg}
\begin{split}
g_{00}^{r}(\omega)\,&=\,
\begin{pmatrix}
g_{L}^{r}(\omega)&0\\
 0&g_{R}^{r}(\omega)\\
\end{pmatrix}\,,\\
g_{-1-1}^{r}(\omega)\,&=\,
\begin{pmatrix}
g_{L}^{r}(\omega\,-\,\omega_{0})& 0\\
 0&g_{R}^{r}(\omega\,-\,\omega_{0})\\
\end{pmatrix}\,,\\
g_{11}^{r}(\omega)\,&=\,
\begin{pmatrix}
g_{L}^{r}(\omega\,+\,\omega_{0})& 0\\
 0&g_{R}^{r}(\omega\,+\,\omega_{0})\\
\end{pmatrix}\,,
\end{split}
\end{equation}
where we approximate $g_{SL}$ to be the leads Green's function, $g_{SL}^{r(a)}(\omega)\approx g_{L}^{r(a)}(\omega)=(\omega-H_{L}\pm i\eta)^{-1}$, and then use
\begin{equation}
\label{djadjhsjhd}
\begin{split}
g^{<}_{L/R}(\omega)\,&=\,{\rm i}\,f_{L/R}(\omega)\,A_{L/R}(\omega)\,,\\
A_{L/R}\,&=\,{\rm i}\,\left[g^{r}_{L/R}\,-\,g^{a}_{L/R} \right]\,,\\
g^{r}_{L/R}(\omega)\,&=\,\frac{\Lambda_{L/R}(\omega)}{2}\,-\,{\rm i}\frac{A_{L/R}(\omega)}{2}\,,\\
g^{a}_{L/R}(\omega)\,&=\,\frac{\Lambda_{L/R}(\omega)}{2}\,+\,{\rm i}\frac{A_{L/R}(\omega)}{2}\,,\\
\end{split}
\end{equation}
with $A$ the spectral function, being Hermitian as well as $\Lambda$. We assume that $g$ is known and in our case calculated numerically.
Then, putting together both the self-energies and disconnected Green's functions, one gets the necessary elements for calculating the current given by Eq.\,(\ref{currentjosephzerothqx})
\begin{equation}
\label{edndjjdjj}
\begin{split}
g^{r}_{-1-1}\Sigma^{<}_{L,-1-1}+g^{<}_{-1-1}\Sigma^{a}_{L,-1-1}&=
\left(
\begin{array}{cc}
g^{r}_{SL}(\omega-\omega_{0})\Sigma^{\prime,<}_{L,-1-1}(\omega)+g^{<}_{SL}(\omega-\omega_{0})\Sigma^{\prime,a}_{L,-1-1}(\omega)& 0\\
 0&0\\
\end{array}
\right) \\
g^{r}_{00}\,\Sigma^{<}_{L,00}\,+\,g^{<}_{00}\,\Sigma^{a}_{L,-1-1}&=
\left(
\begin{array}{cc}
g^{r}_{SL}(\omega)\Sigma^{\prime,<}_{L,00}(\omega)+g^{<}_{SL}(\omega)\Sigma^{\prime,a}_{L,00}(\omega)&0\\
0&0\\
\end{array}
\right) \\
g^{r}_{11}\Sigma^{<}_{L,11}+g^{<}_{11}\Sigma^{a}_{L,11}&=
\left(
\begin{array}{cc}
g^{r}_{SL}(\omega+\omega_{0})\Sigma^{\prime,<}_{L,11}(\omega)+g^{<}_{SL}(\omega+\omega_{0})\Sigma^{\prime,a}_{L,11}(\omega)& 0\\
 0& 0\\
\end{array}
\right)\,. 
\end{split}
\end{equation}
Now, we are left with the self-energies.
Projecting the self-energies on the sideband subspace one gets,
\begin{equation}
\label{selfE}
\begin{split}
\Sigma^{<(a)}_{-1-1}\,&=\,
 V^{\dagger}\,g^{<(a)}_{00}(\omega)\, V\,,
\\
\Sigma^{<(a)}_{00}\,&=\,
V^{\dagger}\,g^{<(a)}_{11}(\omega)\, V\,+\, V\,g^{<(a)}_{-1-1}(\omega)\, V^{\dagger}\,,
\\
\Sigma^{<(a)}_{11}\,&=\,
V\,g^{<(a)}_{00}(\omega)\, V^{\dagger}\,,
\end{split}
\end{equation}
where the Green's functions are given in Eq.\,(\ref{edndjjdjj}), $V$ is given by Eq.\,(\ref{couplingsidebands}), and we rewrite it as
\begin{equation}
\label{Vv}
V\,=\,\eta
\begin{pmatrix}
 0&\lambda_{2}\\
\lambda_{1}& 0\\
\end{pmatrix}\,,\quad
V^{\dagger}\,=\, \eta
\begin{pmatrix}
 0&\lambda_{1}^{\dagger}\\
\lambda_{2}^{\dagger}& 0\\
\end{pmatrix}\,,\quad
\lambda_{1}=
\begin{pmatrix}
0&0\\
0&-v^{*}
\end{pmatrix}\,,\quad
\lambda_{2}=
\begin{pmatrix}
v^{\dagger}&0\\
0&0
\end{pmatrix}
\end{equation}
where $\eta\in[0,1]$ describes the nature of the tunnel junction, and

Then, introducing Eq.\,(\ref{Vv}) and (\ref{disconnectedg}) into Eq.\,(\ref{selfE}), we get the self-energies in their left/right structure 
\begin{equation}
\begin{split}
\Sigma^{<(a)}_{-1-1}\,&=\,\eta^{2}
\left(
\begin{array}{cc}
\lambda^{\dagger}_{1}\,g_{SR}^{<(a)}(\omega)\,\lambda_{1}& 0\\
0&\lambda^{\dagger}_{2}\,g_{SL}^{<(a)}(\omega)\,\lambda_{2}\\
\end{array}
\right)\,,
\\
\Sigma^{<(a)}_{00}\,&=\,\eta^{2}
\left(
\begin{array}{cc}
\lambda^{\dagger}_{1}\,g_{SR}^{<(a)}(\omega+\omega_{0})\,\lambda_{1}&0\\
+\lambda_{2}\,g_{SR}^{<(a)}(\omega-\omega_{0})\,\lambda_{2}^{\dagger}&\\
0&\lambda^{\dagger}_{2}\,g_{SL}^{<(a)}(\omega+\omega_{0})\,\lambda_{2}\\
&+\lambda^{\dagger}_{1}\,g_{SL}^{<(a)}(\omega-\omega_{0})\,\lambda_{1}
\end{array}
\right)\,,
\\
\Sigma^{<(a)}_{11}\,&=\,\eta^{2}
\left(
\begin{array}{cc}
\lambda_{2}\,g_{SR}^{<(a)}(\omega)\,\lambda_{2}^{\dagger}&0\\
 0&\lambda_{1}\,g_{SL}^{<(a)}(\omega)\,\lambda_{1}^{\dagger}\\
\end{array}
\right)\,,
\end{split}
\end{equation}
Matrix elements in previous equations give the prime left self-energies defined in Eq.\,(\ref{edndjjdjj}), then
\begin{equation}
\begin{split}
\Sigma^{\prime,<(a)}_{L,-1-1}\,&=\,\eta^{2}\,\lambda^{\dagger}_{1}\,g_{SR}^{<(a)}(\omega)\,\lambda_{1}\,,\\
\Sigma^{\prime,<(a)}_{L,00}\,&=\,\eta^{2}\left[\lambda^{\dagger}_{1}\,g_{SR}^{<(a)}(\omega+\omega_{0})\,\lambda_{1}
+\lambda_{2}\,g_{SR}^{<(a)}(\omega-\omega_{0})\,\lambda_{2}^{\dagger}\right]\,,\\
\Sigma^{\prime,<(a)}_{L,11}\,&=\,\eta^{2}\,\lambda_{2}\,g_{SR}^{<(a)}(\omega)\,\lambda_{2}^{\dagger}\,.
\end{split}
\end{equation} 
In what follows, we use the definitions given by Eq.\,(\ref{djadjhsjhd}).
Then, we add up the three terms given by Eq.\,(\ref{edndjjdjj}), trace out over the central system subspace, and then take the real part of Eq.\,(\ref{edndjjdjj}),
\begin{equation}
\begin{split}
I_{dc}\,&=\,\frac{2e}{\hbar}\int\frac{d\omega}{2\pi}\Bigg\{ \frac{\eta^{2}}{2}\left[ A_{L}(\omega-\omega_{0})\lambda^{\dagger}_{1}f_{R}(\omega)A_{R}(\omega)\lambda_{1}-f_{L}(\omega-\omega_{0})A_{L}(\omega-\omega_{0})\lambda_{1}^{\dagger}A_{R}(\omega)\lambda_{1}\right]\\
&\times\sigma_{0,{\rm spin}}\otimes\tau_{z,{\rm Nambu}}\\
&+\frac{\eta^{2}}{2}\left[A_{L}(\omega) \lambda_{1}^{\dagger}f_{R}(\omega+\omega_{0})A_{R}(\omega+\omega_{0})\lambda_{1} 
-f_{L}(\omega)A_{L}(\omega) \lambda_{1}^{\dagger}A_{R}(\omega+\omega_{0})\lambda_{1}
\right]\times\\
&\sigma_{0,{\rm spin}}\otimes\tau_{z,{\rm Nambu}}\\
&+\frac{\eta^{2}}{2}\left[A_{L}(\omega)\lambda_{2}f_{R}(\omega-\omega_{0})A_{R}(\omega-\omega_{0})\lambda_{2}^{\dagger}
-f_{L}(\omega)A_{L}(\omega)\lambda_{2}A_{R}(\omega-\omega_{0})\lambda_{2}^{\dagger}\right]\times\\
&\sigma_{0,{\rm spin}}\otimes\tau_{z,{\rm Nambu}}
\\
\,&+ \frac{\eta^{2}}{2}\left[ A_{L}(\omega+\omega_{0})\lambda_{2}f_{R}(\omega)A_{R}(\omega)\lambda_{2}^{\dagger}-f_{L}(\omega+\omega_{0})A_{L}(\omega+\omega_{0})
\lambda_{2}A_{R}(\omega)\lambda_{2}^{\dagger}\right]\times\\
&\sigma_{0,{\rm spin}}\otimes\tau_{z,{\rm Nambu}}\Bigg\}\,.\\
\end{split}
\end{equation}
The first term corresponds to sideband $-1-1$, the second and third to $00$ and the last one to $11$. Notice that previous equations do not contain first order terms in $\eta$.  
Previous equation can be written as
\begin{equation}
\label{hdhdhddhd}
\begin{split}
I_{dc}\,&=\,\frac{2e}{\hbar}\int\frac{d\omega}{2\pi}\frac{\eta^{2}}{2}\sum_{n=0}^{1}\Big\{A_{L}[\omega\,-\,n\,\omega_{0}]\,\lambda_{1}^{\dagger}\,f_{R}[\omega+(1-n)\omega_{0}]\,A_{R}[\omega+(1-n)\omega_{0}]\,\lambda_{1}\\
&-\,
f_{L}[\omega-n\omega_{0}]\,A_{L}[\omega-n\omega_{0}]\,\lambda_{1}^{\dagger}\,A_{R}[\omega+(1-n)\omega_{0}]\,\lambda_{1}\\
&+\,A_{L}[\omega\,+\,n\,\omega_{0}]\,\lambda_{2}\,f_{R}[\omega-(1-n)\omega_{0}]\,A_{R}[\omega-(1-n)\omega_{0}]\,\lambda_{2}^{\dagger}\\
&-\,
f_{L}[\omega+n\omega_{0}]\,A_{L}[\omega+n\omega_{0}]\,\lambda_{1}\,A_{R}[\omega-(1-n)\omega_{0}]\,\lambda_{2}^{\dagger}\Big\}\times\sigma_{0,{\rm spin}}\otimes\tau_{z,{\rm Nambu}}\,,
\end{split}
\end{equation}
where all elements are matrices in Nambu space (label by $\hat{}$ from hereon: $\hat{A}$). So that the spectral function, $\lambda_{1,2}$ and the Fermi distribution function read
\begin{equation}
\begin{split}
\hat{A}_{L}(\omega)\,&=\,
\begin{pmatrix}
A_{11L}(\omega)&A_{12L}(\omega)\\
A_{21L}(\omega)&A_{22L}(\omega)
\end{pmatrix}\,,\quad \hat{f}(\omega)\,=\,
\begin{pmatrix}
f(\omega)&0\\
0&1-f(-\omega)
\end{pmatrix}\,.
\end{split}
\end{equation}
The relevant information information of the system is hidden in the diagonal elements of the spectral function $A$. 
Now, we use the definitions for $\lambda_{1,2}$ given by Eqs.\,(\ref{Vv})
and then trace over Nambu space, 
\begin{equation}
\begin{split}
I_{dc}\,
\,&=\,\frac{2e}{\hbar}\int\frac{d\omega}{2\pi}\Big\{\left[f_{L}(\omega-\omega_{0})-f_{R}(\omega)\right]\,A_{22L}(\omega-\omega_{0})\,v^{T}\,A_{22R}(\omega)\,v^{*}\\
\,&+\,\left[f_{L}(\omega)-f_{R}(\omega+ \omega_{0})\right]\,A_{22L}(\omega)\,v^{T}\,A_{22R}(\omega+ \omega_{0})\,v^{*}\\
\,&+\,\left[f_{R}(\omega-\omega_{0})-f_{L}(\omega)\right]\,A_{11L}(\omega)\,v^{T}\,A_{11R}(\omega- \omega_{0})\,v\\
\,&+\,\left[f_{R}(\omega)-f_{L}(\omega+ \omega_{0})\right]\,A_{11L}(\omega+\omega_{0})\,v^{T}\,A_{11R}(\omega)\,v\,.
\end{split}
\end{equation}
or
\begin{equation}
\begin{split}
I_{dc}\,&=\,\frac{2e}{\hbar}\int\frac{d\omega}{2\pi}\sum_{n=0}^{1}
\Big\{ 
\big[A_{11L}(\omega+n\omega_{0})\,v^{\dagger}\,f_{R}(\omega-(1-n)\omega_{0})\,A_{11R}(\omega-(1-n)\omega_{0})\,v\\
\,&-\,
A_{22L}(\omega-n\omega_{0})\,v^{T}\,f_{R}(\omega+(1-n)\omega_{0})\,A_{22R}(\omega+(1-n)\omega_{0})\,v^{*}\big]
 \\
\,&-\,
\big[f_{L}(\omega+n\omega_{0})\,A_{11L}(\omega+n\omega_{0})\,v^{\dagger}\,A_{11R}(\omega-(1-n)\omega_{0})\,v\\
\,&-\,
f_{L}(\omega-n\omega_{0})\,A_{22L}(\omega-n\omega_{0})\,v^{\dagger}\,A_{22R}(\omega+(1-n)\omega_{0})\,v^{*}\big]
\Big\}\sigma_{0}\,.
\end{split}
\end{equation}
Previous equation can be rewritten in a more elegant way, since the Fermi distribution functions can be factored out,
\begin{equation}
\begin{split}
I_{dc}\,&=\,\frac{2e}{\hbar}\int\frac{d\omega}{2\pi}\sum_{n=0}^{1}\Big\{
A_{11L}(\omega+n\omega_{0})\,v^{\dagger}\,A_{11R}(\omega-(1-n)\omega_{0})\,v\left[ f_{R}(\omega-(1-n)\omega_{0})\,-\,f_{L}(\omega+n\omega_{0})\right]\\
&\,+\,
A_{22L}(\omega-n\omega_{0})\,v^{T}\,A_{22R}(\omega+(1-n)\omega_{0})\,v^{*}\left[f_{L}(\omega-n\omega_{0})\,-\,f_{R}(\omega+(1-n)\omega_{0})\right]
\Big\}\sigma_{0}\,,\\
\end{split}
\end{equation}
and by taking into account that the hopping matrix $v$ is real, one has
\begin{equation}
\label{ksksdj}
\begin{split}
I_{dc}\,&=\,\frac{2e}{\hbar}\int\frac{d\omega}{2\pi}\sum_{n=0}^{1}\Big\{
A_{11L}(\omega+n\omega_{0})\,v^{T}\,A_{11R}(\omega-(1-n)\omega_{0})\,v\left[ f_{R}(\omega-(1-n)\omega_{0})\,-\,f_{L}(\omega+n\omega_{0})\right]\\
&\,+\,
A_{22L}(\omega-n\omega_{0})\,v^{T}\,A_{22R}(\omega+(1-n)\omega_{0})\,v\left[f_{L}(\omega-n\omega_{0})\,-\,f_{R}(\omega+(1-n)\omega_{0})\right]
\Big\}\sigma_{0}\,,\\
\end{split}
\end{equation}

One can also seek for a relation between $A_{11}$ and $A_{22}$. Since the spectral function is computed numerically by using the retarded and advanced Green's functions of the leads, the structure of such Green's functions allows us to relate the $A_{11}$ and $A_{22}$. Thus one has,
\begin{equation}
g^{r}_{11}(\omega)\,=\,-\left[g^{r}_{22}(-\omega)\right]^{*}\,\quad\rightarrow\,\quad A_{11}(\omega)\,=\,-A^{*}_{22}(-\omega)
\end{equation} 
Then, Eq.\,(\ref{ksksdj}), introducing the limits of integration, reads
\begin{equation}
\begin{split}
I_{dc}\,&=\,\frac{2e}{\hbar}\int_{0}^{\hbar\omega_{0}}\,d\,\omega
\sum_{n=0}^{1}\Big\{
A_{22L}^{*}(-\omega-n\omega_{0})\,v^{T}\,A_{22R}^{*}(-\omega+(1-n)\omega_{0})\,v\left[ f_{R}(\omega-(1-n)\omega_{0})\,-\,f_{L}(\omega+n\omega_{0})\right]\\
&\,+\,
A_{22L}(\omega-n\omega_{0})\,v^{T}\,A_{22R}(\omega+(1-n)\omega_{0})\,v\left[f_{L}(\omega-n\omega_{0})\,-\,f_{R}(\omega+(1-n)\omega_{0})\right]\,.
\Big\}
\end{split}
\end{equation}
Notice that previous integral contains a term whose argument  is $-\omega$. Then,
we make the substitution $-\omega\rightarrow \omega$. The distribution functions reads
\begin{displaymath}
\begin{split}
f(-\omega-\omega_{0})\,-\,f(-\omega)\,&=\,f(\omega)\,-\,f(\omega+\omega_{0})\\
f(-\omega)\,-\,f(-\omega+\omega_{0})\,&=\,f(\omega-\omega_{0})\,-\,f(\omega)\,.
\end{split}
\end{displaymath}
Previous considerations lead to
\begin{equation}
\begin{split}
I_{dc}\,&=\,\frac{2e}{\hbar}\Bigg\{\int_{0}^{\omega_{0}}\,d\,\omega
\left[f_{L}(\omega-\omega_{0})\,-\,f_{R}(\omega) \right]\,A_{22L}(\omega-\omega_{0})v^{T}\,A_{22R}(\omega)\,v\,\\
\,&+\,
\,\int_{0}^{\omega_{0}}\,d\,\omega
\left[f_{L}(\omega)\,-\,f_{R}(\omega+\omega_{0}) \right]\,A_{22L}(\omega)v^{T}\,A_{22R}(\omega+\omega_{0})\,v\,\\
\,&+\,
\,\int_{-\omega_{0}}^{0}\,d\,\omega
\left[f_{L}(\omega)\,-\,f_{R}(\omega+\omega_{0}) \right]\,A_{22L}^{*}(\omega)v^{T}\,A_{22R}^{*}(\omega+\omega_{0})\,v\,\,\\
\,&+\,
\,\int_{-\omega_{0}}^{0}\,d\,\omega
\left[f_{L}(\omega-\omega_{0})\,-\,f_{R}(\omega) \right]\,A_{22L}^{*}(\omega-\omega_{0})v^{T}\,A_{22R}^{*}(\omega)\,v\Bigg\}\,.
\end{split}
\end{equation}
Assuming that $f_{L}(\omega)=f_{R}(\omega)$, one gets
\begin{equation}
\begin{split}
I_{dc}\,&=\,\frac{2e}{\hbar}\Bigg\{\int_{0}^{\omega_{0}}\,d\,\omega
\left[f(\omega-\omega_{0})\,-\,f(\omega) \right]\,A_{22L}(\omega-\omega_{0})v^{T}\,A_{22R}(\omega)\,v\,\\
\,&+\,
\,\int_{0}^{\omega_{0}}\,d\,\omega
\left[f(\omega)\,-\,f(\omega+\omega_{0}) \right]\,A_{22L}(\omega)v^{T}\,A_{22R}(\omega+\omega_{0})\,v\,\\
\,&+\,
\,\int_{-\omega_{0}}^{0}\,d\,\omega
\left[f(\omega)\,-\,f(\omega+\omega_{0}) \right]\,A_{22L}^{*}(\omega)v^{T}\,A_{22R}^{*}(\omega+\omega_{0})\,v\,\,\\
\,&+\,
\,\int_{-\omega_{0}}^{0}\,d\,\omega
\left[f(\omega-\omega_{0})\,-\,f(\omega) \right]\,A_{22L}^{*}(\omega-\omega_{0})v^{T}\,A^{*}_{22R}(\omega)\,v\Bigg\}\,.
\end{split}
\end{equation}
One can make a change of variable in the third integral in previous equation, $\omega+\omega_{0}\rightarrow \omega$, then one has that the third term is the conjugated of the first one, 
\begin{equation}
\label{eufbedifh}
\begin{split}
I_{dc}\,&=\,\frac{2e}{\hbar}\Bigg\{\int_{0}^{\omega_{0}}d\,\omega\,2\,{\rm Re}\Big\{
\left[f(\omega-\omega_{0})\,-\,f(\omega) \right]\,A_{22L}(\omega-\omega_{0})v^{T}\,A_{22R}(\omega)\,v\Big\}\,\\
\,&+\,
\,\int_{0}^{\omega_{0}}\,d\,\omega
\left[f(\omega)\,-\,f(\omega+\omega_{0}) \right]\,A_{22L}(\omega)v^{T}\,A_{22R}(\omega+\omega_{0})\,v\,\\
\,&+\,
\,\int_{-\omega_{0}}^{0}\,d\,\omega
\left[f(\omega-\omega_{0})\,-\,f(\omega) \right]\,A_{22L}^{*}(\omega-\omega_{0})v^{T}\,A^{*}_{22R}(\omega)\,v\Bigg\}\,.
\end{split}
\end{equation}
Notice that the last two integrals of previous equation should vanish due to symmetry considerations. 
Let us check it in for the terms $A_{11}$, since this corresponds to the electron spectral function. The conclusion holds also for previous equation.
 One has,
\begin{equation}
\begin{split}
I_{dc}\,&=\,\frac{2e}{\hbar}
\,\int_{-\omega_{0}}^{0}\,d\,\omega
\left[f(\omega)\,-\,f(\omega+\omega_{0}) \right]\,A_{11L}^{*}(\omega+\omega_{0})v^{T}\,A_{11R}^{*}(\omega)\,v\,\,\\
\,&+\,
\,\int_{-\omega_{0}}^{0}\,d\,\omega
\left[f(\omega-\omega_{0})\,-\,f(\omega) \right]\,A_{11L}^{*}(\omega)v^{T}\,A^{*}_{11R}(\omega-\omega_{0})\,v\,\\
\,&+\,
\int_{0}^{\omega_{0}}\,d\,\omega
\left[f(\omega-\omega_{0})\,-\,f(\omega) \right]\,A_{11L}(\omega)v^{T}\,A_{11R}(\omega-\omega_{0})\,v\,\\
\,&+\,
\,\int_{0}^{\omega_{0}}\,d\,\omega
\left[f(\omega)\,-\,f(\omega+\omega_{0}) \right]\,A_{11L}(\omega+\omega_{0})v^{T}\,A_{11R}(\omega)\,v\,.\\
\end{split}
\end{equation}
 \begin{figure}[htb!]
\begin{center}
    \includegraphics[width=.6\textwidth]{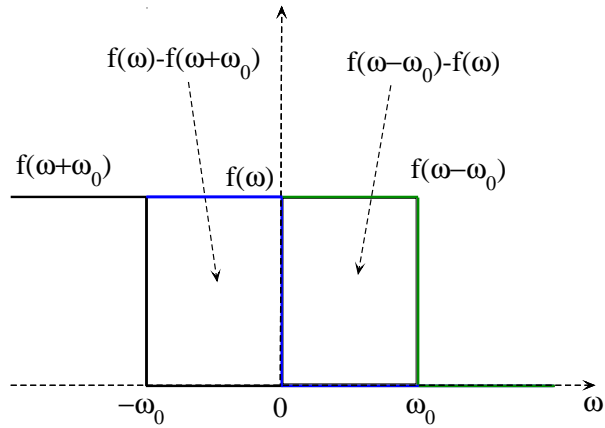}
\caption{Representation of the Fermi functions.}
\label{fig111}
\end{center}
\end{figure}
In the first integral the term $f(\omega)\,-\,f(\omega+\omega_{0})$ is on the interval $[-\omega_{0}, 0]$, then the fourth integral vanish on the interval 
 $[0, \omega_{0}]$. The same analysis can be done between the second and third terms. See Fig.\,\ref{fig111} for more details.
 
Therefore, the terms that survive are only the first and the third. Similar considerations led to the conclusion of vanishing the last two terms in Eq.\,(\ref{eufbedifh}). Therefore, one has
\begin{equation}
\begin{split}
I_{dc}\,&=\,\frac{2e}{\hbar}\Bigg\{
\,\int_{-\omega_{0}}^{0}\,d\,\omega
\left[f(\omega)\,-\,f(\omega+\omega_{0}) \right]\,A_{11L}^{*}(\omega+\omega_{0})v^{T}\,A_{11R}^{*}(\omega)\,v\,\,\\
\,&+\,
\int_{0}^{\omega_{0}}\,d\,\omega
\left[f(\omega-\omega_{0})\,-\,f(\omega) \right]\,A_{11L}(\omega)v^{T}\,A_{11R}(\omega-\omega_{0})\,v\Bigg\}\,,\\
\,&=\frac{2e}{\hbar}\int_{0}^{\omega_{0}}\,d\,\omega
\,2\,{\rm Re}\big\{ \left[f(\omega-\omega_{0})\,-\,f(\omega) \right]\,A_{11L}(\omega)v^{T}\,A_{11R}(\omega-\omega_{0})\,v\,\big\}\,,
\end{split}
\end{equation}
where in the first integral we made a change of variable, $\omega+\omega_{0}\rightarrow \omega$. 
The frequency can now take any arbitrary value, so that the integral limits change.
Then the current in the tunnel limit reads,
\begin{equation}
I_{dc}=\frac{e}{\pi}\,\eta^{2}{\rm Re}\int_{-\infty}^{\infty}\,d\,\omega
\,{\rm Tr}\big\{ \left[f(\omega-\omega_{0})\,-\,f(\omega) \right]\,A_{11L}(\omega)v^{T}\,A_{11R}(\omega-\omega_{0})\,v\,\big\}\,,
\end{equation}
where the trace in previous expression is taken over the spin space. The coupling across the junction is modelled by the hopping matrix $v_{0}=\eta\,v$ between the end sites of each wire, where $\eta\in[0,1]$ is the dimensionless tunneling parameter that controls the junction's normal transparency.

\begin{figure}[t] 
   \centering
   \includegraphics[width=0.3\linewidth,clip]{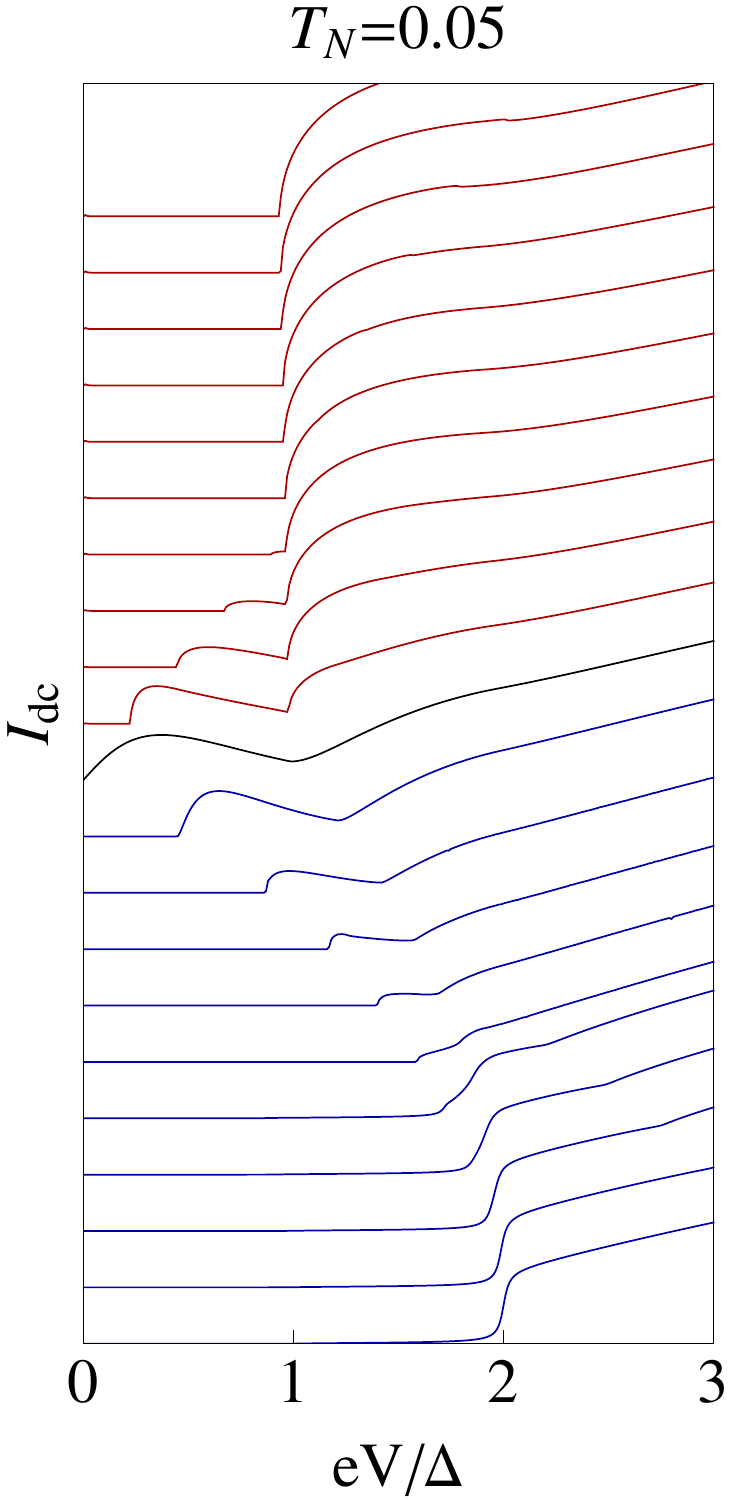} 
     \caption[$I_{dc}(V)$ for increasing Zeeman field $B$ using the tunneling approximation]{(Color online) Time-averaged current $I_{dc}$ as a function of bias $V$ for increasing Zeeman field $B$ using the tunneling approximation of Eq. \ref{Itunnel}. Curves are offset by a constant $2\Delta \mathcal{G}_0/e$, with $\mathcal{G}_0=e^2/h$. Blue and red curves correspond to the non-topological ($B<B_c$) and topological ($B>B_c$) phases respectively. 
}
   \label{fig:Itun}
\end{figure}

Thus, it was possible to calculate the $I_{dc}$ current explicitly in the tunnelling limit, where to leading (second) order in the left-right coupling $v_0$, Eq. (\ref{KeldyshIdc}) reduces, after some algebra, to
\begin{eqnarray}
I_{dc}&\approx&\frac{e}{\pi} \mathrm{Re}\int d\omega \left[f(\omega-\omega_0)-f(\omega)\right]
\mathrm{Tr}\left\{A_{11}^L(\omega) v_0^+ A_{11}^R(\omega-\omega_0)v_0\right\}
\label{Itunnel}
\end{eqnarray}
where $A_{11}^\alpha$ is the particle-particle Nambu $2\times2$ matrix block of the spectral function of the $\alpha=L,R$ decoupled wire,
\[
A^{\alpha}(\omega)
=\begin{pmatrix}
A_{11}^\alpha(\omega) & A_{12}^\alpha(\omega)\\
\left[A_{12}^\alpha(\omega)\right]^+ & -\left[A_{11}^{\alpha}(-\omega)\right]^*
\end{pmatrix}\,
\]
and the trace is taken over spin space. The trace of $A^{\alpha}(\omega)$ is proportional to the local density of states. Fig. \ref{fig:Itun} shows results for the tunnel current using Eq. \ref{Itunnel} for increasing Zeeman fields. Overall, the agreement with the full numerics in Fig. \ref{fig:I0}(a) is very good and, importantly, all the relevant features such as, e. g.,  the closing of the gap, are captured by this tunneling approximation.

\section{Andreev approximation}\label{AppendixB3}

It is conventional, in the study of hybrid superconducting-normal junctions, to assume the limit in which the Fermi energy $\mu$ of the metal under consideration is much greater than the superconducting gap, and any other energy $E$ involved in the problem, $\mu\gg\Delta, E$. This is known as the Andreev approximation. In essence, it allows one to regard the normal system as featureless, with constant Fermi velocity and density of states. In this case, a number of simplifications can be carried out in the computation of equilibrium transport properties. One important consequence of the approximation in the context of our work is that, in a short SNS junction with phase difference $\phi$, and symmetric under time-reversal symmetry ($B=0$ in our case), two degenerate Andreev states will appear of energy $\epsilon(\phi)=|\Delta|\sqrt{1-T_N^2\sin^2(\phi/2)}$ \cite{Beenakker:92}. 
This immediately implies that at perfect normal transparency $T_N=1$, the ABSs will reach zero energy at $\phi=\pi$. In other words, in the Andreev approximation the ABS energy minimum in the non-topological phase will be $\epsilon(\pi)=\delta_\pi=0$.  

While in the topological phase, a zero $\delta_\pi$ is a robust property, protected by parity conservation, a $\delta_\pi=0$ in the non-topological phase is accidental, a direct consequence of the Andreev approximation, and is not protected by any symmetry. In fact, any deviation from the Andreev approximation will induce a finite splitting $\delta_\pi$. In semiconducting nanowires such as the ones considered in this work, this correction is very relevant. Indeed, for the nanowire to undergo a superconducting topological transition at a reasonable Zeeman field $B$, $\mu/\Delta$ must remain relatively small (the wire must be close to depletion, and far from the Andreev approximation), otherwise the critical field $B_c=\sqrt{\mu^2+\Delta^2}$ would be physically unreachable. The splitting $\delta_\pi$, therefore, remains a relevant quantity in the formation and detection of Majorana bound states.

\begin{figure}[t] 
   \centering
   \includegraphics[width=0.5\linewidth,clip]{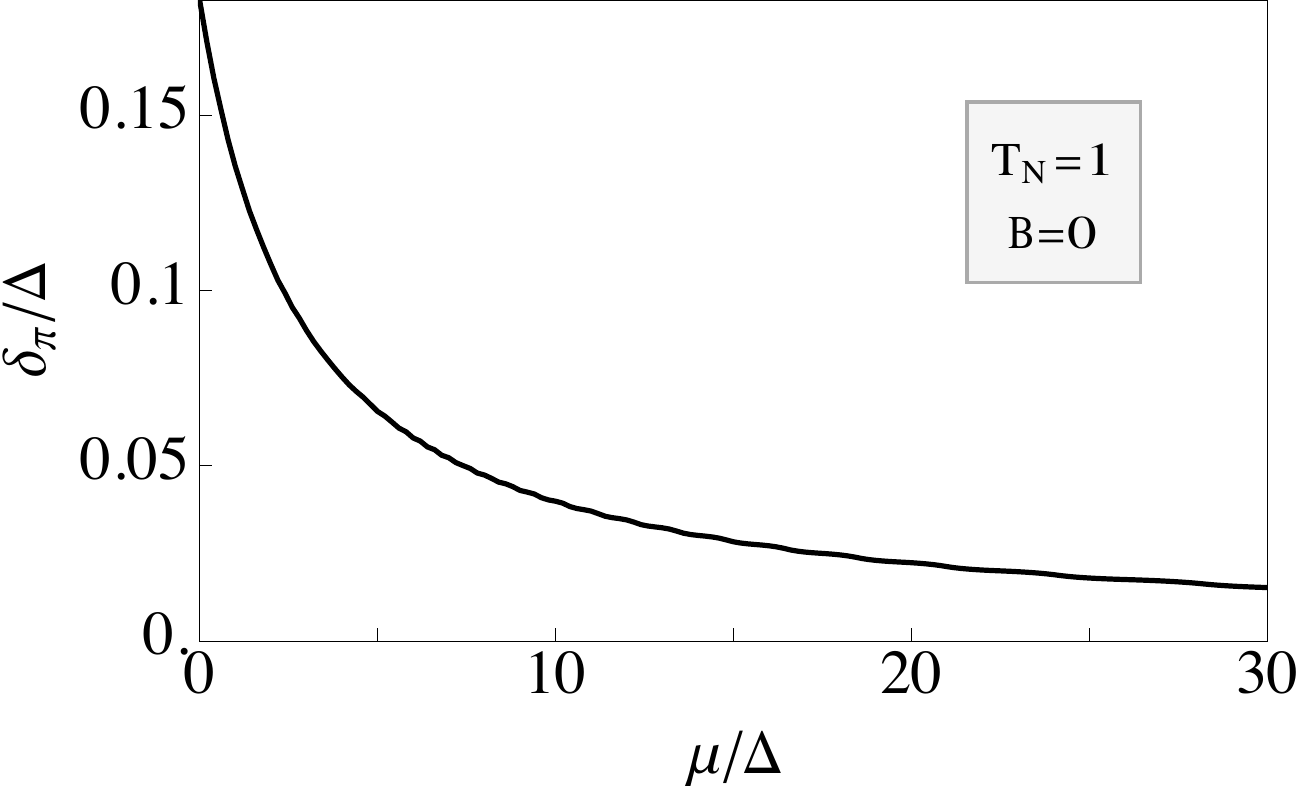} 
     \caption[Minimum Andreev state energy $\delta_\pi$ in a short SNS junction, at $B=0$ and $T_N=1$]{Minimum Andreev state energy $\delta_\pi$ in a short SNS junction, at $B=0$ and $T_N=1$. As $\mu/\Delta$ grows, $\delta_\pi$ decreases to zero, in agreement with predictions within the Andreev approximation.
}
   \label{fig:deltapi}
\end{figure}

The value of $\delta_\pi$ in our system may be computed numerically. The most efficient way is to consider a short SNS junction with finite length superconductors, $T_N=1$, $B=0$ and a phase difference $\phi=\pi$. Since this system is closed, an exact diagonalization of the tight-binding Nambu Hamiltonian yields a minimum eigenvalue that is exactly $\delta_\pi$ if the S leads are long enough (longer than the coherence length). We find that this quantity is finite in the case of wires close to depletion, $\mu\gtrsim\Delta$, and that it vanishes as one approaches the Andreev approximation regime $\mu\gg\Delta$, see Fig. \ref{fig:deltapi}. More specifically, we have found that $\delta_\pi$ scales as $\delta_\pi=c_1 \Delta^2/(\mu+\Delta c_2)$ for some  $c_{1,2}>0$, within very good precision.

\chapter{\bf Appendix for Chapter \ref{Chap2}}
\label{AppA} 
\lhead{Appendix \ref{AppA}. \emph{For Chapter 4}}
In this part we provide some additional technical aspects used in Chapter \ref{Chap2}.
We make a full proximity model in order to justify our calculations within the simplified model description. 
Moreover, we perform an effective model for the normal conductance, where we aim at describing the physics of Fano resonances in the first part of Chapter \ref{Chap2}. At the end, the Majorana localization length is calculated.
\section{Induced superconducting pairing}
\label{inducspe}
\begin{figure}[!ht]
\centering
\includegraphics[width=.8\textwidth]{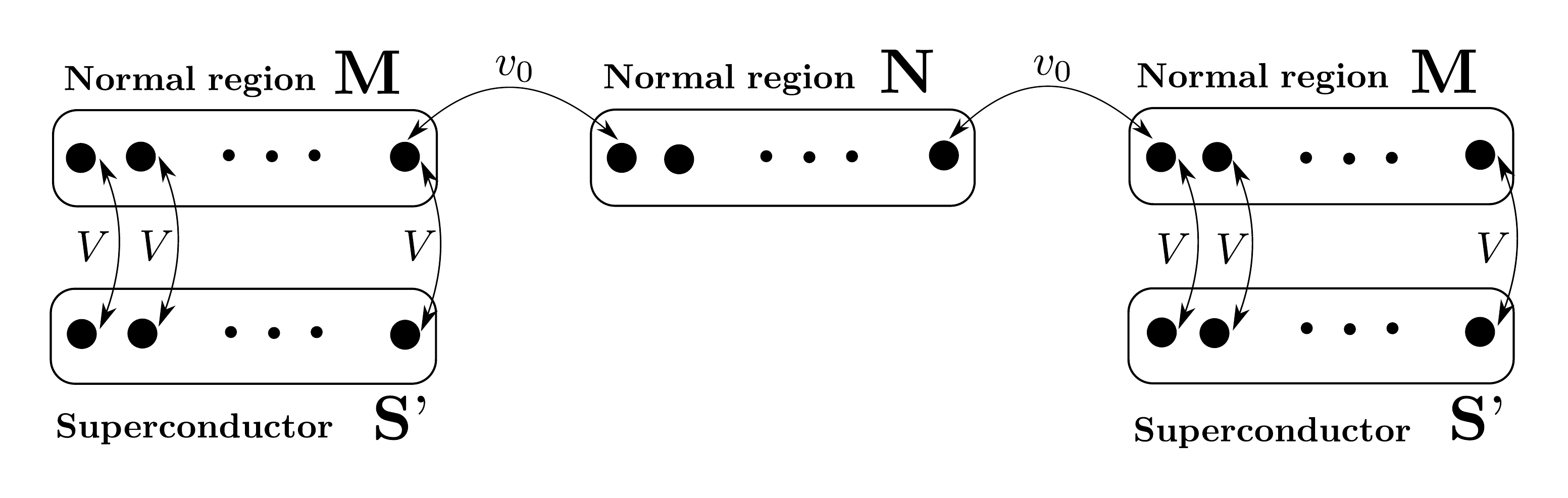} 
\caption[Sketch of the geometry for the appendix]{(Color online) A nanowire is divided in three normal regions denoted by N and M. The sections M are coupled to a superconductor through $V$, while the coupling between N and M is controlled by $v_{0}$.}
\label{fig1App}
\end{figure}

A more realistic model when considering the proximity effect consists on the following description. 
The full NW is divided in three sections: a central normal region (N) and two normal regions (M). See Fig.\,\ref{fig1App}. Each of the M sections describe NW regions coupled to a superconductor which, to distinguish from the previous notation, we denote as S'. As opposed to the previous subsection, the full NW is now a normal system and the proximity effect comes now from the tunneling coupling between the superconductors and the M normal parts of the NW.
In this case, the full (normal) system is described by the following Hamiltonian
\begin{equation}
\label{longsns2}
\hat{h}_{SNS}\,=\,
\begin{pmatrix}
h_{S'_{L}}&h_{S'_{L}M}&0&0&0\\
h_{S'_{L}M}^{\dagger}&h_{M}&h_{MN}&0&0\\
0&h_{MN}^{\dagger}&h_{N}&h_{NM}&0\\
0&0&h_{NM}^{\dagger}&h_{M}&h_{MS'}\\
0&0&0&h_{M S'_{R}}^{\dagger}&h_{S'_{R}}\\
\end{pmatrix}\,,
\end{equation} 
where $h_{M}$ is a normal region of the same dimension as the superconducting one $S$' and
$h_{S'_{i}M}$ is a diagonal matrix in site space that couples the superconductor $S'_{i}$ with the normal lead $M$. This coupling can be parametrized by the parameter $V$.
Then, we introduce the he superconducting pairing, which is written in the same basis as $\hat{h}_{SNS}$, 
\begin{equation}
\label{longpairing}
\begin{split}
\Delta(x)=&
\begin{pmatrix}
\Delta_{S'_{L}}&0&0&0&0\\
0&0&0&0&0\\
0&0&0&0&0\\
0&0&0&0&0\\
0&0&0&0&\Delta_{S'_{R}}
\end{pmatrix}\,,
\end{split}
\end{equation} 
where $\Delta_{S'_{i}}=\Delta_{S'}{\rm e}^{{\rm i}\varphi_{i}}$ with $i=R,L$ describe the bulk s-wave superconducting leads. In a similar way as the one described in Section \ref{SNSjunctionmodel}, we write the full system Hamiltonian in Nambu space as 
\begin{equation}
\hat{H}_{SNS}=\begin{pmatrix}
\hat{h}_{SNS}&\Delta(x)\\
\Delta^{\dagger}(x)&-\hat{h}_{SNS}^{*}
\end{pmatrix}\,.
\end{equation}

\begin{figure}[!ht]
  \label{locMaj}
\centering
\includegraphics[width=.5\textwidth]{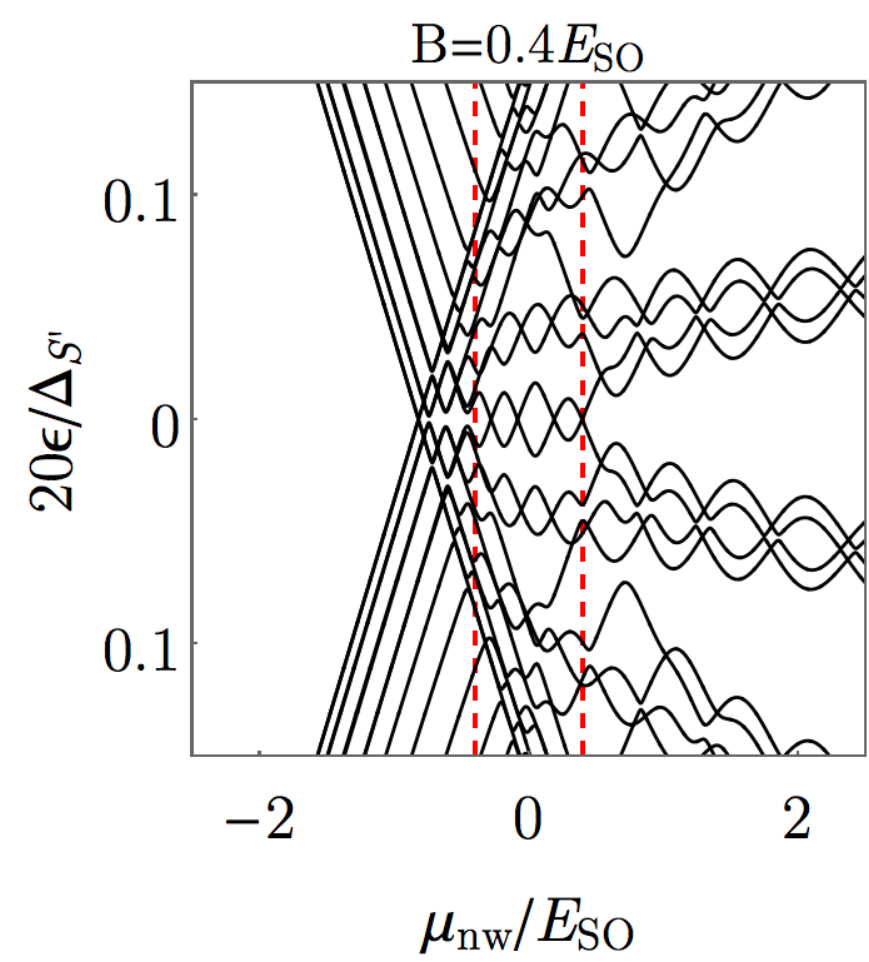} 
\caption[Energy levels as function of the $\mu_{nw}$ for a long junction]{(Color online) Energy levels at $\varphi=0$ as function of the Fermi energy $\mu_{\mathrm{nw}}$ for a long junction $L_{\mathrm{nw}}=4$\,$\mu$m for a fixed Zeeman field. Parameters: $E_{SO}=0.05$meV, $\mu_{leads}=10E_{SO}$, $L_{S}=2\mu$m, $V=20E_{SO}$ and $\Delta_{S'}=20\Delta=5$meV. The rescaled $y$ axis explicitly shows that the relevant energy scale is not the original bulk gap included in the calculation $\Delta_{S'}$ but rather $\Delta$, in agreement with Fig. \ref{fig11}c.}
\label{fig4App}
\end{figure}

As described in the main text, the approximate description of the proximity effect given in Section \ref{SNSjunctionmodel} is a good approximation provided that we are in a large gap limit and that the contact transparency is good. We have benchmarked the approximate solution given in Section \ref{SNSjunctionmodel} against the full proximity model, given here, in various relevant cases and always found good agreement in the correct parameter range. Now, we illustrate this point by showing a calculation using the full proximity effect model of Eq. \ref{longsns2} instead of the approximate model given by Eq.\,(\ref{longsns2}) and fully described in Section \ref{SNSjunctionmodel}. 

In Fig.\,\ref{fig4App} we show results corresponding to the same physical situation we presented in Fig. \ref{fig11}c in the main text, the only difference being that the bulk gap in $S'$ is much larger than the induced gap used in the calculations of Fig. \ref{fig11}c ($\Delta_{S'}=20\Delta$). The overall behaviour of the subgap states in Fig.\,\ref{fig4App} is the same as in Fig. \ref{fig11}c (including the loops in the helical region described in the main text), demonstrating that the simplified model is indeed justified when the bulk gap is the largest energy scale. Importantly, note the rescaled $y$ axis which explicitly shows that the relevant energy scale is not the original bulk gap included in the calculation but the smaller value $\Delta=\Delta_{S'}/20$, in agreement with our previous claim.

\section{Effective model for the conductance}
\label{tightc}
\begin{figure}[!ht]
\centering
\includegraphics[width=.8\textwidth]{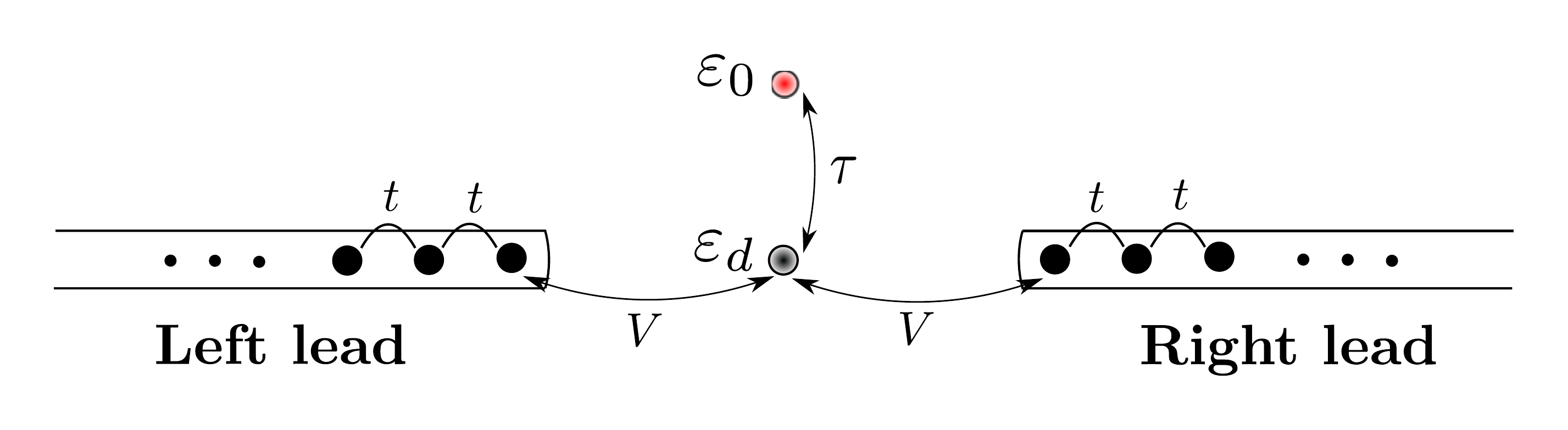} 
\caption[Sketch: model for the conductance]{(Color online) Two tight binding semi-infinite chains (leads) coupled to a central region. We consider a central region consisting of one site with energy $\varepsilon_{d}$ (i.e. a quantum dot) that is weakly coupled to a resonant level $\varepsilon_{0}=\varepsilon_{d}-\varepsilon_{r}$ through $\tau$, where $\varepsilon_{r}$ is a fixed parameter that represents the separation between the quantum dot level and the resonant level. The coupling of the quantum dot to the leads is controlled by $V$.
}\label{fig:sketchGNTN}
\end{figure}

In this part we make use of an effective model to describe the physics of Fano resonances. 
An effective spinless model based on Green's functions is constructed where two semi-infinite tight-binding chains (leads) are coupled through $V$ to a central region $\varepsilon_{d}$, formed by one site, that is additionally weakly coupled through $\tau<<V$ to a resonant level $
\varepsilon_{0}=\varepsilon_{d}-\varepsilon_{r}$, being $\varepsilon_{r}$ a fixed parameter that represents the separation between the quantum dot level and the resonant level (in principle this parameter mimics the role of the Zeeman splitting in our numerics) (see Fig.\,\ref{fig:sketchGNTN}). Consider that $a$ is the lattice constant and $t$ the hopping between sites in the leads.  The normal transmission, $T_{N}$, through a central system formed by one site can be calculated by using the Caroli's formula,
\begin{equation}
\label{transmq}
T_{N}(\omega)\,=\,4\,{\rm Tr}[\Gamma_{L}\,G^{r}\,\Gamma_{R}\,G^{a}]\,,
\end{equation}
where $G^{r(a)}$ is the retarded full system Green's function, and 
\begin{equation}
  \Gamma_{L(R)}(\omega)=\frac{\Sigma_{L(R)}^{r}(\omega)-\Sigma_{L(R)}^{a}(\omega)}{2i}\,,
\end{equation}
takes into account the influence of the leads on the central system through the left(right) L(R) self-energies $\Sigma_{L/R}$.

The full system Green's function can be calculated by using the Dyson's relation,
 \begin{equation}
  G^{r}(\omega)=g_{0}^{r}(\omega)+g_{0}^{r}(\omega)\,\Sigma^{r}(\omega)\,G^{r}(\omega)=(G^{a}(\omega))^{\dagger}\,,\quad
 \text{or}\quad G^{r}(\omega)=\left[[g_{0}^{r}(\omega)]^{-1}-\Sigma^{r}(\omega)\right]^{-1}\,,
 \end{equation}
where $g_{0}^{r}$ is the retarded Green's function of the isolated central region (this central region can for instance be a quantum dot) without the influence of the leads and without the influence of the resonant level. It reads
\begin{equation}
g_{0}^{r}(\omega)\,=\,\frac{1}{\omega\,-\,\varepsilon_{d}\,+\,i\eta}
\end{equation}
where $\varepsilon_{d}$ is the onsite energy of the central region. 

The self-energy $\Sigma^{r}$,
\begin{equation}
\Sigma^{r}(\omega)=\Sigma_{L}^{r}(\omega)+\Sigma_{R}^{r}(\omega)+\Sigma_{res}^{r}(\omega)\,
\end{equation}
contain the effect of the left $\Sigma_{L}(\omega)$ and right $\Sigma_{R}(\omega)$ leads as well as the influence of the resonant level $\Sigma_{res}(\omega)$, respectively. Such self-energies are defined as follows,
\begin{equation}
\Sigma_{L(R)}^{r}(\omega)=t^{\dagger}\,g_{L(R)}^{r}(\omega)\,t
\end{equation}
where $g^{r}_{L(R)}$ is the retarded semi-infinite left (right) lead Green's functions. In principle such lead's Green's functions can be computed considering a recursive approach,
\begin{equation}
g_{L(R)}(\omega)\,=\,\frac{1}{\omega-h-t^{\dagger}\,g_{L(R)}(\omega)\,t}\,,
\end{equation}
$h=2t-\mu$ is the onsite energy in the leads. 
From previous equation one has,
\begin{equation}
|t|^{2}g_{L(R)}\,-\,(\omega-h)\,g_{L(R)}\,+\,1\,=0\,
\end{equation}
therefore,
\begin{equation}
g_{L(R)}(\omega)\,=\,\frac{1}{|t|}\left[\frac{\omega-h}{2|t|}\pm\sqrt{\left(\frac{\omega-h}{2|t|}\right)^{2}\,-\,1}\right]\,.
\end{equation}
Adding a convergence factor to frequency, $\omega\rightarrow\omega\pm i\eta$, one finds the retarded or advanced Green's function.We have the following properties of $g_{L(R)}$,
\begin{widetext}
\begin{equation}
g_{L(R)}(\omega)\, = \begin{cases}
        \,\frac{1}{|t|}\left[\frac{\omega-h}{2|t|}-{\rm sgn}(\omega-h)\sqrt{\left(\frac{\omega-h}{2|t|}\right)^{2}\,-\,1}\right]\,,  &|(\omega-h)/2|t||>1\\
        \,\frac{1}{|t|}\left[\frac{\omega-h}{2|t|}\pm i\sqrt{1\,-\,\left(\frac{\omega-h}{2|t|}\right)^{2}}\right]\,,  & |(\omega-h)/2|t||<1
        \end{cases}
\end{equation}
\end{widetext}
where for the first case the density of states $\rho_{0}=-\frac{1}{\pi}\mathrm{Im}g_{L(R)}$ is zero, while in the second case it exhibits a non zero value. These results allow us to obtain $\Sigma_{L}^{r}(\omega)$. The impurity self-energy $\Sigma_{res}^{r}$ reads,
\begin{equation}
\Sigma_{res}^{r}(\omega)\,=\,\frac{|\tau|^{2}}{\omega\,-\,\varepsilon_{0}\,+\,i\eta},
\end{equation}
where $\tau$ is the coupling of the resonant level to the system.
With these expressions for the different self-energies, we may compute $G^{r}$,
\begin{equation}
G^{r}(\omega)\,=\,\left\{[g_{0}^{r}(\omega)]^{-1}\,-\, \Sigma_{L}^{r}(\omega)\,-\,\Sigma_{R}^{r}(\omega)\,-\,\Sigma_{res}^{r}(\omega)\,\right\}^{-1}\,.
\end{equation}
The normal conductance $G_{N}$ is calculated from the transmission as,
\begin{equation}
G_{N}=\frac{e^2}{h}\int T_{N}(\omega)\left( -\frac{d\,f}{d\,\omega}\right)\, d\,\omega\,
\end{equation}
where by construction we have already in a spinless channel. 
Since we are interested in low temperature physics, $f(\omega)\approx\Theta(\omega_{F}-\omega)$, and $df/d\omega\approx -\delta(\omega_{F}-\omega)$. Therefore,
\begin{equation}
\begin{split}
G_{N}&=\frac{e^{2}}{h}\int T_{N}(\omega)\delta(\omega_{F}-\omega)\, d\,\omega\,\\
G_{N}&=\frac{e^{2}}{h}\, T_{N}(\omega_{F})\,,
\end{split}
\end{equation} 
where $\omega_{F}$ is the Fermi energy which is the zero of energy in our calculations.

 \begin{figure*}[!ht]
\centering
\includegraphics[width=\textwidth,height=.4\textwidth]{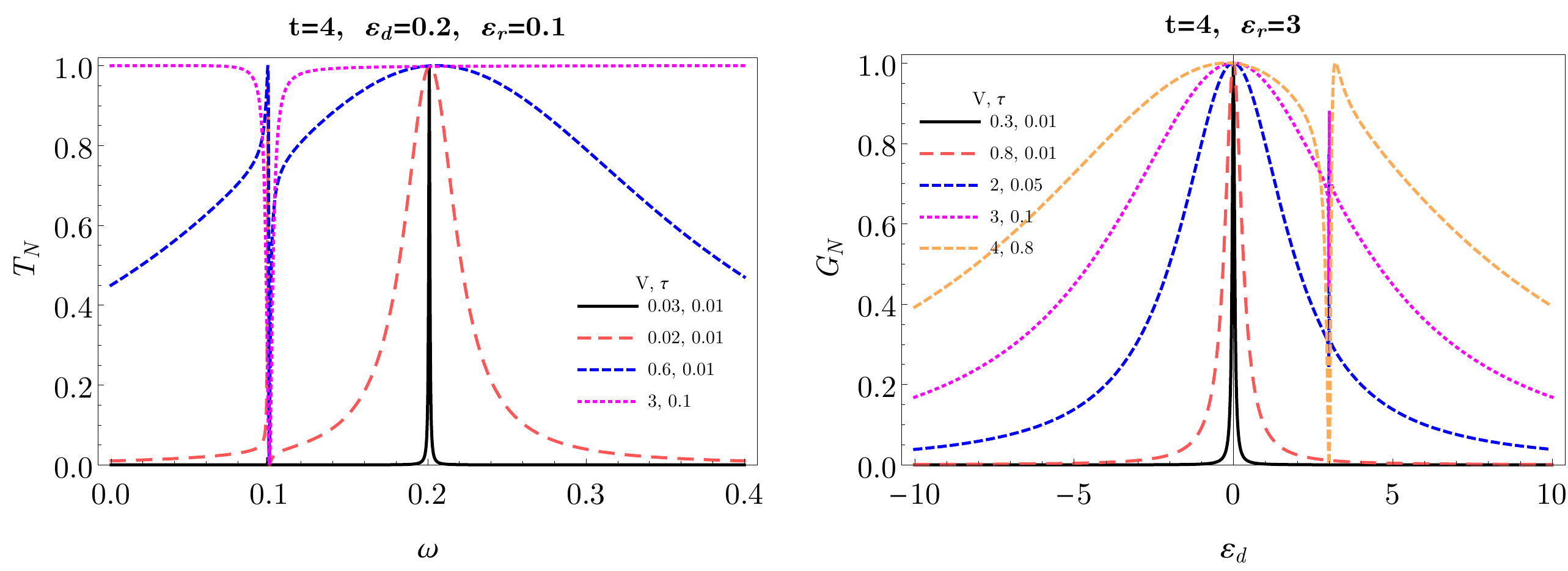} 
\caption[Normal transmission and conductance in a system of two semiinfinite tight-binding  chains]{(Color online) (left) Normal transmission  $T_{N}$ and (right) normal conductance $G_{N}$ across a central region attached to two semi-infinite tight-binding  chains (leads). We consider a central region consisting of one site with energy $\varepsilon_{d}$ (a quantum dot) that is weakly coupled to a resonant level $\varepsilon_{0}=\varepsilon_{d}-\varepsilon_{r}$, where $\varepsilon_{r}$ is a fixed parameter that represents the separation between the quantum dot level and the resonant level.
The plots show $T_{N}$ as a function of the energy $\omega$ and $G_{N}$ as a function of the energy of the quantum dot $\varepsilon_{d}$.
 The hopping among sites in the leads is fixed and strong. By controlling the coupling to the leads $V$ and the one to the resonant level $\tau$ one observes that the normal transmission exhibit a resonant peak at the energy of the quantum dot for weakly coupling, however, by making the coupling to the leads stronger and leaving weak the one to the resonant level, $T_{N}$ develops a dip at the energy of the resonant level.
Likewise, one observes that the normal conductance exhibit a resonant peak when $\varepsilon_{d}=0$, that is the Fermi energy of the leads $\omega_{F}=0$, for weakly couplings, however, by making the coupling to the leads stronger and leaving weak the one to the resonant level, $G_{N}$ develops a dip at the energy of the resonant level.
}\label{fig:GNTN}
\end{figure*}
The aim of this part was to construct an effective model that contains the physics of our numerics, where the transmission (and conductance) develops a resonance in the trivial phase and a dip in the helical phase.
Indeed, by plugging previous equations in the expression for the transmission and conductance, one ends up with the desired result that is plotted in Fig.\,\ref{fig:GNTN}.

In such plots, we consider a strong hopping $t$ between sites in the leads in comparison to the couplings $V$ and $\tau$. 
For weak coupling between leads and the central region a resonant tunnelling peak is obtained at the energy of the central region $\omega=\varepsilon_{d}$. Upon increasing the coupling between the leads and the central region $V$ the resonant peak at $\varepsilon_{d}$ becomes broader and a sharp Fano feature emerges at the resonant impurity $\omega=\varepsilon_{r}$. The new feature has the typical Fano structure of a zero followed by a peak, and arises from the interference of the two possible paths for the carriers, through the very broadened (strongly coupled) site at $\varepsilon_{d}$, and through the weakly coupled resonant level at $\varepsilon_{r}$. For strong enough coupling $V$, the $\varepsilon_{d}$ contributes with a uniform $e^2/h$ background to conductance, while the Fano feature becomes a pure dip to zero.

In conclusion, we have developed an effective model that contains the physics involved in our numerics where a resonance peak is present at the energy of the quantum dot for weakly coupled system. By increasing the coupling of the quantum dot to the leads a Fano feature (dip to zero followed by a peak) appears in conductance at the energy of the resonant level.  
\section{Majorana localization length}
\label{Majorana-length}
The calculation of the Majorana localization length $\ell_{M}$ is carried out by solving the polynomial equation for the wave vector $k$ 
\begin{equation} 
\label{ksol}
k^{2}+4(\mu+C\alpha_{R}^{2})Ck^{2}+8\lambda C^{2}\Delta \alpha_{R} k+4C_{0}C^{2}=0\,,
\end{equation}
where $C=m/\hbar^{2}$ and $C_{0}=\mu^{2}+\Delta^{2}-B^{2}$. Previous equation was derived in Ref. \cite{Lutchyn:PRL10}. We numerically solve Eq.\,(\ref{ksol}), and then we look for real solutions $\{k_{sol}\}$. What we define as the Majorana localisation length is
\begin{equation}
\ell_{M}={\rm Max}\Bigg(\frac{-1}{k_{sol}}\Bigg)\,.
\end{equation}

\begin{figure}[!ht]
  \label{locMaj}
\centering
\includegraphics[width=.6\textwidth]{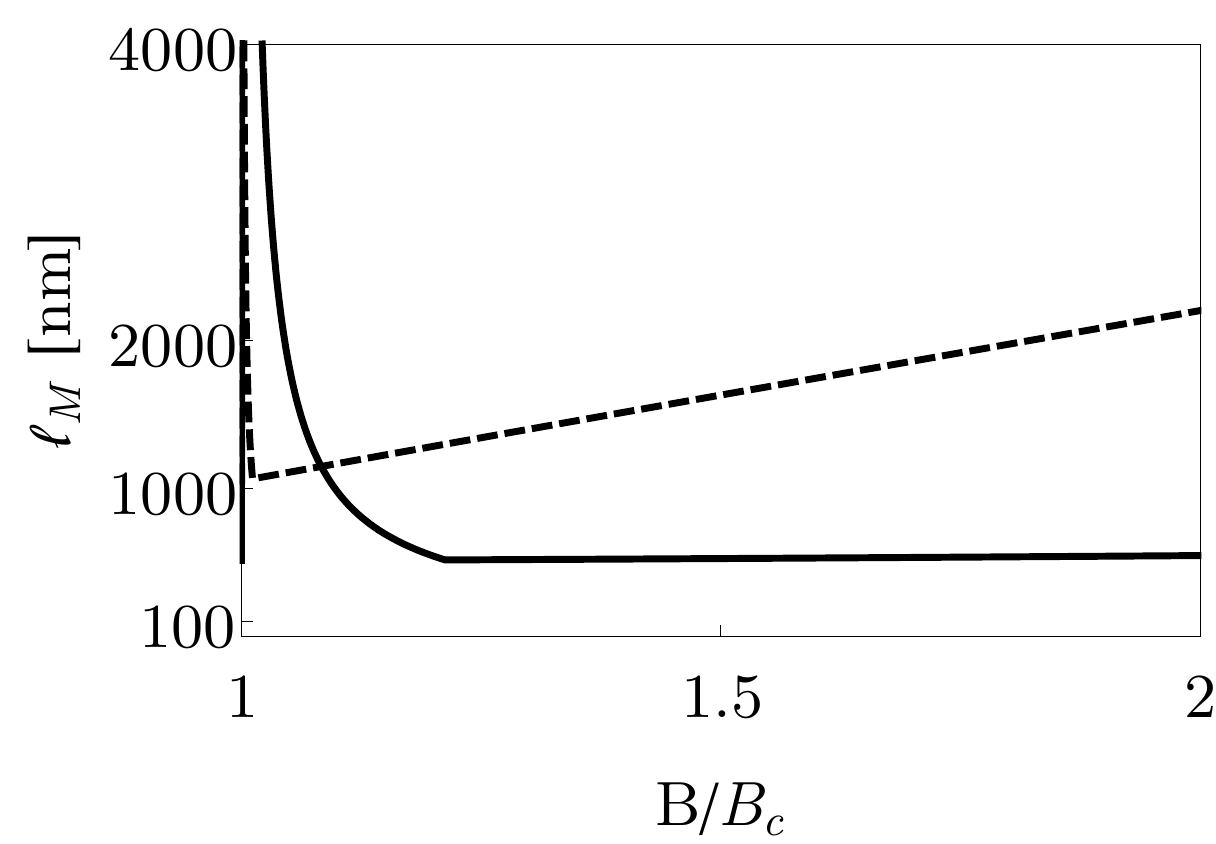} 
\caption[Majorana localization length $\ell_{M}$ as a function of the Zeeman field $B$]{Majorana localization length $\ell_{M}$ as a function of the Zeeman field $B$ for $\alpha_{R}=\alpha_{0}$\, (dashed curve) and $\alpha_{R}=5\alpha_{0}$ (solid curve), where $\alpha_{0}=0.2\mathrm{eV \AA}$. They correspond, to spin-orbit lengths $l_{SO}\approx 200$nm and $l_{SO}\approx 40$nm, respectively. Rest of parameters $\mu=0.5$\,meV, and $\Delta=0.25$\,meV.}
\label{fig3App}
\end{figure}

 Indeed, in Fig.\,\ref{fig3App} one observes that $\ell_{M}$ linearly increases as one increases $B$ for realistic SOC (dashed line), while  it acquires smaller values and remains roughly constant for stronger SOC (solid curve). 

\chapter{\bf Appendix for Chapter \ref{ChapEPs}}
\label{AppChapEPs} 
\lhead{Appendix \ref{AppChapEPs}. \emph{For Chapter 5}}
In this supplemental material we discuss the robustness of the exceptional-point Majorana bound states presented in the main text against finite length and interaction effects.
\section{Methods}
Transport across the NS junction is computed using the nanowire model for a Rashba wire,
\begin{eqnarray} \label{model}
H_\textrm{S}&=&(2t-\mu_S)\sum_{\sigma n}c_{\sigma n}^\dagger c^{\phantom{\dagger}}_{\sigma n}+\sum_{\sigma n}\Delta\, c_{\sigma n}^\dagger c^\dagger_{\bar\sigma n}+\mathrm{H.c} \\
&&-\sum_{\sigma, \langle n,n'\rangle}t\, c_{\sigma n'}^\dagger c^{\phantom{\dagger}}_{\sigma n} -i\sum_{\sigma,\sigma' \langle n,n'\rangle}t^{\mathrm{SO}\phantom{\dagger}}_{n'-n}c_{\sigma' n'}^\dagger \sigma^{y}_{\sigma'\sigma}c^{\phantom{\dagger}}_{\sigma n}\nonumber\\
&&+\sum_{\sigma,\sigma' n}B\,c_{\sigma' n}^\dagger \sigma^{x}_{\sigma'\sigma}c^{\phantom{\dagger}}_{\sigma n}\nonumber
\end{eqnarray} 
with the non-proximized normal section (N) modelled by the same Hamiltonian, albeit with $\Delta=0$ and a $\mu_N$ in place of $\mu_S$. The normal contact transmission $T^{(n)}_N$ of each incoming mode is computed by also setting  $\Delta=0$  on the proximised (S) side, and using the standard Green's function scheme.  One first splits the system into a left lead (with $\mu_N$), a right lead (with $\mu_S$), and a central section (the interface with a non-uniform profile $\mu(x)$ that transitions from $\mu_N$ into $\mu_S$) coupled to the leads through operators $V_{N/S}$. The total conductance $\mathcal{G}$ of the $M$ incoming modes is then given by Caroli's formula \cite{Caroli:JPCSSP71}
\begin{equation}
\mathcal{G}=\sum_n^M T_N^{(n)}=4\mathcal{G}_0\mathrm{Tr}\left[\Gamma_N G \Gamma_S G^\dagger\right]
\end{equation}
where $\mathcal{G}_0=e^2/h$, $G$ is the dressed retarded Green's function of the central region, $\Gamma_{N/S}=(\Sigma_{N/S}+\Sigma_{N/S}^\dagger)/2$ is the decay operator into the left/right leads, $\Sigma_{N/S}=V_{N/S}^\dagger g_{N/S}V_{N/S}$ is the corresponding self energies, and $g_{N/S}$ is the surface Green's function of the decoupled leads.

The poles of the scattering matrix presented in the main text are given, close to the origin of the complex plane, by the eigenvalues of non-Hermitian Hamiltonian $H_S+\Sigma(\omega=0)$, where $H_S$ is the (Hermitian) Hamiltonian of a sufficiently long segment of the wire containing the junction, and $\Sigma$ is the self-energy from the remaining wire (the reservoir), which is computed numerically.

The average normal transmission per mode is defined as $T_N=\mathcal{G}/(M\mathcal{G}_0)$. The values given in the main text were computed for Zeeman $B=0$. $T_N$ depends on the detailed spatial interpolation profile $\mu(x)$ across the interface. An abrupt interface has a smaller transmission than a smooth one, due to the mismatch in Fermi velocity between the two sides. In a real sample, the smoothness of such depletion profile is controlled by geometric parameters of the gating used to deplete the normal side (typically the superconducting side will be difficult to deplete due to screening by the parent superconductor). $T_N$ can also be controlled in a real device by adding a pinch-off gate close to the contact. This possibility is modelled by suppressing a single hopping term $t$ precisely at the contact, where $\Delta(x)$ abruptly jumps from zero to $\Delta$. The combination of mismatch and pinch-off allows to sweep $T_N$ from zero to one.


\chapter{\bf Appendix for Chapter \ref{ChapDensity}}
\label{AppChapDensity} 
\lhead{Appendix \ref{AppChapDensity}. \emph{For Chapter 6}}
In this supplementary material we show how the density-density response function for Rashba nanowires was derived. Then, we describe how the Bogoliubov transformation in a superconducting system was carried out in order to calculate the density-density response function, which is also shown here when the interband pairing is set to zero.
\section{Derivation of the density response in Rashba nanowires}
We introduce the equations for the electron density operator, given by Eqs.\,(\ref{densityoperators}), into the average of the commutator given by Eq.\,(\ref{shortdefnn}), and get
  \begin{equation}
  \label{eqsus}
  \begin{split}
 \chi(1,1')&=-\frac{i}{\hbar}\mathop{\sum_{k_{1,\cdots,4}}}_{\sigma_{1,\cdots,4}}\big<[ 
 \psi^{\dagger}_{k_{1}\sigma_{1}}(r) \psi_{k_{2}\sigma_{2}}(r)c^{\dagger}_{k_{1},\sigma_{1}}(t)c_{k_{2},\sigma_{2}}(t),\psi^{\dagger}_{k_{3}\sigma_{3}}(r') \psi_{k_{4}\sigma_{4}}(r')c^{\dagger}_{k_{3},\sigma_{3}}(t')c_{k_{4},\sigma_{4}}(t')]\big>\,,\\
 &=-\frac{i}{\hbar} \mathop{\sum_{k_{1,\cdots,4}}}_{\sigma_{1,\cdots,4}}
 \psi^{\dagger}_{k_{1}\sigma_{1}}(r) \psi_{k_{2}\sigma_{2}}(r)\psi^{\dagger}_{k_{3}\sigma_{3}}(r') \psi_{k_{4}\sigma_{4}}(r'){\rm e}^{i[(\varepsilon_{k_{1}\sigma_{1}}-\varepsilon_{k_{2}\sigma_{2}})t+(\varepsilon_{k_{3}\sigma_{3}}-\varepsilon_{k_{4}\sigma_{4}})t']/\hbar}\times\\
 &\quad\quad\big<[c^{\dagger}_{k_{1},\sigma_{1}}c_{k_{2},\sigma_{2}},c^{\dagger}_{k_{3},\sigma_{3}}c_{k_{4},\sigma_{4}}]\big>
  \end{split}
 \end{equation}
 
Previous equation is rewritten introducing the expressions for the wave functions $\psi_{k,\sigma}$ from Eq.\,(\ref{eq2}),
 \begin{equation}
 \label{eqq22}
 \begin{split}
 \chi(1,1')&=-\frac{i}{\hbar}\frac{1}{4L^{2}}\mathop{\sum_{k_{1,\cdots,4}}}_{\sigma_{1,\cdots,4}}\,{\rm e}^{-i(k_{1}-k_{2})r}\,{\rm e}^{-i(k_{3}-k_{4})r'}\,{\rm e}^{i[(\varepsilon_{k_{1}\sigma_{1}}-\varepsilon_{k_{2}\sigma_{2}})t+(\varepsilon_{k_{3}\sigma_{3}}-\varepsilon_{k_{4}\sigma_{4}})t']/\hbar}\\
  &\quad\quad\times\big<[c^{\dagger}_{k_{1},\sigma_{1}}c_{k_{2},\sigma_{2}},c^{\dagger}_{k_{3},\sigma_{3}}c_{k_{4},\sigma_{4}}]\big>\eta^{\dagger}_{k_{1}\sigma_{1}}
 \eta_{k_{2}\sigma_{2}}\eta^{\dagger}_{k_{3}\sigma_{3}}\eta_{k_{4}\sigma_{4}}\,.
 \end{split}
 \end{equation}
 Now, we work out the average of the commutator 
 \begin{equation}
 \label{commuta}
 \begin{split}
 \big<[c^{\dagger}_{k_{1},\sigma_{1}}c_{k_{2},\sigma_{2}},c^{\dagger}_{k_{3},\sigma_{3}}c_{k_{4},\sigma_{4}}]\big>&=\big< c^{\dagger}_{k_{1},\sigma_{1}}c_{k_{4},\sigma_{4}}\big>\big<c_{k_{2},\sigma_{2}}c^{\dagger}_{k_{3},\sigma_{3}} \big>-\big< c^{\dagger}_{k_{3},\sigma_{3}}c_{k_{2},\sigma_{2}}\big>\big<c_{k_{4},\sigma_{4}}c^{\dagger}_{k_{1},\sigma_{1}} \big>\\
 &=\big< c^{\dagger}_{k_{1},\sigma_{1}}c_{k_{4},\sigma_{4}}\big>\delta_{k_{2},k_{3}}\delta_{\sigma_{2},\sigma_{3}}-\big< c^{\dagger}_{k_{3},\sigma_{3}}c_{k_{2},\sigma_{2}}\big>\delta_{k_{4},k_{1}}\delta_{\sigma_{4},\sigma_{1}}\\
 &=n_{k_{1}\sigma_{1}}\delta_{k_{1},k_{4}}\delta_{\sigma_{1},\sigma_{4}}\delta_{k_{2},k_{3}}\delta_{\sigma_{2},\sigma_{3}}-n_{k_{2}\sigma_{2}}\delta_{k_{3},k_{2}}\delta_{\sigma_{3},\sigma_{2}}\delta_{k_{4},k_{1}}\delta_{\sigma_{4},\sigma_{1}}\\
 &=(n_{k_{1}\sigma_{1}}-n_{k_{2}\sigma_{2}})\delta_{k_{1},k_{4}}\delta_{\sigma_{1},\sigma_{4}}\delta_{k_{2},k_{3}}\delta_{\sigma_{2},\sigma_{3}}\,,
 \end{split}
 \end{equation}
 where $n_{k\sigma}=\frac{1}{{\rm e}^{\varepsilon_{k\sigma}/k_{B}T}+1}$ is the Fermi distribution function, $k_{B}$ the Boltzman constant and $T$ the system temperature.
Then, we insert Eq.\,(\ref{commuta}) expression into Eq.\,(\ref{eqq22}), and obtain
  \begin{equation}
  \label{eqsus1}
  \begin{split}
 \chi(1,1')
 &=-\frac{i}{\hbar}\frac{1}{L^{2}}\mathop{\sum_{k_{1},k_{2}}}_{\sigma_{1},\sigma_{2}}\,{\rm e}^{-i(k_{1}-k_{2})(r-r')}\,{\rm e}^{i[(\varepsilon_{k_{1}\sigma_{1}}-\varepsilon_{k_{2}\sigma_{2}})(t-t')]/\hbar}(n_{k_{1}\sigma_{1}}-n_{k_{2}\sigma_{2}})\times\\
& \times\eta^{\dagger}_{k_{1}\sigma_{1}}\eta_{k_{2}\sigma_{2}}\eta^{\dagger}_{k_{2}\sigma_{2}}\eta_{k_{1}\sigma_{1}}
 \,.
  \end{split}
 \end{equation}
 The term at the end of previous expression arises from by combined action between the spin-orbit coupling and Zeeman interaction, which can be written from the definition for $\eta_{k\sigma}$ given by Eq.\,(\ref{eq2}), as
 \begin{equation}
 \eta^{\dagger}_{k_{1}\sigma_{1}}\eta_{k_{2}\sigma_{2}}\eta^{\dagger}_{k_{2}\sigma_{2}}\eta_{k_{1}\sigma_{1}}=\frac{1}{2}\bigg[1+\sigma_{1}\sigma_{2}\frac{B^{2}+\alpha_{R}^{2}k_{1}k_{2}}{|v_{k_{1}}||v_{k_{2}}|}\bigg]
 \end{equation}
 The next step is integration over time $t-t'$ and over position $r-r'$, of Eq.\,(\ref{eqsus1}) according to Eq.\,(\ref{densityresponsefunction})
Hence, we obtain
 \begin{equation}
 \label{chiomega}
 \chi(\omega,q)=\frac{1}{2L}\sum_{k\sigma,\sigma'}\bigg[1+\frac{\sigma\sigma'(B^{2}+\alpha_{R}^{2}k(k+q))}{|v_{k}||v_{k+q}|} \bigg]\frac{n_{k\sigma}-n_{k+q\sigma'}}{\hbar\omega+\varepsilon_{k\sigma}-\varepsilon_{k+q\sigma'}+i\hbar\eta}\,,
 \end{equation}
where we have relabeled $k_{1}\rightarrow k$, $\sigma_{1}\rightarrow \sigma$, $\sigma_{2}\rightarrow \sigma'$. The imaginary part of Eq.\,(\ref{chiomega}) gives the structure factor, and notice that we have obtained the same as the one calculated in 
\cite{PhysRevB.89.035131}, which provides a validity check for our calculations.
 
In the static limit, for $\omega=0$, the Lindhard function (or density-density response function) $\chi(\omega,q)$ calculate before is purely real and is given by
 \begin{equation}
   \label{chinnstatic}
 \begin{split}
  \chi(q)&=\frac{1}{2}\frac{1}{2\pi}\sum_{\sigma\sigma'}\int d k\bigg[1+\sigma\sigma'\frac{B^{2}+\alpha_{R}^{2}k(k+q)}{|v_{k}||v_{k+q}|} \bigg]\frac{n_{k,\sigma}-n_{k+q,\sigma'}}{\varepsilon_{k,\sigma}-\varepsilon_{k+q,\sigma'}}\,.
 \end{split}
 \end{equation}
 Notice that the principal part prescription required by the infinitesimal $\eta$ in the denominator of previous equation 
 is not needed here, since the denominator can only vanish simultaneously with the numerator. 
 In the limit $L\rightarrow\infty$, we have replaced the discrete sum over momentum by an integral 
 \begin{equation}
\frac{1}{L^{d}}\sum_{k}(\cdots)\rightarrow \frac{1}{(2\pi)^{d}}\int d{\bf k}(\cdots)\,,
 \end{equation}
 where $d=1,2,3$ is the dimension. 
 
For higher dimensions one needs to calculate the electronic wave functions of the problem with Rashba SO and Zeeman in the desirable dimension and repite all what the steps we have done. 

At zero temperature $T=0$, the Fermi distribution function $n_{k,\sigma}$ becomes a step function
\begin{equation}
n_{k,\sigma}=\lim_{T\rightarrow0}\frac{1}{{\rm e}^{\varepsilon_{k,\sigma}/k_{B}T}+1}=\theta(-\varepsilon_{k,\sigma})=\begin{cases}
    1\,,  &-\varepsilon_{k,\sigma}>0 , \\
    0\,, &-\varepsilon_{k,\sigma}<0\,,
\end{cases}
\end{equation}
where
\begin{equation}
\label{energiess}
\varepsilon_{k,\sigma}=\frac{\hbar^{2}k^{2}}{2m}-\mu+\sigma\sqrt{B^{2}+(\alpha_{R}k)^{2}}\,. 
\end{equation}
Notice that at zero temperature the chemical potential is equal to the Fermi energy $E_{F}=\mu$.
The function $\theta(-\varepsilon_{k,\sigma})$ restricts the $k$ integration to interval $[-k_{F,\sigma},+k_{F,\sigma}]$, with $k_{F}$ defined by Eq.\,(\ref{energiess}) $\varepsilon_{k_{F},\sigma}=0$
\begin{equation}
\pm k_{F,\sigma}=\pm\sqrt{k_{\mu}^{2}+2k_{so}^{2}-\sigma\sqrt{(k_{\mu}^{2}+2k_{so}^{2})^{2}-k_{\mu}^{4}+k_{z}^{4}}}\,,
\end{equation}
where $k_{\mu}=\sqrt{2m\mu/\hbar^{2}}$, $k_{so}=m\alpha_{R}/\hbar^{2}$ and $k_{Z}=\sqrt{2mB/\hbar^{2}}$.
 From Eq.\,(\ref{chinnstatic}), the full density-density response can be expressed as a sum of four terms 
 \begin{equation}
  \chi(q)=\chi^{++}(q)+\chi^{-+}(q)+\chi^{+-}(q)+\chi^{--}(q)=\sum_{\sigma\sigma'}   \chi^{\sigma\sigma'}(q)
  \end{equation}
  where
   \begin{equation}
 \chi^{\sigma\sigma'}(q)=\frac{1}{2}\frac{1}{2\pi}\int d k\,g^{\sigma\sigma'}_{k,q}(B,\alpha_{R})\,\frac{n_{k,\sigma}-n_{k+q,\sigma'}}{\varepsilon_{k,\sigma}-\varepsilon_{k+q,\sigma'}}\,,
 \end{equation} 
 and 
 \begin{equation}
 g^{\sigma\sigma'}_{k,q}(B,\alpha_{R})=\bigg[1+\sigma\sigma'\frac{B^{2}+\alpha_{R}^{2}k(k+q)}{|v_{k}||v_{k+q}|} \bigg]\,.
 \end{equation}
  For a simplification upon integration let us divide each of the $\chi^{\sigma\sigma'}$ into two integrals as follows
  \begin{equation}
   \chi^{\sigma\sigma'}(q)=\frac{1}{2}\frac{1}{2\pi}\int dk\,g^{\sigma\sigma'}_{k,q}(B,\alpha_{R})\,\frac{n_{k,\sigma}}{\varepsilon_{k,\sigma}-\varepsilon_{k+q,\sigma'}}\,
   -
   \frac{1}{2}\frac{1}{2\pi}\int dk\,g^{\sigma\sigma'}_{k,q}(B,\alpha_{R})\,\frac{n_{k+q,\sigma'}}{\varepsilon_{k,\sigma}-\varepsilon_{k+q,\sigma'}}\,.
  \end{equation}
  We know that at zero temperature, the distribution functions become step functions and thus restrict the integration limits,
   \begin{equation}
   \begin{split}
   \chi^{\sigma\sigma'}(q)&=\frac{1}{2}\frac{1}{2\pi}\int dk\,g^{\sigma\sigma'}_{k,q}(B,\alpha_{R})\,\frac{\theta(k_{F,\sigma}-|k|)}{\varepsilon_{k,\sigma}-\varepsilon_{k+q,\sigma'}}\,
   -
   \frac{1}{2}\frac{1}{2\pi}\int d k\,g^{\sigma\sigma'}_{k,q}(B,\alpha_{R})\,\frac{\theta(k_{F,\sigma'}-|k+q|)}{\varepsilon_{k,\sigma}-\varepsilon_{k+q,\sigma'}}\,,\\
   &=\frac{1}{2}\frac{1}{2\pi}\int_{-k_{F,\sigma}}^{{k_{F,\sigma}}} dk\,g^{\sigma\sigma'}_{k,q}(B,\alpha_{R})\,\frac{1}{\varepsilon_{k,\sigma}-\varepsilon_{k+q,\sigma'}}\,
   -
   \frac{1}{2}\frac{1}{2\pi}\int_{-k_{F,\sigma'}-q}^{k_{F,\sigma'}-q} d k\,g^{\sigma\sigma'}_{k,q}(B,\alpha_{R})\,\frac{1}{\varepsilon_{k,\sigma}-\varepsilon_{k+q,\sigma'}}\,,\\
&=  \chi^{\sigma\sigma',(1)}(q) - \chi^{\sigma\sigma',(2)}(q) \,.
   \end{split}
  \end{equation}
To conclude, in this part we have shown how to derive the density-density response function in one-dimensional nanowires with SOC and Zeeman interaction.
\section{Derivation of the density response in superconducting Rashba nanowires}
\label{appdensity}
In this part we derive the density-density response function in superconducting nanowires with SOC and Zeeman interaction.
First, we show how it is found the Bogoliubov transformation as it determines the inclusion of superconducting correlations into the nanowires. Then, we find the density-density response function for a superconducting 1D nanowire when the interband pairing is set to zero, and then in a regime of strong Zeeman field, where only the lower band $\sigma=-$ is taken into account.

\subsection{Derivation of the Bogoliubov transformation at zero interband pairing}
\label{BOGOGOs}
In this part we show how to derive the Bogoliubov transformation when superconducting correlations are induced into the nanowire. 
This transformation consists on finding an unitary operator that diagonalises the system's Hamiltonian.  
The Hamiltonian for our problem in the helical basis described in Secs.\,\ref{Rashbawire} and \ref{appRashba}, see Eq.\,(\ref{basis1}) for instance, is
\begin{equation}
\label{Happ1}
H=\begin{pmatrix}
 \varepsilon_{+}(k)&0&\Delta_{++}(k)&\Delta_{+-}(k)\\
 0&\varepsilon_{-}(k)&-\Delta_{+-}(k)&\Delta_{--}(k)\\
 \Delta_{++}^{\dagger}(k)&-\Delta_{+-}^{\dagger}(k)&-\varepsilon_{+}(-k)&0\\
 \Delta_{+-}^{\dagger}(k)&\Delta_{--}^{\dagger}(k)&0&-\varepsilon_{-}(-k)
 \end{pmatrix}\,,
\end{equation}
where the pairing potentials involved in the problem are
\begin{equation}
\begin{split}
\Delta_{--}(k)&=\frac{i\alpha k \Delta}{\sqrt{B^{2}+\alpha^{2}k^{2}}}\equiv i\Delta_{p}\,,\\
\Delta_{++}(k)&=\frac{-i\alpha k \Delta}{\sqrt{B^{2}+\alpha^{2}k^{2}}}\equiv -i\Delta_{p}\,,\\
\Delta_{+-}(k)&=\frac{B \Delta}{\sqrt{B^{2}+\alpha^{2}k^{2}}}\equiv \Delta_{s}\,, 
\end{split}
\end{equation}
and
\begin{equation}
\Delta_{p}=\frac{\alpha k\Delta}{\sqrt{B^{2}+\alpha^{2}k^{2}}}\,,\quad
\Delta_{s}=\frac{\Delta B}{\sqrt{B^{2}+\alpha^{2}k^{2}}}\,.
\end{equation}
denotes the $p$-wave and $s$-wave nature of the superconducting pairing, respectively.

Notice that the intraband pairing $\Delta_{++}(\Delta_{--})$ contains a phase, which is represented by the imaginary $i$
\begin{equation}
\Delta_{++}(k)\equiv -i\Delta_{p}={\rm e}^{-\frac{\pi}{2}i}\Delta_{p}\,,\quad
\Delta_{--}(k)\equiv i\Delta_{p}={\rm e}^{\frac{\pi}{2}i}\Delta_{p}
\end{equation}
This phase can be gauge away by considering the eigenvalue problem $H\Psi=E\Psi$, where, $T$ being the transpose operation,
\begin{equation}
\Psi=\begin{pmatrix}
u_{1}\,{\rm e}^{-\frac{\pi}{4}i},
u_{2}\,{\rm e}^{\frac{\pi}{4}i},
v_{1}\,{\rm e}^{\frac{\pi}{4}i},
v_{2}\,{\rm e}^{-\frac{\pi}{4}i}
\end{pmatrix}^{T}\,,
\end{equation}
then
\begin{equation}
\begin{split}
(\varepsilon_{+}-E)u_{1}\,{\rm e}^{-\frac{\pi}{4}i}+{\rm e}^{-\frac{\pi}{2}i}\Delta_{p}v_{1}\,{\rm e}^{\frac{\pi}{4}i}+\Delta_{s}v_{2}\,{\rm e}^{-\frac{\pi}{4}i}&=0\,,\\
(\varepsilon_{-}-E)u_{2}\,{\rm e}^{\frac{\pi}{4}i}-\Delta_{s}v_{1}\,{\rm e}^{\frac{\pi}{4}i}+{\rm e}^{\frac{\pi}{2}i}\Delta_{p}v_{2}\,{\rm e}^{-\frac{\pi}{4}i}&=0\,,\\
{\rm e}^{\frac{\pi}{2}i}\Delta_{p}u_{1}\,{\rm e}^{-\frac{\pi}{4}i}-\Delta_{s}u_{2}\,{\rm e}^{\frac{\pi}{4}i}-(\varepsilon_{+}+E)v_{1}\,{\rm e}^{\frac{\pi}{4}i}&=0\,\\
\Delta_{s}u_{1}\,{\rm e}^{-\frac{\pi}{4}i}+{\rm e}^{-\frac{\pi}{2}i}\Delta_{p}u_{2}\,{\rm e}^{\frac{\pi}{4}i}-(\varepsilon_{-}+E)v_{2}\,{\rm e}^{-\frac{\pi}{4}i}&=0\,.
\end{split}
\end{equation}
Therefore, previous equations can be written as
\begin{equation}
\label{systemofEqs}
\begin{split}
(\varepsilon_{+}-E)u_{1}+\Delta_{p}v_{1}+\Delta_{s}v_{2}&=0\,,\\
(\varepsilon_{-}-E)u_{2}-\Delta_{s}v_{1}+\Delta_{p}v_{2}&=0\,,\\
\Delta_{p}u_{1}-\Delta_{s}u_{2}-(\varepsilon_{+}+E)v_{1}&=0\,\\
\Delta_{s}u_{1}+\Delta_{p}u_{2}-(\varepsilon_{-}+E)v_{2}&=0\,.
\end{split}
\end{equation}
Thus, we have gauged away the phase in the intraband pairing potentials. Notice that the dispersion relation we have derived in previous section is not affected by this gauge transformation. We point out here that  $\Delta_{++}$ and $\Delta_{--}$ represent the pairing potentials of two $p$-wave superconductors, while $\Delta_{+-}$ acts as a weak coupling between them, see Sec.\,(\ref{Rashbawire}).

The situation when there is no pairing between states of different band is analysed here
and described by setting $\Delta_{s}=0$. This case resemble a regime with two independent p-wave superconductors described by the two $p$-wave pairing potentials $\Delta_{++}$ and $\Delta_{--}$. 
The eigenvalues in this case are calculated by diagonalising the Hamiltonian given by Eq.\,(\ref{Happ1})
\begin{equation}
E_{\pm}^{2}(k)=\Delta_{p}^{2}(k)+\varepsilon_{\pm}^{2}(k)\,,
\end{equation}
Then, Eqs.\,\ref{systemofEqs} are written as
\begin{equation}
\label{systemofEqs2}
\begin{split}
(\varepsilon_{+}-E)u_{1}+\Delta_{p}v_{1}&=0\,,\\
(\varepsilon_{-}-E)u_{2}+\Delta_{p}v_{2}&=0\,,\\
\Delta_{p}u_{1}-(\varepsilon_{+}+E)v_{1}&=0\,\\
\Delta_{p}u_{2}-(\varepsilon_{-}+E)v_{2}&=0\,,
\end{split}
\end{equation}
where the first and third equations form the Hamiltonian for sector $+$, while second and fourth for sector $-$.
Previous four equations reduce to only two equations
\begin{equation}
\label{uvuv}
u_{1}=-\frac{\Delta_{p}}{\varepsilon_{+}-E}v_{1}\,,\quad
u_{2}=-\frac{\Delta_{p}}{\varepsilon_{-}-E}v_{2}\,,
\end{equation}
which can be solved separately, since the two sectors are completely decoupled. Remain that the coupling between the sectors $+$ and $-$ is described 
by $\Delta_{s}$, whose value is set to zero for simplicity.

\subsubsection*{Sector 1: $\sigma=+$}
From Eqs.\,(\ref{uvuv}), we have
\begin{equation}
\label{eq1s}
\begin{split}
u_{1}&=-\frac{\Delta_{p}}{\varepsilon_{+}-E}v_{1}=\frac{\varepsilon_{+}+E}{\Delta_{p}}v_{1}\,,
\end{split}
\end{equation}
where the energy relation is given by
\begin{equation}
\label{energq1}
E=\pm E_{+}(k)=\pm\sqrt{\Delta_{p}^{2}(k)+\varepsilon_{+}^{2}(k)}\,.
\end{equation}
Moreover, we have the normalization condition
\begin{equation}
u_{1}^{2}+v_{1}^{2}=1\,.
\end{equation}
Then, we insert previous equation into Eq.\,(\ref{eq1s})
\begin{equation}
\frac{(\varepsilon_{+}+E)^{2}}{\Delta_{p}^{2}}v_{1}^{2}+v_{1}^{2}=1\,\Rightarrow\,v_{1}^{2}=\frac{\Delta^{2}_{p}}{\Delta^{2}_{p}+(\varepsilon_{+}+E)^{2}}\,,
\end{equation}
thus
\begin{equation}
\label{eqv1p}
v_{1}=\frac{|\Delta_{p}|}{\sqrt{\Delta^{2}_{p}+(\varepsilon_{+}+E)^{2}}}\,,\quad u_{1}=\frac{\varepsilon_{+}+E}{\sqrt{\Delta^{2}_{p}+(\varepsilon_{+}+E)^{2}}}\,{\rm sgn}(\Delta_{p})\,.
\end{equation}
We can completely neglect the ${\rm sgn}$ function, but for completeness we keep it for now. However, we will show that it won't alter our results.
Now, we need to evaluate previous coherence factors for each energy given by Eq.\,(\ref{energq1}).\\\\
\newline
{\bf (a): For $E=E_{+}=\sqrt{\Delta_{p}^{2}+\varepsilon_{+}^{2}}$}\\\\
From Eq.\,(\ref{eqv1p}), we have for $v_{1}$ and $u_{1}$
\begin{equation}
v_{1}=\frac{1}{\sqrt{2}}\sqrt{1-\frac{\varepsilon_{+}}{E_{+}}}\,,\quad u_{1}=\frac{{\rm sgn}(\Delta_{p})}{\sqrt{2}}\sqrt{1+\frac{\varepsilon_{+}}{E_{+}}}\,.
\end{equation}
Now, we denote previous results as follows
\begin{equation}
\label{defvu}
u_{1}\equiv u_{+}\,{\rm sgn}(\Delta_{p})\,,\quad\quad v_{1}\equiv v_{+}\,.
\end{equation}
The associated vector is therefore
\begin{equation}
\label{u1ma}
\psi_{E_{+}}=
  \begin{pmatrix}
  u_{+}{\rm sgn}(\Delta_{p})\\
  v_{+}
 \end{pmatrix}\,.
\end{equation}
\\\\
\newline
{\bf (b): For $E=-E_{+}=-\sqrt{\Delta_{p}^{2}+\varepsilon_{+}^{2}}$}\\\\
From Eq.\,(\ref{eqv1p}) we have for $v_{1}$ and $u_{1}$,
\begin{equation}
v_{1}=\frac{1}{\sqrt{2}}\sqrt{1+\frac{\varepsilon_{+}}{E_{+}}}\,,\quad 
u_{1}=
\frac{-{\rm sgn}(\Delta_{p})}{\sqrt{2}}\sqrt{1-\frac{\varepsilon_{+}}{E_{+}}}\,.
\end{equation}
Now, in terms of previous definitions $u_{+}$ and $v_{+}$ given by Eqs.\,(\ref{defvu}) we can write 
\begin{equation}
u_{1}\equiv-v_{+} {\rm sgn}(\Delta_{p})\,,\quad\quad v_{1}\equiv u_{+}\,,
\end{equation}
Then, the associated vector is
\begin{equation}
\label{u1m}
\psi_{-E_{+}}=
  \begin{pmatrix}
 - v_{+}{\rm sgn}(\Delta_{p})\\
  u_{+}
 \end{pmatrix}\,.
\end{equation}
\\\\
\newline
{\bf The Bogoliubov transformation }\\\\
This transformation consists on finding an unitary operator that diagonalises the  Hamiltonian for this sector and formed by the first and third equations given by (\ref{systemofEqs2}).
Such unitary operator is a matrix and it is formed by the two vectors calculated previously. 
From Eqs.\,(\ref{u1m}) and (\ref{u1ma}), we neglect the ${\rm sgn}$ function, we have
\begin{equation}
U_{+}^{(1)}=
\begin{pmatrix}
\psi_{E_{+}},\psi_{-E_{+}}
\end{pmatrix}=
\begin{pmatrix}
u_{+} & -v_{+}\\
v_{+}& u_{+}
\end{pmatrix}
\end{equation}
This matrix relates the basis, represented by operators $c$, in which the Hamiltonian is not diagonal with a new one, represented by operators $\alpha$, where it is diagonal.
This new basis is the one of quasiparticles in a superconductor:
\begin{equation}
\begin{pmatrix}
c_{k,+}\\
c_{-k,+}^{\dagger}
\end{pmatrix}=
\begin{pmatrix}
u_{+} & -v_{+}\\
v_{+}& u_{+}
\end{pmatrix}
\begin{pmatrix}
\alpha_{k,+}\\
\alpha_{-k,+}^{\dagger}
\end{pmatrix}\,,
\end{equation}
therefore, the Bogolyubov transformation for operators that belong to the sector $+$ are given by
\begin{equation}
\begin{split}
c_{k,+}=u_{+}\alpha_{k,+}-v_{+}\alpha_{-k,+}^{\dagger}\,,\\
c_{-k,+}^{\dagger}=v_{+}\alpha_{k,+}+u_{+}\alpha_{-k,+}^{\dagger}\,.
\end{split}
\end{equation}
Moreover, the matrix $\Big[U_{+}^{(1)}\Big]^{T}$, $T$ denotes the transpose operation,
\begin{equation}
U_{+}^{(2)}\equiv \Big[U_{+}^{(1)}\Big]^{T}=
\begin{pmatrix}
u_{+} & v_{+}\\
-v_{+}& u_{+}
\end{pmatrix}
\end{equation}
 also diagonalises the Hamiltonian for the sector $+$, and therefore it can also apply for transforming the operators:
\begin{equation}
\begin{split}
c_{k,+}=u_{+}\alpha_{k,+}+v_{+}\alpha_{-k,+}^{\dagger}\,,\\
c_{-k,+}^{\dagger}=-v_{+}\alpha_{k,+}+u_{+}\alpha_{-k,+}^{\dagger}\,.
\end{split}
\end{equation}
\subsubsection*{Sector 2:$\sigma=-$}
From Eqs.\,(\ref{uvuv}), we have
\begin{equation}
\label{esq2}
u_{2}=-\frac{\Delta_{p}}{\varepsilon_{-}-E}\,,\quad v_{2}=\frac{\varepsilon_{-}+E}{\Delta_{p}}v_{2}\,,
\end{equation}
where the energy relation is given by
\begin{equation}
\label{energq2}
E=\pm E_{-}(k)=\pm\sqrt{\Delta_{p}^{2}(k)+\varepsilon_{-}^{2}(k)}\,,
\end{equation}
and the normalization condition is given by
\begin{equation}
u_{2}^{2}+v_{2}^{2}=1\,.
\end{equation}
Then, we insert Eq.\,(\ref{esq2}) into previous equation and get
\begin{equation}
\frac{(\varepsilon_{-}+E)^{2}}{\Delta_{p}^{2}}v^{2}_{2}+v^{2}_{2}=1\,\Rightarrow\,v_{2}^{2}=\frac{\Delta_{p}^{2}}{(\varepsilon_{-}+E)^{2}+\Delta_{p}^{2}}\,.
\end{equation}
Thus,
\begin{equation}
v_{2}=\frac{|\Delta_{p}|}{\sqrt{\Delta_{p}^{2}+(\varepsilon_{-}+E)^{2}}}\,,\quad u_{2}=\frac{(\varepsilon_{-}+E){\rm sgn}(\Delta_{p})}{\sqrt{\Delta_{p}^{2}+(\varepsilon_{-}+E)^{2}}}\,.
\end{equation}
{\bf (a): For $E=E_{-}=\sqrt{\Delta_{p}^{2}+\varepsilon_{-}^{2}}$}\\\\
From Eq.\,(\ref{esq2}) for $v_{2}$ we have
\begin{equation}
v_{2}=\frac{1}{\sqrt{2}}\sqrt{1-\frac{\varepsilon_{-}}{E_{-}}}\,,\quad   u_{2}=\frac{{\rm sgn}(\Delta_{p})}{\sqrt{2}}\sqrt{1+\frac{\varepsilon_{-}}{E_{-}}}\,.
\end{equation}
Again, here we make the following notation
\begin{equation}
u_{2}\equiv u_{-}\,{\rm sgn}(\Delta_{p})\,,\quad\quad v_{2}\equiv v_{-}\,.
\end{equation}
Thus, the associated vector is given by
\begin{equation}
\label{minusv1a}
\psi_{E_{-}}=
\begin{pmatrix}
u_{-}{\rm sgn}(\Delta_{p})\\
v_{-}
\end{pmatrix}\,.
\end{equation}\\\\
{\bf (b): For $E=-E_{-}=-\sqrt{\Delta_{p}^{2}+\varepsilon_{-}^{2}}$}\\\\
From Eq.\,(\ref{esq2}) we have for $v_{2}$
\begin{equation}
v_{2}=\frac{1}{\sqrt{2}}\sqrt{1+\frac{\varepsilon_{-}}{E_{-}}}\,,\quad
u_{2}=-\frac{{\rm sgn}(\Delta_{p})}{\sqrt{2}}\sqrt{1-\frac{\varepsilon_{-}}{E_{-}}}\,.
\end{equation}
Therefore, these coherence factors in terms of previous definitions $u_{-}$ and $v_{-}$, we get
\begin{equation}
u_{2}\equiv -v_{-}\,{\rm sgn}(\Delta_{p})\,,\quad\quad v_{2}\equiv u_{-}\,.
\end{equation}
Thus, the associated vector is given by
\begin{equation}
\label{minusv1b}
\psi_{-E_{-}}=
\begin{pmatrix}
-v_{-}{\rm sgn}(\Delta_{p})\\
u_{-}
\end{pmatrix}\,.
\end{equation}
\\\\
\newline
{\bf The Bogoliubov transformation}\\\\
Here we construct the unitary operator that diagonalizes the Hamiltonian of this sector $-$. 
It is formed by the two vectors given by Eqs.\,(\ref{minusv1a}) and (\ref{minusv1b})calculated previously:
\begin{equation}
U_{-}^{(1)}=
\begin{pmatrix}
\psi_{E_{-}},\psi_{-E_{-}}
\end{pmatrix}=
\begin{pmatrix}
u_{-} & -v_{-}\\
v_{-}& u_{-}
\end{pmatrix}
\end{equation}
As before, this matrix relates the basis, represented by operators $c$, in which the Hamiltonian is not diagonal to a new one, represented by operators $\alpha$, where it is diagonal.
This new basis is the one of quasiparticles in a superconductor:
\begin{equation}
\begin{pmatrix}
c_{k,-}\\
c_{-k,-}^{\dagger}
\end{pmatrix}=
\begin{pmatrix}
u_{-} & -v_{-}\\
v_{-}& u_{-}
\end{pmatrix}
\begin{pmatrix}
\alpha_{k,-}\\
\alpha_{-k,-}^{\dagger}
\end{pmatrix}\,,
\end{equation}
therefore, the Bogoliubov transformation for operators that belong to the sector $-$ are given by
\begin{equation}
\begin{split}
c_{k,-}=u_{-}\alpha_{k,-}-v_{-}\alpha_{-k,-}^{\dagger}\,,\\
c_{-k,-}^{\dagger}=v_{-}\alpha_{k,-}+u_{-}\alpha_{-k,-}^{\dagger}\,.
\end{split}
\end{equation}
As explained before, the matrix $\Big[U_{-}^{(1)}\Big]^{T}$, $T$ denotes the transpose operation,
\begin{equation}
U_{-}^{(2)}\equiv \Big[U_{-}^{(1)}\Big]^{T}=
\begin{pmatrix}
u_{-} & v_{-}\\
-v_{-}& u_{-}
\end{pmatrix}
\end{equation}
diagonalises our Hamiltonian, and therefore it can also apply for transforming the operators $c$ into the new basis $\alpha$
\begin{equation}
\begin{split}
c_{k,-}=u_{-}\alpha_{k,-}+v_{-}\alpha_{-k,-}^{\dagger}\,,\\
c_{-k,-}^{\dagger}=-v_{-}\alpha_{k,-}+u_{-}\alpha_{-k,-}^{\dagger}\,.
\end{split}
\end{equation}
Notice that upon writing the transformations we have omitted the term ${\rm sgn}(\Delta_{p})$. It, however, won't alter our results.
\subsubsection*{The Bogoliubov transformation from previous calculations}
Therefore, for transforming operators $c_{k,\pm}$ into the new basis $\alpha_{k,\pm}$ we can use
\begin{equation}
\label{bogogog1}
\begin{split}
c_{k,+}&=u_{+}\alpha_{k,+}-v_{+}\alpha_{-k,+}^{\dagger}\,,\\
c_{k,-}&=u_{-}\alpha_{k,-}-v_{-}\alpha_{-k,-}^{\dagger}\,,\\
\end{split}
\end{equation}
or the ones related to the transpose of $U$, $U^{T}$,
\begin{equation}
\label{bogogog2}
\begin{split}
c_{k,+}&=u_{+}\alpha_{k,+}+v_{+}\alpha_{-k,+}^{\dagger}\,,\\
c_{k,-}&=u_{-}\alpha_{k,-}+v_{-}\alpha_{-k,-}^{\dagger}\,,\\
\end{split}
\end{equation}
\\\\
{\bf Comment 1:}
For calculating the density response, we can use either Eqs.\,(\ref{bogogog1}) or Eqs.\,(\ref{bogogog2}), since both diagonalize the respective Hamiltonian for each sector.
\subsection*{Density density response at zero interband pairing}
The density-density response function is defined in Eq.\,(\ref{definitionchiq}), where
\begin{equation}
\chi(1,1')=-\frac{i}{\hbar}\,\bar{\chi}(1,1')\,,
\end{equation}
and we here define
\begin{equation}
\label{chichchisc}
\bar{\chi}(1,1')=\big<\big[\hat{n}(1),\hat{n}(1')\big]\big>\,.
\end{equation}
According to the description given in the main text, one writes the electron density operator in terms of field operators, see Eqs.\,(\ref{densityscsc}), which are constructed 
taking into account the one-electron wave functions given by Eq.\,(\ref{eqswave}).
Then, we insert Eqs.\,(\ref{densityscsc}) into  Eq.\,(\ref{chichchisc}), and get
\begin{equation}
\begin{split}
\bar{\chi}(1,1')&=\mathop{\sum_{k_{1}\cdots k_{4}}}_{\sigma_{1}\cdots\sigma_{4}}
\big<\big[\eta_{k_{1}\sigma_{1}}^{\dagger}\eta_{k_{2}\sigma_{2}}\,\phi_{k_{1}}^{\dagger}(r)\phi_{k_{2}}(r)\,c^{\dagger}_{k_{1}\sigma_{1}}(t)c_{k_{2}\sigma_{2}}(t),
\eta_{k_{3}\sigma_{3}}^{\dagger}\eta_{k_{4}\sigma_{4}}\,\phi_{k_{3}}^{\dagger}(r)\phi_{k_{4}}(r)\,c^{\dagger}_{k_{3}\sigma_{3}}(t)c_{k_{4}\sigma_{4}}(t)\big]\big>\,,\\
\end{split}
\end{equation}
and therefore
\begin{equation}
\begin{split}
\bar{\chi}(1,1')&=\frac{1}{L^{2}}\sum_{k_{1}\cdots k_{4}}\,{\rm e}^{i(k_{2}-k_{1})r+i(k_{4}-k_{3})r'}\\
&\Big\{
\eta_{k_{1},+}^{\dagger}\eta_{k_{2},+}\,\eta_{k_{3},+}^{\dagger}\eta_{k_{4},+}
\big<\big[
c^{\dagger}_{k_{1},+}(t)c_{k_{2},+}(t),
c^{\dagger}_{k_{3},+}(t)c_{k_{4},+}(t)\big]\big>\\
&+
\eta_{k_{1},-}^{\dagger}\eta_{k_{2},+}\,\eta_{k_{3},+}^{\dagger}\eta_{k_{4},+}
\big<\big[
c^{\dagger}_{k_{1},-}(t)c_{k_{2},+}(t),
c^{\dagger}_{k_{3},+}(t)c_{k_{4},+}(t)\big]\big>\\
&+
\eta_{k_{1},+}^{\dagger}\eta_{k_{2},-}\,\eta_{k_{3},+}^{\dagger}\eta_{k_{4},+}
\big<\big[
c^{\dagger}_{k_{1},+}(t)c_{k_{2},-}(t),
c^{\dagger}_{k_{3},+}(t)c_{k_{4},+}(t)\big]\big>
\\
&+
\eta_{k_{1},-}^{\dagger}\eta_{k_{2},-}\,\eta_{k_{3},+}^{\dagger}\eta_{k_{4},+}
\big<\big[
c^{\dagger}_{k_{1},-}(t)c_{k_{2},-}(t),
c^{\dagger}_{k_{3},+}(t)c_{k_{4},+}(t)\big]\big>
\\
&+
\eta_{k_{1},+}^{\dagger}\eta_{k_{2},+}\,\eta_{k_{3},-}^{\dagger}\eta_{k_{4},+}
\big<\big[
c^{\dagger}_{k_{1},+}(t)c_{k_{2},+}(t),
c^{\dagger}_{k_{3},-}(t)c_{k_{4},+}(t)\big]\big>
\\
&+
\eta_{k_{1},-}^{\dagger}\eta_{k_{2},+}\,\eta_{k_{3},-}^{\dagger}\eta_{k_{4},+}
\big<\big[
c^{\dagger}_{k_{1},-}(t)c_{k_{2},+}(t),
c^{\dagger}_{k_{3},-}(t)c_{k_{4},+}(t)\big]\big>
\\
&+
\eta_{k_{1},+}^{\dagger}\eta_{k_{2},-}\,\eta_{k_{3},-}^{\dagger}\eta_{k_{4},+}
\big<\big[
c^{\dagger}_{k_{1},+}(t)c_{k_{2},-}(t),
c^{\dagger}_{k_{3},-}(t)c_{k_{4},+}(t)\big]\big>
\\
&+
\eta_{k_{1},-}^{\dagger}\eta_{k_{2},-}\,\eta_{k_{3},-}^{\dagger}\eta_{k_{4},+}
\big<\big[
c^{\dagger}_{k_{1},-}(t)c_{k_{2},-}(t),
c^{\dagger}_{k_{3},-}(t)c_{k_{4},+}(t)\big]\big>
\\
&+
\eta_{k_{1},+}^{\dagger}\eta_{k_{2},+}\,\eta_{k_{3},+}^{\dagger}\eta_{k_{4},-}
\big<\big[
c^{\dagger}_{k_{1},+}(t)c_{k_{2},+}(t),
c^{\dagger}_{k_{3},+}(t)c_{k_{4},-}(t)\big]\big>
\\
&+
\eta_{k_{1},-}^{\dagger}\eta_{k_{2},+}\,\eta_{k_{3},+}^{\dagger}\eta_{k_{4},-}
\big<\big[
c^{\dagger}_{k_{1},-}(t)c_{k_{2},+}(t),
c^{\dagger}_{k_{3},+}(t)c_{k_{4},-}(t)\big]\big>
\\
&+
\eta_{k_{1},+}^{\dagger}\eta_{k_{2},-}\,\eta_{k_{3},+}^{\dagger}\eta_{k_{4},-}
\big<\big[
c^{\dagger}_{k_{1},+}(t)c_{k_{2},-}(t),
c^{\dagger}_{k_{3},+}(t)c_{k_{4},-}(t)\big]\big>
\\
&+
\eta_{k_{1},-}^{\dagger}\eta_{k_{2},-}\,\eta_{k_{3},+}^{\dagger}\eta_{k_{4},-}
\big<\big[
c^{\dagger}_{k_{1},-}(t)c_{k_{2},-}(t),
c^{\dagger}_{k_{3},+}(t)c_{k_{4},-}(t)\big]\big>
\\
&+
\eta_{k_{1},+}^{\dagger}\eta_{k_{2},+}\,\eta_{k_{3},-}^{\dagger}\eta_{k_{4},-}
\big<\big[
c^{\dagger}_{k_{1},+}(t)c_{k_{2},+}(t),
c^{\dagger}_{k_{3},-}(t)c_{k_{4},-}(t)\big]\big>
\\
&+
\eta_{k_{1},-}^{\dagger}\eta_{k_{2},+}\,\eta_{k_{3},-}^{\dagger}\eta_{k_{4},-}
\big<\big[
c^{\dagger}_{k_{1},-}(t)c_{k_{2},+}(t),
c^{\dagger}_{k_{3},-}(t)c_{k_{4},-}(t)\big]\big>
\\
&+
\eta_{k_{1},+}^{\dagger}\eta_{k_{2},-}\,\eta_{k_{3},-}^{\dagger}\eta_{k_{4},-}
\big<\big[
c^{\dagger}_{k_{1},+}(t)c_{k_{2},-}(t),
c^{\dagger}_{k_{3},-}(t)c_{k_{4},-}(t)\big]\big>
\\
&+
\eta_{k_{1},-}^{\dagger}\eta_{k_{2},-}\,\eta_{k_{3},-}^{\dagger}\eta_{k_{4},-}
\big<\big[
c^{\dagger}_{k_{1},-}(t)c_{k_{2},-}(t),
c^{\dagger}_{k_{3},-}(t)c_{k_{4},-}(t)\big]\big>
\Big\}\,,
\end{split}
\end{equation}
where we have summed over $\sigma=\pm$.
There are $16$ terms to be calculated. We need the Bogolyubov transformation to describe excitations in a superconductor, which was derived in previous subsection in Eqs.\,(\ref{bogogog2}),
\begin{equation}
\begin{split}
c_{k,+}&=u_{k,+}\alpha_{k,+}+v_{k,+}\alpha_{-k,+}^{\dagger}\,,\\
c_{k,-}&=u_{k,-}\alpha_{k,-}+v_{k,-}\alpha_{-k,-}^{\dagger}\,,
\end{split}
\end{equation}
where
\begin{equation}
u_{k,\pm}=\frac{1}{\sqrt{2}}\sqrt{1+\frac{\varepsilon_{k,\pm}}{E_{k,\pm}}}\,,\quad
v_{k,\pm}=\frac{1}{\sqrt{2}}\sqrt{1-\frac{\varepsilon_{k,\pm}}{E_{k,\pm}}}\,,
\end{equation}
and $E_{k,\pm}=\sqrt{\Delta_{p}^{2}+\varepsilon_{k,\pm}^{2}}$, $\varepsilon_{k,\pm}=\xi_{k}\pm\sqrt{B^{2}+\alpha^{2}k^{2}}$, $\xi_{k}=\frac{\hbar^{2}k^{2}}{2m}-\mu$.

By working out all the elements, at zero temperature $T=0$, $n_{k,\pm}=0$, we get
\begin{equation}
\label{eqscra}
\begin{split}
\bar{\chi}(1,1')&=\frac{1}{L^{2}}\sum_{k_{1}\cdots k_{4}}\,{\rm e}^{i(k_{2}-k_{1})r+i(k_{4}-k_{3})r'}\\
&\Big\{
\eta_{k_{1},+}^{\dagger}\eta_{k_{2},+}\,\eta_{-k_{1},+}^{\dagger}\eta_{-k_{2},+}
{\rm e}^{i(E_{k_{1},+}+E_{-k_{2},+})(t-t')}u_{k_{1},+}v_{k_{2},+}v_{-k_{1},+}u_{-k_{2},+}\\
&-
\eta_{k_{1},+}^{\dagger}\eta_{k_{2},+}\,\eta_{k_{2},+}^{\dagger}\eta_{k_{1},+}
{\rm e}^{i(E_{k_{1},+}+E_{-k_{2},+})(t-t')}u_{k_{1},+}v_{k_{2},+}v_{k_{2},+}u_{k_{1},+}\\
&-
\eta_{k_{1},+}^{\dagger}\eta_{k_{2},+}\,\eta_{-k_{1},+}^{\dagger}\eta_{-k_{2},+}
{\rm e}^{-i(E_{-k_{1},+}+E_{k_{2},+})(t-t')}v_{k_{1},+}u_{k_{2},+}u_{-k_{1},+}v_{-k_{2},+}\\
&+
\eta_{k_{1},+}^{\dagger}\eta_{k_{2},+}\,\eta_{k_{2},+}^{\dagger}\eta_{k_{1},+}
{\rm e}^{-i(E_{-k_{1},+}+E_{k_{2},+})(t-t')}v_{k_{1},+}u_{k_{2},+}u_{k_{2},+}v_{k_{1},+}\\
&+
\eta_{k_{1},-}^{\dagger}\eta_{k_{2},+}\,\eta_{-k_{1},-}^{\dagger}\eta_{-k_{2},+}
{\rm e}^{i(E_{k_{1},-}+E_{-k_{2},+})(t-t')}u_{k_{1},-}v_{k_{2},+}v_{-k_{1},-}u_{-k_{2},+}\\
&-
\eta_{k_{1},-}^{\dagger}\eta_{k_{2},+}\,\eta_{-k_{1},-}^{\dagger}\eta_{-k_{2},+}
{\rm e}^{-i(E_{-k_{1},-}+E_{k_{2},+})(t-t')}v_{k_{1},-}u_{k_{2},+}u_{-k_{1},-}v_{-k_{2},+}\\
&-
\eta_{k_{1},+}^{\dagger}\eta_{k_{2},-}\,\eta_{k_{2},-}^{\dagger}\eta_{k_{1},+}
{\rm e}^{i(E_{k_{1},+}+E_{-k_{2},-})(t-t')}u_{k_{1},+}v_{k_{2},-}v_{k_{2},-}u_{k_{1},+}\\
&+
\eta_{k_{1},+}^{\dagger}\eta_{k_{2},-}\,\eta_{k_{2},-}^{\dagger}\eta_{k_{1},+}
{\rm e}^{-i(E_{-k_{1},+}+E_{k_{2},-})(t-t')}v_{k_{1},+}u_{k_{2},-}u_{k_{2},-}v_{k_{1},+}\\
&-
\eta_{k_{1},-}^{\dagger}\eta_{k_{2},+}\,\eta_{k_{2},+}^{\dagger}\eta_{k_{1},-}
{\rm e}^{i(E_{k_{1},-}+E_{-k_{2},+})(t-t')}u_{k_{1},-}v_{k_{2},+}v_{k_{2},+}u_{k_{1},-}\\
&+
\eta_{k_{1},-}^{\dagger}\eta_{k_{2},+}\,\eta_{k_{2},+}^{\dagger}\eta_{k_{1},-}
{\rm e}^{-i(E_{-k_{1},-}+E_{k_{2},+})(t-t')}v_{k_{1},-}u_{k_{2},+}u_{k_{2},+}v_{k_{1},-}\\
&+
\eta_{k_{1},+}^{\dagger}\eta_{k_{2},-}\,\eta_{-k_{1},+}^{\dagger}\eta_{-k_{2},-}
{\rm e}^{i(E_{k_{1},+}+E_{-k_{2},-})(t-t')}u_{k_{1},+}v_{k_{2},-}v_{-k_{1},+}u_{-k_{2},-}\\
&-
\eta_{k_{1},+}^{\dagger}\eta_{k_{2},-}\,\eta_{-k_{1},+}^{\dagger}\eta_{-k_{2},-}
{\rm e}^{-i(E_{-k_{1},+}+E_{k_{2},-})(t-t')}v_{k_{1},+}u_{k_{2},-}u_{-k_{1},+}v_{-k_{2},-}\\
&+
\eta_{k_{1},-}^{\dagger}\eta_{k_{2},-}\,\eta_{-k_{1},-}^{\dagger}\eta_{-k_{2},-}
{\rm e}^{i(E_{k_{1},-}+E_{-k_{2},-})(t-t')}u_{k_{1},-}v_{k_{2},-}v_{-k_{1},-}u_{-k_{2},-}\\
&-
\eta_{k_{1},-}^{\dagger}\eta_{k_{2},-}\,\eta_{k_{2},-}^{\dagger}\eta_{k_{1},-}
{\rm e}^{i(E_{k_{1},-}+E_{-k_{2},-})(t-t')}u_{k_{1},-}v_{k_{2},-}v_{k_{2},-}u_{k_{1},-}\\
&-
\eta_{k_{1},-}^{\dagger}\eta_{k_{2},-}\,\eta_{-k_{1},-}^{\dagger}\eta_{-k_{2},-}
{\rm e}^{-i(E_{-k_{1},-}+E_{k_{2},-})(t-t')}v_{k_{1},-}u_{k_{2},-}u_{-k_{1},-}v_{-k_{2},-}\\
&+
\eta_{k_{1},-}^{\dagger}\eta_{k_{2},-}\,\eta_{k_{2},-}^{\dagger}\eta_{k_{1},-}
{\rm e}^{-i(E_{-k_{1},-}+E_{k_{2},-})(t-t')}v_{k_{1},-}u_{k_{2},-}u_{k_{2},-}v_{k_{1},-}
\Big\}\,.
\end{split}
\end{equation}
The frequency dependence is calculated by integrating previous expression in time $t-t'$.
Moreover, the coefficients coming from the Rashba-SO problem in Eq.\,(\ref{eqscra}), the terms with $\eta$, give only two different values that we denote as
\begin{equation}
A_{k_{1},k_{2}}^{\sigma}=\frac{1}{2}\bigg[1+\sigma\frac{B^{2}+\alpha^{2}k_{1}k_{2}}{\sqrt{B^{2}+\alpha^{2}k_{1}^{2}}\sqrt{B^{2}+\alpha^{2}k_{2}^{2}}} \bigg]\,,\quad \sigma=\pm\,.
\end{equation}
After the integration and considering previous definitions we arrive at
\begin{equation}
\begin{split}
\chi(\omega,r,r')&=\frac{1}{L^{2}}\sum_{k_{1},k_{2}}\,{\rm e}^{i(k_{2}-k_{1})(r-r')}\times\Bigg\{\\
&A_{k_{1},k_{2}}^{+}
\bigg[\frac{u_{k_{1},+}v_{k_{2},+}v_{-k_{1},+}u_{-k_{2},+}}{\omega+E_{k_{1},+}+E_{-k_{2},+}+i\eta}
-
\frac{u_{k_{1},+}v_{k_{2},+}v_{k_{2},+}u_{k_{1},+}}{\omega+E_{k_{1},+}+E_{-k_{2},+}+i\eta}\\
&-
\frac{v_{k_{1},+}u_{k_{2},+}u_{-k_{1},+}v_{-k_{2},+}}{\omega-E_{-k_{1},+}-E_{k_{2},+}+i\eta}
+
\frac{v_{k_{1},+}u_{k_{2},+}u_{k_{2},+}v_{k_{1},+}}{\omega-E_{-k_{1},+}-E_{k_{2},+}+i\eta}
\bigg]\\
&+
A_{k_{1},k_{2}}^{-}
\bigg[\frac{u_{k_{1},-}v_{k_{2},+}v_{-k_{1},-}u_{-k_{2},+}}{\omega+E_{k_{1},-}+E_{-k_{2},+}+i\eta}
-
\frac{v_{k_{1},-}u_{k_{2},+}u_{-k_{1},-}v_{-k_{2},+}}{\omega-E_{-k_{1},-}-E_{k_{2},+}+i\eta}\\
&-
\frac{u_{k_{1},+}v_{k_{2},-}v_{k_{2},-}u_{k_{1},+}}{\omega+E_{k_{1},+}+E_{-k_{2},-}+i\eta}
+
\frac{v_{k_{1},+}u_{k_{2},-}u_{k_{2},-}v_{k_{1},+}}{\omega-E_{-k_{1},+}-E_{k_{2},-}+i\eta}
\bigg]\\
&+
A_{k_{1},k_{2}}^{-}
\bigg[-\frac{u_{k_{1},-}v_{k_{2},+}v_{k_{2},+}u_{k_{1},-}}{\omega+E_{k_{1},-}+E_{-k_{2},+}+i\eta}
+
\frac{v_{k_{1},-}u_{k_{2},+}u_{k_{2},+}v_{k_{1},-}}{\omega-E_{-k_{1},-}-E_{k_{2},+}+i\eta}\\
&+
\frac{u_{k_{1},+}v_{k_{2},-}v_{-k_{1},+}u_{-k_{2},-}}{\omega+E_{k_{1},+}+E_{-k_{2},-}+i\eta}
-
\frac{v_{k_{1},+}u_{k_{2},-}u_{-k_{1},+}v_{-k_{2},-}}{\omega-E_{-k_{1},+}-E_{k_{2},-}+i\eta}
\bigg]\\
&+
A_{k_{1},k_{2}}^{+}
\bigg[
\frac{u_{k_{1},-}v_{k_{2},-}v_{-k_{1},-}u_{-k_{2},-}}{\omega+E_{k_{1},-}+E_{-k_{2},-}+i\eta}
-
\frac{u_{k_{1},-}v_{k_{2},-}v_{k_{2},-}u_{k_{1},-}}{\omega+E_{k_{1},-}+E_{-k_{2},-}+i\eta}\\
&-
\frac{v_{k_{1},-}u_{k_{2},-}u_{-k_{1},-}v_{-k_{2},-}}{\omega-E_{-k_{1},-}-E_{k_{2},-}+i\eta}
+
\frac{v_{k_{1},-}u_{k_{2},-}u_{k_{2},-}v_{k_{1},-}}{\omega-E_{-k_{1},-}-E_{k_{2},-}+i\eta}
\bigg]\,.
\Bigg\}
\end{split}
\end{equation}
Now, we take the zero frequency limit and consider $u_{k}=u_{-k}$, $v_{k}=v_{-k}$, $E_{k}=E_{-k}$ and $\varepsilon_{k}=\varepsilon_{-k}$.
Thus, previous equation can be written as
\begin{equation}
\begin{split}
\chi(r,r')&=-\frac{1}{L^{2}}\sum_{k_{1},k_{2}}\,{\rm e}^{i(k_{2}-k_{1})(r-r')}\times\\
&\Bigg\{
\frac{A_{k_{1},k_{2}}^{+}}{E_{k_{1},+}+E_{k_{2},+}}
\big(u_{k_{1},+}v_{k_{2},+} -v_{k_{1},+}u_{k_{2},+}\big)^{2}
+
\frac{A_{k_{1},k_{2}}^{+}}{E_{k_{1},-}+E_{k_{2},-}}
\big(u_{k_{1},-}v_{k_{2},-} -v_{k_{1},-}u_{k_{2},-}\big)^{2}\\
&+
\frac{A_{k_{1},k_{2}}^{-}}{E_{k_{1},-}+E_{k_{2},+}}
\big(u_{k_{1},-}v_{k_{2},+} -v_{k_{1},-}u_{k_{2},+}\big)^{2}
+
\frac{A_{k_{1},k_{2}}^{-}}{E_{k_{1},+}+E_{k_{2},-}}
\big(u_{k_{1},+}v_{k_{2},-} -v_{k_{1},+}u_{k_{2},-}\big)^{2}
\Bigg\}\,.
\end{split}
\end{equation}
Let us denote the term in brackets as $F_{k_{1},k_{2}}$. 
The momentum dependence is calculated as by integrating over space $(r-r')$.
Then,
\begin{equation}
\begin{split}
\chi(q)&=-\frac{1}{L^{2}}\sum_{k_{1},k_{2}}\,F_{k_{1},k_{2}}\,\int{\rm e}^{i(k_{2}-k_{1}-q)(r-r')}d(r-r') =-\frac{1}{L^{2}}\sum_{k_{1},k_{2}}\,F_{k_{1},k_{2}}\, L\delta_{k_{2},k_{1}+q}\\
&=-\frac{1}{L}\sum_{k_{1}}\,F_{k_{1},k_{1}+q}=-\frac{1}{L}\sum_{k}\,F_{k,k+q}\,,\\
&=-\frac{1}{L}\,\frac{L}{2\pi}\int\,dk\,F_{k,k+q}=-\frac{1}{2\pi}\int\,dk\,F_{k,k+q}\,.
\end{split}
\end{equation}
Therefore we get
\begin{equation}
\begin{split}
&\chi(r,r')=\\
&-\frac{1}{2\pi}\int_{-\infty}^{\infty}dk
\Bigg\{
\frac{A_{k,k+q}^{+}}{E_{k,+}+E_{k+q,+}}
\big(u_{k,+}v_{k+q,+} -v_{k,+}u_{k+q,+}\big)^{2}
+
\frac{A_{k,k+q}^{+}}{E_{k,-}+E_{k+q,-}}
\big(u_{k,-}v_{k+q,-} -v_{k,-}u_{k+q,-}\big)^{2}\\
&+
\frac{A_{k,k+q}^{-}}{E_{k,-}+E_{k+q,+}}
\big(u_{k,-}v_{k+q,+} -v_{k,-}u_{k+q,+}\big)^{2}
+
\frac{A_{k,k+q}^{-}}{E_{k,+}+E_{k+q,-}}
\big(u_{k,+}v_{k+q,-} -v_{k,+}u_{k+,-}\big)^{2}
\Bigg\}\,,
\end{split}
\end{equation}
where coherence factors read
\begin{equation}
\begin{split}
\big(u_{k,+}v_{k+q,+} -v_{k,+}u_{k+q,+}\big)^{2}&=\frac{1}{2}\bigg[1-\frac{\varepsilon_{k,+}\varepsilon_{k+q,+}+\Delta_{p}(k)\Delta_{p}(k+q)}{E_{k,+}E_{k+q,+}} \bigg]\\
\big(u_{k,-}v_{k+q,-} -v_{k,-}u_{k+q,-}\big)^{2}&=\frac{1}{2}\bigg[1-\frac{\varepsilon_{k,-}\varepsilon_{k+q,-}+\Delta_{p}(k)\Delta_{p}(k+q)}{E_{k,-}E_{k+q,-}} \bigg]\\
\big(u_{k,-}v_{k+q,+} -v_{k,-}u_{k+q,+}\big)^{2}&=\frac{1}{2}\bigg[1-\frac{\varepsilon_{k,-}\varepsilon_{k+q,+}+\Delta_{p}(k)\Delta_{p}(k+q)}{E_{k,-}E_{k+q,+}} \bigg]\\
\big(u_{k,+}v_{k+q,-} -v_{k,+}u_{k+,-}\big)^{2}&=\frac{1}{2}\bigg[1-\frac{\varepsilon_{k,+}\varepsilon_{k+q,-}+\Delta_{p}(k)\Delta_{p}(k+q)}{E_{k,+}E_{k+q,-}} \bigg]\,.
\end{split}
\end{equation}
\subsection{Regime of strong Zeeman field}
\label{StringBdensity}
In this part we address the situation of high Zeeman field $B>>\mu, \Delta$. In this case, only the lowest band, $-$, is occupied. Therefore, the full Hamiltonian can be projected to the lowest band yielding 
\begin{equation}
\mathcal{H}=\mathcal{H}_{0}+\mathcal{H}_{sc}=\frac{dk}{2\pi}
\varepsilon_{-}(k)\psi^{\dagger}_{-}(k)\psi_{-}(k)+\int \frac{dk}{2\pi} 
\Big[\frac{\Delta_{--}(k)}{2}\psi_{-}^{\dagger}(k)\psi_{-}^{\dagger}(-k)
+h.c\Big]\,,
\end{equation}
The Hamiltonian $\mathcal{H}$ can be written in the BdG form in the $(\psi_{-}(k),\psi_{-}^{\dagger}(-k))$ basis,
\begin{equation}
H_{BdG,-}(k)= \begin{pmatrix}\,
\varepsilon_{-}(k)&\Delta_{--}(k)\\
\Delta_{--}^{\dagger}(k)&-\varepsilon_{-}(-k)
 \end{pmatrix}\,+cte\,,
\end{equation}
where
\begin{equation}
\Delta_{--}(k)=\frac{i\alpha k \Delta}{\sqrt{B^{2}+\alpha^{2}k^{2}}}\equiv i\Delta_{p}(k)\,.
\end{equation}
The eigenvalues are given by
\begin{equation}
\label{energies}
\tilde{E}_{-}(k)=\pm\sqrt{|\Delta_{--}|^{2}+\varepsilon_{-}^{2}(k)}\equiv\pm E_{-}\,.
\end{equation}
Here we proceed similarly to previous subsection in order to calculate the density-density response function.
The response function in momentum and frequency space is given by
\begin{equation}
\chi(q,\omega)=\int d(r-r')\int d(t-t')\,{\rm e}^{-iq(r-r'))}\,{\rm e}^{i(\omega+i\eta)(t-t')}\,\chi(r,r',t,t')\,,
\end{equation}
where, $\chi(r,r',t,t')\equiv\chi(1,1')$,
\begin{equation}
\label{chichchisc00}
\chi(1,1')=-\frac{i}{\hbar}\big<[\hat{n}(1),\hat{n}(1')]\big>\,,
\end{equation}
and $\hat{n}$ is the electron density operator.
According to the description given in the main text, one writes the electron density operator in terms of field operators, see Eqs.\,(\ref{densityscsc}), which are constructed 
taking into account the one-electron wave functions given by Eq.\,(\ref{eqswave}).
Then, we insert Eqs.\,(\ref{densityscsc}) into  Eq.\,(\ref{chichchisc00}), and get
\begin{equation}
\label{densityfull}
\begin{split}
\chi(1,1')&=-\frac{i}{\hbar}\frac{1}{L^{2}}\mathop{\sum_{k_{1}\cdots k_{4}}}_{\sigma_{1}\cdots\sigma_{4}}
\eta^{\dagger}_{k_{1},\sigma_{1}}\eta_{k_{2},\sigma_{2}}
\eta^{\dagger}_{k_{3},\sigma_{3}}\eta_{k_{4},\sigma_{4}}\,
{\rm e}^{i(k_{2}-k_{1})r+i(k_{4}-k_{3})r'}\times\\
&\times\big<\big[c^{\dagger}_{k_{1},\sigma_{1}}(t)c_{k_{2},\sigma_{2}}(t),
c^{\dagger}_{k_{3},\sigma_{3}}(t')c_{k_{4},\sigma_{4}}(t')\big]\big>\,.
\end{split}
\end{equation}
Now, we assume the situation of high Zeeman field, where only the lowest band, $\sigma=-$, is occupied. Therefore,
in Eq.\,(\ref{densityfull}) the unique combination that we need is 
\begin{equation}
\label{densityfull2}
\begin{split}
\chi(1,1')&=-\frac{i}{\hbar}\frac{1}{L^{2}}\sum_{k_{1}\cdots k_{4}}
\eta^{\dagger}_{k_{1},-}\eta_{k_{2},-}
\eta^{\dagger}_{k_{3},-}\eta_{k_{4},-}\,
{\rm e}^{i(k_{2}-k_{1})r+i(k_{4}-k_{3})r'}\times\\
&\times\big<\big[c^{\dagger}_{k_{1},-}(t)c_{k_{2},-}(t),
c^{\dagger}_{k_{3},-}(t')c_{k_{4},-}(t')\big]\big>\,.
\end{split}
\end{equation}
In order to describe the system with superconducting effects, we need to transform our $c$ operators in Eq.\,(\ref{densityfull2}) into the new ones $\alpha$ according to the Bogoliubov transformation given by Eqs.\,(\ref{bogogog1}) or (\ref{bogogog2}),
where \begin{equation}
u_{k}=\sqrt{\frac{1}{2}\bigg[1+\frac{\varepsilon_{-}(k)}{E_{-}(k)} \bigg]}\,,\quad
v_{k}=\sqrt{\frac{1}{2}\bigg[1-\frac{\varepsilon_{-}(k)}{E_{-}(k)} \bigg]}\,.
\end{equation}

Let us define the following quantity 
\begin{equation}
\bar{\chi}(1,1')=\big<\big[c^{\dagger}_{k_{1},\sigma_{1}}(t)c_{k_{2},\sigma_{2}}(t),
c^{\dagger}_{k_{3},\sigma_{3}}(t')c_{k_{4},\sigma_{4}}(t')\big]\big>\,.
\end{equation}
Therefore, in terms of the new quasiparticle operators $\alpha$, we get
\begin{equation}
\label{densityquasi}
\begin{split}
\bar{\chi}(1,1')&=
\big<\big[u_{k_{1}}u_{k_{2}}\alpha^{\dagger}_{k_{1},-}(t)\alpha_{k_{2},-}(t),
u_{k_{3}}u_{k_{4}}\alpha^{\dagger}_{k_{3},-}(t')\alpha_{k_{4},-}(t')
\big]\big>\\
&+\big<\big[u_{k_{1}}u_{k_{2}}\alpha^{\dagger}_{k_{1},-}(t)\alpha_{k_{2},-}(t),
v_{k_{3}}v_{k_{4}}\alpha_{-k_{3},-}(t')\alpha_{-k_{4},-}^{\dagger}(t')
\big]\big>\\
&+\big<\big[u_{k_{1}}v_{k_{2}}\alpha^{\dagger}_{k_{1},-}(t)\alpha_{-k_{2},-}^{\dagger}(t),
v_{k_{3}}u_{k_{4}}\alpha_{-k_{3},-}(t')\alpha_{k_{4},-}(t')\big]\big>\\
&+\big<\big[v_{k_{1}}u_{k_{2}}\alpha_{-k_{1},-}(t)\alpha_{k_{2},-}(t),
u_{k_{3}}v_{k_{4}}\alpha_{k_{3},-}^{\dagger}(t')\alpha_{-k_{4},-}^{\dagger}(t')
\big]\big>\\
&+\big<\big[v_{k_{1}}v_{k_{2}}\alpha_{-k_{1},-}(t)\alpha_{-k_{2},-}^{\dagger}(t),
u_{k_{3}}u_{k_{4}}\alpha^{\dagger}_{k_{3},-}(t')\alpha_{k_{4},-}(t')
\big]\big>\\
&+\big<\big[v_{k_{1}}v_{k_{2}}\alpha_{-k_{1},-}(t)\alpha_{-k_{2},-}^{\dagger}(t),
v_{k_{3}}v_{k_{4}}\alpha_{-k_{3},-}(t')\alpha_{-k_{4},-}^{\dagger}(t')
\big]\big>\,.
\end{split}
\end{equation}
Then, we can work out the average of the commutators in previous equation, consider the time evolution of the operators, $\alpha^{\dagger}_{k,-}(t)={\rm e}^{iE_{k,-}t/\hbar}$,
and obtain for Eq.\,(\ref{densityfull2}), 
\begin{equation}
\begin{split}
\chi(1,1')&=-\frac{i}{\hbar}\frac{1}{L^{2}}\sum_{k_{1}, k_{2}}{\rm e}^{i(k_{2}-k_{1})(r-r')}
\times\\
&\Bigg\{
\eta^{\dagger}_{k_{1},-}\eta_{k_{2},-}
\eta^{\dagger}_{k_{2},-}\eta_{k_{1},-}
\Big[{\rm e}^{i(E_{k_{1},-}-E_{k_{2},-})(t-t')/\hbar}u^{2}_{k_{1}}u^{2}_{k_{2}} (n_{k_{1},-}-n_{k_{2},-})\\
&\quad\quad\quad\quad\quad\quad\quad\quad+{\rm e}^{-i(E_{-k_{1},-}-E_{-k_{2},-})(t-t')/\hbar}v^{2}_{k_{1}}v^{2}_{k_{2}} (n_{-k_{2},-}-n_{-k_{1},-})
\Big]\\
&+\eta^{\dagger}_{k_{1},-}\eta_{k_{2},-}
\eta^{\dagger}_{-k_{1},-}\eta_{-k_{2},-}
\Big[{\rm e}^{i(E_{k_{1},-}-E_{k_{2},-})(t-t')/\hbar}u_{k_{1}}u_{k_{2}}v_{-k_{1}}v_{-k_{2}} (-n_{k_{1},-}+n_{k_{2},-})\\
&\quad\quad\quad\quad\quad\quad\quad\quad+{\rm e}^{-i(E_{-k_{1},-}-E_{-k_{2},-})(t-t')/\hbar}v_{k_{1}}v_{k_{2}}u_{-k_{1}}u_{-k_{2}} (-n_{-k_{2},-}+n_{-k_{1},-})\Big]\\
&+\eta^{\dagger}_{k_{1},-}\eta_{k_{2},-}
\eta^{\dagger}_{-k_{1},-}\eta_{-k_{2},-}
\Big[{\rm e}^{i(E_{k_{1},-}+E_{-k_{2},-})(t-t')/\hbar}u_{k_{1}}v_{k_{2}}v_{-k_{1}}u_{-k_{2}} (1-n_{k_{1},-}-n_{-k_{2},-})\\
&\quad\quad\quad\quad\quad\quad\quad\quad+{\rm e}^{-i(E_{-k_{1},-}+E_{k_{2},-})(t-t')/\hbar}v_{k_{1}}u_{k_{2}}u_{-k_{1}}v_{-k_{2}} (-1+n_{k_{2},-}+n_{-k_{1},-})\Big]\\
&+\eta^{\dagger}_{k_{1},-}\eta_{k_{2},-}
\eta^{\dagger}_{k_{2},-}\eta_{k_{1},-}
\Big[{\rm e}^{i(E_{k_{1},-}+E_{-k_{2},-})(t-t')/\hbar}u_{k_{1}}^{2}v_{k_{2}}^{2} (-1+n_{k_{1},-}+n_{-k_{2},-})\\
&\quad\quad\quad\quad\quad\quad\quad\quad+{\rm e}^{-i(E_{-k_{1},-}+E_{k_{2},-})(t-t')/\hbar}v_{k_{1}}^{2}u_{k_{2}}^{2} (1-n_{k_{2},-}-n_{-k_{1},-})\Big]
\Bigg\}\,.
\end{split}
\end{equation}
Now, we integrate over time and evaluate the zero temperature limit, where $E_{k}>0$, and thus $\lim_{T\rightarrow0}n_{k,-}=0$. Therefore, previous equation reads
\begin{equation}
\begin{split}
\chi(\omega,r,r')&=\frac{1}{L^{2}}\sum_{k_{1}, k_{2}}{\rm e}^{i(k_{2}-k_{1})(r-r')}
\times\\
&\Bigg\{\eta^{\dagger}_{k_{1},-}\eta_{k_{2},-}
\eta^{\dagger}_{-k_{1},-}\eta_{-k_{2},-}
\Bigg[\frac{u_{k_{1}}v_{k_{2}}v_{-k_{1}}u_{-k_{2}} }{\omega+E_{k_{1},-}+E_{-k_{2},-}+i\hbar\eta}
+\frac{v_{k_{1}}u_{k_{2}}u_{-k_{1}}v_{-k_{2}} (-1)}{\omega-E_{-k_{1},-}-E_{k_{2},-}+i\hbar\eta}\Bigg]\\
&+\eta^{\dagger}_{k_{1},-}\eta_{k_{2},-}
\eta^{\dagger}_{k_{2},-}\eta_{k_{1},-}
\Bigg[\frac{u_{k_{1}}^{2}v_{k_{2}}^{2} (-1)}{\omega+E_{k_{1},-}+E_{-k_{2},-}+i\hbar\eta}
+\frac{v_{k_{1}}^{2}u_{k_{2}}^{2} }{\omega-E_{-k_{1},-}-E_{k_{2},-}+i\hbar\eta}\Bigg]
\Bigg\}\,.
\end{split}
\end{equation}
We can further simplify previous equation by considering the zero frequency limit $\omega=0$. 
The coefficients from the Rashba-Zeeman fields reads
\begin{equation}
\begin{split}
\eta^{\dagger}_{k_{1},-}\eta_{k_{2},-}
\eta^{\dagger}_{-k_{1},-}\eta_{-k_{2},-}&=\frac{1}{2}\Bigg[1+\frac{B^{2}+\alpha^{2} k_{1}k_{2}}{\sqrt{B^{2}+\alpha^{2}k_{1}^{2}}\sqrt{B^{2}+\alpha^{2}k_{2}^{2}}}\Bigg]\equiv A_{k_{1},k_{2}}^{+}\,,\\
\eta^{\dagger}_{k_{1},-}\eta_{k_{2},-}
\eta^{\dagger}_{k_{2},-}\eta_{k_{1},-}&=\frac{1}{2}\Bigg[1+\frac{B^{2}+\alpha^{2} k_{1}k_{2}}{\sqrt{B^{2}+\alpha^{2}k_{1}^{2}}\sqrt{B^{2}+\alpha^{2}k_{2}^{2}}}\Bigg]\equiv A_{k_{1},k_{2}}^{+}\,.
\end{split}
\end{equation}
Then, from the relations for the coherence factors $u$ and $v$ and for the dispersion relation $E_{k,-}$, we have that $u_{k}=u_{-k}$, $v_{k}=v_{-k}$ and $E_{k,-}=E_{-k,-}$. Additionally, t
Therefore, taking into account previous comments, the density density response reads,
\begin{equation}
\begin{split}
\chi(r,r')&=\frac{1}{L^{2}}\sum_{k_{1}, k_{2}}{\rm e}^{i(k_{2}-k_{1})(r-r')}
A_{k_{1},k_{2}}^{+}
\Big[u_{k_{1}}v_{k_{2}}v_{k_{1}}u_{k_{2}}+
v_{k_{1}}u_{k_{2}}u_{k_{1}}v_{k_{2}}-u_{k_{1}}^{2}v_{k_{2}}^{2} -v_{k_{1}}^{2}u_{k_{2}}^{2}
\Big]\frac{1}{E_{k_{1},-}+E_{k_{2},-}}
\,,\\
&=-\frac{1}{L^{2}}\sum_{k_{1}, k_{2}}{\rm e}^{i(k_{2}-k_{1})(r-r')}
A_{k_{1},k_{2}}^{+}
\frac{\big(u_{k_{1}}v_{k_{2}}-u_{k_{2}}v_{k_{1}}
\big)^{2}}{E_{k_{1},-}+E_{k_{2},-}}\,.
\end{split}
\end{equation}
Moreover, we can write the momentum dependence of the density-density response 
\begin{equation}
\chi(q)=\int d(r-r')\,{\rm e}^{-iq(r-r')}\chi(r,r')\,.
 \end{equation}
 Thus, we get
 \begin{equation}
 \label{chisc}
 \chi(q)
 =-\frac{1}{L}\sum_{k}
A_{k,k+q}^{+}
\frac{\big(u_{k}v_{k+q}-u_{k+q}v_{k}
\big)^{2}}{E_{k,-}+E_{k+q,-}}\,.
\end{equation}
Previous equation represents the density-density response function in a 1D superconducting wire with Rashba and Zeeman interaction with only the lowest band occupied, also known as the Lindhard function. Previous equation is valid for high Zeeman fields $B>>\mu,\Delta$. Let us write down the coherence coefficient in Eq.\,(\ref{chisc}),
\begin{equation}
\big(u_{k}v_{k+q}-u_{k+q}v_{k}
\big)^{2}=\frac{1}{2}\bigg[1-\frac{\varepsilon_{k,-}\varepsilon_{k+q,-}+\Delta_{p}(k)\Delta_{p}(k+q)}{E_{k,-}E_{k+q,-}} \bigg]\,,
\end{equation}
where we have denoted 
\begin{equation}
\begin{split}
\Delta_{p}(k)&=\frac{\alpha k \Delta}{\sqrt{B^{2}+\alpha^{2}k^{2}}}\,,\\
\varepsilon_{-}(k)&\equiv\varepsilon_{k,-}=\xi_{k}-\sqrt{B^{2}+\alpha^{2}k^{2}}\,,\\
E_{-}(k)&\equiv E_{k,-}=\sqrt{\Delta_{p}^{2}(k)+\varepsilon_{k,-}^{2}}\,.
\end{split}
\end{equation}
As in previous subsection, sum over $k$ in Eq.\,(\ref{chisc}) is transformed into an integral over $k$,
 \begin{equation}
 \label{chisc2}
 \chi(q)
 =-\frac{1}{2\pi}\int_{\infty}^{\infty}
A_{k,k+q}^{+}
\frac{\big(u_{k}v_{k+q}-u_{k+q}v_{k}
\big)^{2}}{E_{k,-}+E_{k+q,-}}\,,
\end{equation}
and the integral is performed numerically. 

\end{appendices}
\addtocontents{toc}{\vspace{1em}} 

\clearpage  





\addtotoc{List of publications}  
\publications{
\addtocontents{toc}{\vspace{1em}}  

\begin{enumerate}

\item Multiple Andreev reflection and critical current in topological superconducting nanowire junctions\\
Pablo San-Jose, {\bf Jorge Cayao}, Elsa Prada and Ram\'{o}n Aguado\\
\href{http://iopscience.iop.org/1367-2630/15/7/075019/}{ New Journal of Physics {\bf 15}, 075019 (2013)}\\
 \href{http://iopscience.iop.org/1367-2630/focus/Focus%20on%20Majorana%20Fermions%20in%20Condensed%20Matter
 }{\textit{Focus issue on Majorana Fermions in Condensed Matter}}
 \item SNS junctions in nanowires with spin-orbit coupling: Role of confinement and helicity on the subgap spectrum\\
 {\bf Jorge Cayao}, Pablo San-Jose, Elsa Prada and Ram\'{o}n Aguado\\
\href{http://journals.aps.org/prb/abstract/10.1103/PhysRevB.91.024514}{Physical Review B {\bf 91}, 024514 (2015)}
 \item Majorana bound states from exceptional points in non-topological superconductors\\
 Pablo San-Jose, {\bf Jorge Cayao}, Elsa Prada and Ram\'{o}n Aguado\\
\href{http://www.nature.com/articles/srep21427}{Scientific reports 6, 21427 (2016).}
\end{enumerate}
}
\clearpage  
\addtotoc{Curriculum Vitae}  
\curriculum{
\addtocontents{toc}{\vspace{1em}}  

\begin{tabular}{ll}
Aug. 4th. 1986 & Born in Lima, Peru \\
Apr. 2000 -- Dec. 2002 & Infantry reserve officer \\ & Colegio Militar Elias Aguirre, Chiclayo, Peru \\
Jan. 2003 -- Sept. 2003 & Academic preparation for the University admission exam \\ & Centro Pre-Universitario de la Universidad Nacional Mayor de San Marcos \\
Apr. 2004 -- Aug. 2005 & Three semesters of Bachelor in Physics \\ & Universidad Nacional Mayor de San Marcos, Lima, Peru\\
Sept. 2005 -- June 2006 & Slovak language and academic preparation \\
 & Comenius University, Bratislava, Slovakia\\
Sept. 2006 -- June 2009 & Bachelor in Physics \\ & Comenius University, Bratislava, Slovakia. \\
Sept. 2009 -- June 2011 & Master in Physics with specialization in Astronomy and Astrophysics \\ & Comenius University, Bratislava, Slovakia\\
Aug. 2011 -- Dec. 2011 & SAP Engineer\\
                                          & Hewlett-Packard (HP) Slovakia\\
Jan. 2012 -- May 2016 & PhD thesis\\  &Advisor: Dr. Ram\'on Aguado \\ & Instituto de Ciencia de Materiales de Madrid (ICMM)\\                  
                                            &Consejo Superior de Investigaciones Cient\'{i}ficas (CSIC), Spain\\
Aug. 2013 -- Sept. 2013 & Visiting Scholar\\ &Group of Prof. Vladimir Bu\v{z}ek, under the supervision of Dr. Peter Sta\v{n}o \\ & Slovak Academy of Sciences. Bratislava, Slovakia\\
Sept. 2014 -- Dec. 2014 & Visiting Scholar\\ &Group of Prof. Daniel Loss, under the supervision of Dr. Peter Sta\v{n}o \\ & Center for Emergent Matter Science, RIKEN. Tokyo, Japan 
\end{tabular}

\clearpage  

\renewcommand\bibname{{\bf References}} 
\addtocontents{toc}{\vspace{0em}}  
\backmatter
\label{References}
\lhead{{ References}}  
\bibliographystyle{unsrt}  
\bibliography{Bibliography}  
\end{document}